\documentclass[11pt,a4paper]{article}
\usepackage{latexsym}
\usepackage{graphicx}
\usepackage{psfrag}
\usepackage{amsmath}
\usepackage{amssymb}
\usepackage{color}
\usepackage{rotating}
\usepackage{multirow}
\usepackage{colortbl}
\usepackage{lscape,epsfig}
\usepackage{a4wide}
\usepackage[]{xcolor}
\usepackage{authblk}

\usepackage[dvips,colorlinks=true,a4paper=true,backref=false,linktocpage=true,citecolor=blue,urlcolor=blue,linkcolor=blue,pdfpagemode=UseOutlines]{hyperref}
\usepackage[sort&compress,square,comma,numbers]{natbib}
\usepackage{hypernat}

\definecolor{orange}{rgb}{1.0,.6,0}
\newcommand{\tbr}{\hspace{0.25mm}{\color{red}$\blacksquare$}} 
\newcommand{\tbg}{{\color{green}$\bigstar$}}

\newcommand{\rC}{C}
\newcommand{\gA}{A}
\newcommand{\oP}{P}

\newcommand{\bda}{\begin{\displaymath}\begin{array}{rl}}
\newcommand{\eda}{\end{array}\end{displaymath}}
\newcommand{\be}{\begin{equation}}
\newcommand{\ee}{\end{equation}}
\newcommand{\bdm}{\begin{displaymath}}
\newcommand{\edm}{\end{displaymath}}
\newcommand{\bea}{\begin{eqnarray}}
\newcommand{\eea}{\end{eqnarray}}

\newcommand{\fs}{\,.}
\newcommand{\co}{\,,}

\newcommand{\ind}{\scriptscriptstyle}

\newcommand{\qbar}{\overline{\rule[0.42em]{0.4em}{0em}}\hspace{-0.45em}q}
\newcommand{\ubar}{\overline{\rule[0.42em]{0.4em}{0em}}\hspace{-0.5em}u}
\newcommand{\dbar}{\,\overline{\rule[0.65em]{0.4em}{0em}}\hspace{-0.6em}d}
\newcommand{\sbar}{\,\overline{\rule[0.42em]{0.4em}{0em}}\hspace{-0.5em}s}
\newcommand{\lbar}{\bar{\ell}}

\newcommand{\lsim}{\,\raisebox{-0.3em}{$\stackrel{\raisebox{-0.1em}{$<$}}{\sim}$
}\,} 
\newcommand{\gsim}{\,\raisebox{-0.3em}{$\stackrel{\raisebox{-0.1em}{$>$}}{\sim}$
}\,}
\newcommand{\lvac}{\langle 0|\,}

\newcommand{\al}{&\!\!\!}

\newcommand{\Ch}{$\chi$} 

\newcommand{\Mpibar}{\rule{0.05cm}{0cm}\overline{\hspace{-0.08cm}M}_{\hspace{-0.04cm}\pi}}
\newcommand{\MKbar}{\rule{0.05cm}{0cm}\overline{\hspace{-0.08cm}M}_{\hspace{-0.04cm}K}}

\newcommand{\bi}{\begin{itemize}}
\newcommand{\ei}{\end{itemize}}

\newcommand{\beq}{\begin{equation}}
\newcommand{\eeq}{\end{equation}}

\newcommand{\Mpi}{M_\pi}
\newcommand{\Fpi}{F_\pi}
\newcommand{\Mka}{M_K}
\newcommand{\Fka}{F_K}

\renewcommand{\dag}{^\dagger}
\newcommand{\<}{\langle}
\renewcommand{\>}{\rangle}

\newcommand{\lonebar}{\ln\frac{\Lambda_1^2}{M_\pi^2}}
\newcommand{\ltwobar}{\ln\frac{\Lambda_2^2}{M_\pi^2}}
\newcommand{\lthreebar}{\ln\frac{\Lambda_3^2}{M_\pi^2}}
\newcommand{\lfourbar}{\ln\frac{\Lambda_4^2}{M_\pi^2}}
\newcommand{\lsixbar}{\ln\frac{\Lambda_6^2}{M_\pi^2}}
\newcommand{\lMbar}{\ln\frac{\Omega_M^2}{M_\pi^2}}
\newcommand{\lFbar}{\ln\frac{\Omega_F^2}{M_\pi^2}}

\newcommand{\MeV}{\,\mathrm{MeV}}
\newcommand{\GeV}{\,\mathrm{GeV}}
\newcommand{\fm}{\,\mathrm{fm}}

\newcommand{\ep}{\epsilon}

\newcommand{\et}{\eta}

\newcommand{\msbar}{{\overline{{\rm MS}}}}
\newcommand{\lms}{\Lambda_\msbar}

\def\mev{{\rm MeV}}
\def\gev{{\rm GeV}}
\def\tev{{\rm TeV}}
\def\fm{{\rm fm}}
 
\def\qbar{\bar{q}}
\def\psibar{\bar{\psi}}

\def\ubar{\bar{u}} 

\def\csw{c_{\rm sw}}

\def\gbar{\bar{g}}

\newcommand{\bd}{\begin{displaymath}}
\newcommand{\ed}{\end{displaymath}}

\newcommand{\eq}[1]{eq.\,(\ref{#1})}
\newcommand{\fig}[1]{Fig.\,\ref{#1}}

\newcommand{\sect}[1]{sect.\,\ref{#1}}

\newcommand{\figurebox}[2]{\fbox{\vbox to#2in{\hbox to #1in{\hfil}\vfil}}}
\newcommand{\bm}[1]{\mbox{\boldmath ${#1}$}}

\newcommand{\gtaeq}{\raisebox{-.6ex}{$\stackrel{\textstyle{>}}{\sim}$}}

\newcommand{\Nf}{N_{\hspace{-0.08 em} f}}
\newcommand{\Tr}{{\rm Tr}\,}

\newcommand{\fKfpicharged}{ \frac{f_{K^\pm}}{f_{\pi^\pm}}}
\newcommand{\fKfpichargedr}{ {f_{K^\pm}}/{f_{\pi^\pm}}}

\newcommand{\mK}{m_{\rm K}}

\newcommand{\fK}{f_{\rm K}}

\newcommand{\half}{\textstyle{1\over2}}

\newcommand{\abar}{\overline{a}}

\newcommand{\rb}[1]{\raisebox{1.5ex}[-1.5ex]{#1}}

\newcommand{\Lo}{\stackrel{\rule[-0.1cm]{0cm}{0cm}\mbox{\tiny LO}}{=}}
\newcommand{\NLo}{\stackrel{\rule[-0.1cm]{0cm}{0cm}\mbox{\tiny NLO}}{=}} 
\newcommand{\epsilonD}{\epsilon}

\definecolor{Gray}{rgb}{0.5,0.5,0.5}
\definecolor{Black}{rgb}{0.0,0.0,0.0}

\def\good{\raisebox{0.35mm}{{\color{green}$\bigstar$}}}
\def\bad{\raisebox{0.35mm}{\hspace{0.65mm}{\color{red}\tiny$\blacksquare$}}} 
\def\soso{\hspace{0.25mm}\raisebox{-0.2mm}{{\color{green}\Large$\circ$}}}
\def\okay{\hspace{0.25mm}\raisebox{-0.2mm}{{\color{green}\large\checkmark}}}

\newcommand{\mr}{\mathrm}

\def\half{{1\over2}}

\def\Tr{\,\mathrm{Tr}}

\def\fm{\mathrm{fm}}
\def\ev{\mathrm{e\kern-0.1em V}}
\def\kev{\mathrm{ke\kern-0.1em V}}
\def\mev{\mathrm{Me\kern-0.1em V}}
\def\gev{\mathrm{Ge\kern-0.1em V}}
\def\tev{\mathrm{Te\kern-0.1em V}}

\let\Re=\re \let\Im=\im

\def\n#1e#2n{{#1}\times 10^{#2}}

\def\bea{\begin{eqnarray}}
\def\eea{\end{eqnarray}}
\def\nn{\nonumber}

\def\cO{\mathcal{O}}

\def\ods2{\mathcal{O}_{\Delta S=2}}
\def\zds2{Z_{\Delta S=2}}

\def\msbar{{\overline{\mathrm{MS}}}}

\def\lat{\mathrm{lat}}

\def\spose#1{\hbox to 0pt{#1\hss}}
\def\ltapprox{\mathrel{\spose{\lower 3pt\hbox{$\mathchar"218$}}
 \raise 2.0pt\hbox{$\mathchar"13C$}}}
\def\gtapprox{\mathrel{\spose{\lower 3pt\hbox{$\mathchar"218$}}
 \raise 2.0pt\hbox{$\mathchar"13E$}}}
\def\inapprox{\mathrel{\spose{\lower 3pt\hbox{$\mathchar"218$}}
 \raise 2.0pt\hbox{$\mathchar"232$}}}

\makeatletter
\def\slash#1{{\mathpalette\c@ncel{#1}}} 
\def\big#1{{\hbox{$\left#1\vbox to1.012\ht\strutbox{}\right.\n@space$}}}
\def\Big#1{{\hbox{$\left#1\vbox to1.369\ht\strutbox{}\right.\n@space$}}}
\def\bigg#1{{\hbox{$\left#1\vbox to1.726\ht\strutbox{}\right.\n@space$}}}
\def\Bigg#1{{\hbox{$\left#1\vbox
to2.083\ht\strutbox{}\right.\n@space$}}}
\makeatother

\newcommand{\nl}{\nonumber \\}

\newcommand{\delv}{{\bf \nabla}}
\newcommand{\delvt}{\tilde{{\bf \nabla}}}

\newcommand{\delfour}{{\Delta^{(4)}}}
\newcommand{\delsq}{\Delta^{(2)}}

\newcommand{\Ev}{\tilde{{\bf E}}}
\newcommand{\Bv}{\tilde{{\bf B}}}
\newcommand{\sigmav}{\mbox{\boldmath$\sigma$}}

\def\spose#1{\hbox to 0pt{#1\hss}}
\def\ltapprox{\mathrel{\spose{\lower 3pt\hbox{$\mathchar"218$}}
\raise 2.0pt\hbox{$\mathchar"13C$}}}
\def\gtapprox{\mathrel{\spose{\lower 3pt\hbox{$\mathchar"218$}}
\raise 2.0pt\hbox{$\mathchar"13E$}}}
\def\inapprox{\mathrel{\spose{\lower 3pt\hbox{$\mathchar"218$}}
\raise 2.0pt\hbox{$\mathchar"232$}}}

\newcommand{\specialcellthree}[3][c]{%
  \begin{tabular}[#1]{@{}c@{}}#2{}#3\end{tabular}}

\newcommand{\alphah}{\alpha_\mathrm{V'}}
\newcommand{\alphav}{\alpha_\mathrm{V}}
\newcommand{\alphap}{\alpha_\mathrm{P}}

\begin{document}

\begin{flushright}
CP3-Origins-2013-040 DNRF90,
DIAS-2013-40\\
FERMILAB-PUB-13-484-T,  FTUAM-13-28\\
IFIC/13-76, IFT-UAM/CSIC-13-106 \\
MITP/13-067, YITP-13-114
\end{flushright}

\vskip 0.5cm
\begin{center}
{\bf \Huge Review of lattice results concerning\\ low energy particle physics}

\vspace{0.5cm}
\today

\vspace{0.5cm}
{\bf \large FLAG Working Group} 

\vspace{0.5cm}
{\small S.~Aoki$^1$, 
Y.~Aoki$^{2,3}$,
C.~Bernard$^4$,
T.~Blum$^{5,3}$,
G.~Colangelo$^6$,
M.~Della~Morte$^7$,
S.~D\"urr$^8$,
A.~X.~El-Khadra$^9$,
H.~Fukaya$^{10}$,
R.~Horsley$^{11}$,
A.~J\"uttner$^{12}$,
T.~Kaneko$^{13}$,
J.~Laiho$^{\star 14}$,
L.~Lellouch$^{15,16}$,
H.~Leutwyler$^6$,
V.~Lubicz$^{17,18}$,
E.~Lunghi$^{19}$,
S.~Necco$^6$,
T.~Onogi$^{10}$,
C.~Pena$^{20}$,
C.~T.~Sachrajda$^{12}$,
S.~R.~Sharpe$^{21}$,
S.~Simula$^{18}$,
R.~Sommer$^{22}$,
R.~S.~Van de Water$^{23}$,
A.~Vladikas$^{24}$,
U.~Wenger$^6$,
H.~Wittig$^{25}$}
\end{center}

\vskip 0.5cm

{\abstract{ We review lattice results related to pion, kaon, $D$- and
    $B$-meson physics with the aim of making them easily accessible to the
    particle physics community. More specifically, we report on the
    determination of the light-quark masses, the form factor $f_+(0)$,
    arising in semileptonic $K \to \pi$ transition at zero momentum
    transfer, as well as the decay constant ratio $f_K/f_\pi$ of decay
    constants and its consequences for the CKM matrix elements $V_{us}$ and
    $V_{ud}$. Furthermore, we describe the results obtained on the lattice
    for some of the low-energy constants of $SU(2)_L\times SU(2)_R$ and
    $SU(3)_L\times SU(3)_R$ Chiral Perturbation Theory and review the
    determination of the $B_K$ parameter of neutral kaon mixing.  The
    inclusion of heavy-quark quantities significantly expands the FLAG
    scope with respect to the previous review. Therefore, we focus here on
    $D$- and $B$-meson decay constants, form factors, and mixing
    parameters, since these are most relevant for the determination of CKM
    matrix elements and the global CKM unitarity-triangle fit.
    In addition we review the status of lattice determinations of the
    strong coupling constant $\alpha_s$.
}}

\newpage

\begin{flushleft}{\small
$^1$ Yukawa Institute for Theoretical Physics, Kyoto University,
Kitashirakawa Oiwakecho, \\ \hskip 0.3 cm Sakyo-ku Kyoto 606-8502, Japan 

$^2$ Kobayashi-Maskawa Institute for the Origin of Particles and the
Universe (KMI), \\ \hskip 0.3 cm
Nagoya University, Nagoya 464-8602, Japan

$^3$ RIKEN BNL Research Center, Brookhaven National Laboratory, Upton, NY 11973, USA

$^4$ Department of Physics, Washington University, Saint Louis, MO 63130, USA

$^5$ Physics Department, University of Connecticut, Storrs, CT 06269-3046, USA and 
  
$^6$ Albert Einstein Center for Fundamental Physics,
Institut f\"ur theoretische Physik, \\ \hskip 0.3 cm Universit\"at Bern, Sidlerstr. 5, 3012
Switzerland 

$^7$ CP3-Origins \&  Danish IAS, University of Southern
    Denmark, Campusvej 55, \\ \hskip 0.3 cm  DK-5230 Odense M, Denmark
    and \\ \hskip 0.3 cm  IFIC (CSIC),  c/ Catedr\'atico Jos\'e Beltr\'an, 2. E-46980, Paterna, Spain

$^8$ Bergische Universit\"at Wuppertal, Gau{\ss}stra{\ss}e\,20, 42119 Wuppertal, Germany
   and \\ \hskip 0.3 cm J\"ulich Supercomputing Center, Forschungszentrum J\"ulich,
   52425 J\"ulich, Germany 

$^9$ Department of Physics, University of Illinois, Urbana, IL 61801, USA
  
\hspace{-0.1cm}$^{10}$  Department of Physics, Osaka University, Toyonaka, Osaka 560-0043 Japan

\hspace{-0.1cm}$^{11}$ School of Physics, University of Edinburgh, Edinburgh EH9 3JZ, UK

\hspace{-0.1cm}$^{12}$ School of Physics \& Astronomy, University of Southampton, SO17 1BJ, UK

\hspace{-0.1cm}$^{13}$ High Energy Accelerator Research Organization (KEK), Ibaraki 305-0801, Japan

\hspace{-0.1cm}$^{14}$ SUPA, Department of Physics and Astronomy,
University of Glasgow, Glasgow, G12~8QQ, \\ \hskip 0.3 cm United Kingdom 

\hspace{-0.1cm}$^{15}$ Aix-Marseille Universit\'e, CNRS, CPT, UMR 7332, 13288 Marseille, France\\
\hspace{-0.1cm}$^{16}$ Universit\'e de Toulon, CNRS, CPT, UMR 7332, 83957 La Garde, France

\hspace{-0.1cm}$^{17}$ Dipartimento di Matematica e Fisica, Universit\`a
                       Roma Tre, Via della Vasca Navale 84, \\ \hskip 0.33 cm 00146 Rome, Italy

\hspace{-0.1cm}$^{18}$ INFN, Sezione di Roma Tre, Via della Vasca Navale 84, 00146 Rome, Italy

\hspace{-0.1cm}$^{19}$ Physics Department, Indiana University, Bloomington, IN 47405,
  USA
  
\hspace{-0.1cm}$^{20}$ Instituto de F\'{\i}sica Te\'orica UAM/CSIC and
Departamento de F\'{\i}sica Te\'orica, 
\\ \hskip 0.3 cm Universidad Aut\'onoma de Madrid, Cantoblanco 28049 Madrid, Spain 

\hspace{-0.1cm}$^{21}$ Physics Department, University of Washington, Seattle, WA 98195-1560, USA 

\hspace{-0.1cm}$^{22}$ NIC @ DESY, Platanenallee~6, 15738~Zeuthen, Germany

\hspace{-0.1cm}$^{23}$ Fermi National Accelerator Laboratory, Batavia, IL 60510, USA

\hspace{-0.1cm}$^{24}$ INFN, Sezione di Tor Vergata, c/o Dipartimento di Fisica,
           Universit\`a di Roma Tor Vergata, \\ \hskip 0.3 cm Via della Ricerca Scientifica 1, 00133 Rome, Italy
  
\hspace{-0.1cm}$^{25}$ PRISMA Cluster of Excellence, Institut f\"ur
Kernphysik and Helmholtz Institute Mainz, \\ \hskip 0.3 cm University of Mainz, 55099 Mainz,
Germany  
  
\vskip 6 cm
$^\star$ Present address: Department of Physics, Syracuse University, Syracuse, New York, USA}
\end{flushleft}

\clearpage
\tableofcontents

\clearpage
\section{Introduction}
\label{sec:introduction}

Flavour physics provides an important opportunity for exploring the
limits of the Standard Model of particle physics and for constraining
possible extensions of theories that go beyond it. As the LHC explores 
a new energy frontier and as experiments continue to extend the precision 
frontier, the importance of flavour physics will grow,
both in terms of searches for signatures of new physics through
precision measurements and in terms of attempts to unravel the
theoretical framework behind direct discoveries of new particles. A
major theoretical limitation consists in the precision with which
strong interaction effects can be quantified. Large-scale numerical
simulations of lattice QCD allow for the computation of these effects
from first principles. The scope of the Flavour Lattice Averaging
Group (FLAG) is to review the current status of lattice results for a
variety of physical quantities in low-energy physics. Set up in
November 2007\footnote{The original group had been set up in the
framework of a European Network on Flavour Physics (Flavianet).}, it
comprises experts in Lattice Field Theory and Chiral Perturbation
Theory. Our aim is to provide an answer to the frequently posed
question ``What is currently the best lattice value for a particular
quantity?", in a way which is readily accessible to
non-lattice-experts. This is generally not an easy question to answer;
different collaborations use different lattice actions
(discretizations of QCD) with a variety of lattice spacings and
volumes, and with a range of masses for the $u-$ and $d-$quarks. Not
only are the systematic errors different, but also the methodology
used to estimate these uncertainties varies between collaborations. In
the present work we summarize the main features of each of the
calculations and provide a framework for judging and combining the
different results. Sometimes it is a single result which provides the
``best" value; more often it is a combination of results from
different collaborations. Indeed, the consistency of values obtained
using different formulations adds significantly to our confidence in
the results.

The first edition of the FLAG review was published in
2011~\cite{Colangelo:2010et}. It was limited to lattice results
related to pion and kaon physics: light-quark masses ($u$-, $d$- and
$s$-flavours), the form factor $f_+(0)$ arising in semileptonic
$K \rightarrow \pi$ transitions at zero momentum transfer and the
decay constant ratio $f_K/f_\pi$, as well as their implications for
the CKM matrix elements $V_{us}$ and $V_{ud}$. Furthermore, results
were reported for some of the low-energy constants of $SU(2)_L \otimes
SU(2)_R$ and $SU(3)_L \otimes SU(3)_R$ Chiral Perturbation Theory and
the $B_K$ parameter of neutral kaon mixing. Results for all of these
quantities have been updated in the present paper. Moreover, the scope
of the present review has been extended by including lattice results
related to $D$- and $B$-meson physics. We focus on $B$- and $D$-meson
decay constants, form factors, and mixing parameters, which are most
relevant for the determination of CKM matrix elements and the global
CKM unitarity-triangle fit. Last but not least, the current status of 
lattice results on the QCD coupling  $\alpha_s$ is also reviewed.
 Bottom- and charm-quark masses, though important parametric
inputs to Standard Model calculations, have not been covered in the
present edition. They will be included in a future FLAG report.

Our plan is to continue providing FLAG updates, in the form of a peer
reviewed paper, roughly on a biannual basis. This effort is
supplemented by our more frequently updated
website \href{http://itpwiki.unibe.ch/flag}{{\tt
http://itpwiki.unibe.ch/flag}}, where figures as well as pdf-files for
the individual sections can be downloaded. The papers reviewed in the
present edition have appeared before the closing date 30 November 2013.

Finally, we draw attention to a particularly important point.  As
stated above, our aim is to make lattice QCD results easily accessible
to non-lattice-experts and we are well aware that it is likely that
some readers will only consult the present paper and not the original
lattice literature. We consider it very important that this paper is
not the only one which gets cited when the lattice results which are
discussed and analysed here are quoted. Readers who find the review
and compilations offered in this paper useful are therefore kindly
requested to also cite the original sources. The bibliography at the
end of this paper should make this task easier. Indeed we hope that
the bibliography will be one of the most widely used elements of the
whole paper.

This review is organized as follows.  In the remainder of
Sec.~\ref{sec:introduction} we summarize the composition and rules of
FLAG, describe the goals of the FLAG effort and general issues that
arise in modern lattice calculations.  For the reader's convenience,
Table~\ref{tab:summary} summarizes the main results (averages and
estimates) of the present review.  In Sec.~\ref{sec:qualcrit} we
explain our general methodology for evaluating the robustness of
lattice results which have appeared in the literature. We also
describe the procedures followed for combining results from different
collaborations in a single average or estimate (see
Sec.~\ref{sec:averages} for our use of these terms). The rest of the
paper consists of sections, each of which is dedicated to a single (or
groups of closely connected) physical quantity(ies). Each of these
sections is accompanied by an Appendix with explicatory notes.

\subsection{FLAG enlargement}

Upon completion of the first review, it was decided to extend the
project by adding new physical quantities and co-authors. FLAG became
more representative of the lattice community, both in terms of the
geographical location of its members and the lattice collaborations to
which they belong. At the time a parallel effort had been carried
out~\cite{Laiho:2009eu,Laiho:webpage}; the two efforts have now merged
in order to provide a single source of information on lattice results
to the particle-physics community.

The experience gained in managing the activities of a medium-sized
group of co-authors taught us that it was necessary to have a more
formal structure and a set of rules by which all concerned had to
abide, in order to make the inner workings of FLAG function smoothly.
The collaboration presently consists of an Advisory Board (AB), an
Editorial Board (EB), and seven Working Groups (WG).  The r\^{o}le of
the Advisory Board is that of general supervision and
consultation. Its members may interfere at any point in the process of
drafting the paper, expressing their opinion and offering advice. They
also give their approval of the final version of the preprint before
it is rendered public.  The Editorial Board coordinates the activities
of FLAG, sets priorities and intermediate deadlines, and takes care of
the editorial work needed to amalgamate the sections written by the
individual working groups into a uniform and coherent review. The
working groups concentrate on writing up the review of the physical
quantities for which they are responsible, which is subsequently circulated
to the whole collaboration for criticisms and suggestions.


\begin{table}[h]
\vspace{-1cm}
\begin{tabular}{|l|l||l|l||l|l||l|l|}
\hline
Quantity \rule[-0.2cm]{0cm}{0.6cm}    & \hspace{-1.5mm}Sect.\hspace{-2mm} &\rule{0cm}{0cm}\hspace{-0.4em} {\color{green}$\blacksquare$}\hspace{-0.4em}&\!$N_f\hspace{-0.2em}=\hspace{-0.2em}2\hspace{-0.2em}+\hspace{-0.2em}1\hspace{-0.2em}+\hspace{-0.2em}1$  &\hspace{-0.2em}{\color{green}$\blacksquare$}\hspace{-0.2em} &  $N_f=2+1$ &\hspace{-0.2em}{\color{green}\hspace{-0em}$\blacksquare$}&$N_f=2$ \\
\hline
\hline
$m_s$(MeV)   &\ref{sec:msmud}&&&3& $93.8(1.5)(1.9)$&2 & $101(3)$ \\
$m_{ud}$(MeV)&\ref{sec:msmud}&&&3& $3.42(6)(7)$&1& $3.6(2)$ \\
$m_s/m_{ud}$ &\ref{sec:msmud}&&&3& $27.46(15)(41)$ &1& $28.1(1.2)$ \\
$m_d$(MeV)   &\ref{subsec:mumd}&&&& $4.68(14)(7)$ && $4.80(23)$ \\
$m_u$(MeV)   &\ref{subsec:mumd}&&&& $2.16(9)(7)$  && $2.40(23)$ \\
$m_u/m_d$    &\ref{subsec:mumd}&&&& $0.46(2)(2)$  && $0.50(4)$ \\
\hline
$f_+^{K\pi}(0)$\rule{0cm}{0.4cm}     &\ref{sec:Direct} &&&2& $0.9661(32)$&1& $0.9560(57)(62)$ \\
$f_{K^+}/f_{\pi^+}$&\ref{sec:Direct} &2& $1.194(5)$ &4& $1.192(5)$ &1& $1.205(6)(17)$ \\
$f_K$(MeV)   &\ref{sec:fKfpi}&&&3& $156.3(0.9)$ &1& $158.1(2.5)$ \\
$f_\pi$(MeV)   &\ref{sec:fKfpi}&&&3& $130.2(1.4)$ &&  \\
\hline
$\Sigma$(MeV) \rule{0cm}{0.4cm}  &\ref{sec:su2}&&&2& $271(15)$ &1& $269(8)$ \\
$F_\pi/F$   & \ref{sec:su2}&1&$1.0760(28)$ &2& $1.0624(21)$ &1& $1.0744(67)$ \\
$\bar\ell_3$ & \ref{sec:su2}&1&$3.70(27)$ &3& $3.05(99)$ &1& $3.41(41)$ \\
$\bar\ell_4$ & \ref{sec:su2}&1&$4.67(10)$ &3& $4.02(28)$ &1& $4.62(22)$ \\
\hline
$\hat{B}_K$ \rule{0cm}{0.4cm}  &\ref{sec:latBK}&&&4& $0.766(10)$ &1& $0.729(25)(17)$ \\
$B_K^{\bar{\mathrm{MS}}}$(2 GeV) &\ref{sec:latBK}&&&4& $0.560(7)$ &1& $0.533(18)(12)$ \\
\hline
\hline
$f_D$(MeV) &\ref{sec:fD}&&&2& $209.2(3.3)$ &1& $208(7)$ \\
$f_{D_s}$(MeV) &\ref{sec:fD}&&&2& $248.6(2.7)$ &1& $250(7)$ \\
$f_{D_s}/f_D$ &\ref{sec:fD}&&&2& $1.187(12)$ &1& $1.20(2)$ \\
\hline
$f_+^{D\pi}(0)$ \rule{0cm}{0.4cm}  &\ref{sec:DtoPiK}&&&1& $0.666(29)$&&  \\
$f_+^{DK}(0)$ &\ref{sec:DtoPiK}&&&1& $0.747(19)$&& \\
\hline
$f_B$(MeV) &\ref{sec:fB}&1&$186(4)$&3& $190.5(4.2)$ &1& $189(8)$ \\
$f_{B_s}$(MeV) &\ref{sec:fB}&1&$224(5)$&3& $227.7(4.5)$ &1& $228(8)$ \\
$f_{B_s}/f_B$ &\ref{sec:fB}&1&$1.205(7)$&2& $1.202(22)$ &1& $1.206(24)$ \\
\hline
$f_{B_d}\sqrt{\rule{0mm}{3.7mm}\rule{5mm}{0mm}}\hspace{-5mm}\hat{B}_{B_d}$(MeV)\rule{0mm}{5mm} &\ref{sec:BMix}&&&1& $216(15)$ && \\
$f_{B_s}\sqrt{\rule{0mm}{3.7mm}\rule{5mm}{0mm}}\hspace{-5mm}\hat{B}_{B_s}$(MeV) &\ref{sec:BMix}&&&1& $266(18)$ && \\
$\hat{B}_{B_d}$ &\ref{sec:BMix}&&&1& $1.27(10)$ && \\
$\hat{B}_{B_s}$ &\ref{sec:BMix}&&&1& $1.33(6)$ && \\
$\xi$ &\ref{sec:BMix} &&&1& $1.268(63)$ && \\
$\hat{B}_{B_s}/\hat{B}_{B_d}$ &\ref{sec:BMix}&&&1&$1.06(11)$ && \\
\hline
$\Delta \zeta^{B\pi}$(ps$^{-1}$)\rule{0cm}{0.4cm}  &\ref{sec:BtoPiK}&&&2& $2.16(50)$ && \\
$f_+^{B\pi}(q^2): \hspace{1mm}a_0^{\rm BCL}$&\ref{sec:BtoPiK}&&&2& $0.453(33)$ && \\
$\textcolor{white}{f_+^{B\pi}(q^2):\hspace{2mm}}  a_1^{\rm BCL}$&&&&2& $-0.43(33)$ &&  \\
$\textcolor{white}{f_+^{B\pi}(q^2):\hspace{2mm}}  a_2^{\rm BCL}$&&&&2& $0.9(3.9)$ &&  \\
\hline
${\cal F}^{B \rightarrow D^*}(1)$\rule{0cm}{0.4cm}   &\ref{sec:BtoD}&&&1& $0.906(4)(12)$ && \\
$R(D)$ &\ref{sec:BtoD} &&&1& $0.316(12)(7)$ && \\ 
\hline
$\alpha_{\overline{\rm MS}}^{(5)}(M_Z) $\rule[-0.25cm]{0cm}{0.7cm} &\ref{s:alpsumm} &&&4&$0.1184(12)$&&\\
\hline 
\end{tabular}
\caption{\label{tab:summary} Summary of the main results of this review, grouped in terms of $\Nf$, the number of dynamical quark flavours in lattice simulations. Quark masses and the quark condensate are given in the $\msbar$ scheme at running scale
   $\mu=2\,\gev$;  the other quantities listed are specified in the quoted sections. The columns marked  {\color{green}$\blacksquare$} indicate the number of results that enter our averages for each quantity. We emphasize that these numbers only give a very rough indication of how thoroughly the quantity in question has been explored on the lattice and recommend to consult the detailed tables and figures in the relevant section for more significant information and for explanations on the source of the quoted errors.}
\end{table}


\clearpage

The most important internal FLAG rules are the following:
\begin{itemize}
\item
members of the AB have a 4-year mandate (to avoid a simultaneous change of
all members, some of the current members of the AB will have a shorter mandate);
\item
the composition of the AB reflects the main geographical areas in
which lattice collaborations are active: one member comes from
America, one from Asia/Oceania and one from Europe ;
\item
the mandate of regular members is not limited in time, but we expect that a
certain turnover will occur naturally;
\item
whenever a replacement becomes necessary this has to keep, and
possibly improve, the balance in FLAG;
\item
in all working groups the three members must belong to three different
lattice collaborations;\footnote{The WG on semileptonic $D$ and $B$
decays has currently four members, but only three of them belong to
lattice collaborations.}
\item
a paper is in general not reviewed (nor colour-coded, as described in
the next section) by one of its authors;
\item
lattice collaborations not represented in FLAG will be asked to
check whether the colour coding of their calculation is correct.
\end{itemize}

The current list of FLAG members and their Working Group assignments is:
\begin{itemize}
\item
Advisory Board (AB):\hfill
S.~Aoki, C.~Bernard, C.~Sachrajda
\item
Editorial Board (EB):\hfill
G.~Colangelo, H.~Leutwyler, A.~Vladikas, U.~Wenger
\item
Working Groups (WG)\\
(each WG coordinator is listed first):
\begin{itemize}
\item Quark masses \hfill L.~Lellouch, T.~Blum, V.~Lubicz
\item $V_{us},V_{ud}$ \hfill A.~J\"uttner, T.~Kaneko, S.~Simula
\item LEC \hfill S.~D\"urr, H.~Fukaya, S.~Necco
\item $B_K$ \hfill H.~Wittig, J.~Laiho, S.~Sharpe
\item $f_{B_{(s)}}$, $f_{D_{(s)}}$, $B_B$ \hfill A.~El-Khadra, Y.~Aoki, M.~Della Morte
\item $B_{(s)}$, $D$ semileptonic and radiative decays \hfill R.~Van de
  Water, E.~Lunghi, C.~Pena, \\
  \hbox{} \hfill J.~Shigemitsu\footnote{J. Shigemitsu has withdrawn from
    FLAG, immediately after completion of the first version of the present
    paper (arXiv:1310.8555 [hep-lat]), of which she is a co-author. She is
    listed here in recognition of her full involvement in the review of
    B(s) and D semileptonic and radiative decays, as well as for her
    valuable contribution of the whole FLAG effort.}
\item $\alpha_s$ \hfill R.~Sommer, R.~Horsley, T.~Onogi
\end{itemize}
\end{itemize}
 
\subsection{General issues and summary of the main results}
The present review aims at two distinct goals:

\vspace{0.1cm}\noindent
a. offer a {\bf description} of the work done on the lattice
concerning low energy particle physics;

\vspace{0.1cm}\noindent
b. draw {\bf conclusions} on the basis of that work, which summarize
the results obtained for the various quantities of physical interest.

\vspace{0.1cm}\noindent
The core of the information about the work done on the lattice is
presented in the form of tables, which not only list the various
results, but also describe the quality of the data that underlie
them. We consider it important that this part of the review represents
a generally accepted description of the work done. For this reason, we
explicitly specify the quality requirements used and provide
sufficient details in the appendices so that the reader can verify
the information given in the tables.

The conclusions drawn on the basis of the available lattice results,
on the other hand, are the responsibility of FLAG alone. We aim at
staying on the conservative side and in several cases reach
conclusions which are more cautious than what a plain average of the
available lattice results would give, in particular when this is
dominated by a single lattice result. An additional issue occurs when
only one lattice result is available for a given quantity.  In such
cases one does not have the same degree of confidence in results and
errors as one has when there is agreement among many different
calculations using different approaches.  Since this degree of
confidence cannot be quantified, it is not reflected in the quoted
errors, but should be kept in mind by the reader. At present, the
issue of having only a single result occurs much more often in
heavy-quark physics than in light-quark physics.  We are confident
that the heavy-quark calculations will soon reach the
state that pertains in light-quark physics.

Several general issues concerning the present review are thoroughly
discussed in Sect. 1.1 of our initial paper~\cite{Colangelo:2010et}
and we encourage the reader to consult the relevant pages. In the
remainder of the present section, we focus on a few important
points. 

Each discretization has its merits, but also its shortcomings. For the
topics covered already in the first edition of the FLAG review, we
have by now a remarkably broad data base, and for most quantities
lattice calculations based on totally different discretizations are
now available. This is illustrated by the dense population of the
tables and figures shown in the first part of this review. Those
calculations which do satisfy our quality criteria indeed lead to
consistent results, confirming universality within the accuracy
reached. In our opinion, the consistency between independent lattice
results, obtained with different discretizations, methods, and
simulation parameters, is an important test of lattice QCD, and
observing such consistency then also provides further evidence that
systematic errors are fully under control.

In the sections dealing with heavy quarks and with $\alpha_s$, the
situation is not the same. Since the $b$-quark mass cannot be resolved
with current lattice spacings, all lattice methods for treating $b$
quarks use effective field theory at some level. This introduces
additional complications not present in the light-quark sector.  An
overview of the issues specific to heavy-quark quantities is given in
the introduction of Sec.~\ref{sec:BDecays}. For $B$ and $D$ meson
leptonic decay constants, there already exist a good number of
different independent calculations that use different heavy-quark
methods, but there are only one or two independent calculations of
semileptonic $B$ and $D$ meson form factors and $B$ meson mixing
parameters. For $\alpha_s$, most lattice methods involve a range of
scales that need to be resolved and controlling the systematic error
over a large range of scales is more demanding. The issues specific to
determinations of the strong coupling are summarized in Sec.~\ref{sec:alpha_s}.

The lattice spacings reached in recent simulations go down to 0.05 fm
or even smaller. In that region, growing autocorrelation times
slow down the sampling of the
configurations~\cite{Antonio:2008zz,Bazavov:2010xr,Schaefer:2010hu,Luscher:2010we,Schaefer:2010qh}. Many
groups check for autocorrelations in a number of observables,
including the topological charge, for which a rapid growth of the
autocorrelation time is observed if the lattice spacing becomes
small. In the following, we assume that the continuum limit can be
reached by extrapolating the existing simulations.

Lattice simulations of QCD currently involve at most four dynamical
quark flavours. Moreover, most of the data concern simulations for
which the masses of the two lightest quarks are set equal. This is
indicated by the notation $\Nf =2+1+1$ which, in this case, denotes a
lattice calculation with four dynamical quark flavours and $m_u = m_d
\neq m_s \neq m_c$. Note that calculations with $\Nf =2$ dynamical
flavours often include strange valence quarks interacting with gluons,
so that bound states with the quantum numbers of the kaons can be
studied, albeit neglecting strange sea quark fluctuations.  The
quenched approximation ($N_f=0$), in which the sea quarks are treated
as a mean field, is no longer used in modern lattice simulations.
Accordingly, we will review results obtained with $N_f=2$, $N_f=2+1$,
and $N_f = 2+1+1$, but omit earlier results with $N_f=0$. On the other
hand, the dependence of the QCD coupling constant $\alpha_s$ on the
number of flavours is a theoretical issue of considerable interest,
and we therefore include results obtained for gluodynamics in the
$\alpha_s$ section. We stress, however, that only results with $N_f 
\ge 3$ are used to determine the physical value of $\alpha_s$ at a high scale.

The remarkable recent progress in the precision of lattice
calculations is due to improved algorithms, better computing resources
and, last but not least, conceptual developments, such as improved
actions which reduce lattice artifacts, actions which preserve
(remnants of) chiral symmetry, understanding finite-size effects,
non-perturbative renormalization, etc. A concise characterization of
the various discretizations that underlie the results reported in the
present review is given in Appendix~\ref{sec_lattice_actions}.

Lattice simulations are performed at fixed values of the bare QCD
parameters (gauge coupling and quark masses) and physical quantities
with mass dimensions (e.g. quark masses, decay constants ...) are
computed in units of the lattice spacing; i.e. they are
dimensionless. Their conversion to physical units requires knowledge
of the lattice spacing at the fixed values of the bare QCD parameters
of the simulations. This is achieved by requiring agreement between
the lattice calculation and experimental measurement of a known
quantity, which ``sets the scale" of a given simulation. A few details
on this procedure are provided in Appendix~\ref{sec_scale}.

Several of the results covered by this review, such as quark masses,
the gauge coupling, and $B$-parameters, are quantities defined in a
given renormalization scheme and scale. The schemes employed are often
chosen because of their specific merits when combined with the lattice
regularization. For a brief discussion of their properties, see
Appendix~\ref{sec_match}. The conversion of the results, obtained in
these so-called intermediate schemes, to more familiar regularization
schemes, such as the $\msbar$-scheme, is done with the aid of
perturbation theory. It must be stressed that the renormalization
scales accessible by the simulations are subject to limitations,
naturally arising in Field Theory computations at finite UV and small
non-zero IR cutoff. Typically, such scales are of the order of the UV
cutoff, or $\Lambda_{\rm QCD}$, depending on the chosen scheme. To
safely match to $\msbar$, a scheme defined in perturbation theory,
Renormalization Group (RG) running to higher scales is performed,
either perturbatively, or non-perturbatively (the latter using
finite-size scaling techniques).

Because of limited computing resources, lattice simulations are often
performed at unphysically heavy pion masses, although results at the
physical point have recently become available.  Further, numerical
simulations must be done at finite lattice spacing.  In order to
obtain physical results, lattice data are generated at a sequence of
pion masses and a sequence of lattice spacings, and then
extrapolated to $M_\pi \approx 135$~MeV and $a \to 0$.  To control the
associated systematic uncertainties, these extrapolations are guided
by effective theory.  For light-quark actions, the lattice-spacing
dependence is described by Symanzik's effective
theory~\cite{Symanzik:1983dc,Symanzik:1983gh}; for heavy quarks, this
can be extended and/or supplemented by other effective theories such
as Heavy-Quark Effective Theory (HQET).  The pion-mass dependence can
be parameterized with Chiral Perturbation Theory ($\chi$PT), which
takes into account the Nambu-Goldstone nature of the lowest
excitations that occur in the presence of light quarks; similarly one
can use Heavy-Light Meson Chiral Perturbation Theory (HM$\chi$PT) to
extrapolate quantities involving mesons composed of one heavy ($b$ or
$c$) and one light quark.  One can combine Symanzik's effective theory
with $\chi$PT to simultaneously extrapolate to the physical pion mass
and continuum; in this case, the form of the effective theory depends
on the discretization.  See Appendix ~\ref{sec_ChiPT} for a brief
description of the different variants in use and some useful
references.

\section{Quality criteria}
\label{sec:qualcrit}

The essential characteristics of our approach to the problem of rating
and averaging lattice quantities reported by different collaborations
have been outlined in our first
publication~\cite{Colangelo:2010et}. Our aim is to help the reader
assess the reliability of a particular lattice result without
necessarily studying the original article in depth. This is a delicate
issue, which may make things appear simpler than they are. However,
it safeguards against the common practice of using lattice results and
drawing physics conclusions from them, without a critical assessment
of the quality of the various calculations. We believe that despite
the risks, it is important to provide some compact information about
the quality of a calculation. However, the importance of the
accompanying detailed discussion of the results presented in the bulk
of the present review cannot be underestimated.
 
\subsection{Systematic errors and colour-coding}
\label{sec:color-code}

In Ref.~\cite{Colangelo:2010et}, we identified a number of sources of
systematic errors, for which a systematic improvement is possible, and
assigned one of three coloured symbols to each calculation: green
star, amber disc or red square. The appearance of a red tag, even in a
single source of systematic error of a given lattice result,
disqualified it from the global averaging. Since results with green
and amber tags entered the averages, and since this policy has been
retained in the present edition, we have decided to substitute the
amber disc by a green unfilled circle. Thus the new colour coding is
as follows: \\
\hspace{-0.1em}\good \hspace{0.2cm} the systematic error has been estimated in a
satisfactory manner and convincingly shown to be under control;\\
\rule{0.1em}{0em}\soso \hspace{0.2cm} a reasonable attempt at estimating the systematic
error has been made, although this could be improved;\\
\rule{0.1em}{0em}\bad \hspace{0.2cm} no or a clearly unsatisfactory attempt at
estimating the systematic error has been made.
We stress once more that only results without a red tag in the
systematic errors are averaged in order to provide a given FLAG
estimate.

The precise criteria used in determining the colour coding is
unavoidably time-dependent; as lattice calculations become more
accurate the standards against which they are measured become
tighter. For quantities related to the light-quark sector, which have
been dealt with in the first edition of the FLAG
review~\cite{Colangelo:2010et}, some of the quality criteria have
remained the same, while others have been tightened up. We will
compare them to those of Ref.~\cite{Colangelo:2010et},
case-by-case, below. For the newly introduced physical quantities, related to
heavy quark physics, the adoption of new criteria was necessary. This
is due to the fact that, in most cases, the discretization of the
heavy quark action follows a very different approach to that of light
flavours. Moreover, the two Working Groups dedicated to heavy flavours
have opted for a somewhat different rating of the extrapolation of
lattice results to the continuum limit. Finally, the strong coupling
being in a class of its own, as far as methods for its computation are
concerned, led to the introduction of dedicated rating criteria for
it.

Of course any colour coding has to be treated with caution; we repeat
that the criteria are subjective and evolving. Sometimes a single
source of systematic error dominates the systematic uncertainty and it
is more important to reduce this uncertainty than to aim for green
stars for other sources of error. In spite of these caveats we hope
that our attempt to introduce quality measures for lattice results
will prove to be a useful guide. In addition we would like to
stress that the agreement of lattice results obtained using
different actions and procedures evident in many of the tables
presented below provides further validation.

For a coherent assessment of the present situation, the quality of the
data plays a key role, but the colour coding cannot be carried over to
the figures. On the other hand, simply showing all data on equal
footing would give the misleading impression that the overall
consistency of the information available on the lattice is
questionable. As a way out, the figures do indicate the quality in a
rudimentary way:
\\ \raisebox{0.35mm}{\hspace{0.65mm}{\color{green}$\blacksquare$}} \hspace{0.2cm}
results included in the
average;\\ \raisebox{0.35mm}{\hspace{0.65mm}{\color{green}$\square$}} \hspace{0.2cm}
results that are not included in the average but pass all quality
criteria;\\ \raisebox{0.35mm}{\hspace{0.65mm}{\color{red}$\square$}} \hspace{0.2cm}
all other results.\\ The reason for not including a given result in
the average is not always the same: the paper may fail one of the
quality criteria, may not be published, be superseded by other results
or not offer a complete error budget.  Symbols other than squares are
used to distinguish results with specific properties and are always
explained in the caption.\\ There are separate criteria for
light-flavour, heavy-flavour, and $\alpha_s$ results. In the following
the criteria for the former two are discussed in detail, while the
criteria for the $\alpha_s$ results will be exposed separately in
sect.~\ref{s:crit}.

\subsubsection{Light-quark physics}
\label{sec:LQCriteria}
The colour code used in the tables is specified as follows:
\begin{itemize}
\item Chiral extrapolation:\\
\good \hspace{0.2cm}  $M_{\pi,\mathrm{min}}< 200$ MeV  \\
\rule{0.05em}{0em}\soso \hspace{0.2cm}  200 MeV $\le M_{\pi,{\mathrm{min}}} \le$ 400 MeV \\
\rule{0.05em}{0em}\bad \hspace{0.2cm}  400 MeV $ < M_{\pi,\mathrm{min}}$ \\
It is assumed that the chiral extrapolation is done with at least a
three-point analysis; otherwise this will be explicitly
mentioned. Note that, compared to Ref.~\cite{Colangelo:2010et}, chiral
extrapolations are now treated in a somewhat more stringent manner and
the cutoff between green star and green open circle (formerly amber
disc), previously set at 250 MeV, is now lowered to 200 MeV.
\item Continuum extrapolation:\\
\good \hspace{0.2cm}  3 or more lattice spacings, at least 2 points below
0.1 fm\\ 
\rule{0.05em}{0em}\soso \hspace{0.2cm}  2 or more lattice spacings, at least 1 point below 0.1
fm \\ 
\rule{0.05em}{0em}\bad \hspace{0.2cm}  otherwise\\
It is assumed that the action is $O(a)$-improved (i.e. the
discretization errors vanish quadratically with the lattice spacing);
otherwise this will be explicitly mentioned. Moreover, for
non-improved actions an additional lattice spacing is required. This
criterion is the same as the one adopted in
Ref.~\cite{Colangelo:2010et}.
\item Finite-volume effects:\\
\good \hspace{0.2cm}  $M_{\pi,\mathrm{min}} L > 4$ or at least 3 volumes \\
\rule{0.05em}{0em}\soso \hspace{0.2cm}  $M_{\pi,\mathrm{min}} L > 3$ and at least 2 volumes \\
\rule{0.05em}{0em}\bad \hspace{0.2cm}  otherwise\\
These ratings apply to calculations in the $p$-regime and it is
assumed that $L_\mathrm{min}\ge $ 2 fm; otherwise this will be
explicitly mentioned and a red square will be assigned.

\item Renormalization (where applicable):\\
\good \hspace{0.2cm}  non-perturbative\\
\rule{0.05em}{0em}\soso \hspace{0.2cm}  1-loop perturbation theory or higher  with a reasonable estimate of truncation errors\\
\rule{0.05em}{0em}\bad \hspace{0.2cm}  otherwise \\
In Ref.~\cite{Colangelo:2010et}, we assigned a red square to all
results which were renormalized at 1-loop in perturbation theory. We
now feel that this is too restrictive, since the error arising from
renormalization constants, calculated in perturbation theory at
1-loop, is often estimated conservatively and reliably.

\item Running (where applicable): \\ 
For scale-dependent quantities, such as quark masses or $B_K$, it is
essential that contact with continuum perturbation theory can be
established.  Various different methods are used for this purpose
(cf.~Appendix \ref{sec_match}): Regularization-independent Momentum
Subtraction (RI/MOM), Schr\"odinger functional, direct comparison with
(resummed) perturbation theory. Irrespective of the particular method used,
the uncertainty associated with the choice of intermediate
renormalization scales in the construction of physical observables
must be brought under control. This is best achieved by performing
comparisons between non-perturbative and perturbative running over a
reasonably broad range of scales. These comparisons were initially
only made in the Schr\"odinger functional (SF) approach, but are now
also being performed in RI/MOM schemes.  We mark the data for which
information about non-perturbative running checks is available and
give some details, but do not attempt to translate this into a
colour-code.
\end{itemize}

The pion mass plays an important r\^{o}le in the criteria relevant for
chiral extrapolation and finite volume.  For some of the
regularizations used, however, it is not a trivial matter to identify
this mass. In the case of twisted-mass fermions, discretization
effects give rise to a mass difference between charged and neutral
pions even when the up- and down-quark masses are equal, with the
charged pion being the heavier of the two. The discussion of the
twisted-mass results presented in the following sections assumes that
the artificial isospin-breaking effects which occur in this
regularization are under control.  In addition, we assume that the mass of
the charged pion may be used when evaluating the chiral extrapolation
and finite volume criteria.  In the case of staggered fermions,
discretization effects give rise to several light states with the
quantum numbers of the pion.\footnote{We refer the interested reader
  to a number of good reviews on the
  subject~\cite{Durr:2005ax,Sharpe:2006re,Kronfeld:2007ek,Golterman:2008gt,Bazavov:2009bb}.}
The mass splitting among these ``taste'' partners represents a
discretization effect of $\cO(a^2)$, which can be significant at big
lattice spacings but shrinks as the spacing is reduced. In the
discussion of the results obtained with staggered quarks given in the
following sections, we assume that these artefacts are under
control. When evaluating the chiral extrapolation criteria, we
conservatively identify $M_{\pi,\mathrm{min}}$ with the root mean
square (RMS) of the mass of all taste partners.  These masses are also
used in sections \ref{sec:vusvud} and \ref{sec:BK} when evaluating the
finite volume criteria, while in sections \ref{sec:qmass},
\ref{sec:LECs}, \ref{sec:DDecays} and \ref{sec:BDecays}, a more
stringent finite volume criterion is applied: $M_{\pi,\mathrm{min}}$
is identified with the mass of the lightest state.

\subsubsection{Heavy-quark physics}

\label{sec:HQCriteria}

This subsection discusses the criteria adopted for the heavy-quark
quantities included in this review, characterized by non-zero charm
and bottom quantum numbers.  There are several different approaches to
treating heavy quarks on the lattice, each with their own issues and
considerations. In general all $b$-quark methods rely on the use of
Effective Field Theory (EFT) at some point in the computation, either
via direct simulation of the EFT, use of the EFT to estimate the size
of cutoff errors, or use of the EFT to extrapolate from the simulated
lattice quark mass up to the physical $b$-quark mass.  Some
simulations of charm-quark quantities use the same heavy-quark methods
as for bottom quarks, but there are also computations that use
improved light-quark actions to simulate charm quarks.  Hence, with
some methods and for some quantities, truncation effects must be
considered together with discretization errors. With other methods,
discretization errors are more severe for heavy-quark quantities than
for the corresponding light-quark quantities.

In order to address these complications, we add a new heavy-quark
treatment category to the ratings system. The purpose of this
criterion is to provide a guideline for the level of action and
operator improvement needed in each approach to make reliable
calculations possible, in principle. In addition, we replace the
rating criteria for the continuum extrapolations of
Sec.~\ref{sec:LQCriteria} with a new empirical approach based on the
size of observed discretization errors in the lattice simulation data.
This accounts for the fact that whether discretization and truncation
effects in a given calculation are sufficiently small as to be
controllable depends not only on the range of lattice spacings used in
the simulations, but also on the simulated heavy-quark masses and on
the level of action and operator improvement. For the other
categories, we adopt the same strict criteria as in
Sec.~\ref{sec:LQCriteria}, with one minor modification, as explained
below.

\begin{itemize}

\item Heavy-quark treatment:\\
A description of the different approaches to treating heavy quarks on
the lattice is given in Appendix~\ref{app:HQactions} including a
discussion of the associated discretization, truncation, and matching
errors.  For truncation errors we use HQET power counting throughout,
since this review is focused on heavy quark quantities involving $B$
and $D$ mesons.  Here we describe the criteria for how each approach
must be implemented in order to receive an acceptable (\okay) rating
for both the heavy quark actions and the weak operators.  Heavy-quark
implementations without the level of improvement described below are
rated not acceptable (\bad). The matching is evaluated together with
renormalization, using the renormalization criteria described in
Sec.~\ref{sec:LQCriteria}.  We emphasize that the heavy-quark
implementations rated as acceptable and described below have been
validated in a variety of ways, such as via phenomenological agreement
with experimental measurements, consistency between independent
lattice calculations, and numerical studies of truncation errors.
These tests are summarized in Sec.~\ref{sec:BDecays}.  \smallskip
\\ {\it Relativistic heavy quark actions:} \\
\noindent 
\okay \hspace{0.2cm}   at least tree-level $O(a)$ improved action and weak operators  \\
This is similar to the requirements for light quark actions. All
current implementations of relativistic heavy quark actions satisfy
these criteria. \smallskip \\
{\it NRQCD:} \\
\noindent 
\okay \hspace{0.2cm}   tree-level matched through $O(1/m_h)$ and improved through $O(a^2)$ \\
The current implementations of NRQCD satisfy these criteria, and also
include tree-level corrections of $O(1/m_h^2)$ in the
action. \smallskip \\
{\it HQET: }\\
\noindent 
\okay \hspace{0.2cm}  tree-level  matched through $O(1/m_h)$ with discretization errors starting at $O(a^2)$ \\
The current implementation of HQET by the ALPHA collaboration
satisfies these criteria with an action and weak operators that are
nonperturbatively matched through $O(1/m_h)$.  Calculations that
exclusively use a static limit action do not satisfy theses criteria,
since the static limit action, by definition, does not include $1/m_h$
terms.  However for SU(3)-breaking ratios such as $\xi$ and $f_{B_s}/f_B$ truncation
errors start at $O((m_s - m_d)/m_h)$. We therefore consider lattice
calculations of such ratios that use a static limit action to still
have controllable truncation errors. \smallskip \\
{\it Light-quark actions for heavy quarks:}  \\
\noindent 
\okay \hspace{0.2cm}  discretization errors starting at $O(a^2)$ or higher \\
This applies to calculations that use the tmWilson action, a
nonperturbatively improved Wilson action, or the HISQ action for charm
quark quantities. It also applies to calculations that use these light
quark actions in the charm region and above together with either the
static limit or with an HQET inspired extrapolation to obtain results
at the physical $b$ quark mass. In these cases, the continuum
extrapolation criteria must be applied to the entire range of heavy
quark masses used in the calculation.

\item Continuum extrapolation:\\
First we introduce the following definitions:
\be
D(a) = \frac{Q(a) - Q(0)}{Q(a)} \, ,
\ee
where $Q(a)$ denotes the central value of quantity $Q$ obtained at
lattice spacing $a$ and $Q(0)$ denotes the continuum extrapolated
value. $D(a)$ is a measure of how far the continuum extrapolated
result is from the lattice data. We evaluate this quantity on the
smallest lattice spacing used in the calculation, $a_{\rm min}$.
\be
\delta(a) = \frac{Q(a) - Q(0)}{\sigma_Q} \, ,
\ee
where $\sigma_Q$ is the combined statistical and systematic (due to
the continuum extrapolation) error.  $\delta(a)$ is a measure of how
well the continuum extrapolated result agrees with the lattice data
within the statistical and systematic errors of the calculation. Again, we
evaluate this quantity on the smallest lattice spacing used in the
calculation, $a_{\rm min}$. \smallskip \\
\smallskip
\noindent \good \hspace{0.15cm} (i) Three or more lattice spacings, and \\
\indent  \hspace{0.49cm} (ii) $a^2_{\rm max} / a^2_{\rm min} \ge 2$, and \\
\indent  \hspace{0.49cm} (iii) $D(a_{\rm min}) \le 2\%$, and \\
\indent \hspace{0.49cm} (iv) $\delta(a_{\rm min}) \le 1$ \\
\rule{0.05em}{0em}\soso \hspace{0.2cm}
(i) Two or more lattice spacings, and \\
\indent \hspace{0.49cm} (ii) $a^2_{\rm max} / a^2_{\rm min} \ge 1.4$, and \\
\indent \hspace{0.49cm} (iii) $D(a_{\rm min}) \le 10\%$, and \\
\indent \hspace{0.49cm} (iv) $\delta(a_{\rm min}) \le 2$ \\
\rule{0.05em}{0em}\bad \hspace{0.2cm}  otherwise \\
For the time being, these new criteria for the quality of the
continuum extrapolation have only been adopted for the heavy-quark
quantities, but their use may be extended to all FLAG quantities in
future reviews.

\item Finite-volume:\\
\good \hspace{0.2cm}  $M_{\pi,\mathrm{min}} L \gtapprox 3.7$ or 2 volumes at fixed parameters \\
\rule{0.05em}{0em}\soso \hspace{0.2cm}  $M_{\pi,\mathrm{min}} L \gtapprox 3$  \\
\rule{0.05em}{0em}\bad \hspace{0.2cm}  otherwise\\
Here the boundary between green star and open circle is slightly
relaxed compared to that in Sec.~\ref{sec:LQCriteria} to account for
the fact that heavy-quark quantities are less sensitive to this
systematic error than light-quark quantities.  A \good\ rating
requires an estimate of the finite volume error either by analysing
data on two or more physical volumes (with all other parameters fixed)
or by using finite volume chiral perturbation theory.  In the case of
staggered sea quarks, $M_{\pi,\mathrm{min}}$ refers to the lightest
(taste Goldstone) pion mass.

\end{itemize}

\subsection{Averages and estimates}\label{sec:averages}

For many observables there are enough independent lattice calculations
of good quality that it makes sense to average them and propose such
an {\em average} as the best current lattice number. In order to
decide whether this is true for a certain observable, we rely on the
colour coding.  We restrict the averages to data for which the colour
code does not contain any red tags. In some cases, the averaging
procedure nevertheless leads to a result which in our opinion does not
cover all uncertainties.  This is related to the fact that procedures
for estimating errors and the resulting conclusions necessarily have
an element of subjectivity, and would vary between groups even with
the same data set.  In order to stay on the conservative side, we may
replace the average by an {\it estimate} (or a {\it range}), which we
consider as a fair assessment of the knowledge acquired on the lattice
at present. This estimate is not obtained with a prescribed
mathematical procedure, but is based on a critical analysis of the
available information.

There are two other important criteria which also play a role in this
respect, but which cannot be colour coded, because a systematic
improvement is not possible. These are: {\em i)} the publication
status, and {\em ii)} the number of flavours $\Nf$. As far as the
former criterion is concerned, we adopt the following policy: we
average only results which have been published in peer reviewed
journals, i.e.~they have been endorsed by referee(s). The only
exception to this rule consists in obvious updates of previously
published results, typically presented in conference proceedings. Such
updates, which supersede the corresponding results in the published
papers, are included in the averages. Nevertheless, all results are
listed and their publication status is identified by the following
symbols:
\begin{itemize}
\item Publication status:\\
\gA  \hspace{0.2cm}published or plain update of published results\\
\oP  \hspace{0.2cm}preprint\\ 
\rC  \hspace{0.2cm}conference contribution
\end{itemize}
Note that updates of earlier results rely, at least partially, on the
same gauge field configuration ensembles. For this reason, we do not
average updates with earlier results. In the present edition, the
publication status on November 30, 2013 is relevant. If the paper
appeared in print after that date this is accounted for in the
bibliography, but does not affect the averages.

In this review we present results from simulations with $N_f=2$,
$N_f=2+1$ and $N_f=2+1+1$ (for $ r_0 \Lambda_\msbar$ also with
$N_f=0$). We are not aware of an {\em a priori} way
to quantitatively estimate the difference between results produced in
simulations with a different number of dynamical quarks. We therefore
average results at fixed $\Nf$ separately; averages of calculations
with different $\Nf$ will not be provided.

To date, no significant differences between results with different
values of $N_f$ have been observed. In the future, as the accuracy
and the control over systematic effects in lattice calculations will
increase, it will hopefully be possible to see a difference between $\Nf
= 2$ and $\Nf = 2 + 1$ calculations and so determine the size of the
Zweig-rule violations related to strange quark loops. This is a very
interesting issue {\em per se}, and one which can be quantitatively 
addressed only with lattice calculations.

\subsection{Averaging procedure and error analysis}
\label{sec:error_analysis}

In \cite{Colangelo:2010et}, the FLAG averages and their errors were
estimated through the following procedure: Having added in quadrature
statistical and systematic errors for each individual result, we
obtained their weighted $\chi^2$ average. This was our central
value. If the fit was of good quality ($\chi_{\rm min}^2/dof \le 1$),
we calculated the net uncertainty $\delta$ from $\chi^2 = \chi_{\rm
  min}^2 + 1$; otherwise, we inflated the result obtained in this way
by the factor $S = \sqrt(\chi^2/dof)$.  Whenever this $\chi^2$
minimization procedure resulted in a total error which was smaller
than the smallest systematic error of any individual lattice result,
we assigned the smallest systematic error of that result to the total
systematic error in the average.

One of the problems arising when forming such averages is that not all
of the data sets are independent; in fact, some rely on the same
ensembles. In particular, the same gauge field configurations,
produced with a given fermion descretization, are often used by
different research teams with different valence quark lattice actions,
obtaining results which are not really independent.  In the present
paper we have modified our averaging procedure, in order to account
for such correlations. To start with, we examine error budgets for
individual calculations and look for potentially correlated
uncertainties. Specific problems encountered in connection with
correlations between different data sets are commented in the text.
If there is any reason to believe that a source of error is correlated
between two calculations, a $100\%$ correlation is assumed. We then
obtain the central value from a $\chi^2$ weighted average, evaluated
by adding statistical and systematic errors in quadrature (just as in
Ref.~\cite{Colangelo:2010et}): for a set of individual measurements
$x_i$ with error $\sigma_i$ and correlation matrix $C_{ij}$, central
value and error of the average are given by:
\begin{eqnarray}
x_{\rm average} \,\, &=& \,\, \sum_i x_i\, \omega_i \,\,,\hspace{0.5cm}  
\omega_i \,\, = \,\, \dfrac{\sigma_i^{-2}}{\sum_j\sigma_j^{-2}}\,\,,\\
\sigma^2_{\rm average} \,\, &=& \,\, \sum_{i,j} \omega_i \,\omega_j \,C_{ij}\,\,. 
\end{eqnarray}
The correlation matrix for the set of correlated lattice results is
estimated with Schmelling's prescription \cite{Schmelling:1994pz}.
When necessary, the statistical and systematic error bars are
stretched by a factor $S$, as specified in the previous paragraph.

\clearpage
\section{Masses of the light quarks}
\label{sec:qmass}
Quark masses are fundamental parameters of the Standard Model. An accurate
determination of these parameters is important for both phenomenological
and theoretical applications. The charm and bottom masses, for instance,
enter the theoretical expressions of several cross sections and decay
rates in heavy-quark expansions. The up-, down- and strange-quark masses
govern the amount of explicit chiral symmetry breaking in QCD. From a
theoretical point of view, the values of quark masses provide information
about the flavour structure of physics beyond the Standard Model. The Review
of Particle Physics of the Particle Data Group contains a review of quark
masses \cite{Manohar_and_Sachrajda}, which covers light as well as heavy
flavours.  The present summary only deals with the light-quark masses (those
of the up, down and strange quarks), but discusses the lattice results for
these in more detail.

Quark masses cannot be measured directly with experiment because
quarks cannot be isolated, as they are confined inside hadrons. On the
other hand, quark masses are free parameters of the theory and, as
such, cannot be obtained on the basis of purely theoretical
considerations. Their values can only be determined by comparing the
theoretical prediction for an observable, which depends on the quark
mass of interest, with the corresponding experimental value. What
makes light-quark masses particularly difficult to determine is the
fact that they are very small (for the up and down) or small (for the
strange) compared to typical hadronic scales. Thus, their impact on
typical hadronic observables is minute and it is difficult to isolate
their contribution accurately.

Fortunately, the spontaneous breaking of SU(3)$_L\otimes$SU(3)$_R$
chiral symmetry provides observables which are particularly sensitive
to the light-quark masses: the masses of the resulting Nambu-Goldstone
bosons (NGB), i.e.~pions, kaons and etas. Indeed, the
Gell-Mann-Oakes-Renner relation~\cite{GellMann:1968rz} predicts that
the squared mass of a NGB is directly proportional to the sum of the
masses of the quark and antiquark which compose it, up to higher-order
mass corrections. Moreover, because these NGBs are light and are
composed of only two valence particles, their masses have a
particularly clean statistical signal in lattice-QCD calculations. In
addition, the experimental uncertainties on these meson
masses are negligible.

\medskip

Three flavour QCD has four free parameters: the strong coupling,
$\alpha_s$ (alternatively $\Lambda_\mathrm{QCD}$) and the up, down and
strange quark masses, $m_u$, $m_d$ and $m_s$. However, present day
lattice calculations are often performed in the isospin limit, and the
up and down quark masses (especially those in the sea) usually get
replaced by a single parameter: the isospin averaged up- and down-quark
mass, $m_{ud}=\frac12(m_u+m_d)$. A lattice determination of these
parameters requires two steps:
\begin{enumerate}
\item Calculations of three experimentally measurable quantities are
  used to fix the three bare parameters. As already discussed, NGB
  masses are particularly appropriate for fixing the light-quark
  masses. Another observable, such as the mass of a member of the
  baryon octet, can be used to fix the overall scale. It is important
  to note that until recently, most calculations were performed at
  values of $m_{ud}$ which were still substantially larger than its
  physical value, typically four times as large. 
  Reaching the physical up- and down-quark mass point required a
  significant extrapolation. This situation is changing fast. The
  PACS-CS~\cite{Aoki:2008sm,Aoki:2009ix,Aoki:2010wm} and
  BMW~\cite{Durr:2010vn,Durr:2010aw} calculations were performed with
  masses all the way down to their physical value (and even below in
  the case of BMW), albeit in very small volumes for PACS-CS. More
  recently, MILC \cite{Bazavov:2011fh} and RBC/UKQCD
  \cite{Arthur:2012opa} have also extended their simulations almost
  down to the physical point, by considering pions with $M_\pi\gsim
  170\,\mev$.\footnote{In the case of MILC, we are referring to the
    staggered root-mean-squared average mass of the taste partners
    (see discussion in Section \ref{sec:color-code}). The mass of the
    corresponding taste-Goldstone-pion in these simulations is the
    physical value.
    }
    Regarding the strange
  quark, modern simulations can easily include them with masses that
  bracket its physical value, and only interpolations are needed.

\item Renormalizations of these bare parameters must be performed to relate
 them to the corresponding cutoff-independent, renormalized
 parameters.\footnote{Throughout this review, the quark masses $m_u$,
   $m_d$ and $m_s$ refer to the $\msbar$ scheme  at running scale
   $\mu=2\,\gev$ and the numerical values are given in MeV units.} These
 are short distance calculations, which may be performed perturbatively.
Experience shows that one-loop calculations are unreliable
 for the renormalization of quark masses: usually at
 least two loops are required to have trustworthy results. Therefore, it
 is best to perform the renormalizations nonperturbatively to avoid
 potentially large perturbative uncertainties due to neglected higher-order 
 terms. {\ However we will include in our averages one-loop results which carry a solid estimate of the systematic uncertainty due to the truncation of the series.}
\end{enumerate}
Of course, in quark mass ratios the renormalization factor cancels, so that
this second step is no longer relevant.

\subsection{Contributions from the electromagnetic interaction}
\label{subsec:electromagnetic interaction}
As mentioned in Section \ref{sec:color-code}, the present review
relies on the hypothesis that, at low energies, the Lagrangian ${\cal
  L}_{\mbox{\tiny QCD}}+{\cal L}_{\mbox{\tiny QED}}$ describes nature
to a high degree of precision. Moreover, we assume that, at the
accuracy reached by now and for the quantities discussed here, the
difference between the results obtained from simulations with three
dynamical flavours and full QCD is small in comparison with the quoted
systematic uncertainties. { This will soon no longer be the case.} The
electromagnetic (e.m.) interaction, on the other hand, cannot be
ignored.  Quite generally, when comparing QCD calculations with
experiment, radiative corrections need to be applied. In lattice simulations, 
where the QCD parameters are fixed in terms of the masses
of some of the hadrons, the electromagnetic contributions to these
masses must be accounted for.\footnote{Since the decomposition of the
  sum ${\cal L}_{\mbox{\tiny QCD}}+{\cal L}_{\mbox{\tiny QED}}$ into
  two parts is not unique, specifying the QCD part requires a
  convention. In order to give results for the quark masses in the
  Standard Model at scale $\mu=2\,\mbox{GeV}$, on the basis of a
  calculation done within QCD, it is convenient to match the two
  theories at that scale. We use this convention throughout the
  present review.  Note that a different convention is used in the
  analysis of the precision measurements carried out in low energy
  pion physics (e.g. \cite{BlochDevaux:2008zz}). When comparing
  lattice results with experiment, it is important to fix the QCD
  parameters in accordance with the convention used in the analysis of
  the experimental data (for a more detailed discussion, see
  \cite{Gasser:2003hk,Rusetsky:2009ic,Gasser:2007de,Leutwyler:2009jg}). }

The electromagnetic interaction plays a crucial role in determinations
of the ratio $m_u/m_d$, because the isospin-breaking effects generated
by this interaction are comparable to those from $m_u\neq m_d$ (see
Subsection \ref{subsec:mumd}). In determinations of the ratio
$m_s/m_{ud}$, the electromagnetic interaction is less important, but
at the accuracy reached, it cannot be neglected. The reason is that,
in the determination of this ratio, the pion mass enters as an input
parameter. Because $M_\pi$ represents a small symmetry breaking
effect, it is rather sensitive to the perturbations generated by QED.

We distinguish the physical mass $M_P$, $P\in\{\pi^+,$ $\pi^0$, $K^+$,
$K^0\}$, from the mass $\hat{M}_P$ within QCD alone. The e.m.\ self-energy
is the difference between the two, $M_P^\gamma\equiv
M_P-\hat{M}_P$. Because the self-energy of the Nambu-Goldstone bosons
diverges in the chiral limit, it is convenient to replace it by the
contribution of the e.m.~interaction to the {\it square} of the mass,
\be \label{eq:DeltaP}
\Delta_{P}^\gamma\equiv M_P^2-\hat{M}_P^2= 2\,M_P M_P^\gamma+O(e^4)\,.\ee 
The main
effect of the e.m.\ interaction is an increase in the mass of the
charged particles, generated by the photon cloud that surrounds
them. The self-energies of the neutral ones are comparatively small,
particularly for the Nambu-Goldstone bosons, which do not have a
magnetic moment. Dashen's theorem \cite{Dashen:1969eg} confirms this
picture, as it states that, to leading order (LO) of the chiral expansion,
the self-energies of the neutral NGBs vanish, while the charged ones
obey $\Delta_{K^+}^\gamma = \Delta_{\pi^+}^\gamma $. It is convenient
to express the self-energies of the neutral particles as well as the
mass difference between the charged and neutral pions within QCD in
units of the observed mass difference, $\Delta_\pi\equiv
M_{\pi^+}^2-M_{\pi^0}^2$:
\be\label{eq:epsilon1}
\Delta_{\pi^0}^\gamma \equiv
\epsilon_{\pi^0}\,\Delta_\pi\co\hspace{0.2cm}\Delta_{K^0}^\gamma \equiv
\epsilon_{K^0}\,\Delta_\pi\co\hspace{0.2cm}\hat{M}_{\pi^+}^2-
\hat{M}_{\pi^0}^2\equiv
\epsilon_m\,\Delta_\pi\fs\ee
In this notation, the self-energies of the charged particles are given
by 
\be\label{eq:epsilon2}
\Delta_{\pi^+}^\gamma=(1+\epsilon_{\pi^0}-\epsilon_m)\,\Delta_\pi\co\hspace{0.5cm}
\Delta_{K^+}^\gamma=(1+\epsilonD+\epsilon_{K^0}-\epsilon_m)\,\Delta_\pi\co\ee
where the dimensionless coefficient $\epsilonD$ parameterizes the
violation of Dashen's theorem,\footnote{\label{fn1}Sometimes,
  e.g. in \cite{Blum:2010ym}, the violation of Dashen's theorem is given in
  terms of a different quantity, $\bar\epsilon\equiv
  (\Delta_{K^+}^\gamma-\Delta_{K^0}^\gamma)/(\Delta_{\pi^+}^\gamma-\Delta_{\pi^0}^\gamma)-1$. This
  parameter is related to $\epsilonD$ used here through
  $\epsilonD=(1-\epsilon_m)\bar\epsilon$. Given the value of
  $\epsilon_m$ (see (\ref{eq:epsilon num})), these two quantities differ
  by only 4\%.}
\be\label{eq:epsilon3}
\Delta_{K^+}^\gamma-\Delta_{K^0}^\gamma-
\Delta_{\pi^+}^\gamma+\Delta_{\pi^0}^\gamma\equiv\epsilonD\,\Delta_\pi\fs\ee
Any determination of the light-quark masses based on a calculation of
the masses of $\pi^+,K^+$ and $K^0$ within QCD requires an estimate
for the coefficients $\epsilonD$, $\epsilon_{\pi^0}$, $\epsilon_{K^0}$
and $\epsilon_m$.

The first determination of the self-energies on the lattice was
carried out by Duncan, Eichten and Thacker \cite{Duncan:1996xy}. Using
the quenched approximation, they arrived at
$M_{K^+}^\gamma-M_{K^0}^\gamma= 1.9\,\mbox{MeV}$. Actually, the
parameterization of the masses given in that paper yields an estimate
for all but one of the coefficients introduced above (since the mass
splitting between the charged and neutral pions in QCD is neglected,
the parameterization amounts to setting $\epsilon_m=0$ ab
initio). Evaluating the differences between the masses obtained at the
physical value of the electromagnetic coupling constant and at $e=0$,
we obtain $\epsilonD = 0.50(8)$, $\epsilon_{\pi^0} = 0.034(5)$ and
$\epsilon_{K^0} = 0.23(3)$. The errors quoted are statistical only: an
estimate of lattice systematic errors is not possible from the limited
results of \cite{Duncan:1996xy}. The result for $\epsilonD$ indicates
that the violation of Dashen's theorem is sizeable: according to this
calculation, the nonleading contributions to the self-energy
difference of the kaons amount to 50\% of the leading term. The result
for the self-energy of the neutral pion cannot be taken at face value,
because it is small, comparable to the neglected mass difference
$\hat{M}_{\pi^+}-\hat{M}_{\pi^0}$. To illustrate this, we note that
the numbers quoted above are obtained by matching the parameterization
with the physical masses for $\pi^0$, $K^+$ and $K^0$. This gives a
mass for the charged pion that is too high by 0.32 MeV. Tuning the
parameters instead such that $M_{\pi^+}$ comes out correctly, the
result for the self-energy of the neutral pion becomes larger:
$\epsilon_{\pi^0}=0.10(7)$ where, again, the error is statistical
only.

In an update of this calculation by the RBC collaboration
\cite{Blum:2007cy} (RBC 07), the electromagnetic interaction is still
treated in the quenched approximation, but the strong interaction is
simulated with $\Nf=2$ dynamical quark flavours. The quark masses are fixed
with the physical masses of $\pi^0$, $K^+$ and $K^0$. The outcome for the
difference in the electromagnetic self-energy of the kaons reads
$M_{K^+}^\gamma-M_{K^0}^\gamma= 1.443(55)\,\mbox{MeV}$. This corresponds to
a remarkably small violation of Dashen's theorem. Indeed, a recent
extension of this work to $\Nf=2+1$ dynamical flavours \cite{Blum:2010ym}
leads to a significantly larger self-energy difference:
$M_{K^+}^\gamma-M_{K^0}^\gamma= 1.87(10)\,\mbox{MeV}$, in good agreement
with the estimate of Eichten et al. Expressed in terms of the coefficient
$\epsilonD$ that measures the size of the violation of Dashen's theorem, it
corresponds to $\epsilonD=0.5(1)$.

The input for the electromagnetic corrections used by MILC is
specified in \cite{Aubin:2004he}. In their analysis of the lattice data, 
$\epsilon_{\pi^0}$, $\epsilon_{K^0}$ and $\epsilon_m$ are set
equal to zero. For the remaining coefficient, which plays a crucial
role in determinations of the ratio $m_u/m_d$, the very conservative
range $\epsilonD=1\pm1$ was used in MILC 04 \cite{Aubin:2004fs}, while
in more recent work, in particular in MILC 09 \cite{Bazavov:2009bb}
and MILC 09A~\cite{Bazavov:2009fk}, this input is replaced by
$\epsilonD=1.2\pm0.5$, as suggested by phenomenological estimates for
the corrections to Dashen's theorem
\cite{Bijnens:1996kk,Donoghue:1996zn}. Results of an evaluation of the
electromagnetic self-energies based on $\Nf=2+1$ dynamical quarks in
the QCD sector and on the quenched approximation in the QED sector are
also reported by MILC \cite{Basak:2008na,Basak:2012zx,Basak:2013iw}. Their preliminary
  result is $\bar\epsilon=0.65(7)(14)(10)$, where the first error is
  statistical, the second systematic, and the third a separate
  systematic for the combined chiral and continuum extrapolation.  The
  estimate of the systematic error does not yet include finite-volume
  effects. With the estimate for $\epsilon_m$ given in
  (\ref{eq:epsilon num}), this result corresponds to $\epsilon=
  0.62(7)(14)(10)$. Similar preliminary results were previously
  reported by the BMW collaboration in conference proceedings
  \cite{Portelli:2010yn,Portelli:2012pn}.

The RM123 collaboration employs a new technique to compute e.m. shifts
in hadron masses in two-flavour QCD: the effects are included at
leading order in the electromagnetic coupling $\alpha$ through simple
insertions of the fundamental electromagnetic interaction in quark
lines of relevant Feynman graphs~\cite{deDivitiis:2013xla}. They find
$\epsilon=0.79(18)(18)$ where the first error is statistical and the
second is the total systematic error resulting from chiral, finite-volume, 
discretization, quenching and fitting errors all added in
quadrature.

The effective Lagrangian that governs the self-energies to next-to-leading
order (NLO) of the chiral expansion was set up in \cite{Urech:1994hd}. The
estimates in \cite{Bijnens:1996kk,Donoghue:1996zn} are obtained by
replacing QCD with a model, matching this model with the effective theory
and assuming that the effective coupling constants obtained in this way
represent a decent approximation to those of QCD. For alternative model
estimates and a detailed discussion of the problems encountered in models
based on saturation by resonances, see
\cite{Baur:1995ig,Baur:1996ya,Moussallam:1997xx}.  In the present review of
the information obtained on the lattice, we avoid the use of models
altogether.

There is an indirect phenomenological determination of $\epsilonD$,
which is based on the decay $\eta\rightarrow 3\pi$ and does not rely
on models.  The result for the quark mass ratio $Q$, defined in
(\ref{eq:Qm}) and obtained from a dispersive analysis of this decay,
implies $\epsilonD = 0.70(28)$ (see Section \ref{subsec:mumd}).  {
  While the values found in older lattice calculations
  \cite{Duncan:1996xy,Blum:2007cy,Blum:2010ym} are a little less than
  one standard deviation lower, the most recent determinations
  \cite{Basak:2008na,Portelli:2010yn,Portelli:2012pn,llconfx12,Basak:2012zx,Basak:2013iw,deDivitiis:2013xla},
  though still preliminary, are in excellent agreement with this
  result and have significantly smaller error bars. However, even in
  the more recent calculations, e.m.\ effects are treated in the
  quenched approximation. Thus, we choose to quote $\epsilonD =
  0.7(3)$, which is essentially the $\eta\rightarrow 3\pi$ result and
  covers generously the range of post 2010 lattice results. Note that
  this value has an uncertainty which is reduced by about 40\%
  compared to the result quoted in the first edition of the FLAG review \cite{Colangelo:2010et}.}

We add a few comments concerning the physics of the self-energies and then
specify the estimates used as an input in our analysis of the data. The
Cottingham formula \cite{Cottingham} represents the self-energy of a
particle as an integral over electron scattering cross sections; elastic as
well as inelastic reactions contribute. For the charged pion, the term due
to elastic scattering, which involves the square of the e.m.~form factor,
makes a substantial contribution. In the case of the $\pi^0$, this term is
absent, because the form factor vanishes on account of charge conjugation
invariance. Indeed, the contribution from the form factor to the
self-energy of the $\pi^+$ roughly reproduces the observed mass difference
between the two particles. Furthermore, the numbers given in
\cite{Socolow:1965zz,Gross:1979ur,Gasser:1982ap} indicate that the
inelastic contributions are significantly smaller than the elastic
contributions to the self-energy of the $\pi^+$. The low energy theorem of
Das, Guralnik, Mathur, Low and Young \cite{Das:1967it} ensures that, in the
limit $m_u,m_d\rightarrow 0$, the e.m.~self-energy of the $\pi^0$ vanishes,
while the one of the $\pi^+$ is given by an integral over the difference
between the vector and axial-vector spectral functions. The estimates for
$\epsilon_{\pi^0}$ obtained in \cite{Duncan:1996xy} are consistent with the
suppression of the self-energy of the $\pi^0$ implied by chiral
SU(2)$\times$SU(2). In our opinion, $\epsilon_{\pi^0}=0.07(7)$ is a
conservative estimate for this coefficient. The self-energy of the $K^0$ is
suppressed less strongly, because it remains different from zero if $m_u$
and $m_d$ are taken massless and only disappears if $m_s$ is turned off as
well. Note also that, since the e.m.~form factor of the $K^0$ is different
from zero, the self-energy of the $K^0$ does pick up an elastic
contribution. The lattice result for $\epsilon_{K^0}$ indicates that the
violation of Dashen's theorem is smaller than in the case of $\epsilonD$.
In the following, we use $\epsilon_{K^0}=0.3(3)$. 

Finally, we consider the mass splitting between the charged and neutral
pions in QCD. This effect is known to be very small, because it is of
second order in $m_u-m_d$. There is a parameter-free prediction, which
expresses the difference $\hat{M}_{\pi^+}^2-\hat{M}_{\pi^0}^2$ in terms of
the physical masses of the pseudoscalar octet and is valid to NLO of the
chiral perturbation series. Numerically, the relation yields
$\epsilon_m=0.04$ \cite{Gasser:1984gg}, indicating that this contribution
does not play a significant role at the present level of accuracy. We
attach a conservative error also to this coefficient: $\epsilon_m=0.04(2)$.
The lattice result for the self-energy difference of the pions, reported in
\cite{Blum:2010ym}, $M_{\pi^+}^\gamma-M_{\pi^0}^\gamma=
4.50(23)\,\mbox{MeV}$, agrees with this estimate: expressed in terms of the
coefficient $\epsilon_m$ that measures the pion mass splitting in QCD, the
result corresponds to $\epsilon_m=0.04(5)$. The corrections of
next-to-next-to-leading order (NNLO) have been worked out
\cite{Amoros:2001cp}, but the numerical evaluation of the 
formulae again meets with the problem that the relevant effective coupling
constants are not reliably known. 

In summary, we use the following estimates for the
e.m.~corrections:
\be\label{eq:epsilon num}\epsilonD={ 0.7(3)}
\co\hspace{0.5cm}\epsilon_{\pi^0}=0.07(7)\co\hspace{0.5cm}
\epsilon_{K^0}=0.3(3)\co\hspace{0.5cm}\epsilon_m=0.04(2)\fs\ee
While the range used for the coefficient $\epsilonD$ affects our
analysis in a significant way, the numerical values of the other
coefficients only serve to set the scale of these contributions. The
range given for $\epsilon_{\pi^0}$ and $\epsilon_{K^0}$ may be overly
generous, but because of the exploratory nature of the 
lattice determinations, we consider it advisable to use a conservative
estimate.

Treating the uncertainties in the four coefficients as statistically
independent and adding errors in quadrature, the numbers in equation
(\ref{eq:epsilon num}) yield the following estimates for the
e.m.~self-energies,
{ 
\bea\label{eq:Mem}&&\hspace{-1cm} M_{\pi^+}^\gamma= 4.7(3)\,
\mbox{MeV}\co\hspace{0.45cm} M_{\pi^0}^\gamma = 0.3(3)\,\mbox{MeV}
\co\hspace{0.5cm} M_{\pi^+}^\gamma-M_{\pi^0}^\gamma=4.4(1)\, \mbox{MeV}\co\\
&&\hspace{-1cm} M_{K^+}^\gamma= 2.5(5)\,\mbox{MeV}\co\hspace{0.35cm}
M_{K^0}^\gamma
=0.4(4)\,\mbox{MeV}\co\hspace{0.3cm}M_{K^+}^\gamma-M_{K^0}^\gamma= 2.1(4)\,
\mbox{MeV}\,,\nonumber \eea
and for the pion and kaon masses occurring in the QCD sector of the
Standard Model, 
\bea\label{eq:MQCD}&&\hspace{-1cm} \hat{M}_{\pi^+}= 134.8(3)\,
\mbox{MeV}\co\hspace{0.2cm} \hat{M}_{\pi^0} = 134.6(3)\,\mbox{MeV}
\co\hspace{0.5cm} \hat{M}_{\pi^+}-\hat{M}_{\pi^0}=\hspace{0.25cm}0.2(1)\,
\mbox{MeV}\co\\ 
&&\hspace{-1.1cm} \hat{M}_{K^+}= 491.2(5)\,\mbox{MeV}\co\hspace{0.12cm}
\hat{M}_{K^0}
=497.2(4)\,\mbox{MeV}\co\hspace{0.3cm}\hat{M}_{K^+}-\hat{M}_{K^0}=-6.1(4)\,
\mbox{MeV}\fs\nonumber \eea
}
The self-energy difference between the charged and neutral pion involves
the same coefficient $\epsilon_m$ that describes the mass difference in QCD
-- this is why the estimate for $ M_{\pi^+}^\gamma-M_{\pi^0}^\gamma$ is so
sharp.

\subsection{Pion and kaon masses in the isospin limit}
\label{subsec:mu not equal md}

As mentioned above, most of the lattice calculations concerning the
properties of the light mesons are performed in the isospin limit of QCD 
($m_u-m_d\rightarrow0$ at fixed $m_u+m_d$). We
denote the pion and kaon masses in that limit by $\Mpibar$ and
$\MKbar$, respectively. Their numerical values can be estimated as
follows. Since the operation $u\leftrightarrow d$ interchanges $\pi^+$
with $\pi^-$ and $K^+$ with $K^0$, the expansion of the quantities
$\hat{M}_{\pi^+}^2$ and $\frac{1}{2}(\hat{M}_{K^+}^2+\hat{M}_{K^0}^2)$
in powers of $m_u-m_d$ only contains even powers. As shown in
\cite{Gasser:1983yg}, the effects generated by $m_u-m_d$ in the mass
of the charged pion are strongly suppressed: the difference
$\hat{M}_{\pi^+}^2-\Mpibar^{\,2}$ represents a quantity of
$O[(m_u-m_d)^2(m_u+m_d)]$ and is therefore small compared to the
difference $\hat{M}_{\pi^+}^2-\hat{M}_{\pi^0}^2$, for which an
estimate was given above. In the case of 
$\frac{1}{2}(\hat{M}_{K^+}^2+\hat{M}_{K^0}^2)-\MKbar^{\,2}$, the
expansion does contain a contribution at NLO, determined by the
combination $2L_8-L_5$ of low energy constants, but the lattice results 
for that combination show that this contribution is very
small, too. Numerically, the effects generated by $m_u-m_d$ in
$\hat{M}_{\pi^+}^2$ and in
$\frac{1}{2}(\hat{M}_{K^+}^2+\hat{M}_{K^0}^2)$ are negligible compared
to the uncertainties in the electromagnetic self-energies. The
estimates for these given in equation (\ref{eq:MQCD}) thus imply
{
\be
\Mpibar = \hat{M}_{\pi^+}=134.8(3)\,\mev\ ,\hspace{1cm} \MKbar=
\sqrt{\frac{1}{2}(\hat{M}_{K^+}^2+\hat{M}_{K^0}^2)}= 494.2(4)\,\mev\ . 
\label{eq:MpiMKiso}
\ee
}
This shows that, for the convention used above to specify the QCD sector of
the Standard Model, and within the accuracy to which this convention can
currently be implemented, the mass of the pion in the isospin limit agrees
with the physical mass of the neutral pion: $\Mpibar-M_{\pi^0}=-0.2(3)$
MeV.

\subsection{Lattice determination of $m_s$ and $m_{ud}$}\label{sec:msmud}

We now turn to a review of the lattice calculations of the light-quark
masses and begin with $m_s$, the isospin averaged up- and down-quark
mass, $m_{ud}$, and their ratio. Most groups quote only $m_{ud}$, not
the individual up- and down-quark masses. We then discuss the ratio
$m_u/m_d$ and the individual determination of $m_u$ and $m_d$.

Quark masses have been calculated on the lattice since the
mid nineties. However early calculations were performed in the quenched
approximation, leading to unquantifiable systematics. Thus in the following,
we only review modern, unquenched calculations, which include the effects of
light sea-quarks.

Tables~\ref{tab:masses2} and \ref{tab:masses3} list the results of
$\Nf=2$ and $\Nf=2+1$ lattice calculations of $m_s$ and
$m_{ud}$. These results are given in the $\msbar$ scheme  at $2\,\gev$,
which is standard nowadays, though some groups are starting to quote
results at higher scales (e.g. \cite{Arthur:2012opa}). The tables also
show the colour-coding of the calculations leading to these
results. The corresponding results for $m_s/m_{ud}$ are given in
Table~\ref{tab:ratio_msmud}.  As indicated earlier in this review, we
treat $\Nf=2$ and $\Nf=2+1$ calculations separately. The latter
include the effects of a strange sea-quark, but the former do not.

\subsubsection{$\Nf=2$ lattice calculations}

We begin with $\Nf=2$ calculations. A quick inspection of
Table~\ref{tab:masses2} indicates that only the most recent
calculations, ALPHA 12~\cite{Fritzsch:2012wq} and ETM
10B~\cite{Blossier:2010cr}, control all systematic effects--the
special case of D\"urr 11 \cite{Durr:2011ed} is discussed below. Only
ALPHA 12~\cite{Fritzsch:2012wq}, ETM 10B~\cite{Blossier:2010cr} and
ETM 07~\cite{Blossier:2007vv} really enter the chiral regime, with
pion masses down to about 270~MeV for ALPHA and
ETM. Because this pion mass is still quite far from the physical pion
mass, ALPHA 12 refrain from determining $m_{ud}$ and give only
$m_s$. All the other calculations have significantly more massive
pions, the lightest being about 430~MeV, in the calculation by CP-PACS
01~\cite{AliKhan:2001tx}.  Moreover, the latter calculation is
performed on very coarse lattices, with lattice spacings $a\ge
0.11\,\fm$ and only one-loop perturbation theory is used to
renormalize the results.

 ETM 10B's~\cite{Blossier:2010cr} calculation of $m_{ud}$ and
  $m_s$ is an update of the earlier twisted-mass determination of ETM
  07~\cite{Blossier:2007vv}. In particular, they have added ensembles with a larger volume and three new 
  lattice spacings, $a = 0.054, 0.067$ and $0.098\,\fm$, allowing for a continuum extrapolation. In addition, it
  presents analyses performed in $SU(2)$ and $SU(3)$ $\chi$PT. 

The new ALPHA 12 \cite{Fritzsch:2012wq} calculation of $m_s$ is an
update of ALPHA 05~\cite{DellaMorte:2005kg}, which pushes computations
to finer lattices and much lighter pion masses. It also importantly
includes a determination of the lattice spacing with the
decay constant $F_K$, whereas ALPHA 05 converted results to physical
units using the scale parameter $r_0$ \cite{Sommer:1993ce}, defined
via the force between static quarks. In particular, the conversion
relied on measurements of $r_0/a$ by QCDSF/UKQCD 04
\cite{Gockeler:2004rp} which differ significantly from the new
determination by ALPHA 12. As in ALPHA 05, in ALPHA 12 both
nonperturbative running and nonperturbative renormalization are
performed in a controlled fashion, using Schr\"odinger functional
methods.

\begin{table}[t]
{\footnotesize{
\begin{tabular*}{\textwidth}[l]{l@{\extracolsep{\fill}}rllllllll}
Collaboration & Ref. & \hspace{0.15cm}\begin{rotate}{60}{publication status}\end{rotate}\hspace{-0.15cm} &
 \hspace{0.15cm}\begin{rotate}{60}{chiral extrapolation}\end{rotate}\hspace{-0.15cm} &
 \hspace{0.15cm}\begin{rotate}{60}{continuum  extrapolation}\end{rotate}\hspace{-0.15cm}  &
 \hspace{0.15cm}\begin{rotate}{60}{finite volume}\end{rotate}\hspace{-0.15cm}  &  
 \hspace{0.15cm}\begin{rotate}{60}{renormalization}\end{rotate}\hspace{-0.15cm} &  
 \hspace{0.15cm}\begin{rotate}{60}{running}\end{rotate}\hspace{-0.15cm}  & 
\rule{0.6cm}{0cm}$m_{ud} $ & \rule{0.6cm}{0cm}$m_s $ \\
&&&&&&&&& \\[-0.1cm]
\hline
\hline
&&&&&&&&& \\[-0.1cm]
{ALPHA 12}& \cite{Fritzsch:2012wq} & \gA & \soso & \good & \good & \good & $\,a,b$
 &  & 102(3)(1) \\ 

{D\"urr 11$^\ddagger$}& \cite{Durr:2011ed} & \gA & \soso & \good & \soso & $-$ & $-$
 & 3.52(10)(9) & 97.0(2.6)(2.5) \\ 

{ETM 10B}& \cite{Blossier:2010cr} & \gA & \soso & \good & \soso & \good & $\,c$
 & 3.6(1)(2) & 95(2)(6) \\ 

{JLQCD/TWQCD 08A}& \cite{Noaki:2008iy} & \gA& \soso&\bad&\bad&\good& $-$& 4.452(81)(38)$\binom{+0}{-227}$ &\rule{0.6cm}{0cm}--\\                  

{RBC 07$^\dagger$} & \cite{Blum:2007cy} & \gA & \bad & \bad & \good  & \good &
$-$       & $4.25(23)(26)$        & 119.5(5.6)(7.4)              \\

{ETM 07} & \cite{Blossier:2007vv} & \gA &  \soso & \bad & \soso & \good &$-$
& $3.85(12)(40)$        & $105(3)(9)$                  \\

\hspace{-0.2cm}{\begin{tabular}{l}QCDSF/\\
UKQCD 06\end{tabular}} & \cite{Gockeler:2006jt} & \gA &  \bad  & \good & \bad &
\good &$-$      & $4.08(23)(19)(23)$ &  $111(6)(4)(6)$ \\

{SPQcdR 05} & \cite{Becirevic:2005ta} & \gA & \bad & \soso & \soso & \good &
$-$& $4.3(4)(^{+1.1}_{-0.0})$ & $101(8)(^{+25}_{-0})$        \\

{ALPHA 05} & \cite{DellaMorte:2005kg} & \gA &  \bad & \soso & \good  & \good &
$\,a$  &                      & 97(4)(18)$^\S$           \\

\hspace{-0.2cm}{\begin{tabular}{l}QCDSF/\\
UKQCD 04\end{tabular}} & \cite{Gockeler:2004rp} & \gA &  \bad  & \good & \bad &
\good & $-$       & $4.7(2)(3)$ & $119(5)(8)$    \\

{JLQCD 02} & \cite{Aoki:2002uc} & \gA &  \bad  & \bad & \soso & \bad & $-$   
& $3.223(^{+46}_{-69})$ & $84.5(^{+12.0}_{-1.7})$        \\

{CP-PACS 01} & \cite{AliKhan:2001tx} & \gA & \bad & \bad & \good & \bad &$-$ &
$3.45(10)(^{+11}_{-18})$ & $89(2)(^{+2}_{-6})^\star$    \\
&&&&&&&&& \\[-0.1cm] 
\hline
\end{tabular*}
\begin{tabular*}{\textwidth}[l]{l@{\extracolsep{\fill}}lllllllll}
\multicolumn{10}{l}{\vbox{\begin{flushleft} 
$^\ddagger$ What is calculated is $m_c/m_s=11.27(30)(26)$. $m_s$ is then obtained using lattice  and phenomenological
      determinations of $m_c$ which rely on perturbation
      theory. Finally, $m_{ud}$ is determined from $m_s$ using BMW 10A,
      10B's $N_f=2+1$ result for $m_s/m_{ud}$ \cite{Durr:2010vn,Durr:2010aw}. Since
      $m_c/m_s$ is renormalization group invariant in QCD, the
      renormalization and running of the quark masses enter indirectly
      through that of $m_c$, a mass that we do not review here.\\
$^\dagger$ The calculation includes quenched e.m. effects.\\
$^\S$ The data used to obtain the bare value of $m_s$ are from UKQCD/QCDSF 04
 \cite{Gockeler:2004rp}.\\
 $^\star$ This value of $m_s$ was obtained
 using the kaon mass as input. If the $\phi$ meson mass is used instead, the\\\hspace{0.3cm}authors find 
 $m_s =90^{+5}_{-11}.$\\\rule{0cm}{0.3cm}\hspace{-0.1cm}
$a$ The masses are renormalized and run nonperturbatively up to
a scale of $100\,\gev$ in the $N_f=2$ SF\\\hspace{0.3cm}scheme. In this
scheme, nonperturbative and NLO running for the quark masses are
shown to agree\\\hspace{0.3cm}well from 100 GeV all the way down to 2
GeV \cite{DellaMorte:2005kg}.\\\rule{0cm}{0.3cm}\hspace{-0.1cm}
$b$ The running and renormalization results of
\cite{DellaMorte:2005kg} are improved in \cite{Fritzsch:2012wq} with
higher statistical and systematic\\\hspace{0.3cm}accuracy.\\\rule{0cm}{0.3cm}\hspace{-0.1cm}
$c$ The masses are renormalized nonperturbatively at scales $1/a\sim 2\div3\,\gev$ in the $N_f=2$ RI/MOM \\\hspace{0.3cm}scheme.  In this
scheme, nonperturbative and N$^3$LO running for the quark masses
are shown to agree\\\hspace{0.3cm}from 4~GeV down 2 GeV to better than 3\%
\cite{Constantinou:2010gr}.\\\rule{0cm}{0.3cm}\hspace{-0.1cm}
\end{flushleft}}}
\end{tabular*}
}}

\vspace{-0.3cm}
\caption{\label{tab:masses2} $\Nf=2$ lattice results for the masses $m_{ud}$ and $m_s$ (MeV, running masses in the $\msbar$ scheme  at scale 2 GeV). The significance of the colours is explained in Sec.~\ref{sec:qualcrit}. If information about nonperturbative running is available, this is indicated in
  the column ``running'', with details given at the bottom of the table.}
\end{table}

The conclusion of our analysis of $\Nf=2$ calculations is that the
results of ALPHA 12 \cite{Fritzsch:2012wq} and ETM 10B
\cite{Blossier:2010cr} (which update and extend ALPHA 05
\cite{DellaMorte:2005kg} and ETM 07 \cite{Blossier:2007vv},
respectively), are the only ones to date which satisfy our selection
criteria. Thus we average those two results for $m_s$, obtaining
101(3)~MeV. Regarding $m_{ud}$, for which only ETM
10B~\cite{Blossier:2010cr} gives a value, we do not offer an average
but simply quote ETM's number. Because ALPHA's result induces an
increase by 7\% of our earlier average for $m_s$
\cite{Colangelo:2010et} while $m_{ud}$ remains unchanged, our average
for $m_s/m_{ud}$ also increases by 7\%. For the latter, however, we
retain the percent error quoted by ETM, who directly estimates this
ratio, and add it in quadrature to the percent error on ALPHA's
$m_s$. Thus, we quote as our estimates:
\be
\label{eq:quark masses Nf=2} \Nf=2 :\hspace{0.4cm}
m_s= 101(3) \,\mbox{MeV}\co \hspace{0.4cm}
m_{ud}= 3.6(2)  \,\mbox{MeV}  \co\hspace{0.4cm}
\frac{m_s}{m_{ud}} = 28.1(1.2)\fs\ee
The errors on these results are 3\%, 6\% and 4\% respectively. The
error is smaller in the ratio than one would get from combining the
errors on $m_{ud}$ and $m_s$, because statistical and systematic
errors cancel in ETM's result for this ratio, most notably those
associated with renormalization and the setting of the scale. It is
worth noting that thanks to ALPHA 12 \cite{Fritzsch:2012wq}, the total
error on $m_s$ has reduced significantly, from 7\% in the last edition
of our report to 3\% now. It is also interesting to remark that ALPHA
12's \cite{Fritzsch:2012wq} central value for $m_s$ is about 1~$\sigma$
larger than that of ETM 10B \cite{Blossier:2010cr} and { nearly
  2~$\sigma$ larger than our present $N_f=2+1$ determination given in
  (\ref{eq:nf3msmud}). Moreover, this larger value for $m_s$ increases
  our $N_f=2$ determination of $m_s/m_{ud}$, making it larger than ETM
  10B's direct measurement, though compatible within errors.}

We have not discussed yet the precise results of D\"urr 11
\cite{Durr:2011ed} which satisfy our selection criteria. This is
because D\"urr 11 pursue an approach which is sufficiently different
from the one of other calculations that we prefer not to include it in
an average at this stage. Following HPQCD~09A, 10 \cite{Davies:2009ih,McNeile:2010ji},
the observable which they actually compute is $m_c/m_s=11.27(30)(26)$,
with an accuracy of 3.5\%. This result is about 1.5 combined standard
deviations below ETM 10B's \cite{Blossier:2010cr} result
$m_c/m_s=12.0(3)$. $m_s$ is subsequently obtained using lattice  and
phenomenological determinations of $m_c$ which rely on perturbation
theory. The value of the charm-quark mass which they use is an average
of those determinations, which they estimate to be
$m_c(2\,\gev)=1.093(13)\,\gev$, with a 1.2\% total uncertainty. Note
that this value is consistent with the PDG average
$m_c(2\,\gev)=1.094(21)\,\gev$  \cite{Beringer:1900zz}, though the latter has a larger
2.0\% uncertainty. D\"urr 11's value of $m_c$ leads to
$m_s=97.0(2.6)(2.5)\,\mev$ given in Table~\ref{tab:masses2}, which has
a total error of 3.7\%. The use of the PDG value for $m_c$
\cite{Beringer:1900zz} would lead to a very similar result. The result
for $m_s$ is perfectly compatible with our estimate given in
(\ref{eq:quark masses Nf=2}) and has a comparable error bar. To
determine $m_{ud}$, D\"urr 11 combine their result for $m_s$ with the
$N_f=2+1$ calculation of $m_s/m_{ud}$ of BMW 10A, 10B
\cite{Durr:2010vn,Durr:2010aw} discussed below. They obtain
$m_{ud}=3.52(10)(9)\,\mev$ with a total uncertainty of less than 4\%,
which is again fully consistent with our estimate of (\ref{eq:quark
  masses Nf=2}) and its uncertainty.

\subsubsection{$\Nf=2+1$ lattice calculations}\label{sec:Nf=2+1}

We turn now to $\Nf=2+1$ calculations. These and the corresponding results
are summarized in Tables~\ref{tab:masses3} and
\ref{tab:ratio_msmud}. Somewhat paradoxically, these calculations are more
mature than those with $\Nf=2$. This is thanks, in large part, to the head
start and sustained effort of MILC, who have been performing $\Nf=2+1$
rooted staggered fermion calculations for the past ten or so years. They
have covered an impressive range of parameter space, with lattice spacings
which, today, go down to 0.045~fm and valence pion masses down to
approximately 180~MeV~\cite{Bazavov:2009fk}.  The most recent updates, MILC
10A~\cite{Bazavov:2010yq} and MILC 09A~\cite{Bazavov:2009fk}, include
significantly more data and use two-loop renormalization.
Since these data sets subsume those of their previous calculations, these
latest results are the only ones that must be kept in any world average.

\begin{table}[!ht]
{\footnotesize{
\begin{tabular*}{\textwidth}[l]{l@{\extracolsep{\fill}}rllllllll}
Collaboration & Ref. & \hspace{0.15cm}\begin{rotate}{60}{publication status}\end{rotate}\hspace{-0.15cm} &
 \hspace{0.15cm}\begin{rotate}{60}{chiral extrapolation}\end{rotate}\hspace{-0.15cm} &
 \hspace{0.15cm}\begin{rotate}{60}{continuum  extrapolation}\end{rotate}\hspace{-0.15cm}  &
 \hspace{0.15cm}\begin{rotate}{60}{finite volume}\end{rotate}\hspace{-0.15cm}  &  
 \hspace{0.15cm}\begin{rotate}{60}{renormalization}\end{rotate}\hspace{-0.15cm} &  
 \hspace{0.15cm}\begin{rotate}{60}{running}\end{rotate}\hspace{-0.15cm}  & 
\rule{0.6cm}{0cm}$m_{ud} $ & \rule{0.6cm}{0cm}$m_s $ \\
&&&&&&&&& \\[-0.1cm]
\hline
\hline
&&&&&&&&& \\[-0.1cm]

{RBC/UKQCD 12$^\ominus$}& \cite{Arthur:2012opa} & \gA & \good & \soso & \good &
\good & $\,a$  &  3.37(9)(7)(1)(2) & 92.3(1.9)(0.9)(0.4)(0.8)\\

{PACS-CS 12$^\star$}& \protect{\cite{Aoki:2012st}} & \gA & \good & \bad & \bad & \good & $\,b$
&  3.12(24)(8) &  83.60(0.58)(2.23) \\

{Laiho 11} & \cite{Laiho:2011np} & \rC & \soso & \good & \good & \soso
& $-$ & 3.31(7)(20)(17)
& 94.2(1.4)(3.2)(4.7)\\

{BMW 10A, 10B$^+$} & \cite{Durr:2010vn,Durr:2010aw} & \gA & \good & \good & \good & \good &
$\,c$ & 3.469(47)(48)& 95.5(1.1)(1.5)\\

{PACS-CS 10}& \cite{Aoki:2010wm} & \gA & \good & \bad & \bad & \good & $\,b$
&  2.78(27) &  86.7(2.3) \\

{MILC 10A}& \cite{Bazavov:2010yq} & \rC & \soso  & \good & \good &
\soso  &$-$& 3.19(4)(5)(16)&\rule{0.6cm}{0cm}-- \\

{HPQCD~10$^\ast$}&  \cite{McNeile:2010ji} &\gA & \soso & \good & \good & $-$
&$-$& 3.39(6)$ $ & 92.2(1.3) \\

{RBC/UKQCD 10A}& \cite{Aoki:2010dy} & \gA & \soso & \soso & \good &
\good & $\,a$  &  3.59(13)(14)(8) & 96.2(1.6)(0.2)(2.1)\\

{Blum~10$^\dagger$}&\cite{Blum:2010ym}& \gA & \soso & \bad & \soso & \good &
$-$ &3.44(12)(22)&97.6(2.9)(5.5)\\

{PACS-CS 09}& \cite{Aoki:2009ix}& \gA &\good   &\bad   & \bad & \good  &  $\,b$
 & 2.97(28)(3) &92.75(58)(95)\\

{HPQCD 09A$^\oplus$}&  \cite{Davies:2009ih}&\gA & \soso & \good & \good & $-$
& $-$& 3.40(7) & 92.4(1.5) \\

{MILC 09A} & \cite{Bazavov:2009fk} & \rC &  \soso & \good & \good & \soso &
$-$ & 3.25 (1)(7)(16)(0) & 89.0(0.2)(1.6)(4.5)(0.1)\\

{MILC 09} & \cite{Bazavov:2009bb} & \gA & \soso & \good & \good & \soso & $-$
& 3.2(0)(1)(2)(0) & 88(0)(3)(4)(0)\\

{PACS-CS 08} & \cite{Aoki:2008sm} &  \gA & \good & \bad & \bad  & \bad & $-$ &
2.527(47) & 72.72(78)\\

{RBC/UKQCD 08} & \cite{Allton:2008pn} & \gA & \soso & \bad & \good & \good &
$-$ &$3.72(16)(33)(18)$ & $107.3(4.4)(9.7)(4.9)$\\

\hspace{-0.2cm}{\begin{tabular}{l}CP-PACS/\\JLQCD 07\end{tabular}} 
& \cite{Ishikawa:2007nn}& \gA & \bad & \good & \good  & \bad & $-$ &
$3.55(19)(^{+56}_{-20})$ & $90.1(4.3)(^{+16.7}_{-4.3})$ \\

{HPQCD 05}
& 
\cite{Mason:2005bj}& \gA & \soso & \soso & \soso & \soso &$-$&
$3.2(0)(2)(2)(0)^\ddagger$ & $87(0)(4)(4)(0)^\ddagger$\\

\hspace{-0.2cm}{\begin{tabular}{l}MILC 04, HPQCD/\\MILC/UKQCD 04\end{tabular}} 
& \cite{Aubin:2004fs,Aubin:2004ck} & \gA & \soso & \soso & \soso & \bad & $-$ &
$2.8(0)(1)(3)(0)$ & $76(0)(3)(7)(0)$\\

\hline
\hline\\[-0.7cm]
\end{tabular*}
\begin{tabular*}{\textwidth}[l]{l@{\extracolsep{\fill}}lllllllll}
\multicolumn{10}{l}{\vbox{\begin{flushleft} 
$^\ominus$ The results are given in the $\msbar$ scheme  at 3 instead
      of 2~GeV: $m_{ud}^\msbar(3\,\gev)=3.05(8)(6)(1)(2)\,\mev$,
      $m_s^\msbar(3\,\gev)=83.5(1.7)(0.8)(0.4)(0.7)\,\mev$, where the
      errors are statistical, chiral, finite-volume and from the
      perturbative matching. We run them
      down to 2~GeV using numerically integrated four-loop running
      \cite{vanRitbergen:1997va,Chetyrkin:1999pq} with
      $N_f=3$ and with the values of $\alpha_s(M_Z)$, $m_b$ and $m_c$
      taken from \cite{Beringer:1900zz}. The running factor is
      1.106. At three loops it is only 0.2\% smaller. We therefore neglect
      the small
      uncertainty associated with this conversion.\\
$^\star$ The calculation
      includes e.m. and $m_u\ne m_d$ effects through
      reweighting.\\ 
\hspace{-0.04cm}$^+$\hspace{-0.05cm} The fermion action used is tree-level improved.\\
$^\ast$ What is calculated is
      $m_c(m_c)=1.273(6)$~GeV, using lattice results and perturbation theory. 
      $m_s$ is then obtained by combing
      this result with HPQCD 09A's $m_c/m_s=11.85(16)$ \cite{Davies:2009ih}.
      Finally, $m_{ud}$
      is determined from $m_s$ with the MILC 09 result for
      $m_s/m_{ud}$. Since $m_c/m_s$ is renormalization group invariant
      in QCD, the renormalization and running of the quark masses
      enter indirectly through that of $m_c$, a mass that we do not review here.\\
$^\dagger$ The calculation includes quenched e.m. effects.\\
$^\oplus$ What is calculated is $m_c/m_s=11.85(16)$. $m_s$ is then obtained by combing
      this result with the determination $m_c(m_c) = 1.268(9)$~GeV
      from \cite{Allison:2008xk}. Finally, $m_{ud}$
      is determined from $m_s$ with the MILC 09 result for
      $m_s/m_{ud}$.\\
$^\ddagger$ The bare numbers are those of MILC 04. The masses are simply rescaled, using the
ratio of the two-loop to\\\hspace{0.3cm}one-loop renormalization factors.
\\\rule{0cm}{0.3cm}\hspace{-0.1cm}
$a$ The masses are renormalized nonperturbatively at a scale of 2~GeV 
in a couple of $N_f=3$ RI/SMOM
\\ \hspace{0.3cm}schemes. A careful study of
perturbative matching uncertainties has been performed by comparing
results
\\\hspace{0.3cm}in the two schemes in the region of 2~GeV to 3~GeV
\cite{Aoki:2010dy}.
\\\rule{0cm}{0.3cm}\hspace{-0.1cm}
$b$ The masses are renormalized and run nonperturbatively up to
a scale of $40\,\gev$ in the $N_f=3$ SF\\ \hspace{0.3cm}scheme. In this
scheme, nonperturbative and NLO running for the quark masses are
shown to agree well\\ \hspace{0.3cm}from 40 GeV all the way down to 3
GeV\cite{Aoki:2010wm}.
\\\rule{0cm}{0.3cm}\hspace{-0.1cm}
$c$ The masses are renormalized and run nonperturbatively up to
a scale of 4 GeV in the $N_f=3$ RI/MOM\\\hspace{0.3cm}scheme.  In this
scheme, nonperturbative and N$^3$LO running for the quark masses
are shown to agree
\\\hspace{0.3cm}from 6~GeV down to 3~GeV to better than 1\%
\cite{Durr:2010aw}.
\end{flushleft}}}
\end{tabular*}
}}
\vspace{-0.7cm}
\caption{$\Nf=2+1$ lattice results for the masses $m_{ud}$ and $m_s$ (see Table~\ref{tab:masses2} for notation).}
\label{tab:masses3}
\end{table}

\begin{table}[!ht]
\vspace{1cm}
{\footnotesize{
\begin{tabular*}{\textwidth}[l]{l@{\extracolsep{\fill}}rllllll}
Collaboration & Ref. & $\Nf$ & \hspace{0.15cm}\begin{rotate}{60}{publication status}\end{rotate}\hspace{-0.15cm}  &
 \hspace{0.15cm}\begin{rotate}{60}{chiral extrapolation}\end{rotate}\hspace{-0.15cm} &
 \hspace{0.15cm}\begin{rotate}{60}{continuum  extrapolation}\end{rotate}\hspace{-0.15cm}  &
 \hspace{0.15cm}\begin{rotate}{60}{finite volume}\end{rotate}\hspace{-0.15cm}  & \rule{0.1cm}{0cm} 
$m_s/m_{ud}$ \\
&&&&&& \\[-0.1cm]
\hline
\hline
&&&&&& \\[-0.1cm]
%

{RBC/UKQCD 12$^\ominus$}& \cite{Arthur:2012opa} &2+1  & \gA & \good & \soso & \good  &  27.36(39)(31)(22)\\

{PACS-CS 12$^\star$}& \cite{Aoki:2012st}       &2+1  & \gA & \good & \bad & \bad 
&  26.8(2.0)\\

{Laiho 11} & \cite{Laiho:2011np}              &2+1  & \rC & \soso & \good & \good & 28.4(0.5)(1.3)\\

{BMW 10A, 10B$^+$}& \cite{Durr:2010vn,Durr:2010aw} &2+1  & \gA & \good & \good & \good & 27.53(20)(8) \\

{RBC/UKQCD 10A}& \cite{Aoki:2010dy}           &2+1  & \gA & \soso & \soso & \good & 26.8(0.8)(1.1) \\

{Blum~10$^\dagger$}&\cite{Blum:2010ym}         &2+1  & \gA & \soso & \bad & \soso & 
28.31(0.29)(1.77)\\

{PACS-CS 09}  & \cite{Aoki:2009ix}            &2+1  &  \gA &\good   &\bad   & \bad    &
31.2(2.7)  \\

{MILC 09A}      & \cite{Bazavov:2009fk}       &2+1  & \rC & \soso & \good & \good &  
27.41(5)(22)(0)(4)  \\

{MILC 09}      & \cite{Bazavov:2009bb}        &2+1  & \gA & \soso & \good & \good  &
27.2(1)(3)(0)(0)  \\

{PACS-CS 08}   & \cite{Aoki:2008sm}           &2+1  & \gA & \good & \bad  & \bad    &
28.8(4)\\

{RBC/UKQCD 08} & \cite{Allton:2008pn}         &2+1  & \gA & \soso & \bad  & \good   &
$28.8(0.4)(1.6)$ \\

\hspace{-0.2cm}{\begin{tabular}{l}MILC 04, HPQCD/\\MILC/UKQCD 04\end{tabular}} 
& \cite{Aubin:2004fs,Aubin:2004ck}            &2+1  & \gA & \soso & \soso & \soso    &
27.4(1)(4)(0)(1)  \\
&&&&&& \\[-0.1cm]
\hline
&&&&&& \\[-0.1cm]

{ETM 10B}& \cite{Blossier:2010cr}            &2  & \gA & \soso & \good  & \soso &   27.3(5)(7) \\

{RBC 07}$^\dagger$ & \cite{Blum:2007cy}       &2  & \gA & \bad  & \bad  & \good   &
28.10(38) \\

{ETM 07}       & \cite{Blossier:2007vv}      &2  & \gA & \soso & \bad  & \soso  & 
$27.3(0.3)(1.2)$ \\

{QCDSF/UKQCD 06}
                 & \cite{Gockeler:2006jt}    &2  & \gA &  \bad & \good & \bad   &
27.2(3.2)\\
&&&&&& \\[-0.1cm]
\hline
\hline
\end{tabular*}
\begin{tabular*}{\textwidth}[l]{l@{\extracolsep{\fill}}lllllll}
\multicolumn{7}{l}{\vbox{\begin{flushleft}
$^\ominus$ The errors are statistical, chiral and finite-volume.\\
$^\star$ The calculation includes e.m. and $m_u\ne m_d$ effects through reweighting.\\
$^+$ The fermion action used is tree-level improved.\\
$^\dagger$
The calculation includes quenched e.m. effects.\\
\end{flushleft}}}
\end{tabular*}
}}
\vspace{-0.7cm}
\caption{Lattice results for the ratio $m_s/m_{ud}$.}
\label{tab:ratio_msmud}
\end{table}

Since our last report \cite{Colangelo:2010et} the situation for
$N_f=2+1$ determinations of light quarks has undergone some
evolution. There are new computations by RBC/UKQCD 12
\cite{Arthur:2012opa}, PACS-CS 12 \cite{Aoki:2012st} and Laiho 11
\cite{Laiho:2011np}. Furthermore, the results of BMW 10A, 10B
\cite{Durr:2010vn,Durr:2010aw} have been published and can now be
included in our averages.

The RBC/UKQCD 12 \cite{Arthur:2012opa} computation improves on the one
of RBC/UKQCD 10A \cite{Aoki:2010dy} in a number of ways. In particular
it involves a new simulation performed at a rather coarse lattice spacing 
of 0.144~fm, but with unitary pion masses down to 171(1)~MeV
and valence pion masses down to 143(1)~MeV in a volume of
$(4.6\,\fm)^3$, compared respectively to 290~MeV, 225~MeV and
$(2.7\,\fm)^3$ in RBC/UKQCD 10A. This provides them with a
significantly better control over the extrapolation to physical
$M_\pi$ and to the infinite-volume limit. As before, they perform
nonperturbative renormalization and running in RI/SMOM schemes. The
only weaker point of the calculation comes from the fact that two of
their three lattice spacings are larger than 0.1~fm and correspond to
different discretizations, while the finest
is only 0.085~fm, making it difficult to convincingly claim full
control over the continuum limit. This is mitigated by the fact that
the scaling violations which they observe on their coarsest lattice
are for many quantities small, around 5\%. 

The Laiho 11 results \cite{Laiho:2011np} are based on MILC
staggered ensembles at the lattice spacings 0.15, 0.09 and 0.06~fm, on
which they propagate domain wall quarks. Moreover they work in volumes
of up to $(4.8\,\fm)^3$. These features give them full control over
the continuum and infinite-volume extrapolations. Their lightest RMS
sea pion mass is 280~MeV and their valence pions have masses down to
210~MeV. The fact that their sea pions do not enter deeply into the
chiral regime penalizes somewhat their extrapolation to physical
$M_\pi$. Moreover, to renormalize the quark masses, they use one-loop
perturbation theory for $Z_A/Z_S-1$ which they combine with $Z_A$
determined nonperturbatively from the axial-vector Ward identity. Although
they conservatively estimate the uncertainty associated with the
procedure to be 5\%, which is the size of their largest one-loop
correction, this represents a weaker point of this calculation.

The new PACS-CS 12 \cite{Aoki:2012st} calculation represents an
important extension of the collaboration's earlier 2010 computation
\cite{Aoki:2010wm}, which already probed pion masses down to
$M_\pi\simeq 135\,\mev$, i.e.\ down to the physical mass point. This
was achieved by reweighting the simulations performed in PACS-CS
08~\cite{Aoki:2008sm} at $M_\pi\simeq 160\,\mev$. If adequately
controlled, this procedure eliminates the need to extrapolate to the
physical mass point and, hence, the corresponding systematic
error. The new calculation now applies similar reweighting techniques
to include electromagnetic and $m_u\ne m_d$ isospin-breaking effects
directly at the physical pion mass. It technically adds to
Blum~10~\cite{Blum:2010ym} and BMW's preliminary results of
\cite{Portelli:2010yn,Portelli:2012pn} by including these effects not
only for valence but also for sea-quarks, as is also done in
\cite{Ishikawa:2012ix}. Further, as in PACS-CS 10
\cite{Aoki:2010wm}, renormalization of quark masses is implemented
nonperturbatively, through the Schr\"odinger functional
method~\cite{Luscher:1992an}. As it stands, the main drawback of the
calculation, which makes the inclusion of its results in a world
average of lattice results inappropriate at this stage, is that for
the lightest quark mass the volume is very small, corresponding to
$LM_\pi\simeq 2.0$, a value for which finite-volume effects will be
difficult to control. Another problem is that the calculation was
performed at a single lattice spacing, forbidding a continuum
extrapolation. Further, it is unclear at this point what might be
the systematic errors associated with the reweighting procedure.

As shown by the colour-coding in Tables~\ref{tab:masses3} and
\ref{tab:ratio_msmud}, the BMW 10A, 10B~\cite{Durr:2010vn,Durr:2010aw}
calculation is still the only one to have addressed all sources of
systematic effects while reaching the physical up- and down-quark mass
by {\it interpolation} instead of by extrapolation. Moreover, their
calculation was performed at five lattice spacings ranging from 0.054
to 0.116~fm, with full nonperturbative renormalization and running
and in volumes of up to (6~fm)$^3$ guaranteeing that the continuum
limit, renormalization and infinite-volume extrapolation are
controlled. It does neglect, however, isospin-breaking effects, which
are small on the scale of their error bars.

Finally we come to another calculation which satisfies our selection
criteria, HPQCD~10~\cite{McNeile:2010ji} (which updates HPQCD~09A~\cite{Davies:2009ih}). 
The strange-quark mass is computed using a
precise determination of the charm-quark mass, $m_c(m_c)=1.273(6)$
GeV~\cite{Allison:2008xk,McNeile:2010ji}, whose accuracy is better
than 0.5\%, and a calculation of the quark-mass ratio
$m_c/m_s=11.85(16)$~\cite{Davies:2009ih}, which achieves a precision
slightly above 1\%.  The determination of $m_s$ via the ratio
$m_c/m_s$ displaces the problem of lattice renormalization in the
computation of $m_s$ to one of renormalization in the continuum for
the determination of $m_c$. To calculate $m_{ud}$
HPQCD~10~\cite{McNeile:2010ji} use the MILC 09 determination of the
quark-mass ratio $m_s/m_{ud}$~\cite{Bazavov:2009bb}.

\begin{figure}[t]
\begin{center}
\psfig{file=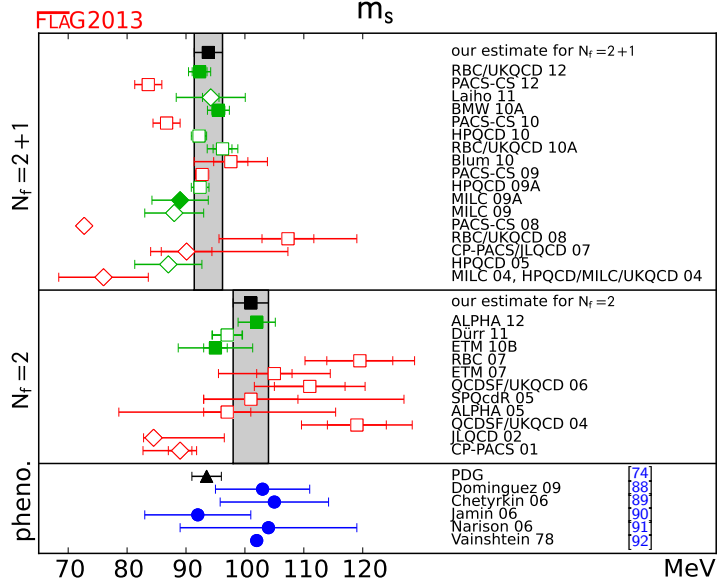,width=11.5cm}
\end{center}
\vspace{-2.47cm}\hspace{11.2cm}\parbox{6cm}{\sffamily\tiny  \cite{Beringer:1900zz}\\

\vspace{-1.4em}\cite{Dominguez:2008jz}\\

\vspace{-1.4em}\cite{Chetyrkin:2005kn}\\

\vspace{-1.4em}\cite{Jamin:2006tj}\\

\vspace{-1.4em}\cite{Narison:2005ny}\\

\vspace{-1.4em}\cite{Vainshtein:1978nn}}

\vspace{0.7cm}
\begin{center}
\caption{ \label{fig:ms}Mass of the strange quark ($\msbar$ scheme ,
  running scale 2 GeV). The central and top panels show the lattice results 
  listed in tables \ref{tab:masses2} and \ref{tab:masses3}. For comparison, the bottom panel collects a few sum rule results and also indicates the current PDG estimate. 
  Diamonds represent results based on perturbative
  renormalization, while squares indicate that, in the relation
  between the lattice regularized and renormalized $\msbar$ masses,
  nonperturbative effects are accounted for. 
The black squares and the grey bands represent our estimates (\ref{eq:quark masses Nf=2}) and
  (\ref{eq:nf3msmud}). The significance of the colours is explained in section \ref{sec:qualcrit}.
}\end{center}

\end{figure}

\begin{figure}[t]

\begin{center}
\psfig{file=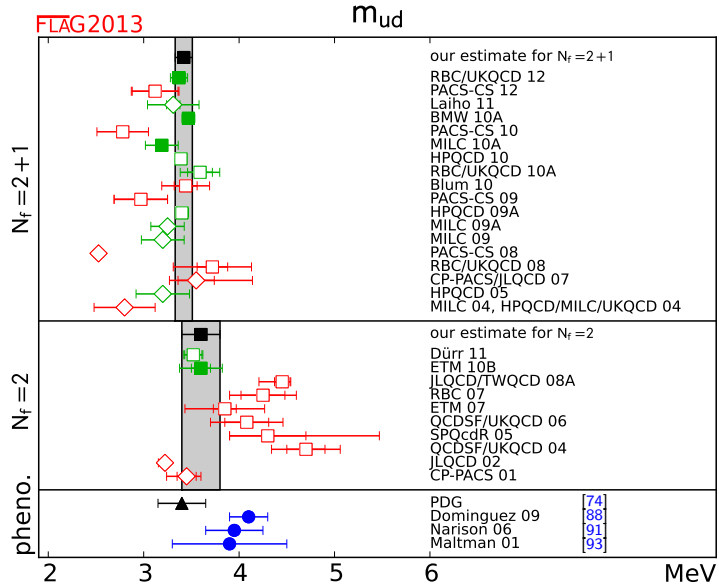,width=11.5cm}
\end{center}
\vspace{-2.16cm}\hspace{10.6cm}\parbox{6cm}{\sffamily\tiny  \cite{Beringer:1900zz}\\

\vspace{-1.35em}\cite{Dominguez:2008jz}\\

\vspace{-1.35em}\cite{Narison:2005ny}\\

\vspace{-1.35em}\cite{Maltman:2001nx}}

\vspace{0.5cm}
\begin{center}
\caption{ \label{fig:mud} Mean mass of the two lightest quarks,
 $m_{ud}=\frac{1}{2}(m_u+m_d)$ (for details see Fig.~\ref{fig:ms}).}\end{center}

\end{figure}

The high precision quoted by HPQCD~10 on the strange-quark mass relies
in large part on the precision reached in the determination of the
charm-quark mass~\cite{Allison:2008xk,McNeile:2010ji}.  This
calculation uses an approach based on the lattice determination of
moments of charm-quark pseudoscalar, vector and axial-vector
correlators. These moments are then combined with four-loop results
from continuum perturbation theory to obtain a determination of the
charm-quark mass in the $\msbar$ scheme . In the preferred case, in
which pseudoscalar correlators are used for the analysis, there are no
lattice renormalization factors required, since the corresponding
axial-vector current is partially conserved in the staggered lattice formalism. 

Instead of combining the result for $m_c/m_s$ of \cite{Davies:2009ih}
with $m_c$ from \cite{McNeile:2010ji}, one can use it with the PDG
\cite{Beringer:1900zz} average $m_c(m_c)=1.275(25)\,\gev$, whose error
is four times as large as the one obtained by HPQCD~10. If one does
so, one obtains $m_s=92.3(2.2)$ in lieu of the value $m_s=92.2(1.3)$
given in Table \ref{tab:masses3}, thereby nearly doubling HPQCD~10's
error. Though we plan to do so in the future, we have not yet
performed a review of lattice determinations of $m_c$. 
Thus, as for
the results of D\"urr 11 \cite{Durr:2011ed} in the $N_f=2$ case, we
postpone its inclusion in our final averages until we have performed
an independent analysis of $m_c$, emphasizing that this novel
strategy for computing the light-quark masses may very well turn out
to be the best way to determine them.

This discussion leaves us with three results for our final average for
$m_s$, those of MILC 09A~\cite{Bazavov:2009fk}, BMW 10A,
10B~\cite{Durr:2010vn,Durr:2010aw} and RBC/UKQCD
12~\cite{Arthur:2012opa}, and the result of
HPQCD~10~\cite{McNeile:2010ji} as an important cross-check. Thus, we
first check that the three other results which will enter our final
average are consistent with HPQCD~10's result. To do this we implement
the averaging procedure described in Sect. \ref{sec:averages} on all
four results. This yields $m_s=93.0(1.0)\,\mev$ with a $\chi^2/dof =
3.0/3=1.0$, indicating overall consistency. Note that in making this
average, we have accounted for correlations in the small statistical
errors of HPQCD~10 and MILC 09A. Omitting HPQCD~10 in our final
average results in an increase by 50\% of the average's uncertainty
and by 0.8~$\sigma$ of its central value. Thus, we obtain
$m_s=93.8(1.5)\,\mev$ with a $\chi^2/dof = 2.26/2=1.13$. 
When repeating the exercise for $m_{ud}$, we replace MILC
09A by the more recent analysis reported in MILC
10A~\cite{Bazavov:2010yq}. A fit of all four results yields
$m_{ud}=3.41(5)\,\mev$ with a $\chi^2/dof = 2.6/3=0.9$ and including
only the same three as above gives $m_{ud}=3.42(6)\,\mev$ with a
$\chi^2/dof = 2.4/2=1.2$. Here the results are barely distinguishable,
indicating full compatibility of all four results. Note that the
outcome of the averaging procedure amounts to a determination of
$m_s$ and $m_{ud}$ of 1.6\%. and 1.8\%, respectively.

The heavy sea-quarks affect the determination of the light-quark
masses only through contributions of order $1/m_c^2$, which moreover
are suppressed by the Okubo-Zweig-Iizuka-rule. We expect these
contributions to be small. However, note that the effect of omitted sea 
quarks on a given quantity is not uniquely defined: the size of the effect 
depends on how the theories with and without these flavours are 
matched. One way to set conventions is to ensure that the bare 
parameters common to both theories are fixed by the same physical 
observables and that the renormalizations are performed in the same 
scheme and at the same scale, with the appropriate numbers of flavours. 

An upper bound on the heavy-quark contributions can be obtained by 
looking at the presumably much larger effect associated with omitting the 
strange quark in the sea. Within errors, the average value 
$m_{ud} = 3.42(6)$ MeV obtained above from the data with $N_f = 2+1$ 
agrees with the result $m_{ud} = 3.6(2)$ MeV for $N_f = 2$ quoted in 
(\ref{eq:quark masses Nf=2}): assuming that the underlying calculations 
more or less follow the above matching prescription, the effects generated 
by the quenching of the strange quark in $m_{ud}$ are within the noise. 
Interpreting the two results as Gaussian distributions, the probability 
distribution of the difference $\Delta m_{ud} \equiv (m_{ud}|_{N_f=2})-
(m_{ud}|_{N_f=3})$ is also Gaussian, with $\Delta m_{ud}=0.18(21)$ MeV. 
The corresponding root-mean-square 
$\langle\Delta m_{ud}^2\rangle^\frac{1}{2}= 0.28$ MeV provides an upper 
bound for the size of the effects due to strange quark quenching; it amounts 
to 8 \% of $m_{ud}$. In the case of $m_s$, the analogous calculation yields 
$\langle\Delta m_s^2\rangle^\frac{1}{2}=7.9$ MeV and thus also amounts 
to an upper bound of about 8 \%. Taking any of these
numbers as an upper bound on the omission of charm effects in the
$N_f=2+1$ results is, we believe, a significant overestimate.

An underestimate of the upper bound on the sea-charm contributions to
$m_s$ can be obtained by transposing, to the $s\bar s$ system, the
perturbative, heavy quarkonium arguments put forward in
\cite{Davies:2010ip} to determine the effect of sea charm on the
$\eta_c$ and $J/\psi$ masses. An estimate using constituent quark
masses \cite{Davies:pc2014} leads very roughly to a 0.05\% effect on
$m_s$, from which \cite{Davies:pc2014} concludes that the error on
$m_s$ and $m_{ud}$ due to the omission of charm is of order 0.1\%.

One could also try to estimate the effect by analysing the relation
between the parameters of QCD$_3$ and those of full QCD in
perturbation theory. The $\beta$- and $\gamma$-functions, which
control the renormalization of the coupling constants and quark
masses, respectively, are known to four loops
\cite{vanRitbergen:1997va,Chetyrkin:1997sg,Chetyrkin:1999pq,Bethke:2009jm}. The
precision achieved in this framework for the decoupling of the $t$-
and $b$-quarks is excellent, but the $c$-quark is not heavy enough: at
the percent level, we believe that the corrections of order $1/m_c^2$ cannot be
neglected and the decoupling formulae of perturbation theory do not
provide a reliable evaluation, because the scale $m_c(m_c)\simeq
1.28\,\gev$ is too low for these formulae to be taken at face
value. Consequently, the accuracy to which it is possible to identify
the running masses of the light quarks of full QCD in terms of those
occurring in QCD$_3$ is limited. For this reason, it is preferable to
characterize the masses $m_u$, $m_d$, $m_s$ in terms of QCD$_4$, where
the connection with full QCD is under good control.

The role of the $c$-quarks in the determination of the
light-quark masses will soon be studied in detail -- some simulations
with 2+1+1 dynamical quarks have already been carried
out~\cite{Baron:2010bv,Bazavov:2011fh}. For the moment, we choose to
consider a crude, and hopefully reasonably conservative, upper bound
on the size of the effects due to the neglected heavy quarks that can
be established within the $\Nf=2+1$ simulations themselves, without
invoking perturbation theory. In \cite{Durr:2008zz} it is found that
when the scale is set by $M_\Xi$, the result for $M_\Lambda$ agrees
well with experiment within the 2.3\% accuracy of the
calculation. Because of the very strong correlations between the
statistical and systematic errors of these two masses, we expect the
uncertainty in the difference $M_\Xi-M_\Lambda$ to also be of order
2\%. To leading order in the chiral expansion this mass difference is
proportional to $m_s-m_{ud}$.  Barring accidental cancellations, we
conclude that the agreement of $\Nf = 2+1$ calculations with
experiment suggests an upper bound on the sensitivity of $m_s$ to
heavy sea-quarks of order 2\%.

Taking this uncertainty into account yields the following averages:
\bea\label{eq:nf3msmud}
\Nf=2+1 :\hspace{1cm} m_{ud}\al=\al 3.42(6)(7)\;\mev 
\co\hspace{1cm}m_s=93.8(1.5)(1.9)\;\mev\, \eea
where the first error comes from the averaging of the lattice results,
and the second is the one that we add to account for the neglect of
sea effects from the charm and more massive quarks. This corresponds
to determinations of $m_{ud}$ and $m_s$ with a precision of and 2.6\%
and 2.7\%, respectively.  These estimates represent the conclusions we
draw from the information gathered on the lattice until now. They are
shown as vertical bands in Figures \ref{fig:ms} and \ref{fig:mud},
together with the $\Nf=2$ results (\ref{eq:quark masses Nf=2}).

In the ratio $m_s/m_{ud}$, one of the sources of systematic error -- the
uncertainties in the renormalization factors -- drops out. Also, we can
compare the lattice results with the leading-order formula of {\Ch}PT, 
\be\label{eq:LO1}\frac{m_s}{m_{ud}}\Lo\frac{\hat{M}_{K^+}^2+
\hat{M}_{K^0}^2-\hat{M}_{\pi^+}^2}{\hat{M}_{\pi^+}^2}\co\ee
which relates the quantity $m_s/m_{ud}$ to a ratio of meson masses in QCD.
Expressing these in terms of the physical masses and the four coefficients
introduced in (\ref{eq:epsilon1})-(\ref{eq:epsilon3}), linearizing the
result with respect to the corrections and inserting the observed mass
values, we obtain 
\be\label{eq:LO1 num} \frac{m_s}{m_{ud}} \Lo 25.9 - 0.1\,
\epsilonD + 1.9\, \epsilon_{\pi^0} - 0.1\, \epsilon_{K^0} -1.8
\,\epsilon_m\fs\ee 
If the coefficients $\epsilonD$, $\epsilon_{\pi^0}$, $\epsilon_{K^0}$
and $\epsilon_m$ are set equal to zero, the right hand side reduces to
the value $m_s/m_{ud}=25.9$ that follows from Weinberg's leading-order
formulae for $m_u/m_d$ and $m_s/m_d$ \cite{Weinberg:1977hb}, in
accordance with the fact that these do account for the
e.m.\ interaction at leading chiral order, and neglect the mass
  difference between the charged and neutral pions in QCD.  Inserting
  the estimates (\ref{eq:epsilon num}) gives the effect of chiral
  corrections to the e.m.\ self-energies and of the mass difference
  between the charged and neutral pions in QCD. With these, the LO
  prediction in QCD becomes:
\be\label{eq:LO
  ms/mud}\frac{m_s}{m_{ud}}\Lo 25.9(1)\fs\ee
The quoted uncertainty does not include an estimate for the
higher-order contributions, but only accounts for the error bars in
the coefficients, which is dominated by the one in the estimate given
for $\epsilon_{\pi^0}$. The fact that the central value remains
  unchanged indicates that chiral corrections to the
  e.m.\ self-energies and mass-difference corrections are small in this
  particular quantity. However, given the high accuracy reached in
  lattice determinations of the ratio $m_s/m_{ud}$, the uncertainties
  associated with e.m.\ corrections are no longer completely
  irrelevant. This is seen by comparing the 0.1 in (\ref{eq:LO
    ms/mud}) with the 0.15 in (\ref{eq:msovmud}). Nevertheless, this
  uncertainty is still smaller than our $\sim 1.\div 1.5\%$ upper bound
  on possible $1/m_c^2$ corrections.

\begin{figure}[t]

\begin{center}
\psfig{file=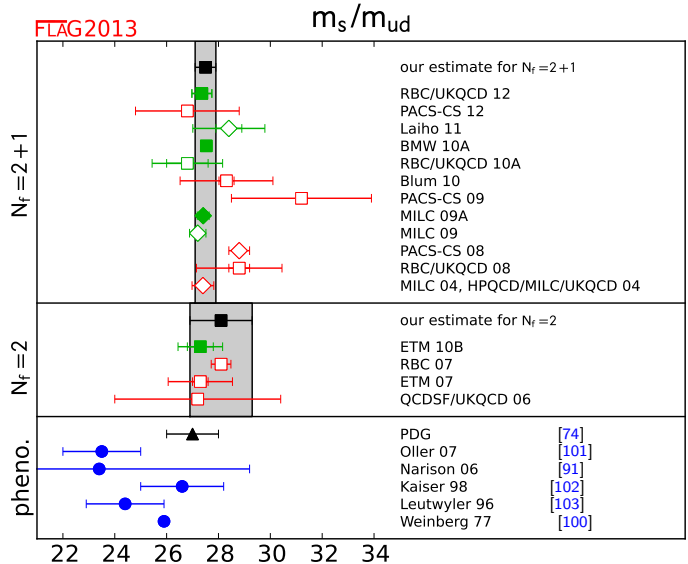,width=11cm}
\end{center}
\vspace{-2.77cm}\hspace{10.5cm}\parbox{6cm}{\sffamily\tiny  \cite{Beringer:1900zz}\\

\vspace{-1.15em}\cite{Oller:2006xb}\\

\vspace{-1.15em}\cite{Narison:2005ny}\\

\vspace{-1.15em}\hspace{-0.48em}\cite{Kaiser}\\

\vspace{-1.15em}\hspace{-0.48em}\cite{Leutwyler:1996qg}\\

\vspace{-1.15em}\cite{Weinberg:1977hb}}

\vspace{1.0cm}
\begin{center}
\caption{ \label{fig:msovmud}Results for the ratio
  $m_s/m_{ud}$. The upper part indicates the lattice results listed in Table
  \ref{tab:ratio_msmud}.  
The lower part shows results obtained from
  $\chi$PT and sum rules, together with the current PDG estimate. }
\end{center}

\end{figure}

The lattice results in Table \ref{tab:ratio_msmud}, which satisfy our
selection criteria, indicate that the corrections generated by the
nonleading terms of the chiral perturbation series are remarkably
small, in the range 3--10\%.  Despite the fact that the SU(3)-flavour-symmetry 
breaking effects in the Nambu-Goldstone boson masses are very
large ($M_K^2\simeq 13\, M_\pi^2$), the mass spectrum of the
pseudoscalar octet obeys the SU(3)$\times$SU(3) formula (\ref{eq:LO1})
very well.

Our average for $m_s/m_{ud}$ is based on the results of MILC 09A, BMW
10A, 10B and RBC/UK\-QCD 12 -- the value quoted by HPQCD~10 does not
represent independent information as it relies on the result for
$m_s/m_{ud}$ obtained by the MILC collaboration. Averaging these
results according to the precription of Section~\ref{sec:error_analysis} gives
$m_s/m_{ud}=27.46(15)$ with $\chi^2/dof=0.2/2$. The fit is dominated
by MILC 09A and BMW 10A, 10B. Since the errors associated with
renormalization drop out in the ratio, the uncertainties are even
smaller than in the case of the quark masses themselves: the above
number for $m_s/m_{ud}$ amounts to an accuracy of 0.5\%.

At this level of precision, the uncertainties in the electromagnetic
and strong isospin-breaking corrections are not completely
negligible. The error estimate in the LO result (\ref{eq:LO ms/mud})
indicates the expected order of magnitude. The uncertainties in $m_s$
and $m_{ud}$ associated with the heavy sea-quarks cancel at least
partly. In view of this, we ascribe a total 1.5\% uncertainty to these
two sources of error. Thus, we are convinced that our final estimate,
\be\label{eq:msovmud} \mbox{$N_f=2+1$ :}
\hspace{1cm}\frac{m_s}{m_{ud}}=27.46(15)(41)\co \ee
is on the conservative side, with a total 1.5\% uncertainty. It is
also fully consistent with the ratio computed from our individual
quark masses in (\ref{eq:nf3msmud}), $m_s/m_{ud}=27.6(6)$, which has a
larger 2.2\% uncertainty. In (\ref{eq:msovmud}) the first
error comes from the averaging of the lattice results, and the second
is the one that we add to account for the neglect of isospin-breaking
and heavy sea-quark effects.

The lattice results show that the LO prediction of {\Ch}PT in
(\ref{eq:LO ms/mud}) receives only small corrections from higher
orders of the chiral expansion: according to (\ref{eq:msovmud}), these
generate a shift of $5.7\pm 1.5 \%$. Our estimate does therefore not
represent a very sharp determination of the higher-order
contributions. 

The ratio $m_s/m_{ud}$ can also be extracted from the masses of the neutral
Nambu-Goldstone bosons: neglecting effects of order $(m_u-m_d)^2$ also
here, the leading-order formula reads 
$m_s/m_{ud}\Lo\frac{3}{2}\hat{M}_\eta^2/\hat{M}_\pi^2-\frac{1}{2}$.
Numerically, this gives $m_s/m_{ud}\Lo 24.2$. The relation has the advantage
that the e.m.\ corrections are expected to be much smaller here, but it is
more difficult to calculate the $\eta$-mass on the lattice. The comparison with
(\ref{eq:msovmud}) shows that, in this case, the contributions of NLO are
somewhat larger: $14\pm 2$\%.

\subsection{Lattice determination of $m_u$ and $m_d$}\label{subsec:mumd}

The determination of $m_u$ and $m_d$ separately requires additional
input.  MILC 09A \cite{Bazavov:2009fk} uses the mass difference
between $K^0$ and $K^+$, from which they subtract electromagnetic
effects using Dashen's theorem with corrections, as discussed in
Section \ref{subsec:electromagnetic interaction}.  The up- and down- 
sea-quarks remain degenerate in their calculation, fixed to the value of
$m_{ud}$ obtained from $M_{\pi^0}$.

To determine $m_u/m_d$, BMW 10A, 10B~\cite{Durr:2010vn,Durr:2010aw}
follow a slightly different strategy. They obtain this ratio from
their result for $m_s/m_{ud}$ combined with a phenomenological
determination of the isospin-breaking quark-mass ratio $Q=22.3(8)$,
defined below in (\ref{eq:Qm}), from $\eta\to3\pi$
decays~\cite{Leutwyler:2009jg} (the decay $\eta\to3\pi$ is very
sensitive to QCD isospin-breaking but fairly insensitive to QED
isospin-breaking).  As discussed in Section \ref{sec:RandQ}, the
central value of the e.m.~parameter $\epsilon$ in (\ref{eq:epsilon
  num}) is taken from the same source.

RM123 11~\cite{deDivitiis:2011eh} actually uses the e.m.~parameter
$\epsilon=0.7(5)$ from the first edition of  the FLAG review \cite{Colangelo:2010et}. However they estimate the effects of
strong isospin-breaking at first nontrivial order, by inserting the
operator $\frac12(m_u-m_d)\int(\bar uu-\bar dd)$ into correlation
functions, while performing the gauge averages in the isospin
limit. Applying these techniques, they obtain $(\hat M_{K^0}^2-\hat
M_{K^+}^2)/(m_d-m_u)=2.57(8)\,\mev$. Combining this result with the
phenomenological $(\hat M_{K^0}^2-\hat M_{K^+}^2)=6.05(63)\times10^3$
determined with the above value of $\epsilon$, they get
$(m_d-m_u)=2.35(8)(24)\,\mev$, where the first error corresponds to the
lattice statistical and systematic uncertainties combined in
quadrature, while the second arises from the uncertainty on
$\epsilon$. Note that below we quote results from RM123 11 for $m_u$,
$m_d$ and $m_u/m_d$. As described in Table~\ref{tab:mu_md_grading}, we
obtain them by combining RM123 11's result for $(m_d-m_u)$ with ETM 10B's
result for $m_{ud}$.

\newpage
\begin{table}[h]
\vspace{1cm}
{\footnotesize{
\begin{tabular*}{\textwidth}[l]{l@{\extracolsep{\fill}}rlllllllll}
Collaboration \al  Ref. \al \hspace{0.15cm}\begin{rotate}{60}{publication status}\end{rotate}\hspace{-0.15cm}  \al 
\hspace{0.15cm}\begin{rotate}{60}{chiral extrapolation}\end{rotate}\hspace{-0.15cm} \al 
\hspace{0.15cm}\begin{rotate}{60}{continuum  extrapolation}\end{rotate}\hspace{-0.15cm}  \al 
\hspace{0.15cm}\begin{rotate}{60}{finite volume}\end{rotate}\hspace{-0.15cm}  \al   
\hspace{0.15cm}\begin{rotate}{60}{renormalization}\end{rotate}\hspace{-0.15cm} \al   
\hspace{0.15cm}\begin{rotate}{60}{running}\end{rotate}\hspace{-0.15cm}  \al  
\rule{0.6cm}{0cm}$m_u$\al 
\rule{0.6cm}{0cm}$m_d$ \al \rule{0.3cm}{0cm} $m_u/m_d$\\
\al \al \al \al \al \al \al \al \al \al  \\[-0.1cm]
\hline
\hline
\al \al \al \al \al \al \al \al \al \al  \\[-0.1cm]
{PACS-CS 12$^\star$} \al \protect{\cite{Aoki:2012st}} \al \gA \al \good \al \bad \al \bad \al \good \al $\,a$
\al  2.57(26)(7) \al  3.68(29)(10) \al 0.698(51)\\

{Laiho 11} \al \cite{Laiho:2011np} \al \rC \al \soso \al \good \al
\good \al \soso \al $-$ \al 1.90(8)(21)(10) \al
4.73(9)(27)(24) \al 0.401(13)(45)\\

{HPQCD~10$^\ddagger$}\al \cite{McNeile:2010ji} \al \gA \al \soso \al \good \al \good \al \good \al
$-$ \al 2.01(14) \al 4.77(15) \al  \\

{BMW 10A, 10B$^+$}\al \cite{Durr:2010vn,Durr:2010aw} \al \gA \al \good \al \good \al \good \al \good \al
$\,b$ \al 2.15(03)(10) \al 4.79(07)(12) \al 0.448(06)(29) \\

{Blum~10$^\dagger$}\al\cite{Blum:2010ym}\al \gA \al \soso \al \bad \al \soso \al \good \al $-$ \al 2.24(10)(34)\al 4.65(15)(32)\al 0.4818(96)(860)\\

{MILC 09A} \al  \cite{Bazavov:2009fk} \al  \rC \al  \soso \al \good \al \good \al \soso \al $-$
\al 1.96(0)(6)(10)(12)
\al  4.53(1)(8)(23)(12)  \al   0.432(1)(9)(0)(39) \\

{MILC 09} \al  \cite{Bazavov:2009bb} \al  \gA \al  \soso \al  \good \al  \good \al  \soso \al 
$-$\al  1.9(0)(1)(1)(1)
\al  4.6(0)(2)(2)(1) \al  0.42(0)(1)(0)(4) \\

\hspace{-0.2cm}{\begin{tabular}{l}MILC 04, HPQCD/\rule{0.1cm}{0cm}\\MILC/UKQCD 04\end{tabular}} \al  \cite{Aubin:2004fs,Aubin:2004ck} \al  \gA \al  \soso \al  \soso \al  \soso \al 
\bad \al$-$\al  1.7(0)(1)(2)(2)
\al  3.9(0)(1)(4)(2)  \al  0.43(0)(1)(0)(8) \\

\al \al \al \al \al \al \al \al \al \al  \\[-0.1cm]
\hline
\al \al \al \al \al \al \al \al \al \al  \\[-0.1cm]

{RM123 13} \al \cite{deDivitiis:2013xla}  \al \gA \al \soso \al \good \al \soso \al \good \al $\,c$
\al 2.40(15)(17) \al  4.80 (15)(17) \al 0.50(2)(3)\\

{RM123 11$^\oplus$} & \cite{deDivitiis:2011eh} \al \gA \al \soso \al \good \al \soso \al \good \al $\,c$
\al { {\em 2.43(11)(23)}} \al { {\em 4.78(11)(23)}} \al { {\em 0.51(2)(4)}}\\

{D\"urr 11$^\ast$}\al \cite{Durr:2011ed} \al \gA \al \soso \al \good \al \soso \al $-$ \al $-$
 \al 2.18(6)(11) \al 4.87(14)(16) \al  \\ 

{RBC 07$^\dagger$} \al  \cite{Blum:2007cy} \al  \gA \al  \bad \al  \bad \al  \good  \al  \good \al  $-$
\al  3.02(27)(19) \al  5.49(20)(34)  \al  0.550(31)\\

\al \al \al \al \al \al \al \al \al \al  \\[-0.1cm]
\hline
\hline\\[-0.7cm]
\end{tabular*}
\begin{tabular*}{\textwidth}[l]{l@{\extracolsep{\fill}}lllllll}
\multicolumn{7}{l}{\vbox{\begin{flushleft}
$^\star$ The calculation includes e.m. and $m_u\ne m_d$ effects
      through reweighting.\\
$^\ddagger$ Values obtained by combining the HPQCD 10 result for $m_s$ with the MILC 09 results for $m_s/m_{ud}$ and\\\hspace{0.3cm}$m_u/m_d$.\\
\hspace{-0.04cm}$^+$\hspace{-0.05cm} The fermion action used is tree-level improved.\\
$^\ast$ Values obtained by combining the D\"urr 11 result for $m_s$
with the BMW 10A, 10B results for $m_s/m_{ud}$ and $m_u/m_d$.\\
{ $^\oplus$ $m_u$, $m_d$ and $m_u/m_d$ are obtained by combining the
result of RM123 11 for $(m_d-m_u)$ \cite{deDivitiis:2011eh} with 
$m_{ud}=3.6(2)\,\mev$ from ETM 10B. $(m_d-m_u)=2.35(8)(24)\,\mev$ in \cite{deDivitiis:2011eh} was
obtained assuming $\epsilonD = 0.7(5)$~\cite{Colangelo:2010et}
and $\epsilon_m=\epsilon_{\pi^0}=\epsilon_{K^0}=0$. In the quoted results, the first
error corresponds to the
lattice statistical and systematic errors combined in
quadrature, while the second arises from the uncertainties associated with
$\epsilonD$.}\\
$^\dagger$
The calculation includes quenched e.m. effects.\\\rule{0cm}{0.3cm}\hspace{-0.1cm}
$a$ The masses are renormalized and run nonperturbatively up to
a scale of $100\,\gev$ in the $N_f=2$ SF\\\hspace{0.3cm}scheme. In this
scheme, nonperturbative and NLO running for the quark masses are
shown to agree\\\hspace{0.3cm}well from 100 GeV all the way down to 2
GeV \cite{DellaMorte:2005kg}.\\\rule{0cm}{0.3cm}\hspace{-0.1cm}
$b$ The masses are renormalized and run nonperturbatively up to
a scale of 4 GeV in the $N_f=3$ RI/MOM\\\hspace{0.3cm}scheme.  In this
scheme, nonperturbative and N$^3$LO running for the quark masses
are shown to agree\\\hspace{0.3cm}from 6~GeV down to 3~GeV to better than 1\%
\cite{Durr:2010aw}. \\\rule{0cm}{0.3cm}\hspace{-0.1cm}
$c$ The masses are renormalized nonperturbatively at scales $1/a\sim 2\div3\,\gev$ in the $N_f=2$ RI/MOM \\\hspace{0.3cm}scheme.  In this
scheme, nonperturbative and N$^3$LO running for the quark masses
are shown to agree\\\hspace{0.3cm}from 4~GeV down 2 GeV to better than 3\%
\cite{Constantinou:2010gr}.\\\rule{0cm}{0.3cm}\hspace{-0.1cm}\end{flushleft}}}
\end{tabular*}
}}
\vspace{-0.7cm}
\caption{Lattice results for $m_u$, $m_d$ (MeV) and for the ratio $m_u/m_d$. The values refer to the 
$\msbar$ scheme  at scale 2 GeV.  The upper part of the table lists results obtained with $\Nf=2+1$, 
while the lower part presents calculations with $N_f = 2$.}
\label{tab:mu_md_grading}
\end{table}

Instead of subtracting electromagnetic effects using phenomenology,
RBC~07~\cite{Blum:2007cy} and Blum~10~\cite{Blum:2010ym} actually
include a quenched electromagnetic field in their calculation. This
means that their results include corrections to Dashen's theorem,
albeit only in the presence of quenched electromagnetism. Since the up-
and down-quarks in the sea are treated as degenerate, very small
isospin corrections are neglected, as in MILC's calculation.

PACS-CS 12 \cite{Aoki:2012st} takes the inclusion of isospin-breaking
effects one step further. Using reweighting techniques, it also
includes electromagnetic and $m_u-m_d$ effects in the sea.

Lattice results for $m_u$, $m_d$ and $m_u/m_d$ are summarized in
Table~\ref{tab:mu_md_grading}. In order to discuss them, we consider
the LO formula
\be\label{eq:LO2}\frac{m_u}{m_d}\Lo\frac{\hat{M}_{K^+}^2-\hat{M}_{K^0}^2+\hat{M}_{\pi^+}^2}
{\hat{M}_{K^0}^2-\hat{M}_{K^+}^2+\hat{M}_{\pi^+}^2} \fs\ee
Using equations (\ref{eq:epsilon1})--(\ref{eq:epsilon3}) to express
the meson masses in QCD in terms of the physical ones and linearizing
in the corrections, this relation takes the form
\be\label{eq:LO2 num}\frac{m_u}{m_d}\Lo 0.558 - 0.084\, \epsilonD - 0.02\,
\epsilon_{\pi^0} + 0.11\, \epsilon_m \fs\ee
Inserting the estimates
(\ref{eq:epsilon num}) and adding errors in quadrature, the LO
prediction becomes
{ 
\be\label{eq:mu/md LO}\frac{m_u}{m_d}\Lo
0.50(3)\fs\ee 
}
Again, the quoted error exclusively accounts for the errors attached
to the estimates (\ref{eq:epsilon num}) for the epsilons --
contributions of nonleading order are ignored. The uncertainty in the
leading-order prediction is dominated by the one in the coefficient
$\epsilonD$, which specifies the difference between the meson
squared-mass splittings generated by the e.m. interaction in the kaon
and pion multiplets. { The reduction in the error on this coefficient
since the previous review \cite{Colangelo:2010et} results in a
reduction of a factor of a little less than 2 in the uncertainty on
the LO value of $m_u/m_d$ given in (\ref{eq:mu/md LO}).}

It is interesting to compare the assumptions made or results obtained
by the different collaborations for the violation of Dashen's
theorem. The input used in MILC 09A is $\epsilonD=1.2(5)$
\cite{Bazavov:2009fk}, 
while the $N_f=2$ computation of RM123 13 finds
$\epsilonD=0.79(18)(18)$~\cite{deDivitiis:2013xla}.  As discussed in
Section \ref{sec:RandQ}, the value of $Q$ used by BMW 10A,
10B~\cite{Durr:2010vn,Durr:2010aw} gives $\epsilonD=0.70(28)$ at NLO
(see (\ref{eq:epsilon eta})). On the other hand, RBC 07
\cite{Blum:2007cy} and Blum~10~\cite{Blum:2010ym} obtain the results
$\epsilonD=0.13(4)$ and $\epsilonD=0.5(1)$. Note that PACS-CS 12
\cite{Aoki:2012st} do not provide results which allow us to determine
$\epsilon$ directly. However, using their result for $m_u/m_d$,
together with (\protect{\ref{eq:LO2 num}}), and neglecting NLO terms,
one finds $\epsilonD=-1.6(6)$, which is difficult to reconcile with
what is known from phenomenology (see Sections
\ref{subsec:electromagnetic interaction} and \ref{sec:RandQ}).  Since
the values assumed or obtained for $\epsilonD$ differ, it does not
come as a surprise that the determinations of $m_u/m_d$ are different.

These values of $\epsilonD$ are also interesting because they allow us
to estimate the chiral corrections to the LO prediction (\ref{eq:mu/md
  LO}) for $m_u/m_d$. Indeed, evaluating the relation (\ref{eq:LO2
  num}) for the values of $\epsilonD$ given above, and neglecting all
other corrections in this equation, yields the LO values
$(m_u/m_d)^\mathrm{LO}=0.46(4)$, 0.547(3), 0.52(1), 0.50(2), 0.49(2)
for MILC 09A, RBC 07, Blum 10, BMW 10A, 10B and RM123 13,
respectively. However, in comparing these numbers to the
nonperturbative results of Table \ref{tab:mu_md_grading} one must be
careful not to double count the uncertainty arising from
$\epsilonD$. One way to obtain a sharp comparison is to consider the
ratio of the results of Table \ref{tab:mu_md_grading} to the LO values $(m_u/m_d)^{\rm LO}$,
in which the uncertainty from $\epsilon$ cancels to good
accuracy. Here we will assume for simplicity that they cancel
completely and will drop all uncertainties related to $\epsilon$. For
$N_f = 2$ we consider RM123 13 \cite{deDivitiis:2013xla}, which
updates RM123 11 and has no red dots.  Since the uncertainties common
to $\epsilon$ and $m_u/m_d$ are not explicitly given in
\cite{deDivitiis:2013xla}, we have to estimate them. For that we use
the leading-order result for $m_u/m_d$, computed with RM123 13's value
for $\epsilon$. Its error bar is the contribution of the uncertainty
on $\epsilon$ to $(m_u/m_d)^{\rm LO}$. To good approximation this
contribution will be the same for the value of $m_u/m_d$ computed in
\cite{deDivitiis:2013xla}. Thus, we subtract it in quadrature from
RM123 13's result in Table~\ref{tab:mu_md_grading} and compute
$(m_u/m_d)/(m_u/m_d)^{\rm LO}$, dropping uncertainties related to
$\epsilon$. We find $(m_u/m_d)/(m_u/m_d)^{\rm LO} = 1.02(6)$. This
result suggests that chiral corrections in the case of $\Nf=2$ are
negligible. For the two most accurate $\Nf=2+1$ calculations, those of
MILC 09A and BMW 10A, 10B, this ratio of ratios is 0.94(2) and
0.90(1), respectively. { Though these two numbers are not fully
  consistent within our rough estimate of the errors, they indicate
  that higher-order corrections to (\ref{eq:mu/md LO}) are negative
  and about 8\% when $\Nf=2+1$. In the following, we will take them to
  be -8(4)\%. The fact that these corrections are seemingly larger and
  of opposite sign than in the $\Nf=2$ case is not understood at this
  point. It could be an effect associated with the quenching of the
  strange quark. It could also be due to the fact that the RM123 13
  calculation does not probe deeply enough into the chiral regime--it
  has $M_\pi\gsim 270\,\mev$--to pick up on important chiral
  corrections. Of course, being less than a two standard deviation
  effect, it may be that there is no problem at all and that
  differences from the LO result are actually small.

Given the exploratory nature of the RBC 07 calculation, its results do
not allow us to draw solid conclusions about the e.m.\,contributions
to $m_u/m_d$ for $\Nf=2$.  As discussed in Section \ref{sec:Nf=2+1},
the $\Nf=2+1$ results of Blum 10 and PACS-CS 12 do not pass our
selection criteria either. We therefore resort to the phenomenological
estimates of the electromagnetic self-energies discussed in Section
\ref{subsec:electromagnetic interaction}, which are validated by
recent, preliminary lattice results. 

Since RM123 13 \cite{deDivitiis:2013xla} includes a lattice estimate of
  e.m.\ corrections, for the $\Nf=2$ final results we simply quote the values
  of $m_u$, $m_d$, and $m_{u}/m_{d}$ from RM123 13 given in
  Table~\ref{tab:mu_md_grading}:
\be
\label{eq:mumdNf2} \hspace{0cm}\Nf = 2:\hspace{0.2cm}m_u
=2.40(23)\,\mev\co\hspace{0.2cm} m_d = 4.80(23) \,\mev\co\hspace{0.2cm}
\frac{m_u}{m_d} = 0.50(4)\ ,
\ee 
with errors of roughly 10\%, 5\% and 8\%, respectively. In these results, the
errors are obtained by combining the lattice statistical and
systematic errors in quadrature.

For $\Nf=2+1$ there is to date no final, published
  computation of e.m.\ corrections. Thus, we take the LO estimate
  for $m_u/m_d$ of (\ref{eq:mu/md LO}) and use the -8(4)\% obtained
  above as an estimate of the size of the corrections from higher
  orders in the chiral expansion. This gives $m_u/m_d=0.46(3)$. The
  two individual masses can then be worked out from the estimate
  (\ref{eq:nf3msmud}) for their mean. Therefore, for $\Nf=2+1$ we
  obtain:
\be
\label{eq:mumd} \hspace{0cm}\Nf = 2+1:\hspace{0.2cm}m_u
=2.16(9)(7)\,\mev\co\hspace{0.2cm} m_d = 4.68(14)(7) \,\mev\co\hspace{0.2cm}
\frac{m_u}{m_d} = 0.46(2)(2)\fs\ee 
In these results, the
first error represents the lattice statistical and systematic errors, combined
in quadrature, while the second arises from the uncertainties
associated with e.m.\ corrections of (\ref{eq:epsilon num}).
The estimates in (\ref{eq:mumd}) have uncertainties of order 5\%, 3\% and 7\%,
respectively.

{
  Naively propagating errors to the end, we
  obtain $(m_u/m_d)_{N_f=2}/(m_u/m_d)_{N_f=2+1}=1.09(10)$. If instead of (\ref{eq:mumdNf2}) we use the results from RM123 11, modified by the e.m. corrections in (\ref{eq:epsilon num}), as was done in our previous review, we obtain $(m_u/m_d)_{N_f=2}/(m_u/m_d)_{N_f=2+1}=1.11(7)(1)$,  confirming again the strong cancellation of
  e.m.\ uncertainties in the ratio. The $N_f=2$ and $2+1$ results are compatible at
  the 1 to 1.5~$\sigma$ level.} 

{ It is interesting to note that in the results above, the errors
  are no longer dominated by the uncertainties in the input used for
  the electromagnetic corrections, though these are still significant
  at the level of precision reached in the $N_f=2+1$ results. This is
  due to the reduction in the error on $\epsilonD$ discussed in
  Section \ref{subsec:electromagnetic interaction}. Nevertheless, the
  comparison of equations (\ref{eq:mu/md LO}) and (\ref{eq:mumd})
  indicates that more than half of the difference between the
  prediction $m_u/m_d=0.558$ obtained from Weinberg's mass formulae
  \cite{Weinberg:1977hb} and the result for $m_u/m_d$ obtained on the
  lattice stems from electromagnetism, the higher orders in the chiral
  perturbation generating a comparable correction.}

In view of the fact that a {\it massless up-quark} would solve the
strong CP-problem, many authors have considered this an attractive
possibility, but the results presented above exclude this possibility:
the value of $m_u$ in (\ref{eq:mumd}) differs from zero by { 20}
standard deviations. We conclude that nature solves the strong
CP-problem differently. This conclusion relies on lattice calculations
of kaon masses and on the phenomenological estimates of the
e.m.~self-energies discussed in Section \ref{subsec:electromagnetic
  interaction}. The uncertainties therein currently represent the
limiting factor in determinations of $m_u$ and $m_d$. As demonstrated
in
\cite{Duncan:1996xy,Blum:2007cy,Blum:2010ym,Basak:2008na,Portelli:2010yn,Portelli:2012pn,llconfx12,Basak:2012zx,Basak:2013iw},
lattice methods can be used to calculate the
e.m.~self-energies. Further progress on the determination of the light-quark 
masses hinges on an improved understanding of the e.m.~effects.

\subsection{Estimates for $R$ and $Q$}\label{sec:RandQ}

The quark-mass ratios
\be\label{eq:Qm}
R\equiv \frac{m_s-m_{ud}}{m_d-m_u}\hspace{0.5cm} \mbox{and}\hspace{0.5cm}Q^2\equiv\frac{m_s^2-m_{ud}^2}{m_d^2-m_u^2}
\ee
compare SU(3)-breaking  with isospin-breaking. The quantity $Q$ is of
particular interest because of a low energy theorem \cite{Gasser:1984pr},
which relates it to a ratio of meson masses,  
\begin{equation}\label{eq:QM}
 Q^2_M\equiv \frac{\hat{M}_K^2}{\hat{M}_\pi^2}\cdot\frac{\hat{M}_K^2-\hat{M}_\pi^2}{\hat{M}_{K^0}^2-
   \hat{M}_{K^+}^2}\co\hspace{1cm}\hat{M}^2_\pi\equiv\mbox{$\frac{1}{2}$}( \hat{M}^2_{\pi^+}+ \hat{M}^2_{\pi^0})
 \co\hspace{0.5cm}\hat{M}^2_K\equiv\mbox{$\frac{1}{2}$}( \hat{M}^2_{K^+}+ \hat{M}^2_{K^0})\fs\end{equation}
Chiral symmetry implies that the expansion of $Q_M^2$ in powers of the
quark masses (i) starts with $Q^2$ and (ii) does not receive any
contributions at NLO:
\be\label{eq:LET Q}Q_M\NLo Q \fs\ee

{
Inserting the estimates for the mass ratios $m_s/m_{ud}$, and
$m_u/m_d$ given for $\Nf=2$ in equations (\ref{eq:quark masses Nf=2})
and (\ref{eq:mumdNf2}) respectively, we obtain
\be\label{eq:RQresNf2} R=40.7(3.7)(2.2)\co\hspace{2cm}Q=24.3(1.4)(0.6)\ ,\ee 
where the errors have been propagated naively and the e.m.\ uncertainty
has been separated out, as discussed in the third paragraph after
(\ref{eq:mu/md LO}). Thus, the meaning of the errors is the same as in
(\ref{eq:mumd}). These numbers agree within errors with those reported
in \cite{deDivitiis:2013xla} where values for $m_s$ and $m_{ud}$ are taken
from ETM 10B~\cite{Blossier:2010cr}.

For $\Nf=2+1$, we use equations
 (\ref{eq:msovmud}) and (\ref{eq:mumd}) and obtain
\be\label{eq:RQres} R=35.8(1.9)(1.8)\co\hspace{2cm}Q=22.6(7)(6)\ ,\ee 
where the meaning of the errors is the same as above. The
$\Nf=2$ and $\Nf=2+1$ results are compatible within 
2$\sigma$, even taking the correlations between e.m. effects into account.
}

It is interesting to use these results to study the size of
chiral corrections in the relations of $R$ and $Q$ to their
expressions in terms of meson masses. To investigate this issue, we
use {\Ch}PT to express the quark-mass ratios in terms of the pion and
kaon masses in QCD and then again use equations
(\ref{eq:epsilon1})--(\ref{eq:epsilon3}) to relate the QCD masses to
the physical ones. Linearizing in the corrections, this leads to {
\bea\label{eq:R epsilon}R\al \Lo\al R_M = 43.9 - 10.8\, \epsilonD +
0.2\, \epsilon_{\pi^0} - 0.2\, \epsilon_{K^0}- 10.7\, \epsilon_m\co\\
Q\al\NLo\al Q_M = 24.3 - 3.0\, \epsilonD + 0.9\,
\epsilon_{\pi^0} - 0.1\, \epsilon_{K^0} + 2.6 \,\epsilon_m
\fs
\label{eq:Q epsilon} 
\eea
}
While
the first relation only holds to LO of the chiral perturbation series, the
second remains valid at NLO, on account of the low energy theorem mentioned
above. The first terms on the right hand side represent the values of $R$
and $Q$ obtained with the Weinberg leading-order formulae for the quark-mass 
ratios \cite{Weinberg:1977hb}. Inserting the estimates
(\ref{eq:epsilon num}), we find that the e.m.~corrections lower the
Weinberg values to { $R_M= 36.7(3.3)$ and $Q_M= 22.3(9)$}, respectively.

{ Comparison of $R_M$ and $Q_M$ with the full results quoted above
  gives a handle on higher-order terms in the chiral
  expansion. Indeed, the ratios $R_M/R$ and $Q_M/Q$ give NLO and NNLO
  (and higher) corrections to the relations $R \Lo R_M$ and $Q\NLo
  Q_M$, respectively. The uncertainties due to the use of the
  e.m.\ corrections of (\ref{eq:epsilon num}) are highly correlated in
  the numerators and denominators of these ratios, and we make the
  simplifying assumption that they cancel in the ratio.  { Thus, for
    $N_f=2$ we evaluate (\ref{eq:R epsilon}) and (\ref{eq:Q epsilon})
    using $\epsilon=0.79(18)(18)$ from RM123 13
    \cite{deDivitiis:2013xla} and the other corrections from
    (\ref{eq:epsilon num}), dropping all uncertainties. We divide them
    by the results for $R$ and $Q$ in (\ref{eq:RQresNf2}), omitting
    the uncertainties due to e.m. We obtain $R_M/R\simeq 0.88(8)$ and
    $Q_M/Q\simeq 0.91(5)$. We proceed analogously for $N_f=2+1$, using
    $\epsilon=0.70(3)$ from (\ref{eq:epsilon num}) and $R$ and $Q$
    from (\ref{eq:RQres}), and find $R_M/R\simeq 1.02(5)$ and
    $Q_M/Q\simeq 0.99(3)$.}  The chiral corrections appear to be small
  for $N_f=2+1$, especially those in the relation of $Q$ to
  $Q_M$. This is less true for $N_f=2$, where the NNLO and higher
  corrections to $Q=Q_M$ could be significant. However, as for other
  quantities which depend on $m_u/m_d$, this difference is not
  significant.}

As mentioned in Section \ref{subsec:electromagnetic interaction}, there is
a phenomenological determination of $Q$ based on the decay $\eta\rightarrow
3\pi$ \cite{Kambor:1995yc,Anisovich:1996tx}. The key point is that the
transition $\eta\rightarrow 3\pi$ violates isospin-conservation. The
dominating contribution to the transition amplitude stems from the mass
difference $m_u-m_d$. At NLO of {\Ch}PT, the QCD part of the amplitude can
be expressed in a parameter free manner in terms of $Q$.  It is well-known
that the electromagnetic contributions to the transition amplitude are
suppressed (a thorough recent analysis is given in \cite{Ditsche:2008cq}).
This implies that the result for $Q$ is less sensitive to the
electromagnetic uncertainties than the value obtained from the masses of
the Nambu-Goldstone bosons.  For a recent update of this determination and
for further references to the literature, we refer to
\cite{Colangelo:2009db}. Using dispersion theory to pin down the 
momentum-dependence of the amplitude, the observed decay rate implies $Q=22.3(8)$
(since the uncertainty quoted in \cite{Colangelo:2009db} does not include
an estimate for all sources of error, we have retained the error estimate
given in \cite{Leutwyler:1996qg}, which is twice as large). The formulae
for the corrections of NNLO are available also in this case
\cite{Bijnens:2007pr} -- the poor knowledge of the effective coupling
constants, particularly of those that are relevant for the dependence on
the quark masses, is currently the limiting factor encountered in the
application of these formulae.

As was to be expected, the central value of $Q$ obtained from
$\eta$-decay agrees exactly with the central value obtained from the
low energy theorem: we have used that theorem to estimate the
coefficient $\epsilonD$, which dominates the e.m.~corrections. Using
the numbers for $\epsilon_m$, $\epsilon_{\pi^0}$ and $\epsilon_{K^0}$
in (\ref{eq:epsilon num}) and adding the corresponding uncertainties
in quadrature to those in the phenomenological result for $Q$, we
obtain
\be\label{eq:epsilon eta}
\epsilonD\NLo 0.70(28)\fs\ee 
The estimate (\ref{eq:epsilon num}) for the size of the coefficient
$\epsilonD$ is taken from here, { as it is confirmed by the most
  recent, preliminary lattice determinations
  \cite{Basak:2008na,Portelli:2010yn,Portelli:2012pn,Basak:2012zx,Basak:2013iw,deDivitiis:2013xla}.}

Our final results for the masses $m_u$, $m_d$, $m_{ud}$, $m_s$ and the mass ratios
$m_u/m_d$, $m_s/m_{ud}$, $R$, $Q$ are collected in Tables \ref{tab:mudms} and
\ref{tab:mumdRQ}. We separate $m_u$, $m_d$, $m_u/m_d$, $R$ and $Q$
from $m_{ud}$, $m_s$ and $m_s/m_{ud}$, because the latter are
completely dominated by lattice results while the former still include
some phenomenological input.

\begin{table}[!thb]\vspace{0.5cm}
{
\begin{tabular*}{\textwidth}[l]{@{\extracolsep{\fill}}cccc}
\hline\hline
$\Nf$ & $m_{ud}$ & $ m_s $ & $m_s/m_{ud}$ \\ 
&&& \\[-2ex]
\hline
&&& \\[-2ex]
2+1 & 3.42(6)(7) & 93.8(1.5)(1.9) & 27.46(15)(41)\\ 
&&& \\[-2ex]
\hline\rule[-0.1cm]{0cm}{0.5cm}
&&& \\[-2ex]
2 & 3.6(2) & 101(3) & 28.1(1.2)\\ 
&&& \\[-2ex]
\hline
\hline
\end{tabular*}
\caption{\label{tab:mudms} Our estimates for the strange and the average
  up-down quark masses in the $\msbar$ scheme  at running scale
  $\mu=2\,\gev$ for $N_f=3$. Numerical values are given in MeV. In the
  results presented here, the first error is the one which we obtain
  by applying the averaging procedure of Section \ref{sec:averages} to the
  relevant lattice results. We have added an uncertainty to the
  $N_f=2+1$ results, which is associated with the neglect of heavy
  sea-quark and isospin-breaking effects, as discussed around
  (\ref{eq:nf3msmud}) and (\ref{eq:msovmud}). This uncertainty is not
  included in the $N_f=2$ results, as it should be smaller than the
  uncontrolled systematic associated with the neglect of strange
  sea-quark effects which we choose not to estimate, as it cannot be
  done so reliably.}  }
\end{table}

\begin{table}[!thb]
{
\begin{tabular*}{\textwidth}[l]{@{\extracolsep{\fill}}cccccc}
\hline\hline
$\Nf$ & $m_u  $ & $m_d $ & $m_u/m_d$ & $R$ & $Q$\\ 
&&&&& \\[-2ex]
\hline
&&&&& \\[-2ex]
2+1 & 2.16(9)(7) & 4.68(14)(7) & 0.46(2)(2) & 35.8(1.9)(1.8) & 22.6(7)(6) \\ 
&&&&& \\[-2ex]
\hline\rule[-0.1cm]{0cm}{0.5cm}
&&&&& \\[-2ex]
2 & 2.40(23) & 4.80(23) & 0.50(4) & 40.7(3.7)(2.2) & 24.3(1.4)(0.6)\\ 
&&&&& \\[-2ex]
\hline
\hline
\end{tabular*}
\caption{\label{tab:mumdRQ} Our estimates for the masses of the
  two lightest quarks and related, strong isospin-breaking
  ratios. Again, the masses refer to the $\msbar$ scheme  at running
  scale $\mu=2\,\gev$ for $N_f=3$ and the numerical values are given
  in MeV. In the results presented here, the first error is the one
  that comes from lattice computations while the second for $N_f=2+1$
  is associated with the phenomenological estimate of e.m.\
  contributions, as discussed after (\protect\ref{eq:mumd}). The
  second error on the $N_f=2$ results for $R$ and $Q$ is also an
  estimate of the e.m.\ uncertainty, this time associated with the
  lattice computation of \protect\cite{deDivitiis:2013xla}, as
  explained after (\protect\ref{eq:RQresNf2}). We present these
  results in a separate table, because they are less firmly
  established than those in Table \protect\ref{tab:mudms}. For
  $N_f=2+1$ they still include information coming from phenomenology,
  in particular on e.m.\ corrections, and for $N_f=2$ the e.m.\
  contributions are computed neglecting the feedback of sea-quarks on
  the photon field.}  }
\end{table}
\newpage

\clearpage
\section{Leptonic and semileptonic kaon and pion decay and 
	$|V_{ud}|$ and $|V_{us}|$}\label{sec:vusvud}
This section summarizes state of the art lattice calculations of the
leptonic kaon and pion decay constants and the kaon semileptonic
decay form factor and provides an analysis in view of the Standard Model.
With respect to the previous edition of the FLAG review \cite{Colangelo:2010et} the data in this section
has been updated, correlations of lattice data are now taken into account
in all the analysis and a subsection on the individual decay constants 
$f_K$ and $f_\pi$  (rather than only the ratio)
has been included. Furthermore, when combining lattice data with experimental
results we now take into account the strong SU(2) isospin correction in chiral 
perturbation theory for the ratio of leptonic decay constants $f_K/f_\pi$.

\subsection{Experimental information concerning $|V_{ud}|$, $|V_{us}|$,
$f_+(0)$ and $\fKfpichargedr$}\label{sec:Exp} 
The following review relies on the fact that precision 
experimental data on kaon decays
very accurately determine the product $|V_{us}|f_+(0)$ and the ratio
$|V_{us}/V_{ud}|f_{K^\pm}/f_{\pi^\pm}$ \cite{Antonelli:2010yf}: 
\be\label{eq:products}
|V_{us}| f_+(0) = 0.2163(5)\co \hspace{1cm} \;
\left|\frac{V_{us}}{V_{ud}}\right|\frac{ f_{K^\pm}}{ f_{\pi^\pm}} \;
=0.2758(5)\fs\ee 
Here and in the following $f_{K^\pm}$ and $f_{\pi^\pm}$ are the isospin-broken 
decay constants, respectively, in QCD
(the
electromagnetic effects have already been subtracted in the experimental
analysis using chiral perturbation theory). We will refer to the decay 
constants in the SU(2) isospin-symmetric limit as $f_{K}$ and $f_{\pi}$.
$|V_{ud}|$ and $|V_{us}|$ are
elements of the Cabibbo-Kobayashi-Maskawa matrix and $f_+(t)$ represents
one of the form factors relevant for the semileptonic decay
$K^0\rightarrow\pi^-\ell\,\nu$, which depends on the momentum transfer $t$
between the two mesons.  What matters here is the value at $t=0$:
$f_+(0)\equiv
f_+^{K^0\pi^-}\hspace{-0.1cm}(t)\,\rule[-0.15cm]{0.02cm}{0.5cm}_{\;t\rightarrow
  0}$. The pion and kaon decay constants are defined by\footnote{The pion
  decay constant represents a QCD matrix element -- in the full Standard
  Model, the one-pion state is not a meaningful notion: the correlation
  function of the charged axial current does not have a pole at
  $p^2=M_{\pi^+}^2$, but a branch cut extending from $M_{\pi^+}^2$ to
  $\infty$. The analytic properties of the correlation function and the
  problems encountered in the determination of $f_\pi$ are thoroughly
  discussed in \cite{Gasser:2010wz}. The ``experimental'' value of $f_\pi$
  depends on the convention used when splitting the sum ${\cal
    L}_{\mbox{\tiny QCD}}+{\cal L}_{\mbox{\tiny QED}}$ into two parts
  (compare section \ref{subsec:electromagnetic interaction}).  The lattice
  determinations of $f_\pi$ do not yet reach the accuracy where this is of
  significance, but at the precision claimed by the Particle Data Group
  \cite{Rosner:2012np}, the numerical value does depend on the convention used
  \cite{Gasser:2003hk,Rusetsky:2009ic,Gasser:2007de,Gasser:2010wz}. }  \bdm
\lvac \dbar\gamma_\mu\gamma_5 \hspace{0.05cm}u|\pi^+(p)\rangle=i
\hspace{0.05cm}p_\mu f_{\pi^+}\co\hspace{1cm} \lvac \sbar\gamma_\mu\gamma_5
\hspace{0.05cm} u|K^+(p)\rangle=i \hspace{0.05cm}p_\mu f_{K^+}\fs\edm In this
normalization, $f_{\pi^\pm} \simeq 130$~MeV, $f_{K^\pm}\simeq 155$~MeV.
 
The measurement of $|V_{ud}|$ based on superallowed nuclear $\beta$
transitions has now become remarkably precise. The result of the 
update of Hardy and Towner \cite{Hardy:2008gy}, which is based on 20
different superallowed transitions, reads\footnote{It is not a trivial
  matter to perform the data analysis at this precision. In particular,
  isospin-breaking effects need to be properly accounted for
  \cite{Towner:2007np,Miller:2008my,Auerbach:2008ut,Liang:2009pf,Miller:2009cg}.
  For a review of recent work on this issue, we refer to
  \cite{Towner:2010bx}.}
\be\label{eq:Vud beta}
|V_{ud}| = 0.97425(22)\fs\ee 

The matrix element $|V_{us}|$ can be determined from semiinclusive 
$\tau$ decays
\cite{Gamiz:2002nu,Gamiz:2004ar,Maltman:2008na,Pich_Kass}. Separating the
inclusive decay $\tau\rightarrow \mbox{hadrons}+\nu$ into nonstrange and
strange final states, e.g. HFAG 12~\cite{Amhis:2012bh} obtain
\be\label{eq:Vus tau}|V_{us}|=0.2173(22) \fs \ee Maltman et
al.~\cite{Maltman:2008ib,Maltman:2008na,Maltman:2009bh} and Gamiz et al.~\cite{Gamiz:2007qs,Gamiz:2013wn}
arrive at very similar values.

In principle, $\tau$ decay offers a clean measurement of $|V_{us}|$, but a
number of open issues yet remain to be clarified. In particular, the
value of $|V_{us}|$ as determined from inclusive $\tau$ decays
differs from the result one obtains from assuming three-flavour SM-unitarity 
by more than three standard deviations~\cite{Amhis:2012bh}. 
It is important to understand this apparent tension better.
The most interesting possibility is that $\tau$ decay involves new physics,
but more work both on the theoretical 
(see e.g.\cite{Beneke:2008ad,Caprini:2009vf,Menke:2009vg,Boyle:2013xw})
and experimental side is required.

The experimental results in equation~(\ref{eq:products}) are for the 
semileptonic decay of a neutral kaon into a negatively charged pion and the
charged pion and kaon leptonic decays, respectively, in QCD. In the case of
the semileptonic decays the corrections for strong
and electromagnetic isospin breaking in chiral perturbation
theory at NLO have allowed for averaging the different experimentally
measured isospin channels~\cite{Antonelli:2010yf}. 
This is quite a convenient procedure as long as lattice QCD does not include
strong or QED isospin-breaking effects. 
Lattice results for $f_K/f_\pi$ are typically quoted for QCD with (squared)
pion and kaon masses of $M_\pi^2=M_{\pi^0}^2$ and $M_K^2=\frac 12
	\left(M_{K^\pm}^2+M_{K^0}^2-M_{\pi^\pm}^2+M_{\pi^0}^2\right)$
for which the leading strong and electromagnetic isospin violations cancel.
While progress
is being made for including strong and electromagnetic isospin breaking in 
the simulations
(e.g.~\cite{Aoki:2008sm,deDivitiis:2011eh,Ishikawa:2012ix,TakuLat12,Portelli:2013jla,Tantalo:2013maa}),
for now contact to experimental results is made
by correcting leading SU(2) isospin breaking 
guided by chiral perturbation theory. 

In the following we will start by presenting the lattice results for
isospin-symmetric QCD. For any Standard Model analysis based on these
results we then utilize chiral perturbation theory to correct for the
leading isospin-breaking effects.


\subsection{Lattice results for $f_+(0)$ and $f_K/f_\pi$}
\begin{table}[t]
\centering 
\vspace{2.8cm}
{\footnotesize\noindent
\begin{tabular*}{\textwidth}[l]{@{\extracolsep{\fill}}llllllll}
Collaboration & Ref. & $\Nf$ & 
\hspace{0.15cm}\begin{rotate}{60}{publication status}\end{rotate}\hspace{-0.15cm}&
\hspace{0.15cm}\begin{rotate}{60}{chiral extrapolation}\end{rotate}\hspace{-0.15cm}&
\hspace{0.15cm}\begin{rotate}{60}{continuum extrapolation}\end{rotate}\hspace{-0.15cm}&
\hspace{0.15cm}\begin{rotate}{60}{finite volume errors}\end{rotate}\hspace{-0.15cm}&\rule{0.3cm}{0cm}
$f_+(0)$ \\
&&&&&&& \\[-0.1cm]
\hline
\hline&&&&&&& \\[-0.1cm]
FNAL/MILC 13C             &\cite{Gamiz:2013xxa} &2+1+1  &\rC&\good&\good&\good& 0.9704(24)(32)\\[-0.5mm]
&&&&&&& \\[-0.1cm]
\hline
&&&&&&& \\[-0.1cm]
RBC/UKQCD 13              & \cite{Boyle:2013gsa}  &2+1  &\gA&\good&\soso&\soso& 0.9670(20)($^{+18}_{-46}$)\\[-0.5mm]
FNAL/MILC 12              & \cite{Bazavov:2012cd} &2+1  &\gA&\soso&\soso&\tbg& 0.9667(23)(33)\\[-0.5mm]
JLQCD 12                  & \cite{Kaneko:2012cta} &2+1  &\rC&\soso&\tbr&\tbg& 0.959(6)(5)\\[-0.5mm]
JLQCD 11                  & \cite{Kaneko:2011rp}  &2+1  &\rC&\soso&\tbr&\tbg& 0.964(6)\\[-1.5mm]
RBC/UKQCD 10              & \cite{Boyle:2010bh}   &2+1  &\gA&\soso&\tbr&\tbg& 0.9599(34)($^{+31}_{-47}$)(14)\rule{0cm}{0.4cm}\\ 
RBC/UKQCD 07              & \cite{Boyle:2007qe}   &2+1  &\gA&\soso&\tbr&\tbg& 0.9644(33)(34)(14)\\
&&&&&&& \\[-0.1cm]
\hline
&&&&&&& \\[-0.1cm]
ETM 10D                   & \cite{Lubicz:2010bv}  &2 &\rC&\soso&\tbg&\soso& 0.9544(68)$_{stat}$\\
ETM 09A 	          & \cite{Lubicz:2009ht}  &2 &\gA&\soso&\soso&\soso& 0.9560(57)(62)\\	
QCDSF 07	          & \cite{Brommel:2007wn} &2 &\rC&\tbr&\tbr&\tbg& 0.9647(15)$_{stat}$ \\
RBC 06  	          & \cite{Dawson:2006qc}  &2 &\gA&\tbr&\tbr&\tbg& 0.968(9)(6)\\	
JLQCD 05 	          & \cite{Tsutsui:2005cj} &2 &\rC&\tbr&\tbr&\tbg& 0.967(6), 0.952(6)\\ 
 &&&&&&& \\[-0.1cm]
\hline
\hline
\end{tabular*}}
\caption{Colour code for the data on $f_+(0)$.\hfill}\label{tab:f+(0)}
\end{table}
The traditional way of determining $|V_{us}|$ relies on using theory for
the value of $f_+(0)$, invoking the Ademollo-Gatto theorem
\cite{Ademollo_Gatto}.  Since this theorem only holds to leading order of
the expansion in powers of $m_u$, $m_d$ and $m_s$, theoretical models are
used to estimate the corrections. Lattice methods have now reached the
stage where quantities like $f_+(0)$ or $f_K/f_\pi$ can be determined to
good accuracy. As a consequence, the uncertainties inherent in the
theoretical estimates for the higher order effects in the value of $f_+(0)$
do not represent a limiting factor any more and we shall therefore not
invoke those estimates. Also, we will use the experimental results based on
nuclear $\beta$ decay and $\tau$ decay exclusively for comparison -- the
main aim of the present review is to assess the information gathered with
lattice methods and to use it for testing the consistency of the SM and its
potential to provide constraints for its extensions.

The database underlying the present review of the semileptonic form factor 
and the ratio of decay constants is
listed in Tables \ref{tab:f+(0)} and \ref{tab:FKFpi}. The properties of the
lattice data play a crucial role for the conclusions to be drawn from these
results: range of $M_\pi$, size of $L M_\pi$, continuum extrapolation,
extrapolation in the quark masses, finite size effects, etc. The key
features of the various data sets are characterized by means of the 
colour code specified in section \ref{sec:color-code}. More detailed information
on individual computations are compiled in appendix~\ref{app:VusVud}.

The quantity $f_+(0)$ represents a matrix element of a strangeness changing
null plane charge, $f_+(0)=(K|Q^{us}|\pi)$. The vector charges obey the
commutation relations of the Lie algebra of SU(3), in particular
$[Q^{us},Q^{su}]=Q^{uu-ss}$. This relation implies the sum rule $\sum_n
|(K|Q^{us}|n)|^2-\sum_n |(K|Q^{su}|n)|^2=1$. Since the contribution from
the one-pion intermediate state to the first sum is given by $f_+(0)^2$,
the relation amounts to an exact representation for this quantity
\cite{Furlan}: \be \label{eq:Ademollo-Gatto} f_+(0)^2=1-\sum_{n\neq \pi}
|(K|Q^{us}|n)|^2+\sum_n |(K|Q^{su}|n)|^2\fs\ee While the first sum on the
right extends over nonstrange intermediate states, the second runs over
exotic states with strangeness $\pm 2$ and is expected to be small compared
to the first.

The expansion of $f_+(0)$ in SU(3) chiral perturbation theory
in powers of $m_u$, $m_d$ and $m_s$ starts with
$f_+(0)=1+f_2+f_4+\ldots\,$ \cite{Gasser:1984gg}.  Since all of the low
energy constants occurring in $f_2$ can be expressed in terms of $M_\pi$,
$M_K$, $M_\eta$ and $f_\pi$ \cite{Gasser:1984ux}, the NLO correction is
known. In the language of the sum rule (\ref{eq:Ademollo-Gatto}), $f_2$
stems from nonstrange intermediate states with three mesons. Like all
other nonexotic intermediate states, it lowers the value of $f_+(0)$:
$f_2=-0.023$ when using the experimental value of $f_\pi$ as input.  
The corresponding expressions have also been derived in
quenched or partially quenched (staggered) chiral perturbation theory
\cite{Becirevic:2005py,Bazavov:2012cd}.  At the same order in the SU(2) expansion
\cite{Flynn:2008tg}, $f_+(0)$ is parameterized in terms of $M_\pi$ and two
\textit{a priori} unknown parameters. The latter can be determined from the
dependence of the lattice results on the masses of the quarks.  Note that
any calculation that relies on the {\Ch}PT formula for $f_2$ is subject to
the uncertainties inherent in NLO results: instead of using the physical
value of the pion decay constant $f_\pi$, one may, for instance, work with
the constant $f_0$ that occurs in the effective Lagrangian and represents
the value of $f_\pi$ in the chiral limit. Although trading $f_\pi$ for
$f_0$ in the expression for the NLO term affects the result only at NNLO,
it may make a significant numerical difference in calculations where the
latter are not explicitly accounted for (the lattice results concerning the
value of the ratio $f_\pi/f_0$ are reviewed in section \ref{sec:su3}).
\begin{table}[t]
\centering
\vspace{1.5cm}{\footnotesize\noindent
\begin{tabular*}{\textwidth}[l]{@{\extracolsep{\fill}}lrlllllll}
Collaboration & Ref. & $\Nf$ &
\hspace{0.15cm}\begin{rotate}{60}{publication status}\end{rotate}\hspace{-0.15cm}&
\hspace{0.15cm}\begin{rotate}{60}{chiral extrapolation}\end{rotate}\hspace{-0.15cm}&
\hspace{0.15cm}\begin{rotate}{60}{continuum extrapolation}\end{rotate}\hspace{-0.15cm}&
\hspace{0.15cm}\begin{rotate}{60}{finite volume errors}\end{rotate}\hspace{-0.15cm}&
\rule{0.2cm}{0cm} $f_K/f_\pi$ &
\rule{0.2cm}{0cm} $f_{K^\pm}/f_{\pi^\pm}$ \\  
&&&&&&& \\[-0.1cm]
\hline
\hline
&&&&&&& \\[-0.1cm]
ETM 13F       &\cite{Dimopoulos:2013qfa}  &2+1+1&\rC&\soso &\good&\soso    		&			1.193(13)(10)			&1.183(14)(10)		\\
HPQCD 13A       &\cite{Dowdall:2013rya}	     &2+1+1&\gA&\good &\soso&\good    		&						&1.1916(15)(16)		\\
MILC 13A        &\cite{Bazavov:2013cp}	     &2+1+1&\gA&\good &\soso&\good    		&						&1.1947(26)(37)		\\
MILC 11        &\cite{Bazavov:2011fh}	     &2+1+1&\rC&\soso &\soso&\soso    		&						&1.1872(42)$^\dagger_{\rm stat.}$		\\
ETM 10E       &\cite{Farchioni:2010tb}&2+1+1&\rC&\soso&\soso&\soso		&1.224(13)$_{\rm stat}$				&						\\
&&&&&&& \\[-0.1cm]                                                                                                              
\hline                                                                                                                          
&&&&&&& \\[-0.1cm]                                                                                                              
RBC/UKQCD 12   &\cite{Arthur:2012opa}    &2+1&\gA&\good    & \soso	 &  \good  	&1.199(12)(14)					&					\\
Laiho 11       &\cite{Laiho:2011np}       &2+1&\rC&\soso    & \soso   &  \soso  	&                                       	&$1.202(11)(9)(2)(5)$$^{\dagger\dagger}$	\\
MILC 10        &\cite{Bazavov:2010hj}&2+1&\rC&\soso&\good&\good			&                             			&1.197(2)($^{+3}_{-7}$)			\\
JLQCD/TWQCD 10 &\cite{Noaki:2010zz}&2+1&\rC&\soso&\tbr&\tbg			&1.230(19)					&                               		\\
RBC/UKQCD 10A  &\cite{Aoki:2010dy}   &2+1&\gA&\soso&\soso&\good			&1.204(7)(25)					&                               		\\
PACS-CS 09     &\cite{Aoki:2009ix}   &2+1&\gA&\good&\tbr&\tbr			&1.333(72)					&                               		\\
BMW 10         &\cite{Durr:2010hr}         &2+1&\gA&\good &\tbg&\tbg			&1.192(7)(6)					&                               		\\
JLQCD/TWQCD 09A&\cite{JLQCD:2009sk}  &2+1&\rC&\soso&\tbr&\tbr			&$1.210(12)_{\rm stat}$				&                      				\\
MILC 09A       &\cite{Bazavov:2009fk}&2+1&\rC&\soso&\tbg&\tbg			&                                               &1.198(2)($^{\hspace{0.01cm}+6}_{-8}$)	\\
MILC 09        &\cite{Bazavov:2009bb}&2+1&\gA&\soso&\tbg&\tbg			&                                               &1.197(3)($^{\;+6}_{-13}$)		\\
Aubin 08       &\cite{Aubin:2008ie}  &2+1&\rC&\soso&\soso&\soso			&                                               &1.191(16)(17)					\\
PACS-CS 08, 08A&\cite{Aoki:2008sm, Kuramashi:2008tb} &2+1&\gA&\tbg&\tbr&\tbr	&1.189(20)					&                                               \\
RBC/UKQCD 08   &\cite{Allton:2008pn} &2+1&\gA&\soso&\tbr&\tbg			&1.205(18)(62)					&                                               \\
HPQCD/UKQCD 07 &\cite{Follana:2007uv}&2+1&\gA&\soso&\tbg&\soso			&1.189(2)(7)					&                                               \\
NPLQCD 06      &\cite{Beane:2006kx}  &2+1&\gA&\soso&\tbr&\tbr			&1.218(2)($^{+11}_{-24}$)			&                                               \\
MILC 04 &\cite{Aubin:2004fs}&2+1&\gA&\soso&\soso&\soso				&						&1.210(4)(13)				\\
&&&&&&& \\[-0.1cm]                                                                                                              
\hline                                                                                                                          
&&&&&&& \\[-0.1cm]                                                                                                              
ALPHA 13       &\cite{Lottini:2013rfa}&2  &\rC&\tbg    &\tbg   &\tbg    	&1.1874(57)(30)					&                                       	\\
BGR 11	       &\cite{Engel:2011aa}  &2  &\gA&\tbg    &\tbr   &\tbr    		&1.215(41)					&                                       	\\
ETM 10D        &\cite{Lubicz:2010bv} &2  &\rC&\soso&\tbg&\soso			&1.190(8)$_{\rm stat}$ 				&                                       	\\
ETM 09         &\cite{Blossier:2009bx}         &2  &\gA&\soso&\tbg&\soso			&1.210(6)(15)(9)				&                                       	\\
QCDSF/UKQCD 07 &\cite{QCDSFUKQCD}    &2  &\rC&\soso&\soso&\tbg			&1.21(3)					&                                       	\\
&&&&&&& \\[-0.1cm]
\hline
\hline
&&&&&&& \\[-0.1cm]
\end{tabular*}}\\[-2mm]
{\tiny $^\dagger$ Result with statistical error only from polynomial interpolation to the physical point.}
{\tiny $^{\dagger\dagger}$ This work is the continuation of Aubin 08.}\\
\vspace{-0.0cm}
\caption{Colour code for the data on the ratio 
of decay constants: $f_K/f_\pi$ is the pure QCD SU(2)-symmetric ratio and 
$f_{K^\pm}/f_{\pi^\pm}$ is in pure QCD with
the SU(2) isospin breaking applied after simulation.
\hfill}\label{tab:FKFpi}

\end{table}
\begin{figure}[ht]
\psfrag{y}{\tiny $\star$}
\hspace{-9mm}\includegraphics[height=6.8cm]{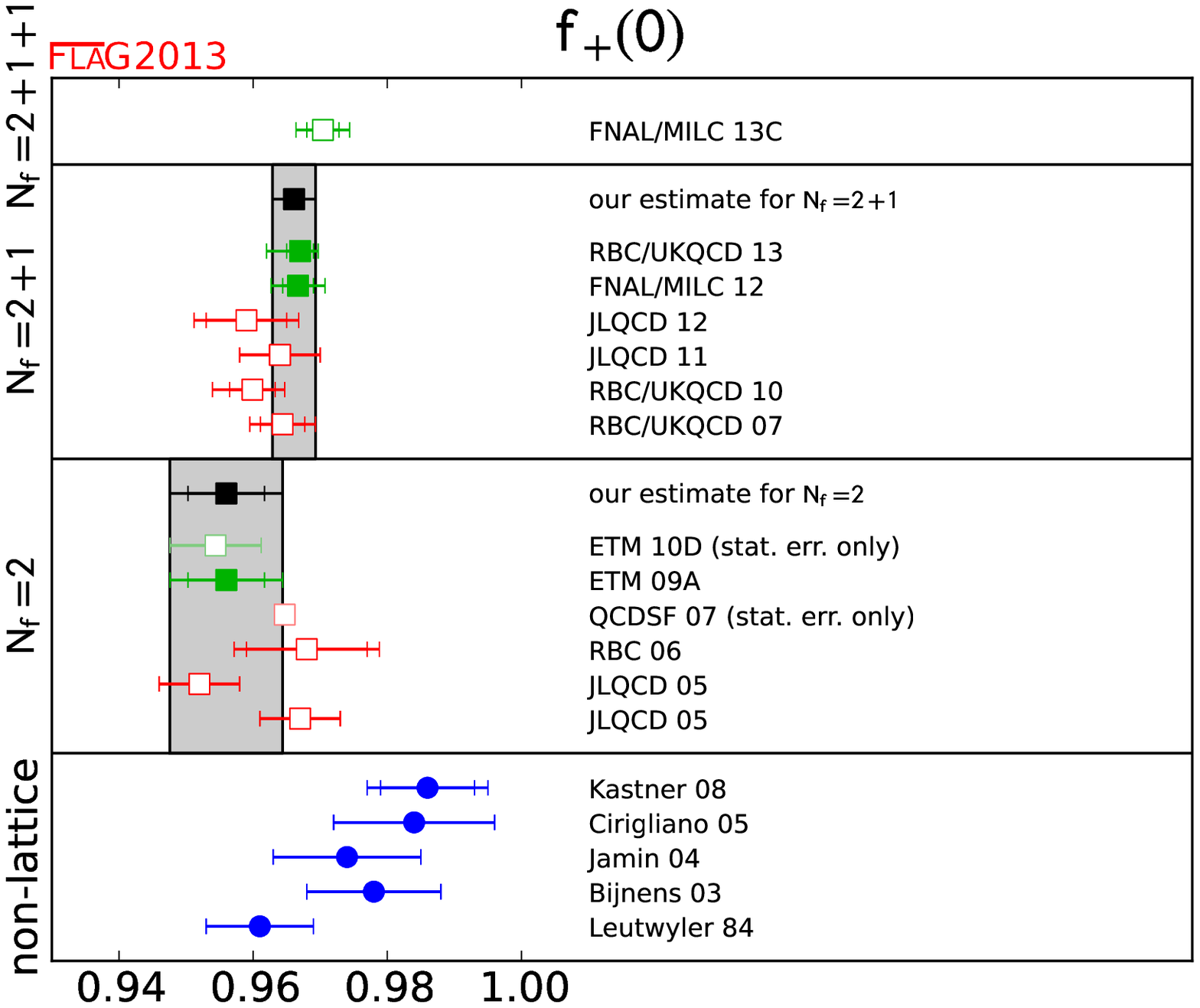}
\hspace{-1cm}
\includegraphics[height=6.8cm]{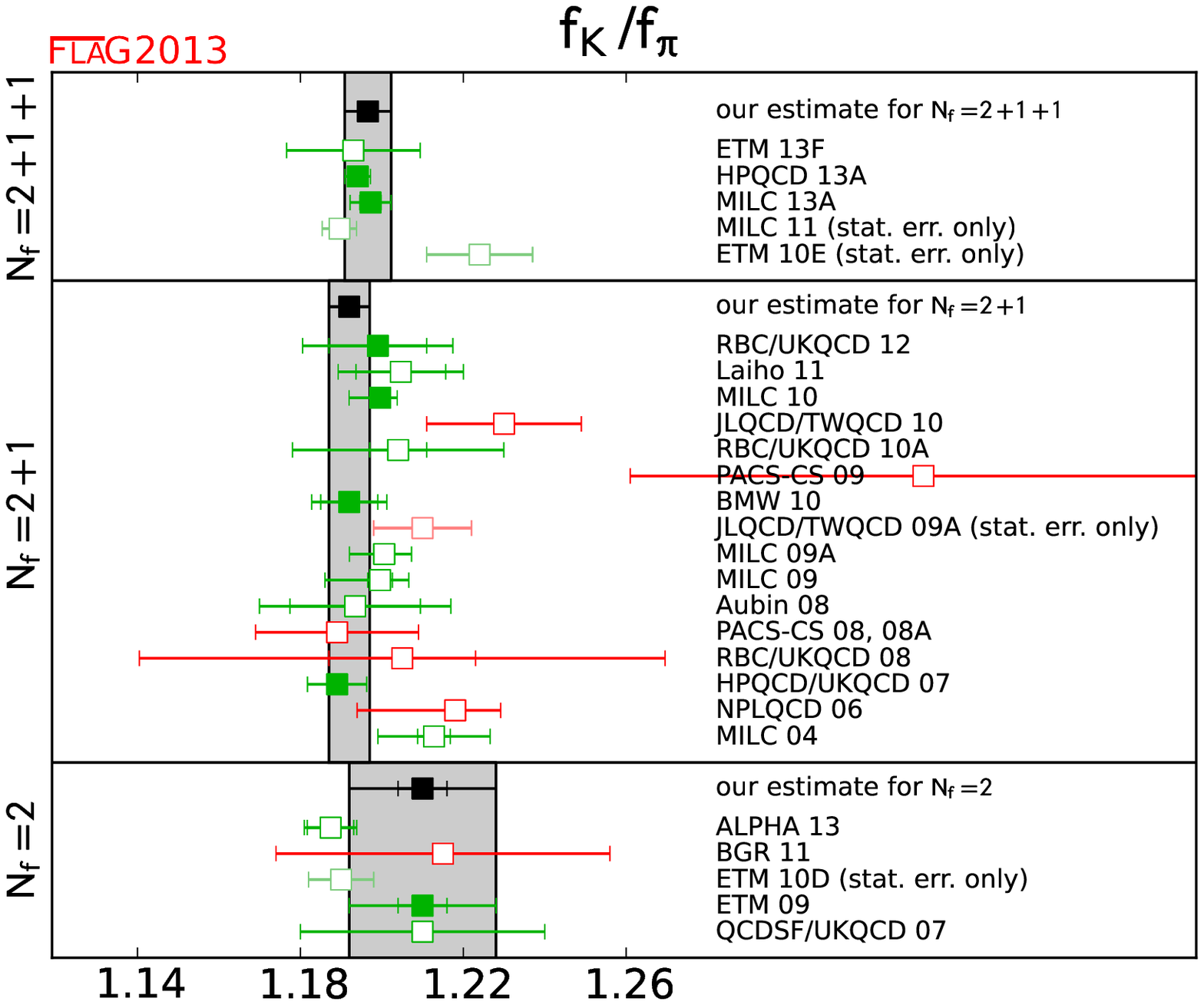}
  
\vspace{-1.99cm}\hspace{5.5cm}\parbox{6cm}{\sffamily\tiny  \cite{Kastner:2008ch}\\

\vspace{-1.13em}\cite{Cirigliano:2005xn}\\

\vspace{-1.13em}\cite{Jamin:2004re}\\

\vspace{-1.13em}\cite{Bijnens:2003uy}\\

\vspace{-1.13em}\cite{Leutwyler:1984je}}
\vspace{1cm}

\caption{\label{fig:lattice data}Comparison of lattice results (squares) 
  for $f_+(0)$ and $f_K/ f_\pi$ with various model estimates based on 
  {\Ch}PT (blue circles).  The black squares and grey bands indicate our estimates. 
  The significance of the colours is explained in section \ref{sec:qualcrit}.}

\end{figure}

The lattice results shown in the left panel of Figure \ref{fig:lattice
  data} indicate that the higher order contributions $\Delta f\equiv
f_+(0)-1-f_2$ are negative and thus amplify the effect generated by $f_2$.
This confirms the expectation that the exotic contributions are small. The
entries in the lower part of the left panel represent various model
estimates for $f_4$. In \cite{Leutwyler:1984je} the symmetry breaking
effects are estimated in the framework of the quark model. The more recent
calculations are more sophisticated, as they make use of the known explicit
expression for the $K_{\ell3}$ form factors to NNLO in {\Ch}PT
\cite{Post:2001si,Bijnens:2003uy}. The corresponding formula for $f_4$
accounts for the chiral logarithms occurring at NNLO and is not subject to
the ambiguity mentioned above.\footnote{Fortran programs for the
  numerical evaluation of the form factor representation in
  \cite{Bijnens:2003uy} are available on request from Johan Bijnens.} 
 The numerical result, however, depends on
the model used to estimate the low energy constants occurring in $f_4$
\cite{Bijnens:2003uy,Jamin:2004re,Cirigliano:2005xn,Kastner:2008ch}. The
figure indicates that the most recent numbers obtained in this way
correspond to a positive rather than a negative value for $\Delta f$.
We note that 
FNAL/MILC~12~\cite{Bazavov:2012cd} have made an attempt at determining some of the
low energy constants appearing in $f_4$ from lattice data.

\subsection{Direct determination of $f_+(0)$ and $f_{K^\pm}/f_{\pi^\pm}$}\label{sec:Direct} 
All lattice results for 
the form factor and the ratio of decay constants that we summarize 
here (Tables~\ref{tab:f+(0)} and~\ref{tab:FKFpi})
have been computed in isospin-symmetric QCD. The reason 
for this unphysical parameter choice 
is that simulations of SU(2) isospin-breaking effects in lattice QCD, while 
ultimately the cleanest way for predicting these effects, are still rare 
and in their 
infancy~\cite{Duncan:1996xy,Basak:2008na,Blum:2010ym,Portelli:2010yn,deDivitiis:2011eh,Portelli:2013jla,Tantalo:2013maa}. 
In the meantime one relies either on chiral perturbation 
theory~\cite{Gasser:1984gg,Aubin:2004fs} to estimate 
the correction to the isospin limit 
or one  calculates the breaking at leading order in $(m_u-m_d)$ in the 
valence quark sector by making a suitable choice of the physical point to 
which the lattice data is extrapolated.
Aubin 08, MILC and Laiho~11 for example 
extrapolate their simulation results for the kaon decay constant
to the physical value of the $up$-quark mass 
(the results for the pion decay constant are extrapolated to the
value of the average light-quark mass $\hat m$).
This then defines their prediction for $f_{K^\pm}/f_{\pi^\pm}$. 
As long as the majority of collaborations present their final results in the 
isospin-symmetric limit (as we will see this comprises the majority of results 
which qualify for inclusion into a FLAG average)
we prefer to provide the overview of world data in Figure~\ref{fig:lattice data}
in this limit.
To this end we compute the isospin-symmetric ratio $f_{K}/f_{\pi}$ 
for Aubin 08, MILC and Laiho 11 using 
NLO chiral perturbation theory~\cite{Gasser:1984gg,Cirigliano:2011tm} where,
\begin{equation}\label{eq:convert}
	\frac{f_K}{f_\pi}=\frac{1}{\sqrt{\delta_{\rm SU(2)}+1}}
		\fKfpicharged\,,
\end{equation}
and where~\cite{Cirigliano:2011tm}\,,
\begin{equation}\label{eq:removeiso}
 \begin{array}{rcl}
	 \delta_{\rm SU(2)}& \approx&
	\sqrt{3}\,\epsilon_{\rm SU(2)}
	\left[-\frac 43 \left(f_{K^\pm}/f_{\pi^\pm}-1\right)+\frac 2{3 (4\pi)^2 f_0^2}
        \left(M_K^2-M_\pi^2-M_\pi^2\ln\frac{M_K^2}{M_\pi^2}\right)
        \right]\,.
  \end{array}
 \end{equation}
We use as input
$\epsilon_{\rm SU(2)}=\sqrt{3}/4/R$ with the FLAG result for $R$ of 
equation~(\ref{eq:RQres}), $F_0=f_0/\sqrt{2}=80(20)$~MeV,
$M_\pi=135$~MeV and $M_K=495$~MeV 
(we decided to choose a conservative uncertainty 
on $f_0$ in order to reflect the magnitude of potential higher order 
corrections)
and obtain  for example
\begin{center}
\begin{tabular}{llll}
\hline\hline\\[-4mm]
		&$f_{K^\pm}/f_{\pi^\pm}$&$\delta_{\rm SU(2)}$&$f_K/f_\pi$\\
\hline\\[-4mm]                           
Aubin 08	&1.202(11)(9)(2)(5)	&-0.0044(8)&1.205(11)(2)(9)(2)(5)\\
MILC 10		&1.197(2)($^{+3}_{-7}$)	&-0.0043(7)&1.200(2)(2)($^{+3}_{-7}$)\\
Laiho 11	&1.191(16)(17)		&-0.0041(9)&1.193(16)(2)(17)		\\
\hline\hline
\end{tabular}
\end{center}
(and similarly also for all other $N_f=2+1$ and $N_f=2+1+1$
results where applicable).
In the last column the first error is statistical and the second is the one 
from the isospin correction (the remaining errors are quoted in the same order
as in the original data).
For $N_f=2$ a dedicated study of the strong-isospin correction
in lattice QCD does exist. 
The result of the RM123 collaboration~\cite{deDivitiis:2011eh}
amounts to $\delta_{\rm SU(2)}=-0.0078(7)$ and we will later use this result for the
correction in the case of $N_f=2$.
We note that this value for the strong-isospin correction is incompatible
with the above results based on SU(3)
chiral perturbation theory.  One would not expect the
strange sea-quark contribution to be responsible for  such a large effect. 
Whether higher order effects in chiral perturbation theory 
or other sources are responsible still needs to be understood.
To remain on the conservative side we attach the difference between
the two- and three-flavour result as an additional uncertainty to the
result based on chiral perturbation theory. For the
further analysis we add both errors in quadrature.

The plots in 
Figure~\ref{fig:lattice data} illustrate our compilation of data for $f_+(0)$
and $f_K/f_\pi$. In both cases the lattice data are largely consistent even when
comparing simulations with different $N_f$.
We now proceed to form the corresponding averages, separately
for the data with $\Nf=2+1+1$, $\Nf=2+1$ and $\Nf=2$ 
dynamical flavours and in the 
following will refer to these averages as the ``direct'' determinations. 
 
For $f_+(0)$ there are currently two computational strategies: 
FNAL/MILC~12 and\linebreak FNAL/MILC~13 use the Ward identity relating the $K\to\pi$ 
form factor at zero momentum transfer to the matrix element $\langle \pi|S|K\rangle$
of the flavour-changing scalar current. Peculiarities of the staggered fermion 
discretisation (see~\cite{Bazavov:2012cd})
which FNAL/MILC is using makes this the favoured
choice. The other collaborations are instead computing the vector current
matrix element $\langle \pi |V_\mu|K\rangle$. Apart from MILC 13C all simulations in 
Table~\ref{tab:f+(0)} involve unphysically heavy quarks and therefore
the lattice data needs to be extrapolated to the physical pion and kaon 
masses corresponding to the $K^0\to\pi^-$ channel. 
We note that all state of the art computations of $f_+(0)$ 
are using partially twisted boundary conditions which allow
to determine the form factor results directly at the relevant kinematical point 
$q^2=0$~\cite{Guadagnoli:2005be,Boyle:2007wg}.

The colour code in Table~\ref{tab:f+(0)} shows that 
for $f_+(0)$, presently only the result of ETM 
(we will be using ETM 09A~\cite{Lubicz:2009ht}) with $\Nf =2$
and the results by the FNAL/MILC and RBC/UKQCD collaborations with 
$\Nf=2+1$ dynamical flavours of fermions, respectively,
are without a red tag. The latter two results,
$f_+(0) =0.9670(20)(^{+18}_{-46})$ (RBC/UKQCD 13) 
and $f_+(0) =0.9667(23)(33)$ (FNAL/MILC 12),
agree very well. This is nice to observe given that the 
two collaborations are using different fermion discretisations 
(staggered fermions in the case of FNAL/MILC
and domain wall fermions in the case of RBC/UKQCD).
Moreover, in the case of FNAL/MILC the form factor has been
determined from the
scalar current matrix element while in the case of RBC/UKQCD it
has been determined from
the matrix element of the vector current. To a certain extent both 
simulations are expected to be 
affected by different systematic effects.

The result FNAL/MILC 12 is from simulations reaching down to a lightest 
RMS pion mass of about 380~MeV (the lightest valence pion mass 
for one of their ensembles is about 260~MeV).
Their combined chiral and continuum extrapolation (results 
for two lattice spacings) is based on NLO staggered chiral
perturbation theory supplemented by the continuum NNLO 
expression~\cite{Bijnens:2003uy} and a phenomenological parameterization
of the breaking of the Ademollo-Gatto theorem at finite lattice spacing
inherent in their approach.
The $p^4$ low energy constants entering the NNLO expression have been fixed in terms
of external input~\cite{Amoros:2001cp}. 

RBC/UKQCD 13 has analysed results on ensembles with pion masses down 
to 170MeV, mapping out nearly the complete range from the SU(3)-symmetric limit
to the physical point. Although no
finite volume or cut-off effects were observed in the simulation
results, the expected residual systematic effects for finite volume
effects in NLO chiral perturbation theory and
an order of magnitude estimate for cutoff effects were included
into the overall error budget. The dominant systematic uncertainty 
is the one due to the extrapolation in the light quark mass to
the physical point which RBC/UKQCD did with the help of a model
motivated and partly based on chiral perturbation theory. The model dependence
is estimated by comparing different ans\"atze for the mass extrapolation.

The ETM collaboration which
uses the twisted-mass discretization provides 
a comprehensive study of the systematics by presenting results for three
lattice spacings \cite{DiVita:2009by} and simulating at light pion masses
(down to $M_\pi=260$~MeV). This allows to constrain the chiral
extrapolation, using both SU(3) \cite{Gasser:1984ux} and SU(2)
\cite{Flynn:2008tg} chiral perturbation theory. Moreover, a rough estimate
for the size of the effects due to quenching the strange quark is given,
based on the comparison of the result for $\Nf=2$ dynamical quark flavours
\cite{Blossier:2009bx} with the one in the quenched approximation, obtained earlier
by the SPQcdR collaboration \cite{Becirevic:2004ya}. 
We note for completeness that ETM extrapolate their lattice results to the
point corresponding to $M_K^2$ and $M_\pi^2$ as defined at the end of 
Section~\ref{sec:Exp}. At the
current level of precision though this is expected to be a tiny effect.

We now compute the $N_f =2+1$ FLAG-average for $f_+(0)$
based on  FNAL/MILC 13 and RBC/UKQCD 12, which we consider uncorrelated, and 
for $\Nf =2$ the only result fulfilling the FLAG criteria is ETM 09A,
\bea\label{eq:fplus_direct}
f_+(0)\al=\al 0.9661(32)    \,, \hspace{1.7cm}(\mbox{direct},\,\Nf=2+1),  \\
f_+(0)\al=\al 0.9560(57)(62)\,, \hspace{1cm}(\mbox{direct},\,\Nf=2).\nonumber \eea
The brackets in the second line indicate the statistical and
systematic errors, respectively.
The dominant source of systematic uncertainty in these simulations of 
$f_+(0)$, the chiral extrapolation,
 will soon be removed by simulations with physical light quark masses
(see FNAL/MILC 13C~\cite{Gamiz:2013xxa} and RBC/UKQCD~\cite{JuettnerRBC:2013})

In the case of the ratio of decay constants
the data sets that meet the criteria formulated in the introduction 
are MILC 13A~\cite{Bazavov:2013cp} and 
HPQCD 13A~\cite{Dowdall:2013rya} with $N_f=2+1+1$, 
MILC 10~\cite{Bazavov:2010hj}, BMW 10~\cite{Durr:2010hr},  
HPQCD/UKQCD 07~\cite{Follana:2007uv} and 
RBC/UKQCD 12~\cite{Arthur:2012opa} 
(which is an update of RBC/UKQCD 10A~\cite{Aoki:2010dy})
 with $\Nf=2+1$ and ETM 09 \cite{Blossier:2009bx} with $\Nf=2$
dynamical flavours.
 
MILC 13A have determined the ratio of decay constants from a 
comprehensive set of ensembles of Highly Improved Staggered Quarks (HISQ) 
which have been taylored to reduce staggered taste breaking effects.
They have generated ensembles for four values of the lattice spacing
(0.06fm-0.15fm, scale set with $f_\pi$) and with
the Goldstone pion masses approximately tuned to the physical point which 
at least on their
finest lattice approximately agrees with the RMS pion mass (i.e. the difference in mass
between different \textit{pion species} which 
originates from staggered taste-splitting). 
Supplementary simulations with slightly heavier Goldstone pion mass
allow to extract the ratio of decay constants
 for the physical value of the light-quark
masses by means of polynomial interpolations. 
In a second step MILC extrapolates the data to the
continuum limit where eventually the ratio $\fKfpichargedr$ is extracted.
The final result of their analysis is 
$\fKfpichargedr=1.1947(26)(33)(17)(2)$ where the errors are statistical,
due to the continuum extrapolation, due to finite volume effects and 
due to electromagnetic effects. MILC has found an increase in the 
central value of the ratio when going from the 2nd finest to their finest 
ensemble
and from this observation they derive the quoted 0.28\% uncertainty in the
continuum extrapolation. They use NLO staggered chiral perturbation theory
to correct for finite volume effects and estimate the uncertainty
in this approach by comparing to the alternative correction
in NLO and NNLO continuum chiral perturbation theory.
Although MILC and HPQCD are independent collaborations, MILC shares its
gauge field ensembles with HPQCD 13A, whose study of $\fKfpichargedr$ is 
 therefore based on the same set of ensembles bar the one for the
finest lattice spacing ($a=$0.09fm-0.15fm, 
scale set with $f_{\pi^+}$ and relative scale
set with the Wilson flow~\cite{Luscher:2010iy,Borsanyi:2012zs})
supplemented by some simulation points with heavier quark masses.
HPQCD employed a global fit based on continuum
NLO SU(3) chiral perturbation theory for the decay constants 
supplemented by a model for higher order terms including 
discretisation and finite volume effects (61 parameters for 
39 data points supplemented by Bayesian priors). 
Their final result is 
$f_{K^\pm}/f_{\pi^\pm}=1.1916(15)(12)(1)(10)$, where the errors are 
statistical, due to the continuum extrapolation, due to finite volume 
effects and the last error contains the combined uncertainties from
the chiral extrapolation, the scale-setting uncertainty, the experimental
input in terms of $f_{\pi^+}$ and from the uncertainty in $m_u/m_d$.

Despite the large overlap in primary lattice data both collaborations 
arrive at surprisingly different error budgets.
In the preparation of this report we interacted with both collaborations trying
to understand the origin of the differences.
HPQCD is using a rather new method to set the relative lattice scale for their
ensembles which together with their more aggressive
binning of the statistical samples, could explain the reduction in 
statistical error by a factor of 1.7 compared to MILC.
Concerning the cutoff dependence, the finest lattice included into MILC's 
analysis is $a=0.06$fm while the finest lattice in HPQCD's case is $a=0.09$fm.
MILC estimates the residual systematic after extrapolating to the continuum
limit by taking the split between
the result of an extrapolation with up to quartic and only up to
 quadratic terms in $a$
as their systematic. HPQCD on the other hand models cutoff
effects within their global fit ansatz up to including terms in $a^8$.
In this way HPQCD arrives at a systematic error due to the continuum limit which
is smaller than MILC's estimate by about a factor 2.8.
HPQCD explains\footnote{Email exchange between HPQCD and FLAG.}
that in their setup,
despite lacking the information from the fine ensemble ($a=0.06$fm),
the approach to the continuum limit is reliably described by the 
chosen fit formula leaving no room for the shift in the result on the
finest lattice observed by MILC. They further explain that their different
way of setting the relative lattice scale leads to reduced
cutoff effects compared to MILC's study.
We now turn to finite volume effects which in the MILC result is the
second largest source of systematic uncertainty.
NLO staggered chiral perturbation theory (MILC) or
continuum chiral perturbation theory (HPQCD) was used for
correcting the lattice data towards the infinite volume limit. MILC then
compared the finite volume correction to the one obtained by
the NNLO expression and took the difference as their estimate for the
residual finite volume error.
In addition they checked the compatibility
of the effective theory predictions (NLO continuum, staggered and NNLO
continuum chiral perturbation theory) against lattice data of different
spacial extent.
The final verdict on the related residual systematic uncertainty 
on $\fKfpichargedr$ made by MILC is larger
by an order of magnitude than the one made by HPQCD.
We note that only HPQCD allows for taste-breaking terms in their fit-model
while MILC postpones such studies to future work.  

The above comparison shows that MILC and HPQCD have studied similar 
sources of systematic uncertainties, e.g. by varying parts of the analysis
procedure or by changing the functional form of a given fit ansatz.
One observation worth mentioning in this context is the way in which 
the resulting variations in the fit result are treated.
MILC tends to include the spread in central 
values from different ans\"atze into the systematic errors.
HPQCD on the other hand determines the final result and attached errors from 
preferred fit-ansatz and then confirms that it agrees within errors
with results from 
other ans\"atze without including the spreads into their error budget.
In this way HPQCD is lifting the calculation of $\fKfpichargedr$
to a new level of precision.  
FLAG is looking forward to independent confirmations of the result for
$\fKfpichargedr$ at the same level of precision. For now we will only
provide a range for the result for $\Nf=2+1+1$ that covers the result of both HPQCD 13A 
and MILC 13A,
\begin{equation}
\fKfpichargedr=1.194(5)\qquad (\textrm{our estimate, direct, }\Nf=2+1+1)
\end{equation}
Concerning simulations with $N_f=2+1$, MILC~10 and HPQCD/UKQCD~07 are based
on staggered fermions, BMW~10 has used improved Wilson fermions
and RBC/UKQCD~12's result is based on the domain wall formulation. 
For $N_f=2$ ETM has simulated twisted-mass fermions.
In contrast to MILC 13A all these latter simulations are for unphysically heavy
quark masses (corresponding to smallest pion masses in the range
 240-260MeV in the case of MILC 10, HPQCD/UKQCD 07 and ETM 09 and
 around 170MeV for RBC/UKQCD 12) and therefore
 slightly more sophisticated extrapolations needed to be controlled.
Various ans\"atze for the mass and cutoff dependence
comprising SU(2) and SU(3) chiral perturbation theory or simply polynomials
were used and compared in order to estimate the model dependence.

We now provide the FLAG average for these data.
While BMW 10 and RBC/UKQCD 12 are entirely independent computations,
subsets of the MILC gauge ensembles used by MILC 10 and HPQCD/UKQCD 07 
are the same. MILC 10 is certainly based on a larger and more
advanced set of gauge configurations than HPQCD/UQKCD 07. This allows
them for a more reliable estimation of systematic effects. 
In this situation we consider only their statistical but 
not their systematic uncertainties to be correlated. 
For $N_f=2$ the FLAG average is just the 
result by ETM 09 and this is illustrated in terms of the
vertical grey band in the r.h.s. panel of Figure~\ref{fig:lattice data}. 
For the purpose of this plot only,
the isospin correction has been removed along the lines laid out earlier.
For the average indicated in the case of $N_f=2+1$ we take
the original data of BMW 10, HPQCD/UKQCD 07 and RBC/UKQCD 12 and use the MILC 10
result as computed above. The 
resulting fit is of good quality, with $f_K/f_\pi=1.194(4)$ and 
$\chi^2/dof=0.4$. 
The systematic errors of the individual data
sets are larger for MILC 10, BMW 10, 
HPQCD/UKQCD 07 and RBC/UKQCD 12, respectively, and 
following again the prescription of section
\ref{sec:error_analysis} we replace the error by the smallest one of these
leading to  
$f_K / f_\pi=  1.194(5)$ for $N_f=2+1$.

Before determining the average for $f_{K^\pm}/f_{\pi^\pm}$ 
which should be used for applications to Standard Model 
phenomenology we apply the isospin
correction individually to all those results which have been published 
in the isospin-symmetric limit, i.e.
BMW 10, HPQCD/UKQCD07 and RBC/UKQCD 12. 
To this end  we invert Equation~(\ref{eq:convert}) and use
\begin{equation}
 \begin{array}{rcl}
	 \delta_{\rm SU(2)} &\approx& \sqrt{3}\,\epsilon_{\rm SU(2)}\,\Big[
	 -\frac 43 (f_K/f_\pi-1)
	 +\frac 2{3(4\pi)^2 f_0^2}\big(M_K^2-M_\pi^2-
	 	M_\pi^2\ln\frac{M_K^2}{M_\pi^2}\big)\Big]\,.
 \end{array}
\end{equation}
The results are:
\begin{center}
\begin{tabular}{llll}
\hline\hline\\[-4mm]
		&$f_K/f_\pi$	&$\delta_{\rm SU(2)}$&$f_{K^\pm}/f_{\pi^\pm}$\\
\hline\\[-4mm]
HPQCD/UKQCD 07	&1.189(2)(7)	&-0.0040(7)&1.187(2) (2)(7)\\
BMW 10		&1.192(7)(6)	&-0.0041(7)&1.190(7) (2)(6)\\
RBC/UKQCD 12	&1.199(12)(14)	&-0.0043(9)&1.196(12)(2)(14)\\
\hline\hline
\end{tabular}
\end{center}
As before, in the last column the first error is statistical and the
second error is due to the isospin correction. 
Using these results we obtain
\begin{equation}
\label{eq:fKfpi_direct_broken} 
\begin{array}{rcll}
f_{K^\pm} / f_{\pi^\pm}&=& 1.192(5), 	  &(\mbox{direct},\, \Nf=2+1)\,,\\
f_{K^\pm} / f_{\pi^\pm}&=& 1.205(6)(17)\,,&(\mbox{direct},\,
\Nf=2)\,, 
\end{array}
\end{equation}
for QCD with broken isospin.

It is instructive to convert the above results for $f_+(0)$ and 
$\fKfpichargedr$
into a corresponding range for the CKM matrix elements $|V_{ud}|$ and
$|V_{us}|$, using the relations (\ref{eq:products}). Consider first the
results for $\Nf=2+1$.  The range for $f_+(0)$ in (\ref{eq:fplus_direct})
is mapped into the interval 
$|V_{us}|=0.2239(7)$, depicted as a horizontal
green band in Figure \ref{fig:VusVersusVud}, while the one for $\fKfpichargedr$
in (\ref{eq:fKfpi_direct_broken}) is converted into $|V_{us}|/|V_{ud}|= 0.2314(11)$,
shown as a tilted 
green band. The smaller green ellipse is the intersection of these two bands.
\begin{figure}[t]\vspace{0.2cm}\centering
\includegraphics[width=12cm]{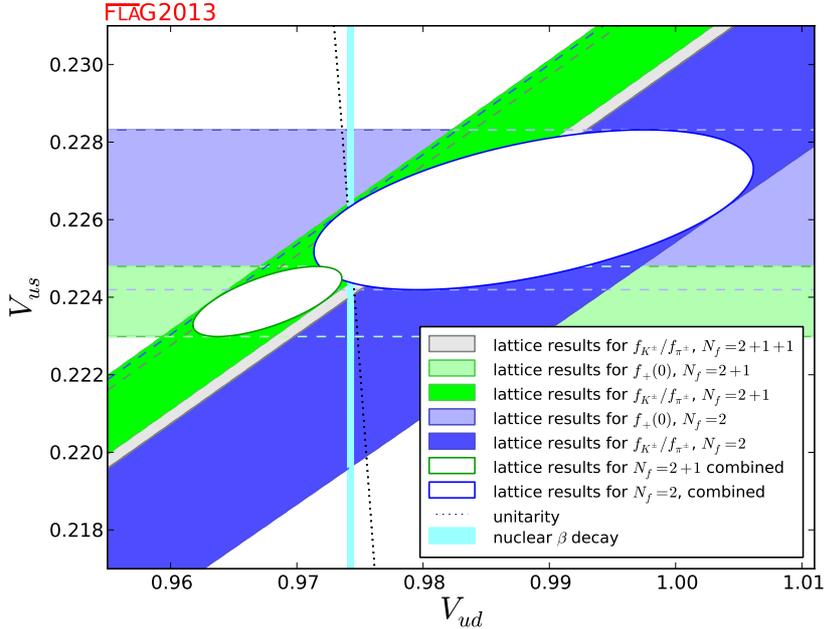}  
\caption{\label{fig:VusVersusVud}The plot compares the information for
  $|V_{ud}|$, $|V_{us}|$ obtained on the lattice with the experimental
  result extracted from nuclear $\beta$ transitions. The dotted arc
  indicates the correlation between $|V_{ud}|$ and $|V_{us}|$ that follows
  if the three-flavour CKM-matrix is unitary. }
\end{figure}
More precisely, it represents the 39\% likelihood contour (note also
that the ellipses shown in Fig. 5 of Ref.~\cite{Colangelo:2010et} have to be
interpreted as 39\% likelihood contours), obtained by treating the
above two results as independent measurements.  Values of $|V_{us}|$,
$|V_{ud}|$ in the region enclosed by this contour are consistent with
the lattice data for $\Nf=2+1$, within one standard deviation. In
particular, the plot shows that the nuclear $\beta$ decay result for
$|V_{ud}|$ is in good agreement with these data.  We note that with
respect to the previous edition of the FLAG review the reanalysis
including new results has moved the ellipse representing QCD with
$N_f=2+1$ slightly down and to the left.

Repeating the exercise for $\Nf=2$ leads to the larger blue
ellipse. The figure indicates a slight tension between the 
$N_f=2$ and $N_f=2+1$ results, which, at the current level of precision
is not visible if considering the $N_f=2$ and $N_f=2+1$ 
results for $f_+(0)$ and $\fKfpichargedr$
in Figure~\ref{fig:lattice data} on their own.
It remains to be seen if this is a first indication of the effect of 
quenching the strange quark.

In the case of $\Nf=2+1+1$ only results for
$\fKfpichargedr$ are without red tags. 
In this case we have therefore only plotted the corresponding band 
for $|V_{us}|$ from $f_{K^\pm}/f_{\pi^\pm}$ corresponding to $|V_{us}|/|V_{ud}|=0.2310(11)$.

\subsection{Testing the Standard Model}\label{sec:testing}

\begin{table}[t]\centering
\noindent
\begin{tabular*}{\textwidth}[l]{@{\extracolsep{\fill}}lrlcl}
Collaboration & Ref. &$\Nf$&from&\rule{0.8cm}{0cm}$|V_{us}|$\\
&&&& \\[-2ex]
\hline \hline &&&&\\[-2ex]
HPQCD 13A &\cite{Dowdall:2013rya}&$2+1+1$&$\fKfpichargedr$ \rule{0cm}{0.45cm} &0.2255(5)(3)\\
MILC 13A &\cite{Bazavov:2013cp}&$2+1+1$&$\fKfpichargedr$ \rule{0cm}{0.45cm} &0.2249(6)(7)\\
&&&& \\[-2ex]
 \hline
&&&& \\[-2ex]
RBC/UKQCD 13 &\cite{Boyle:2013gsa}&$2+1$&$f_+(0)$         \rule{0cm}{0.45cm} &0.2237(7)(7)\\
MILC 12 &\cite{Bazavov:2012cd}&$2+1$&$f_+(0)$         \rule{0cm}{0.45cm} &0.2238(7)(8)\\
MILC 10 &\cite{Bazavov:2010hj}&$2+1$&$\fKfpichargedr$ \rule{0cm}{0.45cm} &0.2249(5)(9)\\
RBC/UKQCD 10A  &\cite{Aoki:2010dy}   &$2+1$&$\fKfpichargedr$ \rule{0cm}{0.45cm} &0.2246(22)(25)\\ 
BMW 10 &\cite{Durr:2010hr}  & $2+1$ \rule{0cm}{0.45cm}& $\fKfpichargedr$ & $0.2259(13)(12)$\\
HPQCD/UKQCD 07 &\cite{Follana:2007uv}\rule{0cm}{0.4cm}& $2+1$ & $\fKfpichargedr$&  $  0.2264(5)(13)$\\
&&&& \\[-2ex]
 \hline
&&&& \\[-2ex]
 ETM 09  &\cite{Blossier:2009bx}\rule{0cm}{0.4cm}&2&$\fKfpichargedr$& $ 0.2231 (11) (31)$\\
 ETM 09A & \cite{Lubicz:2009ht}\rule{0cm}{0.4cm}&2&$f_+(0)$&   $ 0.2263 (14) (15)$\\
&&&& \\[-2ex]
 \hline \hline 
\end{tabular*}
\caption{\label{tab:Vus}Values of $|V_{us}|$ obtained from lattice
  determinations of $f_+(0)$ or $\fKfpichargedr$ with CKM unitarity. The first
  (second) number in brackets represents the statistical (systematic)
  error.} 
  \end{table} 
  
In the Standard Model, the CKM matrix is unitary.  In particular, the
elements of the first row obey \be\label{eq:CKM unitarity}|V_u|^2\equiv
|V_{ud}|^2 + |V_{us}|^2 + |V_{ub}|^2 = 1\fs\ee The tiny contribution from
$|V_{ub}|$ is known much better than needed in the present context: 
{$|V_{ub}|= 4.15 (49) \cdot 10^{-3}$
}~\cite{Beringer:1900zz}. 
In the following, we
first discuss the evidence for the validity of the relation (\ref{eq:CKM
  unitarity}) and only then use it to analyse the lattice data within the 
Standard Model.

In Figure \ref{fig:VusVersusVud}, the correlation between $|V_{ud}|$ and
$|V_{us}|$ imposed by the unitarity of the CKM matrix is indicated by a
dotted arc (more precisely, in view of the uncertainty in $|V_{ub}|$, the
correlation corresponds to a band of finite width, but the effect is too
small to be seen here). 
The plot shows that there is a slight tension with unitarity
in the data for $N_f = 2 + 1$: Numerically, the outcome for the sum of the squares of the first 
row of the CKM matrix reads $|V_u|^2 = 0.987(10)$. Still, it is fair to say that at this level 
the Standard Model passes a nontrivial test that exclusively involves lattice data and 
well-established kaon decay branching ratios. Combining the lattice results for 
$f_+(0)$ and $\fKfpichargedr$ in (\ref{eq:fplus_direct}) and (\ref{eq:fKfpi_direct_broken})
with the $\beta$ decay value of $|V_{ud}|$ quoted in (\ref{eq:Vud beta}), the test sharpens 
considerably: the lattice result for $f_+(0)$ leads to $|V_u|^2 = 0.9993(5)$, 
while the one for $\fKfpichargedr$ implies $|V_u|^2 = 1.0000(6)$, thus confirming 
CKM unitarity at the permille level. 

Repeating the analysis for $N_f = 2$, we find 
$|V_u|^2 = 1.029(35)$ with the lattice data alone. This number is fully compatible 
with 1, in accordance with the fact that the dotted curve penetrates the 
blue contour. Taken by themselves, these results are perfectly consistent 
with the value of $|V_{ud}|$ found in nuclear $\beta$ decay: combining 
this value with the data on $f_+(0)$ yields $|V_u|^2=1.0004(10)$, combining it 
with the data on $\fKfpichargedr$ gives $|V_u|^2= 0.9989(16)$. With respect to the first
edition of the FLAG report the ellipse for $N_f=2$
has moved slightly to the left because we have now taken into account
isospin breaking effects.

For $\Nf=2+1+1$ we can carry out the test of unitarity only with input from 
$\fKfpichargedr$ which leads to $|V_u|^2=0.9998(7)$.

\begin{figure}[t]
\psfrag{y}{\tiny $\star$}
\begin{center}
\vspace{0.5cm} 
\includegraphics[width=13cm]{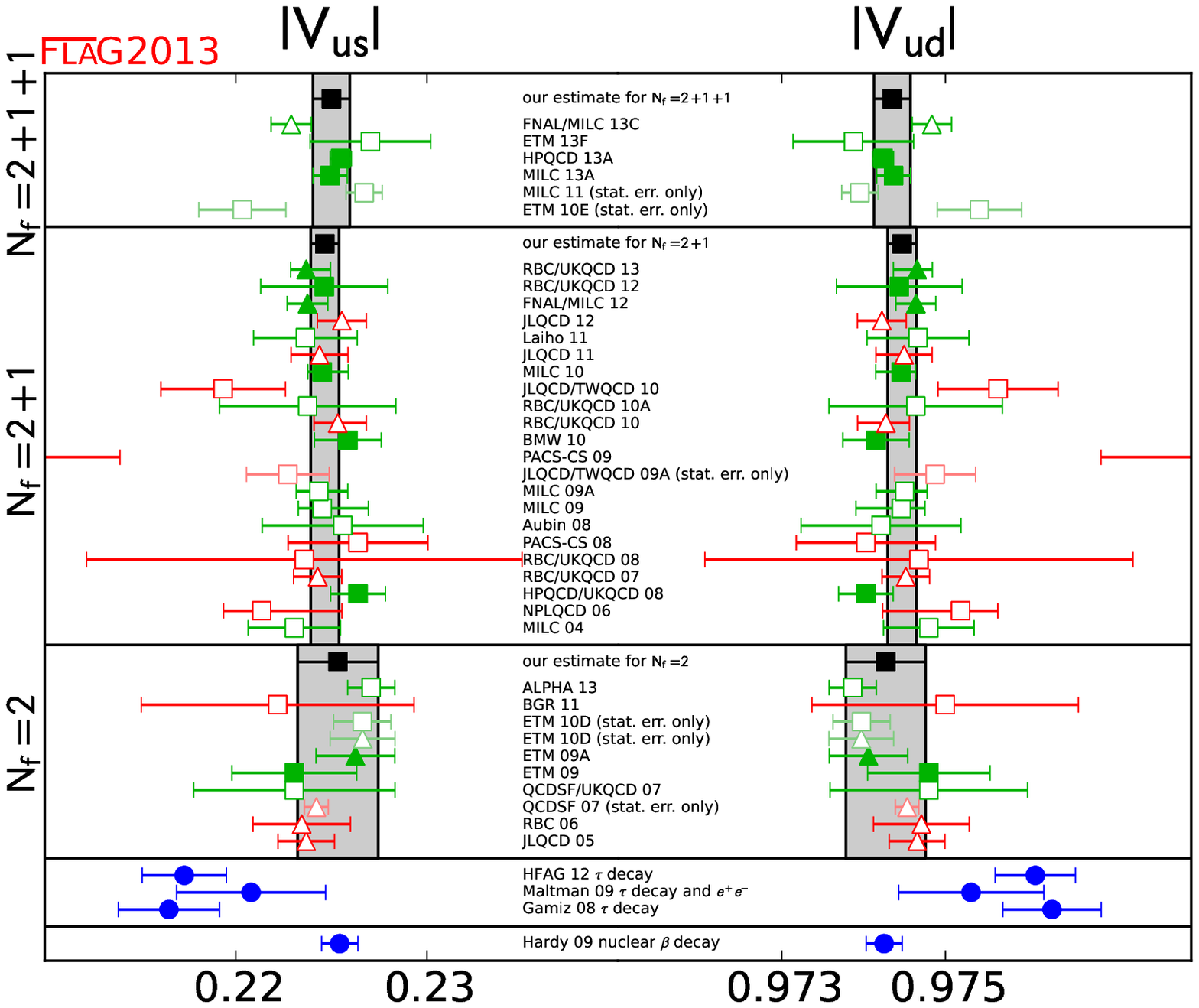}
\end{center}
\mbox{}\\[-4ex]
{\footnotesize $^\star$ Estimates obtained from an analysis of the
  lattice data within the Standard Model, see section \ref{sec:SM}.}

\vspace{-2.72cm}\hspace{9.4cm}\parbox{6cm}{\sffamily\tiny  \cite{Amhis:2012bh}\\

\vspace{-1.34em}\cite{Maltman:2009bh}\\

\vspace{-1.34em}\cite{Gamiz:2007qs}\\

\vspace{-0.95em}\hspace{0em}\cite{Hardy:2008gy}}
\vspace{2cm}

\caption{\label{fig:Vus Vud} Results for $|V_{us}|$ and $|V_{ud}|$ that
  follow from the lattice data for $f_+(0)$ (triangles) and $\fKfpichargedr$
  (squares), on the basis of the assumption that the CKM matrix is
  unitary. 
The black squares and the grey bands represent our estimates,
  obtained by combining these two different ways of measuring $|V_{us}|$
  and $|V_{ud}|$ on a lattice.
  For comparison, the figure also indicates
  the results obtained if the data on nuclear $\beta$ decay and $\tau$
  decay are analysed within the Standard Model. }
\end{figure}  

Note that the above tests also offer a check of the basic hypothesis that
underlies our analysis: we are assuming that the weak interaction between
the quarks and the leptons is governed by the same Fermi constant as the
one that determines the strength of the weak interaction among the leptons
and determines the lifetime of the muon. In certain modifications of the
Standard Model, this is not the case. In those models it need not be true
that the rates of the decays $\pi\rightarrow \ell\nu$,
$K\rightarrow\ell\nu$ and $K\rightarrow \pi\ell \nu$ can be used to
determine the matrix elements $|V_{ud}f_\pi|$, $|V_{us}f_K|$ and
$|V_{us}f_+(0)|$, respectively and that $|V_{ud}|$ can be measured in
nuclear $\beta$ decay. The fact that the lattice data are consistent with
unitarity and with the value of $|V_{ud}|$ found in nuclear $\beta$ decay
indirectly also checks the equality of the Fermi constants.

\subsection{Analysis within the Standard Model} \label{sec:SM} 
 
The Standard Model implies that the CKM matrix is unitary. The precise
experimental constraints quoted in (\ref{eq:products}) and the unitarity
condition (\ref{eq:CKM unitarity}) then reduce the four quantities
$|V_{ud}|,|V_{us}|,f_+(0),\fKfpichargedr$ to a single unknown: any one of these
determines the other three within narrow uncertainties.
 
Figure \ref{fig:Vus Vud} shows that the results obtained for $|V_{us}|$ and
$|V_{ud}|$ from the data on $\fKfpichargedr$ (squares) are quite consistent
with the determinations via $f_+(0)$ (triangles). In order to calculate
the corresponding average values, we restrict ourselves to those
determinations that we have considered best in section \ref{sec:Direct}.
The corresponding results for $|V_{us}|$ are listed in Table~\ref{tab:Vus}
(the error in the experimental numbers used to convert the values of
$f_+(0)$ and $\fKfpichargedr$ into values for $|V_{us}|$ is included in the
statistical error).
  
We consider the fact that the results from the five $\Nf=2+1$ data sets
FNAL/MILC 12~\cite{Bazavov:2012cd}, RBC/UKQCD 13~\cite{Boyle:2013gsa},
RBC/UKQCD 12~\cite{Arthur:2012opa},
BMW 10 \cite{Durr:2010hr}, MILC 10
\cite{Bazavov:2010hj} and HPQCD/UKQCD 07 \cite{Follana:2007uv} are
consistent with each other to be an important reliability test of the
lattice work. Applying the prescription of section
\ref{sec:error_analysis}, where we consider MILC 10, FNAL/MILC 12 and HPQCD/UKQCD 07
on the one hand and RBC/UKQCD 12 and RBC/UKQCD 13 on the other hand,
as mutually statistically correlated 
since the analysis in the two cases starts from partly the same set of gauge-ensembles, 
we arrive at $|V_{us}| = 0.2247(7)$ with $\chi^2/dof=0.8$. 
This result is indicated on the left hand side of
Fig.~\ref{fig:Vus Vud} by the narrow vertical band. The
value for $\Nf=2$,  $|V_{us}|=
0.2253(21)$, with $\chi^2/dof=0.9$, where we have considered ETM 09 and
ETM 09A as statistically correlated is also indicated by a band.
For $\Nf=2+1+1$ we only consider the data for $\fKfpichargedr$ yielding
$|V_{us}|=0.2251(10)$.
The figure shows that the result obtained for the data with
$\Nf=2$, $\Nf=2+1$ and $\Nf=2+1+1$ are perfectly consistent.
 
Alternatively, we can solve the relations for $|V_{ud}|$ instead of
$|V_{us}|$. Again, the result $|V_{ud}|=0.97434(22)$
which follows from the
lattice data with $\Nf=2+1+1$ is perfectly consistent with the values
$|V_{ud}|=0.97447(18)$ and 
$|V_{ud}|=0.97427(49)$ obtained from those with $\Nf=2+1$ and $\Nf=2$, respectively.  
The reduction of
the uncertainties in the result for $|V_{ud}|$ due to CKM unitarity is to
be expected from Figure \ref{fig:VusVersusVud}: the unitarity condition
reduces the region allowed by the lattice results to a nearly vertical
interval.

Next, we determine the value of $f_+(0)$ that follows from the lattice data
within the Standard Model. Using CKM unitarity to convert the lattice
determinations of $\fKfpichargedr$ into corresponding values for $f_+(0)$ and
then combining these with the direct determinations of $f_+(0)$, we find
$f_+(0)= 0.9634(32)$ from the data with $\Nf=2+1$ and $f_+(0)= 0.9595(90)$
for $\Nf=2$. In the case $\Nf=2+1+1$ we obtain $f_+(0)=0.9611(47)$.

Finally, we work out the analogous Standard Model fits for $\fKfpichargedr$,
converting the direct determinations of $f_+(0)$ into corresponding values
for $\fKfpichargedr$ and combining the outcome with the direct determinations of
that quantity. The results read $\fKfpichargedr=1.197(4)$ for $\Nf=2+1$ and
$\fKfpichargedr= 1.192(12) $ for $\Nf=2$, respectively.

\begin{table}[thb]
\centering
\begin{tabular*}{\textwidth}[l]{@{\extracolsep{\fill}}llllll}
\rule[-0.2cm]{0cm}{0.5cm}& Ref. & \rule{0.3cm}{0cm} $|V_{us}|$&\rule{0.3cm}{0cm} $|V_{ud}|$&\rule{0.25cm}{0cm} $f_+(0)$& \rule{0.2cm}{0cm} $\fKfpichargedr$\\
&&&& \\[-2ex]
\hline \hline
&&&& \\[-2ex]
$\Nf= 2+1+1$& &\rule{0cm}{0.4cm}0.2251(10)& 0.97434(22)  & 0.9611(47)   & 1.194(5)\\
&&&& \\[-2ex]
$\Nf= 2+1$&   &\rule{0cm}{0.4cm}0.2247(7)& 0.97447(18)  & 0.9634(32)   & 1.197(4)\\
&&&& \\[-2ex]
\hline
&&&& \\[-2ex]
$\Nf=2$ & &\rule{0cm}{0.4cm}0.2253(21) &0.97427(49)  &0.9595(90) &1.192(12)\\
&&&& \\[-2ex]
\hline\hline
&&&& \\[-2ex]
$\beta$ decay &\cite{Hardy:2008gy}&0.22544(95)& 0.97425(22) & 0.9595(46)&
1.1919(57) \\ 
&&&& \\[-2ex]
$\tau$ decay &\cite{Gamiz:2007qs}&0.2165(26)&0.9763(6)& 0.999(12)&
1.244(16)\\ 
&&&& \\[-2ex]
$\tau$ decay  &\cite{Maltman:2009bh}&0.2208(39)&0.9753(9)& 0.980(18)&
1.218(23)\\ 
&&&& \\[-2ex]
\hline\hline
\end{tabular*}
\caption{\label{tab:Final results}The upper half of the table shows our
  final results for $|V_{us}|$, $|V_{ud}|$,  $f_+(0)$ and $\fKfpichargedr$,
  which are obtained by analysing the lattice 
  data within the Standard Model. For comparison, the lower half lists the
  values that follow if the lattice results are replaced by the
  experimental results on nuclear $\beta$ decay and $\tau$ decay,
  respectively.}
\end{table} 

The results obtained by analysing the lattice data in the framework of the
Standard Model are collected in the upper half of Table \ref{tab:Final
  results}.  In the lower half of this table, we list the analogous
results, found by working out the consequences of CKM-unitarity for the
experimental values of $|V_{ud}|$ and $|V_{us}|$ obtained from nuclear
$\beta$ decay and $\tau$ decay, respectively. The comparison shows that the
lattice result for $|V_{ud}|$ not only agrees very well with the totally
independent determination based on nuclear $\beta$ transitions, but is also
remarkably precise. On the other hand, the values of $|V_{ud}|$, $f_+(0)$
and $\fKfpichargedr$  which follow from the $\tau$ decay data if the Standard
Model is assumed to be valid, are not in good agreement with the lattice
results for these quantities. The disagreement is reduced considerably if
the analysis of the $\tau$ data is supplemented with experimental results
on electroproduction \cite{Maltman:2009bh}: the discrepancy then amounts to
little more than one standard deviation.  
\subsection{Direct determination of $f_K$ and $f_\pi$}\label{sec:fKfpi}
It is useful for flavour physics to provide not only the lattice average of $f_K / f_\pi$, but also the average of  the decay constant $f_K$. 
Indeed, the $\Delta S = 2$ hadronic matrix element for neutral kaon mixing is generally parameterized by $M_K$, $f_K$ and the kaon bag parameter $B_K$. 
The knowledge of both $f_K$ and $B_K$ is therefore crucial for a precise theoretical determination of the CP-violation parameter $\epsilon_K$ and for the constraint on the apex of the CKM unitarity triangle.

The case of the decay constant $f_\pi$ is somehow different, since the experimental value of this quantity is often used for setting the scale in lattice QCD
(see Appendix A.2).
However, the physical  scale can be set in different ways, namely by using as input the mass of the $\Omega$-baryon ($m_\Omega$) or the $\Upsilon$-meson spectrum ($\Delta M_\Upsilon$), which are less sensitive to the uncertainties of the chiral extrapolation in the light-quark mass with respect to $f_\pi$. 
In such cases the value of the decay constant $f_\pi$ becomes a direct prediction of the lattice QCD simulations.
It is therefore interesting to provide also the average of the decay constant $f_\pi$, obtained when the physical scale is set through another hadron observable, in order to check the consistency of different scale setting procedures.

Our compilation of the values of $f_\pi$ and $f_K$ with the corresponding colour code is presented in Table \ref{tab:FK Fpi}.
With respect to the case of $f_K / f_\pi$ we have added two columns indicating which quantity is used to set the physical scale and the possible use of a renormalization constant for the axial current.
Indeed, for several lattice formulations the use of the nonsinglet axial-vector Ward identity allows to avoid the use of any renormalization constant.

One can see that the determinations of $f_\pi$ and $f_K$ suffer from larger uncertainties with respect to the ones of the ratio $f_K / f_\pi$, which is less sensitive to various systematic effects (including the uncertainty of a possible renormalization constant) and, moreover,  is not so 
exposed to the uncertainties of the procedure used to set the physical scale.

According to the FLAG rules three data sets can form the average of $f_\pi$ and $f_K$ for $N_f = 2 + 1$: RBC/UKQCD 12 \cite{Arthur:2012opa} (update of RBC/UKQCD 10A), HPQCD/UKQCD 07 \cite{Follana:2007uv} and MILC 10 \cite{Bazavov:2010hj}, which is the latest update of the MILC program.
\footnote{Since the MILC result is obtained for a charged kaon, we remove the isospin-breaking effect according to the formula $f_K = f_{K^+} (1 - \delta_{\rm SU(2)} / 2)$, valid at NLO in ChPT, with $\delta_{SU(2)}$ for MILC 10 computed using eq.~(\ref{eq:removeiso}).} We consider HPQCD/UKQCD 07 and MILC 10 as 
statistically correlated and use the prescription of \ref{sec:error_analysis} to
form an average.
For $N_f = 2$ the average cannot be formed for $f_\pi$, and only one data set (ETM 09) satisfies the FLAG rules in the case of $f_K$.
Following the discussion around the $N_f=2+1+1$ result for $f_{K^\pm}/f_{\pi^\pm}$ we refrain
from providing a FLAG-average for $f_K$ for this case.

Thus, our estimates (in the isospin-symmetric limit of QCD) read
\begin{eqnarray}
  \label{eq:fPi}
  f_\pi & = & 130.2 ~ (1.4) ~ \mbox{MeV} \qquad \qquad (N_f = 2 + 1), \\ \nonumber \\
   \label{eq:fK}
  f_K & = & 156.3 ~ (0.9) ~ \mbox{MeV} \qquad \qquad (N_f = 2 + 1), \\ \nonumber
  f_K & = & 158.1 ~ (2.5) ~ \mbox{MeV} \qquad \qquad (N_f = 2).
 \end{eqnarray}
The lattice results of Table \ref{tab:FK Fpi} and our estimates (\ref{eq:fPi}-\ref{eq:fK}) are reported in Fig.~\ref{fig:latticedata_decayconstants}. 
The latter ones compare positively within the errors with the latest experimental determinations of $f_\pi$ and $f_K$ from the PDG:
 \begin{eqnarray}
     \label{eq:fps_PDG}
     f_\pi^{(PDG)} = 130.41 ~ (0.20) ~ \mbox{MeV} \qquad , \qquad f_K ^{(PDG)}= 156.1 ~ (0.8) ~ \mbox{MeV} ~ ,
 \end{eqnarray}
which, we recall, do not correspond however to pure QCD results in the isospin-symmetric limit. Moreover the values of $f_\pi$ and $f_K$ quoted by the PDG are 
obtained assuming Eq.~(\ref{eq:products}) for the value of $|V_{ud}|$ and 
adopting the RBC-UKQCD 07 result for $f_+(0)$.

\begin{table}[!htb]
       {\centering
\vspace{1.5cm}{\footnotesize\noindent
\begin{tabular*}{\textwidth}[l]{@{\extracolsep{\fill}}l@{\hspace{1mm}}r@{\hspace{1mm}}l@{\hspace{1mm}}l@{\hspace{1mm}}l@{\hspace{1mm}}l@{\hspace{1mm}}l@{\hspace{3mm}}l@{\hspace{1mm}}l@{\hspace{1mm}}l@{\hspace{5mm}}l@{\hspace{1mm}}l}
Collaboration & Ref. & $\Nf$ &
\hspace{0.15cm}\begin{rotate}{40}{publication status}\end{rotate}\hspace{-0.15cm}&
\hspace{0.15cm}\begin{rotate}{40}{chiral extrapolation}\end{rotate}\hspace{-0.15cm}&
\hspace{0.15cm}\begin{rotate}{40}{continuum extrapolation}\end{rotate}\hspace{-0.15cm}&
\hspace{0.15cm}\begin{rotate}{40}{finite volume errors}\end{rotate}\hspace{-0.15cm}& 
\hspace{0.15cm}\begin{rotate}{40}{renormalization}\end{rotate}\hspace{-0.15cm}&
\hspace{0.05cm}\begin{rotate}{40}{physical scale}\end{rotate}\hspace{-0.15cm}&\rule{0cm}{0cm}
\hspace{0.0cm}\begin{rotate}{40}{SU(2) breaking}\end{rotate}\hspace{-0.15cm}&\rule{0.5cm}{0cm}
$f_\pi$&\rule{0.5cm}{0cm}$f_K$ \\
&&&&&&& \\[-0.1cm]
\hline
\hline
&&&&&&& \\[-0.1cm]
HPQCD 13A&\cite{Dowdall:2013rya}&2+1+1&\gA&\good&\soso&\good&na&$f_\pi$&\checkmark&--&155.37(20)(28)\\[-1mm]
ETM 10E &\cite{Farchioni:2010tb}&2+1+1&\rC&\soso&\soso&\soso&na&$f_\pi$&&--&160(2)\\[-1mm]
&&&&&&& \\[-0.1cm]
\hline
&&&&&&& \\[-0.1cm]
RBC/UKQCD 12   &\cite{Arthur:2012opa}&2+1&\gA&\tbg&\soso&\good&NPR&$m_\Omega$ && $127(3)(3)$&$152(3)(2)$\\
Laiho 11       &\cite{Laiho:2011np}   &2+1&\rC&\soso&\soso&\soso&na&$\dagger$ &\checkmark& $130.53(87)(210)$&$156.8(1.0)(1.7)$\\
MILC 10 &\cite{Bazavov:2010hj}&2+1&\rC&\soso&\good&\good&na&$\dagger$ &\checkmark&129.2(4)(14)&--\\
MILC 10 &\cite{Bazavov:2010hj}&2+1&\rC&\soso&\good&\good&na&$f_\pi$ &\checkmark&--          &156.1(4)($^{+6}_{-9}$)\\
JLQCD/TWQCD 10 &\cite{Noaki:2010zz}&2+1&\rC&\soso&\tbr&\tbg&na&$m_\Omega$&&118.5(3.6)$_{\rm stat}$&145.8(2.7)$_{\rm stat}$\\
RBC/UKQCD 10A  &\cite{Aoki:2010dy} &2+1&\gA&\soso&\soso&\good&NPR&$m_\Omega$&&124(2)(5)&149(2)(3)\\
PACS-CS 09     &\cite{Aoki:2009ix} &2+1&\gA&\good&\tbr&\tbr  &NPR &$m_\Omega$&&124.6(8.6)(0.9) & 166.1(3.4)(1.2)\\
JLQCD/TWQCD 09A&\cite{JLQCD:2009sk} &2+1&\rC&\soso&\tbr&\tbr &na&$f_\pi$&&--&157.3(5.5)$_{\rm stat}$\\
MILC 09A &\cite{Bazavov:2009fk}&2+1&\rC&\soso&\tbg&\tbg &na&$\Delta M_\Upsilon$ &\checkmark&128.0(0.3)(2.9)&          153.8(0.3)(3.9)\\
MILC 09A &\cite{Bazavov:2009fk}&2+1&\rC&\soso&\tbg&\tbg &na&$f_\pi$&\checkmark&--&156.2(0.3)(1.1)\\
MILC 09 &\cite{Bazavov:2009bb}&2+1&\gA&\soso&\tbg&\tbg &na&$\Delta M_\Upsilon$&\checkmark&128.3(0.5)($^{+2.4}_{-3.5}$)&154.3(0.4)($^{+2.1}_{-3.4}$) \\
MILC 09 &\cite{Bazavov:2009bb}&2+1&\gA&\soso&\tbg&\tbg &na&$f_\pi$&\checkmark&&156.5(0.4)($^{+1.0}_{-2.7}$)\\
Aubin 08       &\cite{Aubin:2008ie} &2+1&\rC&\soso&\soso&\soso&na&$\Delta M_\Upsilon$     &\checkmark& 129.1(1.9)(4.0)   & 153.9(1.7)(4.4)  \\
PACS-CS 08, 08A&\cite{Aoki:2008sm, Kuramashi:2008tb} &2+1&\gA&\tbg&\tbr&\tbr&1lp&$m_\Omega$&&134.0(4.2)$_{\rm stat}$& 159.4(3.1)$_{\rm stat}$\\
RBC/UKQCD 08   &\cite{Allton:2008pn} &2+1&\gA&\soso&\tbr&\tbg&NPR&$m_\Omega$&&124.1(3.6)(6.9) &        149.6(3.6)(6.3)\\
HPQCD/UKQCD 07 &\cite{Follana:2007uv}&2+1&\gA&\soso&\tbg&\soso&na&$\Delta M_\Upsilon$&& 132(2)                &   157(2)\\
MILC 04 &\cite{Aubin:2004fs}&2+1&\gA&\soso&\soso&\soso&na&$\Delta M_\Upsilon$&\checkmark&129.5(0.9)(3.5)     &     156.6(1.0)(3.6)\\[-1mm]
&&&&&&& \\[-0.1cm]
\hline
&&&&&&& \\[-0.1cm]
TWQCD 11       &\cite{Chiu:2011bm}   &2 &\oP&\tbg&\tbr&\tbr&na&$r_0^\ast$&&127.3(1.7)(2.0)$^{\ast\ast}$&--\\
ETM 09         &\cite{Blossier:2009bx}         &2 &\gA&\soso&\tbg&\soso&na&$f_\pi$&& --&158.1(0.8)(2.0)(1.1)$^{\dagger\dagger}$\\
JLQCD/TWQCD 08A&\cite{Noaki:2008iy} &2  &\gA  &\soso &\tbr   &\tbr &na&$r_0$&&119.6(3.0)($^{+6.5}_{-1.0}$)$^{\ast\ast}$&--\\[-1mm]
&&&&&&& \\[-0.1cm]
\hline
\hline
&&&&&&& \\[-0.1cm]
\end{tabular*}}\\[-2mm]
}

\begin{minipage}{\linewidth}
\footnotesize The label "na" indicates the lattice calculations which do not require the use of any renormalization constant for the axial current, while the label "NPR" ("1lp") signals the use of a renormalization constant calculated nonperturbatively (at one-loop order in perturbation theory).  
\begin{itemize}
{\footnotesize 
\item[$^{\dagger}$] The ratios of lattice spacings within the ensembles were determined using the quantity $r_1$. 
The conversion
to physical units was made basing on ref.~\cite{Davies:2009tsa} and we note that such a determination depends on the
experimental value of the pion decay constant\\[-5mm]
\item[$^{\dagger\dagger}$] Errors are (stat+chiral)($a\neq 0$)(finite size).
\\[-5mm]
\item[$^\ast$] The ratio $f_\pi/M_\pi$ was used as experimental input to fix the
light-quark mass.
\\[-5mm]
\item[$^{\ast\ast}$] $L_{\rm min}<2$fm in these simulations.
\\[-5mm]
}
\end{itemize}
\end{minipage}
\caption{Colour code for the lattice data on $f_\pi$ and $f_K$ 
together with information on the way the lattice spacing was
converted to physical units and on whether or not an isospin-breaking
correction has been applied (using chiral perturbation theory)
to the quoted result. The numerical values are listed in MeV units. 
\hfill}\label{tab:FK Fpi}

\end{table}

\begin{figure}[!htb]
\begin{center}
\includegraphics[height=10cm]{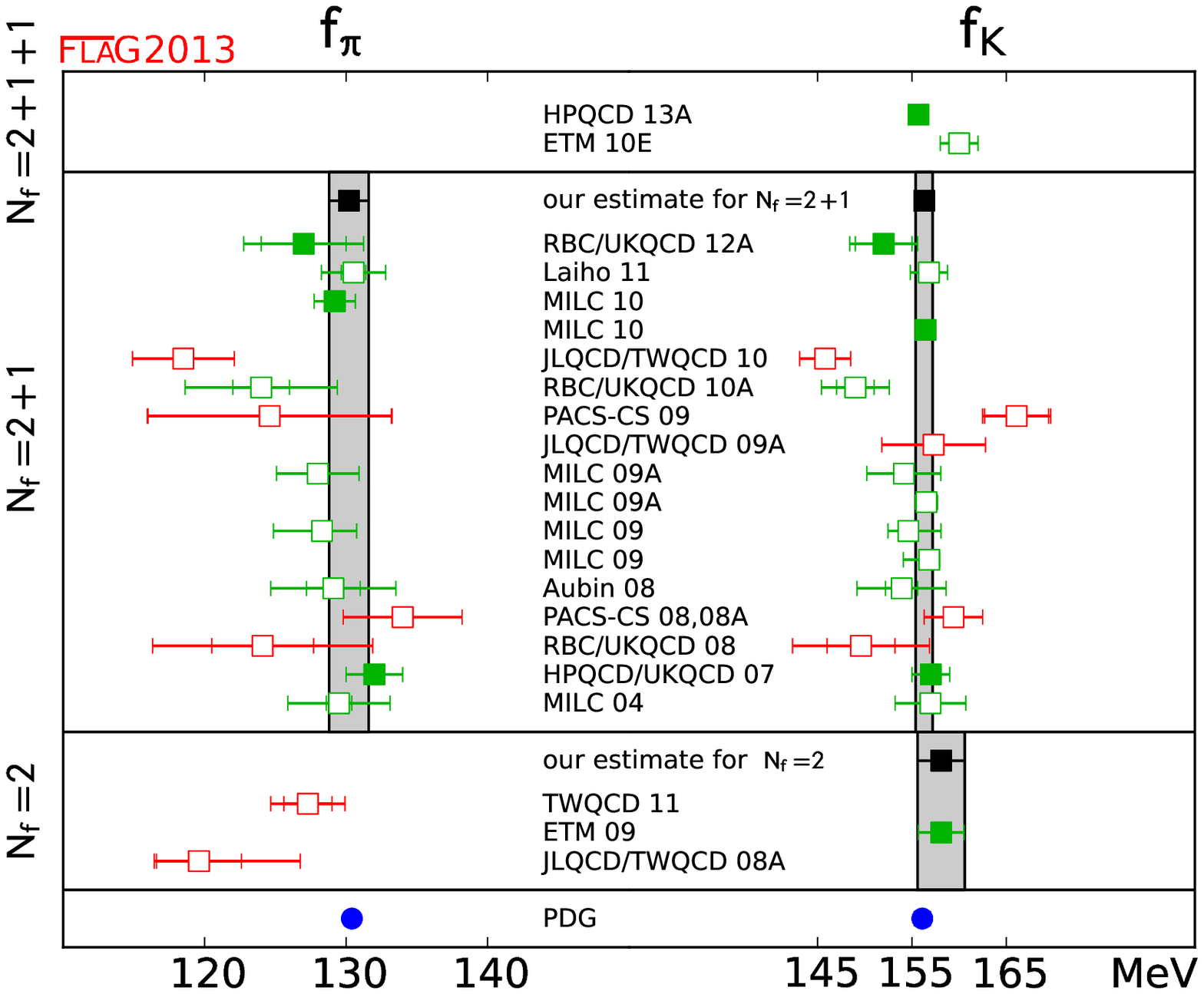}
\end{center}
  \vspace{-2.08cm}\hspace{8.2cm}\parbox{6cm}{\sffamily\tiny  \cite{Beringer:1900zz}}
\vspace{1cm}
\caption{\label{fig:latticedata_decayconstants}
Values of $f_\pi$ and $f_K$.
The black squares and grey bands indicate our estimates (\ref{eq:fPi}) and (\ref{eq:fK}).
The blue dots represent the experimental values quoted by the PDG (\ref{eq:fps_PDG}).}

\end{figure}


\clearpage


\section{Low-energy constants\label{sec:LECs}}


In the study of the quark-mass dependence of QCD observables calculated on the
lattice it is common practice to invoke Chiral Perturbation Theory ({\Ch}PT).
For a given quantity this framework predicts the nonanalytic quark-mass
dependence  and it provides symmetry relations among different observables.
These relations are best expressed with the help of a set of linearly
independent and universal (i.e. process-independent) low-energy constants
(LECs), which appear as coefficients of the polynomial terms (in $m_q$ or
$\Mpi^2$) in different observables.
If one expands around the SU(2) chiral limit, in the Chiral Effective
Lagrangian there appear two LECs at order $p^2$
\be
F\equiv F_\pi\,\rule[-0.3cm]{0.01cm}{0.7cm}_{\;m_u,m_d\rightarrow 0} \; , \qquad
B\equiv\frac{\Sigma}{F^2}\quad\mbox{where}\quad\Sigma\equiv-\<\ubar u\>\,\Big|_{\;m_u,m_d\rightarrow 0} \; ,
\ee
and seven at order $p^4$, indicated by $\bar \ell_i$ with $i=1,\ldots,7$.
In the analysis of the SU(3) chiral limit there are also just two LECs at order
$p^2$
\be
F_0\equiv   F_\pi\,\rule[-0.3cm]{0.01cm}{0.7cm}_{\;m_u,m_d,m_s\rightarrow 0} \; , \qquad
B_0\equiv\frac{\Sigma_0}{F_0^2}\quad\mbox{where}\quad\Sigma_0\equiv-\<\ubar  u\>\,\Big|_{\;m_u,m_d,m_s\rightarrow 0} \; ,
\ee
but ten at order $p^4$, indicated by the capital letter $L_i(\mu)$ with
$i=1,\ldots,10$.
These constants are independent of the quark masses%
\footnote{More precisely, they are independent of the 2 or 3 light
quark masses which are explicitly considered in the respective framework.
However, all low-energy constants depend on the masses of the remaining quarks
$s,c,b,t$ or $c,b,t$ in the SU(2) and SU(3) framework, respectively.},
but they become scale dependent after renormalization (sometimes a superscript
$r$ is added).
The SU(2) constants $\bar\ell_i$ are scale independent, since they are defined
at $\mu=M_\pi$ (as indicated by the bar).
For the precise definition of these constants and their scale dependence we
refer the reader to \cite{Gasser:1983yg,Gasser:1984gg}.

First of all, lattice calculations can be used to test if chiral symmetry is
indeed broken as SU$(\Nf)_L \times $SU$(\Nf)_R \to $SU$(\Nf)_{L+R}$ by
measuring nonzero chiral condensates and by verifying the validity of the GMOR
relation $M_\pi^2\propto m$ close to the chiral limit.
If the chiral extrapolation of quantities calculated on the lattice is made
with the help of {\Ch}PT, apart from determining the observable at the physical
value of the quark masses one also obtains the relevant LECs.
This is a very important by-product for two reasons:
\begin{enumerate}
\itemsep-2pt
\item
All LECs up to order $p^4$ (with the exception of $B$ and $B_0$, since only the
product of these times the quark masses can be estimated from phenomenology)
have either been determined by comparison to experiment or estimated
theoretically.
A lattice determination of the better known ones thus provides a test of the
{\Ch}PT approach.
\item
The less well known LECs are those which describe the quark-mass dependence of
observables -- these cannot be determined from experiment, and therefore the
lattice provides unique quantitative information.
This information is essential for improving phenomenological {\Ch}PT
predictions in which these LECs play a role.
\end{enumerate}
We stress that this program is based on the nonobvious assumption that {\Ch}PT
is valid in the region of masses used in the lattice simulations under
consideration.

The fact that, at large volume, the finite-size effects, which occur if a
system undergoes spontaneous symmetry breakdown, are controlled by the
Nambu-Goldstone modes, was first noted in solid state physics, in connection
with magnetic systems \cite{Fisher:1985zz,Brezin:1985xx}.
As pointed out in \cite{Gasser:1986vb} in the context of QCD, the thermal
properties of such systems can be studied in a systematic and model-independent
manner by means of the corresponding effective field theory, provided the
temperature is low enough.
While finite volumes are not of physical interest in particle physics, lattice
simulations are necessarily carried out in a finite box.
As shown in \cite{Gasser:1987ah,Gasser:1987zq,Hasenfratz:1989pk}, the ensuing
finite-size effects can also be studied on the basis of the effective theory --
{\Ch}PT in the case of QCD -- provided the simulation is close enough to the
continuum limit, the volume is sufficiently large and the explicit breaking of
chiral symmetry generated by the quark masses is sufficiently small.
Indeed, {\Ch}PT represents also a useful tool for the analysis of the
finite-size effects in lattice simulations.

In the following two subsections we summarize the lattice results for the SU(2)
and SU(3) LECs, respectively.
In either case we first discuss the $O(p^2)$ constants and then proceed to
their $O(p^4)$ counterparts.
The $O(p^2)$ LECs are determined from the chiral extrapolation of masses and
decay constants or, alternatively, from a finite-size study of correlators in
the $\ep$-regime.
At order $p^4$ some LECs affect two-point functions while other appear only in
three- or four-point functions; the latter need to be determined from form
factors or scattering amplitudes.
The {\Ch}PT analysis of the (non-lattice) phenomenological quantities is
nowadays%
\footnote{Some of the $O(p^6)$ formulae presented below have been derived in an
unpublished note by three of us (GC, SD and HL) and J\"urg Gasser. We thank him
for allowing us to publish them here.}
based on $O(p^6)$ formulae.
At this level the number of LECs explodes and we will not discuss any of these.
We will, however, discuss how comparing different orders and different
expansions (in particular $x$ versus $\xi$-expansion, see below) can help to
assess the theoretical uncertainties of the LECs determined on the lattice.


\subsection{SU(2) Low-Energy Constants \label{sec:su2}}


\subsubsection{Quark-mass dependence of pseudoscalar masses and decay constants\label{sec_MF}}

The expansions%
\footnote{Here and in the following, we stick to the notation used in the
papers where the {\Ch}PT formulae were established, i.e.\ we work with
$F_\pi\equiv f_\pi/\sqrt{2}=92.2(1)\MeV$ and $F_K\equiv f_K/\sqrt{2}$. The
occurrence of different normalization conventions is not convenient, but
avoiding it by reformulating the formulae in terms of $f_\pi$, $f_K$ is not a
good way out. Since we are using different symbols, confusion cannot arise.
\label{foot:fpi}}
of $M_\pi^2$ and $F_\pi$ in powers of the quark mass are known to
next-to-next-to-leading order in the SU(2) chiral effective theory.
In the isospin limit, $m_u=m_d=m$, the explicit expressions may be written
in the form \cite{Colangelo:2001df}
\begin{eqnarray}
M_\pi^2 & = & M^2\left\{1-\frac{1}{2}x\ln\frac{\Lambda_3^2}{M^2}
  +\frac{17}{8}x^2 \left(\ln\frac{\Lambda_M^2}{M^2}  \right)^2 +x^2 k_M
  +O(x^3)             \right\},
\label{eq:MF}
\\
F_\pi & = & F\left\{1+x\ln\frac{\Lambda_4^2}{M^2} -\frac{5}{4}x^2
  \left(\ln\frac{\Lambda_F^2}{M^2}  \right)^2 +x^2k_F   +O(x^3)
\right\}.
\nonumber
\end{eqnarray}
Here the expansion parameter is given by
\begin{equation}
x=\frac{M^2}{(4\pi F)^2},\;\;\;\;\;\;\;\;\;\;M^2=2Bm=\frac{2\Sigma m}{F^2},
\label{eq:xM2}
\end{equation}
but there is another option as discussed below.
The scales $\Lambda_3,\Lambda_4$ are related to the effective coupling
constants $\bar\ell_3,\bar\ell_4$ of the chiral Lagrangian at running scale
$M_\pi\equiv\Mpi^\mr{phys}$ by
\begin{equation}
\bar\ell_n=\ln\frac{\Lambda_n^2}{M_\pi^2},\;\;\;\;\;\;\;\;\;\;\;n=1,...,7.
\end{equation}
Note that in Eq.\,(\ref{eq:MF}) the logarithms are evaluated at $M^2$, not at
$M_\pi^2$.
The coupling constants $k_M,k_F$ in Eq.\,(\ref{eq:MF}) are mass-independent.
The scales of the squared logarithms can be expressed in terms of the
$O(p^4)$ coupling constants as
\begin{eqnarray}
  \ln\frac{\Lambda_M^2}{M^2} & = &
  \frac{1}{51}\left(28\ln\frac{\Lambda_1^2}{M^2}
    +32\ln\frac{\Lambda_2^2}{M^2}    -9 \ln\frac{\Lambda_3^2}{M^2}+49
  \right),
\\
  \ln\frac{\Lambda_F^2}{M^2} & = &
  \frac{1}{30}\left(14\ln\frac{\Lambda_1^2}{M^2}
    +16\ln\frac{\Lambda_2^2}{M^2}    +6 \ln\frac{\Lambda_3^2}{M^2}
    - 6 \ln\frac{\Lambda_4^2}{M^2}      +23  \right).
\nonumber
\end{eqnarray}
Hence by analysing the quark-mass dependence of $M_\pi^2$ and $F_\pi$ with
Eq.\,(\ref{eq:MF}), possibly truncated at NLO, one can determine%
\footnote{Notice that one could analyse the quark-mass dependence entirely in
terms of the parameter $M^2$ defined in Eq.\,(\ref{eq:xM2}) and determine
equally well all other LECs. Using the determination of the quark masses
described in Sec.~\ref{sec:qmass} one can then extract $B$ or $\Sigma$.}
the $O(p^2)$ LECs $B$ and $F$, as well as the $O(p^4)$ LECs $\bar \ell_3$ and
$\bar \ell_4$.
The quark condensate in the chiral limit is given by $\Sigma=F^2B$.
With precise enough data at several small enough pion masses, one could in
principle also determine $\Lambda_M$, $\Lambda_F$ and $k_M$, $k_F$.
To date this is not yet possible.
The results for the LO and NLO constants will be presented in
Sec.~\ref{sec:SU2results}.

Alternatively, one can invert Eq.\,(\ref{eq:MF}) and express $M^2$ and $F$ as
an expansion in
\be
\xi \equiv \frac{M_\pi^2}{16 \pi^2 F_\pi^2} \; \; ,
\label{eq:xi}
\ee
and the corresponding expressions then take the form
\bea
\label{eq:MpiFpi}
M^2&=& M_\pi^2\,\left\{
1+\frac{1}{2}\,\xi\,\lthreebar-
\frac{5}{8}\,\xi^2 \left(\!\lMbar\!\right)^2+
\xi^2 c_{\ind M}+O(\xi^3)\right\} \co
\\
F&=& F_\pi\,\left\{1-\xi\,\lfourbar-\frac{1}{4}\,\xi^2
\left(\!\lFbar\!\right)^2
+\xi^2 c_{\ind F}+O(\xi^3)\right\} \fs \nn
\eea
The scales of the quadratic logarithms are determined by
$\Lambda_1,\ldots,\Lambda_4$ through
\bea
\lMbar&=&\frac{1}{15}\left(28\,\lonebar+32\,\ltwobar-
33\,\lthreebar-12\,\lfourbar +52\right) \co \\
\lFbar&=&\frac{1}{3}\,\left(-7\,\lonebar-8\,\ltwobar+
18\,\lfourbar- \frac{29}{2}\right)\nonumber \fs
\eea


\subsubsection{Two-point correlation functions in the epsilon-regime\label{sec_eps}}

The finite-size effects encountered in lattice calculations can be used to
determine some of the LECs of QCD.
In order to illustrate this point, we focus on the two lightest quarks, take
the isospin limit $m_u=m_d=m$ and consider a box of size $L_s$ in the three
space directions and size $L_t$ in the time direction.
If $m$ is sent to zero at fixed box size, chiral symmetry is restored.
The behaviour of the various observables in the symmetry-restoration region is
controlled by the parameter $\mu\equiv m\,\Sigma\,V$, where $V=L_s^3L_t$ is the
four-dimensional volume of the box.
Up to a sign and a factor of two, the parameter $\mu$ represents the minimum of
the classical action that belongs to the leading-order effective Lagrangian of
QCD.

For $\mu\gg1$, the system behaves qualitatively as if the box was infinitely
large.
In that region, the $p$-expansion, which counts $1/L_s$, $1/L_t$ and $M$ as
quantities of the same order, is adequate.
In view of $\mu=\frac{1}{2}F^2 M^2V $, this region includes configurations with
$ML\gsim\! 1$, where the finite-size effects due to pion loop diagrams are
suppressed by the factor $e^{-ML}$.

If $\mu$ is comparable to or smaller than 1, however, the chiral perturbation
series must be reordered.
The $\ep$-expansion achieves this by counting $1/L_s, 1/L_t$ as quantities of
$O(\ep)$, while the quark mass $m$ is booked as a term of $O(\ep^4)$.
This ensures that the symmetry-restoration parameter $\mu$ represents a term of
order $O(\ep^0)$, so that the manner in which chiral symmetry is restored can
be worked out.

As an example, we consider the correlator of the axial charge carried by the
two lightest quarks, $q(x)=\{u(x),d(x)\}$.
The axial current and the pseudoscalar density are given by
\be
A_\mu^i(x)=
\qbar(x)\mbox{$\frac{1}{2}$} \tau^i\,\gamma_\mu\gamma_5\,q(x)\,,
\hspace{1cm}P^i(x) = \qbar(x)\mbox{$\frac{1}{2}$} \tau^i\,\mr{i} \gamma_5\,q(x)\,,
\ee
where $\tau^1, \tau^2,\tau^3$, are the Pauli matrices in flavour space.
In Euclidean space, the correlators of the axial charge and of the space
integral over the pseudoscalar density are given by
\begin{eqnarray}\label{eq:correlators}
\delta^{ik}C_{AA}(t)\al = \al L_s^3\int \hspace{-0.12cm}d^3\hspace{-0.04cm}\vec{x}\;\langle A_4^i(\vec{x},t)
A_4^k(0)\rangle\,,
\\
\delta^{ik}C_{PP}(t)\al  =\al L_s^3\int \hspace{-0.12cm}d^3\hspace{-0.04cm}\vec{x}\;\langle P^i(\vec{x},t)
P^k(0)\rangle\,.\nonumber
\end{eqnarray}
{\Ch}PT yields explicit finite-size scaling formulae for these quantities
\cite{Hasenfratz:1989pk,Hansen:1990un,Hansen:1990yg}.
In the $\epsilon$-regime, the expansion starts with
\begin{eqnarray}  \label{aa-eps}
C_{AA}(t) \al = \al \frac{F^2L_s^3}{L_t}\left[a_A+
  \frac{L_t}{F^2L_s^3}\,b_A\,h_1\hspace{-0.1cm}\left(\frac{t}{L_t}  \right)
+O(\epsilon^4)\right],
\\
C_{PP}(t) \al = \al
\Sigma^2L_s^6\left[a_P+\frac{L_t}{F^2L_s^3}\,b_P\,h_1\hspace{-0.1cm}\left(\frac{t}{L_t}  \right)
+O(\epsilon^4)\right],\nonumber
\end{eqnarray}
where the coefficients $a_A$, $b_A$, $a_P$, $b_P$ stand for quantities of
$O(\ep^0)$.
They can be expressed in terms of the variables $L_s$, $L_t$ and $m$ and
involve only the two leading low-energy constants $F$ and $\Sigma$.
In fact, at leading order only the combination $\mu=m\,\Sigma\,L_s^3 L_t$
matters, the correlators are $t$-independent and the dependence on $\mu$ is
fully determined by the structure of the groups involved in the SSB pattern.
In the case of SU(2)$\times$SU(2) $\rightarrow$ SU(2), relevant for QCD in the
symmetry restoration region with two light quarks, the coefficients can be
expressed in terms of Bessel functions.
The $t$-dependence of the correlators starts showing up at $O(\ep^2)$, in the
form of a parabola, viz.\ $h_1(\tau)=\frac{1}{2}\left[\left(\tau-\frac{1}{2}
\right)^2-\frac{1}{12} \right]$.
Explicit expressions for $a_A$, $b_A$, $a_P$, $b_P$ can be found in
\cite{Hasenfratz:1989pk,Hansen:1990un,Hansen:1990yg}, where some of the
correlation functions are worked out to NNLO.
By matching the finite-size scaling of correlators computed on the lattice with
these predictions one can extract $F$ and $\Sigma$.
A way to deal with the numerical challenges genuine to the $\ep$-regime has
been described \cite{Giusti:2004yp}.

The fact that the representation of the correlators to NLO is not
``contaminated'' by higher-order unknown LECs, makes the $\ep$-regime
potentially convenient for a clean extraction of the LO couplings.
The determination of these LECs is then affected by different systematic
uncertainties with respect to the standard case; simulations in this regime
yield complementary information which can serve as a valuable cross-check
to get a comprehensive picture of the low-energy properties of QCD.

The effective theory can also be used to study the distribution of the
topological charge in QCD \cite{Leutwyler:1992yt} and the various
quantities of interest may be defined for a fixed value of this charge. The
expectation values and correlation functions then not only depend on the
symmetry restoration parameter $\mu$, but also on the topological charge
$\nu$. The dependence on these two variables can explicitly be calculated.
It turns out that the two-point correlation functions considered above
retain the form (\ref{aa-eps}), but the coefficients $a_A$, $b_A$, $a_P$,
$b_P$ now depend on the topological charge as well as on the symmetry
restoration parameter (see
\cite{Damgaard:2001js,Damgaard:2002qe,Aoki:2009mx} for explicit
expressions).

A specific issue with $\ep$-regime calculations is the scale setting.
Ideally one would perform a $p$-regime study with the same bare parameters to
measure a hadronic scale (e.g.\ the proton mass).
In the literature, sometimes a gluonic scale (e.g.\ $r_0$) is used to avoid
such expenses.
Obviously the issues inherent in scale setting are aggravated if the
$\ep$-regime simulation is restricted to a fixed sector of topological charge.

It is important to stress that in the $\epsilon$-expansion higher-order
finite-volume corrections might be significant, and the physical box size (in
fm) should still be large in order to keep these contributions under control.
The criteria for the chiral extrapolation and finite-volume effects are
obviously different with respect to the $p$-regime.
For these reasons we have to adjust the colour coding defined in
Sect.\,\ref{sec:color-code} (see \ref{sec:SU2results} for more details).

Recently, the effective theory has been extended to the ``mixed regime''
where some quarks are in the $p$-regime and some in the $\ep$-regime
\cite{Bernardoni:2008ei,Hernandez:2012tw}.
In \cite{Damgaard:2008zs} a technique is proposed to smoothly connect the
$p$- and $\ep$-regimes.
In \cite{Aoki:2011pza} the issue is reconsidered with a counting rule which is
essentially the same as in the $p$-regime.
In this new scheme, the theory remains IR finite even in the chiral limit,
while the chiral-logarithmic effects are kept present.


\subsubsection{Energy levels of the QCD Hamiltonian in a box and $\delta$-regime\label{sec_su2_delta}}

At low temperature, the properties of the partition function are governed
by the lowest eigenvalues of the Hamiltonian.
In the case of QCD, the lowest levels are due to the Nambu-Goldstone bosons and
can be worked out with {\Ch}PT \cite{Leutwyler:1987ak}.
In the chiral limit the level pattern follows the one of a quantum-mechanical
rotator, i.e.\ $E_\ell=\ell(\ell+1)/(2\,\Theta)$ with $\ell = 0, 1,2,\ldots$.
For a cubic spatial box and to leading order in the expansion in inverse powers
of the box size $L_s$, the moment of inertia is fixed by the value of the pion
decay constant in the chiral limit, i.e.\ $\Theta=F^2L_s^3$.

In order to analyse the dependence of the levels on the quark masses and on the parameters that
specify the size of the box, a reordering of the chiral series is required,
the so-called $\delta$-expansion; the region where the properties of the
system are controlled by this expansion is referred to as the
$\delta$-regime. Evaluating the chiral perturbation series in this regime,
one finds that the expansion of the partition function goes in even inverse
powers of $FL_s$, that the rotator formula for the energy levels holds up
to NNLO and the expression for the moment of inertia is now also known up
to and including terms of order $(FL_s)^{-4}$
\cite{Hasenfratz:2009mp,Niedermayer:2010mx,Weingart:2010yv}. Since the
level spectrum is governed by the value of the pion decay constant in the
chiral limit, an evaluation of this spectrum on the lattice can be used to
measure $F$. More generally, the evaluation of various observables in the
$\delta$-regime offers an alternative method for a determination of some of
the low-energy constants occurring in the effective Lagrangian. At present,
however, the numerical results obtained in this way
\cite{Hasenfratz:2006xi,Bietenholz:2010az} are not yet competitive with
those found in the $p$- or $\epsilon$-regimes.


\subsubsection{Other methods for the extraction of the Low-Energy Constants\label{sec_su2_extra}}

An observable that can be used to extract the LECs is the topological
susceptibility
\begin{equation}
\chi_t=\int d^4\!x\; \langle \omega(x) \omega(0)\rangle,
\end{equation}
where $\omega(x)$ is the topological charge density,
\begin{equation}
\omega(x)=\frac{1}{32\pi^2}
\epsilon^{\mu\nu\rho\sigma}{\rm Tr}\left[F_{\mu\nu}(x)F_{\rho\sigma}(x)\right].
\end{equation}
At infinite volume, the expansion of $\chi_t$ in powers of the quark masses
starts with \cite{DiVecchia:1980ve}
\begin{equation}\label{chi_t}
\chi_t=\overline{m}\,\Sigma \,\{1+O(m)\}\,,\hspace{2cm}
\overline{m}\equiv\left(
\frac{1}{m_u}+\frac{1}{m_d}+\frac{1}{m_s}+\ldots
\right)^{-1}.
\end{equation}
The condensate $\Sigma$ can thus be extracted from the properties of the
topological susceptibility close to the chiral limit.
The behaviour at finite volume, in particular in the region where the symmetry
is restored, is discussed in \cite{Hansen:1990yg}.
The dependence on the vacuum angle $\theta$ and the projection on sectors of
fixed $\nu$ have been studied in \cite{Leutwyler:1992yt}.
For a discussion of the finite-size effects at NLO, including the dependence on
$\theta$, we refer to \cite{Mao:2009sy,Aoki:2009mx}.

The role that the topological susceptibility plays in attempts to determine
whether there is a large paramagnetic suppression when going from the $\Nf=2$
to the $\Nf=2+1$ theory has been highlighted in Ref.\,\cite{Bernard:2012fw}.
And the potential usefulness of higher moments of the topological charge
distribution to determine LECs has been investigated in \cite{Bernard:2012ci}.

Another method for computing the quark condensate has been proposed in
\cite{Giusti:2008vb}, where it is shown that starting from the Banks-Casher
relation \cite{Banks:1979yr} one may extract the condensate from suitable
(renormalizable) spectral observables, for instance the number of Dirac
operator modes in a given interval.
For those spectral observables higher-order corrections can be systematically
computed in terms of the chiral effective theory.
A recent paper based on this strategy is ETM 13 \cite{Cichy:2013gja}.
As an aside let us remark that corrections to the Banks-Casher relation that
come from a finite quark mass, a finite four-dimensional volume and (with
Wilson-type fermions) a finite lattice spacing can be parameterized in a
properly extended version of the chiral framework \cite{Necco:2013sxa}.

An alternative strategy is based on the fact that at LO in the $\ep$-expansion
the partition function in a given topological sector $\nu$ is equivalent to the
one of a chiral Random Matrix Theory (RMT) \cite{Shuryak:1992pi,
Verbaarschot:1993pm,Verbaarschot:1994qf,Verbaarschot:2000dy}.
In RMT it is possible to extract the probability distributions of individual
eigenvalues \cite{Nishigaki:1998is,Damgaard:2000ah,Basile:2007ki} in terms of
two dimensionless variables $\zeta=\lambda\Sigma V$ and $\mu=m\Sigma V$, where
$\lambda$ represents the eigenvalue of the massless Dirac operator and $m$
is the sea quark mass.
More recently this approach has been extended to the Hermitian (Wilson) Dirac
operator \cite{Kieburg:2013tca} which is easier to study in numerical simulations.
Hence, if it is possible to match the QCD low-lying spectrum of the Dirac
operator to the RMT predictions, then one may extract%
\footnote{By introducing an imaginary isospin chemical potential, the framework
can be extended such that the low-lying spectrum of the Dirac operator is also
sensitive to the pseudoscalar decay constant $F$ at LO \cite{Akemann:2006ru}.}
the chiral condensate $\Sigma$.
One issue with this method is that for the distributions of individual
eigenvalues higher-order corrections are still not known in the effective
theory, and this may introduce systematic effects which are hard%
\footnote{Higher-order systematic effects in the matching with RMT have been
investigated in \cite{Lehner:2010mv,Lehner:2011km}.}
to control.
Another open question is that, while it is clear how the spectral density is
renormalized \cite{DelDebbio:2005qa}, this is not the case for the individual
eigenvalues, and one relies on assumptions.
There have been many lattice studies \cite{Fukaya:2007yv,Lang:2006ab,
DeGrand:2006nv,Hasenfratz:2007yj,DeGrand:2007tm} which
investigate the matching of the low-lying Dirac spectrum with RMT.
In this review the results of the LECs obtained in this way%
\footnote{The results for $\Sigma$ and $F$ lie in the same range as the
determinations reported in Tables~\ref{tabsigma} and \ref{tabf}.}
are not included.


\subsubsection{Pion form factors\label{sec:su2_form}}

The scalar and vector form factors of the pion are defined by the matrix
elements
\bea
\langle \pi^i(p_2) |\, \qbar\, q  \, | \pi^j(p_1) \rangle \al = \al
\delta^{ij} F_S^\pi(t) \co
\\
\langle \pi^i(p_2) | \,\qbar\, \mbox{$\frac{1}{2}$}\tau^k \gamma^\mu q\,| \pi^j(p_1)
\rangle \al = \al
\mr{i} \,\epsilon^{ikj} (p_1^\mu + p_2^\mu) F_V^\pi(t) \co\nonumber
\eea
where the operators contain only the lightest two quark flavours, i.e.\
$\tau^1$, $\tau^2$, $\tau^3$ are the Pauli matrices, and $t\equiv (p_1-p_2)^2$
denotes the momentum transfer.

The vector form factor has been measured by several experiments for
timelike as well as for spacelike values of $t$.
The scalar form factor is not directly measurable, but it can be evaluated
theoretically from data on the $\pi \pi$ and $\pi K$ phase shifts
\cite{Donoghue:1990xh} by means of analyticity and unitarity, i.e.\ in a
model-independent way.
Lattice calculations can be compared with data or model-independent theoretical
evaluations at any given value of $t$.
At present, however, most lattice studies concentrate on the region close to
$t=0$ and on the evaluation of the slope and curvature which are defined as
\begin{eqnarray}
  F^\pi_V(t) & = & 1+\mbox{$\frac{1}{6}$}\langle r^2 \rangle^\pi_V t +
c_V\hspace{0.025cm} t^2+\ldots \;\co
\\
  F^\pi_S(t) & = & F^\pi_S(0) \left[1+\mbox{$\frac{1}{6}$}\langle r^2
    \rangle^\pi_S t + c_S\, t^2+ \ldots \right] \; \; . \nn
\end{eqnarray}
The slopes are related to the mean-square vector and scalar radii which are the
quantities on which most experiments and lattice calculations concentrate.

In chiral perturbation theory, the form factors are known at NNLO
\cite{Bijnens:1998fm}.
The corresponding formulae are available in fully analytical form and are
compact enough that they can be used for the chiral extrapolation of the data
(as done, for example in \cite{Frezzotti:2008dr,Kaneko:2008kx}).
The expressions for the scalar and vector radii and for the $c_{S,V}$
coefficients at two-loop level read
\bea
\langle r^2 \rangle^\pi_S &=& \frac{1}{(4\pi\Fpi)^2} \left\{6 \lfourbar-\frac{13}{2}
-\frac{29}{3}\,\xi \left(\!\ln\frac{\Omega_{r_S}^2}{M_\pi^2} \!\right)^2+ 6
\xi \, k_{r_S}+O(\xi^2)\right\} \co\nn
\\
\langle r^2 \rangle^\pi_V &=& \frac{1}{(4\pi\Fpi)^2} \left\{ \lsixbar-1
+2\,\xi \left(\!\ln\frac{\Omega_{r_V}^2}{M_\pi^2} \!\right)^2+6 \xi \,k_{r_V}+O(\xi^2)\right\}\co
\label{formula_rsqu}
\\
c_S &=&\frac{1}{(4\pi\Fpi\Mpi)^2} \left\{\frac{19}{120}  + \xi \left[ \frac{43}{36} \left(\!
      \ln\frac{\Omega_{c_S}^2}{M_\pi^2} \!\right)^2 + k_{c_S} \right]
\right\} \co \nn
\\
c_V &=&\frac{1}{(4\pi\Fpi\Mpi)^2} \left\{\frac{1}{60}+\xi \left[\frac{1}{72} \left(\!
      \ln\frac{\Omega_{c_V}^2}{M_\pi^2} \!\right)^2 + k_{c_V} \right]
\right\} \co \nn
\eea
where
\bea
\ln\frac{\Omega_{r_S}^2}{M_\pi^2}&=&\frac{1}{29}\,\left(31\,\lonebar+34\,\ltwobar -36\,\lfourbar
  +\frac{145}{24}\right)  \co \nn\\
\ln\frac{\Omega_{r_V}^2}{M_\pi^2}&=&\frac{1}{2}\,\left(\lonebar-\ltwobar+\lfourbar+\lsixbar
-\frac{31}{12}\right) \co  \\
\ln\frac{\Omega_{c_S}^2}{M_\pi^2}&=&\frac{43}{63}\,\left(11\,\lonebar+14\,\ltwobar+18\,\lfourbar
  -\frac{6041}{120}\right)
\co \nn \\
\ln\frac{\Omega_{c_V}^2}{M_\pi^2}&=&\frac{1}{72}\,\left(2\lonebar-2\ltwobar-\lsixbar
-\frac{26}{30}\right) \co \nn
\eea
and $k_{r_S},k_{r_V}$ and $k_{c_S},k_{c_V}$ are independent of the quark
masses.
Their expression in terms of the $\ell_i$ and of the $O(p^6)$ constants
$c_M,c_F$ is known but will not be reproduced here.

The difference between the quark-line connected and the full (i.e.\ containing
the connected and the disconnected piece) scalar pion form factor has been
investigated by means of Chiral Perturbation Theory in \cite{Juttner:2011ur}.
It is expected that the technique used can be applied to a large class of
observables relevant in QCD-phenomenology.

As a point of practical interest let us remark that there are no finite-volume
correction formulae for the mean-square radii $\<r^2\>_{V,S}$ and the curvatures
$c_{V,S}$.
The lattice data for $F_{V,S}(t)$ need to be corrected, point by point in $t$,
for finite-volume effects.
In fact, if a given $t$ is realized through several inequivalent $p_1\!-\!p_2$
combinations, the level of agreement after the correction has been applied is
indicative of how well higher-order effects are under control.


\subsubsection{Lattice determinations\label{sec:SU2results}}

In this section we summarize the lattice results for the SU(2) couplings in
a set of tables (\ref{tabsigma}--\ref{tabl6}) and figures
(\ref{fig_sigma}--\ref{fig_l3l4l6}).
The tables present our usual colour coding which summarizes the main aspects
related to the treatment of the systematic errors of the various
calculations.

A delicate issue in the lattice determination of chiral LECs (in particular at
NLO) which cannot be reflected by our colour coding is a reliable assessment of
the theoretical error that comes from the chiral expansion.
We add a few remarks on this point:
\begin{enumerate}
\item
Using \emph{both} the $x$ and the $\xi$ expansion is a good way to test how
the ambiguity of the chiral expansion (at a given order) affects the numerical
values of the LECs that are determined from a particular set of data.
For instance, to determine $\bar{\ell}_4$ (or $\Lambda_4$) from lattice data
for $\Fpi$ as a function of the quark mass, one may compare the fits based on
the parameterization $\Fpi=F\{1+x\ln(\Lambda_4^2/M^2)\}$
[see Eq.\,(\ref{eq:MF})]
with those obtained from $\Fpi=F/\{1-\xi\ln(\Lambda_4^2/\Mpi^2)\}$
[see Eq.\,(\ref{eq:MpiFpi})].
The difference between the two results provides an estimate of the uncertainty
due to the truncation of the chiral series.
Which central value one chooses is in principle arbitrary, but we find it
advisable to use the one obtained with the $\xi$ expansion%
\footnote{There are theoretical arguments suggesting that the $\xi$ expansion
is preferable to the $x$ expansion, based on the observation that the
coefficients in front of the squared logs in (\ref{eq:MF}) are somewhat larger
than in (\ref{eq:MpiFpi}). This can be traced to the fact that a part of every
formula in the $x$ expansion is concerned with locating the position of the
pion pole (at the previous order) while in the $\xi$ expansion the knowledge of
this position is built in exactly. Numerical evidence supporting this view is
presented in \cite{Noaki:2008iy}.},
in particular because it makes the comparison with phenomenological
determinations (where it is standard practice to use the $\xi$ expansion) more
meaningful.
\item
Alternatively one could try to estimate the influence of higher chiral orders
by reshuffling irrelevant higher-order terms.
For instance, in the example mentioned above one might use
$\Fpi=F/\{1-x\ln(\Lambda_4^2/M^2)\}$ as a different functional form at NLO.
Another way to establish such an estimate is through introducing by hand
``analytical'' higher-order terms (e.g.\ ``analytical NNLO'' as done, in the
past, by MILC \cite{Bazavov:2009bb}).
In principle it would be preferable to include all NNLO terms or none, such
that the structure of the chiral expansion is preserved at any order (this is
what ETM \cite{Baron:2009wt} and JLQCD/TWQCD \cite{Noaki:2008iy} have done
for SU(2) {\Ch}PT and MILC for SU(3) {\Ch}PT \cite{Bazavov:2009fk}).
There are different opinions in the field as to whether it is advisable to
include terms to which the data are not sensitive.
In case one is willing to include external (typically: non-lattice)
information, the use of priors is a theoretically well founded option
(e.g.\ priors for NNLO LECs if one is interested in LECs at LO/NLO).
\item
Another issue concerns the $s$-quark mass dependence of the LECs $\bar{\ell}_i$
or $\Lambda_i$ of the SU(2) framework.
As far as variations of $m_s$ around $m_s^\mr{phys}$ are concerned (say for
$0<m_s<1.5m_s^\mr{phys}$ at best) the issue can be studied in SU(3) ChPT, and this
has been done in a series of papers \cite{Gasser:1984gg,Gasser:2007sg,Gasser:2009hr}.
However, the effect of sending $m_s$ to infinity, as is the case in $\Nf=2$ lattice
studies of SU(2) LECs, cannot be addressed in this way.
A unique way to analyse this difference is to compare the numerical values of LECs
determined in $\Nf=2$ lattice simulations to those determined in $\Nf=2+1$ lattice
simulations (see e.g.\ \cite{Durr:2013koa} for a discussion).
\item
Last but not least let us recall that the determination of the LECs is affected
by discretization effects, and it is important that these are removed by means
of a continuum extrapolation.
In this step invoking an extended version of the chiral Lagrangian
\cite{Rupak:2002sm,Aoki:2003yv,Bar:2003mh} may be useful%
\footnote{This means that for any given lattice formulation one
needs to determine additional low-energy constants, often denoted $W_i$.
For certain formulations, e.g.\ the twisted-mass approach, first steps in
this direction have already been taken \cite{Herdoiza:2013sla}.}
in case one aims for a global fit of lattice data involving several
$\Mpi$ and $a$ values and several chiral observables.
\end{enumerate}

In the tables and figures we summarize the results of various lattice
collaborations for the SU(2) LECs at LO ($F$ or $F/F_\pi$, $B$ or $\Sigma$) and
at NLO ($\bar\ell_1-\bar\ell_2$, $\bar\ell_3$, $\bar\ell_4$, $\bar\ell_5$,
$\bar\ell_6$).
Throughout we group the results into those which stem from $\Nf=2+1+1$
calculations, those which come from $\Nf=2+1$ calculations and those which stem
from $\Nf=2$ calculations (since, as mentioned above, the LECs are logically
distinct even if the current precision of the data is not sufficient to resolve
the differences).
Furthermore, we make a distinction whether the results are obtained from
simulations in the $p$-regime or whether alternative methods ($\ep$-regime,
spectral quantities, topological susceptibility, etc.) have been used (this
should not affect the result).
For comparison we add, in each case, a few phenomenological determinations with
high standing.

A generic comment applies to the issue of the scale setting.
In the past none of the lattice studies with $\Nf\geq2$ involved simulations in
the $p$-regime at the physical value of $m_{ud}$.
Accordingly, the setting of the scale $a^{-1}$ via an experimentally measurable
quantity did necessarily involve a chiral extrapolation, and as a result of
this dimensionful quantities used to be particularly sensitive to this
extrapolation uncertainty, while in dimensionless ratios such as $F_\pi/F$,
$F/F_0$, $B/B_0$, $\Sigma/\Sigma_0$ this particular problem is much reduced
(and often finite lattice-to-continuum renormalization factors drop out).
Now, there is a new generation of lattice studies
\cite{Aoki:2009ix,Durr:2010vn,Durr:2010aw,Borsanyi:2012zv,Bazavov:2012xda,Bazavov:2012cd}
which does involve simulations at physical pion masses.
In such studies even the uncertainty that the scale setting has on dimensionful
quantities is much mitigated.

It is worth repeating here that the standard colour-coding scheme of our
tables is necessarily schematic and cannot do justice to every calculation.
In particular there is some difficulty in coming up with a fair adjustment
of the rating criteria to finite-volume regimes of QCD.
For instance, in the $\epsilon$-regime%
\footnote{Also in case of \cite{Fukaya:2009fh} and \cite{Fukaya:2010na} the
colour-coding criteria for the $\epsilon$-regime have been applied.}
we re-express the ``chiral extrapolation'' criterion in terms of
$\sqrt{2m_\mr{min}\Sigma}/F$, with the same threshold values (in MeV)
between the three categories as in the $p$-regime.
Also the ``infinite volume'' assessment is adapted to the $\ep$-regime, since
the $\Mpi L$ criterion does not make sense here; we assign a green star if at
least 2 volumes with $L>2.5\fm$ are included, an open symbol if at least 1 volume
with $L>2\fm$ is invoked and a red square if all boxes are smaller than $2\fm$.
Similarly, in the calculation of form factors and charge radii the tables do
not reflect whether an interpolation to the desired $q^2$ has been performed or
whether the relevant $q^2$ has been engineered by means of
``partially-twisted boundary conditions'' \cite{Boyle:2008yd}.
In spite of these limitations we feel that these tables give an adequate
overview of the qualities of the various calculations.


\begin{table}[!tbp] 
\vspace*{3cm}
\centering
\footnotesize
\begin{tabular*}{\textwidth}[l]{l@{\extracolsep{\fill}}rlllllll}
Collaboration & Ref. & $\Nf$ &
\hspace{0.15cm}\begin{rotate}{60}{publication status}\end{rotate}\hspace{-0.15cm} &
\hspace{0.15cm}\begin{rotate}{60}{chiral extrapolation}\end{rotate}\hspace{-0.15cm}&
\hspace{0.15cm}\begin{rotate}{60}{continuum  extrapolation}\end{rotate}\hspace{-0.15cm} &
\hspace{0.15cm}\begin{rotate}{60}{finite volume}\end{rotate}\hspace{-0.15cm} &
\hspace{0.15cm}\begin{rotate}{60}{renormalization}\end{rotate}\hspace{-0.15cm} & \rule{0.4cm}{0cm}$\Sigma^{1/3}$  \\
&&&&&&&& \\[-0.1cm]
\hline
\hline
&&&&&&&& \\[-0.1cm]
ETM 13         & \cite{Cichy:2013gja}     &2+1+1& \gA & \soso & \good & \soso & \good & 274(08)(08)  \\
&&&&&&&& \\[-0.1cm]
\hline
&&&&&&&& \\[-0.1cm]
BMW 13         & \cite{Durr:2013goa}      & 2+1 & \oP & \good & \good & \good & \good & 271(4)(1)      \\
Borsanyi 12    & \cite{Borsanyi:2012zv}   & 2+1 & \gA & \good & \good & \soso & \good & 272.3(1.2)(1.4)\\
MILC 10A       & \cite{Bazavov:2010yq}    & 2+1 & \rC & \soso & \good & \good & \soso & 281.5(3.4)$\binom{+2.0}{-5.9}$(4.0) \\
JLQCD/TWQCD 10 & \cite{Fukaya:2010na}     & 2+1 & \gA & \good & \bad  & \soso & \good & 234(4)(17)    \\
RBC/UKQCD 10A  & \cite{Aoki:2010dy}       & 2+1 & \gA & \soso & \soso & \good & \good & 256(5)(2)(2)  \\
JLQCD 09       & \cite{Fukaya:2009fh}     & 2+1 & \gA & \good & \bad  & \soso & \good & 242(4)$\binom{+19}{-18}$\\
MILC 09A       & \cite{Bazavov:2009fk}    & 2+1 & \rC & \soso & \good & \good & \soso & 279(1)(2)(4)  \\
MILC 09A       & \cite{Bazavov:2009fk}    & 2+1 & \rC & \soso & \good & \good & \soso & 280(2)$\binom{+4}{-8}$(4)\\
MILC 09        & \cite{Bazavov:2009bb}    & 2+1 & \gA & \soso & \good & \good & \soso & 278(1)$\binom{+2}{-3}$(5)\\
TWQCD 08       & \cite{Chiu:2008jq}       & 2+1 & \gA & \soso & \bad  & \bad  & \good & 259(6)(9)     \\
JLQCD/TWQCD 08B& \cite{Chiu:2008kt}       & 2+1 & \rC & \soso & \bad  & \bad  & \good & 253(4)(6)     \\
PACS-CS 08     & \cite{Aoki:2008sm}       & 2+1 & \gA & \good & \bad  & \bad  & \bad  & 312(10)       \\
PACS-CS 08     & \cite{Aoki:2008sm}       & 2+1 & \gA & \good & \bad  & \bad  & \bad  & 309(7)        \\
RBC/UKQCD 08   & \cite{Allton:2008pn}     & 2+1 & \gA & \soso & \bad  & \good & \good & 255(8)(8)(13) \\
&&&&&&&& \\[-0.1cm]
\hline
&&&&&&&& \\[-0.1cm]
Brandt 13      & \cite{Brandt:2013dua}    &  2  & \gA & \soso & \good & \good & \good & 261(13)(1)   \\
ETM 13         & \cite{Cichy:2013gja}     &  2  & \gA & \soso & \good & \soso & \good & 277(06)(12)  \\
ETM 12         & \cite{Burger:2012ti}     &  2  & \gA & \soso & \good & \soso & \good & 299(26)(29)  \\
Bernardoni 11  & \cite{Bernardoni:2011kd} &  2  & \rC & \soso & \bad  & \bad  & \soso & 306(11)      \\
TWQCD 11       & \cite{Chiu:2011bm}       &  2  & \gA & \soso & \bad  & \bad  & \good & 235(8)(4)    \\
TWQCD 11A      & \cite{Chiu:2011dz}       &  2  & \gA & \soso & \bad  & \bad  & \good & 259(6)(7)    \\
Bernardoni 10  & \cite{Bernardoni:2010nf} &  2  & \gA & \soso & \bad  & \bad  & \good & 262$\binom{+33}{-34}\binom{+4}{-5}$ \\
JLQCD/TWQCD 10 & \cite{Fukaya:2010na}     &  2  & \gA & \good & \bad  & \bad  & \good & 242(5)(20)   \\
ETM 09C        & \cite{Baron:2009wt}      &  2  & \gA & \soso & \good & \soso & \good & 270(5)$\binom{+3}{-4}$ \\
ETM 08         & \cite{Frezzotti:2008dr}  &  2  & \gA & \soso & \soso & \soso & \good & 264(3)(5)    \\
CERN 08        & \cite{Giusti:2008vb}     &  2  & \gA & \soso & \bad  & \soso & \good & 276(3)(4)(5) \\
JLQCD/TWQCD 08A& \cite{Noaki:2008iy}      &  2  & \gA & \soso & \bad  & \bad  & \good & 235.7(5.0)(2.0)$\binom{+12.7}{-0.0}$\\
JLQCD/TWQCD 07A& \cite{Aoki:2007pw}       &  2  & \gA & \soso & \bad  & \bad  & \good & 252(5)(10)   \\
&&&&&&&& \\[-0.1cm]
\hline
&&&&&&&& \\[-0.1cm]
ETM 09B        & \cite{Jansen:2009tt}     &  2  & \rC & \good & \soso & \bad  & \good & 239.6(4.8) \\
Hasenfratz 08  & \cite{Hasenfratz:2008ce} &  2  & \gA & \soso & \bad  & \soso & \good & 248(6)     \\
JLQCD/TWQCD 07 & \cite{Fukaya:2007pn}     &  2  & \gA & \good & \bad  & \bad  & \good & 239.8(4.0) \\
&&&&&&&& \\[-0.1cm]
\hline
\hline
\end{tabular*}
\normalsize
\vspace*{-4mm}
\caption{\label{tabsigma}
Quark condensate $\Sigma\equiv |\langle\ubar u\rangle|_{m_u,m_d\to0}$: colour code and numerical values
in MeV (compare Fig.~\ref{fig_sigma}).}
\end{table}

\begin{figure}[!tb]
\centering
\includegraphics[width=12.0cm]{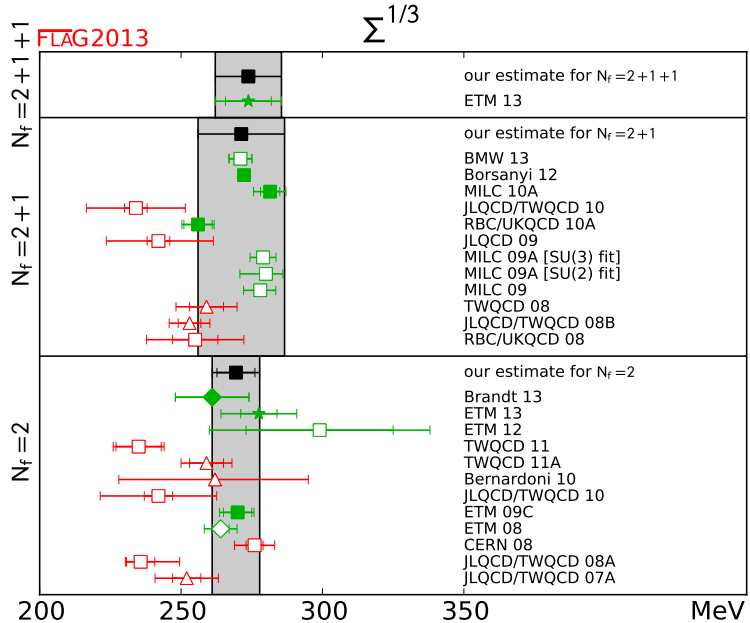}
\vspace*{-4mm}
\caption{\label{fig_sigma}
Quark condensate $\Sigma\equiv |\langle\ubar u\rangle|_{m_u,m_d\to0}$
($\overline{\rm MS}$-scheme, scale $\mu=2$ GeV).
Squares and left triangles indicate determinations from correlators in the
$p$- and $\epsilon$-regimes, respectively. Up triangles refer to extractions
from the topological susceptibility, diamonds to determinations from the pion
form factor, and star symbols refer to the spectral density method.
The black squares and grey bands indicate our estimates.
The meaning of the colours is explained in section \ref{sec:qualcrit}.}
\end{figure}

We begin with a discussion of the lattice results for the SU(2) LEC $\Sigma$.
We present the results in Table~\ref{tabsigma} and Figure \ref{fig_sigma}.
We add that results which include only a statistical error are listed in the
table but omitted from the plot.
Regarding the $\Nf=2$ computations there are five entries without a red tag
(ETM 08, ETM 09C, ETM 12, ETM 13, Brandt 13).
We form the average based on ETM 09C, ETM 13 (here we deviate from our
``superseded'' rule, since the latter work has a much bigger error) and
Brandt 13.
Regarding the $\Nf=2+1$ computations there are three published papers
(RBC/UKQCD 10A, MILC 10A and Borsanyi 12) which make it into the $\Nf=2+1$
average and a preprint (BMW 13) which will be included in a future update.
We also remark that among the three works included RBC/UKQCD 10A is
inconsistent with the other two (MILC 10A and Borsanyi 12).
For the time being we inflate the error of our $\Nf=2+1$ average such
that it includes all three central values it is based on.
This yields
\beq
\Sigma\big|_{\Nf=2}=269(08)\MeV
\;,\quad
\Sigma\big|_{\Nf=2+1}=271(15)\MeV
\;,
\eeq
where the errors include both statistical and systematic uncertainties.
In accordance with our guidelines we plead with the reader to cite
\cite{Baron:2009wt,Cichy:2013gja,Brandt:2013dua} (for $\Nf=2$) or
\cite{Aoki:2010dy,Bazavov:2010yq,Borsanyi:2012zv} (for $\Nf=2+1$)
when using these numbers.
Finally, for $\Nf=2+1+1$ there is only one calculation, and we recommend
to use the result of \cite{Cichy:2013gja} as given in Table~\ref{tabsigma}.
Another look at Figure~\ref{fig_sigma} confirms that these values
are well consistent with each other.


\begin{table}[!tbp] 
\vspace*{3cm}
\centering
\footnotesize
\begin{tabular*}{\textwidth}[l]{l@{\extracolsep{\fill}}rllllllll}
Collaboration & Ref. & $\Nf$ &
\hspace{0.15cm}\begin{rotate}{60}{publication status}\end{rotate}\hspace{-0.15cm}&
\hspace{0.15cm}\begin{rotate}{60}{chiral extrapolation}\end{rotate}\hspace{-0.15cm}&
\hspace{0.15cm}\begin{rotate}{60}{continuum  extrapolation}\end{rotate}\hspace{-0.15cm} &
\hspace{0.15cm}\begin{rotate}{60}{finite volume}\end{rotate}\hspace{-0.15cm} &
\hspace{0.15cm}\begin{rotate}{60}{renormalization}\end{rotate}\hspace{-0.15cm} &\rule{0.4cm}{0cm}
 $F$  &\rule{0.1cm}{0cm} $F_\pi/F$\\
&&&&&&&&& \\[-0.1cm]
\hline
\hline
&&&&&&&&& \\[-0.1cm]
ETM 11         & \cite{Baron:2011sf}      &2+1+1& \rC & \soso & \good & \soso & \good & 85.60(4)       & {\sl 1.077(2)} \\
ETM 10 $\dag$  & \cite{Baron:2010bv}      &2+1+1& \gA & \soso & \soso & \soso & \good & 85.66(6)(13)   & 1.076(2)(2)  \\
&&&&&&&&& \\[-0.1cm]
\hline
&&&&&&&&& \\[-0.1cm]
BMW 13         & \cite{Durr:2013goa}      & 2+1 & \oP & \good & \good & \good & \good & 88.0(1.3)(0.3) & 1.055(7)(2)                 \\
Borsanyi 12    & \cite{Borsanyi:2012zv}   & 2+1 & \gA & \good & \good & \soso & \good & 86.78(05)(25)  & 1.0627(06)(27)              \\
NPLQCD 11      & \cite{Beane:2011zm}      & 2+1 & \gA & \soso & \soso & \good &       &                & 1.062(26)$\binom{+42}{-40}$ \\
MILC 10A       & \cite{Bazavov:2010yq}    & 2+1 & \rC & \soso & \good & \good & \soso & 87.5(1.0)$\binom{+0.7}{-2.6}$ & {\sl 1.06(3)}\\
MILC 10        & \cite{Bazavov:2010hj}    & 2+1 & \rC & \soso & \good & \good & \soso & 87.0(4)(5)     & {\sl 1.060(8)}              \\
MILC 09A       & \cite{Bazavov:2009fk}    & 2+1 & \rC & \soso & \good & \good & \soso & 86.8(2)(4)     & 1.062(1)(3)                 \\
MILC 09        & \cite{Bazavov:2009bb}    & 2+1 & \gA & \soso & \good & \good &       &                & 1.052(2)$\binom{+6}{-3}$    \\
PACS-CS 08     & \cite{Aoki:2008sm}       & 2+1 & \gA & \good & \bad  & \bad  & \bad  & 89.4(3.3)      & 1.060(7)                    \\
RBC/UKQCD 08   & \cite{Allton:2008pn}     & 2+1 & \gA & \soso & \bad  & \good & \good & 81.2(2.9)(5.7) & 1.080(8)                    \\
&&&&&&&&& \\[-0.1cm]
\hline
&&&&&&&&& \\[-0.1cm]
Brandt 13      & \cite{Brandt:2013dua}    &  2  & \gA & \soso & \good & \good & \good & 84(8)(2)       & 1.080(16)(6)                \\
QCDSF 13       & \cite{Horsley:2013ayv}   &  2  & \oP & \good & \good & \bad  & \good & 86(1)          & {\sl 1.07(1)}               \\
Bernardoni 11  & \cite{Bernardoni:2011kd} &  2  & \rC & \soso & \bad  & \bad  & \soso & 79(4)          & {\sl 1.17(5)}               \\
TWQCD 11       & \cite{Chiu:2011bm}       &  2  & \gA & \good & \bad  & \bad  & \good & 83.39(35)(38)  & {\sl 1.106(6)}              \\
ETM 09C        & \cite{Baron:2009wt}      &  2  & \gA & \soso & \good & \soso &       &                & 1.0755(6)$\binom{+08}{-94}$ \\
ETM 08         & \cite{Frezzotti:2008dr}  &  2  & \gA & \soso & \soso & \soso & \good & 86.6(7)(7)     & 1.067(9)(9)                 \\
JLQCD/TWQCD 08A& \cite{Noaki:2008iy}      &  2  & \gA & \soso & \bad  & \bad  & \good & 79.0(2.5)(0.7)$\binom{+4.2}{-0.0}$ & {\sl 1.17(4)} \\
&&&&&&&&& \\[-0.1cm]
\hline
&&&&&&&&& \\[-0.1cm]
ETM 09B $^\S$  & \cite{Jansen:2009tt}     &  2  & \rC & \good & \soso & \bad  & \good & 90.2(4.8)      & {\sl 1.02(5)} \\
Hasenfratz 08  & \cite{Hasenfratz:2008ce} &  2  & \gA & \soso & \bad  & \soso & \good & 90(4)          & {\sl 1.02(4)} \\
JLQCD/TWQCD 07 & \cite{Fukaya:2007pn}     &  2  & \gA & \good & \bad  & \bad  & \good & 87.3(5.6)      & {\sl 1.06(6)} \\
&&&&&&&&& \\[-0.1cm]
\hline
&&&&&&&&& \\[-0.1cm]
Colangelo 03   & \cite{Colangelo:2003hf}  &     &     &       &       &       &       & 86.2(5)        & 1.0719(52) \\
&&&&&&&&& \\[-0.1cm]
\hline
\hline
\end{tabular*}
\flushleft
$\dag$ The values of $M_{\pi^+} L$ correspond to a green tag in the FV-column,
while those of $M_{\pi^0} L$ imply a red one; 
since both masses play a role in finite-volume effects, we opt for open green.
\\[0mm]
$^\S$ Result for $r_0 F$ converted into a value for $F$ via $r_0=0.49\,\fm$
(despite ETM quoting smaller values of $r_0$).
\normalsize
\vspace*{-2mm}
\caption{\label{tabf}
Results for the leading-order SU(2) low-energy constant $F$ (in MeV) and for the ratio $F_\pi/F$.
Numbers in slanted fonts have been calculated by us (see text for details). Horizontal
lines establish the same grouping as in Table~\ref{tabsigma}.}
\end{table}

\begin{figure}[!tb]
\begin{center}
\includegraphics[width=12.0cm]{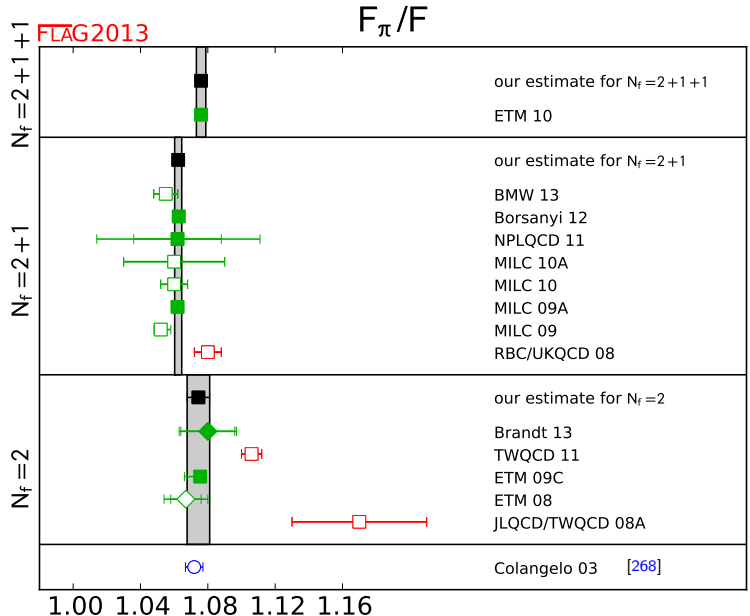}
\end{center}
\vspace{-2.0cm}\hspace{11.0cm}\parbox{6cm}{\sffamily\tiny  \cite{Colangelo:2003hf}}
\begin{center}
\vspace{0.25cm}
\caption{\label{fig_f}
Comparison of the results for the ratio of the physical pion decay constant
$F_\pi$ and  the leading-order SU(2) low-energy constant $F$.
The meaning of the symbols is the same as in Figure~\ref{fig_sigma}. }
\end{center}
\vspace{-0.3cm}
\end{figure}

The next quantity considered is $F$, i.e.\ the pion decay constant in the SU(2)
chiral limit ($m_{ud}\to0$ at fixed physical $m_s$) in the Bernese normalization.
As argued on previous occasions we tend to give preference to $\Fpi/F$ (here the
numerator is meant to refer to the physical-pion-mass point) wherever it is
available, since often some of the systematic uncertainties are mitigated.
We collect the results in Table~\ref{tabf} and Figure~\ref{fig_f}.
In those cases where the collaboration provides only $F$, the ratio is computed
on the basis of the phenomenological value of $\Fpi$, and the corresponding
entries in Table~\ref{tabf} are in slanted fonts.
Among the $\Nf=2$ determinations only three (ETM 08, ETM 09C and Brandt 13) are
without red tags.
Since the first two are by the same collaboration, only the latter two enter
the average.
Among the  $\Nf=2+1$ determinations three values (MILC 09A as an obvious
update of MILC 09, NPLQCD 11 and Borsanyi 12) make it into the average.
Finally, there is a single $\Nf=2+1+1$ determination (ETM 10) which forms the
current best estimate in this category.

Given this input our averaging procedure yields
\beq
\frac{\Fpi}{F}\big|_{\Nf=2}=1.0744(67)
\;,\quad
\frac{\Fpi}{F}\big|_{\Nf=2+1}=1.0624(21)
\;,
\label{results_fpi_f}
\eeq
where the errors include both statistical and systematic uncertainties.
We plead with the reader to cite
\cite{Baron:2009wt,Brandt:2013dua} (for $\Nf=2$) or
\cite{Bazavov:2009fk,Beane:2011zm,Borsanyi:2012zv} (for $\Nf=2+1$)
when using these numbers.
Finally, for $\Nf=2+1+1$ we recommend to use the result of
\cite{Baron:2010bv}; see Table~\ref{tabf} for the numerical value.
From these numbers (or from a look at Figure~\ref{fig_f}) it is obvious that
the $\Nf=2+1$ and $\Nf=2+1+1$ results are not quite consistent.
From a theoretical viewpoint this is rather surprising, since the only difference
(the presence of absence of a dynamical charm quark) is expected to have a rather
insignificant effect on this ratio (which, in addition, would be monotonic in
$\Nf$, contrary to what is seen in Figure~\ref{fig_f}).
In our view this indicates that --~in spite of the conservative attitude
taken in this report~-- the theoretical uncertainties in at least one of the
two cases is likely underestimated.
We hope that a future release of the FLAG report can clarify the issue.


\begin{table}[!tbp] 
\vspace*{3cm}
\centering
\footnotesize
\begin{tabular*}{\textwidth}[l]{l@{\extracolsep{\fill}}rlllllll}
Collaboration & Ref. & $\Nf$ &
\hspace{0.15cm}\begin{rotate}{60}{publication status}\end{rotate}\hspace{-0.15cm} &
\hspace{0.15cm}\begin{rotate}{60}{chiral extrapolation}\end{rotate}\hspace{-0.15cm}&
\hspace{0.15cm}\begin{rotate}{60}{continuum  extrapolation}\end{rotate}\hspace{-0.15cm} &
\hspace{0.15cm}\begin{rotate}{60}{finite volume}\end{rotate}\hspace{-0.15cm} &\rule{0.3cm}{0cm} $\bar\ell_3$ & \rule{0.3cm}{0cm}$\bar\ell_4$\\
&&&&&&&& \\[-0.1cm]
\hline
\hline
&&&&&&&& \\[-0.1cm]
ETM 11         & \cite{Baron:2011sf}      &2+1+1& \rC & \soso & \good & \soso & 3.53(5)     & 4.73(2)  \\
ETM 10         & \cite{Baron:2010bv}      &2+1+1& \gA & \soso & \soso & \soso & 3.70(7)(26) & 4.67(3)(10) \\
&&&&&&&& \\[-0.1cm]
\hline
&&&&&&&& \\[-0.1cm]
BMW 13         & \cite{Durr:2013goa}      & 2+1 & \oP & \good & \good & \good & 2.5(5)(4)    & 3.8(4)(2)    \\
RBC/UKQCD 12   & \cite{Arthur:2012opa}    & 2+1 & \gA & \good & \soso & \good & 2.91(23)(07) & 3.99(16)(09) \\
Borsanyi 12    & \cite{Borsanyi:2012zv}   & 2+1 & \gA & \good & \good & \soso & 3.16(10)(29) & 4.03(03)(16) \\
NPLQCD 11      & \cite{Beane:2011zm}      & 2+1 & \gA & \soso & \soso & \good & 4.04(40)$\binom{+73}{-55}$ & 4.30(51)$\binom{+84}{-60}$ \\
MILC 10A       & \cite{Bazavov:2010yq}    & 2+1 & \rC & \soso & \good & \good & 2.85(81)$\binom{+37}{-92}$ & 3.98(32)$\binom{+51}{-28}$ \\
MILC 10        & \cite{Bazavov:2010hj}    & 2+1 & \rC & \soso & \good & \good & 3.18(50)(89) & 4.29(21)(82) \\
RBC/UKQCD 10A  & \cite{Aoki:2010dy}       & 2+1 & \gA & \soso & \soso & \good & 2.57(18)     & 3.83(9) \\
MILC 09A       & \cite{Bazavov:2009fk}    & 2+1 & \rC & \soso & \good & \good & 3.32(64)(45) & 4.03(16)(17) \\
MILC 09A       & \cite{Bazavov:2009fk}    & 2+1 & \rC & \soso & \good & \good & 3.0(6)$\binom{+9}{-6}$& 3.9(2)(3) \\
PACS-CS 08     & \cite{Aoki:2008sm}       & 2+1 & \gA & \good & \bad  & \bad  & 3.47(11)     & 4.21(11) \\
PACS-CS 08     & \cite{Aoki:2008sm}       & 2+1 & \gA & \good & \bad  & \bad  & 3.14(23)     & 4.04(19) \\
RBC/UKQCD 08   & \cite{Allton:2008pn}     & 2+1 & \gA & \soso & \bad  & \good & 3.13(33)(24) & 4.43(14)(77) \\
&&&&&&& \\[-0.1cm]
\hline
&&&&&&& \\[-0.1cm]
G\"ulpers 13   & \cite{Gulpers:2013uca}   &  2  & \oP & \soso & \bad  & \good &               & 4.76(13)(--)  \\
Brandt 13      & \cite{Brandt:2013dua}    &  2  & \gA & \soso & \good & \good & 3.0(7)(5)     & 4.7(4)(1)     \\
QCDSF 13       & \cite{Horsley:2013ayv}   &  2  & \oP & \good & \good & \bad  &               & 4.2(1)        \\
Bernardoni 11  & \cite{Bernardoni:2011kd} &  2  & \rC & \soso & \bad  & \bad  & 4.46(30)(14)  & 4.56(10)(4)   \\
TWQCD 11       & \cite{Chiu:2011bm}       &  2  & \gA & \good & \bad  & \bad  & 4.149(35)(14) & 4.582(17)(20) \\
ETM 09C        & \cite{Baron:2009wt}      &  2  & \gA & \soso & \good & \soso & 3.50(9)$\binom{+09}{-30}$&4.66(4)$\binom{+04}{-33}$ \\
JLQCD/TWQCD 09 & \cite{JLQCD:2009qn}      &  2  & \gA & \soso & \bad  & \bad  &               & 4.09(50)(52) \\
ETM 08         & \cite{Frezzotti:2008dr}  &  2  & \gA & \soso & \soso & \soso & 3.2(8)(2)     & 4.4(2)(1) \\
JLQCD/TWQCD 08A& \cite{Noaki:2008iy}      &  2  & \gA & \soso & \bad  & \bad  & 3.38(40)(24)$\binom{+31}{-00}$& 4.12(35)(30)$\binom{+31}{-00}$ \\
CERN-TOV 06    & \cite{DelDebbio:2006cn}  &  2  & \gA & \soso & \good & \bad  & 3.0(5)(1)     & \\
&&&&&&& \\[-0.1cm]
\hline
&&&&&&& \\[-0.1cm]
Colangelo 01   & \cite{Colangelo:2001df}  &&&&&&         & 4.4(2) \\
Gasser 84      & \cite{Gasser:1983yg}     &&&&&& 2.9(2.4)& 4.3(9) \\
&&&&&&& \\[-0.1cm]
\hline
\hline
\end{tabular*}
\normalsize
\caption{\label{tabl3and4}
Results for the SU(2) NLO couplings $\bar\ell_3$ and $\lbar_4$. The MILC 10
results are obtained by converting the SU(3) LECs, while the MILC 10A results
are obtained with a direct SU(2) fit. For comparison, the last two lines show
results from phenomenological analyses.}
\end{table}

\begin{figure}[!p]
\begin{center}
\includegraphics[width=9.0cm]{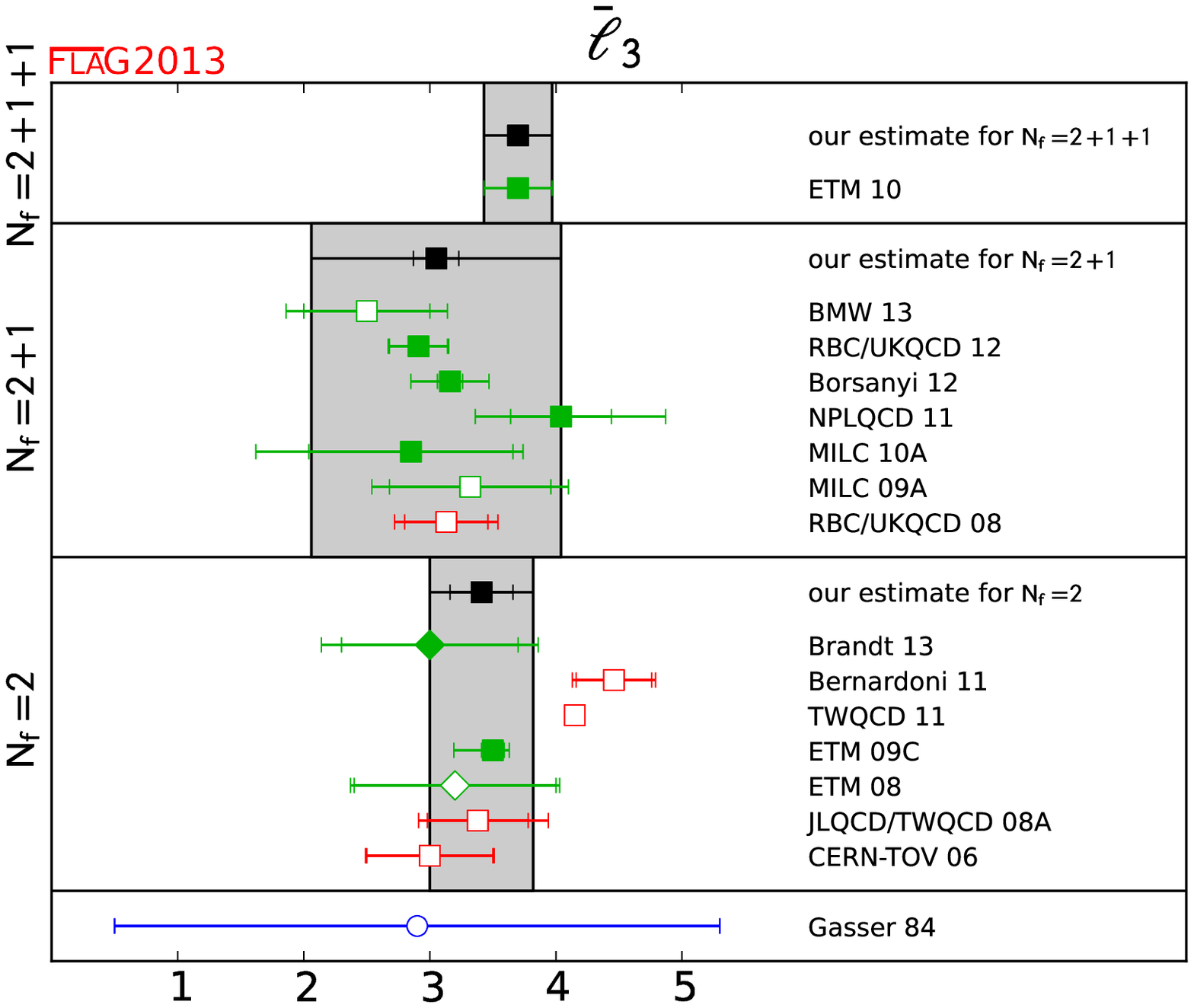}\\[-4mm]
\end{center}
\vspace{-1.68cm}\hspace{10.3cm}\parbox{6cm}{\sffamily\tiny\cite{Gasser:1983yg}}
\vspace{0.1cm}
\begin{center}
\includegraphics[width=9.0cm]{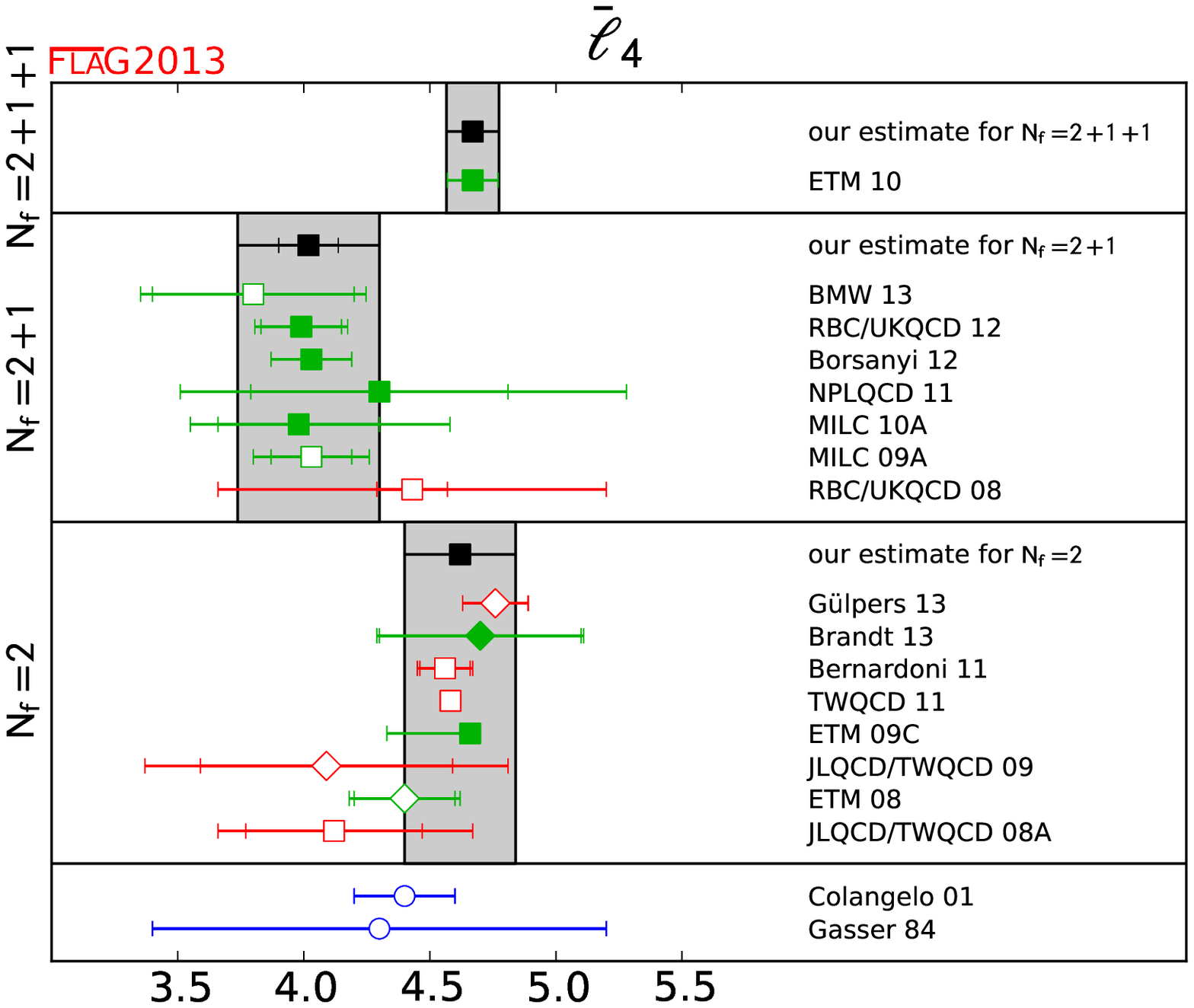}\\[-4mm]
\end{center}
\vspace{-1.85cm}\hspace{10.5cm}\parbox{6cm}{\sffamily\tiny\cite{Colangelo:2001df}\\
\vspace{-1.8em}\hspace{0.17em}\cite{Gasser:1983yg}}
\vspace{0.1cm}
\begin{center}
\includegraphics[width=9.0cm]{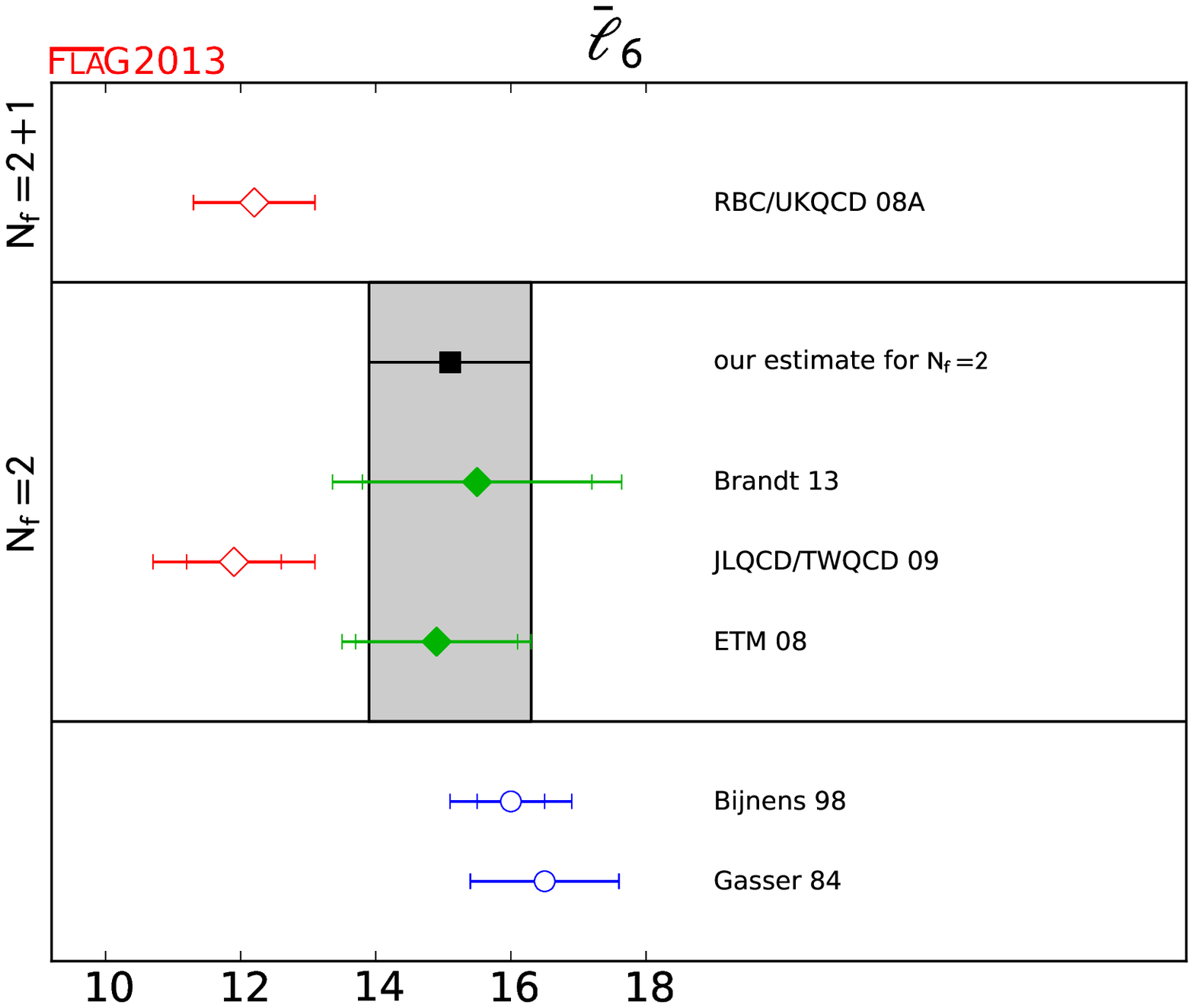}
\end{center}
\vspace{-2.2cm}\hspace{9.7cm}\parbox{6cm}{\sffamily\tiny\cite{Bijnens:1998fm}\\
\\
\vspace{0.02em}\hspace{0.17em}\cite{Gasser:1983yg}}
\vspace{0.5cm}
\begin{center}
\caption{\label{fig_l3l4l6}
Effective coupling constants $\bar \ell_3$, $\bar\ell_4$ and $\bar\ell_6$.
Squares indicate determinations from correlators in the $p$-regime, diamonds
refer to determinations from the pion form factor.}
\end{center}
\end{figure}

We move on to a discussion of the lattice results for the NLO LECs
$\bar\ell_3$ and $\bar\ell_4$.
We remind the reader that on the lattice the former LEC is obtained as a result
of the tiny deviation from linearity seen in $\Mpi^2$ versus $Bm_{ud}$, whereas
the latter LEC is extracted from the curvature in $\Fpi$ versus $Bm_{ud}$.
The available determinations are presented in Table~\ref{tabl3and4} and
Figure~\ref{fig_l3l4l6}.
Among the $\Nf=2$ determinations ETM 08, ETM 09C and Brandt 13 are published
and without red tags, and our rules imply that the latter two determinations
enter our average.
The colour coding of the $\Nf=2+1$ results looks very promising; there is
a significant number of lattice determinations without any red tag.
At first sight it seems that RBC/UKQCD 10A, MILC 10A, NPLQCD 11, Borsanyi 12 and
RBC/UKQCD 12 make it into the average.
Unfortunately, $\bar\ell_3$ and $\bar\ell_4$ of RBC/UKQCD 10A have no systematic
error; therefore we exclude this work from the $\Nf=2+1$ average.
Among the $\Nf=2+1+1$ determinations only ETM 10 qualifies for an average.

Given this input our averaging procedure yields
\beq
\bar\ell_3\big|_{\Nf=2}=3.41(41)
\;,\quad
\bar\ell_3\big|_{\Nf=2+1}=3.05(99)
\;,
\label{results_l3}
\eeq
\beq
\bar\ell_4\big|_{\Nf=2}=4.62(22)
\;,\quad
\bar\ell_4\big|_{\Nf=2+1}=4.02(28)
\;,
\label{results_l4}
\eeq
where the errors include both statistical and systematic uncertainties.
Again we plead with the reader to cite
\cite{Baron:2009wt,Brandt:2013dua} (for $\Nf=2$) or
\cite{Bazavov:2010yq,Beane:2011zm,Borsanyi:2012zv,Arthur:2012opa}
(for $\Nf=2+1$) when using these numbers.
For $\Nf=2+1+1$ we stay with the recommendation to use the results of
\cite{Baron:2010bv}, see Table~\ref{tabl3and4} for the numerical values.

Let us add two remarks.
On the input side our procedure%
\footnote{There are two naive procedures to symmetrize an asymmetric systematic
error: ($i$) keep the central value untouched and enlarge the smaller error,
($ii$) shift the central value by half of the difference between the two original
errors and enlarge/shrink both errors by the same amount. Our procedure ($iii$)
is to average the results of ($i$) and ($ii$). In other words a result
$c(s)\binom{u}{\ell}$ with $\ell>u$ is changed into $c+(u-\ell)/4$ with
statistical error $s$ and a symmetric systematic error $(u+3\ell)/4$.
The case $\ell<u$ is handled accordingly.}
symmetrizes the asymmetric error of ETM 09C with a slight adjustment of the
central value.
On the output side the error of the $\bar\ell_3$ average for $\Nf=2$ and
of the $\bar\ell_3,\bar\ell_4$ averages for $\Nf=2+1$, according to the FLAG
procedure, got inflated by hand to cover all central values.
From these numbers (or from a look at Figure~\ref{fig_l3l4l6}) it is clear that
the lattice results for $\bar\ell_3$ do not show any obvious $\Nf$-dependence
-- thanks, chiefly, to our conservative error treatment strategy.
On the other hand, in the case of $\bar\ell_4$ even our practice of inflating
the error of the $\Nf=2+1$ average did not manage to avoid some mild
inconsistency between the $\Nf=2+1$ average on one side and either the $\Nf=2$
or the $\Nf=2+1+1$ average on the other side.
Again, the dependence of the average on the number of active flavours is not
monotonic, and this raises a decent amount of suspicion that some of the
systematic errors might still be underestimated.

More specifically, it seems that again the $\Nf=2+1+1$ value by ETM shows some
tension relative to the average $\Nf=2+1$ value quoted above, in close analogy to
what happened for $F$ or $\Fpi/F$; see the discussion around (\ref{results_fpi_f}).
Since both $F$ and $\bar\ell_4$ are determined from the quark-mass dependence of
the pseudoscalar decay constant, perhaps the formulas in
Refs.\,\cite{Bar:2010jk,Colangelo:2010cu} for dealing with cut-off and
finite-volume effects with twisted-mass data might prove useful in future analysis.

From a more phenomenological viewpoint there is a notable difference between
$\bar\ell_3$ and $\bar\ell_4$ in Figure~\ref{fig_l3l4l6}.
For $\bar\ell_4$ the precision of the phenomenological determination achieved in Colangelo 01
\cite{Colangelo:2001df} represents a significant improvement compared to Gasser 84 \cite{Gasser:1983yg}.
Picking any $\Nf$, the lattice average of $\bar\ell_4$ is consistent with
both of the phenomenological values and comes with an error which is
roughly comparable to the uncertainty of the result in Colangelo 01 \cite{Colangelo:2001df}.
By contrast, for $\bar\ell_3$ the error of the lattice determination is
significantly smaller than the error of the estimate given in Gasser 84 \cite{Gasser:1983yg}.
In other words, here the lattice really provides some added value.


\begin{table}[!tbp] 
\vspace*{3cm}
\footnotesize
\begin{tabular*}{\textwidth}[l]{l@{\extracolsep{\fill}}rllllllll}
Collaboration & Ref. & $\Nf$ &
\hspace{0.15cm}\begin{rotate}{60}{publication status}\end{rotate}\hspace{-0.15cm} &
\hspace{0.15cm}\begin{rotate}{60}{chiral extrapolation}\end{rotate}\hspace{-0.15cm}&
\hspace{0.15cm}\begin{rotate}{60}{continuum  extrapolation}\end{rotate}\hspace{-0.15cm} &
\hspace{0.15cm}\begin{rotate}{60}{finite volume}\end{rotate}\hspace{-0.15cm} &
\rule{0.3cm}{0cm}$\<r^2\>_V^\pi$ & \rule{0.3cm}{0cm}$c_V$ & \rule{0.3cm}{0cm}$\bar\ell_6$ \\
&&&&&&&&& \\[-0.1cm]
\hline
\hline
&&&&&&&&& \\[-0.1cm]
RBC/UKQCD 08A  & \cite{Boyle:2008yd}     & 2+1 & \gA & \soso & \bad  & \good & 0.418(31) &     & 12.2(9)\\
LHP 04         & \cite{Bonnet:2004fr}    & 2+1 & \gA & \soso & \bad  & \soso & 0.310(46) &     &    \\
&&&&&&&&& \\[-0.1cm]
\hline
&&&&&&&&& \\[-0.1cm]
Brandt 13      & \cite{Brandt:2013dua}   &  2  & \gA & \soso & \good & \good & 0.481(33)(13) &              & 15.5(1.7)(1.3) \\
JLQCD/TWQCD 09 & \cite{JLQCD:2009qn}     &  2  & \gA & \soso & \bad  & \bad  & 0.409(23)(37) & 3.22(17)(36) & 11.9(0.7)(1.0) \\
ETM 08         & \cite{Frezzotti:2008dr} &  2  & \gA & \soso & \soso & \soso & 0.456(30)(24) & 3.37(31)(27) & 14.9(1.2)(0.7) \\
QCDSF/UKQCD 06A& \cite{Brommel:2006ww}   &  2  & \gA & \soso & \good & \soso & 0.441(19)(56)(29) &     &     \\[-0.1cm]
&&&&&&&&& \\[-0.1cm]
\hline
&&&&&&&&& \\[-0.1cm]
Bijnens 98     & \cite{Bijnens:1998fm}   &&&&&& 0.437(16) & 3.85(60) & 16.0(0.5)(0.7) \\
NA7 86         & \cite{Amendolia:1986wj} &&&&&& 0.439(8)  &          &                \\
Gasser 84      & \cite{Gasser:1983yg}    &&&&&&           &          & 16.5(1.1)      \\
&&&&&&&&& \\[-0.1cm]
\hline
\hline
\end{tabular*}
\\[3cm]
\begin{tabular*}{\textwidth}[l]{l@{\extracolsep{\fill}}rlllllll}
Collaboration & Ref. & $\Nf$ &
\hspace{0.15cm}\begin{rotate}{60}{publication status}\end{rotate}\hspace{-0.15cm} &
\hspace{0.15cm}\begin{rotate}{60}{chiral extrapolation}\end{rotate}\hspace{-0.15cm}&
\hspace{0.15cm}\begin{rotate}{60}{continuum  extrapolation}\end{rotate}\hspace{-0.15cm} &
\hspace{0.15cm}\begin{rotate}{60}{finite volume}\end{rotate}\hspace{-0.15cm} &
\rule{0.3cm}{0cm}$\<r^2\>_S^\pi$ & \rule{0.3cm}{0cm}$\bar\ell_1-\bar\ell_2$ \\
&&&&&&&& \\[-0.1cm]
\hline
\hline
&&&&&&&& \\[-0.1cm]
G\"ulpers 13   & \cite{Gulpers:2013uca} &  2  & \oP & \soso & \bad  & \good & 0.637(23)(--) &                \\
JLQCD/TWQCD 09 & \cite{JLQCD:2009qn}    &  2  & \gA & \soso & \bad  & \bad  & 0.617(79)(66) & -2.9(0.9)(1.3) \\
&&&&&&&& \\[-0.1cm]
\hline
&&&&&&&& \\[-0.1cm]
Colangelo 01 & \cite{Colangelo:2001df}     &&&&&& 0.61(4) & -4.7(6) \\
&&&&&&&& \\[-0.1cm]
\hline
\hline
&&&&&&&& \\[-0.1cm]
\end{tabular*}
\normalsize
\vspace*{-4mm}
\caption{\label{tabl6}
{\it Top panel: vector form factor of the pion.} Lattice results for the charge
radius $\<r^2\>_V^\pi$ (in $\mathrm{fm}^2$), the curvature $c_V$ (in $\mathrm{GeV}^{-4}$)
and the effective coupling constant $\bar\ell_6$ are compared with the experimental value
obtained by NA7 and some phenomenological estimates.
{\it Bottom panel: scalar form factor of the pion.} Lattice results for the scalar
radius $\< r^2 \>_S^\pi$ (in $\mathrm{fm}^2$) and the combination $\bar\ell_1-\bar\ell_2$
are compared with a dispersive calculation of these quantities \cite{Colangelo:2001df}.}
\end{table}

We finish with a discussion of the lattice results for $\bar\ell_6$ and
$\bar\ell_1-\bar\ell_2$.
The LEC $\lbar_6$ determines the leading contribution in the chiral expansion
of the pion charge radius -- see (\ref{formula_rsqu}).
Hence from a lattice study of the vector form factor of the pion with
several $\Mpi$ one may extract the radius $\<r^2\>_V^\pi$, the curvature $c_V$
(both at the physical pion-mass point) and the LEC $\bar\ell_6$ in one go.
Similarly, the leading contribution in the chiral expansion of the scalar
radius of the pion determines $\bar\ell_4$ -- see (\ref{formula_rsqu}).
This LEC is also present in the pion-mass dependence of $\Fpi$, as we have seen.
The difference $\lbar_1-\lbar_2$, finally, may be obtained  from the momentum
dependence of the vector and scalar pion form factors, based on the two-loop
formulae of \cite{Bijnens:1998fm}.
The top part of Table~\ref{tabl6} collects the results obtained from the vector
form factor of the pion (charge radius, curvature and $\lbar_6$).
Regarding this low-energy constant two $\Nf=2$ calculations are published works
without a red tag; we thus arrive at the estimate
\beq
\bar\ell_6\big|_{\Nf=2}=15.1(1.2)
\eeq
which is represented as a grey band in the last panel of Fig.\,\ref{fig_l3l4l6}.
Here we plead with the reader to cite \cite{Frezzotti:2008dr,Brandt:2013dua}
when using this number.

The experimental information concerning the charge radius is excellent and
the curvature is also known very accurately, based on $e^+e^-$ data and
dispersion theory.
The vector form factor calculations thus present an excellent testing ground
for the lattice methodology.
The table shows that most of the available lattice results pass the test.
There is, however, one worrisome point.
For $\lbar_6$ the agreement seems less convincing than for the charge radius,
even though the two quantities are closely related.
So far we have no explanation, but we urge the groups to pay special attention
to this point.
Similarly, the bottom part of Table~\ref{tabl6} collects the results obtained
for the scalar form factor of the pion and the combination $\lbar_1-\lbar_2$
that is extracted from it.

Perhaps the most important physics result of this section is that the lattice
simulations confirm the approximate validity of the Gell-Mann-Oakes-Renner
formula and show that the square of the pion mass indeed grows in proportion
to $m_{ud}$.
The formula represents the leading term of the chiral perturbation series and
necessarily receives corrections from higher orders.
At first nonleading order, the correction is determined by the effective
coupling constant $\lbar_3$.
The results collected in Table~\ref{tabl3and4} and in the top panel of
Figure~\ref{fig_l3l4l6} show that $\lbar_3$ is now known quite well.
They corroborate the conclusion drawn already in Ref.\,\cite{Durr:2002zx}:
the lattice confirms the estimate of $\bar\ell_3$ derived in \cite{Gasser:1983yg}.
In the graph of $M_\pi^2$ versus $m_{ud}$, the values found on the lattice for
$\lbar_3$ correspond to remarkably little curvature: the Gell-Mann-Oakes-Renner
formula represents a reasonable first approximation out to values of $m_{ud}$
that exceed the physical value by an order of magnitude.

As emphasized by Stern and collaborators \cite{Fuchs:1991cq,Stern:1993rg,
DescotesGenon:1999uh}, the analysis in the framework of $\chi$PT is coherent
only if ($i$) the leading term in the chiral expansion of $M_\pi^2$
dominates over the remainder and ($ii$) the ratio $m_s/m_{ud}$ is close to
the value $25.6$ that follows from Weinberg's leading-order formulae.
In order to investigate the possibility that one or both of these conditions
might fail, the authors proposed a more general framework, referred to as
``Generalized {\Ch}PT'', which includes {\Ch}PT as a special case.
The results found on the lattice demonstrate that QCD does satisfy both of the
above conditions -- in the context of QCD, the proposed generalization of
the effective theory does not appear to be needed.
There is a modified version, however, referred to as ``Resummed {\Ch}PT''
\cite{Bernard:2010ex}, which is motivated by the possibility that the Zweig
rule violating couplings $L_4$ and $L_6$ might be larger than expected.
The available lattice data do not support this possibility, but they do not
rule it out either (see section \ref{sec:SU3results} for details).


\subsection{SU(3) Low-Energy Constants \label{sec:su3}}


\subsubsection{Quark-mass dependence of pseudoscalar masses and decay constants}

In the isospin limit, the relevant SU(3) formulae take the form
\cite{Gasser:1984gg}
\bea
\Mpi^2\!\!&\!\!\NLo\!\!&\!\! 2B_0m_{ud}
\Big\{
1+\mu_\pi-\frac{1}{3}\mu_\et+\frac{B_0}{F_0^2}
\Big[16m_{ud}(2L_8\!-\!L_5)+16(m_s\!+\!2m_{ud})(2L_6\!-\!L_4)\Big]
\Big\}\;,\nn
\\
M_{\!K}^2\!\!&\!\!\NLo\!\!&\!\! B_0(m_s\!\!+\!m_{ud})
\Big\{
1\!+\!\frac{2}{3}\mu_\et\!+\!\frac{B_0}{F_0^2}
\Big[8(m_s\!\!+\!m_{ud})(2L_8\!-\!L_5)\!+\!16(m_s\!\!+\!2m_{ud})(2L_6\!-\!L_4)\Big]
\Big\}\;,\quad\nonumber
\\
\Fpi\!\!&\!\!\NLo\!\!&\!\!F_0
\Big\{
1-2\mu_\pi-\mu_K+\frac{B_0}{F_0^2}
\Big[8m_{ud}L_5+8(m_s\!+\!2m_{ud})L_4\Big]
\Big\}\;,
\\
\Fka\!\!&\!\!\NLo\!\!&\!\!F_0
\Big\{
1-\frac{3}{4}\mu_\pi-\frac{3}{2}\mu_K-\frac{3}{4}\mu_\et+\frac{B_0}{F_0^2}
\Big[4(m_s\!+\!m_{ud})L_5+8(m_s\!+\!2m_{ud})L_4\Big]
\Big\}\;,\nonumber
\eea
where $m_{ud}$ is the common up and down quark mass (which may be different
from the one in the real world), and $B_0=\Sigma_0/F_0^2$, $F_0$ denote the
condensate parameter and the pseudoscalar decay constant in the SU(3) chiral
limit, respectively.
In addition, we use the notation
\beq
\mu_P=\frac{M_P^2}{32\pi^2F_0^2}
\ln\!\Big(\frac{M_P^2}{\mu^2}\Big)\;.
\label{def_muP}
\eeq
At the order of the chiral expansion used in these formulae, the quantities
$\mu_\pi$, $\mu_K$, $\mu_\eta$ can equally well be evaluated with the
leading-order expressions for the masses,
\beq
M_\pi^2\Lo 2B_0\,m_{ud}\;,\quad
M_K^2\Lo B_0(m_s\!+\!m_{ud})\;,\quad
M_\et^2\Lo \mbox{$\frac{2}{3}$}B_0(2m_s\!+\!m_{ud})
\;.
\eeq
Throughout, $L_i$ denotes the renormalized low-energy constant/coupling (LEC)
at scale $\mu$, and we adopt the convention which is standard in phenomenology,
$\mu=770\MeV$.
The normalization used for the decay constants is specified in
footnote \ref{foot:fpi}.


\subsubsection{Charge radius}

The SU(3) formula for the slope of the pion vector form factor reads
\cite{Gasser:1984ux}
\beq
\<r^2\>_V^\pi\;\Lo\;-\frac{1}{32\pi^2F_0^2}
\Big\{
3+2\ln(\frac{\Mpi^2}{\mu^2})+\ln(\frac{\Mka^2}{\mu^2})
\Big\}
+\frac{12L_9}{F_0^2}
\;,
\eeq
while the expression $\<r^2\>_S^\mathrm{oct}$ for the octet part of the
scalar radius does not contain any NLO low-energy constant at the one-loop order
\cite{Gasser:1984ux} (cf.\ \ref{sec:su2_form} for the situation in SU(2)).


\subsubsection{Partially quenched formulae}

The term ``partially quenched QCD'' is used in two ways.
For heavy quarks ($c,b$ and sometimes $s$) it usually means that these flavours
are included in the valence sector, but not into the functional determinant.
For the light quarks ($u,d$ and sometimes $s$) it means that they are
present in both the valence and the sea sector of the theory, but with
different masses (e.g.\ a series of valence quark masses is evaluated on an
ensemble with a fixed sea quark mass).

The program of extending the standard (unitary) SU(3) theory to the
(second version of) ``partially quenched QCD''
has been completed at the two-loop (NNLO) level for masses and decay
constants~\cite{Bijnens:2006jv}.
These formulae tend to be complicated, with the consequence that a
state-of-the-art analysis with $O(2000)$ bootstrap samples on $O(20)$
ensembles with $O(5)$ masses each [and hence $O(200'000)$ different fits] will
require significant computational resources for the global fits.
For an up-to-date summary of recent developments in Chiral Perturbation Theory
relevant to lattice QCD we refer to~\cite{Bijnens:2011tb}.

The theoretical underpinning of how ``partial quenching'' is to be treated in
the (properly extended) chiral framework is given in \cite{Bernard:2013kwa}.
Specifically for partially quenched QCD with staggered quarks it is shown that
a transfer matrix can be constructed which is not Hermitian but bounded, and
can thus be used to construct correlation functions in the usual way.


\subsubsection{Lattice determinations\label{sec:SU3results}}

\begin{table}[!tb] 
\vspace*{3cm}
\centering
\footnotesize
\begin{tabular*}{\textwidth}[l]{l@{\extracolsep{\fill}}rclllllllllll}
 & Ref. & $\Nf$ &
\hspace{0.15cm}\begin{rotate}{60}{publication status}\end{rotate}\hspace{-0.15cm} &
\hspace{0.15cm}\begin{rotate}{60}{chiral extrapolation}\end{rotate}\hspace{-0.15cm} &
\hspace{0.15cm}\begin{rotate}{60}{continuum  extrapolation}\end{rotate}\hspace{-0.15cm} &
\hspace{0.15cm}\begin{rotate}{60}{finite volume}\end{rotate}\hspace{-0.15cm} &
\hspace{0.15cm}\begin{rotate}{60}{renormalization}\end{rotate}\hspace{-0.15cm} &
\rule{0.3cm}{0cm}$F_0$ & \rule{0.1cm}{0cm} $F/F_0$ & \rule{0.2cm}{0cm}$B/B_0$ & \hspace{2.5cm} \\
&&&&&&&&&&&& \\[-0.1cm]
\hline
\hline
&&&&&&&&&&&& \\[-0.1cm]
JLQCD/TWQCD 10 & \cite{Fukaya:2010na}  &  3  & \gA & \bad  & \bad  & \bad  & \good & 71(3)(8)       &                           & \\
&&&&&&&&&&&& \\[-0.2cm]
\hline \\[-0.1cm]
MILC 10        & \cite{Bazavov:2010hj} & 2+1 & \rC & \soso & \good & \good & \soso & 80.3(2.5)(5.4) &                           & \\
MILC 09A       & \cite{Bazavov:2009fk} & 2+1 & \rC & \soso & \good & \good & \soso & 78.3(1.4)(2.9) & {\sl 1.104(3)(41)}        & {\sl 1.21(4)$\binom{+5}{-6}$}    \\
MILC 09        & \cite{Bazavov:2009bb} & 2+1 & \gA & \soso & \good & \good & \soso &                & 1.15(5)$\binom{+13}{-03}$ & {\sl 1.15(16)$\binom{+39}{-13}$} \\
PACS-CS 08     & \cite{Aoki:2008sm}    & 2+1 & \gA & \good & \bad  & \bad  & \bad  & 83.8(6.4)      & 1.078(44)                 & 1.089(15)\\
RBC/UKQCD 08   & \cite{Allton:2008pn}  & 2+1 & \gA & \soso & \bad  & \good & \good & 66.1(5.2)      & 1.229(59)                 & 1.03(05) \\
&&&&&&&&&&& \\[-0.1cm]
\hline
\hline
\end{tabular*}
\newline
\vspace*{3cm}
\begin{tabular*}{\textwidth}[l]{l@{\extracolsep{\fill}}rclllllll}
 & Ref. & $\Nf$ &
\hspace{0.15cm}\begin{rotate}{60}{publication status}\end{rotate}\hspace{-0.15cm} &
\hspace{0.15cm}\begin{rotate}{60}{chiral extrapolation}\end{rotate}\hspace{-0.15cm} &
\hspace{0.15cm}\begin{rotate}{60}{continuum  extrapolation}\end{rotate}\hspace{-0.15cm} &
\hspace{0.15cm}\begin{rotate}{60}{finite volume}\end{rotate}\hspace{-0.15cm} &
\hspace{0.15cm}\begin{rotate}{60}{renormalization}\end{rotate}\hspace{-0.15cm} &
\rule{0.3cm}{0cm}$\Sigma_0^{1/3}$ & \rule{0.1cm}{0cm} $\Sigma/\Sigma_0$ \\
&&&&&&&&& \\[-0.1cm]
\hline
\hline
&&&&&&&&& \\[-0.1cm]
JLQCD/TWQCD 10 & \cite{Fukaya:2010na}  &  3  & \gA & \bad  & \bad  & \bad  & \good & 214(6)(24)                  & {\sl 1.31(13)(52)} \\
&&&&&&&&& \\[-0.2cm]
\hline \\[-0.1cm]
MILC 09A       & \cite{Bazavov:2009fk} & 2+1 & \rC & \soso & \good & \good & \soso & 245(5)(4)(4)                & {\sl 1.48(9)(8)(10)}       \\
MILC 09        & \cite{Bazavov:2009bb} & 2+1 & \gA & \soso & \good & \good & \soso & 242(9)$\binom{+05}{-17}$(4) & 1.52(17)$\binom{+38}{-15}$ \\
PACS-CS 08     & \cite{Aoki:2008sm}    & 2+1 & \gA & \good & \bad  & \bad  & \bad  & 290(15)                     & 1.245(10)     \\
RBC/UKQCD 08   & \cite{Allton:2008pn}  & 2+1 & \gA & \soso & \bad  & \good & \good &                             & 1.55(21)      \\
&&&&&&&&& \\[-0.1cm]
\hline
\hline
\end{tabular*}
\normalsize
\vspace*{-2mm}
\caption{\label{tab:SU3_overview}
Lattice results for the low-energy constants $F_0$, $B_0$ and
$\Sigma_0\equiv F_0^2B_0$, which specify the effective SU(3) Lagrangian at
leading order (MeV units). The ratios $F/F_0$, $B/B_0$, $\Sigma/\Sigma_0$, which
compare these with their SU(2) counterparts, indicate the strength of the
Zweig-rule violations in these quantities (in the large-$N_c$ limit, they
tend to unity). Numbers in slanted fonts are calculated by us, from the
information given in the quoted references.}
\end{table}

To date, there are three comprehensive SU(3) papers with results based on
lattice QCD with $\Nf\!=\!2\!+\!1$ dynamical flavours \cite{Allton:2008pn,
Aoki:2008sm,Bazavov:2009bb}, and one more with results based on
$\Nf\!=\!2\!+\!1\!+\!1$ dynamical flavours \cite{Dowdall:2013rya}.
It is an open issue whether the data collected at
$m_s\!\simeq\!m_s^\mathrm{phys}$ allow for an unambiguous determination of
SU(3) low-energy constants (cf.\ the discussion in \cite{Allton:2008pn}).
To make definite statements one needs data at considerably smaller $m_s$, and
so far only MILC has some \cite{Bazavov:2009bb}.
We are aware of a few papers with a result on one SU(3) low-energy constant
each \cite{Beane:2006kx,Boyle:2008yd,Shintani:2008qe,Aoki:2010dy} which we
list for completeness.
Some particulars of the computations are listed in Table~\ref{tab:SU3_overview}.

\begin{table}[!p] 
\vspace*{3cm}
\centering
\footnotesize
\begin{tabular*}{\textwidth}[l]{l@{\extracolsep{\fill}}rcllllllll}
{} & Ref. & $\Nf$ &
\hspace{0.15cm}\begin{rotate}{60}{publication status}\end{rotate}\hspace{-0.15cm} &
\hspace{0.15cm}\begin{rotate}{60}{chiral extrapolation}\end{rotate}\hspace{-0.15cm} &
\hspace{0.15cm}\begin{rotate}{60}{continuum  extrapolation}\end{rotate}\hspace{-0.15cm} &
\hspace{0.15cm}\begin{rotate}{60}{finite volume}\end{rotate}\hspace{-0.15cm} &
\rule{0.1cm}{0cm}$10^3L_4$ &$\rule{0.1cm}{0cm}10^3L_6$ & \hspace{-0.3cm} $10^3(2L_6\!-\!L_4)$ \\
\\[-3mm]
\hline
\hline
\\[-3mm]
HPQCD 13A       & \cite{Dowdall:2013rya} &2+1+1& \gA & \good & \good & \good & 0.09(34)                  & 0.16(20)     & 0.22(17)\\
\hline
\\[-3mm]
JLQCD/TWQCD 10A & \cite{Fukaya:2010na}   &  3  & \gA & \bad  & \bad  & \bad  &                           & 0.03(7)(17)  & \\
\hline
\\[-3mm]
MILC 10         & \cite{Bazavov:2010hj}  & 2+1 & \rC & \soso & \good & \good & -0.08(22)$\binom{+57}{-33}$ & {\sl-0.02(16)$\binom{+33}{-21}$} & 0.03(24)$\binom{+32}{-27}$ \\
MILC 09A        & \cite{Bazavov:2009fk}  & 2+1 & \rC & \soso & \good & \good & 0.04(13)(4)               & 0.07(10)(3)          & 0.10(12)(2)          \\
MILC 09         & \cite{Bazavov:2009bb}  & 2+1 & \gA & \soso & \good & \good & 0.1(3)$\binom{+3}{-1}$    & 0.2(2)$\binom{+2}{-1}$ & 0.3(1)$\binom{+2}{-3}$ \\
PACS-CS 08      & \cite{Aoki:2008sm}     & 2+1 & \gA & \good & \bad  & \bad  & -0.06(10)(-)              & {\sl0.02(5)(-)}      & 0.10(2)(-)           \\
RBC/UKQCD 08    & \cite{Allton:2008pn}   & 2+1 & \gA & \soso & \bad  & \good & 0.14(8)(-)                & 0.07(6)(-)           & 0.00(4)(-)           \\
\hline
\\[-3mm]
Bijnens 11      & \cite{Bijnens:2011tb}  &     &     &       &       &       & 0.75(75)                  & 0.29(85)             & {\sl-0.17(1.86)}       \\
Gasser 85       & \cite{Gasser:1984gg}   &     &     &       &       &       & -0.3(5)                   & -0.2(3)              & {\sl-0.1(8)}         \\
\\[-3mm]
\hline
\hline
\\
{} & Ref. & $\Nf$ &
& 
& 
& 
& 
\rule{0.1cm}{0cm} $10^3L_5$ &\rule{0.05cm}{0cm}  $10^3L_8$ &\hspace{-0.3cm} $10^3(2L_8\!-\!L_5)$ \\
\\[-3mm]
\hline
\hline
\\[-3mm]
HPQCD 13A       & \cite{Dowdall:2013rya} &2+1+1& \gA & \good & \good & \good & 1.19(25)             & 0.55(15)        & -0.10(20) \\
\hline
\\[-3mm]
MILC 10         & \cite{Bazavov:2010hj}  & 2+1 & \rC & \soso & \good & \good & 0.98(16)$\binom{+28}{-41}$ & {\sl0.42(10)$\binom{+27}{-23}$} & -0.15(11)$\binom{+45}{-19}$\\
MILC 09A        & \cite{Bazavov:2009fk}  & 2+1 & \rC & \soso & \good & \good & 0.84(12)(36)         & 0.36(5)(7)      & -0.12(8)(21)     \\
MILC 09         & \cite{Bazavov:2009bb}  & 2+1 & \gA & \soso & \good & \good & 1.4(2)$\binom{+2}{-1}$ & 0.8(1)(1)     & 0.3(1)(1)        \\
PACS-CS 08      & \cite{Aoki:2008sm}     & 2+1 & \gA & \good & \bad  & \bad  & 1.45(7)(-)           & {\sl0.62(4)(-)} & -0.21(3)(-)      \\
RBC/UKQCD 08    & \cite{Allton:2008pn}   & 2+1 & \gA & \soso & \bad  & \good & 0.87(10)(-)          & 0.56(4)(-)      & 0.24(4)(-)       \\
\hline
\\[-3mm]
Bijnens 11      & \cite{Bijnens:2011tb}  &     &     &       &       &       & 0.58(13)             & 0.18(18)        & {\sl-0.22(38)}   \\
Gasser 85       & \cite{Gasser:1984gg}   &     &     &       &       &       & 1.4(5)               & 0.9(3)          & {\sl0.4(8)}      \\
\\[-3mm]
\hline
\hline
\\
{} & Ref. & $\Nf$ &
& 
& 
& 
& 
\rule{0.1cm}{0cm} $10^3L_5$ &\rule{0.1cm}{0cm} $10^3L_9$ &\rule{0.1cm}{0cm} $10^3L_{10}$ \\
\\[-3mm]
\hline
\hline
\\[-3mm]
RBC/UKQCD 09    & \cite{Boyle:2009xi}                      & 2+1 & \gA & \soso & \bad  & \soso &                           &              & -5.7(11)(07)          \\
RBC/UKQCD 08A   & \cite{Boyle:2008yd}                      & 2+1 & \gA & \soso & \bad  & \good &                           & 3.08(23)(51) &                       \\
NPLQCD 06       & \cite{Beane:2006kx}                      & 2+1 & \gA & \soso & \bad  & \bad  & 1.42(2)$\binom{+18}{-54}$ &              &                       \\
\hline
\\[-3mm]
JLQCD 08A       & \cite{Shintani:2008qe}                   &  2  & \gA & \soso & \bad  & \bad  &                           &              & -5.2(2)$\binom{+5}{-3}$ \\
\hline
\\[-3mm]
Bijnens 11      & \cite{Bijnens:2011tb}   &     &     &       &       &       & 0.58(13)                &              &                       \\
Bijnens 02      & \cite{Bijnens:2002hp}   &     &     &       &       &       &                         & 5.93(43)     &                       \\
Davier 98       & \cite{Davier:1998dz}    &     &     &       &       &       &                         &              & -5.13(19)             \\
Gasser 85       & \cite{Gasser:1984gg}    &     &     &       &       &       & 1.4(5)                  & 6.9(7)       & -5.5(7)               \\
\\[-3mm]
\hline
\hline
\end{tabular*}
\normalsize
\vspace*{-4mm}
\caption{\label{tab:SU3_NLO}
Low-energy constants that enter the effective
SU(3) Lagrangian at NLO (running scale $\mu\!=\!770\MeV$ -- the values in
\cite{Dowdall:2013rya,Bazavov:2010hj,Bazavov:2009bb,Bazavov:2009fk,Gasser:1984gg} are
evolved accordingly). The MILC 10 entry for $L_6$ is obtained from their results for
$2L_6\!-\!L_4$ and $L_4$ (and similarly for other entries in slanted fonts).
The JLQCD 08A result [which is for $\ell_5(770\MeV)$ despite the paper saying
$L_{10}(770\MeV)$] has been converted to $L_{10}$ with the standard one-loop formula,
assuming that the difference between $\bar\ell_5(m_s\!=\!m_s^\mr{phys})$ [needed
in the formula] and $\bar\ell_5(m_s\!=\!\infty)$ [computed by JLQCD] can be ignored.}
\end{table}



\begin{figure}[!tbh]
\includegraphics[width=7.5cm]{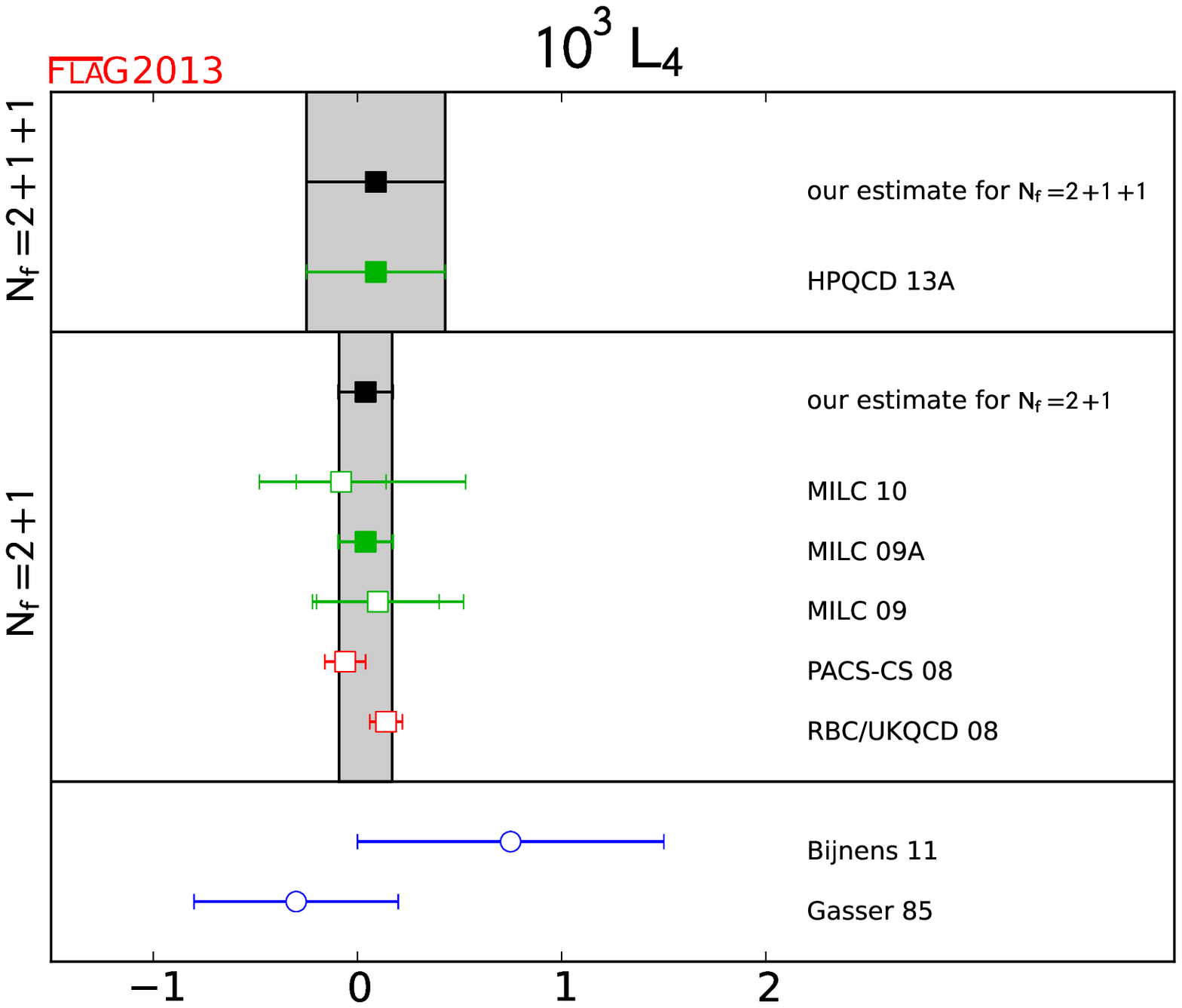}
\includegraphics[width=7.5cm]{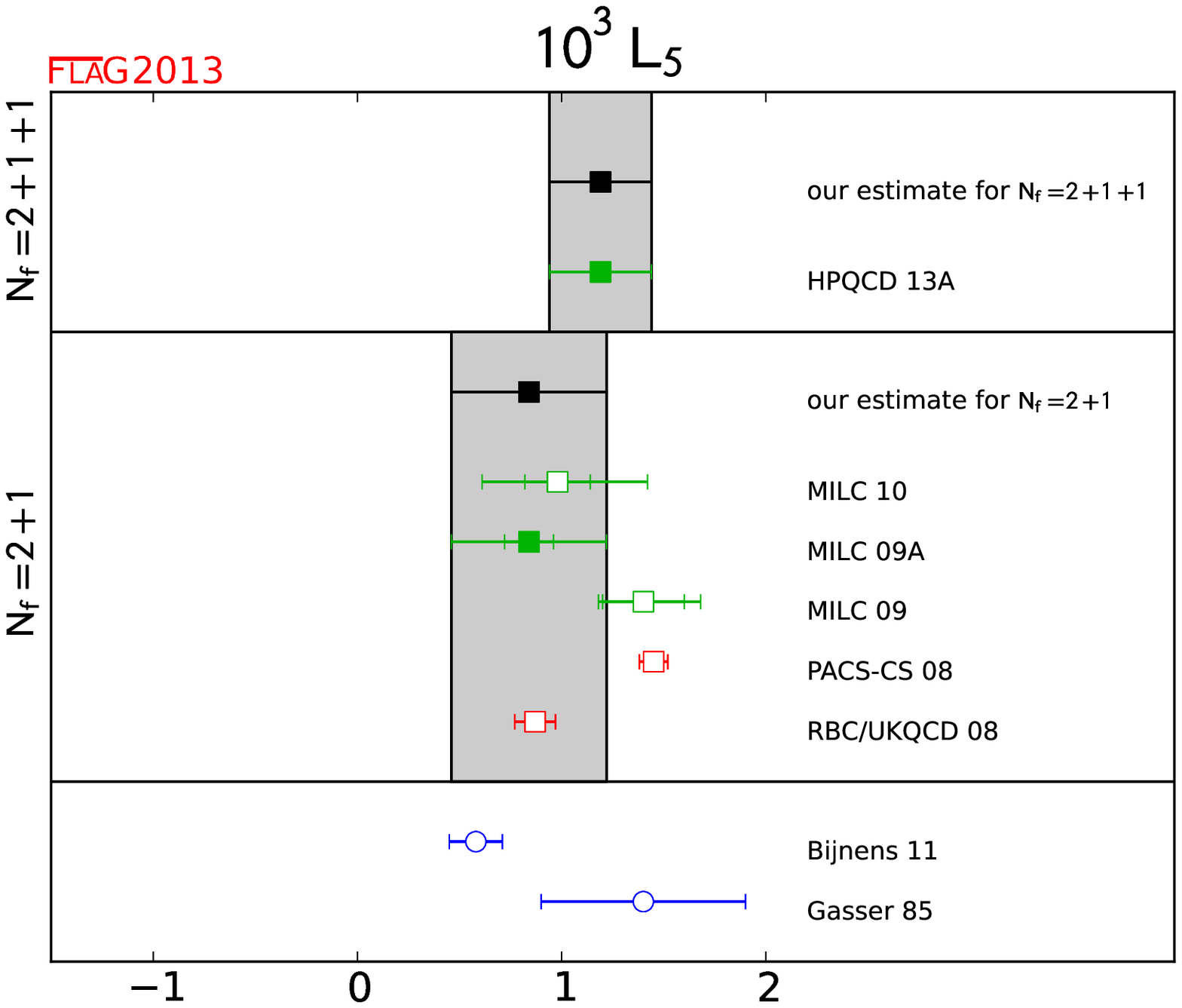}
\\
\includegraphics[width=7.5cm]{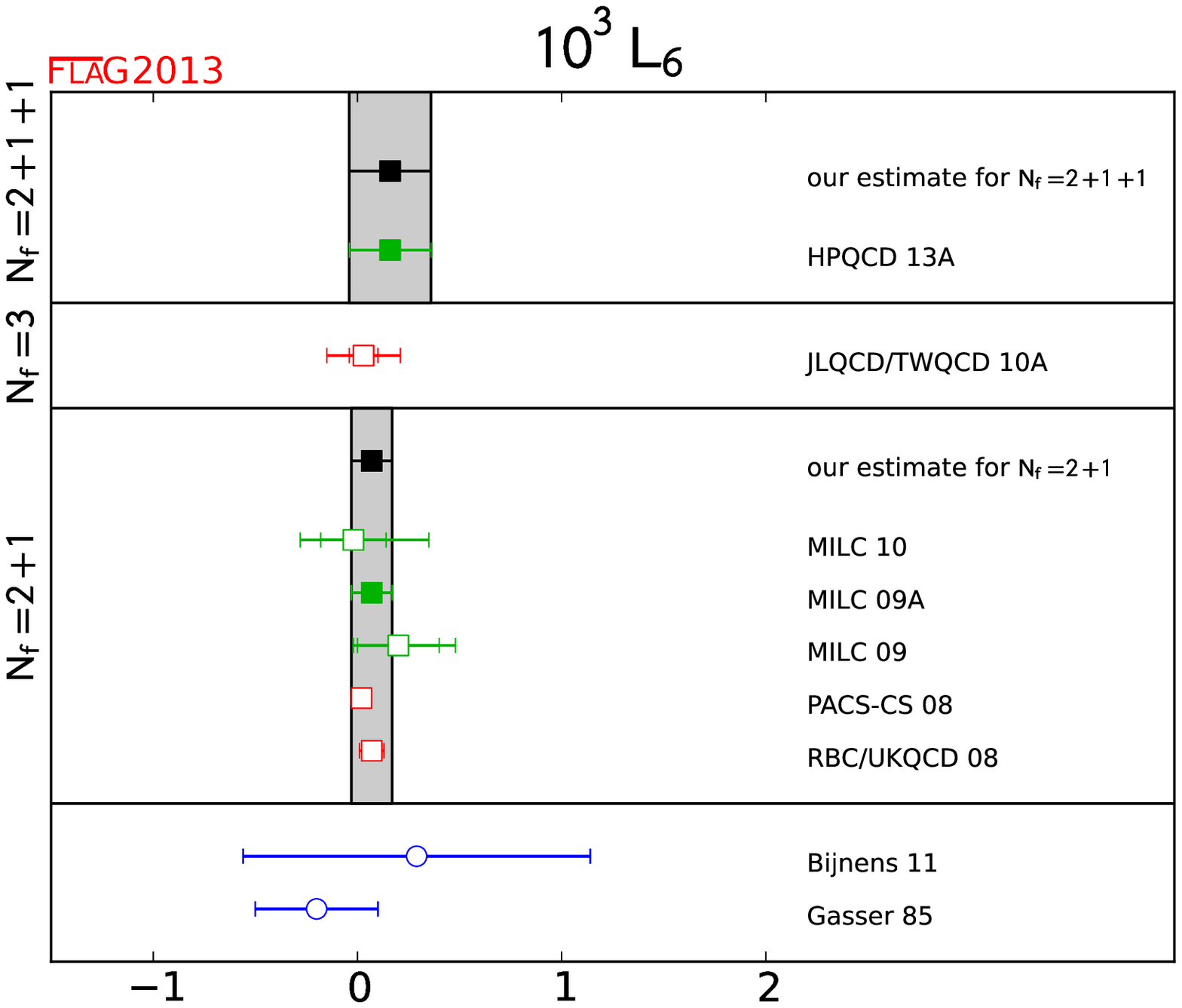}
\includegraphics[width=7.5cm]{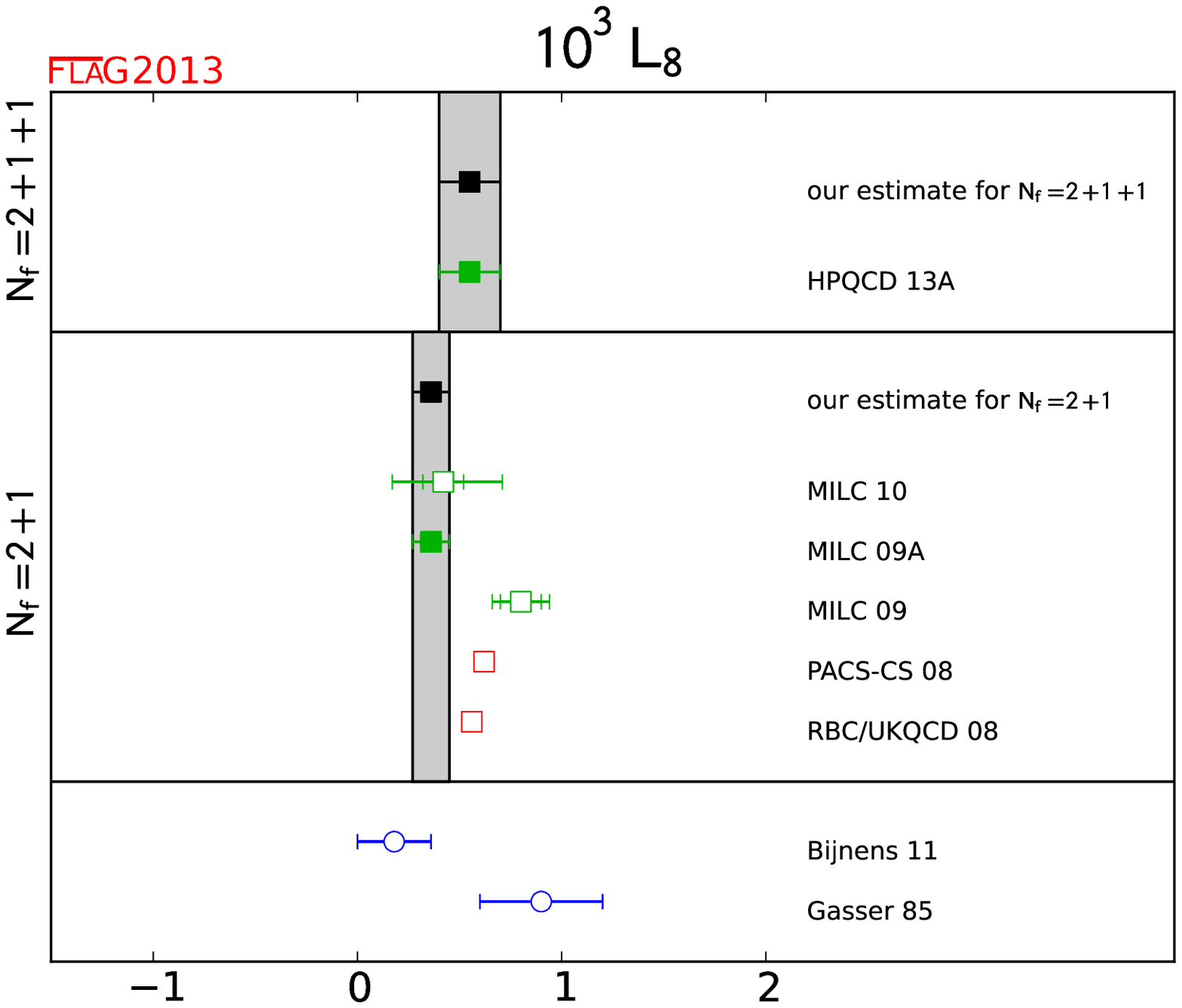}
\\
\includegraphics[width=7.5cm]{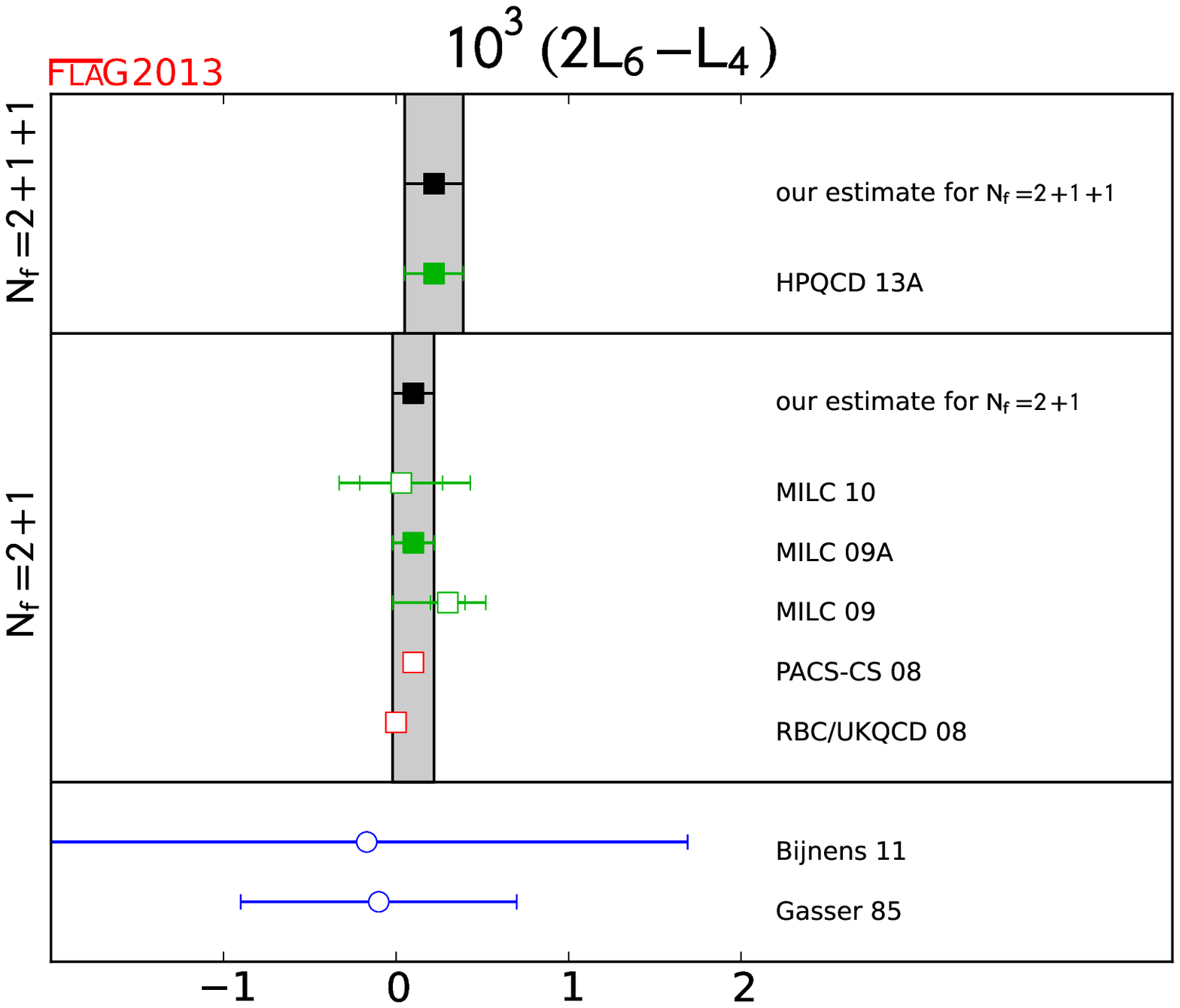}
\includegraphics[width=7.5cm]{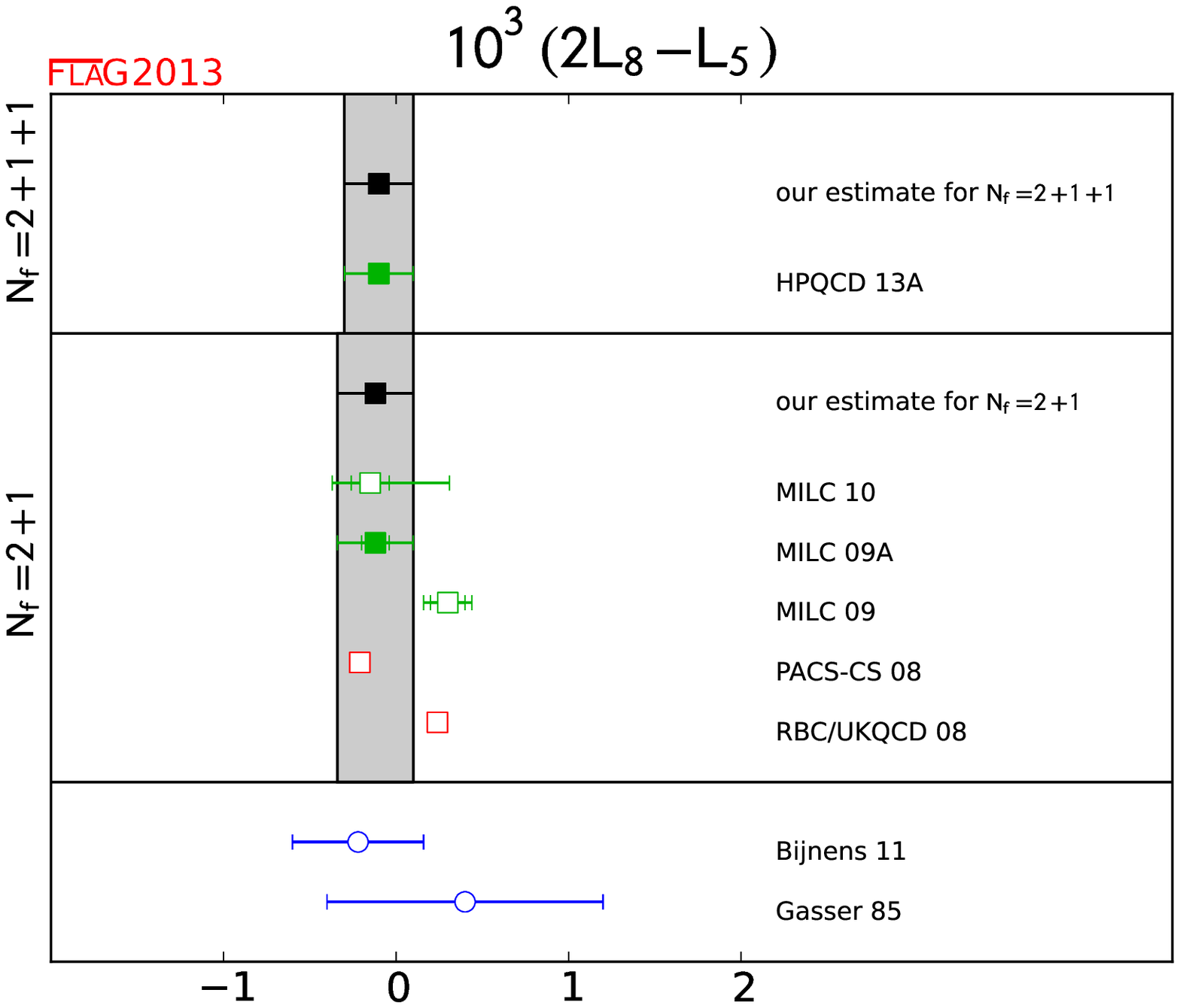}
\caption{\label{fig:plot_LECs_SU3}
Low-energy constants that enter the effective SU(3) Lagrangian at NLO.
The grey bands and black dots labeled as ``our estimate'' coincide with the
results of MILC 09A \cite{Bazavov:2009fk} for $\Nf=2+1$ and HPQCD 13A
\cite{Dowdall:2013rya} for $\Nf=2+1+1$, respectively.}
\end{figure}

Results for the SU(3) low-energy constants of leading order are found in
Table~\ref{tab:SU3_overview} and analogous results for some of the
effective coupling constants that enter the chiral SU(3) Lagrangian at NLO
are collected in Table~\ref{tab:SU3_NLO}.  From PACS-CS \cite{Aoki:2008sm}
only those results are quoted which have been \emph{corrected} for
finite-size effects (misleadingly labeled ``w/FSE'' in their tables).
For staggered data our colour-coding rule states that $\Mpi$ is to be
understood as $\Mpi^\mr{RMS}$.
The rating of \cite{Bazavov:2009bb,Bazavov:2010hj} is based on the
information regarding the RMS masses given in \cite{Bazavov:2009fk}.

A graphical summary of the lattice results for the coupling constants $L_4$,
$L_5$, $L_6$ and $L_8$, which determine the masses and the decay constants of
the pions and kaons at NLO of the chiral SU(3) expansion, is displayed in
Figure~\ref{fig:plot_LECs_SU3}, along with the two phenomenological
determinations quoted in the above tables.
The overall consistency seems fairly convincing.
In spite of this apparent consistency, there is a point which needs to be
clarified as soon as possible. Some collaborations (RBC/UKQCD and PACS-CS)
find that they are having difficulties in fitting their partially quenched
data to the respective formulas for pion masses above $\simeq$ 400 MeV.
Evidently, this indicates that the data are stretching the regime of
validity of these formulas. To date it is, however, not clear which subset
of the data causes the troubles, whether it is the unitary part extending
to too large values of the quark masses or whether it is due to
$m^\mathrm{val}/m^\mathrm{sea}$ differing too much from one.
In fact, little is known, in the framework of partially quenched {\Ch}PT, about
the \emph{shape} of the region of applicability in the $m^\mathrm{val}$ versus
$m^\mathrm{sea}$ plane for fixed $\Nf$.
This point has also been emphasized in \cite{Durr:2013koa}.

To date only the computations MILC 09A \cite{Bazavov:2009fk} (as an obvious
update of MILC 09) and HPQCD 13A \cite{Dowdall:2013rya} are free of red tags.
Since they use different $\Nf$ (in the former case $\Nf=2+1$, in the latter
case $\Nf=2+1+1$) we stay away from averaging them.
Hence the situation remains unsatisfactory in the sense that for each $\Nf$
only a single determination of high standing is available.
Accordingly, we stay with the recommendation to use the results of
MILC 09A \cite{Bazavov:2009fk} and HPQCD 13A \cite{Dowdall:2013rya}
for $\Nf=2+1$ and $\Nf=2+1+1$, respectively, as given in Table~\ref{tab:SU3_NLO}.
These numbers are shown as grey bands in Figure~\ref{fig:plot_LECs_SU3}.

In the large-$N_c$ limit, the Zweig-rule becomes exact, but the quarks have
$N_c=3$.
The work done on the lattice is ideally suited to disprove or confirm the
approximate validity of this rule for QCD.
Two of the coupling constants entering the effective SU(3) Lagrangian at NLO
disappear when $N_c$ is sent to infinity: $L_4$ and $L_6$.
The upper part of Table~\ref{tab:SU3_NLO} and the left panels of
Figure~\ref{fig:plot_LECs_SU3} show that the lattice results for these are
quite coherent.
At the scale $\mu=M_\rho$, $L_4$ and $L_6$ are consistent with zero, indicating
that these constants do approximately obey the Zweig-rule.
As mentioned above, the ratios $F/F_0$, $B/B_0$ and $\Sigma/\Sigma_0$ also test
the validity of this rule.
Their expansion in powers of $m_s$ starts with unity and the contributions of
first order in $m_s$ are determined by the constants $L_4$ and $L_6$, but they
also contain terms of higher order.
Apart from measuring the Zweig-rule violations, an accurate determination
of these ratios will thus also allow us to determine the range of $m_s$ where
the first few terms of the expansion represent an adequate approximation.
Unfortunately, at present, the uncertainties in the lattice data on these
ratios are too large to draw conclusions, both concerning the relative size of
the subsequent terms in the chiral perturbation series and concerning the
magnitude of the Zweig-rule violations.
The data seem to confirm the {\it paramagnetic inequalities}
\cite{DescotesGenon:1999uh}, which require $F/F_0>1$, $\Sigma/\Sigma_0>1$, and
it appears that the ratio $B/B_0$ is also larger than unity, but the numerical
results need to be improved before further conclusions can be drawn.

In principle, the matching formulae in \cite{Gasser:1984gg} can be used to
calculate%
\footnote{For instance, for the MILC data this yields $\bar{l}_3=3.32(64)(45)$
and $\bar{l}_4=4.03(16)(17)$ \cite{Bazavov:2009fk}.}
the SU(2) couplings $\bar{l}_i$ from the SU(3) couplings $L_j$.
This procedure, however, yields less accurate results than a direct
determination within SU(2), as it relies on the expansion in powers of $m_s$,
where the omitted higher-order contributions generate comparatively large
uncertainties.
We plead with every collaboration performing $\Nf=2+1$ simulations to
\emph{directly} analyse their data in the SU(2) framework.
In practice, lattice simulations are performed at values of $m_s$ close to
the physical value and the results are then corrected for the difference of
$m_s$ from its physical value.
If simulations with more than one value of $m_s$ have been performed, this can
be done by interpolation.
Alternatively one can use the technique of \emph{reweighting} (for a review see
e.g.\ \cite{Jung:2010jt}) to shift $m_s$ to its physical value.

\clearpage

\clearpage
\section{Kaon $B$-parameter $B_K$}
\label{sec:BK}
\subsection{Indirect CP-violation and $\epsilon_{K}$}

The mixing of neutral pseudoscalar mesons plays an important role in
the understanding of the physics of CP-violation. In this section we
will only focus on $K^0 - \bar K^0$ oscillations, which probe the
physics of indirect CP-violation. We collect here the basic formulae;
for extended reviews on the subject see, among others,
Refs.~\cite{Branco:1999fs,Buchalla:1995vs,Buras:1998raa}. Indirect
CP-violation arises in $K_L \rightarrow \pi \pi$ transitions through
the decay of the $\rm CP=+1$ component of $K_L$ into two pions (which
are also in a $\rm CP=+1$ state). Its measure is defined as
\be 
\epsilon_{K} \,\, = \,\, \dfrac{{\cal A} [ K_L \rightarrow
(\pi\pi)_{I=0}]}{{\cal A} [ K_S \rightarrow (\pi\pi)_{I=0}]} \,\, ,
\ee
with the final state having total isospin zero. The parameter
$\epsilon_{K}$ may also be expressed in terms of $K^0 - \bar K^0$
oscillations. In particular, to lowest order in the electroweak
theory, the contribution to these oscillations arises from so-called
box diagrams, in which two $W$-bosons and two ``up-type" quarks
(i.e. up, charm, top) are exchanged between the constituent down and
strange quarks of the $K$-mesons. The loop integration of the box
diagrams can be performed exactly. In the limit of vanishing external
momenta and external quark masses, the result can be identified with
an effective four-fermion interaction, expressed in terms of the
``effective Hamiltonian"
\be
  {\cal H}_{\rm eff}^{\Delta S = 2} \,\, = \,\,
  \frac{G_{\rm{F}}^2 M_{{W}}^2}{16\pi^2} {\cal F}^0 Q^{\Delta S=2}
  \,\, + \,\, {\rm h.c.} \,\,.
\ee
In this expression, $G_{\rm{F}}$ is the Fermi coupling, $M_{{W}}$ the
$W$-boson mass, and
\be
   Q^{\Delta S=2} =
   \left[\bar{s}\gamma_\mu(1-\gamma_5)d\right]
   \left[\bar{s}\gamma_\mu(1-\gamma_5)d\right]
   \equiv O_{\rm VV+AA}-O_{\rm VA+AV} \,\, 
\ee
is a dimension-six, four-fermion operator. The function ${\cal F}^0$
is given by
\be
{\cal F}^0 \,\, = \,\, \lambda_c^2 S_0(x_c) \, + \, \lambda_t^2
S_0(x_t) \, + \, 2 \lambda_c  \lambda_t S_0(x_c,x_t)  \,\, , 
\ee
where $\lambda_a = V^\ast_{as} V_{ad}$, and $a=c\,,t$ denotes a flavour
index. The quantities $S_0(x_c),\,S_0(x_t)$ and $S_0(x_c,x_t)$ with
$x_c=m_c^2/M_{{W}}^2$, $x_t=m_t^2/M_{{W}}^2$ are the Inami-Lim functions
\cite{Inami:1980fz}, which express the basic electroweak loop contributions
without QCD corrections. The contribution of the up quark, which is taken to
be massless in this approach, has been taken into account by imposing the
unitarity constraint $\lambda_u + \lambda_c + \lambda_t = 0$.  For future
reference we note that the dominant contribution comes from the term
$\lambda_t^2 S_0(x_t)$.  This factor is proportional to $|V_{cb}|^4$ if one
enforces the unitarity of the CKM matrix.  The dependence on a high power of
$V_{cb}$ is important from a phenomenological point of view because it implies
that uncertainties in $V_{cb}$ are magnified when considering $\epsilon_K$.

When strong interactions are included, $\Delta{S}=2$ transitions can
no longer be discussed at the quark level. Instead, the effective
Hamiltonian must be considered between mesonic initial and final
states. Since the strong coupling constant is large at typical
hadronic scales, the resulting weak matrix element cannot be
calculated in perturbation theory. The operator product expansion
(OPE) does, however, factorize long- and short-distance effects. For
energy scales below the charm threshold, the $K^0-\bar K^0$ transition
amplitude of the effective Hamiltonian can be expressed as
\begin{eqnarray}
\label{eq:Heff}
\langle \bar K^0 \vert {\cal H}_{\rm eff}^{\Delta S = 2} \vert K^0
\rangle  \,\, = \,\, \frac{G_{\rm{F}}^2 M_{{W}}^2}{16 \pi^2}  
\Big [ \lambda_c^2 S_0(x_c) \eta_1  \, + \, \lambda_t^2 S_0(x_t)
  \eta_2 \, + \, 2 \lambda_c  \lambda_t S_0(x_c,x_t) \eta_3
  \Big ]  \nn \\ 
\times 
%
%
  \left(\frac{\gbar(\mu)^2}{4\pi}\right)^{-\gamma_0/(2\beta_0)}
  \exp\bigg\{ \int_0^{\gbar(\mu)} \, dg \, \bigg(
  \frac{\gamma(g)}{\beta(g)} \, + \, \frac{\gamma_0}{\beta_0g} \bigg)
  \bigg\} 
   \langle \bar K^0 \vert  Q^{\Delta S=2}_{\rm R} (\mu) \vert K^0
   \rangle \,\, + \,\, {\rm h.c.} \,\, ,
\end{eqnarray}
where $\gbar(\mu)$ and $Q^{\Delta S=2}_{\rm R}(\mu)$ are the
renormalized gauge coupling and four-fermion operator in some
renormalization scheme. The factors $\eta_1, \eta_2$ and $\eta_3$
depend on the renormalized coupling $\gbar$, evaluated at the various
flavour thresholds $m_t, m_b, m_c$ and $ M_{{W}}$, as required by
the OPE and RG-running procedure that separates high- and low-energy
contributions. Explicit expressions can be found
in~\cite{Buchalla:1995vs} and references therein, except that $\eta_1$
and $\eta_3$ have been recently calculated to NNLO in
Refs.~\cite{Brod:2011ty} and \cite{Brod:2010mj}, respectively. We
follow the same conventions for the RG-equations as in
Ref.~\cite{Buchalla:1995vs}. Thus the Callan-Symanzik function and the
anomalous dimension $\gamma(\gbar)$ of $Q^{\Delta S=2}$ are defined by
\be
\label{eq:four_quark_operator_anomalous_dimensions}
\dfrac{d \gbar}{d \ln \mu} = \beta(\gbar)\,,\qquad
\dfrac{d Q^{\Delta S=2}_{\rm R}}{d \ln \mu} =
-\gamma(\gbar)\,Q^{\Delta S=2}_{\rm R} \,\,, 
\ee
with perturbative expansions
\begin{eqnarray}
\label{eq:four_quark_operator_anomalous_dimensions_perturbative}
\beta(g)  &=&  -\beta_0 \dfrac{g^3}{(4\pi)^2} \,\, - \,\, \beta_1
\dfrac{g^5}{(4\pi)^4} \,\, - \,\, \cdots 
\\
\gamma(g)  &=&  \gamma_0 \dfrac{g^2}{(4\pi)^2} \,\, + \,\,
\gamma_1 \dfrac{g^4}{(4\pi)^4} \,\, + \,\, \cdots \,.\nn
\end{eqnarray}
We stress that $\beta_0, \beta_1$ and $\gamma_0$ are universal,
i.e. scheme-independent. $K^0-\bar K^0$ mixing is usually considered
in the naive dimensional regularization (NDR) scheme of $\msbar$, and
below we specify the perturbative coefficient $\gamma_1$ in that
scheme:
\begin{eqnarray}
\label{eq:RG-coefficients}
& &\beta_0 = 
         \left\{\frac{11}{3}N_c-\frac{2}{3}\Nf\right\}, \qquad
   \beta_1 = 
         \left\{\frac{34}{3}N_c^2-\Nf\left(\frac{13}{3}N_c-\frac{1}{N_c}
         \right)\right\}, \\[0.3ex]
& &\gamma_0 = \frac{6(N_c-1)}{N_c}, \qquad
         \gamma_1 = \frac{N_c-1}{2N_c} 
         \left\{-21 + \frac{57}{N_c} - \frac{19}{3}N_c + \frac{4}{3}\Nf
         \right\}\,.\nn
\end{eqnarray}
Note that for QCD the above expressions must be evaluated for $N_c=3$
colours, while $\Nf$ denotes the number of active quark flavours. As
already stated, Eq.~(\ref{eq:Heff}) is valid at scales below the charm
threshold, after all heavier flavours have been integrated out,
i.e. $\Nf = 3$.

In Eq.~(\ref{eq:Heff}), the terms proportional to $\eta_1,\,\eta_2$
and $\eta_3$, multiplied by the contributions containing
$\gbar(\mu)^2$, correspond to the Wilson coefficient of the OPE,
computed in perturbation theory. Its dependence on the renormalization
scheme and scale $\mu$ is canceled by that of the weak matrix element
$\langle \bar K^0 \vert Q^{\Delta S=2}_{\rm R} (\mu) \vert K^0
\rangle$. The latter corresponds to the long-distance effects of the
effective Hamiltonian and must be computed nonperturbatively. For
historical, as well as technical reasons, it is convenient to express
it in terms of the $B$-parameter $B_{{K}}$, defined as
\be
   B_{{K}}(\mu)= \frac{{\left\langle\bar{K}^0\left|
         Q^{\Delta S=2}_{\rm R}(\mu)\right|K^0\right\rangle} }{
         {\frac{8}{3}\fK^2\mK^2}} \,\, .
\ee
The four-quark operator $Q^{\Delta S=2}(\mu)$ is renormalized at scale
$\mu$ in some regularization scheme, for instance,
NDR-$\msbar$. Assuming that $B_{{K}}(\mu)$ and the anomalous
dimension $\gamma(g)$ are both known in that scheme, the
renormalization group invariant (RGI) $B$-parameter $\hat{B}_{K}$ 
is related to $B_{{K}}(\mu)$ by the exact formula
\be
  \hat{B}_{{K}} = 
  \left(\frac{\gbar(\mu)^2}{4\pi}\right)^{-\gamma_0/(2\beta_0)}
  \exp\bigg\{ \int_0^{\gbar(\mu)} \, dg \, \bigg(
  \frac{\gamma(g)}{\beta(g)} \, + \, \frac{\gamma_0}{\beta_0g} \bigg)
  \bigg\} 
\, B_{{K}}(\mu) \,\,\, .
\ee
At NLO in perturbation theory the above reduces to
\be
   \hat{B}_{{K}} =
   \left(\frac{\gbar(\mu)^2}{4\pi}\right)^{- \gamma_0/(2\beta_0)}
   \left\{ 1+\dfrac{\gbar(\mu)^2}{(4\pi)^2}\left[
   \frac{\beta_1\gamma_0-\beta_0\gamma_1}{2\beta_0^2} \right]\right\}\,
   B_{{K}}(\mu) \,\,\, .
\label{eq:BKRGI_NLO}
\ee
To this order, this is the scale-independent product of all
$\mu$-dependent quantities in Eq.~(\ref{eq:Heff}).

Lattice QCD calculations provide results for $B_K(\mu)$. These results,
however, are usually obtained in intermediate schemes other than the continuum
$\msbar$ scheme used to calculate the Wilson coefficients appearing in
Eq.~(\ref{eq:Heff}). Examples of intermediate schemes are the RI/MOM scheme
\cite{Martinelli:1994ty} (also dubbed the ``Rome-Southampton method'') and the
Schr\"odinger functional (SF) scheme \cite{Luscher:1992an}, which both allow
for a nonperturbative renormalization of the four-fermion operator, using an
auxiliary lattice simulation. In this way $B_K(\mu)$ can be calculated with
percent-level accuracy, as described below.

In order to make contact with phenomenology, however, and in
particular to use the results presented above, one must convert from
the intermediate scheme to the $\msbar$ scheme or to the RGI quantity
$\hat{B}_{K}$. This conversion relies on one or two-loop
perturbative matching calculations, the truncation errors in which
are, for many recent calculations, the dominant source of error in
$\hat{B}_{{K}}$~\cite{Aubin:2009jh,Kim:2011qg,Bae:2011ff,Laiho:2011np,Arthur:2012opa}.
While this scheme-conversion error is not, strictly speaking, an error
of the lattice calculation itself, it must be included in results for
the quantities of phenomenological interest, namely
$B_K(\msbar,2\,{\rm GeV})$ and $\hat{B}_{K}$. We note that this
error can be minimized by matching between the intermediate scheme and
$\msbar$ at as large a scale $\mu$ as possible (so that the coupling
constant which determines the rate of convergence is
minimized). Recent calculations have pushed the matching $\mu$ up to
the range $3-3.5\,$GeV. This is possible because of the use of
nonperturbative RG running determined on the
lattice~\cite{Durr:2011ap,Arthur:2012opa}. The Schr\"odinger
functional offers the possibility to run nonperturbatively to scales
$\mu\sim M_{{W}}$ where the truncation error can be safely
neglected. However, so far this has been applied only for two flavours
of Wilson quarks~\cite{Dimopoulos:2007ht}.

Perturbative truncation errors in Eq.~(\ref{eq:Heff}) also affect the
Wilson coefficients $\eta_1$, $\eta_2$ and~$\eta_3$. It turns out that
the largest uncertainty comes from that in
$\eta_1$~\cite{Brod:2011ty}. Although it is now calculated at NNLO,
the series shows poor convergence. The net effect is that the
uncertainty in $\eta_1$ is larger than that in present lattice
calculations of $B_K$.

The ``master formula'' for $\epsilon_{K}$, which connects the
experimentally observable quantity $\epsilon_{K}$ to the matrix
element of ${\cal H}_{\rm eff}^{\Delta S = 2}$,
is~\cite{Buras:1998raa,Anikeev:2001rk,Nierste:2009wg,Buras:2010pz}
\be
\epsilon_{K} \,\,\, = \,\,\, \exp(i \phi_\epsilon) \,\,
\sin(\phi_\epsilon) \,\, \Big [ \frac{\Im [ \langle \bar K^0 \vert
{\cal H}_{\rm eff}^{\Delta S = 2} \vert K^0 \rangle ]} {\Delta m_K }
\,\,\, + \,\,\, \rho \frac{\Im(A_0)}{\Re(A_0)} \,\, \Big ] \,\,\, ,
\label{eq:epsK}
\ee
for $\lambda_u$ real and positive; the phase of $\epsilon_{K}$ is
given by
\be
\phi_\epsilon \,\,\, = \,\,\, \arctan \frac{\Delta m_{K}}{\Delta
  \Gamma_{K}/2} \,\,\, . 
\ee
The quantities $\Delta m_K\equiv m_{K_L}-m_{K_S}$ and $\Delta\Gamma_K\equiv \Gamma_{K_S}-\Gamma_{K_L}$ are the mass- and
decay width-differences between long- and short-lived neutral Kaons,
while $A_0$ is the amplitude of the Kaon decay into a two-pion state
with isospin zero. The experimentally measured values of the above quantities
are\,\cite{Beringer:1900zz}:
\begin{eqnarray}
\vert \epsilon_{K} \vert \,\, &=& \,\, 2.228(11) \times 10^{-3} \,\,\, ,
\nn \\
\phi_\epsilon \,\, &=& \,\, 43.52(5)^\circ \,\,\, ,
 \\
\Delta m_{K} \,\, &=& \,\, 3.4839(59) \times 10^{-12}\, {\rm MeV} \,\,\, ,
\nn \\
\Delta \Gamma_{K}  \,\, &=& \,\ 7.3382(33) \times 10^{-12} \,{\rm MeV} \,\,\,.\nn
\end{eqnarray}
The second term in the square brackets of Eq.~(\ref{eq:epsK}), has
been discussed and estimated, e.g., in
Refs.~\cite{Buras:2008nn,Buras:2010pz}. It can best be thought of as
$\xi + (\rho-1)\xi$, with $\xi={\rm Im}(A_0)/{\rm Re}(A_0)$. The $\xi$
term is the contribution of direct CP violation to $\epsilon_K$. Using
the estimate of $\xi$ from Ref.~\cite{Buras:2008nn} (obtained from the
experimental value of $\epsilon'/\epsilon$) this gives a
$\sim-6.0(1.5)\%$ correction.\footnote{%
A very recent lattice calculation of ${\rm Im}(A_2)$ by the RBC/UKQCD
collaboration opens up the possibility of a more accurate
determination of $\xi$ using the measured value of
$\epsilon'$~\cite{Blum:2011ng,Blum:2012uk}. This lattice calculation
uses only a single lattice spacing, so we do not quote the resulting
value here, but note that it is consistent with that obtained in
Ref.~\cite{Buras:2008nn}, with errors estimated to be significantly
smaller.}
The $(\rho-1)\xi$ term arises from long-distance contributions to the
imaginary part of $K^0 -\bar K^0$ mixing\,\cite{Buras:2010pz}
[contributions which are neglected in Eq.~(\ref{eq:Heff})]. Using the
estimate $\rho=0.6\pm 0.3$~\cite{Buras:2010pz}, this gives a
contribution of about $+2\%$ with large errors. Overall these
corrections combine to give a $(4 \pm 2)\%$ reduction in the
prediction for $\epsilon_K$. Although this is a small correction, we
note that its contribution to the error of $\epsilon_K$ is larger than
that arising from the value of $B_{K}$ reported below.

\subsection{Lattice computation of $B_{{K}}$}\label{sec:latBK}

Lattice calculations of $B_{{K}}$ are affected by the same
systematic effects discussed in previous sections. However, the issue
of renormalization merits special attention. The reason is that the
multiplicative renormalizability of the relevant operator $Q^{\Delta
S=2}$ is lost once the regularized QCD action ceases to be invariant
under chiral transformations. For Wilson fermions, $Q^{\Delta S=2}$
mixes with four additional dimension-six operators, which belong to
different representations of the chiral group, with mixing
coefficients that are finite functions of the gauge coupling. This
complicated renormalization pattern was identified as the main source
of systematic error in earlier, mostly quenched calculations of
$B_{{K}}$ with Wilson quarks. It can be bypassed via the
implementation of specifically designed methods, which are either
based on Ward identities~\cite{Becirevic:2000cy} or on a modification
of the Wilson quark action, known as twisted-mass
QCD~\cite{Frezzotti:2000nk,Dimopoulos:2006dm}.

An advantage of staggered fermions is the presence of a remnant $U(1)$
chiral symmetry. However, at nonvanishing lattice spacing, the
symmetry among the extra unphysical degrees of freedom (tastes) is
broken. As a result, mixing with other dimension-six operators cannot
be avoided in the staggered formulation, which complicates the
determination of the $B$-parameter. The effects of the broken taste
symmetry are usually treated via an effective field theory, such as
staggered Chiral Perturbation Theory (S$\chi$PT).

Fermionic lattice actions based on the Ginsparg-Wilson
relation~\cite{Ginsparg:1981bj} are invariant under the chiral group,
and hence four-quark operators such as $Q^{\Delta S=2}$ renormalize
multiplicatively. However, depending on the particular formulation of
Ginsparg-Wilson fermions, residual chiral symmetry breaking effects
may be present in actual calculations. For instance, in the case of
domain wall fermions, the finiteness of the extra 5th dimension
implies that the decoupling of modes with different chirality is not
exact, which produces a residual nonzero quark mass in the chiral
limit. Whether or not a significant mixing with dimension-six
operators is induced as well must be investigated on a case-by-case
basis.

In this section we focus on recent results for $B_{{K}}$, obtained
for $\Nf=2$ and $2+1$ flavours of dynamical quarks. A compilation of
results is shown in Table~\ref{tab_BKsumm} and
Fig.\,\ref{fig_BKsumm}. An overview of the quality of systematic error
studies is represented by the colour coded entries in
Table~\ref{tab_BKsumm}. In Appendix~\ref{app-BK} we gather the
simulation details and results from different collaborations, the
values of the most relevant lattice parameters, and comparative tables
on the various estimates of systematic errors.

Some of the groups whose results are listed in Table~\ref{tab_BKsumm}
do not quote results for both $B_{{K}}(\overline{\rm MS},2\,{\rm
GeV})$ -- which we denote by the shorthand $B_{{K}}$ from now on --
and $\hat{B}_{{K}}$. This concerns
Refs.~\cite{Aoki:2004ht,Constantinou:2010qv} for $\Nf=2$
and\,\cite{Laiho:2011np,Arthur:2012opa} for~2+1 flavours. In these
cases we perform the conversion ourselves by evaluating the
proportionality factor in Eq.\,(\ref{eq:BKRGI_NLO}) at $\mu=2\,\gev$,
using the following procedure: For $\Nf=2+1$ we use the value
$\alpha_s(M_{{Z}})=0.1184$ from the PDG\,\cite{Beringer:1900zz} and
run it across the quark thresholds at $m_b=4.19$\,GeV and
$m_c=1.27$\,GeV, and then run up in the three-flavour theory to
$\mu=2\,\gev$. All running is done using the four-loop RG
$\beta$-function. The resulting value of $\alpha_s(2\,\gev)$ is then
used to evaluate $\hat{B}_{{K}}/B_{{K}}$ in one-loop
perturbation theory, which gives $\hat{B}_{{K}}/B_{{K}}=1.369$
in the three-flavour theory.

In two-flavour QCD one can insert the updated nonperturbative
estimate for the $\Lambda$-parameter by the ALPHA
Collaboration\,\cite{Fritzsch:2012wq},
i.e.\,$\Lambda^{(2)}=310(20)$\,MeV, into the NLO expressions for
$\alpha_s$. The resulting value of the perturbative conversion factor
$\hat{B}_K/B_K$ for $\Nf=2$ is then equal to~1.386. However, since the
running coupling in the $\msbar$ scheme enters at several stages in
the entire matching and running procedure, it is difficult to use this
estimate of $\alpha_s$ consistently without a partial reanalysis of
the data in Refs.~\cite{Aoki:2004ht,Constantinou:2010qv}. We have
therefore chosen to apply the conversion factor of~1.369 not only to
results obtained for $\Nf=2+1$ flavours but also to the two-flavour
theory (in cases where only one of $\hat{B_K}$ and $B_K$ are
quoted). This is a change from the convention used in the previous edition of the FLAG review \cite{Colangelo:2010et}. We note
that the difference between 1.386 and 1.369 will produce an ambiguity
of the order of~1\%, which is well below the overall uncertainties in
Refs.~\cite{Aoki:2004ht,Constantinou:2010qv}. We have indicated
explicitly in Table~\ref{tab_BKsumm} in which way the conversion
factor 1.369 has been applied to the results
of~Refs.~\cite{Aoki:2004ht,Constantinou:2010qv,Laiho:2011np,Arthur:2012opa}.

Note that in this section the colour code for chiral extrapolations is
interpreted differently. We recall that the criteria are:
\begin{itemize}
\item[] Chiral extrapolation:\\
\good \hspace{0.2cm}  $M_{\pi,\mathrm{min}}< 200$ MeV  \\
\rule{0.05em}{0em}\soso \hspace{0.2cm}  200 MeV $\le
M_{\pi,{\mathrm{min}}} \le$ 400 MeV \\
\rule{0.05em}{0em}\bad \hspace{0.2cm}  $M_{\pi,\mathrm{min}}> 400$ MeV
\end{itemize}
Many calculations of $B_K$ employ partially quenched $\chi$PT, and in
this case it is the mass of the valence pion which enters in chiral
logarithms and leads to the most significant dependence on quark
masses. Therefore, whenever a specific calculation employs partially
quenched pions, the above colour code is applied with respect to the
minimum valence pion mass.\footnote{%
This approach is supported by the results of the calculations using
partial quenching (see in particular Refs.~\cite{Laiho:2011np} and
\cite{Aoki:2010pe}), which find that the dependence on sea-quark
masses is weaker than that on the valence-quark masses (which itself
is very mild).}
As before, it is assumed that the chiral extrapolation is done with at
least a three-point analysis -- otherwise this will be explicitly
mentioned in a footnote. In case of nondegeneracies among the
different pion states $M_{\pi,\mathrm{min}}$ stands for a
root-mean-squared (RMS) pion mass.

\begin{table}[t]
\begin{center}
\mbox{} \\[3.0cm]
\footnotesize
\begin{tabular*}{\textwidth}[l]{l @{\extracolsep{\fill}} r l l l l l l l l l}
Collaboration & Ref. & $\Nf$ & 
\hspace{0.15cm}\begin{rotate}{60}{publication status}\end{rotate}\hspace{-0.15cm} &
\hspace{0.15cm}\begin{rotate}{60}{continuum extrapolation}\end{rotate}\hspace{-0.15cm} &
\hspace{0.15cm}\begin{rotate}{60}{chiral extrapolation}\end{rotate}\hspace{-0.15cm}&
\hspace{0.15cm}\begin{rotate}{60}{finite volume}\end{rotate}\hspace{-0.15cm}&
\hspace{0.15cm}\begin{rotate}{60}{renormalization}\end{rotate}\hspace{-0.15cm}  &
\hspace{0.15cm}\begin{rotate}{60}{running}\end{rotate}\hspace{-0.15cm} & 
\rule{0.3cm}{0cm}$B_{{K}}(\overline{\rm MS},2\,{\rm GeV})$ 
& \rule{0.3cm}{0cm}$\hat{B}_{{K}}$ \\
&&&&&&&&&& \\[-0.1cm]
\hline
\hline
&&&&&&&&&& \\[-0.1cm]

SWME 13 & \cite{Bae:2013lja} & 2+1 & \rC & \good & \soso &
\good & \soso$^\ddagger$ & $-$ & 0.539(3)(25) & 0.738(5)(34) \\[0.5ex]

RBC/UKQCD 12
& \cite{Arthur:2012opa} & 2+1 & \gA & \soso & \good &
     \soso & \good & $\,a$ & 0.554(8)(14)$^1$ & 0.758(11)(19) \\[0.5ex]  

Laiho 11 & \cite{Laiho:2011np} & 2+1 & \rC & \good & \soso &
     \soso & \good & $-$ & 0.5572(28)(150)& 0.7628(38)(205)$^2$ \\[0.5ex]  

SWME 11A & \cite{Bae:2011ff} & 2+1 & \gA & \good & \soso &
\soso & \soso$^\ddagger$ & $-$ & 0.531(3)(27) & 0.727(4)(38) \\[0.5ex]

BMW 11 & \cite{Durr:2011ap} & 2+1 & \gA & \good & \good & \good & \good
& $\,b$ & 0.5644(59)(58) & 0.7727(81)(84) \\[0.5ex]

RBC/UKQCD 10B & \cite{Aoki:2010pe} & 2+1 & \gA & \soso & \soso & \good &
\good & $\,c$ & 0.549(5)(26) & 0.749(7)(26) \\[0.5ex] 

SWME 10 & \cite{Bae:2010ki} & 2+1 & \gA & \good & \soso & \soso & \soso
& $-$ & 0.529(9)(32) &  0.724(12)(43) \\[0.5ex] 

Aubin 09 & \cite{Aubin:2009jh} & 2+1 & \gA & \soso & \soso &
     \soso & \tbg & $-$ & 0.527(6)(21)& 0.724(8)(29) \\[0.5ex]  

RBC/UKQCD 07A, 08 \rule{1em}{0em}& \cite{Antonio:2007pb,Allton:2008pn} & 2+1 & \gA
                              & \tbr & \soso & \tbg     & \tbg & $-$ &
0.524(10)(28) & 0.720(13)(37) \\[0.5ex]  
HPQCD/UKQCD 06  & \cite{Gamiz:2006sq} & 2+1 & \gA
                              & \tbr & \soso$^\ast$ & \tbg     & \tbr &
$-$ & 0.618(18)(135)& 0.83(18) \\[0.5ex]  
&&&&&&&&&& \\[-0.1cm]
\hline
&&&&&&&&&& \\[-0.1cm]
ETM 10A & \cite{Constantinou:2010qv} & 2 & \gA & \good & \soso & \soso
& \good&  $\,d$ &   0.533(18)(12)$^1$  & 0.729(25)(17) \\[0.5ex]
JLQCD 08 & \cite{Aoki:2008ss} & 2 & \gA  & \tbr      & \soso      &
\tbr          &\tbg    & $-$ & 0.537(4)(40) &
0.758(6)(71)\\[0.5ex]  
RBC 04   & \cite{Aoki:2004ht} & 2 & \gA & \tbr      & \tbr      &
\tbr$^\dagger$ & \tbg      &$-$ & 0.495(18)    & 0.678(25)$^2$
\\[0.5ex]  
UKQCD 04 & \cite{Flynn:2004au} & 2  & \gA  & \tbr      & \tbr      &
\tbr$^\dagger$ & \tbr      & $-$ & 0.49(13)     & 0.68(18)
\\[0.5ex]  
&&&&&&&&&& \\[-0.1cm]
\hline
\hline
\end{tabular*}
\begin{tabular*}{\textwidth}[l]{l@{\extracolsep{\fill}}lllllllll}
  \multicolumn{10}{l}{\vbox{\begin{flushleft} 
        $^\ddagger$ The renormalization is performed using perturbation
        theory at one loop, with a conservative 
	\\\hspace{0.2cm} 
        estimate of the uncertainty. 
\\\rule{0cm}{0.4cm}\hspace{-0.1cm} 
	%
	%
        $^\ast$ This result has been obtained with only two ``light'' sea
        quark masses. 
\\ \rule{0cm}{0.4cm}\hspace{-0.1cm} 
        $^\dagger$ These results have been obtained at  
	$(M_\pi L)_{\rm min} > 4$ in a lattice box 
        with a spatial extension  $L < 2$~fm.
\\ \rule{0cm}{0.3cm}\hspace{-0.1cm} 
        $a$ $B_K$ is renormalized non-perturbatively at a scale of 1.4 GeV
        in two RI/SMOM schemes for $\Nf = 3$, and 
	\\\hspace{0.2cm}
        then run to 3 GeV using a non-perturbatively determined step-scaling
        function. 
	\\\hspace{0.2cm}
	Conversion to $\msbar$ is at one-loop order at 3 GeV.
        \\\rule{0cm}{0.3cm}\hspace{-0.1cm}
        $b$ $B_K$ is renormalized and run non-perturbatively to a scale of
        $3.4\,\gev$ in the RI/MOM scheme.
	\\\hspace{0.2cm}
	Non-perturbative and NLO
        perturbative running agrees down to scales of $1.8\,\gev$ within
        statistical
	\\\hspace{0.2cm}
	uncertainties of about 2\%.
        \\\rule{0cm}{0.4cm}\hspace{-0.1cm}
        $c$ $B_K$ is renormalized non-perturbatively at a scale of 2\,GeV
        in two RI/SMOM schemes for $\Nf = 3$, and then 
	\\\hspace{0.2cm}
        run to 3 GeV using a non-perturbatively determined step-scaling
        function. Conversion to $\msbar$ is at 
	\\\hspace{0.2cm}
        one-loop order at 3 GeV.
        \\\rule{0cm}{0.3cm}\hspace{-0.1cm}
        $d$ $B_K$ is renormalized non-perturbatively at scales $1/a \sim 2
        \div 3\,\gev$ in the $\Nf = 2$ RI/MOM scheme. In this
        \\\hspace{0.2cm}
        scheme, non-perturbative and NLO
        perturbative running are shown to agree from 4 GeV down 2 GeV to
        \\\hspace{0.2cm} 
        better than 3\%
        \cite{Constantinou:2010gr,Constantinou:2010qv}.  
        \\\rule{0cm}{0.4cm}\hspace{-0.1cm}
        $^1$ $B_{K}(\msbar, 2\,\gev)$ is obtained from the estimate for
        $\hat{B}_{{K}}$ using the conversion factor 1.369. 
\\ \rule{0cm}{0.4cm}\hspace{-0.1cm}  
        $^2$ $\hat{B}_{{K}}$ is obtained from the estimate for
        $B_{K}(\msbar, 2\,\gev)$ using the conversion factor 1.369. 
\\ \rule{0cm}{0.4cm}\hspace{-0.1cm}  
\end{flushleft}}}
\end{tabular*}
\vspace{-0.5cm}
\caption{Results for the Kaon $B$-parameter together with a summary of
  systematic errors. If information about non-perturbative running is
  available, this is indicated in the column ``running", with details
  given at the bottom of the table.\label{tab_BKsumm}}
\end{center}
\end{table}

Since the first publication of the FLAG review\,\cite{Colangelo:2010et}
several new or updated results for the Kaon $B$-parameter have been reported
for $\Nf=2+1$, i.e. BMW\,11\,\cite{Durr:2011ap}, SWME\,11A\,\cite{Bae:2011ff},
SWME\,13\,\cite{Bae:2013lja}, Laiho\,11\,\cite{Laiho:2011np}, and
RBC/UKQCD\,12\,\cite{Arthur:2012opa}. No new results for two-flavour QCD have
appeared recently. There is a first, preliminary calculation with
$\Nf=2+1+1$~\cite{Carrasco:2011gr} from the ETM collaboration. We do not
include this result in the following discussion, however, because the
interpretation of $B_{K}$ with active charm involves several subtleties that
have yet to be addressed.\footnote{%
For example, the master formula Eq.~(\ref{eq:epsK}) no longer holds as
written because contributions containing two insertions of $\Delta
S=1$ weak Hamiltonians connected by dynamical charm quarks no longer
lead to a short-distance $\Delta S=2$ matrix element. }
We briefly discuss the main features of the most recent calculations
below.

The BMW Collaboration has produced a new result for
$B_{{K}}$\,\cite{Durr:2011ap}, using their ensembles of
HEX-smeared, tree-level O($a$) improved Wilson
fermions\,\cite{Durr:2010aw}. To this end the four finest lattice
spacings, with $a$ ranging from $0.054 - 0.093$\,fm, are
employed. Simulations are performed close to the physical pion mass,
or even below that value (for the two largest lattice spacings). The
smearing of the link variables results in a significant suppression of
the effects of chiral symmetry breaking, since the coefficients
multiplying the dimension-six operators of different chirality are
found to be very small, in some cases even compatible with zero. The
quoted value for $\hat{B}_{{K}}$ is obtained from a combined chiral
and continuum extrapolation. In order to investigate the systematics
associated with the chiral behaviour, several different cuts on the
maximum pion mass are performed. Another important ingredient in
BMW\,11~\cite{Durr:2011ap} is the nonperturbative determination of
the continuum step scaling function for scales varying between~1.8
and~3.5\,GeV. In this way, the perturbative matching to the RGI
$B$-parameter can be performed at $\mu=3.5$\,GeV, a value where
perturbation theory at NLO is found to yield a good description of the
scale dependence.

The SWME\,11,\,11A,\,13 results~\cite{Kim:2011qg,Bae:2011ff,Bae:2013lja} are
obtained using a mixed action: HYP-smeared valence staggered quarks on the
Asqtad improved, rooted staggered MILC ensembles. Compared to 
the previous edition of the FLAG review~\cite{Colangelo:2010et}, one
major update is the addition of a fourth, finer, lattice spacing. This allows
for a more extensive analysis of the continuum extrapolation, leading to more
reliable estimates of the associated error (which is the second-largest error
at 1.1\%). A second major update, implemented only in SWME\,13, is the
addition of several ensembles with a range of sea-quark masses allowing a
simultaneous extrapolation in $a^2$ and the sea-quark masses.  A third change
in SWME\,13 is the use of larger volumes. Other updates include the use of
correlated fits in the chiral extrapolation, the inclusion of finite-volume
corrections in the chiral fits, and a significant reduction in statistical
errors due to the use of an order of magnitude more sources on each
lattice. The dominant error remains that from the use of one-loop perturbative
matching between lattice and $\msbar$ schemes. This error is estimated
conservatively assuming a missing two-loop matching term of size
$1\times\alpha(1/a)^2$, i.e. with no factors of $1/(4\pi)$ included. The other
methods for estimating this error described earlier in this review lead to
smaller estimates~\cite{Kim:2009tia}. This procedure is, in this review,
deemed conservative enough to merit inclusion in the global average described
below. The resulting matching error is 4.4\%.

The Laiho\,11 result~\cite{Laiho:2011np} uses a mixed action, with
HYP-smeared domain wall valence quarks on the Asqtad MILC
ensembles. Compared to the earlier result obtained by this
collaboration (Aubin 09~\cite{Aubin:2009jh}), the main improvement
consists in the implementation of an RI/MOM scheme based on
nonexceptional momenta in the nonperturbative renormalization of
$B_{{K}}$, as well as the addition of a third lattice spacing. The
largest error is still the matching factor between the lattice and
$\msbar$ schemes. This error is $2.4\%$ out of a total quoted error of
$2.8\%$. The present calculation uses five additional ensembles over
that of the previous edition of the FLAG review \cite{Colangelo:2010et}, leading to a reduction of the chiral/continuum
extrapolation error and to the statistical error.

The RBC/UKQCD Collaboration employ domain wall fermions to determine
$B_{{K}}$. The main feature of their latest update,
Ref.~\cite{Arthur:2012opa}, is the addition of two ensembles with
unitary pion masses as low as~171\,MeV and a minimum partially
quenched pion mass of~143\,MeV. In order to keep the numerical effort
of simulating near-physical pion masses at a manageable level, the new
ensembles are generated at a larger lattice spacing. Moreover, in
order to control the larger residual chiral symmetry breaking effects
which are incurred on coarser lattices, a modified fermion action, the
Dislocation Suppressing Determinant Ratio
(DSDR)\,\cite{Vranas:1999rz,Vranas:2006zk,Fukaya:2006vs,Renfrew:2009wu},
is used in the simulations. As in their earlier
publication\,\cite{Aoki:2010pe}, RBC/UKQCD employ nonperturbative
renormalization factors computed for a variety of RI/MOM schemes with
nonexceptional momenta. Owing to the addition of ensembles with
larger lattice spacing, the matching between lattice regularization
and the intermediate RI/MOM schemes is performed at the lower scale of
$1.4\,\gev$. When combined with the nonperturbative determinations of
the continuum step scaling functions, the perturbative conversion to
the $\msbar$ or RGI schemes can be done at $\mu=3\,\gev$. The use of
near-physical valence pion masses at a spatial volume of
$L\approx4.6$\,fm implies a rather small value of $M_{\pi,\rm
min}L\approx3.3$. However, the entire set of results collected in
Refs.~\cite{Aoki:2010pe,Arthur:2012opa} comprises several volumes with
$L>2.7$\,fm . The combined analysis of all data should allow for a
reliable determination of $B_{{K}}$ with controlled finite-volume
effects. It is noted in Ref.~\cite{Arthur:2012opa} that the inclusion
of the lighter pion masses essentially halves the uncertainty in
$B_{{K}}$ due to the chiral/continuum extrapolation. The largest
systematic uncertainty remains the perturbative truncation error of
2.1\%. As regards the effects of residual chiral symmetry breaking
induced by the finite extent of the 5th dimension in the domain wall
fermion formulation, it is noted in Ref.~\cite{Aoki:2007xm} that the
mixing of $Q^{\Delta{S}=2}$ with operators of opposite chirality is
negligibly small.

Summarizing the new developments, one must note that the biggest
improvements since the previous edition of the FLAG review \cite{Colangelo:2010et} concern the chiral extrapolation and the
issue of renormalization. Ensembles at near-physical pion masses have
significantly reduced the uncertainty associated with chiral fits,
while nonperturbative running is about to become routine. One must
realize that, despite this improvement, perturbative matching is still
applied only at moderately large scales. Most collaborations therefore
identify the largest uncertainty to arise from neglecting higher
orders in the perturbative relation to the RGI or $\msbar$ schemes.

We now describe our procedure for obtaining global averages. The rules
of section \ref{sec:color-code} stipulate that results which are free
of red tags and are published in a refereed journal may enter an
average. Papers which at the time of writing are still unpublished but
are obvious updates of earlier published results can also be taken
into account.

In the previous edition of the FLAG review \cite{Colangelo:2010et} the
results by SWME were excluded from the average, since 
the renormalization factors were estimated in one-loop perturbation
theory only. However, in this review such calculations are included as
long as the estimate of the matching error is sufficiently
conservative. Thus the result of SWME\,13\,\cite{Bae:2013lja} 
(which is a update of the earlier published calculations
of Refs.~\cite{Kim:2011qg,Bae:2011ff})
now qualifies
for inclusion, despite the fact that nonperturbative information on
the renormalization factors is not available. Ref.~\cite{Laiho:2011np},
Laiho\,11 has appeared only as conference proceedings, but since it extends
the study of Ref.~\cite{Aubin:2009jh} it will be included in our average.

Thus, for $N_f=2+1$ our global average is based on the results of
BMW\,11~\cite{Durr:2011ap}, SWME\,13~\cite{Bae:2013lja},
Laiho\,11~\cite{Laiho:2011np} and
RBC/UKQCD\,12~\cite{Arthur:2012opa}. Our procedure is as follows: in
a first step statistical and systematic errors of each individual
result for the RGI $B$-parameter, $\hat{B}_{{K}}$, are combined in
quadrature. Next, a weighted average is computed from the set of
results. For the final error estimate we take correlations between
different collaborations into account. To this end we note that we
consider the statistical and finite-volume errors of SWME\,13 and
Laiho\,11 to be correlated, since both groups use the Asqtad ensembles
generated by the MILC Collaboration. Laiho\,11 and RBC/UKQCD\,12A both
use domain wall quarks in the valence sector and also employ similar
procedures for the nonperturbative determination of matching
factors. Hence, we treat the quoted renormalization and matching
uncertainties by the two groups as correlated. After constructing the
global covariance matrix according to
Schmelling~\cite{Schmelling:1994pz}, we arrive at
\begin{equation}\label{eq:BKNf3}
  \Nf=2+1:\hspace{1cm} \hat{B}_{{K}} = 0.7661(99)\, ,
\end{equation}
with a reduced $\chi^2$-value of 0.387. The error is dominated by
systematic uncertainties.\footnote{%
We can approximately quantify this as follows. A weighted average of
BMW\,11, Laiho\,11 and RBC/UKQCD\,12A using only statistical errors
gives $\hat{B}_{K}=0.7640(33)$. Taking $0.0033$ as the total
statistical error, a total systematic error of $0.0093$ is needed to
obtain the combined total error of $0.0099$ quoted in the text. (We
exclude the SWME\,13 result from this calculation as it is only
consistent with the other results when its relatively large systematic
error is included.) We note that this estimate of the total systematic
error is larger than the smallest individual systematic error
($0.0084$ from BMW\,11). }
By applying the NLO conversion
factor $\hat{B}_{{K}}/B_{{K}}^\msbar (2\,{\rm GeV})=1.369$, this
translates into
\begin{equation}
  \Nf=2+1:\hspace{1cm} B_{{K}}^\msbar (2\,{\rm GeV}) = 0.5596(72)\, .
\end{equation}
Thus, the accuracy of the current global estimate stands at an impressive
1.3\%, which represents a significant improvement over the 2.7\% uncertainty
quoted in the previous edition of the FLAG review
($\hat{B}_{K}=0.738(20)$). The two results are, however, completely
consistent.

\begin{figure}[ht]
\centering
\includegraphics[width=13cm]{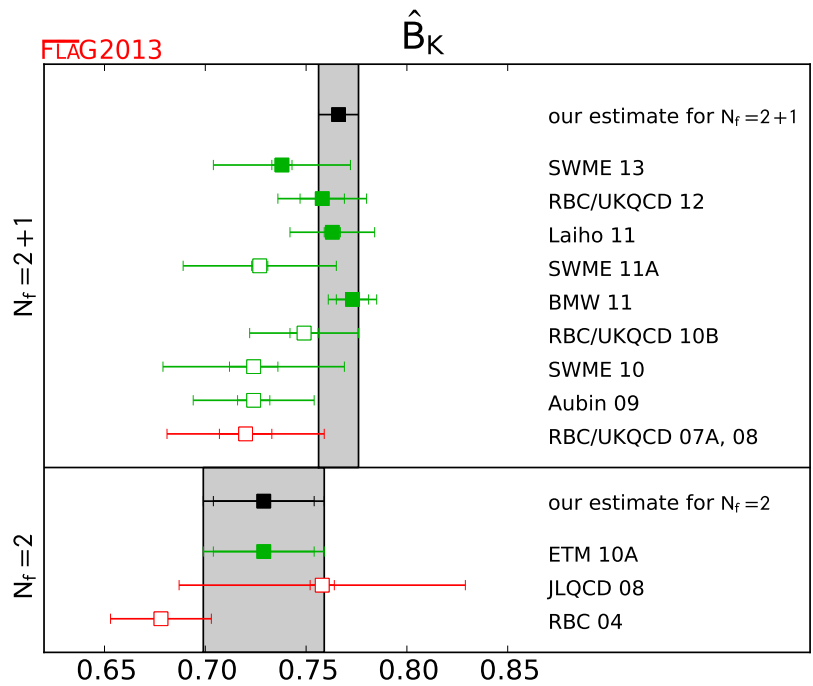}
\caption{Lattice results for the renormalization group invariant $B$-parameter (compare Table \ref{tab_BKsumm}). 
The black squares and grey bands indicate our global averages (\ref{eq:BKNf3}) and (\ref{eq:BKNf2}).
 Our $\Nf=2$ estimate coincides
  with the ETM\,10A result. The significance of the colours is explained in section \ref{sec:qualcrit}.
\label{fig_BKsumm}}
\end{figure}

Passing over to describing the results computed for $\Nf=2$ flavours,
we note that the situation is unchanged since the publication of
the previous edition of the FLAG review \cite{Colangelo:2010et}. In particular, the result of
ETM\,10A~\cite{Constantinou:2010qv} is the only one which allows for
an extensive investigation of systematic uncertainties. In fact, it is
the only published $\Nf=2$ calculation involving data computed at
three values of the lattice spacing. Being the only result without red
tags, it can therefore be identified with the currently best global
estimate for two-flavour QCD, i.e.
\begin{equation}\label{eq:BKNf2}
\Nf=2:\hspace{1cm}\hat{B}_{{K}} = 0.729(25)(17)\, ,
\hspace{1cm}B_{{K}}^\msbar (2\,{\rm GeV}) = 0.533(18)(12)\,.
\end{equation}
The result in the $\msbar$ scheme has been obtained by applying the
same conversion factor of 1.369 as in the three-flavour theory.

The grey bands in Fig.\,\ref{fig_BKsumm} represent the global
estimates for $\Nf=2$ and $\Nf=2+1$. It appears that $B_{{K}}$ may be
slightly smaller in two-flavour QCD, but in view of the relatively
large uncertainty of the $\Nf=2$ result, the difference is hardly
significant.

\clearpage

\input{HQ/macros_static.sty}
\clearpage
\section{$D$-meson decay constants and form factors}
\label{sec:DDecays}

Leptonic and semileptonic decays of charmed $D$ and $D_s$ mesons occur
via charged $W$-boson exchange, and are sensitive probes of $c \to d$
and $c \to s$ quark flavour-changing transitions.  Given experimental
measurements of the branching fractions combined with sufficiently
precise theoretical calculations of the hadronic matrix elements, they
enable the determination of the CKM matrix elements $|V_{cd}|$ and
$|V_{cs}|$ (within the Standard Model) and a precise test of the
unitarity of the second row of the CKM matrix.  Here we summarize the
status of lattice-QCD calculations of the charmed leptonic decay
constants and semileptonic form factors.  Significant progress has
been made in computing $f_{D_{(s)}}$ and the $D\to \pi (K) \ell \nu$
form factors in the last few years, largely due to the introduction of
highly-improved lattice-fermion actions that enable the simulation of
$c$-quarks with the same action as for the $u$, $d$, and $s$-quarks.

The charm-quark methods discussed in this review have been validated
in a number of ways.  Because several groups use the same action for
charm and bottom quarks, tests of charm-quark methods are also
relevant for the $B$-physics results discussed in
Sec.~\ref{sec:BDecays}, and are therefore summarized in the
introduction of that section.  Finally, we note that we limit our
review to results based on modern simulations with reasonably light
pion masses (below approximately 500~MeV). This excludes results
obtained from the earliest unquenched simulations, which typically had
two flavours in the sea, and which were limited to heavier pion masses
because of the constraints imposed by the computational resources and
methods available at that time.

Following our review of lattice-QCD calculations of $D_{(s)}$-meson
leptonic decay constants and semileptonic form factors, we then
interpret our results within the context of the Standard Model.  We
combine our best-determined values of the hadronic matrix elements
with the most recent experimentally-measured branching fractions to
obtain $|V_{cd(s)}|$ and test the unitarity of the second row of the
CKM matrix.

\subsection{Leptonic decay constants $f_D$ and $f_{D_s}$}
\label{sec:fD}

In the Standard Model the decay constant $f_{D_{(s)}}$ of a
pseudoscalar $D$ or $D_s$ meson is related to the branching ratio for
leptonic decays mediated by a $W$ boson through the formula
\be
{\mathcal{B}}(D_{(s)} \to \ell\nu_\ell)= {{G_F^2|V_{cq}|^2 \tau_{D_{(s)}}}\over{8 \pi}} f_{D_{(s)}}^2 m_\ell^2 
m_{D_{(s)}} \left(1-{{m_\ell^2}\over{m_{D_{(s)}}^2}}\right)^2\;,
 \label{eq:Dtoellnu}
\ee
where $V_{cd}$ ($V_{cs}$) is the appropriate CKM matrix element for a
$D$ ($D_s$) meson.  The branching fractions have been experimentally
measured by CLEO, Belle and Babar with a precision around 5-6$\%$ for
the $D_s$-meson; the uncertainties are twice as large for the Cabibbo
suppressed $D$-meson decay modes~\cite{Beringer:1900zz}.  When
combined with lattice results for the decay constants, they allow for
determinations of $|V_{cs}|$ and $|V_{cd}|$.

In lattice-QCD calculations the decay constants $f_{D_{(s)}}$ are extracted from 
Euclidean  matrix elements of the axial current
\be
\langle 0| A^{\mu}_{cq} | D_q(p) \rangle = f_{D_q}\;p_{D_q}^\mu  \;,
\ee
with $q=d,s$ and $ A^{\mu}_{cq} =\bar{c}\gamma_\mu \gamma_5
q$. Results for $N_f=2,\; 2+1$ and $2+1+1$ dynamical flavours are
summarized in Table~\ref{tab_FDsummary} and Figure~\ref{fig:fD}.

\begin{table}[htb]
\begin{center}
\mbox{} \\[3.0cm]
\footnotesize
\begin{tabular*}{\textwidth}[l]{@{\extracolsep{\fill}}l@{\hspace{1mm}}r@{\hspace{1mm}}l@{\hspace{1mm}}l@{\hspace{1mm}}l@{\hspace{1mm}}l@{\hspace{1mm}}l@{\hspace{1mm}}l@{\hspace{1mm}}l@{\hspace{1mm}}l@{\hspace{1mm}}l@{\hspace{1mm}}l}
Collaboration & Ref. & $\Nf$ & 
\hspace{0.15cm}\begin{rotate}{60}{publication status}\end{rotate}\hspace{-0.15cm} &
\hspace{0.15cm}\begin{rotate}{60}{continuum extrapolation}\end{rotate}\hspace{-0.15cm} &
\hspace{0.15cm}\begin{rotate}{60}{chiral extrapolation}\end{rotate}\hspace{-0.15cm}&
\hspace{0.15cm}\begin{rotate}{60}{finite volume}\end{rotate}\hspace{-0.15cm}&
\hspace{0.15cm}\begin{rotate}{60}{renormalization/matching}\end{rotate}\hspace{-0.15cm}  &
\hspace{0.15cm}\begin{rotate}{60}{heavy quark treatment}\end{rotate}\hspace{-0.15cm} & 
\rule{0.4cm}{0cm}$f_D$ & \rule{0.4cm}{0cm}$f_{D_s}$  & 
 \rule{0.3cm}{0cm}$f_{D_s}/f_D$ \\[0.2cm]
\hline
\hline
&&&&&&&&&&& \\[-0.1cm]
ETM 13F & \cite{Dimopoulos:2013qfa} & 2+1+1 & \rC & \soso & \soso  &  \soso & \good  &  \okay &
202(8)   & 242(8) &  1.199(25) \\[0.5ex]

FNAL/MILC 13$^\nabla$ & \cite{Bazavov:2013nfa} & 2+1+1 & \rC & \good    & \good    & \good     
&\good & \okay  & 212.3(0.3)(1.0)   & 248.7(0.2)(1.0) & 1.1714(10)(25)\\[0.5ex]

FNAL/MILC 12B & \cite{Bazavov:2012dg} & 2+1+1 & \rC & \good    & \good    & \good     
&\good & \okay  & 209.2(3.0)(3.6)   & 246.4(0.5)(3.6) & 1.175(16)(11)\\[0.5ex]

&&&&&&&&&&& \\[-0.1cm]
\hline
&&&&&&&&&&& \\[-0.1cm]
HPQCD 12A &\cite{Na:2012iu} & 2+1 & \gA &\soso  &\soso &\good &\good &\okay 
& 208.3(1.0)(3.3) & 246.0(0.7)(3.5) & 1.187(4)(12)\\[0.5ex]

FNAL/MILC 11& \cite{Bazavov:2011aa} & 2+1 & \gA & \soso &\soso &\good  & 
 \soso & \okay & 218.9(11.3) & 260.1(10.8)&   1.188(25)   \\[0.5ex]  

PACS-CS 11 & \cite{Namekawa:2011wt} & 2+1 & \gA & \tbr & \good & \tbr  & 
\soso & \okay & 226(6)(1)(5) & 257(2)(1)(5)&  1.14(3)   \\[0.5ex] 

HPQCD 10A & \cite{Davies:2010ip} & 2+1 & \gA & \good  & \soso  & 
\good & \good & \okay & 213(4)$^{*}$ & 248.0(2.5)  \\[0.5ex]

HPQCD/UKQCD 07 & \cite{Follana:2007uv} & 2+1 &  \gA & \good & \soso & 
\good & \good  & \okay & 207(4) & 241 (3)& 1.164(11)  \\[0.5ex] 

FNAL/MILC 05 & \cite{Aubin:2005ar} & 2+1 & \gA &\soso &   \soso    &
\good      & \soso    &  \okay       & 201(3)(17) & 249(3)(16)  & 1.24(1)(7) \\[0.5ex]

&&&&&&&&&&& \\[-0.1cm]
\hline
&&&&&&&&&&& \\[-0.1cm]
ETM 13B$^\square$ & \cite{Carrasco:2013zta} & 2 & \oP & \good & \soso  &  \good & \good  &  \okay &
208(7)   & 250(7) &  1.20(2) \\[0.5ex]

ETM 11A & \cite{Dimopoulos:2011gx} & 2 & \gA & \good & \soso  &  \good & \good  &  \okay &
212(8)   & 248(6) &  1.17(5) \\[0.5ex]

ETM 09 & \cite{Blossier:2009bx} & 2 & \gA & \soso & \soso  &  \good & \good  &  \okay & 
197(9)   & 244(8) &  1.24(3) \\[0.5ex]

&&&&&&&&&&& \\[-0.1cm]
\hline
\hline
\end{tabular*}
\begin{tabular*}{\textwidth}[l]{l@{\extracolsep{\fill}}lllllllll}
  \multicolumn{10}{l}{\vbox{\begin{flushleft} 
$^{\nabla}$ Update of FNAL/MILC 12B.\\
$^{*}$ This result is obtained by using the central value for $f_{D_s}/f_D$ from HPQCD/UKQCD~07 
and increasing the error to account for the effects from the change in the physical value of $r_1$. \\
$^{\square}$ Update of ETM 11A and ETM 09.
\end{flushleft}}}
\end{tabular*}

\vspace{-0.5cm}
\caption{Decay constants of the $D$ and $D_{s}$ mesons (in MeV) and their ratio.}
\label{tab_FDsummary}
\end{center}
\end{table}
%
%
\begin{figure}[tb]
\hspace{-0.8cm}\includegraphics[width=0.58\linewidth]{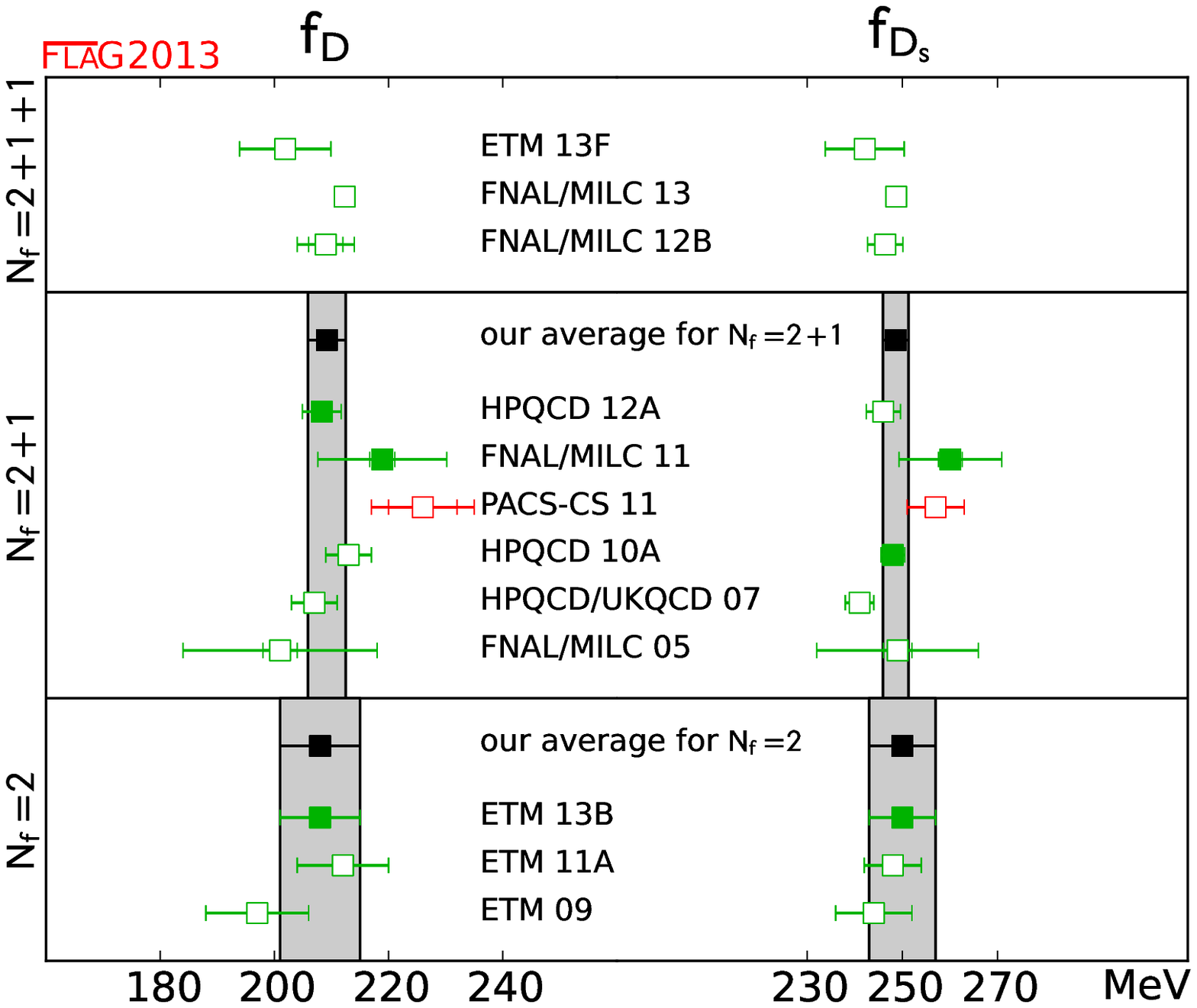} \hspace{-1cm}
\includegraphics[width=0.58\linewidth]{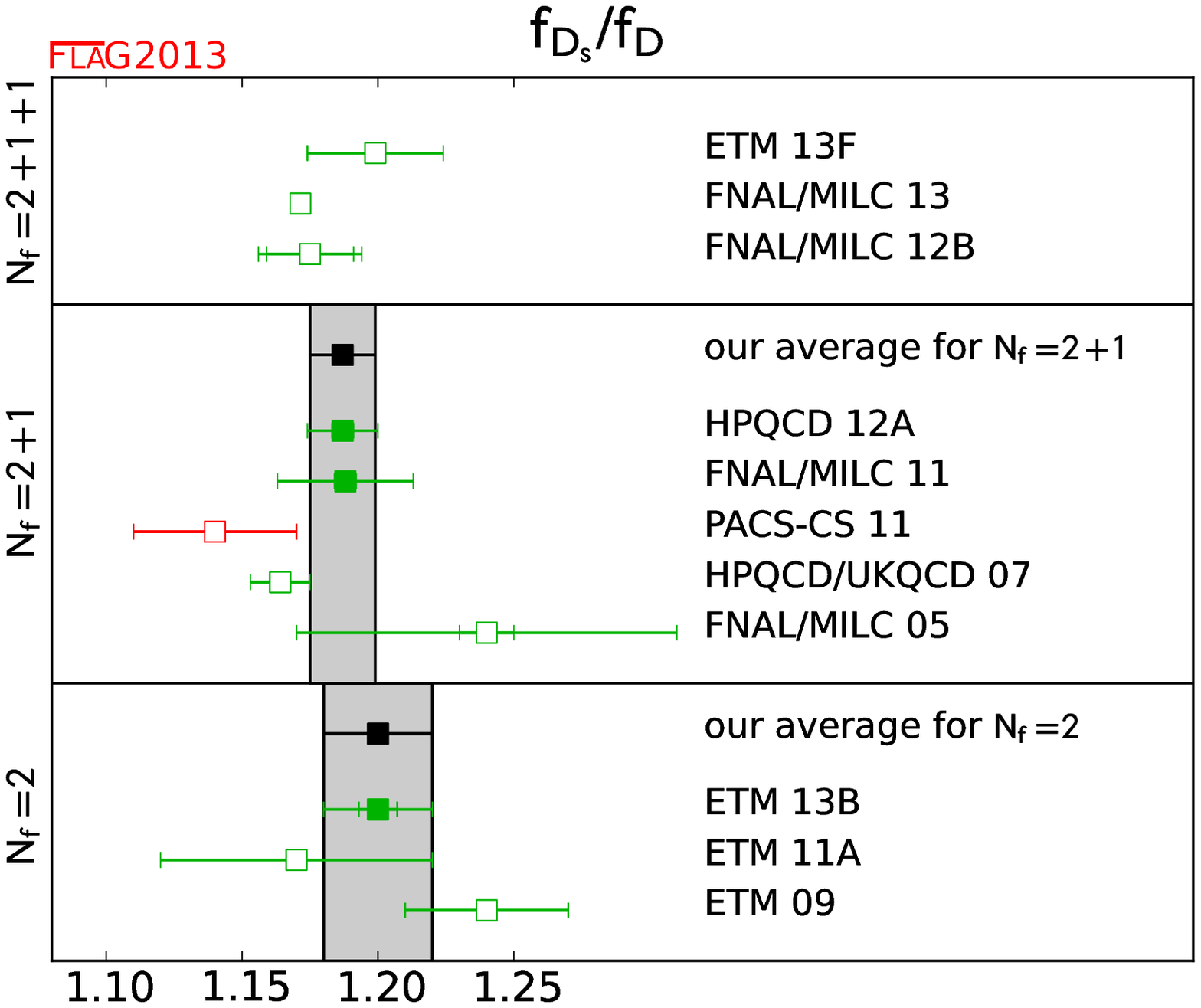}

\vspace{-2mm}
\caption{Decay constants of the $D$ and $D_s$ mesons [values in Table
    \ref{tab_FDsummary} and Eqs.~(\ref{eq:Nf2av}),
    (\ref{eq:Nf3av})]. The significance of the colours is explained in
  section \ref{sec:qualcrit}. The black squares and grey bands
  indicate our averages. Errors in FNAL/MILC~13 are smaller than the
  symbols.}\label{fig:fD}
\end{figure}

The ETM collaboration has published results for $D$ and $D_s$ meson
decay constants with two dynamical flavours, using the twisted-mass
fermionic action at maximal twist with the tree-level improved
Symanzik gauge action.  In this setup the decay constants can be
extracted from an absolutely normalized current and they are
automatically ${\mathcal{O}}(a)$ improved. In ETM~09 three lattice
spacings between $0.1$ and $0.07$ fm are considered with pion masses
down to 270 MeV. Heavy~meson~$\chi$PT formulae plus terms linear in
$a^2$ have been used for the continuum/chiral extrapolations, which
have been performed in two different ways in order to estimate
sytematic effects.  In the first approach $f_{D_s}\sqrt{m_{D_s}}$ and
$\frac{f_{D_s}\sqrt{m_{D_s}}}{ f_{D}\sqrt{m_{D}}}$ are fitted, whereas
in the second case the ratios $\frac{f_{D_s}\sqrt{m_{D_s}}}{f_K}$ and
$\frac{f_{D_s}\sqrt{m_{D_s}}}{f_K} \times
\frac{f_\pi}{f_{D}\sqrt{m_{D}}}$ are analysed. As expected, the
pion-mass dependence of $f_{D_s}\sqrt{m_{D_s}}$ turns out to be very
mild. In addition the double ratio $\frac{f_{D_s}\sqrt{m_{D_s}}}{f_K}
\times \frac{f_\pi}{f_{D}\sqrt{m_{D}}}$ shows little dependence on the
pion mass as well as on the lattice spacing. Cutoff effects on the
contrary are rather large on the decay constants, with the difference
between the physical-mass result at the finest lattice spacing and in
the continuum being approximately 5\%.  ETM~11A contains an update of
the results in ETM~09 obtained by enlarging the statistics on some of
the ensembles and by including a finer lattice resolution with
$a\approx 0.054$ fm, which implies a reduction of cutoff effects by a
factor two.  Moreover in ETM~11A the continuum extrapolations are
performed after interpolating the results at different lattice
spacings to fixed values of the heavy-quark mass. In the case of the
SU(3) breaking ratio $f_{D_s}/f_D$, the uncertainty associated with
the chiral extrapolation is estimated by comparing fits either
following heavy~meson~$\chi$PT or assuming a simple linear dependence
on the light-quark mass.  These results have been further updated in
ETM~13B~\cite{Carrasco:2013zta} by using optimized smearing
interpolating fields in order to suppress excited states contributions
and by changing the chiral extrapolation. The ensembles used are the
same as in ETM~11A.  Values at the physical point are obtained by
first extrapolating $f_{D_s} \sqrt{m_{D_s}}$ linearly in $m_l^2$ and
in $a^2$ and then by extrapolating the double ratio
$(f_{D_s}/f_D)/(f_K/f_\pi)$ using HM$\chi$PT. The value of $f_K/f_\pi$
is taken from the $N_f=2+1$ average in~\cite{Colangelo:2010et}, in
order to avoid correlations with estimates obtained by the ETM
collaboration.
 
As results from just one collaboration exist in the literature, the
$N_f=2$ averages are simply given by the values in ETM~13B, which read
\begin{equation}
N_f=2: \quad\!\! f_D=(208\pm 7) \;{\rm MeV}, \quad\!  f_{D_s}= (250\pm 7)\;{\rm MeV}, \quad\! {{f_{D_s}}\over{f_D}}=1.20\pm 0.02\;.
\label{eq:Nf2av}
\end{equation}

The ALPHA Collaboration presented preliminary results on $f_{D_{(s)}}$
with two dynamical flavours at the Lattice 2013
Conference~\cite{Heitger:2013fds}.  The proceedings however appeared
after the deadline for consideration in this review and therefore are
not discussed here.

Several collaborations have produced results with $N_f=2+1$ dynamical
flavours. The most precise determinations come from a sequence of
publications by
HPQCD/UKQCD~\cite{Na:2012iu,Davies:2010ip,Follana:2007uv}. In all
cases configurations generated by MILC with Asqtad rooted staggered
quarks in the sea and a one-loop tadpole improved Symanzik gauge
action have been analysed (see~\cite{Bazavov:2009bb} and references
therein).  The main differences are in the ensembles utilized and in
the absolute scale setting.  The relative scale is always set through
$r_1$ derived from the static quark-antiquark potential.

In HPQCD/UKQCD~07~\cite{Follana:2007uv} three lattice spacings,
$a\approx 0.15,\; 0.12$ and $0.09$ fm, with RMS pion masses between
542 and 329 MeV, have been considered. This gives rather large values
for the charm-quark mass in lattice units, $0.43 <am_c< 0.85$, and
indeed lattice artifacts are estimated to be the second largest
systematic uncertainty in the computation.  The main systematic error
is resulting from the absolute scale setting, which had previously
been performed through the $\Upsilon$ spectrum, using NRQCD for the
$b$ quark. The estimate reads $r_1=0.321(5)$ fm.

In 2010, HPQCD obtained a more precise determination of
$r_1=0.3133(23)$, based on several different physical inputs
(including $f_\pi$, $f_K$ and the $\Upsilon$ spectrum) and improved
continuum limit extrapolations. It is worth noting that the new $r_1$
is about 1.5$\sigma$ lower than the older value. The publications
HPQCD~10A~\cite{Davies:2010ip} and HPCQD~12A~\cite{Na:2012iu} update
the computations of $f_{D_s}$ and $f_D$, respectively, using the new
scale determination. These results enter our final averages. The
change in the scale requires a retuning of the bare quark masses and a
change in the conversion of dimensionless quantitities, measured in
units of $r_1$, to physical ones, measured in MeV.

In HPQCD~10A, $f_{D_s}$ is calculated on ensembles with $a\approx
0.06$ and $0,045$ fm and with RMS pion masses ranging between 542 and
258 MeV.  The chiral and continuum extrapolations have been performed
simultaneously by employing polynomials quadratic in the sea-quark
mass $\delta_q = \frac{m_{q,sea} -m_{q,phys}}{m_{q,phys}}$, with $q =
s, l$, and through the eighth power of the charm-quark mass, including
cross terms of the form $\delta_q (am_c)^n$.  The valence strange- and
charm-quark masses are fixed to their physical values obtained from
matching to the $\eta_s$ and $\eta_c$ masses.  The fits are robust
against variations, such as the exclusion of ensembles with the
coarsest and finest lattice spacings, or a change in the functional
form such that terms up to $(am_c)^4$ only are kept.  The largest
source of uncertainty in HPQCD~10A still comes from the value of $r_1$
and it amounts to 0.6\%.  The published error includes a 0.1\%
contribution coming from an estimate of electromagnetic effects
obtained using a potential model.

The process of switching to the improved determination of $r_1$ is
 finally completed in HPQCD~12A~\cite{Na:2012iu}, where new values of
 $f_D$ and the ratio $f_{D_s}/f_D$ are reported.  The statistics is
 enlarged at the $a \approx 0.12$ fm and $a \approx 0.09$ fm lattices
 and for the latter a more chiral point, with light-quark masses
 halved with respect to HPQCD/UKQCD~07, is added.  The three-point
 function for $D \to \pi$ at zero recoil momentum (calculated for a
 different project) is used to perform simultaneous fits to two- and
 three-point functions. This turns out to be beneficial in reducing
 the statistical errors on the hadron masses and decay constant matrix
 elements.  Chiral and continuum extrapolations are carried out at the
 same time adopting partially~quenched~heavy~meson~$\chi$PT augmented
 by $(am_c)^2$ and $(am_c)^4$ terms. Given the rather large values of
 $am_c$ between 0.4 and 0.6, the continuum extrapolation gives the
 largest systematical uncertainty, amounting to roughly 1\% out of the
 total 1.7\% and 1.1\% total errors on $f_D$ and on $f_{D_s}/f_D$
 respectively.  Finally, the HPQCD collaboration also calculates the
 ratio $f_+^{D \to \pi}(0)/f_D$ using the result for the semileptonic
 form factor from~\cite{Na:2011mc} and find good agreement with the
 experimental ratio which is independent of $|V_{cd}|$.  Summarizing
 the computations by HPQCD: concerning $f_D$, HPQCD~12A supersedes
 HPQCD/UKQCD~07 and HPQCD~10A because of the more chiral points
 considered but does not supersede HPQCD~10A for $f_{D_s}$ as finer
 resolutions are included in the latter, which contains the
 collaboration's most precise result for the $D_s$ meson decay
 constant.

The PACS-CS Collaboration published in 2011 a computation of the $D$
and $D_s$ decay constants with 2+1 flavours of nonperturbatively
${\mathcal{O}}(a)$ improved Wilson fermions and the Iwasaki gauge
action~\cite{Namekawa:2011wt}.  For the charm quark the Tsukuba heavy
quark action is used.  The parameters in the action and the
renormalization constants of the charm-light and charm-strange axial
currents are computed in a mixed setup, partly nonperturbatively
(typically the massless contribution) and partly relying on one-loop
perturbation theory, see Appendix~A for details.  This leaves residual
cutoff and matching effects of ${\mathcal{O}}(\alpha^2_s
a\Lambda_{QCD},\, (a\Lambda)^2,\, \alpha_s^2)$ in the computation,
which, in addition is carried out at one value of the lattice spacing
only ($a\approx 0.09$ fm). Quark masses are quite low, yielding $m_\pi
=152(6)$ MeV and the ensemble is reweighted to the physical point
using the technique in~\cite{Aoki:2009ix}. However, measurements are
performed on only one set of configurations with $L/a=32$, such that
$m_\pi L$ is around 2.2. For this reason, and for the limitation to a
single lattice spacing, the PACS-CS~11 results do not enter our
averages.

The Fermilab Lattice and MILC collaborations have presented several
computations of $D_{(s)}$ meson decay constants with $2+1$ flavours of
dynamical quarks \cite{Aubin:2005ar,Bazavov:2011aa}. Their first
published results are in Ref.~\cite{Aubin:2005ar} (FNAL/MILC~05),
which were later updated and superseded in Ref.~\cite{Bazavov:2011aa}
(FNAL/MILC~11).  The MILC Asqtad ensembles, as for the HPQCD results,
have been used in both cases.  For the charm quark the Fermilab action
is adopted, with mostly nonperturbative (mNPR) renormalization of the
axial currents (see Appendix A for details).  In FNAL/MILC~05 three
lattice spacings with $a\approx 0.18,\; 0.12$ and $0.09$ fm, according
to the original estimate $r_1=0.321(5)$ fm, have been considered. RMS
pion masses are slightly larger than 400 MeV.  Chiral and continuum
extrapolations are performed at the same time by using the $\chi$PT
expressions at NLO for staggered quarks.  Discretization effects and
the chiral fits are the largest sources of systematic errors in $f_D$
and in $f_{D_s}$, each effect being responsible for a systematic
between 4\% and 6\%. Cutoff effects are significantly smaller in the
ratio $f_{D_s}/f_D$, whose systematic uncertainty (around 5\%) is
dominated by the chiral extrapolation.

These uncertainties are reduced in FNAL/MILC~11. The same setup
concerning lattice actions and renormalization is used as in
FNAL/MILC~05 but lighter pion masses (down to 320 MeV for the RMS
values) are included in the analysis and the extremely coarse
$0.18$~fm ensembles are replaced by finer $0.15$~fm ones.  The scale
is set through $r_1=0.3120(22)$ fm, as obtained from an average of
previous MILC and HPQCD determinations. One-loop
rooted~staggered~partially~quenched~$\chi$PT plus leading order in the
heavy-quark expansion formulae are used for the chiral and continuum
extrapolations. The expressions parameterize also the effects of
hyperfine and flavour splittings.  Discretization effects are estimated
using a combination of heavy-quark and Symanzik effective theories to
be around $3\%$ for $f_{D_{(s)}}$ and negligible for the ratio.  At
this level of accuracy the truncation errors in the small correction
factor inherent in the mNPR method are not negligible anymore; the
authors conservatively estimate the two-loop and higher-order
perturbative truncation errors to the full size of the known one-loop
term, i.e. roughly 1$\%$ for the decay constants.

As shown in Table~\ref{tab_FDsummary} the $N_f=2+1$ computations which
fulfill our quality criteria and can enter the averages are HPCQD~12A
and FNAL/MILC~11 for $f_D$ and the SU(3) breaking ratio
$f_{D_{s}}/f_D$, and HPQCD~10A and FNAL/MILC~11 for $f_{D_{s}}$.
Because FNAL/MILC and HPQCD use a largely overlapping set of
configurations, we treat the statistical errors as 100\% correlated
and finally quote
\begin{equation}
N_f=2+1: \quad\!\!\! f_D=(209.2\pm 3.3) \;{\rm MeV}, \quad\!\!\!   f_{D_s}=(248.6\pm 2.7) \;{\rm MeV}, \quad\!\!\! {{f_{D_s}}\over{f_D}}=1.187\pm 0.012\;.
\label{eq:Nf3av}
\end{equation}

The first computation of $f_D$ and $f_{D_s}$ with $N_f=2+1+1$ sea
quarks is presented in Ref.~\cite{Bazavov:2012dg} (FNAL/MILC~12B),
published as a proceeding contribution to the Lattice 2012
Conference. The calculation is performed on configurations generated
by the MILC Collaboration using HISQ sea quarks and a one-loop tadpole
improved Symanzik gauge action~\cite{Bazavov:2012xda}.  Light, strange
and charm valence quarks are also in the HISQ regularization.  Four
lattice resolutions in the range $a\approx 0.15 - 0.06$ fm are
considered.  RMS pion masses vary between 306 and 144 MeV and include
ensembles at each lattice spacing with Goldstone pions at the physical
point.  The dominant systematic uncertainties are due to the scale
setting (through $f_\pi$) and the continuum extrapolation, and they
are both estimated to be at the percent level.  The results have been
updated in FNAL/MILC~13~\cite{Bazavov:2013nfa}.  New measurements at
the finest lattice spacing have been included in the analysis and the
statistics have been significantly increased in each ensemble.  In
addition, heavy-meson, rooted, all-staggered chiral perturbation
theory (HMrAS$\chi$PT), as introduced in Ref.~\cite{Bernard:2013qwa}
to treat both the light and charm quarks as staggered, has been used
at NLO in performing chiral and continuum extrapolations.  The
configurations used in these computations have been generated using
both the RHMC and the RHMD algorithms.  The latter is an inexact
algorithm, where the accept/reject step at the end of the
molecular-dynamics trajectory is skipped. In
Ref.~\cite{Bazavov:2012xda} results for the plaquette, the bare
fermion condensates and a few meson masses, using both algorithms, are
compared and found to agree within statistical uncertainties.

The ETM collaboration has also reported results with $2+1+1$ dynamical
flavours at the Lattice 2013 Conference~\cite{Dimopoulos:2013qfa}.
The configurations have been generated using the Iwasaki action in the
gauge and the Wilson twisted mass action for sea quarks.  The charm
and strange valence quarks are discretized as Osterwalder-Seiler
fermions~\cite{Osterwalder:1977pc}.  Three different lattice spacings
in the range $0.09 - 0.06$ fm have been analysed with pion masses as
low as 210 MeV in lattices of linear spatial extent of about 2.5 to 3
fm. As in the $N_f=2$ computation in ETM~13B, the chiral and continuum
extrapolations are performed first for $f_{D_s}$, including terms
linear and quadratic in $m_l$ and one term linear in $a^2$ in the
parameterization, and then for the double ratio
$(f_{D_s}/f_D)/(f_K/f_\pi)$ using continuum HM$\chi$PT.  The main
systematic uncertaintes are due to the continuum and chiral
extrapolation for $f_{D_s}$ and to the error on $f_K/f_\pi$, which is
also provided in these proceedings and discussed in
Sec.~\ref{sec:vusvud} of this review, for $f_D$.

As a final remark, since the accuracy of the lattice determinations of the
$D$ meson decay constant is rapidly improving, it will become important
in the future, especially when comparing to experimental numbers,
to distinguish between $f_{D^+}$ and the average of $f_{D^+}$ and $f_{D^0}$.
The current status is summarized as follows: FNAL/MILC results
concern $f_{D^+}$, whereas HPQCD, PACS-CS and ETMC numbers correspond to the average
of the decay constants for $D^+$ and $D^0$.

 \subsection{Semileptonic form factors for $D\to \pi \ell \nu$ and $D\to K \ell \nu$}
 \label{sec:DtoPiK}

The form factors for semileptonic $D\to \pi \ell\nu$ and $D\to
K \ell \nu$ decay, when combined with experimental measurements of the
decay widths, enable determinations of the CKM matrix elements
$|V_{cd}|$ and $|V_{cs}|$ via:
\begin{eqnarray}
	\frac{d\Gamma(D\to P\ell\nu)}{dq^2} = \frac{G_F^2 |V_{cx}|^2}{24 \pi^3}
	\,\frac{(q^2-m_\ell^2)^2\sqrt{E_P^2-m_P^2}}{q^4m_{D}^2} \,
	\bigg[ \left(1+\frac{m_\ell^2}{2q^2}\right)m_{D}^2(E_P^2-m_P^2)|f_+(q^2)|^2 & \nonumber\\
+ \frac{3m_\ell^2}{8q^2}(m_{D}^2-m_P^2)^2|f_0(q^2)|^2 & \!\!\!\! \bigg]\,, \label{eq:DtoPiKFull}
\end{eqnarray}
where $x = d, s$ is the daughter light quark, $P= \pi, K$ is the
daughter light pseudoscalar meson, and $q = (p_D - p_P)$ is the
momentum of the outgoing lepton pair.  The vector and scalar form
factors $f_+(q^2)$ and $f_0(q^2)$ parameterize the hadronic matrix
element of the heavy-to-light quark flavour-changing vector current
$V_\mu = i \overline{x} \gamma_\mu c$:
\begin{equation}
\langle P| V_\mu | D \rangle  = f_+(q^2) \left( {p_D}_\mu+ {p_P}_\mu - \frac{m_D^2 - m_P^2}{q^2}\,q_\mu \right) + f_0(q^2) \frac{m_D^2 - m_P^2}{q^2}\,q_\mu \,,
\end{equation}
and satisfy the kinematic constraint $f_+(0) = f_0(0)$ at zero
momentum-transfer.  Because the contribution to the decay width from
the scalar form factor is proportional to $m_\ell^2$, it can be
neglected for $\ell = e, \mu$, and Eq.~(\ref{eq:DtoPiKFull})
simplifies to
\begin{equation}
\frac{d\Gamma \!\left(D \to P \ell \nu\right)}{d q^2} = \frac{G_F^2}{24 \pi^3} |\vec{p}_{P}|^3 {|V_{cx}|^2 |f_+^{DP} (q^2)|^2} \,. \label{eq:DtoPiK}
\end{equation}

In practice, most lattice-QCD calculations of $D\to \pi \ell\nu$ and
$D\to K \ell \nu$ focus on providing the value of the vector form
factor at a single value of the momentum transfer, $f_+(q^2=0)$, which
is sufficient to obtain $|V_{cd}|$ and $|V_{cs}|$.  Because the decay
rate cannot be measured directly at zero momentum transfer, comparison
of these lattice-QCD results with experiment requires a slight
extrapolation of the experimental measurement.  Some lattice-QCD
calculations also provide determinations of the $D\to \pi \ell\nu$ and
$D\to K \ell \nu$ form factors over the full kinematic range $0 < q^2
< q^2_{\rm max} = (m_D - m_P)^2$, thereby allowing a comparison of the
shapes of the lattice simulation and experimental data.  This
nontrivial test in the $D$ system provides a strong check of
lattice-QCD methods that are also used in the $B$-meson system.

Lattice-QCD calculations of the $D\to \pi \ell\nu$ and $D\to
K \ell \nu$ form factors typically use the same light-quark and
charm-quark actions as those of the leptonic decay constants $f_D$ and
$f_{D_s}$.  Therefore many of the same issues arise, e.g. chiral
extrapolation of the light-quark mass(es) to the physical point and
discretization errors from the charm quark, and matching the lattice
weak operator to the continuum, as discussed in the previous section.
Two strategies have been adopted to eliminate the need to renormalize
the heavy-light vector current in recent calculations of
$D\to \pi \ell\nu$ and $D\to K \ell \nu$, both of which can be applied
to simulations in which the same relativistic action is used for the
light $(u,d,s)$ and charm quarks.  The first method was proposed by
Be{\'c}irevi{\'c} and Haas in Ref.~\cite{Becirevic:2007cr}, and
introduces double-ratios of lattice three-point correlation functions
in which the vector current renormalization cancels.  Discretization
errors in the double ratio are of ${\mathcal O}((am_h)^2)$ provided
that the vector-current matrix elements are ${\mathcal O}(a)$
improved.  The vector and scalar form factors $f_+(q^2)$ and
$f_0(q^2)$ are obtained by taking suitable linear combinations of
these double ratios.  The second method was introduced by the HPQCD
Collaboration in Ref.~\cite{Na:2010uf}.  In this case, the quantity
$(m_{c} - m_{x} ) \langle P | S | D \rangle$, where $m_{x}$ and $m_c$
are the bare lattice quark masses and $S = \bar{x}c$ is the lattice
scalar current, does not get renormalized.  The desired form factor at
zero momentum transfer can be obtained by (i) using a Ward identity to
relate the matrix element of the vector current to that of the scalar
current, and (ii) taking advantage of the kinematic identity at zero
momentum transfer $f_+(0) = f_0(0)$, such that $f_+(q^2=0) = (m_{c} -
m_{x} ) \langle P | S | D \rangle / (m^2_D - m^2_P)$.

Additional complications enter for semileptonic decay matrix elements
due to the nonzero momentum of the outgoing pion or kaon.  Both
statistical errors and discretization errors increase at larger
momenta, so results for the lattice form factors are most precise at
$q^2_{\rm max}$.  However, because lattice calculations are performed
in a finite spatial volume, the pion or kaon three-momentum can only
take discrete values in units of $2\pi/L$ when periodic boundary
conditions are used.  For typical box sizes in recent lattice $D$- and
$B$-meson form-factor calculations, $L \sim 2.5$--3~fm; thus the
smallest nonzero momentum in most of these analyses ranges from
$p_P \equiv |\vec{p}_P| \sim 400$--$500$~MeV.  The largest momentum in lattice
heavy-light form-factor calculations is typically restricted to
$ p_P \leq 4\pi/L$ 
For $D \to \pi \ell \nu$ and $D \to
K \ell \nu$, $q^2=0$ corresponds to $p_\pi \sim 940$~MeV and $p_K \sim
1$~GeV, respectively, and the full recoil-momentum region is within
the range of accessible lattice momenta.\footnote{This situation
differs from that of calculations of the $K\to\pi\ell\nu$ form factor,
where the physical pion recoil momenta are smaller than $2 \pi/L$.
For $K\to\pi\ell\nu$ it is now standard to use nonperiodic
(``twisted") boundary
conditions~\cite{Bedaque:2004kc,Sachrajda:2004mi} to simulate directly
at $q^2=0$; see Sec.~\ref{sec:Direct}.  Some collaborations have also
begun to use twisted boundary conditions for $D$
decays~\cite{DiVita:2011py,Koponen:2011ev,Koponen:2012di,Koponen:2013tua}.}  Therefore
the interpolation to $q^2=0$ is relatively insensitive to the fit
function used to parameterize the momentum dependence, and the
associated systematic uncertainty in $f_+(0)$ is small. In contrast,
determinations of the form-factor shape can depend strongly on the
parameterization of the momentum dependence, and the systematic
uncertainty due to the choice of model function is often difficult to
quantify.  This is becoming relevant for $D \to \pi \ell \nu$ and $D
\to K \ell \nu$ decays as collaborations are beginning to present
results for $f_+(q^2)$ and $f_0(q^2)$ over the full kinematic range.
The parameterization of the form-factor shape is even more important
for semileptonic $B$ decays, for which the momentum range needed to
connect to experiment is often far from $q^2_{max}$.

A class of functions based on general field-theory properties, known
as $z$-expansions, has been introduced to allow model-independent
parameterizations of the $q^2$ dependence of semileptonic form factors
over the entire kinematic range (see, e.g.,
Refs.~\cite{Boyd:1994tt,Bourrely:2008za}).  The use of such functions
is now standard for the analysis of $B \to \pi \ell\nu$ transitions
and the determination of
$|V_{ub}|$~\cite{Bailey:2008wp,Ha:2010rf,Lees:2012vv,Amhis:2012bh}; we
therefore discuss approaches for parameterizing the $q^2$ dependence
of semileptonic form factors, including $z$-expansions, in
Sec.~\ref{sec:BtoPiK}.  Here we briefly summarize the aspects most
relevant to calculations of $D \to \pi \ell \nu$ and $D \to K \ell
\nu$.  In general, all semileptonic form factors can be expressed as a
series expansion in powers of $z^n$ times an overall multiplicative
function that accounts for any sub-threshold poles and branch cuts,
where the new variable $z$ is a nonlinear function of $q^2$.  The
series coefficients $a_n$ depend upon the physical process (as well as
the choice of the prefactors), and can only be determined empirically
by fits to lattice or experimental data.  Unitarity establishes strict
upper bounds on the size of the $a_n$'s, while guidance from
heavy-quark power counting provides even tighter constraints.
Recently the HPQCD Collaboration introduced a variation on this
approach, which they refer to as a ``modified $z$-expansion," that
they use to simultaneously extrapolate their lattice simulation data
to the physical light-quark masses and the continuum limit, and to
interpolate/extrapolate their lattice data in $q^2$.  They do so by
allowing the coefficients $a_n$ to depend on the light-quark masses,
squared lattice spacing, and, in some cases the charm-quark mass and
pion or kaon energy.  Because the modified $z$-expansion is not
derived from an underlying effective field theory, there are several
potential concerns with this approach that have yet to be studied in
the literature.  The most significant is that there is no theoretical
derivation relating the coefficients of the modified $z$-expansion to
those of the physical coefficients measured in experiment; it
therefore introduces an unquantified model dependence in the
form-factor shape.  Further, if Bayesian methods are used to constrain
the parameters of the modified $z$-expansion, there is no {\it a
  priori} way to obtain priors for their natural size.  The
``modified" $z$-expansion is now being utilized by collaborations
other than HPQCD and for quantities other than $D \to \pi \ell \nu$
and $D \to K \ell \nu$~\cite{Horgan:2013hoa,SanfilippoLat13}.  We
advise treating results that utilize the ``modified" $z$-expansion to
obtain form-factor shapes and CKM matrix elements with caution,
however, since the systematics of this approach warrant further study.

\subsubsection{Results for $f_+(0)$}

We now review the status of lattice calculations of the $D \to \pi
\ell \nu$ and $D \to K \ell \nu$ form factors at $q^2=0$.  As in the
first version of this review, although we also describe ongoing
calculations of the form-factor shapes, we do not rate these
calculations.

The most advanced $N_f = 2$ lattice-QCD calculation of the
$D \to \pi \ell \nu$ and $D \to K \ell \nu$ form factors is by the ETM
Collaboration~\cite{DiVita:2011py}.  This still preliminary work uses
the twisted-mass Wilson action for both the light and charm quarks,
with three lattice spacings down to $a \approx 0.068$~fm and (charged)
pion masses down to $m_\pi \approx 270$~MeV.  The calculation employs
the ratio method of Ref.~\cite{Becirevic:2007cr} to avoid the need to
renormalize the vector current, and extrapolates to the physical
light-quark masses using $SU(2)$ heavy-light meson $\chi$PT formulated
for twisted-mass fermions.  ETM simulate with nonperiodic boundary
conditions for the valence quarks to access arbitrary momentum values
over the full physical $q^2$ range, and interpolate to $q^2=0$ using
the Be{\'c}irevi{\'c}-Kaidalov ansatz~\cite{Becirevic:1999kt}.  The
statistical errors in $f_+^{D\pi}(0)$ and $f_+^{DK}(0)$ are 9\% and
7\%, respectively, and lead to rather large systematic uncertainties
in the fits to the light-quark mass and energy dependence (7\% and
5\%, respectively).  Another significant source of uncertainty is from
discretization errors (5\% and 3\%, respectively).  On the finest
lattice spacing used in this analysis $am_c \sim 0.17$, so ${\mathcal
O}((am_c)^2)$ cutoff errors are expected to be about 5\%.  This can be
reduced by including the existing $N_f = 2$ twisted-mass ensembles
with $a \approx 0.051$~fm discussed in Ref.~\cite{Baron:2009wt}. Work
is in progress by the ETM Collaboration to compute $f_+^{D\pi}(0)$ and
$f_+^{DK}(0)$ using the same methods on the $N_f = 2+1+1$ twisted-mass
Wilson lattices~\cite{Baron:2010bv}.  This calculation will include
dynamical charm-quark effects and use three lattice spacings down to
$a\approx 0.06$~fm.

The first published $N_f = 2+1$ lattice-QCD calculation of the
$D \to \pi \ell \nu$ and $D \to K \ell \nu$ form factors is by the
Fermilab Lattice, MILC, and HPQCD Collaborations~\cite{Aubin:2004ej}.
(Because only two of the authors of this work are in HPQCD, and to
distinguish it from other more recent works on the same topic by
HPQCD, we hereafter refer to this work as ``FNAL/MILC.")  This work
uses Asqtad-improved staggered sea quarks and light ($u,d,s$) valence
quarks and the Fermilab action for the charm quarks, with a single
lattice spacing of $a \approx 0.12$ fm.  At this lattice spacing, the
staggered taste splittings are still fairly large, and the minimum RMS
pion mass is $\approx 510$~MeV.  This calculation renormalizes the
vector current using a mostly nonperturbative approach, such that the
perturbative truncation error is expected to be negligible compared to
other systematics.  The Fermilab Lattice and MILC Collaborations
present results for the $D \to \pi \ell \nu$ and $D \to K \ell \nu$
semileptonic form factors over the full kinematic range, rather than
just at zero momentum transfer.  In fact, the publication of this
result predated the precise measurements of the $D\to K \ell\nu$ decay
width by the FOCUS~\cite{Link:2004dh} and Belle
experiments~\cite{Abe:2005sh}, and predicted the shape of
$f_+^{DK}(q^2)$ quite accurately.  This bolsters confidence in
calculations of the $B$-meson semileptonic decay form factors using
the same methodology.  Work is in progress~\cite{Bailey:2012sa} to
reduce both the statistical and systematic errors in $f_+^{D\pi}(q^2)$
and $f_+^{DK}(q^2)$ through increasing the number of configurations
analysed, simulating with lighter pions, and adding lattice spacings
as fine as $a \approx 0.045$~fm.  In parallel, the Fermilab Lattice
and MILC collaborations are initiating a new calculation of
$D \to \pi \ell \nu$ and $D \to K \ell \nu$ using the HISQ action for
all valence and sea quarks~\cite{Gamiz:2012fh}; this calculation will
focus on obtaining the form factors at zero momentum transfer using
the scalar form-factor method~\cite{Na:2010uf} to avoid the need for
current renormalization and (partially) twisted boundary
conditions~\cite{Sachrajda:2004mi,Bedaque:2004ax} to simulate directly
at $q^2=0$.

\begin{table}[h]
\begin{center}
\mbox{} \\[3.0cm]
\footnotesize
\begin{tabular*}{\textwidth}[l]{l @{\extracolsep{\fill}} r l l l l l l l c c}
Collaboration & Ref. & $\Nf$ & 
\hspace{0.15cm}\begin{rotate}{60}{publication status}\end{rotate}\hspace{-0.15cm} &
\hspace{0.15cm}\begin{rotate}{60}{continuum extrapolation}\end{rotate}\hspace{-0.15cm} &
\hspace{0.15cm}\begin{rotate}{60}{chiral extrapolation}\end{rotate}\hspace{-0.15cm}&
\hspace{0.15cm}\begin{rotate}{60}{finite volume}\end{rotate}\hspace{-0.15cm}&
\hspace{0.15cm}\begin{rotate}{60}{renormalization}\end{rotate}\hspace{-0.15cm}  &
\hspace{0.15cm}\begin{rotate}{60}{heavy-quark treatment}\end{rotate}\hspace{-0.15cm}  &
$f_+^{D\pi}(0)$ & $f_+^{DK}(0)$\\
&&&&&&&&& \\[-0.1cm]
\hline
\hline
&&&&&&&&& \\[-0.1cm]
%
HPQCD 11 & \cite{Na:2011mc} & 2+1 & \gA  & \soso & \soso & \good & \good &  \okay & 0.666(29) &\\[0.5ex]
HPQCD 10B & \cite{Na:2010uf} & 2+1 & \gA  & \soso & \soso & \good & \good &  \okay & & 0.747(19)  \\[0.5ex]
FNAL/MILC 04 & \cite{Aubin:2004ej} & 2+1 & \gA  & \tbr & \tbr & \good & \soso & \okay & 0.64(3)(6)& 0.73(3)(7)
\\[0.5ex]
&&&&&&&&& \\[-0.1cm]
\hline\\[0.5ex]
ETM 11B & \cite{DiVita:2011py} & 2 & \rC  & \soso & \soso & \good & \good &  \okay & 0.65(6)(6) & 0.76(5)(5)\\[0.5ex]
&&&&&&&&& \\[-0.1cm]
\hline
\hline
\end{tabular*}
\caption{$D \to \pi\ell\nu$ and $D\to K\ell\nu$ semileptonic form factors at zero momentum transfer.\label{tab_DtoPiKsumm2}}
\end{center}
\end{table}

The most precise published calculations of the
$D \to \pi \ell \nu$~\cite{Na:2011mc} and $D \to
K \ell \nu$~\cite{Na:2010uf} form factors are by the HPQCD
Collaboration.  These analyses also use the $N_f = 2+1$
Asqtad-improved staggered MILC configurations at two lattice spacings
$a \approx 0.09$ and 0.12~fm, but use the HISQ action for the valence
$u,d,s$, and $c$ quarks.  In these mixed-action calculations, the HISQ
valence light-quark masses are tuned so that the ratio $m_l/m_s$ is
approximately the same as for the sea quarks; the minimum RMS sea-pion
mass is $\approx 390$~MeV.  They calculate the form factors at zero
momentum transfer by relating them to the matrix element of the scalar
current, which is not renormalized. They use the ``modified
$z$-expansion'' to simultaneously extrapolate to the physical
light-quark masses and continuum and interpolate to $q^2 = 0$, and
allow the coefficients of the series expansion to vary with the light-
and charm-quark masses.  The form of the light-quark dependence is
inspired by $\chi$PT, and includes logarithms of the form $m_\pi^2
{\rm log} (m_\pi^2)$ as well as polynomials in the valence-, sea-, and
charm-quark masses.  Polynomials in $E_{\pi(K)}$ are also included to
parameterize momentum-dependent discretization errors. The
coefficients of each term are constrained using Gaussian priors with
widths inspired by $\chi$PT power counting for the light-quark mass
terms and by HISQ power-counting for the others.
The number of terms is increased until the result for $f_+(0)$
stabilizes, such that the quoted fit error for $f_+(0)$ includes both
statistical uncertainties and those due to most systematics.  The
largest uncertainties in these calculations are from statistics and
charm-quark discretization errors.

The HPQCD Collaboration is now extending their work on $D$-meson
semileptonic form factors to determining their shape over the full
kinematic range~\cite{Koponen:2011ev}, and recently obtained results
for the $D \to K \ell \nu$ form factors $f+(q^2)$ and
$f_0(q^2)$~\cite{Koponen:2012di}.  This analysis uses a subset of the
ensembles included in their earlier work, with two sea-quark masses at
$a \approx 0.12$~fm and one sea-quark mass at $a \approx 0.09$~fm, but
with approximately three times more statistics on the coarser
ensembles and ten times more statistics on the finer ensemble.  As
above, the scalar current is not renormalized.  The spatial vector
current renormalization factor is obtained by requiring that
$f_+(0)^{H\to H} = 1$ for $H = D, D_s, \eta_s$, and $\eta_c$.  The
renormalization factors for the flavour-diagonal currents agree for
different momenta as well as for charm-charm and strange-strange
external mesons within a few percent, and are then used to renormalize
the flavour-changing charm-strange and charm-light currents.  The
charm-strange temporal vector current is normalized by matching to the
scalar current $f_0(q^2_{\rm max})$.  Also as above, they
simultaneously extrapolate to the physical light-quark masses and
continuum and interpolate/extrapolate in $q^2$ using the modified
$z$-expansion.  In this case, however, they only allow for light-quark
mass and lattice-spacing dependence in the series coefficients, but
not for charm-quark mass or kaon energy dependence, and constrain the
parameters with Bayesian priors.  It is not clear, however, that only
three sea-quark ensembles at two lattice spacings are sufficient to
resolve the quark-mass and lattice spacing dependence, even within the
context of constrained fitting.  The quoted error in the zero-recoil
form factor $f_+(0) = 0.745(11)$ is significantly smaller than in
their 2010 work, but we are unable to understand the sources of this
improvement with the limited information provided in
Ref.~\cite{Koponen:2012di}. The preprint does not provide an error
budget, nor any information on how the systematic uncertainties are
estimated. Thus we cannot rate this calculation, and do not include it
in the summary table and plot.

Table~\ref{tab_DtoPiKsumm2} summarizes the existing $N_f =2$ and $N_f
= 2+1$ calculations of the $D \to \pi \ell \nu$ and $D \to K \ell \nu$
semileptonic form factors.  The quality of the systematic error
studies is indicated by the symbols.  Additional tables in
appendix~\ref{app:DtoPi_Notes} provide further details on the
simulation parameters and comparisons of the error estimates.  Recall
that only calculations without red tags that are published in a
refereed journal are included in the FLAG average.  Of the
calculations described above, only those of HPQCD~10B,11 satisfy all
of the quality criteria.  Therefore our average of the
$D \to \pi \ell \nu$ and $D \to K \ell \nu$ semileptonic form factors
from $N_f = 2+1$ lattice QCD is
\begin{equation}
	N_f=2+1: \qquad f_+^{D\pi}(0) =  0.666(29) \,, \qquad f_+^{DK}(0) = 0.747(19)  \,. \label{eq:Nf=2p1Dsemi}
\end{equation}
Figure~\ref{fig:DtoPiK} plots the existing $N_f =2$ and $N_f = 2+1$
results for $f_+^{D\pi}(0)$ and $f_+^{DK}(0)$; the grey bands show our
average of these quantities.  Section~\ref{sec:Vcd} discusses the
implications of these results for determinations of the CKM matrix
elements $|V_{cd}|$ and $|V_{cs}|$ and tests of unitarity of the
second row of the CKM matrix.

\begin{figure}[h]
\begin{center}
\includegraphics[width=0.7\linewidth]{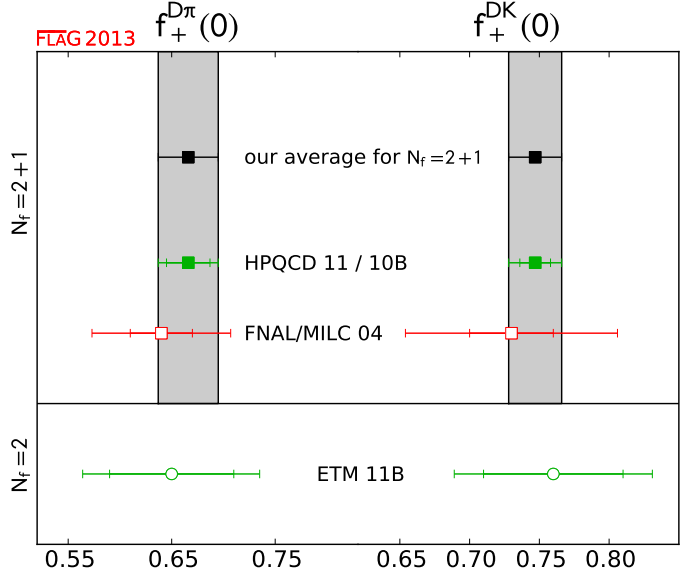}

\vspace{-2mm}
\caption{$D\to\pi \ell\nu$ and $D\to K\ell\nu$ semileptonic form
  factors at zero momentum transfer. The HPQCD result for
  $f_+^{D\pi}(0)$ is from HPQCD 11, the one for $f_+^{DK}(0)$
  represents HPQCD 10B (see Table
  \ref{tab_DtoPiKsumm2}). \label{fig:DtoPiK}}
 \end{center}
\end{figure}

\subsection{Determinations of $|V_{cd}|$ and $|V_{cs}|$ and test of  second-row CKM unitarity}
\label{sec:Vcd}

We now interpret the lattice-QCD results for the $D_{(s)}$ meson decay
constants and semileptonic form factors as determinations of the CKM
matrix elements $|V_{cd}|$ and $|V_{cs}|$ in the Standard Model.

For the leptonic decays, we use the latest experimental averages from
Rosner and Stone for the Particle Data Group~\cite{Rosner:2012np}
(where electromagnetic corrections of $\sim 1\%$ have been removed):
\begin{equation}
	f_D |V_{cd}| = 46.40(1.98)~{\rm MeV} \,, \qquad f_{D_s} |V_{cs}| = 253.1(5.3)~{\rm MeV} \,.
\end{equation}
We combine these with the average values of $f_D$ and $f_{D_s}$ from
the individual $N_f = 2$ and $N_f = 2+1$ lattice-QCD calculations that
satisfy the FLAG criteria, and summarize the results for the CKM
matrix elements $|V_{cd}|$ and $|V_{cs}|$ in
Table~\ref{tab:VcdVcsIndividual}.  For our preferred values we use the
averaged $N_f=2$ and $N_f = 2+1$ results for $f_D$ and $f_{D_s}$ in
Eqs.~(\ref{eq:Nf2av}) and~(\ref{eq:Nf3av}).  We obtain
\begin{eqnarray}
	|V_{cd}| = 0.2218(35)(95) \,, \quad  |V_{cs}| = 1.018 (11)(21) \,, \quad && ({\rm leptonic~decays}, N_f=2+1) \, \\
	|V_{cd}| = 0.2231(95)(75) \,, \quad  |V_{cs}| =  1.012(21)(28) \,, \quad && ({\rm leptonic~decays}, N_f=2) \,
\end{eqnarray}
where the errors shown are from the lattice calculation and experiment
(plus non-lattice theory), respectively.  For the $N_f = 2+1$
determinations, the uncertainties from the lattice-QCD calculations of
the decay constants are two to three times smaller than the
experimental uncertainties in the branching fractions; the lattice
central values and errors are dominated by those of the HPQCD
calculations.  Although the $N_f=2$ and $N_f = 2+1$ results for
$|V_{cs}|$ are slightly larger than one, they are both consistent with
unity within errors.

\begin{table}[tb]
\begin{center}
\noindent
\begin{tabular*}{\textwidth}[l]{@{\extracolsep{\fill}}lrlcr}
Collaboration & Ref. &$\Nf$&from&\rule{0.8cm}{0cm}$|V_{cd}|$ or $|V_{cs}|$\\
&&&& \\[-2ex]
\hline \hline &&&&\\[-2ex]
HPQCD 12A & \cite{Na:2012iu} & 2+1 & $f_{D}$  & 0.2228(36)(95) \\
FNAL/MILC 11  & \cite{Bazavov:2011aa} & 2+1 & $f_{D}$  &  0.2120(109)(91)  \\
HPQCD 11 & \cite{Na:2011mc} & 2+1 & $D \to \pi \ell \nu$  & 0.2192(95)(45) \\
ETM 13B  & \cite{Carrasco:2013zta} & 2 & $f_{D}$  &  0.2231(95)(75)  \\
&&&& \\[-2ex]
 \hline
&&&& \\[-2ex]
HPQCD 10A & \cite{Davies:2010ip} & 2+1 & $f_{D_s}$  & 1.021(10)(21)  \\
FNAL/MILC 11 & \cite{Bazavov:2011aa} & 2+1 & $f_{D_s}$  &  0.9731(404)(202) \\
HPQCD 10B & \cite{Na:2010uf} & 2+1 & $D \to K \ell \nu$  & 0.9746(248)(67) \\
ETM 13B & \cite{Carrasco:2013zta} & 2 & $f_{D_s}$  &  1.012(21)(28) \\
&&&& \\[-2ex]
 \hline \hline 
\end{tabular*}
\caption{Determinations of $|V_{cd}|$ (upper panel) and $|V_{cs}|$
  (lower panel) obtained from lattice calculations of $D$-meson
  leptonic decay constants and semileptonic form factors. The errors
  shown are from the lattice calculation and experiment (plus
  non-lattice theory), respectively. \label{tab:VcdVcsIndividual}}
\end{center}
\end{table}

For the semileptonic decays, we use the latest experimental averages
from the Heavy Flavour Averaging Group~\cite{Amhis:2012bh}:\footnote{We
note that HFAG currently averages results for neutral and charged $D$
meson decays without first removing the correction due to the Coulomb
attraction between the charged final-state particles for the neutral
$D$ meson decays.}
\begin{equation}
	f_+^{D\pi}(0) |V_{cd}| = 0.146(3) \,, \qquad f_+^{DK}(0) |V_{cs}| = 0.728(5)  \,.
\end{equation}
For each of $f_+^{D\pi}(0)$ and $f_+^{DK}(0)$, there is only a single
$N_f = 2+1$ lattice-QCD calculation that satisfies the FLAG criteria.
Using these results, which are given in Eq.~(\ref{eq:Nf=2p1Dsemi}), we
obtain our preferred values for $|V_{cd}|$ and $|V_{cs}|$:
\begin{eqnarray}
	|V_{cd}| = 0.2192(95)(45)  \,, \quad  |V_{cs}| = 0.9746(248)(67)   \,, \quad  ({\rm semileptonic~decays}, N_f=2+1) 
\end{eqnarray}
where the errors shown are from the lattice calculation and experiment
(plus non-lattice theory), respectively.

Table~\ref{tab:VcdVcsSummary} summarizes the results for $|V_{cd}|$
and $|V_{cs}|$ from leptonic and semileptonic decays, and compares
them to determinations from neutrino scattering (for $|V_{cd}|$ only)
and CKM unitarity.  These results are also plotted in
Fig.~\ref{fig:VcdVcs}.  The determinations of $|V_{cd}|$ all agree
within uncertainties, but the errors in the direct determinations from
leptonic and semileptonic decays are approximately ten times larger
than the indirect determination from CKM unitarity.  The determination
of $|V_{cs}|$ from $N_f = 2+1$ lattice-QCD calculations of leptonic
decays is noticeably larger than that from both semileptonic decays
and CKM unitarity.  The disagreement between $|V_{cs}|$ from leptonic
and semileptonic decays is slight (only 1.2$\sigma$ assuming no
correlations), but the disagreement between $|V_{cs}|$ from leptonic
decays and CKM unitarity is larger at 1.9$\sigma$.  This tension with
CKM unitarity is driven primarily by the HPQCD calculation of
$f_{D_s}$ in Ref.~\cite{Davies:2010ip}, but we note that the ETM
$N_f=2$ calculation of $f_{D_s}$ in Ref.~\cite{Carrasco:2013zta} leads
to the same high central value of $|V_{cs}|$, just with larger
uncertainties.  Further, the recent preliminary lattice-QCD
calculation of $f_{D_s}$ using $N_f = 2+1+1$ configurations with
dynamical HISQ quarks by Fermilab/MILC~\cite{Bazavov:2013nfa} agrees
with the HPQCD result and quotes smaller uncertainties due to the
inclusion of data at the physical light-quark mass, so it will be
interesting to see how this tension evolves with improved experimental
measurements and more independent lattice-QCD results with competitive
errors.

The $N_f=2+1$ averages for $|V_{cd}|$ and $|V_{cs}|$ in
Fig.~\ref{fig:VcdVcs} are obtained by averaging the results in
Table~\ref{tab:VcdVcsIndividual} including correlations.  We assume
that the statistical errors are 100\% correlated between all of the
calculations because they use the MILC Asqtad gauge configurations.
We also assume that the heavy-quark discretization errors are 100\%
correlated between the HPQCD calculations of leptonic and semileptonic
decays because they use the same charm-quark action, and that the
scale-setting uncertainties are 100\% correlated between the HPQCD
results as well.  Finally, we include the 100\% correlation between
the experimental inputs for the two extractions of $|V_{cd(s)}|$ from
leptonic decays.  We obtain
\begin{eqnarray}
	|V_{cd}| = 0.2191(83) \,, \quad |V_{cs}| = 0.996(21)  \,, \quad  ({\rm our\ average}, N_f=2+1) \label{eq:VcdsNf2p1}
\end{eqnarray}
where the errors include both theoretical and experimental
uncertainties, and the error on $|V_{cs}|$ has been increased by
$\sqrt{\chi^2/{\rm dof}}=1.03$.

Using the determinations of $|V_{cd}|$ and $|V_{cs}|$ in
Eq.~(\ref{eq:VcdsNf2p1}), we can test the unitarity of the second row
of the CKM matrix.  We obtain
\begin{equation}
	|V_{cd}|^2 + |V_{cs}|^2 + |V_{cb}|^2 - 1 = 0.04(6) \,
\end{equation}
which agrees with the Standard Model at the percent level.  Given the
current level of precision, this result does not depend on the value
used for $|V_{cb}|$, which is of ${\mathcal{O}}(10^{-2})$ [see
Eq.~(\ref{eq:VcbNf2p1})].

\begin{table}[tb]
\begin{center}
\noindent
\begin{tabular*}{\textwidth}[l]{@{\extracolsep{\fill}}lcrcc}
& from & Ref. &\rule{0.8cm}{0cm}$|V_{cd}|$ & \rule{0.8cm}{0cm}$|V_{cs}|$\\
&& \\[-2ex]
\hline \hline &&\\[-2ex]
$N_f = 2+1$&  $f_D$ \& $f_{D_s}$ && 0.2218(101) & 1.018(24) \\
$N_f = 2$ &  $f_D$ \& $f_{D_s}$ && 0.2231(121) & 1.012(35) \\
&& \\[-2ex]
 \hline
&& \\[-2ex]
$N_f = 2+1$ & $D \to \pi \ell\nu$ and $D\to K \ell\nu$ && 0.2192(105) & 0.9746(257) \\
&& \\[-2ex]
 \hline
&& \\[-2ex]
PDG & neutrino scattering & \cite{Beringer:1900zz} & 0.230(11)&  \\
Rosner 12 ({\it for the} PDG) & CKM unitarity & \cite{Rosner:2012np} & 0.2245(12) & 0.97345(22) \\
&& \\[-2ex]
 \hline \hline 
\end{tabular*}
\caption{Comparison of determinations of $|V_{cd}|$ and $|V_{cs}|$
  obtained from lattice methods with non-lattice determinations and
  the Standard Model prediction assuming CKM
  unitarity. \label{tab:VcdVcsSummary}}
\end{center}
\end{table}

\begin{figure}[h]

\begin{center}
\includegraphics[width=0.7\linewidth]{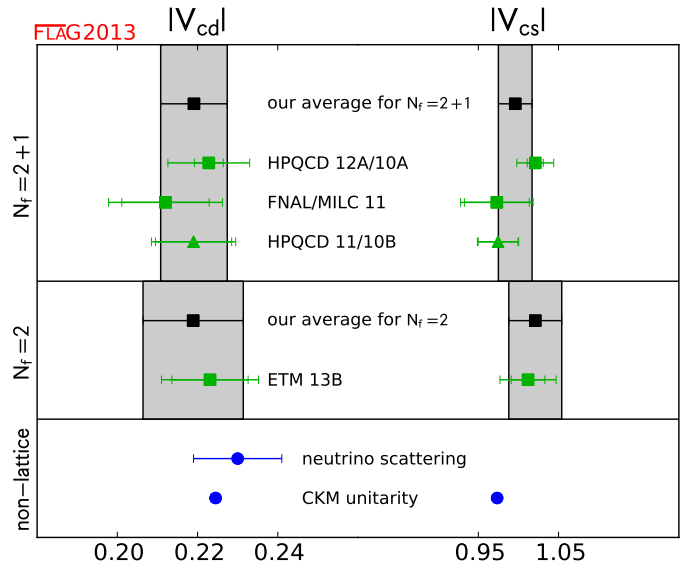}

\vspace{-2mm}
\caption{Comparison of determinations of $|V_{cd}|$ and $|V_{cs}|$
  obtained from lattice methods with non-lattice determinations and
  the Standard Model prediction based on CKM unitarity.  When two
  references are listed on a single row, the first corresponds to the
  lattice input for $|V_{cd}|$ and the second to that for $|V_{cs}|$.
  The results denoted by squares are from leptonic decays, while those
  denoted by triangles are from semileptonic
  decays.  \label{fig:VcdVcs}}
\end{center}
\end{figure}
\clearpage

\clearpage
\section{$B$-meson decay constants, mixing parameters and form factors}
\label{sec:BDecays}

Leptonic and semileptonic decays of bottom $B$ and $B_s$ mesons probe
the quark-flavour changing transitions $b \to u$ and $b \to c$.
Tree-level semileptonic $B$ decays with light charged leptons ($\ell =
e,\mu$) in the final state, such as $B \to\pi\ell\nu$ and $B \to
D^{(*)}\ell\nu$, enable determinations of the CKM matrix elements
$|V_{ub}|$ and $|V_{cb}|$ within the Standard Model.  Semileptonic $B$
decays that occur via loops in the Standard Model, such as $B\to
K^{(*)} \ell^+ \ell^-$, provide sensitive probes of physics
beyond-the-Standard Model because contributions from new heavy
particles in the loops may be comparable to the Standard Model
``background."  Further, because $B$ mesons are sufficiently massive,
they can decay to final states involving $\tau$-leptons.  Tree-level
decays such as $B\to\tau \nu$ and $B\to D^{(*)}\tau\nu$ are promising
new-physics search channels because they can receive significant
contributions from charged-Higgs bosons.

Mixing of neutral $B^0_d$ and $B^0_s$ mesons occurs in the Standard
Model via one-loop box diagrams containing up-type quarks ($u,c,t$)
and charged $W$ bosons.  Because the Standard Model contributions are
proportional to the CKM factors $|V_{u(c,t)q}V_{u(c,t)b}^*|^2$ (where
$q=d,s$) and the quark masses $m_{u(c,t)}^2$, neutral $B$-meson mixing
is dominated by intermediate top quarks.  Thus experimental
measurements of the neutral $B^0_{d(s)}$-meson oscillation
frequencies, $\Delta M_{d(s)}$ combined with sufficiently precise
theoretical calculations of the hadronic mixing matrix elements (often
presented as dimensionless ``bag" parameters), enable the
determination of the CKM matrix elements $|V_{td}|$ and $|V_{ts}|$
within the Standard Model.  Conversely, neutral $B$-meson mixing
places stringent constraints on the scale of generic new heavy
particles that can enter the loops in beyond-the-Standard Model
scenarios.  Finally, neutral meson mixing is also sensitive to the
phase of the CKM matrix $(\rho, \eta)$.  Thus the ratio of oscillation
frequencies $\Delta M_d / \Delta M_s$ places a tight constraint on the
apex of the CKM unitarity triangle that is complementary to those from
other observables.

Lattice-QCD calculations of $b$ quarks have an added complication not
present for charm and light quarks: at the lattice spacings that are
currently used in numerical simulations, the $b$ quark mass is of
order one in lattice units. Therefore a direct treatment of $b$ quarks
with the fermion actions commonly used for light quarks will result in
large cutoff effects, and all current lattice-QCD calculations of $b$
quark quantities make use of effective field theory at some stage.
The two most widely used general approaches for lattice $b$ quarks are
(i) direct application of effective field theory treatments such as
HQET or NRQCD, which allow for a systematic expansion in $1/m_b$; or
(ii) the interpretation of a relativistic quark action in a manner
suitable for heavy quarks using an extended Symanzik improvement
program to suppress cutoff errors.  This introduces new systematic
uncertainties that are not present in light-quark calculations, either
from truncation of the effective theory, or from more complicated
lattice-spacing dependence.  Further, because with these approaches
the light and bottom quarks are simulated with different fermion
actions, it is in general not possible to construct absolutely
normalized bottom-light currents; this leads to systematic
uncertainties due to matching the lattice operators to the continuum
that can be significant. A third approach is to use an improved
light-quark action to calculate the quantity of interest over a range
of heavy-quark masses with $am_h < 1$, and then use heavy-quark
effective theory and/or knowledge of the static limit to extrapolate
or interpolate to the physical $b$-quark mass. Such methods can avoid
some of the aforementioned complications, but require simulations at
very small lattice spacings in order to keep discretization errors
under control. Appendix~\ref{app:HQactions} reviews the methods used
to treat $b$ quarks on the lattice in more detail.

Here we summarize the status of lattice-QCD calculations of the bottom
leptonic decay constants, neutral meson mixing parameters, and
semileptonic form factors. We limit our review to results based on
modern simulations with reasonably light pion masses (below
approximately 500~MeV). This excludes results obtained from the
earliest unquenched simulations, which typically had two flavours in
the sea, and which were limited to heavier pion masses because of the
constraints imposed by the computational resources and methods
available at that time.  Fewer collaborations have presented results
for these quantities than for the light-quark sector ($u$, $d$, $s$),
and the calculations tend to be on coarser lattice spacings with
heavier pions.  Therefore, for some quantities, there is only a single
lattice calculation that satisfies the criteria to be included in our
average.  Several collaborations, however, are currently pursuing the
needed matrix-element calculations with different lattice $b$-quark
actions, finer lattice spacings, and lighter pions, so we expect the
appearance of many new results with controlled errors in the next year
or two.

We also note that the heavy-quark methods discussed in this review
have been validated in a number of ways.  Because several groups use
the same action for charm and bottom quarks, tests of such methods
with charm quarks are relevant for $B$ physics results, and are
therefore included in the following discussion.  Calculations of
hadron masses with one or more heavy (charm or bottom) valence quark
provide phenomenological tests of the heavy-quark action.  Such
calculations have been performed with NRQCD, HQET, Fermilab, RHQ,
Tsukuba, HISQ, Overlap, twisted-mass Wilson, and other ${\cal O}(a)$
improved Wilson heavy quarks for the hyperfine splittings in the
$D_{(s)}$ and $B_{(s)}$ meson systems
\cite{Davies:2010ip,Bernard:2010fr,Gregory:2010gm,Namekawa:2011wt,Blossier:2011dk,McNeile:2011ng,Bali:2011dc,Mohler:2011ke,Dowdall:2012ab,Bernardoni:2012ti,Basak:2012py,Bali:2012ua,Moir:2013ub,Kalinowski:2013wsa},
and for the low-lying charmonium
\cite{Burch:2009az,Namekawa:2011wt,Bali:2011dc,Mohler:2011ke,DeTar:2012xk,Donald:2012ga,Becirevic:2013gy,Basak:2012py,Liu:2012ze},
bottomonium
\cite{Gray:2005ur,Meinel:2009rd,Meinel:2010pv,Dowdall:2011wh,Daldrop:2011aa,Aoki:2012xaa,Lewis:2012ir},
and $B_c$
\cite{Allison:2004be,Gregory:2009hq,Gregory:2010gm,Dowdall:2012ab,McNeile:2012qf}
systems.  All of them are in good agreement with experimental
measurements.  Hyperfine splittings are sensitive to higher-order
terms in the heavy-quark action and therefore provide particularly
good tests of such terms.  The comparison of lattice-QCD calculations
of hadronic matrix elements for leptonic and radiative decays in
charmonium \cite{Donald:2012ga,McNeile:2012qf} with experimental
measurements provides CKM-free tests of heavy-HISQ currents.  The
comparison of lattice-QCD calculations of the shape of the
semileptonic form factors for $D \to \pi (K) \ell \nu$
\cite{Aubin:2004ej} with experimental measurements provides CKM
independent tests of charm-quark currents with the Fermilab action.
In two of the above mentioned tests, the lattice-QCD calculations were
predictions, in one case predating the experimental discovery of the
$B_c$ mass, and in the other predating experimental measurements of
the shape of the semilleptonic $D$-meson form factors with comparable
precision.  Truncation errors in HQET have been studied by comparing
simulations of the effective field theory with corresponding quenched
simulations using a nonperturbatively improved Wilson action with
heavy quark masses in the charm-mass region in large
volumes~\cite{Blossier:2010mk} and up to the $b$-quark mass in small
volumes~\cite{Heitger:2004gb}.  Moreover, the consistency between
independent determinations of the bottom
\cite{Blossier:2009hg,McNeile:2010ji,Dimopoulos:2011gx,Blossier:2011dk,Bernardoni:2012ti,Carrasco:2012de,Lee:2013mla}
and charm
\cite{Allison:2008xk,Davies:2009ih,Blossier:2010cr,McNeile:2010ji,Jansen:2011vr,Durr:2011ns,Namekawa:2011wt}
quark masses using NRQCD, HQET, Tsukuba, HISQ, twisted-mass Wilson,
and other ${\cal O}(a)$ improved Wilson heavy quarks, as well as their
agreement with non-lattice determinations
 \cite{Beringer:1900zz} further validate lattice heavy-quark methods.

Following our review of lattice-QCD calculations of $B_{(s)}$-meson
leptonic decay constants, neutral meson mixing parameters, and
semileptonic form factors, we then interpret our results within the
context of the Standard Model.  We combine our best-determined values
of the hadronic matrix elements with the most recent
experimentally-measured branching fractions to obtain $|V_{(u)cb}|$
and compare these results to those obtained from inclusive
semileptonic $B$ decays.

\subsection{Leptonic decay constants $f_B$ and $f_{B_s}$}
\label{sec:fB}

The $B$ and $B_s$ meson decay constants are relevant for decays of
charged $B$-mesons to a lepton-neutrino pair via the charged current
interaction, as well as for rare leptonic decays of neutral $B_{d(s)}$
mesons to a charged-lepton pair via a flavour-changing neutral current
(FCNC) interaction.

In the Standard Model the decay rate for $B^+ \to \ell^+ \nu_{\ell}$
is given by a formula identical to the one for $D$ decays in
Eq.~(\ref{eq:Dtoellnu}) but with $D_{(s)}$ replaced by $B$ and the
relevant CKM matrix element $V_{cq}$ replaced by $V_{ub}$:
\be
\Gamma ( B \to \ell \nu_{\ell} ) =  \frac{ m_B}{8 \pi} G_F^2  f_B^2 |V_{ub}|^2 m_{\ell}^2 
           \left(1-\frac{ m_{\ell}^2}{m_B^2} \right)^2 \;. \label{eq:B_leptonic_rate}
\ee
The only charged-current $B$ meson decay that has been observed so far
is $B \to \tau \nu_{\tau}$, which has been measured by the Belle and
Babar collaborations with a combined precision of
20\% \cite{Beringer:1900zz}. This measurement can therefore be used to
determine $|V_{ub}|$ when combined with lattice-QCD predictions of the
corresponding decay constant.

The decay of a neutral $B_{d(s)}$ meson to a charged lepton pair 
is  loop-suppressed in the Standard Model. The corresponding expression for the 
branching fraction has the  form 
\be
B ( B_q \to \ell^+ \ell^-) = \tau_{B_q} \frac{G_F^2}{\pi} \, Y \,
\left(  \frac{\alpha}{4 \pi \sin^2 \Theta_W} \right)^2
m_{B_q} f_{B_q}^2 |V_{tb}^*V_{tq}|^2 m_{\ell}^2 
           \sqrt{1- 4 \frac{ m_{\ell}^2}{m_B^2} }\;,
\ee
where the light quark $q=s$ or $d$, and the loop function $Y$ includes
 NLO QCD and electro-weak corrections \cite{Buchalla:1993bv}. Evidence
 for $B_s \to \mu^+ \mu^-$ decay was recently seen at LHCb at the
 $3.5 \sigma$ level, with a branching fraction of
 $BR(B_s \to \mu^+ \mu^-) =
 (3.2^{+1.5}_{-1.2}) \,10^{-9}$ \cite{Aaij:2012nna}.

The decay constants $f_{B_q}$ (with $q=u,d,s$) parameterize the matrix
elements of the corresponding axial-vector currents, $A^{\mu}_{bq}
= \bar{b}\gamma^{\mu}\gamma^5q$, analogously to the definition of
$f_{D_q}$ in section~\ref{sec:fD}:
\be
\langle 0| A^{\mu} | B_q(p) \rangle = p_B^{\mu} f_{B_q} \;.
\ee
For heavy-light mesons, it is convenient to define and analyse the quantity 
\be
 \Phi_{B_q} \equiv f_{B_q} \sqrt{m_{B_q}} \;,
\ee
which approaches a constant (up to logarithmic corrections) in the
$m_B \to \infty$ limit.
In the following discussion we denote lattice data for $\Phi$($f$)
 obtained at a heavy quark mass $m_h$ and light valence-quark mass
 $m_{\ell}$ as $\Phi_{h\ell}$($f_{hl}$), to differentiate them from
 the corresponding quantities at the physical $b$ and light-quark
 masses.

The SU(3) breaking ratio $f_{B_s}/f_B$ is an interesting quantity to
study with lattice QCD, since most systematic errors partially cancel
in this ratio, including discretization errors, heavy-quark mass
tuning effects, and renormalization errors, among others. The SU(3)
breaking ratio is however sensitive to the chiral extrapolation. So
one can, in principle, combine a lattice-QCD calculation of the SU(3)
breaking ratio that includes a careful study of the chiral
extrapolation, with a different lattice-QCD calculation of $f_{B_s}$
(which is relatively insensitive to chiral extrapolation errors) that
includes a careful study of all other systematic errors to obtain a
more precise result for $f_B$ than would be possible from either
lattice-QCD calculation alone.  Indeed, this strategy is used by both
the ETM and HPQCD collaborations, as described below.

A number of different heavy-quark formulations are being used to
obtain results for $B_{q}$ meson decay constants from numerical
simulations with $N_f=2$, $N_f=2+1$, and $N_f=2+1+1$ sea quarks. They
are summarized in Tables~\ref{tab:FBssumm}~and~\ref{tab:FBratsumm} and
in Figure~\ref{fig:fB}. Additional details about the underlying
simulations and systematic error estimates are given in
Appendix~\ref{app:fB_Notes}.

\begin{table}[htb]
\begin{center}
\mbox{} \\[3.0cm]
\footnotesize
\begin{tabular*}{\textwidth}[l]{l @{\extracolsep{\fill}} r l l l l l l l l l l l l}
Collaboration \al Ref. \al $\Nf$ \al
\hspace{0.1cm}\begin{rotate}{60}{publication status}\end{rotate}\hspace{-0.1cm} \al
\hspace{0.1cm}\begin{rotate}{60}{continuum extrapolation}\end{rotate}\hspace{-0.1cm} \al
\hspace{0.1cm}\begin{rotate}{60}{chiral extrapolation}\end{rotate}\hspace{-0.1cm} \al
\hspace{0.1cm}\begin{rotate}{60}{finite volume}\end{rotate}\hspace{-0.1cm} \al
\hspace{0.1cm}\begin{rotate}{60}{renormalization/matching}\end{rotate}\hspace{-0.1cm}  \al
\hspace{0.1cm}\begin{rotate}{60}{heavy quark treatment}\end{rotate}\hspace{-0.1cm} \al
 $f_{B^+}$ \al $f_{B^0}$   \al $f_{B}$ \al $f_{B_s}$  \\
&&&&&&&&&&&&\\[-0.1cm]
\hline
\hline
&&&&&&&&&&&& \\[-0.1cm]

ETM 13E \al \cite{Carrasco:2013naa} \al 2+1+1 \al \rC \al \soso \al \soso \al \soso 
\al \soso \al  \okay \al  $-$ \al $-$ \al 196(9) \al 235(9) \\[0.5ex]

HPQCD 13 \al \cite{Dowdall:2013tga} \al 2+1+1 \al \gA \al \good \al \good \al \good \al \soso
\al \okay \al  184(4) \al 188(4) \al186(4) \al 224(5)  \\[0.5ex]

&&&&&&&&&& \\[-0.1cm]
\hline
&&&&&&&&&& \\[-0.1cm]

RBC/UKQCD 13A \al \cite{Witzel:2013sla} \al 2+1 \al \rC \al \soso \al \soso \al \good 
  \al \soso \al \okay \al $-$ \al $-$ \al  191(6)$_{\rm stat}^\diamond$ \al 233(5)$_{\rm stat}^\diamond$ \\[0.5ex]

HPQCD 12 & \cite{Na:2012sp} \al 2+1 \al \gA \al \soso \al \soso \al \good \al \soso
\al \okay \al $-$ \al $-$ \al 191(9) \al 228(10)  \\[0.5ex]

HPQCD 12 \al \cite{Na:2012sp} \al 2+1 \al \gA \al \soso \al \soso \al \good \al \soso
\al \okay \al $-$ \al $-$ \al 189(4)$^\triangle$ \al  $-$  \\[0.5ex]

HPQCD 11A \al \cite{McNeile:2011ng} \al 2+1 \al \gA \al \good \al \soso \al
 \good \al \good \al \okay \al $-$ \al $-$ \al $-$ \al 225(4)$^\nabla$ \\[0.5ex] 

FNAL/MILC 11 \al \cite{Bazavov:2011aa} \al 2+1 \al \gA \al \soso \al \soso \al
     \good \al \soso \al \okay \al 197(9) \al $-$ \al $-$ \al 242(10) \al  \\[0.5ex]  

HPQCD 09 \al \cite{Gamiz:2009ku} \al 2+1 \al \gA \al \soso \al \soso \al \good \al
\soso \al \okay \al $-$ \al $-$ \al 190(13)$^\bullet$ \al 231(15)$^\bullet$  \\[0.5ex] 

\al&&&&&&&&& \\[-0.1cm]
\hline
&&&&&&&&&& \\[-0.1cm]

ALPHA 13 \al \cite{Bernardoni:2013oda} \al 2 \al \rC  \al \good   \al \good   \al
\good    \al\good  \al \okay   \al $-$ \al $-$ \al 187(12)(2) \al  224(13)  \\[0.5ex] 

ETM 13B, 13C \al \cite{Carrasco:2013zta,Carrasco:2013iba} \al 2 \al \oP$^\dagger$ \al \good \al \soso \al \good
\al \soso \al  \okay \al  $-$ \al $-$ \al 189(8) \al 228(8) \\[0.5ex]

ALPHA 12A\al \cite{Bernardoni:2012ti} \al 2 \al \rC  \al \good      \al \good      \al
\good          \al\good  \al \okay   \al $-$ \al $-$ \al 193(9)(4) \al  219(12)   \\[0.5ex] 

ETM 12B \al \cite{Carrasco:2012de} \al 2 \al \rC \al \good \al \soso \al \good
\al \soso \al  \okay \al  $-$ \al $-$ \al 197(10) \al 234(6) \\[0.5ex]

ALPHA 11\al \cite{Blossier:2011dk} \al 2 \al \rC  \al \good      \al \soso      \al
\good          \al\good  \al \okay  \al $-$ \al $-$ \al 174(11)(2) \al  $-$  \\[0.5ex]  

ETM 11A \al \cite{Dimopoulos:2011gx} \al 2 \al \gA \al \soso \al \soso \al \good
\al \soso \al  \okay \al $-$ \al $-$ \al 195(12) \al 232(10) \\[0.5ex]

ETM 09D \al \cite{Blossier:2009hg} \al 2 \al \gA \al \soso \al \soso \al \soso
\al \soso \al  \okay \al $-$ \al $-$ \al 194(16) \al 235(12) \\[0.5ex]
&&&&&&&&&& \\[-0.1cm]
\hline
\hline
\end{tabular*}
\begin{tabular*}{\textwidth}[l]{l@{\extracolsep{\fill}}lllllllll}
  \multicolumn{10}{l}{\vbox{\begin{flushleft} 
	$^\diamond$Statistical errors only. \\
        $^\triangle$Obtained by combining $f_{B_s}$ from HPQCD 11A with $f_{B_s}/f_B$ calculated in this work.\\
        $^\nabla$This result uses one ensemble per lattice spacing with light to strange sea-quark mass 
        ratio $m_{\ell}/m_s \approx 0.2$. \\
        $^\bullet$This result uses an old determination of  $r_1=0.321(5)$ fm from Ref.~\cite{Gray:2005ur} that 
        has since been superseded. \\
        $^\dagger$Update of ETM 11A and 12B. 
\end{flushleft}}}
\end{tabular*}
\vspace{-0.5cm}
\caption{Decay constants of the $B$, $B^+$, $B^0$ and $B_{s}$ mesons (in MeV). Here $f_B$ stands for the mean value of $f_{B^+}$ and $f_{B^0}$, extrapolated (or interpolated) in the mass of the light valence-quark to the physical value of $m_{ud}$.}
\label{tab:FBssumm}
\end{center}
\end{table}

\begin{figure}[htb]
\hspace{-0.8cm}\includegraphics[width=0.58\linewidth]{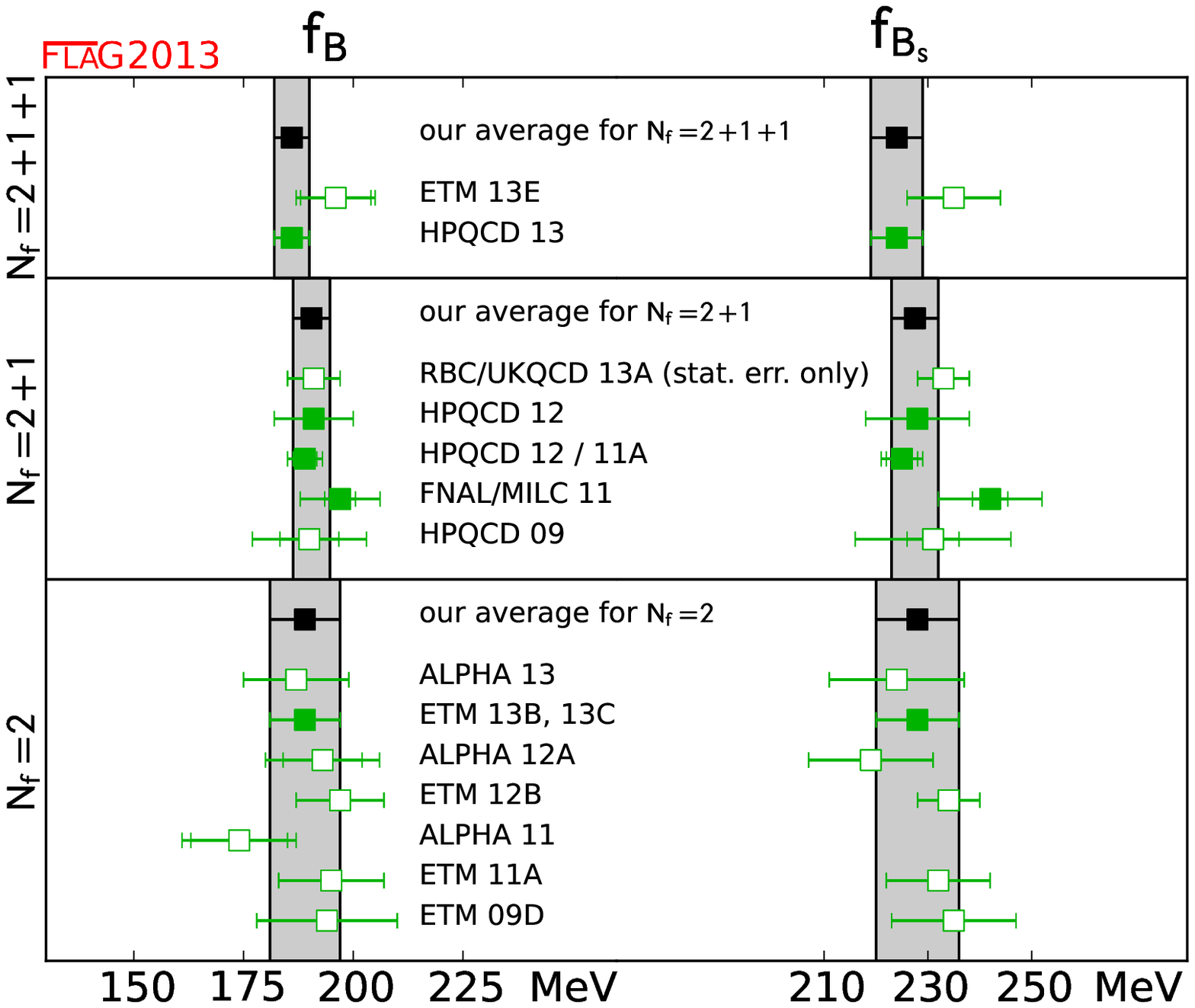}\hspace{-0.95cm}
\includegraphics[width=0.58\linewidth]{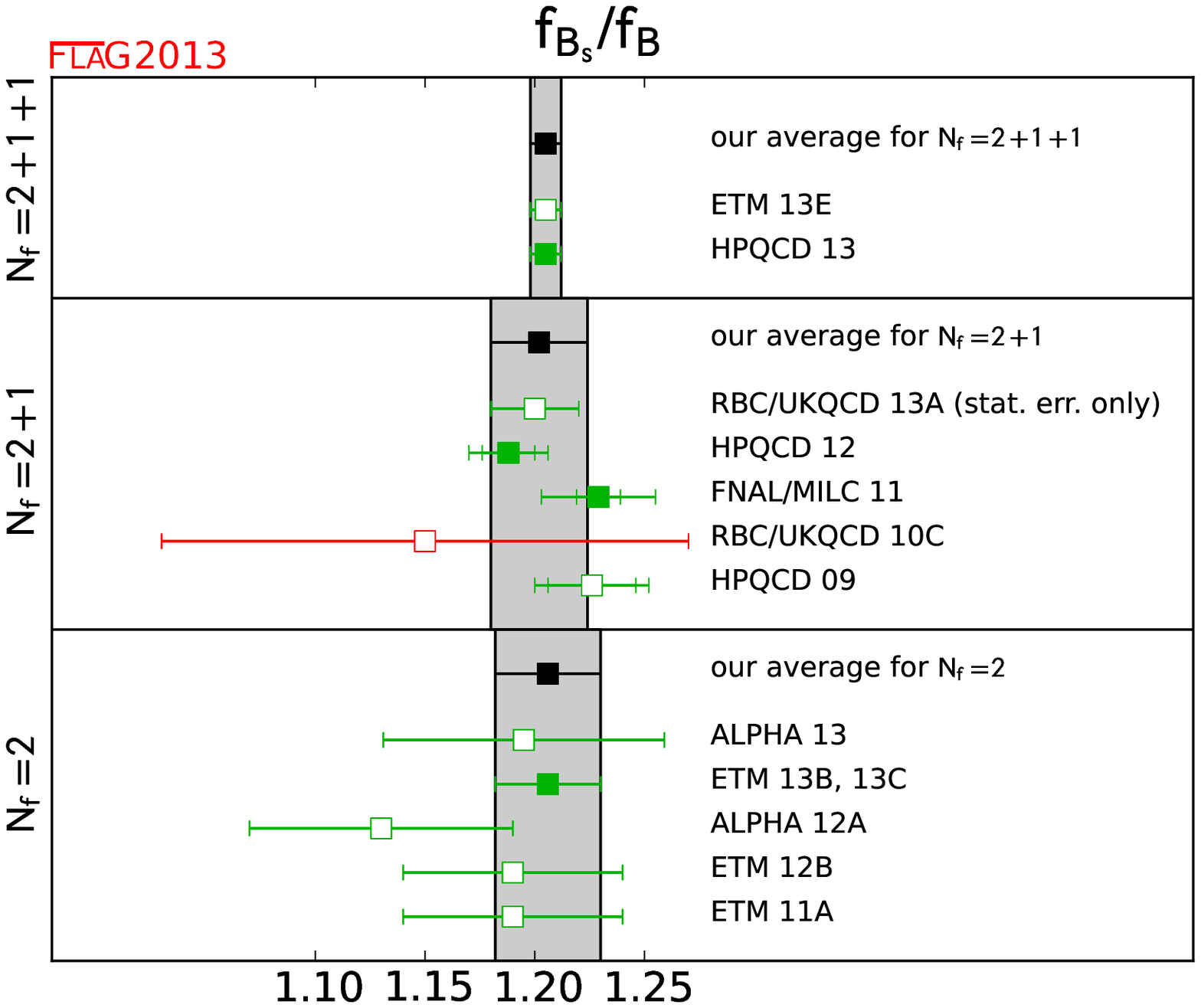}
 
\vspace{-2mm}
\caption{Decay constants of the $B$ and $B_s$ mesons. The values are taken from Table \ref{tab:FBssumm} 
(the $f_B$ entry for FNAL/MILC~11 represents $f_{B^+}$). The
significance of the colours is explained in
section \ref{sec:qualcrit}. The black squares and grey bands indicate
our averages in Eqs.~(\ref{eq:fbav2}), (\ref{eq:fbav21}) and (\ref{eq:fbav211}).}\label{fig:fB}
\end{figure}

\begin{table}[htb]
\begin{center}
\mbox{} \\[3.0cm]
\footnotesize
\begin{tabular*}{\textwidth}[l]{l @{\extracolsep{\fill}} r l l l l l l l l l l l l}
Collaboration & Ref. & $\Nf$ & 
\hspace{0.15cm}\begin{rotate}{60}{publication status}\end{rotate}\hspace{-0.15cm} &
\hspace{0.15cm}\begin{rotate}{60}{continuum extrapolation}\end{rotate}\hspace{-0.15cm} &
\hspace{0.15cm}\begin{rotate}{60}{chiral extrapolation}\end{rotate}\hspace{-0.15cm}&
\hspace{0.15cm}\begin{rotate}{60}{finite volume}\end{rotate}\hspace{-0.15cm}&
\hspace{0.15cm}\begin{rotate}{60}{renormalization/matching}\end{rotate}\hspace{-0.15cm}  &
\hspace{0.15cm}\begin{rotate}{60}{heavy quark treatment}\end{rotate}\hspace{-0.15cm} & 
 $f_{B_s}/f_{B^+}$  & $f_{B_s}/f_{B^0}$  & $f_{B_s}/f_{B}$  \\
&&&&&&&&&& \\[-0.1cm]
\hline
\hline
&&&&&&&&&& \\[-0.1cm]

ETM 13E & \cite{Carrasco:2013naa} & 2+1+1 & \rC & \good & \soso & \soso
& \soso &  \okay &  $-$ & $-$ & 1.201(25) \\[0.5ex]

HPQCD 13 & \cite{Dowdall:2013tga} & 2+1+1 & \gA & \good & \good & \good & \soso
& \okay & 1.217(8) & 1.194(7) & 1.205(7)  \\[0.5ex]

&&&&&&&&&& \\[-0.1cm]
\hline
&&&&&&&&&& \\[-0.1cm]

RBC/UKQCD 13A & \cite{Witzel:2013sla} & 2+1 & \rC & \soso & \soso & \good 
  & \soso & \okay & $-$ & $-$ &  1.20(2)$_{\rm stat}^\diamond$ \\[0.5ex]

HPQCD 12 & \cite{Na:2012sp} & 2+1 & \gA & \soso & \soso & \good & \soso
& \okay & $-$ & $-$ & 1.188(18) \\[0.5ex]

FNAL/MILC 11 & \cite{Bazavov:2011aa} & 2+1 & \gA & \soso & \soso &
     \good& \soso & \okay & 1.229(26) & $-$ & $-$ \\[0.5ex]  
     
RBC/UKQCD 10C & \cite{Albertus:2010nm} & 2+1 & \gA & \tbr & \tbr & \good 
  & \soso & \okay & $-$ & $-$ & 1.15(12) \\[0.5ex]

HPQCD 09 & \cite{Gamiz:2009ku} & 2+1 & \gA & \soso & \soso & \good &
\soso & \okay & $-$ & $-$ & 1.226(26)  \\[0.5ex] 

&&&&&&&&&& \\[-0.1cm]
\hline
&&&&&&&&&& \\[-0.1cm]
ALPHA 13 & \cite{Bernardoni:2013oda} & 2 & \rC  & \good  & \good  &
\good   &\good  & \okay   & $-$ & $-$ & 1.195(61)(20)  &  \\[0.5ex] 

ETM 13B, 13C & \cite{Carrasco:2013zta,Carrasco:2013iba} & 2 & \oP$^\dagger$ & \good & \soso & \good
& \soso &  \okay &  $-$ & $-$ & 1.206(24)  \\[0.5ex]

ALPHA 12A & \cite{Bernardoni:2012ti} & 2 & \rC & \good & \good & \good
& \good &  \okay & $-$ & $-$ & 1.13(6)  \\ [0.5ex]

ETM 12B & \cite{Carrasco:2012de} & 2 & \rC & \good & \soso & \good
& \soso &  \okay & $-$ & $-$ & 1.19(5) \\ [0.5ex]

ETM 11A & \cite{Dimopoulos:2011gx} & 2 & \gA & \soso & \soso & \good
& \soso &  \okay & $-$ & $-$ & 1.19(5) \\ [0.5ex]
&&&&&&&&&& \\[-0.1cm]
\hline
\hline
\end{tabular*}
\begin{tabular*}{\textwidth}[l]{l@{\extracolsep{\fill}}lllllllll}
  \multicolumn{10}{l}{\vbox{\begin{flushleft}
 	 $^\diamond$Statistical errors only. \\
          $^\dagger$Update of ETM 11A and 12B. 
\end{flushleft}}}
\end{tabular*}
\vspace{-0.5cm}
\caption{Ratios of decay constants of the $B$ and $B_s$ mesons (for details see Table \ref{tab:FBssumm}).}
\label{tab:FBratsumm}
\end{center}
\end{table}

The ETM collaboration has presented a series of calculations of the
$B$-meson decay constants based on simulations with $N_f=2$ sea quarks 
\cite{Blossier:2009hg,Dimopoulos:2011gx,Carrasco:2012de,Carrasco:2013zta,Carrasco:2013iba}.
Three lattice spacings in the range $a \approx 0.067 - 0.098 \,{\rm fm}$ are 
used in ETM~09D \cite{Blossier:2009hg}. In ETM~11A, ETM~12B, and ETM~13B, 13C 
\cite{Dimopoulos:2011gx,Carrasco:2012de,Carrasco:2013zta,Carrasco:2013iba} 
additional ensembles at $a\approx 0.054$~fm are included.
The valence and sea quarks are
simulated with two different versions of the twisted-mass Wilson
fermion action. In ETM~09D and ETM~11A the heavy-quark masses are 
in the charm region and above while keeping $am_h \lsim 0.5$. 
ETM~12B includes slightly heavier masses than ETM~09D and ETM~11A, 
while ETM~13B, 13C includes masses as heavy as $am_h \sim 0.85$ at the
largest two lattice spacings.
In ETM~11A two methods are used to
obtain $f_{B_{(s)}}$ from their heavy Wilson data: the ratio and the
interpolation methods. In the interpolation method they supplement
their heavy Wilson data with a static limit calculation.  In the ratio
method (see Appendix~\ref{app:HQactions}) they construct ratios
(called $z_{(s)}$) from a combination of the decay constants
$f_{h\ell(s)}$ and the heavy-quark pole masses that are equal to unity
in the static limit. Ratios of pole-to-$\msbar$ mass conversion
factors are included at NLO in continuum perturbation theory.  
ETM~09D, ETM~12B and ETM~13B, 13C use only the ratio method.  
Finally, ETM analyses the SU(3)
breaking ratio $\Phi_{hs}/\Phi_{h\ell}$ (or the ratio of ratios,
$z_s/z$) and combines it with $\Phi_{hs}$ or ($z_s$) to obtain $f_B$,
instead of directly extracting it from their $\Phi_{h\ell}$ (or $z$)
data.  In ETM~11A, ETM~12B, and ETM~13B, 13C 
the data are interpolated to a fixed set of reference masses
on all ensembles, and subsequently extrapolated to the continuum and
to the physical light-quark masses in a combined fit.  The static
limit calculation for the interpolation method in ETM~11A is done at
two intermediate lattice spacings, $a \approx 0.085, 0.067$ fm.
The results from the interpolation method have
larger (statistical and systematic) errors than those from the ratio
method, since statistical and systematic errors tend to cancel in the
ratios. The observed discretization effects (as measured by the
percentage difference between the lattice data at the smallest lattice
spacing and the continuum extrapolated results) are smaller than what
would be expected from power-counting estimates. Over the range of
heavy quark masses used in their simulations ETM finds discretization
errors $\lsim 3\%$ for $\Phi_{hs}$ and $\lsim 1.5\%$ for the ratio
$z_s$. As a result, the dominant error on $f_{B_s}$ is the statistical
(combined with the chiral and continuum extrapolation and heavy quark
interpolation) uncertainty, whereas the dominant error on the SU(3)
breaking ratio is due to the chiral extrapolation.

The ALPHA collaboration calculates the $B$- and $B_s$-meson decay
constants at the physical $b$-quark mass using nonperturbative lattice
HQET through ${\cal O}(1/m_h)$ on ensembles with $N_f=2$
nonperturbatively ${\cal O}(a)$ improved Wilson quarks at three
lattice spacings in the range $a \approx 0.048 - 0.075 \,{\rm fm}$.
The parameters of the HQET action and the static-current
renormalization are determined nonperturbatively in a separate
matching calculation using small physical volumes ($L\simeq 0.4 $~fm)
with Schr\"odinger functional boundary conditions together with a
recursive finite-size scaling procedure to obtain the nonperturbative
parameters at the large physical volumes used in the simulations.  In
ALPHA~11 \cite{Blossier:2011dk} ensembles with pion masses in the
range $m_{\pi} \approx 440 - 270$~MeV are used.  ALPHA~12A 
\cite{Bernardoni:2012ti} and ALPHA~13 \cite{Bernardoni:2013oda} 
include an ensemble at a lighter sea-quark mass corresponding to 
$m_{\pi} \approx 190$~MeV.  ALPHA~11
presents results for $f_B$ only, while ALPHA~12A  also presents
a preliminary result for $f_{B_s}$, and ALPHA~13 presents the collaboration's
final results for $f_B$, $f_{B_s}$, and $f_{B_s}/f_B$.
The combined statistical and
extrapolation errors are of order $5-6\%$ in these calculations, and are
larger than the chiral fit uncertainty. Truncation errors which are
${\mathcal O}(\Lambda_{\rm QCD}/m_h)^2$ are not included in this error
budget.  Simple power-counting would suggest that they are $ \approx
1-4\%$.  However, the results from both the ETM collaboration
discussed above and the HPQCD collaboration (from their heavy HISQ
analysis) discussed below, as well as results obtained by ALPHA in the
quenched approximation \cite{Blossier:2010mk} indicate that ${\mathcal
O}(\Lambda_{\rm QCD}/m_h)^2$ effects are probably quite small for
heavy-light decay constants at the physical $b$-quark mass.

In summary, for the $N_f=2$ case, only ETM's results qualify for averaging, since
ALPHA's results have appeared in conference proceedings only so far. 
Since ETM~13B, 13C updates the published ETM~11A results, we use it for
our average:
\be
N_f=2:  \hspace{0.2em} f_B = (189 \pm 8) \MeV, \hspace{0.2em}
f_{B_s} = (228 \pm 8) \MeV, \hspace{0.2em}
f_{B_s}/f_B = 1.206 \pm 0.024 \,.
\label{eq:fbav2}
\ee

For the $N_f=2+1$ case there are currently four published papers
describing lattice-QCD calculations of $f_{B_{(s)}}$ performed by two
different groups: FNAL/MILC and HPQCD.  The HPQCD collaboration has
published several calculations of the $B$ meson decay constants with
NRQCD $b$ quarks \cite{Gamiz:2009ku,Na:2012sp}.  In
Ref.~\cite{Gamiz:2009ku} (HPQCD~09) they use Asqtad light valence
quarks, and include ensembles at two lattice spacings $a\approx 0.12,
0.09$~fm and sea quarks with minimum RMS sea-pion masses $m_{\pi, {\rm
RMS}} \approx 400$~MeV
 equal to the light sea-quark masses.  In Ref.~\cite{Na:2012sp} HISQ
 light valence quarks are employed instead.  This analysis uses the
 same Asqtad ensembles as in HPQCD~09 but includes an additional
 ensemble at $a \approx 0.09$~fm at a lighter sea-quark mass, so that
 the minimum RMS sea pion mass is approximately $320$~MeV. The HISQ
 light valence masses are matched to the Asqtad sea-quark masses via
 the ratio $m_{\ell}/m_s$. The dominant systematic error in both
 calculations is due to using one-loop mean-field improved lattice
 perturbation theory for the current renormalization.  In both
 calculations, HPQCD performs a combined chiral and continuum
 extrapolation of the data, in the first case using NLO (full QCD)
 heavy meson rooted staggered $\chi$PT (HMrS$\chi$PT) and in the
 latter case using NLO continuum partially quenched HM$\chi$PT,
 supplemented in both cases by NNLO analytic and generic
 discretization terms.  HPQCD finds a significant reduction in
 discretization errors in their calculation with HISQ light valence
 quarks, as compared to their calculation with Asqtad valence quarks.
 Indeed, in HPQCD~12 the continuum extrapolated results overlap within
 errors with the data at finite lattice spacing.

Another calculation of the $B_s$-meson decay constant is presented by
the HPQCD collaboration in Ref.~\cite{McNeile:2011ng} , this time
using the HISQ action for the strange and heavy valence quarks,
i.e. the heavy HISQ method. This analysis includes Asqtad ensembles
over a large range of lattice spacings, $a\approx 0.15 - 0.045$~fm and
heavy-quark masses in the range $am_h \approx 0.2 - 0.85$. Only one
sea-quark ensemble per lattice spacing is included in this analysis,
all with a sea-quark to strange-quark mass ratio of
$m_{\ell}/m_s \approx 0.2$, yielding a minimum RMS sea pion mass of
approximately $330$~MeV.  The sea-quark mass dependence is assumed to
be negligible, which is based on the analysis of $f_{D_s}$ in
Ref.~\cite{Davies:2010ip}.  HPQCD uses an HQET-type expansion in
$1/m_H$ (where $m_H$ is the mass of an $h$-flavoured meson) with
coefficients that are polynomials in $am_h$, $a\Lambda$, and $am_s$ to
perform a combined fit to all their data, including terms up to
$1/m_H^3$, $(am_h)^6$, $(a\Lambda)^6$, and $(am_s)^6$.  The continuum
extrapolated fit curve is then used to obtain the decay constant at
the physical $B_s$ meson mass, which requires another small
extrapolation. As can be seen in Figure~1 of
Ref.~\cite{McNeile:2011ng}, discretization errors (as measured by the
percentage difference between the lattice data and the continuum fit
curve) are smaller for a given value of $am_h$ when $m_H$ is
larger. This somewhat counterintuitive result for an action that
formally contains discretization errors of ${\cal O}(am_h)^2$ is
likely due to coefficients in the form of powers of $v/c$ that
suppress these errors.  After statistical (and extrapolation) errors,
the largest sources of uncertainty in this analysis are discretization
and heavy-quark extrapolation errors. They are estimated by varying
the fit Ansatz and by excluding data at the largest and smallest
lattice spacings as well as data at the largest values of $am_h$.

The Fermilab Lattice and MILC collaborations present a lattice-QCD
calculation of the $D$- and $B$-meson decay constants in
Ref.~\cite{Bazavov:2011aa}, which uses the Fermilab method for the
heavy ($b$ and $c$) valence quarks together with Asqtad light and
strange valence quarks on a subset of the MILC Asqtad $N_f=2+1$
ensembles. The current renormalizations are calculated using a mostly
nonperturbative renormalization (mNPR) method. Their estimate of the
perturbative errors for the small perturbative correction factors
calculated at one-loop in mean field improved lattice perturbation
theory are comparable to the size of actual one-loop corrections.  The
simulations include lattice spacings in the range $a \approx 0.15 -
0.09$~fm and a minimum RMS pion mass of approximately $320$~MeV. In
this calculation lattice data at $9-12$ valence light-quark masses are
generated for each sea-quark ensemble.  The chiral and continuum
extrapolated results are obtained from combined chiral and continuum
fits. The chiral fit function uses NLO partially quenched HMrS$\chi$PT
including $1/m_h$ terms and supplemented by NNLO analytic terms. Also
included are light-quark as well as heavy-quark discretization
terms. The dominant uncertainties after statistical errors are due to
heavy-quark discretization effects, heavy-quark mass tuning, and
correlator fit errors.  A calculation of the $B$ and $D$ meson decay
constants using Fermilab heavy quarks on the full set of Asqtad
ensembles is still in progress \cite{Neil:2011ku}.
 
The RBC/UKQCD collaboration has presented a result for the SU(3)
breaking ratio in Ref.~\cite{Albertus:2010nm} using a static-limit
action on $N_f=2+1$ domain wall ensembles at a single lattice spacing
$a \approx 0.11$~fm with a minimum pion mass of approximately
$430$~MeV.  They use both HYP and APE smearing for the static action
and one-loop mean field improved lattice perturbation theory to
renormalize and improve the static-limit current. Their static-limit
action and current do not, however, include $1/m_h$
effects. Ref.~\cite{Albertus:2010nm} includes an estimate of this
effect via power counting as ${\cal O}\left((m_s-m_d)/m_b\right)$ in
the error budget.  The statistical errors in this work are
significantly larger ($\sim5-8 \%$), as are the chiral extrapolation
errors ($\sim7 \%$), due to the rather large pion masses used in this
work. With data at only one lattice spacing, discretization errors
cannot be estimated from the data. A power counting estimate of this
error of $3\%$ is included in the systematic error budget.  
An update of this work was presented at the Lattice 2013 conference
\cite{Ishikawa:2013faa}, where the new analysis includes ensembles 
at two lattice spacings and with smaller pion masses, 
as well as calculations of the decay constants themselves. However,
Ref.~\cite{Ishikawa:2013faa} did not appear until after the closing deadline 
and is therefore not included in this review. 
The RBC/UKQCD collaboration has also presented 
preliminary calculations of the $B$-meson decay constants using the 
RHQ action (another relativistic heavy-quark action) 
\cite{Witzel:2012pr,Witzel:2013sla} on $N_f=2+1$ domain wall ensembles 
at two lattice spacings, $a \approx 0.086, 0.11$~fm with sea-pion masses 
in the range $m_{\pi} \approx 420 - 290$~MeV. The parameters of the RHQ 
action are tuned nonperturbatively, and the axial vector current is renormalized 
using the mNPR method. Results are quoted  with statistical errors only 
\cite{Witzel:2013sla} after a combined chiral-continuum extrapolation using 
SU(2) HM$\chi$PT and a term linear in $a^2$. A complete systematic error 
analysis is still in progress.

In summary, for the
$N_f=2+1$ case there currently are four different results for the $B$
and $B_s$ meson decay constants and three different results for the
SU(3) breaking ratio that satisfy the quality criteria (see 
Tables~\ref{tab:FBssumm}~and~\ref{tab:FBratsumm},). However, they
all use overlapping subsets of MILC Asqtad ensembles. We therefore
treat the statistical errors between the results as 100\% correlated.
Furthermore, one of the results for $f_B$ in HPQCD~12 \cite{Na:2012sp}
is obtained by combining HPQCD~12's result for the ratio $f_{B_s}/f_B$
using NRQCD $b$ quarks with HPQCD~11A's result for $f_{B_s}$. However,
no itemized error budget is given for the so-combined $f_B$ result. In
order to include sensible correlations between the two HPQCD results
for $f_B$, we construct an itemized error budget for the combined
$f_B$ from the individual itemized error budgets, by adding the
itemized errors in quadrature.  This is conservative, because the
resulting total uncertainty on the combined $f_B$ is slightly larger
than the quoted uncertainty in Ref.~\cite{Na:2012sp}, $4.3$ MeV
compared to $4$ MeV.  We then treat the chiral extrapolation errors,
the light-quark discretization errors, the scale setting errors, and
renormalization errors as 100\% correlated between the two $f_B$
results in HPQCD~12. Finally, the HPQCD~09 result was obtained using a
value for the scale $r_1$ that has since been superseded. We drop this
result from the average, since it is effectively updated by HPQCD~12.
We find:
\be
N_f=2+1: \hspace{0.2em} f_B = (190.5 \pm 4.2) \MeV, \hspace{0.2em}
f_{B_s} = (227.7 \pm 4.5) \MeV,
\hspace{0.2em} f_{B_s}/f_B = 1.202 \pm 0.022 \,.
\label{eq:fbav21}
\ee
The uncertainties on the averages for $f_{B_s}$ and for the SU(3)
breaking ratio $f_{B_s}/f_B$ include PDG rescaling factors of $1.1$ and $1.3$, 
respectively.
 
Finally, the first published results for $B$ meson decay constants with
$N_f=2+1+1$ sea quarks are presented by the HPQCD
collaboration \cite{Dowdall:2013tga} (HPQCD~13) using the MILC HISQ
ensembles at three lattice spacings, $a\approx 0.15, 0.12, 0.09$~fm,
where at each lattice spacing one ensemble with Goldstone pions at the
physical value is included.  HPQCD~13 uses NRQCD $b$ quarks and HISQ
light valence quarks.  The combined chiral interpolation and continuum
extrapolation is performed using NLO (full QCD) HM$\chi$PT,
supplemented by generic discretization terms of $O(a^2,a^4)$. HPQCD
also performs a continuum extrapolation of the data at the physical
point only, with results that are in good agreement with the
extrapolated results obtained from the full data set.  The dominant
systematic error in this calculation is due to using one-loop
mean-field improved lattice perturbation theory for the current
renormalization.  In HPQCD~13 it is estimated at $1.4$\%, almost a
factor of 3 smaller than in HPQCD~12, after reorganizing the
perturbative series similar to the mNPR method, and using the fact
that the heavy-heavy NRQCD temporal vector current is absolutely
normalized and that the light-light HISQ vector current has a small
one-loop correction.  The next largest uncertainties are due to
heavy-quark truncation effects and statistics and scale setting.  In
this work the scale is set using the $\Upsilon(2S-1S)$ splitting
calculated in Ref.~\cite{Dowdall:2011wh} without using $r_1$ to set
the relative scale between ensembles at different lattice spacings, as
was done in previous HPQCD work.  

Most recently, the ETM collaboration presented their new results for $B$-meson
decay constants on their $N_f=2+1+1$ ensembles in 
ETM~13E \cite{Carrasco:2013naa}. This work uses the same methods as ETM's $N_f=2$
$B$-meson decay constant analyses. In particular, different versions of 
twisted-mass Wilson actions are used for sea and valence quarks. The decay 
constants are calculated with the ratio method using heavy-quark 
masses in the charm region and above while keeping $am_h \lsim 0.8$.
ETM~13E includes ensembles with lattice spacings in the range 
$a \approx 0.062-0.089$~fm and with sea-pion masses in the range 
$m_{\pi} \approx 211-443$~MeV which are used for combined chiral-continuum
extrapolations. As before, the ratio data for $z_s$ show small discretization effects. 
Somewhat larger discretization effects are observed, however, for the decay-constant 
data at the charm-quark mass, since the smallest lattice spacing for the $N_f=2+1+1$ 
ensembles is larger than for $N_f=2$.

In summary, for the $N_f=2+1+1$ case, the only published results are from HPQCD~13,
which therefore form our average:
\be
N_f=2+1+1: \hspace{0.2em} f_B = (186 \pm 4) \MeV, \hspace{0.2em}
f_{B_s} = (224 \pm 5) \MeV,
\hspace{0.2em} f_{B_s}/f_B = 1.205 \pm 0.007 \,.
\label{eq:fbav211}
\ee

A comparison of all $N_f=2$ , $N_f=2+1$ and $N_f=2+1+1$ lattice-QCD results for
$f_B$, $f_{B_s}$, and their ratio is shown in Figure~\ref{fig:fB}. The
averages presented in Eqs.~(\ref{eq:fbav2}),~(\ref{eq:fbav21})~and~(\ref{eq:fbav211})
are represented by the grey bands in the figures.

A final comment concerns which light valence-quark mass is used for
the chiral extrapolations (or interpolations) to the physical
point. First, we note that all the results discussed in this review
use simulations with degenerate up and down sea-quark masses. However,
since the observed sea-quark mass dependence is much smaller than the
valence-quark mass dependence, the dominant contribution to
differences between $B^+$- and $B^0$-meson quantities is due to the
light valence quarks. Almost all the results quoted in this review are
obtained from chiral extrapolations to the average of the up- and
down-quark masses, and therefore correspond to the average of the
$B^0$- and $B^+$-meson decay constants. The exceptions are FNAL/MILC~11
and HPQCD~13 which both quote results for the $B^+$ meson decay
constant from chiral extrapolations (interpolations) of the light
valence-quark to the physical up-quark mass.  HPQCD~13 also quotes
results for the $B^0$ meson decay constant from chiral interpolations
to the physical down-quark mass as well as results for the average of
the $B^+$ and $B^0$ mesons.  The $N_f = 2+1$ and $N_f = 2$ averages
presented in Eqs.~(\ref{eq:fbav2}), (\ref{eq:fbav21}) and (\ref{eq:fbav211}) are 
for the average of the $B^+$- and $B^0$-meson decay constant, $f_B$, and the
corresponding ratio, $f_{B_s}/f_B$.  Given the errors quoted in the
results that enter our averages, we currently include the FNAL/MILC~11
results for the $B^+$ meson in Eq.~(\ref{eq:fbav21}).  As the
precision with which $B$-meson decay constants are obtained continues
to improve, especially given the availability of physical mass
ensembles, future reviews will need to distinguish between these
cases. Indeed HPQCD~13 finds a $2$\% difference between the $B^+$ and
$B^0$ decay constants, which is the same size as the total uncertainty
in this calculation. We strongly recommend that future lattice-QCD
calculations of $B$-meson decay constants quote results for the $B^+$
and $B^0$ mesons separately.

\subsection{Neutral $B$-meson mixing matrix elements}
\label{sec:BMix}

Neutral $B$-meson mixing is induced in the Standard Model through
one-loop box diagrams to lowest order in the electroweak theory,
similar to those for neutral kaon mixing.  The effective Hamiltonian
is given by
\begin{equation}
  {\cal H}_{\rm eff}^{\Delta B = 2, {\rm SM}} \,\, = \,\,
  \frac{G_F^2 M_{\rm{W}}^2}{16\pi^2} ({\cal F}^0_d {\cal Q}^d_1 + {\cal F}^0_s {\cal Q}^s_1)\,\, +
   \,\, {\rm h.c.} \,\,,
   \label{eq:HeffB}
\end{equation}
with
\begin{equation}
 {\cal Q}^q_1 =
   \left[\bar{b}\gamma_\mu(1-\gamma_5)q\right]
   \left[\bar{b}\gamma_\mu(1-\gamma_5)q\right],
   \label{eq:Q1}
\end{equation}
where $q=d$ or $s$. The short-distance function ${\cal F}^0_q$ in
Eq.~(\ref{eq:HeffB}) is much simpler compared to the kaon mixing case
due to the hierarchy in the CKM matrix elements. Here, only one term
is relevant,
\begin{equation}
 {\cal F}^0_q = \lambda_{tq}^2 S_0(x_t)
\end{equation}
where
\begin{equation}
 \lambda_{tq} = V^*_{tq}V_{tb},
\end{equation}
and where $S_0(x_t)$ is an Inami-Lim function with $x_t=m_t^2/M_W^2$,
which describes the basic electroweak loop contributions without QCD
\cite{Inami:1980fz}. The transition amplitude for $B_q^0$ with $q=d$
or $s$ can be written as
\begin{eqnarray}
\label{eq:BmixHeff}
&&\langle \bar B^0_q \vert {\cal H}_{\rm eff}^{\Delta B = 2} \vert B^0_q\rangle  \,\, = \,\, \frac{G_F^2 M_{\rm{W}}^2}{16 \pi^2}  
\Big [ \lambda_{tq}^2 S_0(x_t) \eta_{2B} \Big ]  \nn \\ 
&&\times 
  \left(\frac{\gbar(\mu)^2}{4\pi}\right)^{-\gamma_0/(2\beta_0)}
  \exp\bigg\{ \int_0^{\gbar(\mu)} \, dg \, \bigg(
  \frac{\gamma(g)}{\beta(g)} \, + \, \frac{\gamma_0}{\beta_0g} \bigg)
  \bigg\} 
   \langle \bar B^0_q \vert  Q^q_{\rm R} (\mu) \vert B^0_q
   \rangle \,\, + \,\, {\rm h.c.} \,\, ,
   \label{eq:BBME}
\end{eqnarray}
where $Q^q_{\rm R} (\mu)$ is the renormalized four-fermion operator
(usually in the NDR scheme of $\msbar$). The running coupling
($\gbar$), the $\beta$-function ($\beta(g)$), and the anomalous
dimension of the four quark operator ($\gamma(g)$) are defined in
Eqs.~(\ref{eq:four_quark_operator_anomalous_dimensions})~and~(\ref{eq:four_quark_operator_anomalous_dimensions_perturbative}).
The product of $\mu$ dependent terms on the second line of
Eq.~(\ref{eq:BBME}) is, of course, $\mu$ independent (up to truncation
errors if perturbation theory is used).  The explicit expression for
the short-distance QCD correction factor $\eta_{2B}$(calculated to
NLO) can be found in Ref.~\cite{Buchalla:1995vs}.

For historical reasons the $B$-meson mixing matrix elements are often
parameterized in terms of bag parameters defined as
\begin{equation}
 B_{B_q}(\mu)= \frac{{\left\langle\bar{B}^0_q\left|
   Q^q_{\rm R}(\mu)\right|B^0_q\right\rangle} }{
         {\frac{8}{3}f_{B_q}^2\mB^2}} \,\, .
         \label{eq:bagdef}
\end{equation}
The RGI $B$ parameter $\hat{B}$ is defined, as in the case of the kaon,
and expressed to two-loop order as
\begin{equation}
 \hat{B}_{B_q} = 
   \left(\frac{\gbar(\mu)^2}{4\pi}\right)^{- \gamma_0/(2\beta_0)}
   \left\{ 1+\dfrac{\gbar(\mu)^2}{(4\pi)^2}\left[
   \frac{\beta_1\gamma_0-\beta_0\gamma_1}{2\beta_0^2} \right]\right\}\,
   B_{B_q}(\mu) \,\,\, ,
\label{eq:BBRGI_NLO}
\end{equation}
with $\beta_0$, $\beta_1$, $\gamma_0$, and $\gamma_1$ defined in
Eq.~(\ref{eq:RG-coefficients}). 

Nonzero transition amplitudes result in a mass difference between the
$CP$ eigenstates of the neutral $B$ meson system. Writing the mass
difference for a $B_q^0$ meson as $\Delta m_q$, its Standard Model
prediction is
\begin{equation}
 \Delta m_q = \frac{G^2_Fm^2_W m_{B_q}}{6\pi^2} \,
  |\lambda_{tq}|^2 S_0(x_t) \eta_{2B} f_{B_q}^2 \hat{B}_{B_q}.
\end{equation}
Experimentally the mass difference is measured as oscillation
frequency of the $CP$ eigenstates. The frequencies are measured
precisely with an error of less than a percent. Many different
experiments have measured $\Delta m_d$, but the current average
\cite{Beringer:1900zz} is dominated by measurements from the
$B$-factory experiments Belle and Babar, and from the LHC experiment
LHC$b$. For $\Delta m_s$ the experimental average is based on results
from the Tevatron experiment CDF and from the LHC experiment LHC$b$
\cite{Beringer:1900zz}.  With these experimental results and
lattice-QCD calculations of $f_{B_q}^2\hat{B}_{B_q}$ at hand,
$\lambda_{tq}$ can be determined.  In lattice-QCD calculations the
flavour $SU(3)$ breaking ratio
\begin{equation}
 \xi^2 = \frac{f_{B_s}^2B_{B_s}}{f_{B_d}^2B_{B_d}}
 \label{eq:xidef}
\end{equation} 
can be obtained more precisely than the individual $B_q$-mixing matrix
elements because statistical and systematic errors cancel in part.
With this ratio $|V_{td}/V_{ts}|$ can be determined, which can be used
to constrain the apex of the CKM triangle.

Neutral $B$-meson mixing, being loop-induced in the Standard Model is
also a sensitive probe of new physics. The most general $\Delta B=2$
effective Hamiltonian that describes contributions to $B$-meson mixing
in the Standard Model and beyond is given in terms of five local
four-fermion operators:
\be
  {\cal H}_{\rm eff}^{\Delta B = 2} = \sum_{i=1}^5 {\cal C}_i {\cal Q}_i \;,
\ee
where ${\cal Q}_1$ is defined in Eq.~(\ref{eq:Q1}) and where
\bd
{\cal Q}^q_2 =  \left[\bar{b}(1-\gamma_5)q\right]
   \left[\bar{b}(1-\gamma_5)q\right], \qquad
{\cal Q}^q_3 =  \left[\bar{b}^{\alpha}(1-\gamma_5)q^{\beta}\right]
   \left[\bar{b}^{\beta}(1-\gamma_5)q^{\alpha}\right],
 \ed
  \be
{\cal Q}^q_4 =  \left[\bar{b}(1-\gamma_5)q\right]
   \left[\bar{b}(1+\gamma_5)q\right], \qquad
{\cal Q}^q_5 =  \left[\bar{b}^{\alpha}(1-\gamma_5)q^{\beta}\right]
   \left[\bar{b}^{\beta}(1+\gamma_5)q^{\alpha}\right], 
   \label{eq:Q25}
\ee 
with the superscripts $\alpha,\beta$ denoting colour indices, which
are shown only when they are contracted across the two bilinears. The
short-distance Wilson coefficients ${\cal C}_i$ depend on the
underlying theory and can be calculated perturbatively.  In the
Standard Model only matrix elements of ${\cal Q}^q_1$ contribute to
$\Delta m_q$, and combinations of matrix elements of ${\cal Q}^q_1$,
${\cal Q}^q_2$, and ${\cal Q}^q_3$ contribute to the width difference
$\Delta \Gamma_q$ \cite{Lenz:2006hd,Beneke:1996gn}.  Matrix elements
of ${\cal Q}^q_4$ and ${\cal Q}^q_5$ are needed for calculating the
contributions to $B_q$-meson mixing from beyond the Standard Model
theories.

In this section we report on results from lattice-QCD calculations for
the neutral $B$-meson mixing parameters $\hat{B}_{B_d}$,
$\hat{B}_{B_s}$, $f_{B_d}\sqrt{\hat{B}_{B_d}}$,
$f_{B_s}\sqrt{\hat{B}_{B_s}}$ and the $SU(3)$ breaking ratios
$B_{B_s}/B_{B_d}$ and $\xi$ defined in Eqs.~(\ref{eq:bagdef}),
(\ref{eq:BBRGI_NLO}), and (\ref{eq:xidef}).  The results are
summarized in Tables~\ref{tab_BBssumm} and \ref{tab_BBratsumm} and in
Figures~\ref{fig:fBsqrtBB2} and \ref{fig:xi}. Additional details about
the underlying simulations and systematic error estimates are given in
Appendix~\ref{app:BMix_Notes}.  Some collaborations do not provide the
RGI quantities $\hat{B}_{B_q}$ but quote instead
$B_B(\mu)^{\overline{MS},NDR}$. In such cases we convert the results
to the RGI quantities quoted in Table~\ref{tab_BBssumm} using
Eq.~(\ref{eq:BBRGI_NLO}). More details on the conversion factors are
provided below in the descriptions of the individual results. One
group also reports results for $B$-meson matrix elements of the other
operators ${\cal Q}_{2-5}$ in Ref.~\cite{Bouchard:2011xj}, which is a
conference proceedings.

\begin{table}[htb]
\begin{center}
\mbox{} \\[3.0cm]
\footnotesize
\begin{tabular*}{\textwidth}[l]{l @{\extracolsep{\fill}} r l l l l l l l l l l l }
Collaboration \al Ref. \al $\Nf$ \al
\hspace{0.10cm}\begin{rotate}{60}{publication status}\end{rotate}\hspace{-0.10cm} \al
\hspace{0.10cm}\begin{rotate}{60}{continuum extrapolation}\end{rotate}\hspace{-0.1cm} \al
\hspace{0.10cm}\begin{rotate}{60}{chiral extrapolation}\end{rotate}\hspace{-0.10cm}\al
\hspace{0.10cm}\begin{rotate}{60}{finite volume}\end{rotate}\hspace{-0.15cm}\al
\hspace{0.10cm}\begin{rotate}{60}{renormalization/matching}\end{rotate}\hspace{-0.1cm}  \al
\hspace{0.10cm}\begin{rotate}{60}{heavy quark treatment}\end{rotate}\hspace{-0.1cm} \al 
\rule{0.12cm}{0cm}
\parbox[b]{1.2cm}{$f_{\rm B_d}\sqrt{\hat{B}_{\rm B_d}}$} \al
\rule{0.012cm}{0cm}
\parbox[b]{1.2cm}{$f_{\rm B_d}\sqrt{\hat{B}_{\rm B_s}}$} \al
\rule{0.12cm}{0cm}
$\hat{B}_{\rm B_d}$ \al 
\rule{0.12cm}{0cm}
$\hat{B}_{\rm B_{\rm s}}$ \\
&&&&&&&&&& \\[-0.1cm]
\hline
\hline
&&&&&&&&&& \\[-0.1cm]

FNAL/MILC 11A \al \cite{Bouchard:2011xj} \al 2+1 \al \rC \al \good \al \soso \al
     \good \al \soso
	\al \okay & 250(23)$^\dagger$ \al 291(18)$^\dagger$ \al $-$ \al $-$\\[0.5ex]

HPQCD 09 \al \cite{Gamiz:2009ku} \al 2+1 \al \gA \al \soso \al \soso$^\nabla$ \al \good \al
\soso 
\al \okay & 216(15)$^\ast$ \al 266(18)$^\ast$ \al 1.27(10)$^\ast$ \al 1.33(6)$^\ast$ \\[0.5ex] 

HPQCD 06A \al \cite{Dalgic:2006gp} \al 2+1 \al \gA \al \tbr \al \tbr \al \good \al 
\soso
	\al \okay & $-$ \al  281(21) \al $-$ \al 1.17(17) \\

&&&&&&&&&& \\[-0.1cm]
\hline
&&&&&&&&&& \\[-0.1cm]
ETM 13B \al \cite{Carrasco:2013zta} \al 2 \al \oP \al \good \al \soso \al \good \al
    \good \al \okay & 216(6)(8) \al 262(6)(8) \al  1.30(5)(3) \al 1.32(5)(2) \\[0.5ex]

ETM 12A, 12B \al \cite{Carrasco:2012dd,Carrasco:2012de} \al 2 \al \rC \al \good \al \soso \al \good \al
    \good \al \okay & $-$ \al $-$ \al  1.32(8)$^\diamond$ \al 1.36(8)$^\diamond$ \\[0.5ex]
&&&&&&&&&& \\[-0.1cm]
\hline
\hline
\end{tabular*}
\begin{tabular*}{\textwidth}[l]{l@{\extracolsep{\fill}}lllllllll}
  \multicolumn{10}{l}{\vbox{\begin{flushleft} 
	$^\dagger$ Reported $f_B^2B$ at $\mu=m_b$ is converted to RGI by
	multiplying the two-loop factor
	1.517.\\
	  $^\nabla$Wrong-spin contributions are not included in the rS$\chi$PT fits. \\
        $^\ast$This result uses an old determination of  $r_1=0.321(5)$ fm from Ref.~\cite{Gray:2005ur} that has since been superseded. \\
        $^\diamond$  Reported $B$ at $\mu=m_b=4.35$ GeV is converted to
     RGI by multiplying the two-loop factor 1.521.\\
\end{flushleft}}}
\end{tabular*}

\vspace{-0.5cm}
\caption{Neutral $B$ and $B_{\rm s}$ meson mixing matrix
 elements (in MeV) and bag parameters.}
\label{tab_BBssumm}
\end{center}
\end{table}

\begin{figure}[htb]
\hspace{-0.8cm}\includegraphics[width=0.57\linewidth]{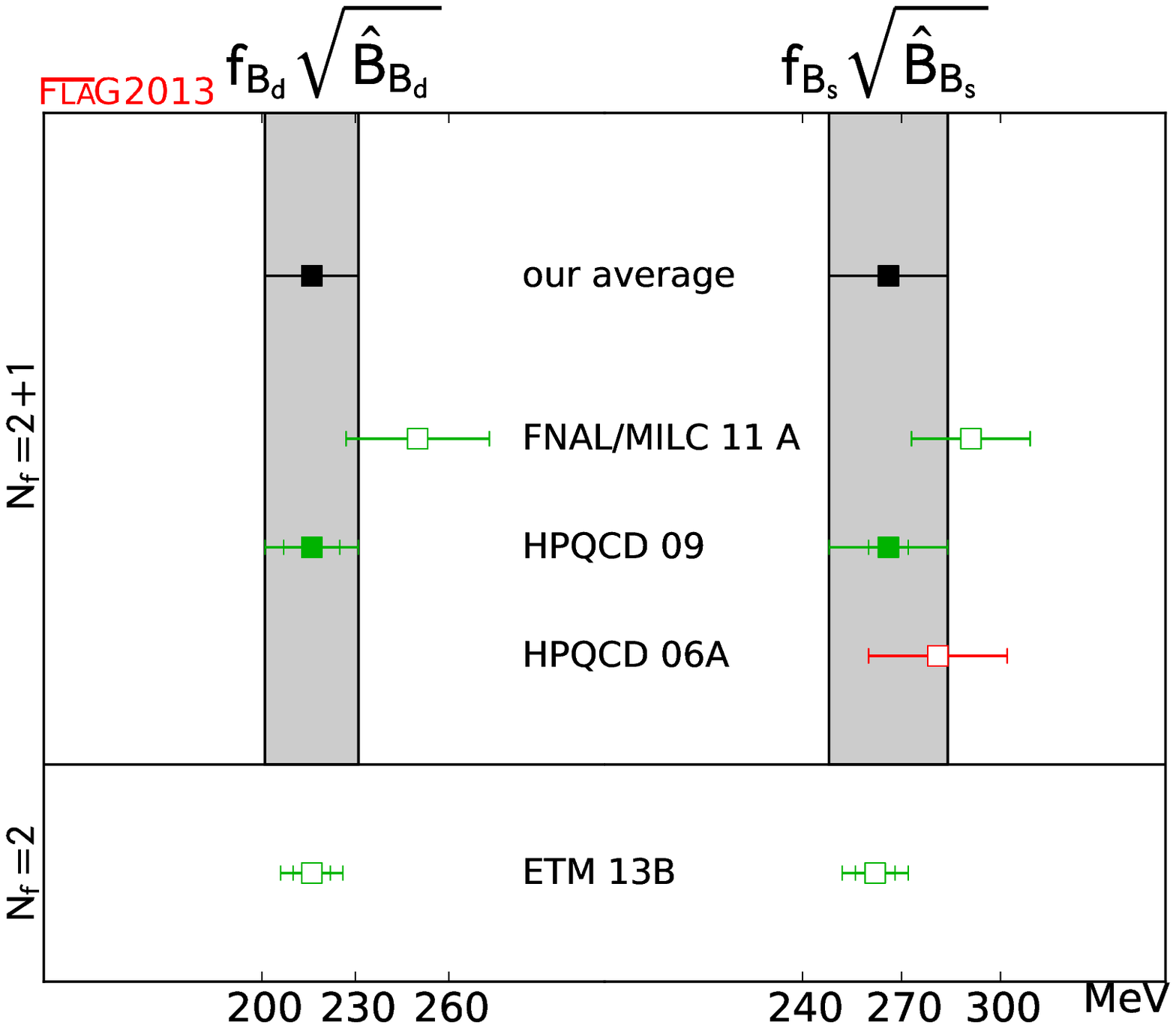}\hspace{-0.8cm}
\includegraphics[width=0.57\linewidth]{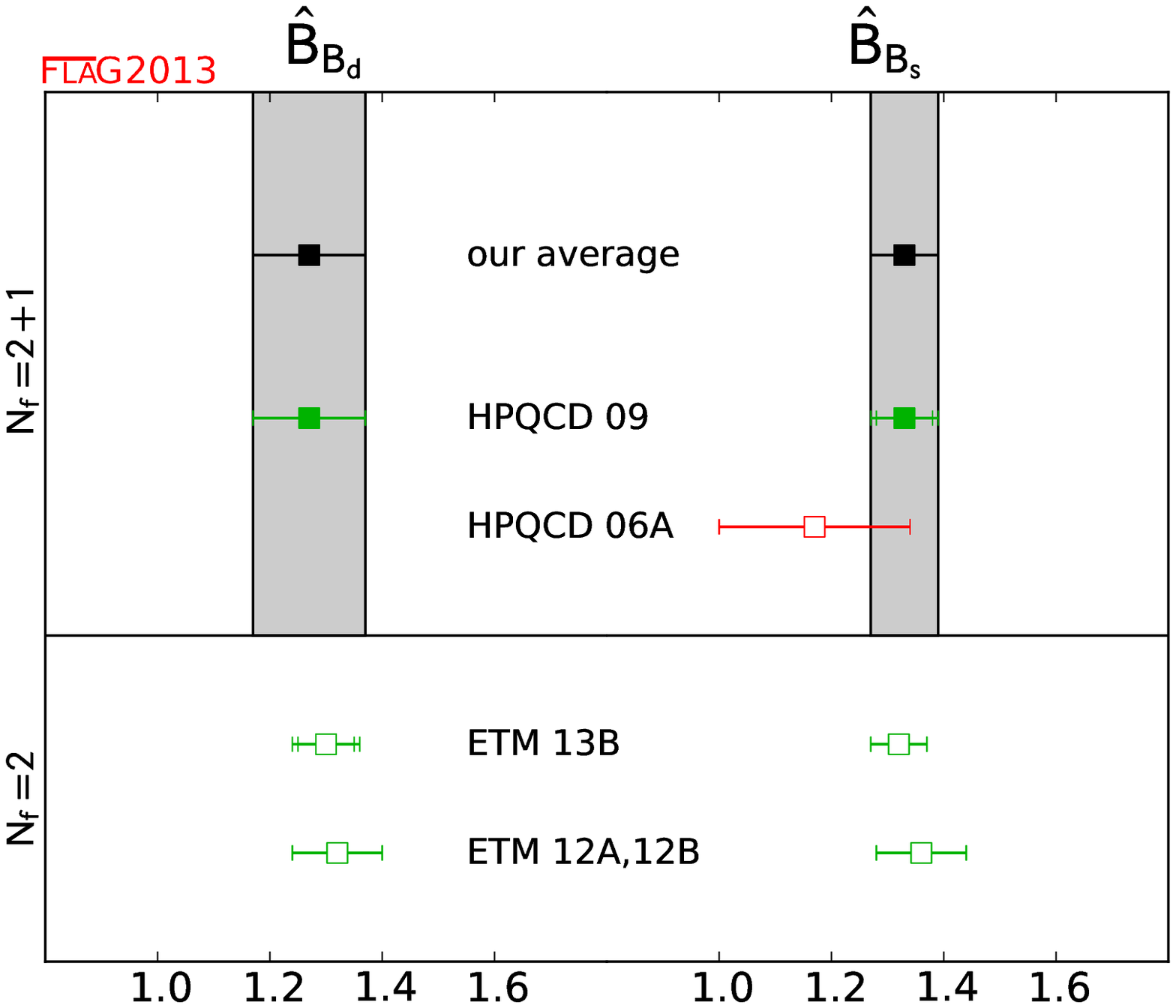}

\vspace{-2mm}
\caption{Neutral $B$ and $B_{\rm s}$ meson mixing matrix
 elements and bag parameters [values in Table \ref{tab_BBssumm} and
 Eqs.~(\ref{eq:avfBB}), (\ref{eq:avBB})]. 
 \label{fig:fBsqrtBB2}}
\end{figure}

\begin{table}[htb]
\begin{center}
\mbox{} \\[3.0cm]
\footnotesize
\begin{tabular*}{\textwidth}[l]{l @{\extracolsep{\fill}} r l l l l l l l l l}
Collaboration & Ref. & $\Nf$ & 
\hspace{0.15cm}\begin{rotate}{60}{publication status}\end{rotate}\hspace{-0.15cm} &
\hspace{0.15cm}\begin{rotate}{60}{continuum extrapolation}\end{rotate}\hspace{-0.15cm} &
\hspace{0.15cm}\begin{rotate}{60}{chiral extrapolation}\end{rotate}\hspace{-0.15cm}&
\hspace{0.15cm}\begin{rotate}{60}{finite volume}\end{rotate}\hspace{-0.15cm}&
\hspace{0.15cm}\begin{rotate}{60}{renormalization/matching}\end{rotate}\hspace{-0.15cm}  &
\hspace{0.15cm}\begin{rotate}{60}{heavy quark treatment}\end{rotate}\hspace{-0.15cm} & 
\rule{0.12cm}{0cm}$\xi$ &
 \rule{0.12cm}{0cm}$B_{\rm B_{\rm s}}/B_{\rm B_d}$ \\
&&&&&&&&&& \\[-0.1cm]
\hline
\hline
&&&&&&&&&& \\[-0.1cm]

FNAL/MILC 12 & \cite{Bazavov:2012zs} & 2+1 & \gA & \soso & \soso &
     \good & \soso & \okay & 1.268(63) & 1.06(11) \\[0.5ex]

RBC/UKQCD 10C
 & \cite{Albertus:2010nm} & 2+1 & \gA & \tbr & \tbr & \good
  & \soso & \okay & 1.13(12) & $-$ \\[0.5ex]

HPQCD 09 & \cite{Gamiz:2009ku} & 2+1 & \gA & \soso & \soso$^\nabla$ & \good &
\soso & \okay & 1.258(33) & 1.05(7) \\[0.5ex] 

&&&&&&&&&& \\[-0.1cm]

\hline

&&&&&&&&&& \\[-0.1cm]

ETM 13B & \cite{Carrasco:2013zta} & 2 & \oP & \good & \soso & \good & \good
			     & \okay & 1.225(16)(14)(22) & 1.007(15)(14) \\

ETM 12A, 12B & \cite{Carrasco:2012dd,Carrasco:2012de} & 2 & \rC & \good & \soso & \good & \good
			     & \okay & 1.21(6) & 1.03(2) \\

&&&&&&&&&& \\[-0.1cm]
\hline
\hline
\end{tabular*}
\begin{tabular*}{\textwidth}[l]{l@{\extracolsep{\fill}}lllllllll}
  \multicolumn{10}{l}{\vbox{\begin{flushleft} 
  $^\nabla$Wrong-spin contributions are not included in the rS$\chi$PT fits. \\
\end{flushleft}}}
\end{tabular*}

\vspace{-0.5cm}
\caption{Results for $SU(3)$ breaking ratios of neutral $B_{d}$ and 
 $B_{s}$ meson mixing matrix elements and bag parameters.}
\label{tab_BBratsumm}
\end{center}
\end{table}

\begin{figure}[htb]
\begin{center}
\includegraphics[width=0.7\linewidth]{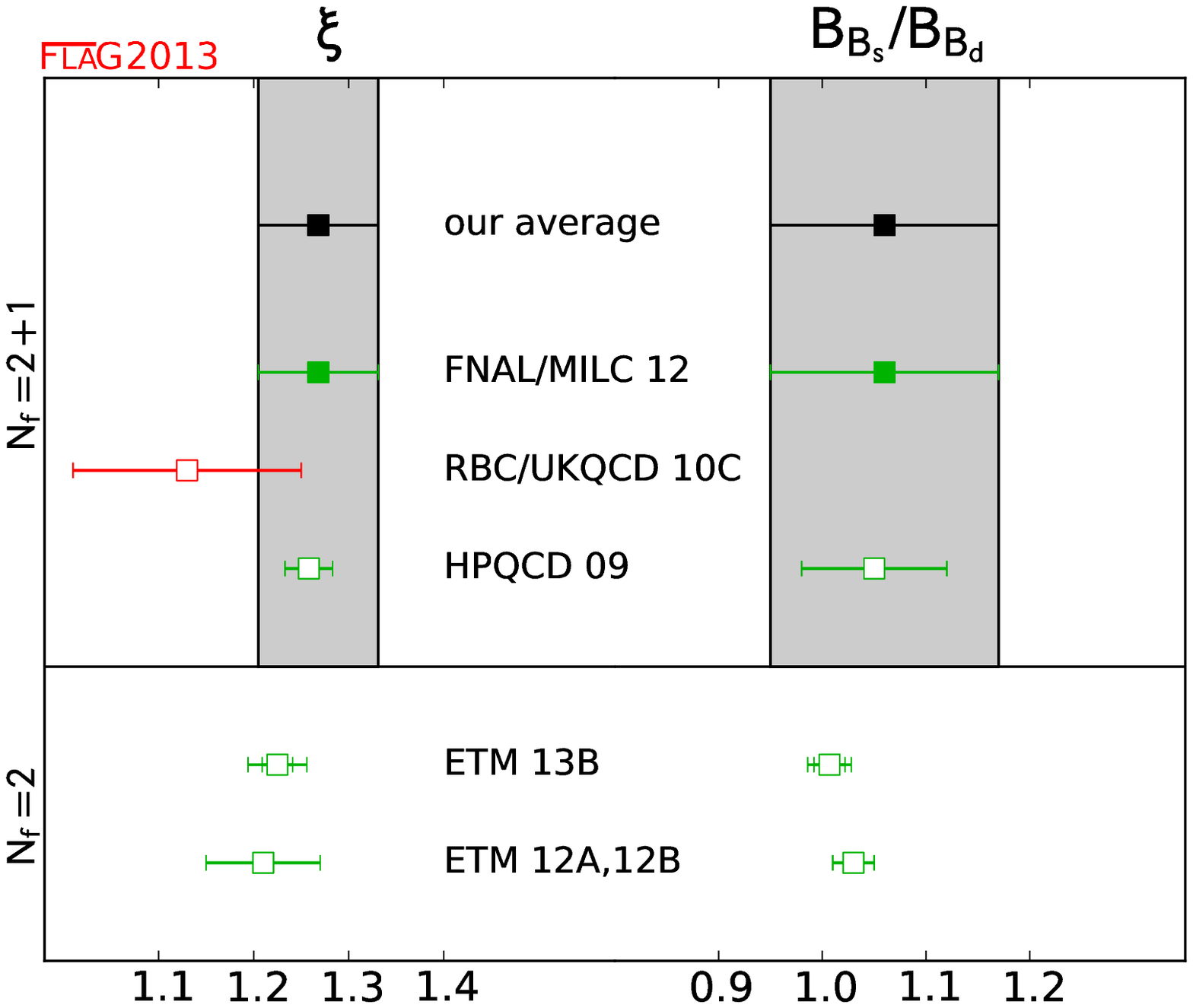}

\vspace{-2mm}
\caption{The $SU(3)$ breaking quantities $\xi$ and $B_{B_s}/B_{B_d}$ [values in Table \ref{tab_BBratsumm} and Eq.~(\ref{eq:avxiBB})].}\label{fig:xi} 
\end{center}
\end{figure}

The ETM collaboration has presented their first results for $B$-mixing
quantities with $N_f=2$ sea quarks in
Refs.~\cite{Carrasco:2012dd,Carrasco:2012de} (ETM~12A,~12B) using
ensembles at three lattice spacings in the range $a \approx 0.065 -
0.098 \,{\rm fm}$ with a minimum pion mass of $270$~MeV. 
Additional ensembles at $a \approx 0.054 \,{\rm fm}$ are included
in ETM~13B \cite{Carrasco:2013zta}.
 The valence
and sea quarks are simulated with two different versions of the
twisted-mass Wilson fermion action. The heavy-quark masses are in the
charm region and above while keeping $am_h \lsim 0.6$ for ETM~12A and 12B.
Larger masses up to $am_h \lsim 0.85$ are used for ETM~13B.
In this series of calculations
the ratio method first developed for $B$-meson decay constants
(see Appendix~\ref{app:HQactions} and Section~\ref{sec:fB}) is
extended to $B$-meson mixing quantities.  ETM again constructs ratios
of $B$-mixing matrix elements (now called $\omega_{d(s)}$) that are
equal to unity in the static limit, including also an analogous ratio
for $\xi$. The renormalization of the four-quark operator is
calculated nonperturbatively in the RI'/MOM scheme.  As an
intermediate step for the interpolation to the physical $b$-quark
mass, these ratios include perturbative matching factors to match the
four-quark operator from QCD to HQET; 
these include tree-level and leading log contributions
in ETM~12A and 12B, and additionally next-to-leading-log
contributions in ETM~13B.
Similar to their decay constant analysis, ETM
analyses the $SU(3)$ breaking ratio of ratios,
$\omega_{s}/\omega_{\ell}$, and combines it with $\omega_{s}$ to
obtain $B_{B_d}$. The data are interpolated to a fixed set of
heavy-quark reference masses on all ensembles, and subsequently
extrapolated to the continuum and to the physical light-quark masses
in a combined fit. The interpolation to the physical $b$-quark mass is
linear or quadratic in the inverse of the heavy-quark mass.  
While ETM~13B reports RGI bag parameters,
ETM~12A and 12B report only 
$B_B(m_b)^{\overline{MS},NDR}$ at $m_b=4.35$ GeV.
Taking
$\alpha_s(M_Z)=0.1184$~\cite{Bethke:2009jm}, we apply an RGI
conversion factor of $\hat{B}_B/B_B(m_b)^{\overline{MS},NDR}=1.521$ to
obtain the $\hat{B}_B$ values quoted in Table~\ref{tab_BBssumm}.  The
observed discretization effects (as measured by the percentage
difference between the lattice data at the smallest lattice spacing
and the continuum extrapolated results) are $\lsim 1\%$ over the range
of heavy-quark masses used in their simulations. As a result, the
dominant error on the bag parameters and on the ratio of bag
parameters is the combined statistical uncertainty, whereas the
dominant error on the $SU(3)$ breaking ratio $\xi$ is due to the
chiral extrapolation. 
Because these studies appear either in conference
proceedings or preprint only, the results do not enter our averages.

For the $N_f=2+1$ case there are three
collaborations that have presented results for $B-\bar{B}$ mixing
matrix elements: HPQCD, RBC/UKQCD, and FNAL/MILC.  The first published
results are by the HPQCD collaboration
\cite{Dalgic:2006gp,Gamiz:2009ku} and use NRQCD $b$ quarks and Asqtad
light valence quarks on $N_f=2+1$ MILC Asqtad ensembles. In HPQCD 06A
\cite{Dalgic:2006gp} results are presented for $B_s$-mixing quantities
only, using one lattice spacing and two light sea-quark masses with a
minimum RMS pion mass of $510$~MeV. The observed sea-quark mass
dependence is much smaller than the rather large statistical
errors. This calculation uses one-loop mean-field improved lattice
perturbation theory for the operator renormalization. Discretization
errors cannot be estimated from the data with only one lattice
spacing, but are estimated using power counting arguments to be
smaller than the dominant statistical and renormalization errors. With
only one lattice spacing and given the rather large minimum RMS pion
mass, this result does not enter our averages.  These shortcomings are
removed in HPQCD~09 \cite{Gamiz:2009ku} with two lattice spacings, ($a
\approx 0.09, 0.12$~fm) and four or two sea-quark masses per lattice
spacing with a minimum RMS pion mass of about $400$~MeV. The
calculation is also extended to include both $B_d$ and $B_s$ mixing
quantities and thus also the $SU(3)$ breaking ratios.  A combined
chiral and continuum extrapolation of the data is performed, using NLO
HMrS$\chi$PT, supplemented by NNLO analytic and generic discretization
terms of ${\cal O}(\alpha_s a^2, a^4)$.  The dominant systematic error
is due to using one-loop mean-field improved lattice perturbation
theory for the operator renormalization and matching, the same as in
HPQCD~06. It is estimated as $4\%$ and $2.5\%$, respectively,
consistent with power counting.  The statistical, chiral, and
continuum extrapolation uncertainties are also prominent sources of
uncertainty, followed by heavy-quark truncation and scale setting
errors. The dominant error on $\xi$ is due to statistics and chiral
extrapolation. Finally, we note that this work uses an old
determination of $r_1=0.321(5)$ fm from Ref.~\cite{Gray:2005ur} to set
the scale, that has since been superseded, and that differs from the
new value by about two standard deviations. Dimensionless quantities
are, of course, affected by a change in $r_1$ only through the inputs,
which are a subdominant source of uncertainty. The scale uncertainty
itself is also subdominant in the error budget, and this change
therefore does not affect HPQCD 09's results for
$f_{B_q}\sqrt{B_{B_q}}$ outside of the total error.

The RBC/UKQCD collaboration has presented a result for
the $SU(3)$ breaking ratio $\xi$ in Ref.~\cite{Albertus:2010nm} using
a static-limit action on $N_f=2+1$ domain wall ensembles at a single
lattice spacing $a \approx 0.11$~fm with a minimum pion mass of
approximately $430$~MeV.  They use both HYP and APE smearing for the
static-limit action and one-loop mean field improved lattice
perturbation theory to renormalize the static-limit four-quark
operators. Effects of ${\cal O}(1/m_h)$ are not included in the
static-limit action and operators, but Ref.~\cite{Albertus:2010nm}
includes an estimate of this effect via power counting as ${\cal
  O}\left((m_s-m_d)/m_b\right)$ in the error budget.  The statistical
errors in this work are significant ($\sim5-6 \%$), as are the chiral
extrapolation errors ($\sim7 \%$, estimated from the difference
between fits using NLO $SU(2)$ HM$\chi$PT and a linear fit function),
due to the rather large pion masses used in this in this work. With
data at only one lattice spacing, discretization errors cannot be
estimated from the data, but a power counting estimate of this error
of $4\%$ is included in the systematic error budget. With only one
lattice spacing this result does not enter our averages.
The RBC/UKQCD collaboration 
reported at Lattice 2013 \cite{Ishikawa:2013faa} that they are
extending this study, using
HYP and HYP2 smearings for the static-limit action,
smaller pion masses, larger volumes and two lattice spacings.
The conference proceedings~\cite{Ishikawa:2013faa}, however, 
did not appear until after the
closing deadline and is therefore not included in this review.

Another calculation of the $SU(3)$ breaking ratio
$\xi$ is presented by the Fermilab Lattice and MILC collaborations in
Ref.~\cite{Bazavov:2012zs} (FNAL/MILC~12).  The calculation uses the
Fermilab method for the $b$ quarks together with Asqtad light and
strange valence quarks on a subset of the MILC Asqtad $N_f=2+1$
ensembles, including lattice spacings in the range $a \approx 0.09 -
0.12$~fm and a minimum RMS pion mass of approximately $320$~MeV.  This
analysis includes partially-quenched lattice data at six valence
light-quark masses for each sea-quark ensemble.  The operator
renormalizations are calculated using one-loop mean-field improved
lattice perturbation theory, which does not result in a significant
source of uncertainty for the $SU(3)$ breaking ratios.  The combined
chiral and continuum extrapolations use a chiral fit function based on
NLO partially quenched HMrS$\chi$PT supplemented by NNLO analytic
terms. Also included are light-quark discretization terms of ${\cal
  O}(\alpha_s^2 a^2, a^4)$. The combined statistical, light-quark
discretization, and chiral extrapolation error dominates the error
budget together with an uncertainty that is described as the error due
to the omission of ``wrong-spin contributions" (see below).  First
results for the $B$ mixing matrix elements from an ongoing FNAL/MILC
calculation of all $B$ meson mixing quantities on the full set of
Asqtad ensembles are presented in \cite{Bouchard:2011xj}, including
the matrix elements of all five operators that contribute to $B$ meson
mixing in the Standard Model and beyond .  The dominant uncertainties
on the matrix elements are due to the combined statistical, chiral
extrapolation, and light-quark discretization error and due to the
one-loop matching.  FNAL/MILC 11A reports results for
$f_{B_q}\sqrt{B_{B_q}}$ evaluated at $\mu=m_b$ in the $\overline{\rm
  MS}$ NDR scheme.  Taking $\alpha_s(M_Z)=0.1184$~\cite{Bethke:2009jm}
and $m_b = 4.19$~GeV~\cite{Beringer:1900zz}, we apply an RGI
conversion factor of $\hat{B}_B/B_B(m_b)^{\overline{MS},NDR}=1.517$ to
obtain the values for the RGI quantities listed in
Table~\ref{tab_BBssumm}. Ref.~\cite{Bouchard:2011xj} presents a
complete error budget, but since the paper is a conference
proceedings, its results are not included in our averages.

For the $N_f=2$ case there are no published results,
so we do not quote an average for this case.  For $N_f=2+1$ only the
results of HPQCD 09 and FNAL/MILC~12 enter our averages.  First, we
must consider the issue of the so-called ``wrong-spin contributions,"
described in Ref.~\cite{Bazavov:2012zs} and explained in detail in
Ref.~\cite{Bernard:2013dfa}.  With staggered light quarks,
interactions between different unphysical species (``tastes'') of
quarks induce mixing between the operator ${\mathcal Q}_1^q$ in
Eq.~(\ref{eq:Q1}) and the operators ${\mathcal Q}_2^q$ and ${\mathcal
  Q}_3^q$ in Eq.~(\ref{eq:Q25}) at ${\mathcal O}(a^2)$.  These
additional contributions to the matrix element $f_{B_q}
\sqrt{B_{B_q}}$ are discretization errors that vanish in the continuum
limit.  The contributions of ${\mathcal Q}_1^q$--${\mathcal Q}_5^q$
have been derived at next-to-leading order in
HMrS$\chi$PT~\cite{Bernard:2013dfa}.  The result is that, in the
chiral expansion of the matrix elements of ${\mathcal Q}_1^q$, the
matrix elements of ${\mathcal Q}_{2,3}^q$ appear with ${\mathcal
  O}(a^2)$ coefficients that depend upon the light-quark masses.
These contributions can be accounted for in the chiral-continuum
extrapolation by fitting the numerical results for the matrix elements
of the three operators simultaneously.  Further, if the matrix
elements of all five basis operators in Eqs.~(\ref{eq:Q1})
and~(\ref{eq:Q25}) are computed on the lattice, then no additional
low-energy constants are required to describe the wrong-spin
contributions effects in the chiral-continuum extrapolation.  In
principle, instead of using HMrS$\chi$PT as described above, it is
possible to account for the wrong-spin terms via the inclusion of
generic mass-dependent terms such as $O(a^2 m^2_{\pi})$ in the
combined chiral-continuum extrapolation, provided that the lattice
spacing and light-quark masses are small enough.

Both HPQCD~09 and FNAL/MILC~11A use chiral fit functions based on NLO
HMrS$\chi$PT.  Since, however, these works predate
Refs.~\cite{Bazavov:2012zs,Bernard:2013dfa}, the wrong-spin terms are
not included in their chiral extrapolations.  The calculation in
FNAL/MILC~12 also does not include the matrix elements of all three
operators, so here the effect of the wrong-spin contributions is
treated as a systematic error, which is estimated using the lattice
data described in Ref.~\cite{Bouchard:2011xj}. As discussed above, the
estimated uncertainty of 3\% for $\xi$ is a dominant contribution to
the error budget in Ref.~\cite{Bazavov:2012zs}.  Because, however,
HPQCD~09 does not include the wrong-spin contributions in its chiral
extrapolations, we must consider how they affect the results. First,
the chiral fit functions used in HPQCD~09 and in FNAL/MILC~12 are very
similar with similar (though not identical) choices for prior
widths. The main difference is that the generic light-quark
discretization term of ${\cal O}(\alpha_s a^2)$ included in HPQCD 09
is a little less constrained than the ${\cal O}(\alpha_s^2 a^2)$ term
included in FNAL/MILC~12.  It is therefore possible that the chiral
extrapolation in HPQCD~09 accounts for the wrong-spin contributions
via the generic discretization terms.  Furthermore, for
$f_{B_q}\sqrt{B_{B_q}}$ the chiral extrapolation error, while not
insignificant, is not a dominant source of error in the HPQCD
calculation.  For $\xi$, however, the chiral extrapolation error is a
dominant source of uncertainty, and the FNAL/MILC~12 analysis
indicates that the omission of the wrong-spin contributions from
HMrS$\chi$PT fits may also be a significant source of error.  We
therefore make the conservative choice of excluding HPQCD 09's result
for $\xi$ from our average, but keeping HPQCD~09's results for
$f_{B_q}\sqrt{B_{B_q}}$ and $B_{B_q}$ in our averages.  As a result,
we now have only one calculation that enters our averages for each
quantity. Our averages are ($N_f=2+1$):
\begin{eqnarray} 
 && f_{B_d}\sqrt{\hat{B}_{B_d}} =  216(15)\;\; {\rm MeV}\,,\hspace{2em}
 f_{B_s}\sqrt{\hat{B}_{B_s}}  =  266(18)\;\; {\rm MeV}\,,  \label{eq:avfBB}\\
 &&\hat{B}_{B_d}  = 1.27(10)\,, \hspace{7em}
 \hat{B}_{B_s} =  1.33(6)\,, \label{eq:avBB}\\
 &&\xi  =  1.268(63)\,, \hspace{7.8em}
 B_{B_s}/B_{B_d}  =  1.06(11). \label{eq:avxiBB}
\end{eqnarray}
Finally, we note that the above results are all correlated with each other:  the numbers in
(\ref{eq:avfBB}) and (\ref{eq:avBB}) are from HPQCD~09 \cite{Gamiz:2009ku}, while those in (\ref{eq:avxiBB})  
are from FNAL/MILC~12 \cite{Bazavov:2012zs} --
the same Asqtad MILC ensembles are used in these 
simulations. The results are also correlated with the averages obtained in 
Section~\ref{sec:fB} and shown in
Eq.~(\ref{eq:fbav21}), because the calculations of $B$-meson decay constants and  
mixing quantities 
are performed on the same (or on similar) sets of ensembles, and results obtained by a 
given collaboration 
use the same actions and setups. These correlations must be considered when 
using our averages as inputs to UT fits. In the future, as more independent 
calculations enter the averages, correlations between the lattice-QCD inputs to the UT 
fit will become less significant.

\subsection{Semileptonic form factors for $B$ decays to light flavours}
\label{sec:BtoPiK}

The Standard Model differential rate for the decay $B_{(s)}\to
P\ell\nu$ involving a quark-level $b\to u$ transition is given, at
leading order in the weak interaction, by a formula identical to the
one for $D$ decays in Eq.~(\ref{eq:DtoPiKFull}) but with $D \to
B_{(s)}$ and the relevant CKM matrix element $|V_{cq}| \to |V_{ub}|$:
\begin{eqnarray}
	\frac{d\Gamma(B_{(s)}\to P\ell\nu)}{dq^2} = \frac{G_F^2 |V_{ub}|^2}{24 \pi^3}
	\,\frac{(q^2-m_\ell^2)^2\sqrt{E_P^2-m_P^2}}{q^4m_{B_{(s)}}^2}
	\bigg[& \!\!\!\!\!\!\!\!\!\!\! \left(1+\frac{m_\ell^2}{2q^2}\right)m_{B_{(s)}}^2(E_P^2-m_P^2)|f_+(q^2)|^2 \nonumber\\
&~~~~\,+\,\frac{3m_\ell^2}{8q^2}(m_{B_{(s)}}^2-m_P^2)^2|f_0(q^2)|^2
\bigg]\,. \label{eq:B_semileptonic_rate}
\end{eqnarray}
Again, for $\ell=e,\mu$ the contribution from the scalar form factor
$f_0$ can be neglected, and one has a similar expression to
Eq.~(\ref{eq:DtoPiK}), which in principle allows for a direct
extraction of $|V_{ub}|$ by matching theoretical predictions to
experimental data.  However, while for $D$ (or $K$) decays the entire
physical range $0 \leq q^2 \leq q^2_{\rm max}$ can be covered with
moderate momenta accessible to lattice simulations, in
$B \to \pi \ell\nu$ decays one has $q^2_{\rm max} \sim 26~{\rm GeV}^2$
and only part of the full kinematic range is reachable.  As a
consequence, obtaining $|V_{ub}|$ from $B\to\pi\ell\nu$ is more
complicated then obtaining $|V_{cd(s)}|$ from semileptonic $D$-meson
decays.  The standard procedure involves the matching of theoretical
predictions and experimental data for the integrated decay rate over a
limited $q^2$ range,
\begin{gather}\label{eq:Deltazeta}
	\Delta \zeta = \frac{1}{|V_{ub}|^2} \int_{q^2_{1}}^{q^2_{2}} \left( \frac{d \Gamma}{d q^2} \right) dq^2\,.
\end{gather}
This requires knowledge of the relevant form factor(s) within the
integration interval. In practice, lattice computations are restricted
to small values of the momentum transfer (see Sec.~\ref{sec:DtoPiK})
where statistical and momentum-dependent discretization errors can be
controlled,\footnote{The variance of hadron correlation functions at
nonzero momentum is dominated at large Euclidean times by
zero-momentum multiparticle states~\cite{DellaMorte:2012xc}; therefore
the noise-to-signal grows more rapidly than for the vanishing momentum
case.} which in existing calculations roughly cover the upper third of
the kinematically allowed $q^2$ range.  Experimental results normally
cover the whole interval, but are more precise in the low-$q^2$
region.  Therefore, both experimental and lattice data for the $q^2$
dependence have to be parameterized by fitting data to a specific
ansatz, either separately or jointly (with the relative normalization
$|V_{ub}|^2$ as a free parameter). A good control of the systematic
uncertainty induced by the choice of parameterization is hence crucial
to obtain a precise determination of $|V_{ub}|$.

\subsubsection{Parameterizations of heavy-to-light semileptonic form factors}
\label{sec:zparam}

All form factors are analytic functions of $q^2$ outside physical
poles and inelastic threshold branch points; in the case of
$B\to\pi\ell\nu$, the only pole expected below the $B\pi$ production
region, starting at $q^2 = t_+ = (m_B+m_\pi)^2$, is the $B^*$.  A
simple ansatz for the $q^2$ dependence of the $B\to\pi\ell\nu$
semileptonic form factors that incorporates vector-meson dominance is
the Be\'cirevi\'c-Kaidalov (BK)
parameterization~\cite{Becirevic:1999kt}:
\begin{gather}
f_+(q^2) = \frac{f(0)}{(1-q^2/m_{B^*}^2)(1-\alpha q^2/m_{B^*}^2)}\,,~~~~~~~~~~~~
f_0(q^2) = \frac{f(0)}{1-\frac{1}{\beta}\,q^2/m_{B^*}^2}\,. \label{eq:BKparam}
\end{gather}
Because the BK ansatz has few free parameters, it has been used
extensively to parameterize the shape of experimental
branching-fraction measurements and theoretical form-factor
calculations.  A variant of this parameterization proposed by Ball and
Zwicky (BZ) adds extra pole factors to the expressions in
Eq.~(\ref{eq:BKparam}) in order to mimic the effect of multiparticle
states~\cite{Ball:2004ye}.  Another variant (RH) has been proposed by
Hill in~\cite{Hill:2005ju}.  Although all of these parameterizations
capture some known properties of form factors, they do not manifestly
satisfy others.  For example, perturbative QCD scaling constrains the
high-$q^2$ behaviour to be $f_+(q^2)\sim 1/q^2$ up to logarithmic
corrections~\cite{Lepage:1980fj,Akhoury:1993uw,Lellouch:1995yv}, and
angular momentum conservation constrains the asymptotic behaviour near
thresholds --- e.g. ${\rm Im}\,f_+(q^2) \sim (q^2-t_+)^{3/2}$ (see
e.g.~\cite{Bourrely:2008za}).  Further, they do not allow for an easy
quantification of systematic uncertainties.

A more systematic approach that improves upon the use of simple models
for the $q^2$ behaviour exploits the positivity and analyticity
properties of two-point functions of vector currents to obtain optimal
parameterizations of form
factors~\cite{Bourrely:1980gp,Boyd:1994tt,Lellouch:1995yv,Boyd:1997qw,Arnesen:2005ez,Becher:2005bg}.
Any form factor $f$ can be shown to admit a series expansion of the
form
\begin{gather}
f(q^2) = \frac{1}{B(q^2)\phi(q^2,t_0)}\,\sum_{n=0}^\infty a_n(t_0)\,z(q^2,t_0)^n\,,
\end{gather}
where the squared momentum transfer is replaced by the variable
\begin{gather}
z(q^2,t_0) = \frac{\sqrt{t_+-q^2}-\sqrt{t_+-t_0}}{\sqrt{t_+-q^2}+\sqrt{t_+-t_0}}\,.
\end{gather}
This is a conformal transformation, depending on an arbitrary real
parameter $t_0<t_+$, that maps the $q^2$ plane cut for $q^2 \geq t_+$
onto the disk $|z(q^2,t_0)|<1$ in the $z$ complex plane. The function
$B(q^2)$ is called the {\it Blaschke factor}, and contains poles and
cuts below $t_+$ --- for instance, in the case of $B\to\pi$ decays
\begin{gather}
B(q^2)=\frac{z(q^2,t_0)-z(m_{B^*}^2,t_0)}{1-z(q^2,t_0)z(m_{B^*}^2,t_0)}=z(q^2,m_{B^*}^2)\,.
\end{gather}
Finally, the quantity $\phi(q^2,t_0)$, called the {\em outer
function}, is an analytic function that does not introduce further
poles or branch cuts.  The crucial property of this series expansion
is that the sum of the squares of the coefficients
\begin{gather}
\sum_{n=0}^\infty a_n^2 = \frac{1}{2\pi i}\oint \frac{dz}{z}\,|B(z)\phi(z)f(z)|^2\,,
\end{gather}
is a finite quantity. Therefore, by using this parameterization an
absolute bound to the uncertainty induced by truncating the series can
be obtained.  The criteria involved in the optimal choice of $\phi$
then aim at obtaining a bound that is useful in practice, while
(ideally) preserving the correct behaviour of the form factor at high
$q^2$ and around thresholds.

The simplest form of the bound would correspond to $\sum_{n=0}^\infty
a_n^2=1$.  {\it Imposing} this bound yields the following ``standard''
choice for the outer function
\begin{gather}
\label{eq:comp_of}
\begin{split}
\phi(q^2,t_0)=&\sqrt{\frac{1}{32\pi\chi_{1^-}(0)}}\,
\left(\sqrt{t_+-q^2}+\sqrt{t_+-t_0}\right)\\
&\times\,\left(\sqrt{t_+-q^2}+\sqrt{t_+-t_-}\right)^{3/2}
\left(\sqrt{t_+-q^2}+\sqrt{t_+}\right)^{-5}
\,\frac{t_+-q^2}{(t_+-t_0)^{1/4}}\,,
\end{split}
\end{gather}
where $\chi_{1^-}(0)$ is the derivative of the transverse component of
the polarisation function (i.e. the Fourier transform of the vector
two-point function) $\Pi_{\mu\nu}(q)$ at Euclidian momentum
$Q^2=-q^2=0$. It is computed perturbatively, using operator product
expansion techniques, by relating the $B\to\pi\ell\nu$ decay amplitude
to $\ell\nu\to B\pi$ inelastic scattering via crossing symmetry and
reproducing the correct value of the inclusive rate $\ell\nu\to X_b$.
We will refer to the series parameterization with the outer function
in Eq.~(\ref{eq:comp_of}) as Boyd, Grinstein, and Lebed (BGL).  The
perturbative and OPE truncations imply that the bound is not strict,
and one should take it as
\begin{gather}
\sum_{n=0}^N a_n^2 \lesssim 1\,,
\end{gather}
where this holds for any choice of $N$.  Since the values of $|z|$ in
the kinematical region of interest are well below~1 for judicious
choices of $t_0$, this provides a very stringent bound on systematic
uncertainties related to truncation for $N\geq 2$. On the other hand,
the outer function in Eq.~(\ref{eq:comp_of}) is somewhat unwieldy and,
more relevantly, spoils the correct large $q^2$ behaviour and induces
an unphysical singularity at the $B\pi$ threshold.

A simpler choice of outer function has been proposed by Bourrely,
Caprini and Lellouch (BCL) in~\cite{Bourrely:2008za}, which leads to a
parameterization of the form
\begin{gather}
\label{eq:bcl}
f_+(q^2)=\frac{1}{1-q^2/m_{B^*}^2}\,\sum_{n=0}^N a_n(t_0) z(q^2,t_0)^n\,.
\end{gather}
This satisfies all the basic properties of the form factor, at the price
of changing the expression for the bound to
\begin{gather}
\sum_{j,k=0}^N B_{jk}(t_0)a_j(t_0)a_k(t_0) \leq 1\,.
\end{gather}
The constants $B_{jk}$ can be computed and shown to be
$|B_{jk}|\lesssim \mathcal{O}(10^{-2})$ for judicious choices of
$t_0$; therefore, one again finds that truncating at $N\geq 2$
provides sufficiently stringent bounds for the current level of
experimental and theoretical precision.  It is actually possible to
optimize the properties of the expansion by taking
\begin{gather}
t_0 = t_{\rm opt} = (m_B-m_\pi)(\sqrt{m_B}-\sqrt{m_\pi})^2\,,
\end{gather}
which for physical values of the masses results in the semileptonic
domain being mapped onto the symmetric interval $|z| \ltapprox 0.279$
(where this range differs slightly for the $B^{\pm}$ and $B^0$ decay
channels), minimizing the maximum truncation error.  If one also
imposes that the asymptotic behaviour ${\rm Im}\,f_+(q^2) \sim
(q^2-t_+)^{3/2}$ near threshold is satisfied, then the highest-order
coefficient is further constrained as
\begin{gather}
\label{eq:red_coeff}
a_N=-\,\frac{(-1)^N}{N}\,\sum_{n=0}^{N-1}(-1)^n\,n\,a_n\,.
\end{gather}
Substituting the above constraint on $a_N$ into Eq.~(\ref{eq:bcl})
leads to the constrained BCL parameterization
\begin{gather}
\label{eq:bcl_c}
f_+(q^2)=\frac{1}{1-q^2/m_{B^*}^2}\,\sum_{n=0}^{N-1} a_n\left[z^n-(-1)^{n-N}\,\frac{n}{N}\,z^N\right]\,,
\end{gather}
which is the standard implementation of the BCL parameterization used
in the literature.

Parameterizations of the BGL and BCL kind (to which we will refer
collectively as ``$z$-parameterizations'') have already been adopted
by the Babar and Belle collaborations to report their results, and
also by the Heavy Flavour Averaging Group (HFAG). Some lattice
collaborations, such as FNAL/MILC and ALPHA, have already started to
report their results for form factors in this way.  The emerging trend
is to use the BCL parameterization as a standard way of presenting
results for the $q^2$ dependence of semileptonic form factors. Our
policy will be to quote results for $z$-parameterizations when the
latter are provided in the paper (including the covariance matrix of
the fits); when this is not the case, but the published form factors
include the full correlation matrix for values at different $q^2$, we
will perform our own fit to the constrained BCL ansatz
in~Eq.(\ref{eq:bcl_c}); otherwise no fit will be quoted.

\subsubsection{Form factors for $B\to\pi\ell\nu$ and $B_s\to K\ell\nu$}

The semileptonic decay processes $B\to\pi\ell\nu$ and $B_s\to
K\ell\nu$ enable determinations of the CKM matrix element $|V_{ub}|$
within the Standard Model via Eq.~(\ref{eq:B_semileptonic_rate}).
Results for the $B\to\pi\ell\nu$ form factors have been published by
the HPQCD~\cite{Dalgic:2006dt} and FNAL/MILC~\cite{Bailey:2008wp}
Collaborations, in both cases for $N_f=2+1$ dynamical quark flavours.
Work is also underway by ALPHA~\cite{Bahr:2012vt,Bahr:2012qs} (on
$N_f=2$ nonperturbatively $\mathcal{O}(a)$ improved Wilson
configurations), FNAL/MILC~\cite{Zhou:2012sna,Du:2013kea} (updating
the published analysis), HPQCD~\cite{Bouchard:2012tb,Bouchard:2013zda}
(with HISQ valence light quarks), and the RBC/UKQCD
Collaborations~\cite{Kawanai:2012id,Kawanai:2013qxa} (with $N_f=2+1$
DWF). These calculations, however, are so far described only in
conference proceedings which do not provide quotable results, so they
will not be discussed in this report.  No unquenched computation of
$B_s\to K\ell\nu$ form factors is currently available.  Preliminary
results by the HPQCD Collaboration are reported
in~\cite{Bouchard:2012tb,Bouchard:2013zda}, while work in progress by
the FNAL/MILC Collaboration is discussed
in~\cite{Zhou:2012sna,Liu:2013sya}.

Both the HPQCD and the FNAL/MILC computations of the $B\to\pi\ell\nu$
amplitudes use ensembles of gauge configurations with $N_f=2+1$
flavours of rooted staggered quarks produced by the MILC Collaboration
at two different values of the lattice spacing ($a\sim0.12,~0.09~{\rm
fm}$). The relative scale is fixed in both cases through $r_1/a$,
while the absolute scale is set through the $\Upsilon$ $2S$--$1S$
splitting for HPQCD and $f_\pi$ (with uncertainty estimated from the
same $\Upsilon$ splitting) for FNAL/MILC.  The spatial extent of the
lattices is $L\simeq 2.4~{\rm fm}$, save for the lightest mass point
($a\sim 0.09~{\rm fm}$) for which $L\simeq 2.9~{\rm fm}$. The lightest
RMS pion mass is around $400~{\rm MeV}$. Lattice-discretization
effects are estimated within HMrS$\chi$PT in the FNAL/MILC
computation, while HPQCD quotes the results at $a\sim 0.12~{\rm fm}$
as central values and uses the $a\sim 0.09~{\rm fm}$ results to quote
an uncertainty.

The main difference between the computations lies in the treatment of
heavy quarks. HPQCD uses the NRQCD formalism, with a one-loop matching
of the relevant currents to the ones in the relativistic
theory. FNAL/MILC employs the clover action with the Fermilab
interpretation, with a mostly nonperturbative renormalization of the
relevant currents, within which light-light and heavy-heavy currents
are renormalized nonperturbatively and one-loop perturbation theory is
used for the relative normalization.  (See Table~\ref{tab_BtoPisumm2};
full details about the computations are provided in tables in
Appendix~\ref{app:BtoPi_Notes}.)

Chiral extrapolations are an important source of systematic
uncertainty, since the pion masses at which the computations are
carried out are relatively heavy. In order to control deviations from
the expected $\chi$PT behaviour, FNAL/MILC supplements $SU(3)$
HMrS$\chi$PT formulae with higher-order powers in $E_\pi$ to extend
the form factor parameterization up to $E_\pi \sim 1~{\rm
GeV}$. Chiral extrapolation effects do indeed make the largest
contribution to their systematic error budget. HPQCD performs chiral
extrapolations using HMrS$\chi$PT formulae, and estimates systematic
uncertainties by comparing the result with the ones from fits to a
linear behaviour in the light-quark mass, continuum HM$\chi$PT, and
partially quenched HMrS$\chi$PT formulae (including also data with
different sea and valence light quark masses).  This is again the
dominant contribution to the error budget of the computation, along
with the matching of the heavy-light current.

HPQCD provides results for both $f_+(q^2)$ and $f_0(q^2)$. In this
case, the parameterization of the $q^2$ dependence of form factors is
somewhat intertwined with chiral extrapolations: a set of fiducial
values $\{E_\pi^{(n)}\}$ is fixed for each value of the light-quark
mass, and $f_{+,0}$ are interpolated to each of the $E_\pi^{(n)}$;
chiral extrapolations are then performed at fixed $E_\pi$. The
interpolation is performed using a BZ ansatz.  The $q^2$ dependence of
the resulting form factors in the chiral limit is then described by
means of a BZ ansatz, which is cross-checked against BK, RH, and BGL
parameterizations. FNAL/MILC presents results for $f_+(q^2)$ only, and
provides as its preferred description a three-parameter fit to the BGL
form in a companion paper~\cite{Bernard:2009ke}; this result is quoted
in Table~\ref{tab_BtoPisumm2}.  HPQCD, on the other hand, does not
provide the correlation matrix for the values of $f_{+}(q^2)$ in the
chiral limit, and therefore no independent fit to a
$z$-parameterization is possible.

Results for the integrated decay rate $\Delta \zeta^{B\pi}$, which is
defined in equation (\ref{eq:Deltazeta}) and depends on the chosen
interval of integration, are available in both cases (see
Table \ref{tab_BtoPisumm2} and Fig. \ref{fig:BtoPi}). We quote the
average ($q_1=4\, \mbox{GeV}$, $q_2=q_{max}$):
\begin{gather}\label{eq:DeltazetaBtoPi}
N_f=2+1: \qquad \Delta \zeta^{B\pi} = 2.16(50)~{\rm ps}^{-1}\,,
\end{gather}
where we have conservatively assumed that the calculations are 100\%
correlated because neither FNAL/MILC nor HPQCD provide itemized error
budgets for $\Delta
\zeta^{B\pi}$.\footnote{These calculations are based on an overlapping set of gauge-field ensembles, so
their statistical errors are highly correlated.  They use different
heavy-quark actions, renormalization methods, and chiral extrapolation
fit functions, however, so we expect their systematic errors to be
largely uncorrelated.  Therefore we expect that assuming 100\%
correlation will lead to a conservative (over)estimate for the errors
in $\Delta \zeta^{B\pi}$.}

The results for $f_+(q^2)$ in HPQCD~06 and FNAL/MILC~08A can also be
combined into a single fit to our preferred BCL $z$-parameterization,
Eq.~(\ref{eq:bcl_c}).  While FNAL/MILC~08A provides the full
correlation matrix between $f_+(q^2)$ values, this information is not
available for HPQCD data; we thus perform a simultaneous fit including
all the $f_+(q^2)$ values from FNAL/MILC~08A and only one point from
HPQCD~06.  The value of $f_+$ from HPQCD~06 that we choose to include
in the fit is the one at the lowest quoted momentum transfer for which
no extrapolation in the energy of the final state pion is involved in
the computation, $q^2_{\rm min}=17.35~{\rm GeV}^2$.  Since in
FNAL/MILC~08A $q^2_{\rm min}=18.4~{\rm GeV}^2$, this extends the
covered kinematical range, and, together with the smaller relative
error of the HPQCD datum, results in the latter having a significant
weight in the fit.  The HPQCD and FNAL/MILC computations are
correlated by the use of an overlapping set of gauge-field ensembles
for the evaluation of observables.  We therefore treat the combined
statistical plus chiral-extrapolation errors as 100\% correlated
between the two calculations in the fit.  We treat the other
systematic uncertainties as uncorrelated because they are mostly
associated with the choice of $b$-quark action, which is different in
the two calculations.

\begin{table}[h]
\begin{center}
\mbox{} \\[3.0cm]
\footnotesize
\begin{tabular*}{\textwidth}[l]{l @{\extracolsep{\fill}} c @{\hspace{2mm}} c l l l l l l @{\hspace{0mm}} c @{\hspace{2mm}} c m{1.7cm} @{\hspace{0mm}} c}
Collaboration & Ref. & $\Nf$ & 
\hspace{0.15cm}\begin{rotate}{60}{publication status}\end{rotate}\hspace{-0.15cm} &
\hspace{0.15cm}\begin{rotate}{60}{continuum extrapolation}\end{rotate}\hspace{-0.15cm} &
\hspace{0.15cm}\begin{rotate}{60}{chiral extrapolation}\end{rotate}\hspace{-0.15cm}&
\hspace{0.15cm}\begin{rotate}{60}{finite volume}\end{rotate}\hspace{-0.15cm}&
\hspace{0.15cm}\begin{rotate}{60}{renormalization}\end{rotate}\hspace{-0.15cm}  &
\hspace{0.15cm}\begin{rotate}{60}{heavy-quark treatment}\end{rotate}\hspace{-0.15cm}  &
\rule{0.3cm}{0cm}$\Delta \zeta^{B\pi} $ & \multicolumn{2}{c}{ \hspace{-3mm} \specialcellthree[b]{$z$-parameterization \\[0.0cm] \parbox[c]{3.0cm}{type \hfill $\left\{ a_0, a_1, a_2 \right\}$}} } & \parbox[b]{0.9cm}{\centering cov. \\[-0.5mm] \vspace{0mm}matrix} \\
&&&&&&&&&& \\[-0.1cm]
\hline
\hline
&&&&&&&&&& \\[-0.6cm]
FNAL/MILC 08A & \cite{Bailey:2008wp} & 2+1 & \gA  & \soso & \soso & \good & \soso & \okay &
$\quad $2.21$^{+0.47}_{-0.42}$$^\dagger$ & BGL$^\ddagger$ & \begin{center}$\big\{0.0216(27)$, $-0.038(19)$, $-0.113(27)\big\}$\end{center} & yes$^\S$  \\[-0.5cm]
HPQCD 06 & \cite{Dalgic:2006dt} & 2+1 & \gA  & \soso & \soso & \good
& \soso & \okay &
$\quad $2.07(41)(39) & --  & \begin{center}--\end{center} & no \\[-0.4cm]
&&&&&&&&&&&& \\[-0.1cm]
\hline
\hline
\end{tabular*}
\begin{tabular*}{\textwidth}[l]{l@{\extracolsep{\fill}}llllllllllll}
  \multicolumn{13}{l}{\vbox{\begin{flushleft}
  $^\dagger$ Value based on the calculation of Ref.~\cite{Bailey:2008wp} (private communication with the FNAL/MILC collaboration). \\
  $^\ddagger$ Result of BGL fit to FNAL/MILC data in Ref.~\cite{Bailey:2008wp} using $\chi_{1^-}(0) = 6.88919 \times 10^{-4}$ and given in~\cite{Bernard:2009ke}. \\
 $^\S$ Covariance matrix $C_{ij} = {\rm cov}(a_i,a_j)$ given in Table~IV of Ref.~\cite{Bernard:2009ke}.\\  
\end{flushleft}}}
\end{tabular*}
\caption{Results for the $B \to \pi\ell\nu$ semileptonic form factor.  The quantity $\Delta\zeta$ is defined in Eq.~(\ref{eq:Deltazeta}); the quoted values correspond to $q_1=4$~GeV, $q_2=q_{max}$, and are given in $\mbox{ps}^{-1}$. The ``cov.~matrix" entry indicates whether or not the correlations, either between the lattice form-factor data at different values of $q^2$, or between the coefficients of a $z$-parameterization, are provided.  This information is needed to use the lattice results in a combined fit to obtain $|V_{ub}|$. \label{tab_BtoPisumm2}}
\end{center}
\end{table}

\begin{figure}[tbp]
\begin{center}
\includegraphics[width=0.55\linewidth]{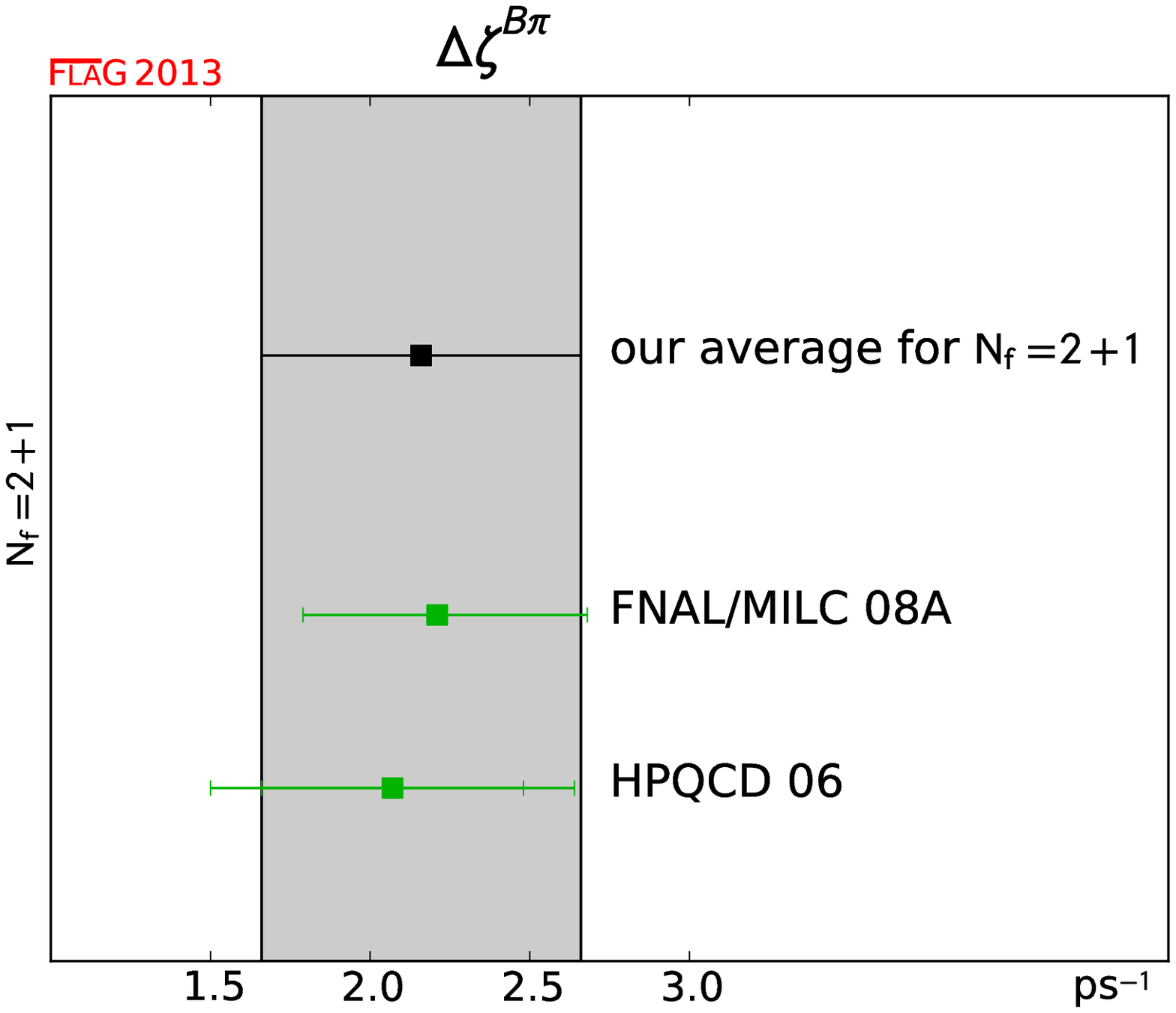}
\vspace{-2mm}
\caption{Integrated width of the decay $B\to\pi\ell\nu$ divided by $|V_{ub}|^2$ [values in Table \ref{tab_BtoPisumm2} and Eq.~(\ref{eq:DeltazetaBtoPi})].}\label{fig:BtoPi}
\end{center}
\end{figure}

We fit the two sets of lattice data for $f_+(q^2)$ together to the BCL
parameterization in Eq.~(\ref{eq:bcl_c}) and assess the systematic
uncertainty due to truncating the series expansion by considering fits
to different orders in $z$.  Figure~\ref{fig:LQCDzfit} plots the
FNAL/MILC and HPQCD data points for $(1-q^2/m_{B^*}^2) f_+(q^2)$
versus $z$; the data is highly linear, and only a simple two-parameter
fit is needed for a good $\chi^2/{\rm d.o.f.}$. (Note that a fit to
the constrained BCL form in Eq.~(\ref{eq:bcl_c}) with two free
parameters corresponds to a polynomial through ${\mathcal O}(z^2)$,
etc.)  Further, we cannot constrain the coefficients of the
$z$-expansion beyond this order, as evidenced by the error on the
coefficient $a_2$ being significantly greater than 100\% for a
three-parameter fit.  Because the FNAL/MILC synthetic data points are
all from the output of the same chiral-continuum extrapolation, they
are strongly correlated, so inverting the full $12\times 12$
correlation matrix is problematic.  We address these correlations in
the FNAL/MILC data in several ways and make sure that the outcome of
the fit is stable: we thin the data set to either six (every other) or
four (every third) points, and imposing singular value decomposition
(SVD) cuts of various severities in the construction of the
pseudoinverse.  The results (central values and errors) for the fit
parameters are all very consistent irrespective of the treatment of
correlations.

We quote as our preferred result the outcome of the three-parameter
${\mathcal O}(z^3)$ BCL fit using a thinned FNAL/MILC dataset that
includes every second data point starting at $q^2 = 18.4~{\rm GeV}^2$
in addition to the HPQCD point at $q^2=17.35~{\rm GeV}^2$:
 \begin{gather}
N_f=2+1:  \qquad 
a_0 = 0.453(33)\,,~~~~
a_1 = -0.43(33)\,,~~~~
a_2 = 0.9(3.9)\,;\label{eq:BtoPiLatBCLFit}\\[1.0ex]
\nonumber \qquad\qquad
{\rm cov}(a_i,a_j)=\left(\begin{array}{rrr}
 1.00 & -0.55 & -0.63 \\
-0.55 &  1.00 &  0.59 \\
-0.63 &  0.59 &  1.00
\end{array}\right)\,,
\end{gather}
where the above uncertainties encompass both the lattice errors and
the systematic error due to truncating the series in $z$.  This can be
used as the averaged FLAG result for the lattice-computed form factor
$f_+(q^2)$.  The coefficient $a_3$ can be obtained from the values for
$a_0$--$a_2$ using Eq.~(\ref{eq:red_coeff}).  We emphasize that future
lattice-QCD calculations of semileptonic form factors should publish
their full statistical and systematic correlation matrices to enable
others to use the data fully.

\begin{figure}[tbp]
\begin{center}
\includegraphics[width=0.55\linewidth]{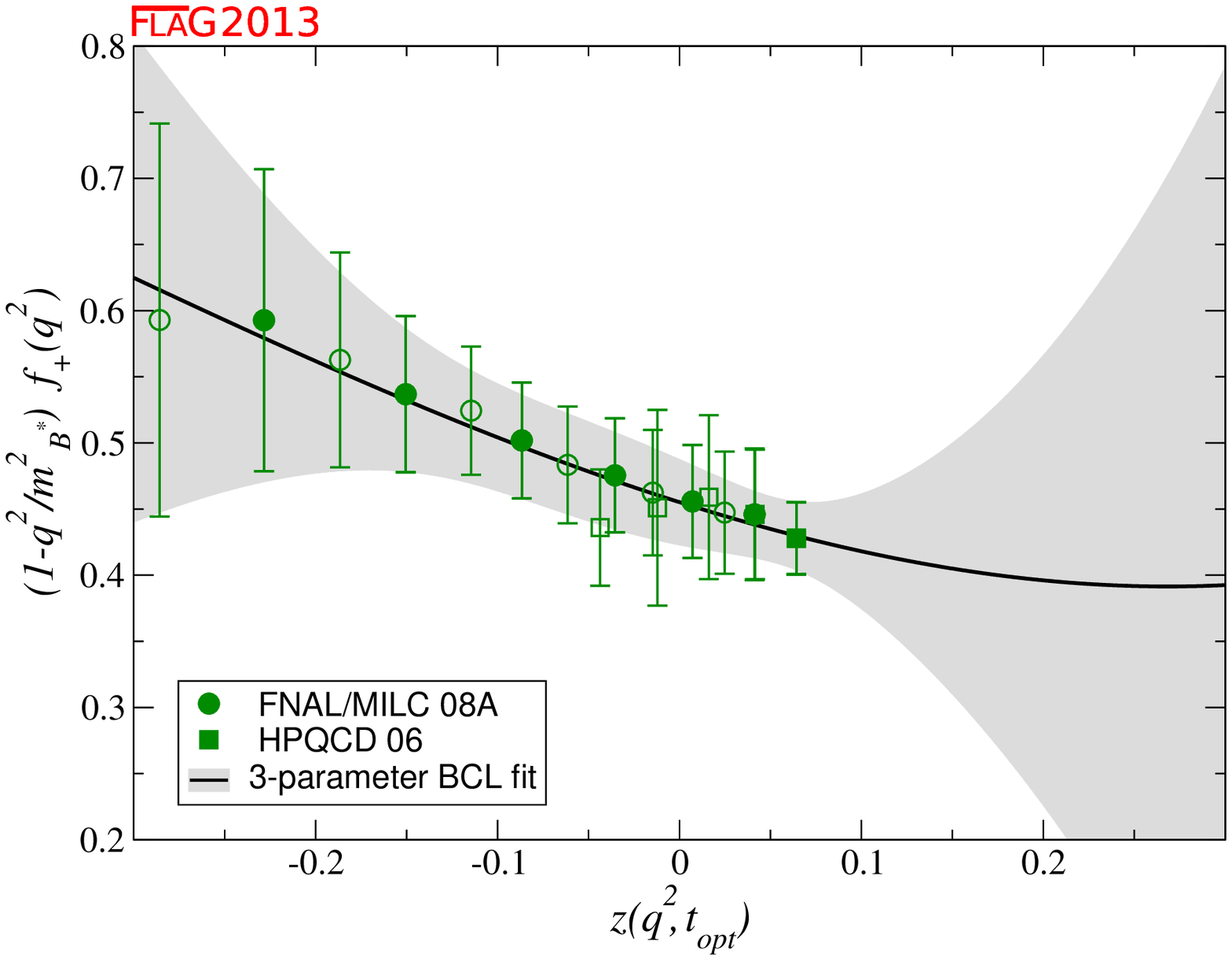}
\vspace{-2mm}
\caption{The form factors $(1 - q^2/m_{B^*}^2) f_+(q^2)$ versus $z$.  The filled symbols denote data points included in the fit, while the open symbols show points that are not included in the fit (either because of unknown correlations or strong correlations).  The grey band displays our preferred three-parameter BCL fit to the plotted data with errors.}\label{fig:LQCDzfit}
\end{center}
\end{figure}

\subsubsection{Form factors for rare and radiative $B$ semileptonic decays to light flavours}

Lattice-QCD input is also available for some exclusive semileptonic
decay channels involving neutral-current $b\to s$ transitions at the
quark level. Being forbidden at tree level in the SM, these processes
allow for stringent tests of potential new physics; simple examples
are $B\to K^*\gamma$ and $B\to K^{(*)}\ell^+\ell^-$, where the $B$
meson (and therefore the kaon) can be either neutral or charged.

The corresponding SM effective
weak Hamiltonian is considerably more complicated than the one for the
tree-level processes discussed above: after neglecting top quark
effects, as many as ten dimension-six operators formed by the product
of two hadronic currents or one hadronic and one leptonic current
appear.\footnote{See e.g.~\cite{Antonelli:2009ws} and references
therein.}  Three of the latter, coming from penguin and box diagrams,
dominate at short distances, and within a reasonable approximation one
can keep these contributions only. Long-distance hadronic physics is
then again encoded in matrix elements of current operators (vector,
tensor, and axial-vector) between one-hadron states, which in turn can
be parameterized in terms of a number of form factors
(see~\cite{Liu:2009dj} for a complete description).  In addition, the
lattice computation of the relevant form factors in channels with a
vector meson in the final state faces extra challenges on top of those
already present when the decay product is a Goldstone boson: the state
is unstable and the extraction of the relevant matrix element from
correlation functions is significantly more complicated; and $\chi$PT
cannot be used as a guide to extrapolate results at unphysically heavy
pion masses to the chiral limit.  As a result, the current lattice
methods and simulations that allow for control over systematic errors
for kaon and pion final states leave uncontrolled systematic errors in
calculations of weak decay form factors into unstable vector meson
final states, such as the $K^*$ or $\rho$ mesons.

Several collaborations are calculating form factors for $B \to
K^{(*)}$ transitions in the Standard Model and beyond on the MILC
$N_f=2+1$ rooted Aqstad staggered gauge configurations.  Two new
results have appeared since the initial April closing date for this
review.  We summarize their content briefly here, but a full
discussion of the calculations, including their rating, is postponed
to the next major update of the FLAG review. The HPQCD Collaboration
has published in Ref.~\cite{Bouchard:2013eph} a determination of the
three form factors for $B\to K\ell^+\ell^-$ with NRQCD $b$ quarks and
HISQ valence light quarks.  In this work, they parameterize the form
factors over the full kinematic range using a model-independent
$z$-expansion as in Sec.~\ref{sec:zparam}, and provide the series
coefficients and covariance matrix.  HPQCD also published a companion
paper~\cite{Bouchard:2013mia} in which they calculate the
Standard Model predictions for the differential branching fractions
and other observables and compare to experiment.  Horgan {\it et al.}
have obtained the seven form factors governing $B \to K^* \ell^+
\ell^-$ (as well as those for $B_s \to \phi\, \ell^+ \ell^-$) in
Ref.~\cite{Horgan:2013hoa} using NRQCD $b$ quarks and Asqtad staggered
light quarks.  In this work, they use a ``modified" $z$-expansion to
simultaneously extrapolate to the physical light-quark masses and
continuum and extrapolate in $q^2$ to the full kinematic range.  As
discussed in Sec.~\ref{sec:DtoPiK}, the ``modified" $z$-expansion is
not based on an underlying effective theory, and the associated
uncertainties have yet to be fully studied.  Horgan {\it et al.} use
their form-factor results to calculate the differential branching
fractions and angular distributions and discuss the implications for
phenomenology in a companion paper~\cite{Horgan:2013pva}.  Finally,
the FNAL/MILC Collaboration has reported preliminary results for the
three $B \to K\ell^+\ell^-$ form factors using Fermilab bottom quarks
and Astqad light quarks in Refs.~\cite{Zhou:2012sna,Liu:2013sya}.

\subsection{Semileptonic form factors for $B \to D \ell \nu$, $B \to D^*  \ell \nu$, and $B \to D \tau \nu$}
\label{sec:BtoD}

The semileptonic processes $ B \rightarrow D \ell \nu$ and
$B \rightarrow D^* \ell \nu$ ($\ell = e, \mu$) have been studied
extensively by experimentalists and theorists over the years.  They
allow for the determination of the CKM matrix element $|V_{cb}|$, an
extremely important parameter of the Standard Model. $|V_{cb}|$
appears in many quantities that serve as inputs into CKM Unitarity
Triangle analyses and reducing its uncertainties is of paramount
importance.  For example, when $\epsilon_K$, the measure of indirect
$CP$-violation in the neutral kaon system, is written in terms of the
parameters $\rho$ and $\eta$ that specify the apex of the unitarity
triangle, a factor of $|V_{cb}|^4$ multiplies the dominant term.  As a
result, the errors coming from $|V_{cb}|$ (and not those from $B_K$)
are now the dominant uncertainty in the Standard Model (SM) prediction
for this quantity.  Decay rates for $B \rightarrow D^{(*)} \ell \nu$
processes can be parameterized as
\begin{eqnarray}
    \frac{d\Gamma_{B^-\to D^{0} \ell^-\bar{\nu}}}{dw} & = &
        \frac{G^2_\mu m^3_{D}}{48\pi^3}(m_B+m_{D})^2(w^2-1)^{3/2}  |\eta_\mathrm{EW}|^2|V_{cb}|^2 |\mathcal{G}(w)|^2,
    \label{eq:vxb:BtoD} \\
    \frac{d\Gamma_{B^-\to D^{0*}\ell^-\bar{\nu}}}{dw} & = &
        \frac{G^2_\mu m^3_{D^*}}{4\pi^3}(m_B-m_{D^*})^2(w^2-1)^{1/2}  |\eta_\mathrm{EW}|^2|V_{cb}|^2\chi(w)|\mathcal{F}(w)|^2 ,
    \label{eq:vxb:BtoDstar}
\end{eqnarray}
where $w \equiv v_B \cdot v_{D^{(*)}}$, $v_P=p_P/m_P$ are the
four-velocities of the mesons, and $\eta_\mathrm{EW}=1.0066$
 is the one-loop electroweak correction~\cite{Sirlin:1981ie}. The
 function $\chi(w)$ in Eq.~(\ref{eq:vxb:BtoDstar}) depends upon the
 recoil $w$ and the meson masses, and reduces to unity at zero
 recoil~\cite{Antonelli:2009ws}.  These formulas do not include terms
 that are proportional to the lepton mass squared which can be
 neglected for $\ell = e, \mu$.

Most unquenched lattice calculations for $B \rightarrow D^* \ell \nu$ and
$B \rightarrow D l \nu$ decays to date focus on the form
factors at zero recoil ~\cite{Bailey:2010gb,Bernard:2008dn} ${\cal
F}^{B \rightarrow D^*}(1)$ and ${\cal G}^{B \rightarrow D}(1)$.  These
can then be combined with experimental input to extract $|V_{cb}|$.
The main reasons for concentrating on the zero recoil point are that
(i) the decay rate then depends on a single form factor, and (ii) for
$B \rightarrow D^*\ell\nu$, there are no ${\cal O}(\Lambda_{QCD}/m_Q)$
contributions due to Luke's theorem.  Further, the zero recoil form
factor can be computed via a double ratio in which most of the current
renormalization cancels and heavy-quark discretization errors are
suppressed by an additional power of $\Lambda_{QCD}/m_Q$.

Some recent work on $B \rightarrow D^{(*)}\ell\nu$ transitions
has started to explore the dependence of the relevant form factors on the
momentum transfer, but these results are not yet published.
The methodology for this is similar to the one employed
in $B\to\pi\ell\nu$ transitions; we refer the reader to Section~\ref{sec:BtoPiK}
for a detailed discussion.
Also recently, first results have appeared for $B_s \rightarrow D_s\ell\nu$
amplitudes, again including information about the momentum transfer dependence;
this will allow for an independent determination of $|V_{cb}|$ as soon as
experimental data are available for these transitions.

\subsubsection{ $B_{(s)} \rightarrow D_{(s)}$ decays}

Until recently, the only unquenched lattice result for the $B \rightarrow D \ell \nu$
form factor ${\cal G}^{B \rightarrow D}(1)$ at zero recoil had appeared in
a 2004 conference proceeding by FNAL/MILC ~\cite{Okamoto:2004xg}.
This calculation employs MILC $N_f = 2 +1$ configurations at a single
lattice spacing, again with Fermilab bottom and charm quarks and
Asqtad staggered light quarks.  Three values of the light-quark mass are used
and results extrapolated linearly to the chiral limit.  The
preliminary result is ${\cal G}^{B \rightarrow D}(1) = 1.074(18)(16)$.

The FNAL/MILC study of $B \rightarrow D \ell \nu$ transitions is now being greatly
updated by considering several lattice spacings and quark masses, as well
as transitions outside the zero recoil limit. Preliminary results have been
published in conference proceedings~\cite{Qiu:2013ofa}, following the
strategy previously outlined in~\cite{Qiu:2012xi}. This work employs ensembles at
four values of the lattice spacing ranging between approximately $0.045~{\rm fm}$
and $0.12~{\rm fm}$, and four values of the light-quark mass corresponding to pions
with RMS masses ranging between $330~{\rm MeV}$ and $470~{\rm MeV}$.

The quantities directly studied are the form factors $h_\pm$
defined by
\begin{equation}
\frac{\langle D(p_D)| i\bar c \gamma_\mu b| B(p_B)\rangle}{\sqrt{m_D m_B}} =
h_+(w)(v_B+v_D)_\mu\,+\,h_-(w)(v_B-v_D)_\mu\,,
\end{equation}
which are related to the standard vector and scalar form factors by
\begin{equation}
f_+(q^2) = \frac{1}{2\sqrt{r}}\,\left[(1+r)h_+(w)-(1-r)h_-(w)\right]\,,~~~~
f_0(q^2) = \sqrt{r}\left[\frac{1+w}{1+r}\,h_+(w)\,+\,\frac{1-w}{1-r}\,h_-(w)\right]\,,
\end{equation}
with $r=m_D/m_B$. (Recall that
$q^2=(p_B-p_D)^2=m_B^2+m_D^2-2wm_Bm_D$.)  The hadronic form factor
relevant for experiment, $\mathcal{G}(w)$, is then obtained from the
relation $\mathcal{G}(w)=4rf_+(q^2)/(1+r)$. The form factors are
obtained from double ratios of three-point functions in which the
flavour-conserving current renormalization factors cancel. The
remaining matching factor $\rho_{V^\mu_{cb}}$ is is estimated with
one-loop lattice perturbation theory.

In order to obtain $h_\pm(w)$ the results are fitted to an ansatz that
contains the light-quark mass and lattice spacing dependence predicted
by next-to-leading order rSHMChPT, and the leading dependence on $m_c$
predicted by the heavy quark expansion ($1/m_c^2$ for $h_+$ and
$1/m_c$ for $h_-$). The $w$-dependence, which allows for an
interpolation in $w$, is given by analytic terms up to $(1-w)^2$, as
well as a contribution from the log proportional to $g^2_{D^*D\pi}$.
The total systematic error is $2.1\%$ for $h_+$ and $22\%$ for $h_-$
(note that $h_-$ is of ${\cal O}(1-w)$ in the recoil parameter, while
$h_+$ is of ${\cal O}(1)$), where the error budget is dominated by the
heavy-quark discretization (estimated from HQET) in the case of $h_+$,
and by the perturbative current matching factor for $h_-$.

Synthetic data points at three values of $w$ that cover the simulated
range are generated for $h_\pm(w)$, from which the form factors
$f_{+,0}$ are reconstructed and their $q^2$-dependence fitted to a
$z$-parameterization of the BGL form~\cite{Boyd:1994tt},
cf. Section~\ref{sec:BtoPiK}.  The values of the series coefficients
and their correlations are not given in the conference proceedings,
but are left for a forthcoming full publication.  From the fit result
one can extract, in particular, the value of the relevant hadronic
form factor at zero recoil
\begin{equation}
{\cal G}^{B \rightarrow D}(1) = 1.081(25)\,.
\end{equation}

Another recent work~\cite{Atoui:2013zza} provides the first study of
$B_s \rightarrow D_s\ell\nu$ transitions with $N_f=2$ flavours of
dynamical quarks, using the publicly available ETMC configurations
obtained with the twisted-mass QCD action at maximal twist.  Four
values of the lattice spacing, ranging between $0.054~{\rm fm}$ and
$0.098~{\rm fm}$, are considered, with physical box lengths ranging
between $1.7~{\rm fm}$ and $2.7~{\rm fm}$.  At two values of the
lattice spacing two different physical volumes are available.
Charged-pion masses range between $\approx 270~{\rm MeV}$ and $\approx
490~{\rm MeV}$, with two or three masses available per lattice spacing
and volume, save for the $a \approx 0.054~{\rm fm}$ point at which
only one light mass is available for each of the two volumes. The
strange and heavy valence quarks are also treated with maximally
twisted-mass QCD.

The quantities of interest are again the form factors $h_\pm$ defined above.
In order to control discretization effects from the heavy quarks, a strategy
similar to the one employed by the ETM Collaboration in their studies of
$B$-meson decay constants (cf. Section~\ref{sec:fB}) is employed: the value of
${\cal G}(w)$ is computed at a fixed value of $m_c$ and several values of
a heavier quark mass $m_h^{(k)}=\lambda^k m_c$, where $\lambda$ is a fixed
scaling parameter, and step-scaling functions are built as
\begin{equation}
\Sigma_k(w) = \frac{{\cal G}(w,\lambda^{k+1} m_c,m_c,a^2)}{{\cal G}(w,\lambda^k m_c,m_c,a^2)}\,.
\end{equation}
Each ratio is extrapolated to the continuum limit,
$\sigma_k(w)=\lim_{a \to 0}\Sigma_k(w)$.  One then exploits the fact
that the $m_h \to \infty$ limit of the step-scaling is fixed --- in
particular, it is easy to find from the heavy-quark expansion that
$\lim_{m_h\to\infty}\sigma(1)=1$. In this way, the physical result at
the $b$-quark mass can be reached by interpolating $\sigma(w)$ between
the charm region (where the computation can be carried out with
controlled systematics) and the known static limit value.

In practice, the values of $m_c$ and $m_s$ are fixed at each value of
the lattice spacing such that the experimental kaon and $D_s$ masses
are reached at the physical point, as determined
in~\cite{Blossier:2010cr}.  For the scaling parameter $\lambda=1.176$
is chosen, and eight step-scaling steps are performed, reaching
$m_h/m_c=1.176^9\simeq 4.30$, approximately corresponding to the ratio
of the physical $b$ and $c$ masses in the $\overline{\rm MS}$ scheme
at $2~{\rm GeV}$.  All observables are obtained from ratios that do
not require (re)normalization.  The ansatz for the continuum and
chiral extrapolation of $\Sigma_k$ contains a constant and linear
terms in $m_{\rm sea}$ and $a^2$.  Twisted boundary conditions in
space are used for valence-quark fields for better momentum
resolution.  Applying this strategy the form factors are finally
obtained at four reference values of $w$ between $1.004$ and $1.062$,
and, after a slight extrapolation to $w=1$, the result is quoted
\begin{equation}
{\cal G}^{B_s \rightarrow D_s}(1) = 1.052(46)\,.
\end{equation}

The authors also provide values for the form factor relevant for the
meson states with light valence quarks, obtained from a similar
analysis to the one described above for the $B_s\rightarrow D_s$ case.
Values are quoted from fits with and without a linear $m_{\rm
  sea}/m_s$ term in the chiral extrapolation. The result in the former
case, which safely covers systematic uncertainties, is
\begin{equation}
{\cal G}^{B \rightarrow D}(1)=1.033(95)\,.
\end{equation}
Given the identical strategy, and the small sensitivity of the ratios
used in their method to the light valence- and sea-quark masses, we
assign this result the same ratings in Table~\ref{tab_BtoDStarsumm2}
as those for their calculation of ${\cal G}^{B_s \rightarrow D_s}(1)$.
Currently the precision of this calculation is not competitive with
that of FNAL/MILC 13A, but this is due largely to the small number of
configurations analysed by Atoui et al.  The viability of their method
has been clearly demonstrated, however, which leaves significant room
for improvement on the errors of both the $B \to D$ and $B_s \to D_s$
form factors with this approach by including either additional
two-flavour data or analysing more recent ensembles with $N_f>2$.

Finally, Atoui et al. also study the scalar and tensor form factors, as well as the
momentum transfer dependence of $f_{+,0}$. The value of the ratio $f_0(q^2)/f_+(q^2)$
is provided at a reference value of $q^2$ as a proxy for the slope of ${\cal G}(w)$
around the zero-recoil limit.

\subsubsection{$B \rightarrow D^*$ decays}

The most precise computation of the zero-recoil form
factors needed for the determination of $|V_{cb}|$ from exclusive $B$
semileptonic decays comes from the $B \rightarrow D^* \ell \nu$ form
factor at zero recoil, ${\cal F}^{B \rightarrow D^*}(1)$, calculated
by the Fermilab Lattice and MILC
Collaborations~\cite{Bailey:2010gb,Bernard:2008dn}.  This work uses
the MILC $N_f = 2 + 1$ ensembles.  The bottom and charm quarks are
simulated using the clover action with the Fermilab interpretation and
light quarks are treated via the Asqtad staggered fermion action.  At
zero recoil ${\cal F}^{B \rightarrow D^*}(1)$ reduces to a single form
factor, $h_{A_1}(1)$, coming from the axial-vector current
\begin{equation}
\langle D^*(v,\epsilon^\prime)| {\cal A}_\mu | \overline{B}(v) \rangle = i \sqrt{2m_B 2 m_{D^*}} \; {\epsilon^\prime_\mu}^\ast h_{A_1}(1),
\end{equation}
where $\epsilon^\prime$ is the polarization of the $D^*$.  Reference
~\cite{Bernard:2008dn} introduces a new ratio of three-point
correlators which directly gives $|h_{A_1}(1)|$:
\begin{equation}
{\cal R}_{A_1} = \frac{\langle D^*|\bar{c} \gamma_j \gamma_5 b | \overline{B} 
\rangle \; \langle \overline{B}| \bar{b} \gamma_j \gamma_5 c | D^* \rangle}
{\langle D^*|\bar{c} \gamma_4 c | D^* 
\rangle \; \langle \overline{B}| \bar{b} \gamma_4 b | \overline{B} \rangle} 
= |h_{A_1}(1)|^2.
\end{equation}
In reference ~\cite{Bernard:2008dn} simulation data are obtained on
MILC ensembles with three lattice spacings, $a \approx 0.15$, 0.12,
and 0.09~fm, for 2, 4 or 3 different light-quark masses respectively.
Results are then extrapolated to the physical, continuum/chiral, limit
employing staggered $\chi$PT.

The $D^*$ meson is not a stable particle in QCD and decays
 predominantly into a $D$ plus a pion.  Nevertheless, heavy-light
 meson $\chi$PT can be applied to extrapolate lattice simulation
 results for the $B\to D^*\ell\nu$ form factor to the physical
 light-quark mass.  The $D^*$ width is quite narrow, 0.096 MeV for the
 $D^{*\pm}(2010)$ and less than 2.1MeV for the $D^{*0}(2007)$, making
 this system much more stable and long lived than the $\rho$ or the
 $K^*$ systems. The fact that the $D^* - D$ mass difference is close
 to the pion mass leads to the well known ``cusp'' in ${\cal
 R}_{A_1}$ just above the physical pion
 mass~\cite{Randall:1993qg,Savage:2001jw,Hashimoto:2001nb}. This cusp
 makes the chiral extrapolation sensitive to values used in the
 $\chi$PT formulas for the $D^*D \pi$ coupling $g_{D^*D\pi}$.  The
 error budget in reference ~\cite{Bernard:2008dn} includes a separate
 error of 0.9\% coming from the uncertainty in $g_{D^*D \pi}$ in
 addition to general chiral extrapolation errors in order to take this
 sensitivity into account.

The final value presented in ~\cite{Bernard:2008dn}, ${\cal F}^{B
  \rightarrow D^*}(1) = h_{A_1}(1) = 0.921(13)(20)$, where the first
error is statistical, and the second the sum of systematic errors
added in quadrature, has a total error of $2.6$\%.  This result is
updated in Ref.~\cite{Bailey:2010gb} after increasing statistics and
adding data from $a \approx 0.06$ fm lattices, and even further in
Ref.~\cite{Qiu:2013ofa} adding data from an $a \approx 0.045$~fm
ensemble.  The latest value is
\begin{equation}
{\cal F}^{B \rightarrow D^*}(1) = 0.906(4)_{\rm stat}(12)_{\rm sys}\,,
\end{equation}
with the total error reduced to $1.4$\%.  The largest systematic
uncertainty comes from discretization errors followed by effects of
higher-order corrections in the chiral perturbation theory ansatz.

\begin{table}[h] 
\begin{center}
\mbox{} \\[3.0cm]
\footnotesize\hspace{-0.2cm}
\begin{tabular*}{\textwidth}[l]{l @{\extracolsep{\fill}} r l l l l l l l c l}
Collaboration & Ref. & $\Nf$ & 
\hspace{0.15cm}\begin{rotate}{60}{publication status}\end{rotate}\hspace{-0.15cm} &
\hspace{0.15cm}\begin{rotate}{60}{continuum extrapolation}\end{rotate}\hspace{-0.15cm} &
\hspace{0.15cm}\begin{rotate}{60}{chiral extrapolation}\end{rotate}\hspace{-0.15cm}&
\hspace{0.15cm}\begin{rotate}{60}{finite volume}\end{rotate}\hspace{-0.15cm}&
\hspace{0.15cm}\begin{rotate}{60}{renormalization}\end{rotate}\hspace{-0.15cm}  &
\hspace{0.15cm}\begin{rotate}{60}{heavy-quark treatment}\end{rotate}\hspace{-0.15cm}  & \multicolumn{2}{l}{\hspace{1cm}form factor}\\
&&&&&&&&&& \\[-0.1cm]
\hline
\hline
&&&&&&&&&& \\[-0.1cm]
FNAL/MILC 13B & \cite{Qiu:2013ofa} & 2+1 & \rC$^\triangledown$ & \good &  \soso &  \good & \soso & \okay&${\mathcal F}^{B\to D^*}(1)$   & 0.906(4)(12) \\[0.5ex]  
FNAL/MILC 10 & \cite{Bailey:2010gb} & 2+1 & \rC$^\S$ & \good & \soso & \good & \soso & \okay&${\mathcal F}^{B\to D^*}(1)$   & 0.9017(51)(87)(83)(89)(30)(33) $^\ddagger$  \\[0.5ex]  
FNAL/MILC 08 & \cite{Bernard:2008dn} & 2+1 & \gA & \good & \soso & \good & \soso & \okay&${\mathcal F}^{B\to D^*}(1)$  & 0.921(13)(8)(8)(14)(6)(3)(4) \\[0.5ex]
&&&&&&&&& \\[-0.1cm]
\hline
&&&&&&&&& \\[-0.1cm]
FNAL/MILC 13B & \cite{Qiu:2013ofa} & 2+1 & \rC & \good &  \soso & \good & \soso & \okay & ${\mathcal G}^{B\to D}(1)$ & 1.081(25) \\[0.5ex]
FNAL/MILC 04A & \cite{Okamoto:2004xg} & 2+1 & \rC & \tbr & \tbr & \soso$^\ast$ & \soso$^\dagger$ & \okay & ${\mathcal G}^{B\to D}(1)$ &1.074(18)(16) \\[0.5ex]
&&&&&&&&& \\[-0.1cm]
\hline
&&&&&&&&& \\[-0.1cm]
FNAL/MILC 12A & \cite{Bailey:2012jg} & 2+1 & \gA & \soso & \soso & \good & \soso & \okay &$R(D)$& 0.316(12)(7) \\[0.5ex] 
&&&&&&&&& \\[-0.1cm]
\hline
&&&&&&&&& \\[-0.1cm]
Atoui 13 & \cite{Atoui:2013zza} & 2 & \oP & \good & \good & \good & --- & \okay & ${\mathcal G}^{B\to D}(1)$  & 1.033(95) \\[0.5ex]
&&&&&&&&& \\[-0.1cm]
\hline
&&&&&&&&& \\[-0.1cm]
Atoui 13 & \cite{Atoui:2013zza} & 2 & \oP & \good & \good & \good & --- & \okay & ${\mathcal G}^{B_s\to D_s}(1)$  & 1.052(46) \\
&&&&&&&&& \\[-0.1cm]
\hline
\hline
\end{tabular*}
\begin{tabular*}{\textwidth}[l]{l@{\extracolsep{\fill}}llllllllll}
  \multicolumn{11}{l}{\vbox{\begin{flushleft} 
  $^\triangledown$ Update of FNAL/MILC 08 for Lattice 2013.  \\ 
  $^\S$ Update of FNAL/MILC 08 for CKM 2010.  \\ 
  $^\ddagger$ Value of ${\mathcal F}(1)$ presented in Ref.~\cite{Bailey:2010gb} includes 0.7\% correction $\eta_{EW}$.  This correction is unrelated to the\\\hspace{0.5em} lattice calculation and has been removed here.\\
   $^\ast$ No explicit estimate of FV error, but expected to be small.\rule{0em}{1em} \\
 $^\dagger$  No explicit estimate of perturbative truncation error in vector current renormalization factor, but expected\\ \hspace{0.5em} to be small because of mostly-nonperturbative approach. \\
\end{flushleft}}}
\end{tabular*}
\caption{Lattice results for the $B \to D^* \ell\nu$, $B\to D\ell\nu$, and $B_s \to D_s \ell \nu$ semileptonic form factors and $R(D)$. \label{tab_BtoDStarsumm2}}
\end{center}
\end{table}

\subsubsection{$B \rightarrow D^{(*)} \tau \nu$ decays}

Another interesting semileptonic process is $B \rightarrow
D^{(*)} \tau \nu$.  Here the mass of the outgoing charged lepton
cannot be neglected in the decay rate formula, so that both vector and
scalar form factors come into play.  Recently Babar announced their
first observations of the semileptonic decays of $B$ mesons into third
generation leptons at a rate in slight excess over SM expectations.
Since the lepton mass is now large enough for the branching fraction
${\cal B}(B \rightarrow D \tau \nu)$ to be sensitive to the scalar
form factor $f_0(q^2)$, this could be a hint for some New Physics
scalar exchange contribution.  Accurate SM predictions for the ratio
\begin{equation}
R(D^{(*)}) = {\cal B}(B \rightarrow D^{(*)} \tau \nu) /
{\cal B}(B \rightarrow D^{(*)} \ell \nu)\hspace{1cm}\mbox{with}\;\ell=e,\mu
\end{equation}
have therefore become important and timely.  FNAL/MILC has published
the first unquenched lattice determination of $R(D)$
~\cite{Bailey:2012jg}.  They use a subset of the MILC ensembles from
the ongoing $B \rightarrow D\ell\nu$ semileptonic project
~\cite{Qiu:2012xi}, namely two light-quark masses each on $a \approx
0.12$ and $0.09$~fm lattices, and find,
\begin{equation}
R(D) = 0.316(12)(7).
\end{equation}
This SM prediction is about $\sim 1.7\sigma$ lower than 
the Babar measurement.

\subsubsection{Ratios of $B$ and $B_s$ semileptonic decay form factors}

In addition to $B \rightarrow D\ell\nu$ semileptonic decays there is
also interest in $B_s \rightarrow D_s \ell\nu$ semileptonic decays. In
particular, $[B_s \rightarrow D_s \ell\nu] / [B \rightarrow
D \ell\nu]$ semileptonic form factor ratios can be used to obtain
ratios of $B_q$ meson ($q = d,s$) fragmentation fractions, $f_s /
f_d$.  This latter ratio enters into LHCb's analysis of
$B_s \rightarrow \mu^+ \mu^-$ decays.  There is now one unquenched
calculation by FNAL/MILC of ratios of the scalar form factors
$f_0^{(q)}(q^2)$ \cite{Bailey:2012rr}:
\begin{equation}
f_0^{(s)}(M_\pi^2) / f_0^{(d)}(M_K^2) = 1.046(44)(15),  
\qquad 
f_0^{(s)}(M_\pi^2) / f_0^{(d)}(M_\pi^2) = 1.054(47)(17),
\end{equation}
where the first error is statistical and the second systematic.  These
results lead to fragmentation fraction ratios $f_s/f_d$ that are
consistent with LHCb's measurements via other methods.

\subsubsection{ Summary}

In Table~\ref{tab_BtoDStarsumm2} we summarize the existing results for
the $B \rightarrow D^* \ell \nu$, $B \rightarrow D \ell \nu$, and $B_s
\rightarrow D _s\ell \nu$ form factors at zero recoil, ${\cal F}^{B
  \rightarrow D^*}(1)$, ${\cal G}^{B \rightarrow D}(1)$, and ${\cal
  G}^{B_s \rightarrow D_s}(1)$, as well as for the ratio $R(D) = {\cal
  B}(B \rightarrow D \tau \nu) / {\cal B}(B \rightarrow D l \nu)$.
Further details of the lattice calculations are provided in
Appendix~\ref{app:BtoD_Notes}.  Selecting those results that are
published in refereed journals (or are straightforward updates
thereof) and have no red tags, our averages for ${\cal F}^{B
  \rightarrow D^*}(1)$ and $R(D)$ are
\begin{equation}
N_f = 2 + 1 :  \quad {\cal F}^{B \rightarrow D^*} = 0.906(4)(12), 
\quad R(D) = 0.316(12)(7). \label{eq:Nf2p1_BtoDResults}
\end{equation}

\begin{figure}[!h]
\begin{center}
\includegraphics[width=0.65\linewidth]{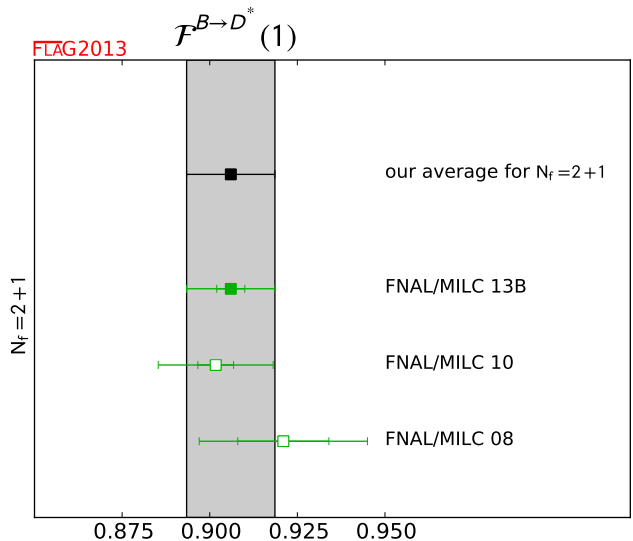}
\vspace{-2mm}
\caption{$B \to D^* \ell \nu$ semileptonic form factor at zero recoil [values in Table \ref{tab_BtoDStarsumm2} and Eq.~(\ref{eq:Nf2p1_BtoDResults})].}\label{fig:BtoDStar}
\end{center}
\end{figure}

\subsection{Determination of $|V_{ub}|$}
\label{sec:Vub}

\noindent
We now use the lattice-determined Standard Model transition amplitudes
for leptonic (Sec.~\ref{sec:fB}) and semileptonic
(Sec.~\ref{sec:BtoPiK}) $B$-meson decays to obtain exclusive
determinations of the CKM matrix element $|V_{ub}|$. The relevant
formulae are Eqs.~(\ref{eq:B_leptonic_rate})
and~(\ref{eq:B_semileptonic_rate}). Among leptonic channels the only
input comes from $B\to\tau\nu_\tau$, since the rates for decays to $e$
and $\mu$ have not yet been measured.  In the semileptonic case we
only consider $B\to\pi\ell\nu_\ell$ transitions (experimentally
measured for $\ell=e,\mu$), since no theoretical prediction for
hadronic effects in other $b\to u$ transitions is currently available
that satisfies FLAG requirements for controlled systematics.

The branching fraction for the decay $B\to\tau\nu_\tau$ has been
measured by the Belle and Babar collaborations with both
semileptonic~\cite{Aubert:2009wt,Hara:2010dk} and hadronic
tagging~\cite{Lees:2012ju,Adachi:2012mm} methods.  The uncertainties
in these measurements are still dominated by statistical errors, and
none of them individually are of 5$\sigma$ significance.  When
combined, however, they cross the threshold needed to establish
discovery of this mode.  Until recently, the various
largely-independent measurements have agreed well within errors.
Earlier this year, however, the Belle collaboration published the
single-most precise measurement of $B\to\tau\nu$ using the hadronic
tagging method with an improved efficiency and the full
dataset~\cite{Adachi:2012mm}, and obtained a result which is more than
2$\sigma$ below the previous
average~\cite{Beringer:1900zz,Amhis:2012bh}.  The errors on the
analogous measurement from Babar~\cite{Lees:2012ju} are not
competitive due to the smaller available data set, and the Babar
result has not yet been published.

Both Belle and Babar quote averages of the hadronic and the
semileptonic tagging modes that we can use to obtain $|V_{ub}|$.  In
the case of Belle, the average $BR(B^+ \to\tau^+ \nu_\tau)=(0.96 \pm
0.26)\times 10^{-4}$~\cite{Adachi:2012mm} includes slight correlations
between systematics with the two tagging methods, but does not include
a rescaling factor due to the fact that the hadronic and semileptonic
measurements are inconsistent at the $\sim 1.5\sigma$ level.  The
Babar average $BR(B^+ \to\tau^+ \nu_\tau)=(1.79 \pm 0.48)\times
10^{-4}$~\cite{Lees:2012ju} neglects correlations.  By combining these
values with the mean $B^+$-meson lifetime $\tau_{B^+}=1.641(8)~{\rm
ps}$ quoted by the PDG, and our averages
$f_B = (189\pm 8)~{\rm MeV}~(N_f=2)$,
$f_B= 190.5 \pm 4.2~{\rm MeV}~(N_f=2+1)$,
and $f_B= 186 \pm 4~{\rm MeV}~(N_f=2+1+1)$
for the $B$-meson decay constants, we obtain
\begin{gather}
\begin{split}
\mbox{Belle}~B\to\tau\nu_\tau: \qquad  & |V_{ub}| = 3.90(53)(17) \times 10^{-3}\,,~~~~~~~N_f=2\,,\\
\mbox{Belle}~B\to\tau\nu_\tau: \qquad  & |V_{ub}| = 3.87(52)(9) \times 10^{-3} \,,~~~~~~~~N_f=2+1\,,\\
\mbox{Belle}~B\to\tau\nu_\tau: \qquad  & |V_{ub}| = 3.96(54)(9) \times 10^{-3} \,,~~~~~~~~N_f=2+1+1\,;\\
& \\
\mbox{Babar}~B\to\tau\nu_\tau: \qquad  & |V_{ub}| =  5.32(71)(23) \times 10^{-3}\,,~~~~~~~N_f=2\,,\\
\mbox{Babar}~B\to\tau\nu_\tau: \qquad  & |V_{ub}| = 5.28(71)(12) \times 10^{-3} \,,~~~~~~~N_f=2+1\,,\\
\mbox{Babar}~B\to\tau\nu_\tau: \qquad  & |V_{ub}| = 5.41(73)(12) \times 10^{-3} \,,~~~~~~~N_f=2+1+1\,.
\end{split}
\end{gather}
where the first error comes from experiment and the second comes from
the uncertainty in $f_B$.  We can also average all four results for
$BR(B^+ \to\tau^+ \nu_\tau)$ from Belle and Babar.  The measurements
using hadronic and semileptonic tagging are statistically independent;
further, because the measurements are dominated by statistical errors,
the correlations between systematic errors in the two approaches can
be reasonably neglected.  We obtain $BR(B^+ \to\tau^+ \nu_\tau) =
(1.12 \pm 0.28) \times 10^{-4}$, where we have applied a
$\sqrt{(\chi^2/{\rm d.o.f.})} \sim 1.3$ rescaling factor because the
Belle hadronic tagging measurement differs significantly from the
other three.  Using this value for the branching fraction, and again
combining with the $N_f=2$, $N_f=2+1$ and $N_f=2+1+1$ lattice-QCD averages for
$f_B$ from Eqs.~(\ref{eq:fbav2})--(\ref{eq:fbav211}), our preferred
determinations of $|V_{ub}|$ from leptonic $B\to\tau\nu$ decay are
\begin{gather}
\begin{split}
\mbox{Belle + Babar}~B\to\tau\nu_\tau:  & \qquad |V_{ub}| = 4.21(53)(18) \times 10^{-3}\,,~~~~~~~N_f=2\,,\\
\mbox{Belle + Babar}~B\to\tau\nu_\tau:  & \qquad |V_{ub}| =  4.18(52)(9) \times 10^{-3} \,,~~~~~~~~~N_f=2+1\,,\\
\mbox{Belle + Babar}~B\to\tau\nu_\tau:  & \qquad |V_{ub}| =  4.28(53)(9) \times 10^{-3} \,,~~~~~~~~~N_f=2+1+1\,.
\end{split}
\end{gather}


In semileptonic decays, the experimental value of $|V_{ub}|f_+(q^2)$
can be extracted from the measured branching fractions of
$B^0\to\pi^-\ell^+\nu$ decays by applying
Eq.~(\ref{eq:B_semileptonic_rate}); $|V_{ub}|$ can then be determined
by performing fits to the constrained BCL $z$-parameterization of the
form factor $f_+(q^2)$ given in Eq.~(\ref{eq:bcl_c}).  This can be
done in two ways: one option is to perform separate fits to lattice
(cf. Sec.~\ref{sec:BtoPiK}) and experimental results, and extract the
value of $|V_{ub}|$ from the ratio of the respective $a_0$
coefficients; a second option is to perform a simultaneous fit to
lattice and experimental data, leaving their relative normalization
$|V_{ub}|$ as a free parameter. We adopt the second strategy because
it more optimally combines the lattice and experimental information
and minimizes the uncertainty in $|V_{ub}|$.  As experimental input we
take the latest untagged 12-bin Babar data~\cite{Lees:2012vv} and
13-bin Belle data~\cite{Ha:2010rf}, and we assume no correlation
between experimental and lattice data.  As in the fit to lattice data
only in Sec.~\ref{sec:BtoPiK}, we assume that the statistics plus
chiral-extrapolation errors are 100\% correlated between the FNAL/MILC
08A and HPQCD 06 data, and we reduce the correlations in the FNAL/MILC
data by keeping only every second data point.

Figure~\ref{fig:LQCDExpzfit} shows both the lattice and experimental
data for $(1 - q^2/m_{B^*}^2) f_+(q^2)$ versus $z$.  For illustration,
the experimental data are divided by the value of $|V_{ub}|$ obtained
from the preferred fit.  Both the lattice-QCD and experimental data
are linear and display no visible signs of curvature; further, the
slopes of the lattice and experimental data sets appear consistent.  A
simple three-parameter constrained BCL fit (i.e.~through
${\mathcal O}(z^2)$ plus $|V_{ub}|$) is sufficient to describe the
combined data sets with a good $\chi^2/{\rm d.o.f.}$, however, the
addition of the experimental points enables a better determination of
higher-order terms in the $z$-expansion than from the lattice-only
fit.  In order to address the potential systematic uncertainty due to
truncating the series in $z$, we continue to add terms to the fit
until the result for $|V_{ub}|$ stabilizes, i.e.~the central
value settles and the errors stop increasing.  We find that this
happens at ${\mathcal O}(z^3)$, and take the value of $|V_{ub}|$ from
this combined fit of the lattice-QCD and experimental data as our
preferred result:
\begin{gather}
\begin{split}
\mbox{global lattice + Babar:} & \qquad |V_{ub}| = 3.37(21) \times 10^{-3} \,,~~~~~~~N_f=2+1\,,\\
\mbox{global lattice + Belle:} & \qquad |V_{ub}| = 3.47(22) \times 10^{-3} \,,~~~~~~~N_f=2+1\,.
\end{split}
\end{gather}
We do not quote a result for a combined lattice + Babar + Belle fit,
since we are unable to properly take into account possible
correlations between experimental results.  Again, we emphasize the
importance of publishing statistical and systematic correlation
matrices in future lattice-QCD work on semileptonic form factors, so
that the lattice results can be fully used to obtain CKM matrix
elements and for other phenomenological applications.

\begin{figure}[thb]
\hspace{0.6cm}\includegraphics[width=0.47\linewidth]{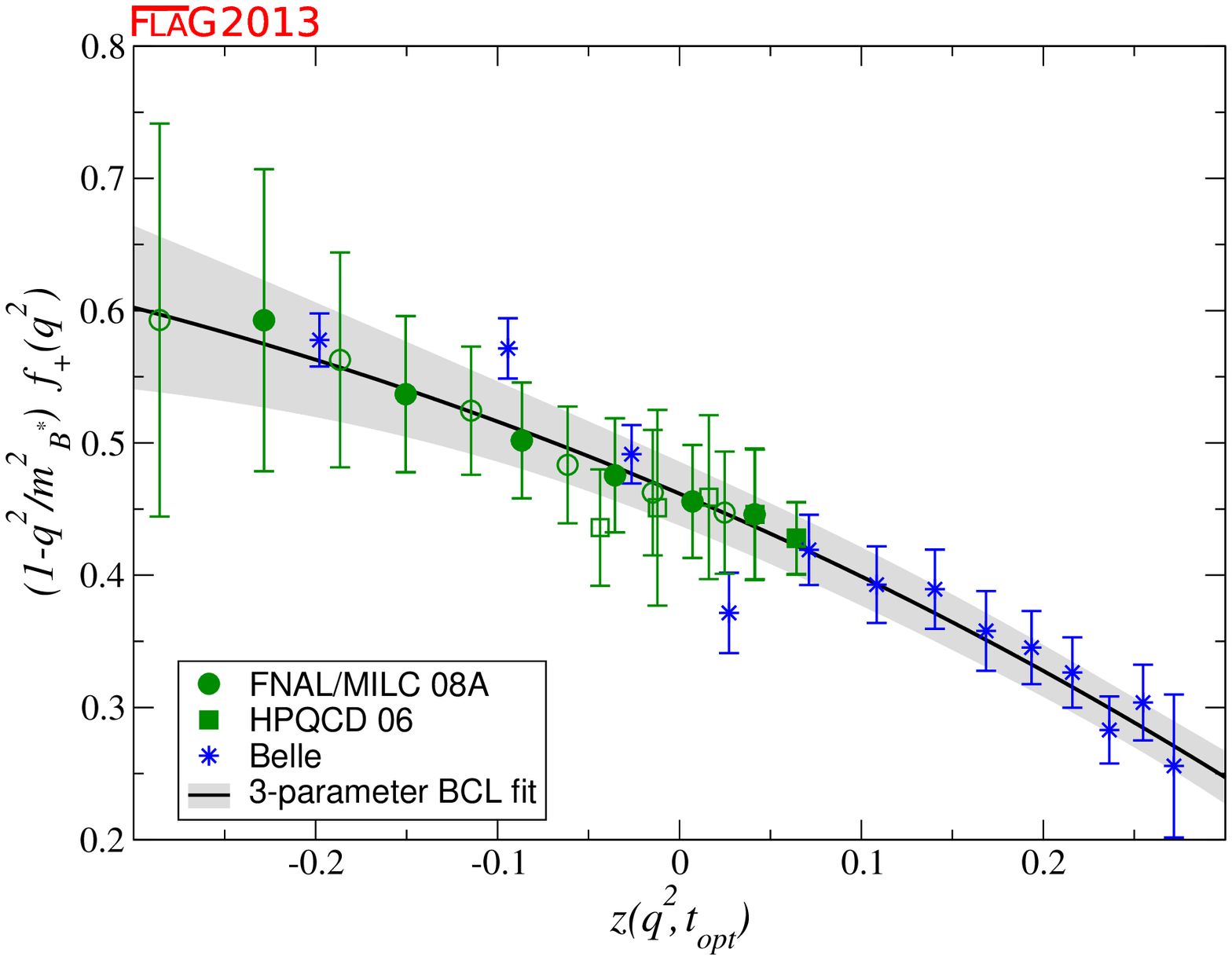}\hspace{0.7cm}
\includegraphics[width=0.47\linewidth]{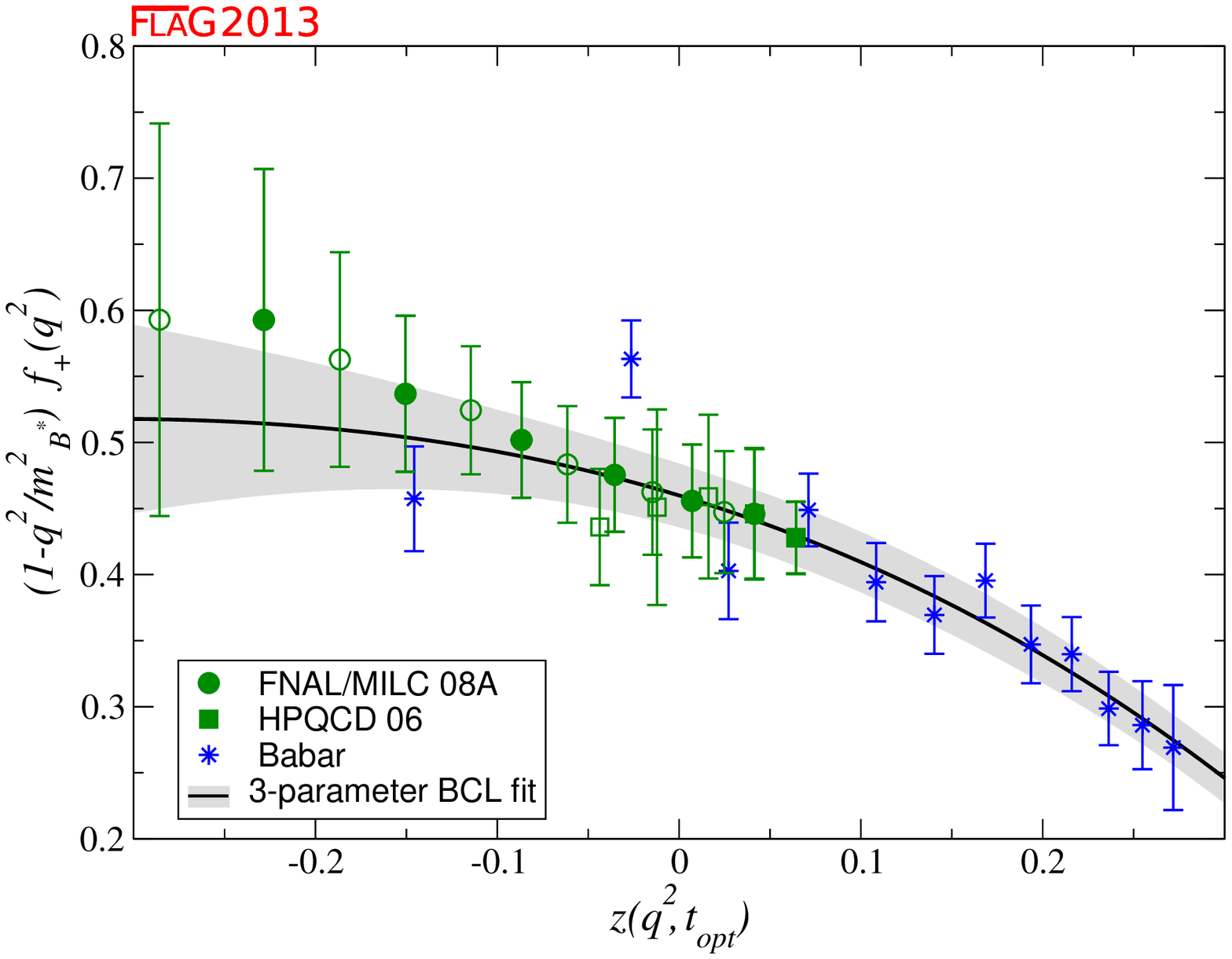}

\vspace{-0.3cm}
\caption{Lattice and experimental data for $(1 - q^2/m_{B^*}^2)
  f_+(q^2)$ versus $z$.  The filled green symbols denote lattice-QCD
  points included in the fit, while the open green symbols show those
  that are not included in the fit (either because of unknown
  correlations or strong correlations).  The blue stars show the
  experimental data divided by the value of $|V_{ub}|$ obtained from
  the fit.  The grey band in the left (right) plots shows the
  preferred three-parameter BCL fit to the lattice-QCD and Belle
  (Babar) data with errors.}\label{fig:LQCDExpzfit}
\end{figure}

Our results for $|V_{ub}|$ are summarized in
Table~\ref{tab:VubSummary} and Figure~\ref{fig:VubVcbSummary}, where
we also show the inclusive determinations from HFAG for comparison.
The spread of values for $|V_{ub}|$ does not yield a clear picture.
We observe the well-known $\sim 3\sigma$ tension between
determinations of $|V_{ub}|$ from exclusive and inclusive semileptonic
decays.  The determination of $|V_{ub}|$ from leptonic $B\to\tau\nu$
decay lies in between the inclusive and exclusive determinations, but
the experimental errors in ${\rm BR}(B\to\tau\nu)$ are so large that
it agrees with both within $\sim 1.5\sigma$.  If, however, we consider
separately the different experimental measurements of ${\rm
BR}(B\to\tau\nu)$, the Belle measurement from hadronic tagging leads
to a value of $|V_{ub}|$ that agrees well with the one from exclusive
$B\to\pi\ell\nu$ decay, while the remaining Belle and Babar
measurements lead to values of $|V_{ub}|$ that are larger than both
the latter and inclusive determinations.  The exclusive determination
of $|V_{ub}|$ will improve in the next few years with better
lattice-QCD calculations of the $B\to\pi\ell\nu$ form factor, while
the improvement in $|V_{ub}|$ from $B\to\tau\nu$ decays will have to
wait longer for the Belle~II experiment, which aims to begin running
in 2016, to collect a larger data set than is currently available.

\begin{table}[h]
\begin{center}
\noindent
\begin{tabular*}{\textwidth}[l]{@{\extracolsep{\fill}}lll}
& from & $|V_{ub}|\times 10^3$\\
&& \\[-2ex]
\hline \hline &&\\[-2ex]
our result for $N_f = 2$ &  $B\to\tau\nu$  & $4.21(53)(18)$ \\
our result for $N_f = 2+1$&   $B\to\tau\nu$ & $4.18(52)(9)$ \\
our result for $N_f = 2+1+1$&   $B\to\tau\nu$ & $4.28(53)(9)$ \\
&& \\[-2ex]
 \hline
&& \\[-2ex]
our result for $N_f = 2+1$ & $B\to\pi\ell\nu$ (Babar) & $3.37(21)$\\
our result for $N_f = 2+1$ &  $B\to\pi\ell\nu$ (Belle)  & $3.47(22)$\\
&& \\[-2ex]
 \hline\hline
&& \\[-2ex]
Bauer 01  \cite{Bauer:2001rc} & $B\to X_u \ell\nu$  & $4.62(20)(29)$\\
Lange 05  \cite{Lange:2005yw} & $B\to X_u \ell\nu$ & $4.40(15)(^{+19}_{-21})$ \\
Andersen 05  \cite{Andersen:2005mj}, Gardi 08 \cite{Gardi:2008bb}&$B\to X_u \ell\nu$ & $4.45(15)(^{+15}_{-16})$ \\
Gambino 07 \cite{Gambino:2007rp}  & $B\to X_u \ell\nu$  & $4.39(15)(^{+12}_{-14})$\\
Aglietti 07  \cite{Aglietti:2007ik} & $B\to X_u \ell\nu$ & $4.03(13)(^{+18}_{-12})$\\
&& \\[-2ex]
 \hline
&& \\[-2ex]
HFAG inclusive average  \cite{Amhis:2012bh} & $B\to X_u \ell\nu$ & $4.40(15)(20)$\\
&& \\[-2ex]
\hline \hline && \\[-2ex]
\end{tabular*}
\caption{Comparison of exclusive determinations of $|V_{ub}|$ (upper panel)
and inclusive determinations (lower panel).  For $B\to\tau\nu$, the
two uncertainties shown come from experiment (plus non-lattice theory)
and from the lattice calculation, respectively. Each inclusive
determination corresponds to a different theoretical treatment of the
same experimental partial branching fractions compiled by the Heavy
Flavour Averaging Group~\cite{HFAG_Semi_12}; the errors shown are
experimental and theoretical, respectively.}
\label{tab:VubSummary}
\end{center}
\end{table}

\subsection{Determination of $|V_{cb}|$}
\label{sec:Vcb}

We now interpret the lattice-QCD results for the $B\to D^{(*)}\ell\nu$
form factors as determinations of the CKM matrix element $|V_{cb}|$ in
the Standard Model.

For the experimental branching fractions at zero recoil, we use the
latest experimental averages from the Heavy Flavour Averaging
Group~\cite{Amhis:2012bh}:\footnote{We note that HFAG currently
averages results for neutral and charged $B$ meson decays without
first removing the correction due to the Coulomb attraction between
the charged final-state particles for the neutral $B$ meson decays.}
\begin{equation}
	 {\mathcal F}^{B\to D^*}(1) \eta_{\rm EW} |V_{cb}| =  35.90(45) \,, \qquad {\mathcal G}^{B\to D}(1) \eta_{EW} |V_{cb}|  = 42.64(1.53)  \,.
\end{equation}
For ${\mathcal F}^{B\to D^*}(1)$, there is only a single $N_f = 2+1$
lattice-QCD calculation that satisfies the FLAG criteria, while there
is currently no such calculation of ${\mathcal G}^{B \to D}(1)$.
Using the
result given in Eq.~(\ref{eq:Nf2p1_BtoDResults}), we obtain our
preferred value for $|V_{cb}|$:
\begin{equation}
	B \to D^* \ell \nu: \quad |V_{cb}| = 39.36(56)(50) \times 10^{-3}  \,, \qquad  N_f=2+1 \label{eq:VcbNf2p1}
\end{equation}
where the errors shown are from the lattice calculation and experiment
(plus non-lattice theory), respectively.  Table~\ref{tab:VcbCompare}
compares the determination of $|V_{cb}|$ from exclusive $B\to
D^* \ell\nu$ decays to that from inclusive $B \to X_c \ell \nu$
decays, where $X_c$ denotes all possible charmed hadronic final
states.  The results, also shown in Fig.~\ref{fig:VubVcbSummary},
differ by approximately 2.7$\sigma$.  The exclusive determination of
$|V_{cb}|$ will improve significantly over the next year or two with
new lattice-QCD calculations of the $B \to D^{(*)} \ell \nu$ form
factors at nonzero recoil.

\begin{table}[h]
\begin{center}
\noindent
\begin{tabular*}{\textwidth}[l]{@{\extracolsep{\fill}}lclc}
& Ref.& from  & $|V_{cb}| \times 10^3$\\
&& \\[-2ex]
\hline \hline &&\\[-2ex]
our average for $N_f = 2+1$ & \cite{Bailey:2010gb} &$B\to D^* \ell \nu$ & 39.36(56)(50) \\
&& \\[-2ex]
 \hline
&& \\[-2ex]
Inclusive (Gambino 13)  & \cite{Gambino:2013rza}& $B\to X_c \ell\nu$ & 42.42(86) \\
&& \\[-2ex]
 \hline \hline 
\end{tabular*}
\caption{Determinations of $|V_{cb}|$ obtained from semileptonic $B$ decay.
  The errors shown in the first row indicate those from lattice
  and experimental (plus non-lattice theory) uncertainties, respectively,
  while the error shown in the second row is the total (experimental plus
  theoretical) uncertainty. \label{tab:VcbCompare}}
\end{center}
\end{table}
\begin{figure}[h]
\hspace{-0.7cm}\includegraphics[width=0.57\linewidth]{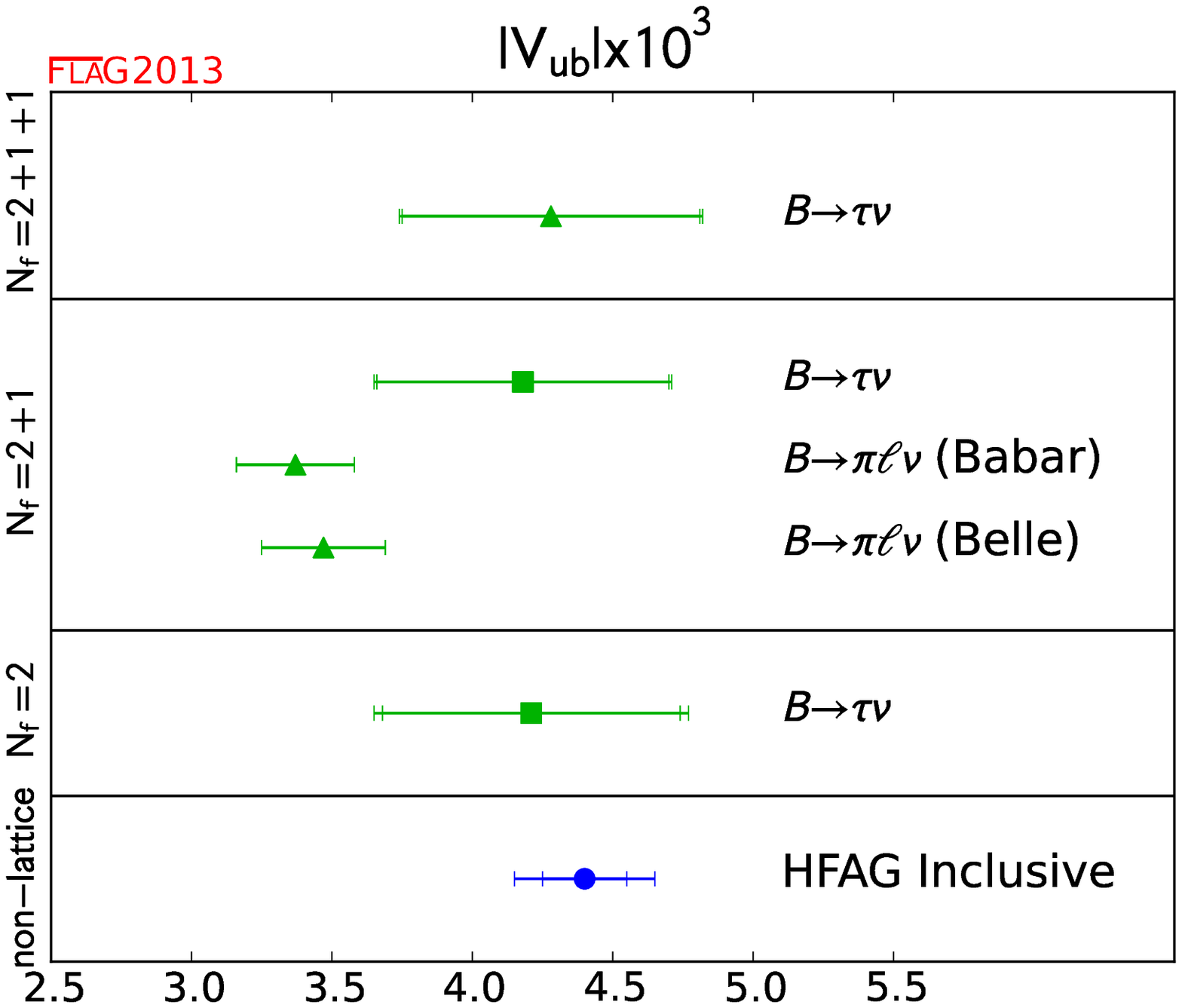}\hspace{-0.7cm}
\includegraphics[width=0.57\linewidth]{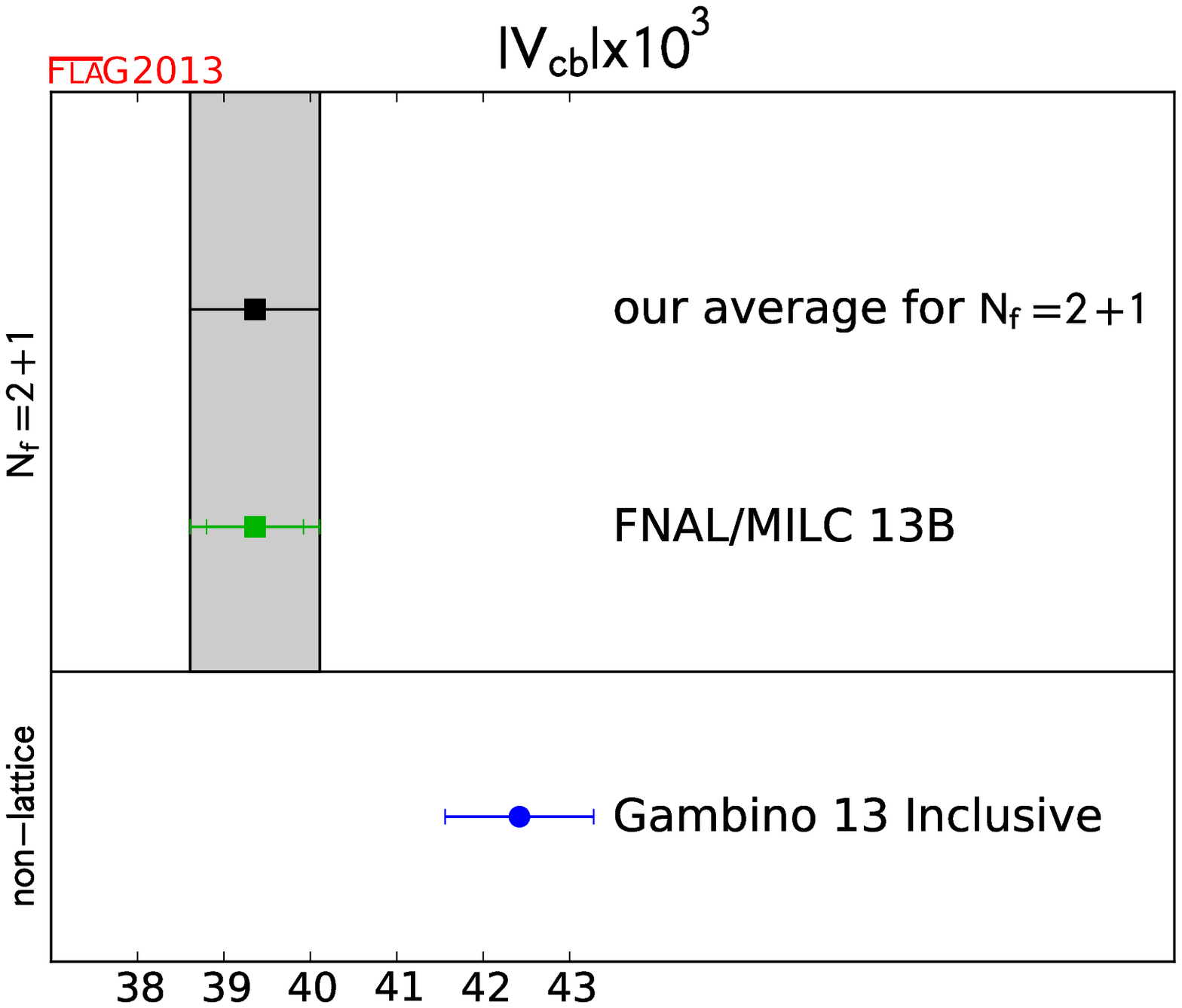}

\vspace{-2mm}
\caption{Comparison of the results for $|V_{ub}|$ and $|V_{cb}|$
  obtained from lattice methods with non-lattice determinations based
  on inclusive semileptonic $B$ decays. In the left plot, the results
  denoted by squares are from leptonic decays, while those denoted by
  triangles are from semileptonic decays. The grey band indicates our
  $N_f=2+1$ average. \label{fig:VubVcbSummary}}
\end{figure}
\clearpage

\clearpage
\section{The strong coupling $\alpha_{\rm s}$}
\label{sec:alpha_s}


\subsection{Introduction}


\label{introduction}


The strong coupling $\gbar(\mu)$ defined at scale $\mu$, plays a key
r\^ole in the understanding of QCD and in its application for collider
physics. For example, the parametric uncertainty from $\alpha_s$ is one of
the dominant sources of uncertainty in the Standard Model prediction for
the $H \to b\bar{b}$ partial width, and the largest source of uncertainty
for $H \to gg$.  Thus higher precision determinations  $\alpha_s$ are
needed to maximize the potential of experimental measurements at the LHC,
and for high-precision Higgs studies at future
colliders~\cite{Heinemeyer:2013tqa,Adams:2013qkq,Dawson:2013bba}.  The
value of $\alpha_s$ also yields one of the essential boundary conditions
for completions of the standard model at high energies. 

In order to determine the running coupling at scale $\mu$
\begin{eqnarray}
   \alpha_s(\mu) = { \gbar^2(\mu) \over 4\pi} \,,
\end{eqnarray}
we should first ``measure'' a short distance quantity ${\cal O}$ at scale
$\mu$ either experimentally or by lattice calculations and then 
match it with a perturbative expansion in terms of a running coupling,
conventionally taken as $\alpha_{\overline{\rm MS}}(\mu)$,
\begin{eqnarray}
   {\cal O}(\mu) = c_1 \alpha_{\overline{\rm MS}}(\mu)
              +  c_2 \alpha_{\overline{\rm MS}}(\mu)^2 + \cdots \,.
\label{eq:alpha_MSbar}
\end{eqnarray}
The essential difference between continuum determinations of
$\alpha_s$ and lattice determinations is the origin of the values of
$\cal O$ in \eq{eq:alpha_MSbar}.

The basis of continuum determinations are 
experimentally measurable cross sections from which $\cal O$ is
defined. These cross sections have to be sufficiently inclusive 
and at sufficiently high scales such that perturbation theory 
can be applied. Often hadronisation corrections have to be used
to connect the observed hadronic cross sections to the perturbative
ones. Experimental data at high $\mu$, where perturbation theory
is progressively more precise, usually has increasing experimental errors, 
and it is  not easy to find processes which allow one
to follow the $\mu$ dependence of a single $\cal O(\mu)$ over
a range where $\alpha_s(\mu)$ changes significantly and precision is 
maintained. 

In contrast, in lattice gauge theory, one can design $\cal O(\mu)$ as
Euclidean short distance quantities which are not directly related to
experimental observables. This allows us to follow the $\mu$
dependence until the perturbative regime is reached and
non-perturbative ``corrections'' are negligible.  The only
experimental input for lattice computations of $\alpha_s$ is the
hadron spectrum which fixes the overall energy scale of the theory and
the quark masses. Therefore experimental errors are completely
negligible and issues such as hadronisation do not occur.  We can
construct many short-distance quantities that are easy to calculate
nonperturbatively in lattice simulations with small statistical
uncertainties.  We can also simulate at parameter values that do not
exist in nature (for example with unphysical quark masses between
bottom and charm) to help control systematic uncertainties.  These
features mean that very precise results for $\alpha_s$ can be achieved
with lattice gauge theory computations.  Further, as in the continuum,
the many different methods available to determine $\alpha_s$ in
lattice calculations with different associated systematic
uncertainties enable valuable cross-checks.  Practical limitations are
discussed in the next section, but a simple one is worth mentioning
here. Experimental results (and therefore the continuum
determinations) of course have all quarks present, while in lattice
gauge theories only the light ones are included and one then is forced
to use the matching at thresholds, as discussed in the following
sub-section.

It is important to keep in mind that the dominant source of
uncertainty in most present day lattice-QCD calculations of $\alpha_s$
is from the truncation of either continuum or lattice perturbation
theory.  Perturbative truncation errors are of a different nature than
most other lattice (or experimental) systematics, in that they often
cannot be estimated from studying the data itself.  Further, the size
of higher-order coefficients in the perturbative series can sometimes
turn out to be larger than naive expectations based on power-counting
from the behaviour of lower-order terms.  Therefore for the purposes of
this review we choose to be cautious in the range presented in
Sec.~\ref{s:alpsumm} for $\alpha^{(5)}_{\overline{\rm MS}}(M_Z)$ from
lattice calculations.

The various phenomenological approaches to determining the running
coupling, $\alpha^{(5)}_{\overline{\rm MS}}(M_Z)$ are summarized by the
Particle Data Group \cite{Beringer:1900zz}. 
The PDG review lists $4$ categories of phenomenological results
used to obtain the running coupling using hadronic
$\tau$ decays, hadronic final states of $e^+e^-$ annihilation,
deep inelastic lepton--nucleon scattering and electroweak precision data.
Excluding lattice results, the PDG quotes a
weighted average of
\begin{eqnarray}
   \alpha^{(5)}_{\overline{\rm MS}}(M_Z) = 0.1183(12) \,.
\label{PDG_nolat}
\end{eqnarray}
For a general overview of the status of the various phenomenological
and lattice approaches see e.g.\ \cite{Bethke:2011tr}.  We note
that perturbative truncation errors are also the dominant
source of uncertainty in several of the phenomenological
determinations of $\alpha_s$.  In particular, the extraction of
$\alpha_s$ from $\tau$ data, which is the most precise and has the
largest impact on the non-lattice average in eq.~(\ref{PDG_nolat}) is
especially sensitive to the treatment of higher-order perturbative
terms.  This is important to keep in mind when comparing our chosen
range for $\alpha^{(5)}_{\overline{\rm MS}}(M_Z)$ from lattice
determinations in eq.~(\ref{eq:alpmz}) with the non-lattice average
from the PDG.

\subsubsection{Scheme and scale dependence of $\alpha_s$ and $\Lambda_{\rm QCD}$}

Despite the fact that the notion of the QCD coupling is 
initially a perturbative concept, the associated $\Lambda$-parameter
is non-perturbatively defined
\begin{eqnarray}
   \Lambda 
      \equiv \mu\,(b_0\gbar^2(\mu))^{-b_1/(2b_0^2)} 
              e^{-1/(2b_0\gbar^2(\mu))}
             \exp\left[ -\int_0^{\gbar(\mu)}\,dx 
                        \left( {1\over \beta(x)} + {1 \over b_0x^3} 
                                                 - {b_1 \over b_0^2x}
                        \right) \right] \,,
\label{eq:Lambda}
\end{eqnarray}
where $\beta$ is the full renormalization group function in the scheme
which defines $\gbar$, and $b_0$ and $b_1$ are the first two
scheme-independent coefficients of the perturbative expansion
$\beta(x) \sim -b_0 x^3 -b_1 x^5 + \ldots$.  Thus the
$\Lambda$-parameter is renormalization scheme dependent but in an
exactly computable way, and lattice gauge theory is an ideal method to
relate it to the low-energy properties of QCD.

The change in the coupling from one scheme, $S$, to another (taken here
to be the $\overline{\rm MS}$ scheme) is perturbative,
\begin{eqnarray}
   g_{\overline{\rm MS}}^2(\mu) 
      = g_{\rm S}^2(\mu) (1 + c^{(1)}_g g_{\rm S}^2(\mu) + \ldots ) \,,
\label{eq:g_conversion}
\end{eqnarray}
where $c^{(i)}_g$ are the finite renormalization coefficients.  The
scale $\mu$ must be taken high enough for the error in keeping only
the first few terms in the expansion to be small.  The conversion to
the $\Lambda$-parameter in the $\overline{\rm MS}$ scheme is given by
\begin{eqnarray}
   \Lambda_{\overline{\rm MS}} 
      = \Lambda_{\rm S} \exp\left[ c_g^{(1)}/(2b_0)\right] \,.
\end{eqnarray}

By convention $\alpha_\msbar$ is usually quoted at a scale $\mu=M_Z$
where the appropriate effective coupling is the one in the
five-flavour theory: $\alpha^{(5)}_{\overline{\rm MS}}(M_Z)$.  In
order to obtain it from a lower-flavour result, one connects effective
theories with different number of flavour as discussed by Bernreuther
and Wetzel~\cite{Bernreuther:1981sg}.  For example one considers the
$\msbar$ scheme, matches the three-flavour theory to the four-flavour
theory at a scale given by the charm quark mass, runs with the
four-loop beta-function of the four-flavour theory to a scale given by
the b-quark mass and there matches to the five-flavour theory, after
which one runs up to $\mu=M_Z$.  For the matching relation at a given
quark threshold we use the mass $m_\star$ which satisfies $m_\star=
\overline{m}_\msbar(m_\star)$, where $\overline{m}$ is the running
mass (analogous to the running coupling). Then
\begin{eqnarray}
\label{e:convnfm1}
 \gbar^2_{N_f-1}(m_\star) =  \gbar^2_{N_f}(m_\star)\times 
       [1+t_2\,\gbar^4_{N_f}(m_\star)+t_3\,\gbar^6_{N_f}(m_\star)+ \ldots]
\label{e:grelation}
\end{eqnarray}
with \cite{Chetyrkin:2005ia}
\begin{eqnarray}
  t_2 &=&  {1 \over (4\pi^2)^2} {11\over72}\\
  t_3 &=&  {1 \over (4\pi^2)^3} \left[- {82043\over27648}\zeta_3 + 
                     {564731\over124416}-{2633\over31104}(N_f-1)\right]\, 
\end{eqnarray}
(where $\zeta_3$ is the Riemann zeta-function) provides the matching
at the thresholds in the $\msbar$-scheme.  While $t_2$, $t_3$ are
numerically small coefficients, the charm threshold scale is also
relatively low and so there could be some non-perturbative
uncertainties in the matching procedure, which are difficult to
estimate.  Obviously there is no perturbative matching formula across
the strange ``threshold''; here matching is entirely non-perturbative.
Model dependent extrapolations of $\gbar^2_{N_f}$ from $N_f=0,2$ to
$N_f=3$ were done in the early days of lattice gauge theory. We will
include these in our listings of results but not in our estimates,
since such extrapolations are based on untestable assumptions.

\subsubsection{Overview of the review of $\alpha_s$}

We begin by explaining lattice-specific difficulties in \sect{s:crit}
and the FLAG quality criteria designed to assess whether the
associated systematic uncertainties can be controlled and estimated in
a reasonable manner.  We then discuss, in \sect{s:SF} -- \sect{s:glu},
the various lattice approaches. For completeness, we present results
from calculations with $N_f = 0, 2, 3$, and 4 flavours.  Finally, in
section~\ref{s:alpsumm}, we present averages together with our best
estimates for $\alpha_{\overline{\rm MS}}^{(5)}$. These are determined
from three- and four-flavour QCD simulations. The earlier $N_f = 0, 2$
works obtained results for $N_f = 3$ by extrapolation in
$N_f$. Because this is not a theoretically controlled procedure, we do
not include these results in our averages.  For the $\Lambda$
parameter, we also give results for other number of flavours,
including $N_f=0$. Even though the latter numbers should not be used
for phenomenology, they represent valuable non-perturbative
information concerning field theories with variable numbers of quarks.


\subsection{Discussion of criteria for computations entering the averages}


\label{s:crit}


As in the PDG review, we only use calculations of $\alpha_s$ published
in peer-reviewed journals, and that use NNLO or higher-order
perturbative expansions, to obtain our final range in
Sec.~\ref{s:alpsumm}.  We also, however, introduce further quality
criteria designed to assess the ability to control important
systematics which we describe here.  Some of these criteria, 
  e.g. that for the continuum extrapolation, are associated with
lattice-specific systematics and have no continuum analogue.  Other
criteria, e.g. that for the renormalization scale, could in
principle be applied to non-lattice determinations but are not
considered in the PDG average.  Expecting that lattice calculations
will continue to improve significantly in the near future, our goal in
reviewing the state of the art here is to be conservative and avoid
prematurely choosing an overly small range.

In lattice calculations, we generally take ${\cal O}$ to be some
combination of physical amplitudes or Euclidean correlation functions
which are free from UV and IR divergences and have a well-defined
continuum limit.  Examples include the force between static quarks and
$2$-point functions of quark bilinear currents.

In comparison to values of observables ${\cal O}$ determined
experimentally, those from lattice calculations require two more
steps.  The first step concerns setting the scale $\mu$ in \mbox{GeV},
where one needs to use some experimentally measurable low energy scale
as input. Ideally one employs a hadron mass. Alternatively convenient
intermediate scales such as $\sqrt{t_0}$, $w_0$, $r_0$, $r_1$,
\cite{Luscher:2010iy,Borsanyi:2012zs,Sommer:1993ce,Bernard:2000gd} can
be used if their relation to an experimental dimensionful observable
is established. The low energy scale needs to be computed at the same
bare parameters where ${\cal O}$ is determined, at least as long as
one does not use the step scaling method (see below).  This induces a
practical difficulty given present computing resources.  In the
determination of the low energy reference scale the volume needs to be
large enough to avoid finite size effects. On the other hand, in order
for the perturbative expansion of eq.~(\ref{eq:alpha_MSbar}) to be
reliable, one has to reach sufficiently high values of $\mu$,
i.e.\ short enough distances. To avoid uncontrollable discretisation
effects the lattice spacing $a$ has to be accordingly small.  This
means
\begin{eqnarray}
   L \gg \mbox{hadron size}\sim \Lambda_{\rm QCD}^{-1}\quad 
   \mbox{and} \quad  1/a \gg \mu \,,
   \label{eq:scaleproblem}
\end{eqnarray}
(where $L$ is the box size) and therefore
\begin{eqnarray} 
   L/a \ggg \mu/\Lambda_{\rm QCD} \,.
   \label{eq:scaleproblem2}
\end{eqnarray}
The currently available computer power, however, limits $L/a$, 
typically to
$L/a = 20-64$. 
Unless one accepts compromises in controlling  discretisation errors
or finite size effects, this means one needs to set 
the scale $\mu$ according to
\begin{eqnarray}
   \mu \lll L/a \times \Lambda_{\rm QCD} & \sim 5-20\, \mbox{GeV} \,.
\end{eqnarray}
Therefore, $\mu$ can be $1-3\, \mbox{GeV}$ at most.
This raises the concern whether the asymptotic perturbative expansion
truncated at $1$-loop, $2$-loop, or $3$-loop in eq.~(\ref{eq:alpha_MSbar})
is sufficiently accurate. There is a finite size scaling method,
usually called step scaling method, which solves this problem by identifying 
$\mu=1/L$ in the definition of ${\cal O}(\mu)$, see \sect{s:SF}. 

For the second step after setting the scale $\mu$ in physical units
($\mbox{GeV}$), one should compute ${\cal O}$ on the lattice,
${\cal O}_{\rm lat}(a,\mu)$ for several lattice spacings and take the continuum
limit to obtain the left hand side of eq.~(\ref{eq:alpha_MSbar}) as
\begin{eqnarray}
   {\cal O}(\mu) \equiv \lim_{a\rightarrow 0} {\cal O}_{\rm lat}(a,\mu) 
              \mbox{  with $\mu$ fixed}\,.
\end{eqnarray}
This is necessary to remove the discretisation error.

Here it is assumed that the quantity ${\cal O}$ has a continuum limit,
which is regularisation-independent up to discretisation errors.
The method discussed in \sect{s:WL}, which is based on the perturbative
expansion of a lattice-regulated, divergent short-distance quantity
$W_{\rm lat}(a)$ differs in this respect and must be
treated separately.

In summary, a controlled determination of $\alpha_s$ 
needs to satisfy the following:
\begin{enumerate}

   \item The determination of $\alpha_s$ is based on a
         comparison of a
         short distance quantity ${\cal O}$ at scale $\mu$ with a well--defined
         continuum limit without UV and IR divergences to a perturbative
         expansion formula in eq.~(\ref{eq:alpha_MSbar}).

   \item The scale $\mu$ is large enough so that the perturbative expansion
         in eq.~(\ref{eq:alpha_MSbar}) is precise, i.e. it has good 
         {\em asymptotic} convergence.
         \label{pt_converg}

   \item If ${\cal O}$ is defined by physical quantities in infinite volume,  
         one needs to satisfy \eq{eq:scaleproblem2}.
         \label{constraints}

   \item[] Non-universal quantities need a separate discussion, see
        \sect{s:WL}.

\end{enumerate}

Conditions \ref{pt_converg}. and \ref{constraints}. give approximate lower and
upper bounds for $\mu$ respectively. It is important to see whether there is a
window to satisfy \ref{pt_converg}. and \ref{constraints}. at the same time.
If it exists, it remains to examine whether a particular lattice
calculation is done inside the window or not. 

Obviously, an important issue for the reliability of a calculation is
whether the scale $\mu$ that can be reached lies in a regime where
perturbation theory can be applied with confidence. However, the value
of $\mu$ does not provide an unambiguous criterion. For instance, the
Schr\"odinger Functional, or SF-coupling (sect.~\ref{s:SF}) is
conventionally identified with $\mu=1/L$, but one could also choose
$\mu=2/L$. Instead of $\mu$ we therefore define an effective
$\alpha_{\rm eff}$.  For schemes such as SF (see sect.~\ref{s:SF}) or
$qq$ (see sect.~\ref{s:qq}) this is directly the coupling constant of
the scheme. For other schemes such as the vacuum polarisation we use
the perturbative expansion \eq{eq:alpha_MSbar} for the observable
${\cal O}$ to define
\begin{eqnarray}
   \alpha_{\rm eff} =  {\cal O}/c_1 \,.
   \label{eq:alpeff}
\end{eqnarray}
If there is an $\alpha_s$-independent term it should first be subtracted.
Note that this is nothing but defining an effective,
regularisation-independent coupling,
a physical renormalization scheme.

Let us now comment further on the use of the perturbative series.
Since it is only an asymptotic expansion, the remainder $R_n({\cal
  O})={\cal O}-\sum_{i\leq n}c_i \alpha_s^i$ of a truncated
perturbative expression ${\cal O}\sim\sum_{i\leq n}c_i \alpha_s^i$
cannot just be estimated as a perturbative error $k\,\alpha_s^{n+1}$.
The error is non-perturbative. Often one speaks of ``non-perturbative
contributions'', but non-perturbative and perturbative cannot be
strictly separated due to the asymptotic nature of the series (see
e.g. \cite{Martinelli:1996pk}).

Still, we do have some general ideas concerning the 
size of non-perturbative effects. The known ones such as instantons
or renormalons decay for large $\mu$ like inverse powers of $\mu$
and are thus roughly of the form 
\begin{eqnarray}
   \exp(-\gamma/\alpha_s) \,,
\end{eqnarray}
with some positive constant $\gamma$. Thus we have,
loosely speaking,
\begin{eqnarray}
   {\cal O} = c_1 \alpha_s + c_2 \alpha_s^2 + \ldots + c_n\alpha_s^n
                  + \mathrm{O}(\alpha_s^{n+1}) 
                  + \mathrm{O}(\exp(-\gamma/\alpha_s)) \,.
   \label{eq:Owitherr}
\end{eqnarray}
For small $\alpha_s$, the $\exp(-\gamma/\alpha_s)$ is negligible.
Similarly the perturbative estimate for the magnitude of
relative errors in \eq{eq:Owitherr} is small; as an
illustration for $n=3$ and $\alpha_s = 0.2$ the relative error
is $\sim 0.8\%$ (assuming coefficients $|c_n /c_1 | \sim 1$).

For larger values of $\alpha_s$ non-perturbative effects can become
significant in eq.~(\ref{eq:Owitherr}). An instructive example comes
from the values obtained from $\tau$
decays, for which $\alpha_s\approx 0.3$. Here, different applications
of perturbation theory (fixed order, FOPT, and contour improved, CIPT)
each look reasonably asymptotically convergent but the difference does
not seem to decrease much with the order (see, e.g., the contribution
of Pich in \cite{Bethke:2011tr}). In addition non-perturbative terms
in the spectral function may be non-negligible even after the
integration up to $m_\tau$ (Golterman in \cite{Bethke:2011tr}). All of
this is because $\alpha_s$ is not really small.

Since the size of the non-perturbative effects is very hard to
estimate one should try to avoid such regions of the coupling.  In a
fully controlled computation one would like to verify the perturbative
behaviour by changing $\alpha_s$ over a significant range instead of
estimating the errors as $\sim \alpha_s^{n+1}$ .  Some computations
try to take non-perturbative power `corrections' to the perturbative
series into account by including such terms in a fit to the $\mu$
dependence. We note that this is a delicate procedure, both because
the separation of non-perturbative and perturbative is theoretically
not well defined and because in practice a term like, e.g.,
$\alpha_s(\mu)^3$ is hard to distinguish from a $1/\mu^2$ term when
the $\mu$-range is restricted and statistical and systematic errors
are present. We consider it safer to restrict the fit range to the
region where the power corrections are negligible compared to the
estimated perturbative error.

The above considerations lead us to the following special quality
criteria for the determination of $\alpha_s$. 

\begin{itemize}
   \item Renormalization scale         
         \begin{itemize}
            \item[\good] all points relevant in the analysis have
             $\alpha_{\rm eff} < 0.2$
            \item[\soso] all points have $\alpha_{\rm eff} < 0.4$
                         and at least one $\alpha_{\rm eff} \le 0.25$
            \item[\bad]  otherwise                                   
         \end{itemize}
   \item Perturbative behaviour
         
         \begin{itemize}
            \item[\good]  verified  over a range of a factor $2$
                         in $\alpha_{\rm eff}$ (without power
                         corrections)  
            \item[\soso] agreement with perturbation theory 
                 over a range of a factor
                 $1.5$ in $\alpha_{\rm eff}$ 
                 (possibly fitting with power corrections)
            \item[\bad]  otherwise
        \end{itemize}

   \item Continuum extrapolation 
        
        At a reference point of $\alpha_{\rm eff} = 0.3$ (or less) we require
         \begin{itemize}
            \item[\good] three lattice spacings with
                         $\mu a < 1/2$ and full $O(a)$
                         improvement, \\
                         or three lattice spacings with
                         $\mu a \leq 1/4$ and $2$-loop $O(a)$
                         improvement, \\
                         or $\mu a \leq 1/8$ and $1$-loop $O(a)$
                         improvement 
            \item[\soso] three lattice spacings with $\mu a < 1.5$
                         reaching down to $\mu a =1$ and full
                         $O(a)$ improvement, \\
                         or three lattice spacings with
                         $\mu a \leq 1/4$ and 1-loop $O(a)$
                         improvement        
            \item[\bad]  otherwise  
         \end{itemize}

\end{itemize}  
       
We here assume that the two-loop relation between the used
coupling and $\alpha_{\rm \overline{MS}}$ is always known such that
the three-loop beta-function is known in the scheme considered.
Therefore we have no separate criterion for the order of
perturbation theory. Similarly we assume that quark mass effects
of light quarks (including strange) are negligible in the effective
coupling itself where large, perturbative, $\mu$ is considered.

We also need to specify what is meant by $\mu$. For SF we mean
$\mu=1/L$, for $qq$ it is $\mu=2/r$, for schemes with observables in
momentum space we take the magnitude of the momentum.  Finally, for
moments of heavy quark currents with quark masses $m_\mathrm{h}$ we
use $\mu=2m_\mathrm{h}$.  We note again that the above criteria cannot
be applied when regularisation dependent quantities
$W_\mathrm{lat}(a)$ are used instead of ${\cal O}(\mu)$. These cases
are specifically discussed in \sect{s:WL}.

The usual criterion for the chiral extrapolation and the control over
finite volume effects is missing here for the following reason. These
criteria would apply only to the setting of the scale.  Usually this
has been determined in preceding papers of the collaboration
determining the coupling constant (or indeed by another
collaboration). However, the determination of the scale does not need
to be very precise, since using the lowest order $\beta$-function
shows that a 3\% error in the scale determination corresponds to a
$\sim 0.5\%$ error in $\alpha_s(M_Z)$.  So as long as systematic
errors from chiral extrapolation and finite volume effects are below
3\% we do not need to be concerned about those. This covers
practically all cases. When, exceptionally, it matters we include the
precision of the scale setting in our discussion.

A popular scale choice is the intermediate $r_0$ scale, although one
should also bear in mind that its determination from physical
observables has also to be taken into account.  The phenomenological
value of $r_0$ was originally determined as $r_0 \approx
0.49\,\mbox{fm}$ through potential models describing quarkonia
\cite{Sommer:1993ce}. Recent determinations from 2-flavour QCD are
$r_0 = 0.420(14) - 0.450(14)\,\mbox{fm}$ by the ETM collaboration
\cite{Baron:2009wt,Blossier:2009bx}, using as input $f_\pi$ and $f_K$
and carrying out various continuum extrapolations. On the other hand,
the ALPHA collaboration \cite{Fritzsch:2012wq} determined $r_0 =
0.503(10)\,\mbox{fm}$ with input from $f_K$, and the QCDSF
Collaboration \cite{Bali:2012qs} cites $0.501(10)(11)\,\mbox{fm}$ from
the mass of the nucleon (no continuum limit).  Recent determinations
from three-flavour QCD are consistent with $r_1 = 0.313(3)\,\mbox{fm}$
and $r_0 = 0.472(5)\,\mbox{fm}$
\cite{Davies:2009tsa,Bazavov:2010hj,Bazavov:2011nk}.  Due to the
uncertainty in these estimates, and as many results are based directly
on $r_0$ to set the scale, we shall often give both the dimensionless
number $r_0 \Lambda_{\overline{\rm MS}}$, as well as
$\Lambda_{\overline{\rm MS}}$. In case $r_1 \Lambda_{\overline{\rm
    MS}}$ is given in the publications, we use $r_0 /r_1 = 1.508$
\cite{Bazavov:2011nk} to convert, neglecting the error on this ratio.

The attentive reader will have noticed that bounds such as $\mu a <
1.5$ and $\alpha_\mathrm{eff}<0.25$ which we require for \soso\ are
not very stringent. There is a considerable difference between
\soso\ and \good.  We have chosen the above bounds since not too many
computations would satisfy more stringent ones at present.
Nevertheless, we believe that the \soso\ criteria already give
reasonable bases for estimates of systematic errors. In the future, we
expect that we will be able to tighten our criteria for inclusion in
the average, and that many more computations will reach the present
\good\ rating in one or more categories. 

In principle one should also
account for electro-weak radiative corrections. However, both in the
determination of $\alpha_\mathrm{s}$ at intermediate scales $\mu$ and
in the running to high scales, we expect electro-weak effects to be
much smaller than the presently reached precision. Such effects are
therefore not further discussed.

{

\subsection{$\alpha_s$ from the Schr\"odinger Functional}
\label{s:SF}
\subsubsection{General considerations}


The method of step-scaling functions avoids the scale problem,
\eq{eq:scaleproblem}. It is in principle independent of the particular
boundary conditions used and was first developed with periodic
boundary conditions in a two-dimensional
model~\cite{Luscher:1991wu}. However, at present all applications in
QCD use Schr\"odinger functional boundary
conditions~\cite{Luscher:1992an,Sint:1993un}.  An important reason is
that these boundary conditions avoid zero modes for the quark fields
and quartic modes \cite{Coste:1985mn} in the perturbative expansion in
the gauge fields.  Furthermore the corresponding renormalization
scheme is well studied in perturbation
theory~\cite{Luscher:1993gh,Sint:1995ch,Bode:1999sm} with the
three-loop $\beta$-function and two-loop cutoff effects (for the
standard Wilson regularisation) known.

Let us first briefly review the step scaling strategy.  The essential
idea is to split the determination of the running coupling at large
$\mu$ and of a hadronic scale into two lattice calculations and
connect them by `step scaling'.  In the former part, we determine the
running coupling constant in a finite-volume scheme, in practice a
`Schr\"odinger Functional (SF) scheme' in which the renormalization
scale is set by the inverse lattice size $\mu = 1/L$. In this
calculation, one takes a high renormalization scale while keeping the
lattice spacing sufficiently small as
\begin{eqnarray}
   \mu \equiv 1/L \sim 10\,\ldots\, 100\,\mbox{GeV}\,, \qquad a/L \ll 1 \,.
\end{eqnarray}
In the latter part, one chooses a certain 
$\gbar^2_\mathrm{max}=\gbar^2(1/L_\mathrm{max})$, 
typically such that $L_\mathrm{max}$ is around $0.5\,\fm$. With a 
common discretisation, one then determines $L_\mathrm{max}/a$ and
(in a large volume $L \ge 2 - 3\,\mbox{fm} $)
a hadronic scale such as a hadron mass, $\sqrt{t_0}/a$ or $r_0/a$
at the same bare parameters. In this way one gets numbers for
$L_\mathrm{max}/r_0$ and by changing the lattice spacing $a$ 
carries out a continuum limit extrapolation of that ratio. 
 
In order to connect $\gbar^2(1/L_\mathrm{max})$ to $\gbar^2(\mu)$ at
high $\mu$, one determines the change of the coupling in the continuum
limit when the scale changes from $L$ to $L/2$, starting from
$L=L_{\rm max}$ and arriving at $\mu = 2^k /L_{\rm max}$. This part of
the strategy is called step scaling. Combining these results yields
$\gbar^2(\mu)$ at $\mu = 2^k {r_0 \over L_\mathrm{max}} r_0^{-1}$,
where $r_0$ stands for the particular chosen hadronic scale.

In order to have a perturbatively well-defined scheme,
the SF scheme uses Dirichlet boundary condition at time 
$t = 0$ and $t = T$. These break translation invariance and permit
${\rm O}(a)$ counter terms at the boundary through quantum corrections. 
Therefore, the leading discretisation error is ${\rm O}(a)$.
In practice, improving the lattice action is achieved by adding
$1$-loop or $2$-loop perturbative counter terms at the boundaries
whose coefficients are denoted as $c_t,\tilde c_t$.
A better precision in this step yields a better 
control over discretisation errors, which is important, as can be
seen, e.g., in \cite{Takeda:2004xha,Necco:2001xg}.
The finite $c^{(i)}_g$, \eq{eq:g_conversion}, are 
known for $i=1,2$ 
\cite{Sint:1995ch,Bode:1999sm}.


\subsubsection{Discussion of computations}


In Table~\ref{tab_SF3} we give results from various determinations
\begin{table}[!htb]
   \vspace{3.0cm}
   \footnotesize
   \begin{tabular*}{\textwidth}[l]{l@{\extracolsep{\fill}}rlllllllll}
      Collaboration & Ref. & $\Nf$ &
      \hspace{0.15cm}\begin{rotate}{60}{publication status}\end{rotate}
                                                       \hspace{-0.15cm} &
      \hspace{0.15cm}\begin{rotate}{60}{renormalization scale}\end{rotate}
                                                       \hspace{-0.15cm} &
      \hspace{0.15cm}\begin{rotate}{60}{perturbative behaviour}\end{rotate}
                                                       \hspace{-0.15cm} &
      \hspace{0.15cm}\begin{rotate}{60}{continuum extrapolation}\end{rotate}
                               \hspace{-0.25cm} & 
                         scale & $\Lambda_\msbar[\MeV]$ & $r_0\Lambda_\msbar$ \\
      & & & & & & & & \\[-0.1cm]
      \hline
      \hline
      & & & & & & & & \\
      ALPHA 10A & \cite{Tekin:2010mm} & 4 & \gA &\good & \good &\good 
                    & \multicolumn{3}{l}{only running of $\alpha_s$ in Fig.~4}
                    \\  
      Perez 10 & \cite{PerezRubio:2010ke} & 4 & \oP &\good & \good &\soso  
                    & \multicolumn{3}{l}{only step scaling function in Fig.~4}
                    \\           
      & & & & & & & & & \\[-0.1cm]
      \hline
      & & & & & & & & & \\[-0.1cm]
      PACS-CS 09A& \cite{Aoki:2009tf} & 2+1 
                    & \gA &\good &\good &\soso
                    & $m_\rho$ & $371(13)(8)(^{+0}_{-27})$$^{\#}$
                    & $0.888(30)(18)(^{+0}_{-65})$$^\dagger$
                    \\ 
                    &&&\gA &\good &\good &\soso 
                    & $m_\rho$  & $345(59)$$^{\#\#}$
                    & $0.824(141)$$^\dagger$
                    \\ 
      & & & & & & & & \\[-0.1cm]
      \hline  \\[-1.0ex]
      & & & & & & & & \\[-0.1cm]
      ALPHA 12 $^*$ & \cite{Fritzsch:2012wq} & 2 
                    & \gA &\good &\good &\good
                    &  $f_{\rm K}$ & $310(20)$ &  $0.789(52)$
                    \\
      ALPHA 04 & \cite{DellaMorte:2004bc} & 2 
                    & \gA &\bad &\good &\good
                    & $r_0$$^\S$  & $245(16)(16)^\S$ & $0.62(2)(2)^\S$
                    \\
      ALPHA 01A & \cite{Bode:2001jv} & 2 
                    &\gA & \good & \good & \good 
                    &\multicolumn{3}{l}{only running of $\alpha_s$  in Fig.~5}
                    \\
      & & & & & & & & \\[-0.1cm]
      \hline  \\[-1.0ex]
      & & & & & & & & \\[-0.1cm]
      CP-PACS 04 $^\&$  & \cite{Takeda:2004xha} & 0 
                    & \gA & \good & \good & \soso  
                    & \multicolumn{3}{l}{only tables of $g^2_{\rm SF}$}
                    \\
      ALPHA 98 $^{\dagger\dagger}$ & \cite{Capitani:1998mq} & 0 
                    & \gA & \good & \good & \soso 
                    &  $r_0=0.5\fm$ & $238(19)$ & 0.602(48) 
                    \\
      L\"uscher 93  & \cite{Luscher:1993gh} & 0 
                    & \gA & \good & \good & \soso
                    & $r_0=0.5\fm$ & 233(23)  & 0.590(60)$^{\S\S}$ 
                    \\
      &&&&&&& \\[-0.1cm]
      \hline
      \hline
\end{tabular*}
\begin{tabular*}{\textwidth}[l]{l@{\extracolsep{\fill}}lllllll}
\multicolumn{8}{l}{\vbox{\begin{flushleft} 
$^{\#}$    Result with a constant (in $a$) continuum extrapolation
         of the combination $L_\mathrm{max}m_\rho$. \\
$^\dagger$ In conversion to $r_0\Lambda_{\overline{\rm MS}}$, $r_0$ is
         taken to be $0.472\,\mbox{fm}$. \\
$^{\#\#}$     Result with a linear continuum extrapolation
         in $a$ of the combination $L_\mathrm{max}m_\rho$. \\
$^*$      Supersedes ALPHA 04. \\
$^\S$     The $N_f=2$ results were based on values for $r_0/a$
         which have later been found to be too small by
          \cite{Fritzsch:2012wq}. The effect will be of the order of
          10--15\%, presumably an increase in $\Lambda r_0$.
          We have taken this into account by a $\bad$ in the 
          renormalization scale. \\
$^\&$     This investigation was a precursor for PACS-CS 09A
          and confirmed two step scaling functions as well as the
          scale setting of ALPHA~98. \\
$^{\dagger\dagger}$  Uses data of L\"uscher~93 and therefore supersedes it. \\
$^{\S\S}$ Converted from $\alpha_\msbar(37r_0^{-1})=0.1108(25)$.
\end{flushleft}}}
\end{tabular*}
\vspace{-0.3cm}
\normalsize
%
%
\caption{Results for the $\Lambda$-parameter from computations using 
         step scaling of the SF-coupling. Entries without values for $\Lambda$
         computed the running and established perturbative behaviour
         at large $\mu$. 
         }
\label{tab_SF3}
\end{table}
of the $\Lambda$-parameter. For a clear assessment of the $N_f$
dependence, the last column also shows results that refer to a common
hadronic scale, $r_0$. As discussed above, the renormalization scale
can be chosen large enough such that $\alpha_s < 0.2$ and the
perturbative behaviour can be verified.  Consequently only $\good$ is
present for these criteria.  With dynamical fermions, results for the
step scaling functions are always available for at least $a/L = \mu a
=1/4,1/6, 1/8$.  All calculations have a non-perturbatively
$\mathrm{O}(a)$ improved action in the bulk. For the discussed
boundary $\mathrm{O}(a)$ terms this is not so. In most recent
calculations 2-loop $\mathrm{O}(a)$ improvement is employed together
with at least three lattice spacings.\footnote{With two-loop
  $\mathrm{O}(a)$ improvement we here mean $c_\mathrm{t}$ including
  the $g_0^4$ term and $\tilde c_\mathrm{t}$ with the $g_0^2$
  term. For gluonic observables such as the running coupling this is
  sufficient for cutoff effects being suppressed to $\mathrm{O}(g^6
  a)$.} This means a \good\ for the continuum extrapolation.  In the
other contributions only 1-loop $c_t$ was available and we arrive at \soso. We
note that the discretisation errors in the step scaling functions are
usually found to be very small, at the percent level or
below. However, the overall desired precision is very high as well,
and the results in CP-PACS 04~\cite{Takeda:2004xha} show that
discretisation errors at the below percent level cannot be taken for
granted.  In particular with staggered fermions (unimproved except for
boundary terms) few percent effects are seen in
Perez~10~\cite{PerezRubio:2010ke}.

In the work by PACS-CS 09A~\cite{Aoki:2009tf}, the continuum
extrapolation in the scale setting is performed using a constant
function in $a$ and with a linear function.
Potentially the former leaves a considerable residual discretisation 
error. We here use, as discussed with the collaboration, 
the continuum extrapolation linear in $a$,
as given in the second line  of PACS-CS 09A results in Table~\ref{tab_SF3}.

A single computation, PACS-CS 09A~\cite{Aoki:2009tf}, quotes also
$\alpha_\msbar(M_Z)$. 
We take the linear continuum extrapolation as discussed above:
\begin{eqnarray}
 \alpha_\msbar^{(5)}(M_Z)=0.118(3)\,, 
\end{eqnarray}
where the conversion from a three-flavour result to five-flavours
was done perturbatively (see \sect{s:crit}).
Other results do not have a sufficient number of quark flavours
(ALPHA~10A \cite{Tekin:2010mm}, Perez~10~\cite{PerezRubio:2010ke})
or do not yet contain the conversion of the
scale to physical units. Thus no value for $\alpha_\msbar^{(5)}(M_Z)$ is quoted.
 
More results for $\alpha_\msbar^{(5)}(M_Z)$ using step scaling functions 
can be expected soon. Their precision is likely to be much better than what
we were able to report on here. A major reason is the use of the gradient flow
\cite{Luscher:2010iy} in definitions of finite volume schemes 
\cite{Fodor:2012td,Fritzsch:2013je}.


\subsection{$\alpha_s$ from the potential at short distances}


\subsubsection{General considerations}
\label{s:qq}


The basic method was introduced in \cite{Michael:1992nj} and developed in
\cite{Booth:1992bm}. The force or potential between an infinitely
massive quark and antiquark pair defines an effective coupling
constant via
\begin{eqnarray}
   F(r) = {d V(r) \over dr} 
        = C_F {\alpha_\mathrm{qq}(r) \over r^2} \,.
\label{force_alpha}
\end{eqnarray}
The coupling can be evaluated non-perturbatively from the potential
through a numerical differentiation, see below. In perturbation theory
one also defines couplings in different schemes $\alpha_{\bar{V}}$,
$\alpha_V$ via 
\begin{eqnarray}
   V(r) = - C_F {\alpha_{\bar{V}}(r) \over r} \,, 
   \qquad \mbox{or} \quad
   \tilde{V}(Q) = - C_F {\alpha_V(Q) \over Q^2} \,,
\label{pot_alpha}
\end{eqnarray}
where one fixes the unphysical constant in the potential
by $\lim_{r\to\infty}V(r)=0$ and $\tilde{V}(Q)$ is the
Fourier transform of $V(r)$. Non-perturbatively, the subtraction
of a constant in the potential introduces an additional 
renormalization constant, the value of $V(r_\mathrm{ref})$ at some 
distance $r_\mathrm{ref}$.  Perturbatively, it entails a 
renormalon ambiguity. In perturbation theory, these definitions
are all simply related to each other, and their perturbative
expansions are known including the $\alpha_s^4$ 
and $\alpha_s^4 \log\alpha_s$  terms
\cite{Fischler:1977yf,Billoire:1979ih,Peter:1997me,Schroder:1998vy,
Brambilla:1999qa,Smirnov:2009fh,Anzai:2009tm,Brambilla:2009bi}.
 
The potential $V(r)$ is determined from ratios of Wilson loops,
$W(r,t)$, which behave as
\begin{eqnarray}
   \langle W(r, t) \rangle 
      = |c_0|^2 e^{-V(r)t} + \sum_{n\not= 0} |c_n|^2 e^{-V_n(r)t} \,,
\end{eqnarray}
where $t$ is taken as the temporal extension of the loop, $r$ is the
spatial one and $V_n$ are excited-state potentials.  To improve the
overlap with the ground state, and to suppress the effects of excited
states, $t$ is taken large. Also various additional techniques are
used, such as a variational basis of operators (spatial paths) to help
in projecting out the ground state.  Furthermore some lattice
discretisation effects can be reduced by averaging over Wilson loops
related by rotational symmetry in the continuum.

In order to reduce discretisation errors it is of advantage 
to define the numerical derivative giving the force as
\begin{eqnarray}
   F(r_\mathrm{I}) = { V(r) - V(r-a) \over a } \,,
\end{eqnarray}
where $r_\mathrm{I}$ is chosen so that at tree level the force is the
continuum force. $F(r_\mathrm{I})$ is then a `tree level improved' quantity
and similarly the tree-level improved potential can be defined
\cite{Necco:2001gh}.

Finally, as was noted in sect.~\ref{s:crit}, a determination
of the force can also be used to determine the $r_0$ scale,
by defining it from the static force by
\begin{eqnarray}
   r_0^2 F(r_0) = {1.65} \,.
\end{eqnarray}


\subsubsection{Discussion of computations}


In Table~\ref{tab_short_dist}, we list results of determinations
\begin{table}[htb]
   \vspace{3.0cm}
   \footnotesize
   \begin{tabular*}{\textwidth}[l]{l@{\extracolsep{\fill}}rlllllllll}
      Collaboration & Ref. & $N_f$ &
      \hspace{0.15cm}\begin{rotate}{60}{publication status}\end{rotate}
                                                       \hspace{-0.15cm} &
      \hspace{0.15cm}\begin{rotate}{60}{renormalization scale}\end{rotate}
                                                       \hspace{-0.15cm} &
      \hspace{0.15cm}\begin{rotate}{60}{perturbative behaviour}\end{rotate}
                                                       \hspace{-0.15cm} &
      \hspace{0.15cm}\begin{rotate}{60}{continuum extrapolation}\end{rotate}
                               \hspace{-0.25cm} & 
                         scale & $\Lambda_\msbar[\MeV]$ & $r_0\Lambda_\msbar$ \\
      & & & & & & & & & \\[-0.1cm]
      \hline
      \hline
      & & & & & & & & & \\[-0.1cm]
      Bazavov 12   & \cite{Bazavov:2012ka}   & 2+1       & \gA & \soso$^\dagger$
                   & \soso\  & \soso$^\#$
                   & $r_0 = 0.468\,\mbox{fm}$ 
                   & $295(30)$\,$^\star$ 
                   & $0.70(7)$$^{**}$                                   \\
      & & & & & & & & & \\[-0.1cm]
      \hline
      & & & & & & & & & \\[-0.1cm]
      ETM 11C      & \cite{Jansen:2011vv}    & 2         & \gA & \soso  
                   & \soso  & \soso
                   & $r_0 = 0.42\,\mbox{fm}$
                   & $315(30)$$^\S$ 
                   & $0.658(55)$                                        \\
      & & & & & & & & & \\[-0.1cm]
      \hline
      & & & & & & & & & \\[-0.1cm]
      Brambilla 10 & \cite{Brambilla:2010pp} & 0         & \gA & \soso 
                   & \soso\ & \soso$^{\dagger\dagger}$ &  &  &
                   $0.637(^{+32}_{-30})$$^{\dagger\dagger +}$                  \\
      UKQCD 92     & \cite{Booth:1992bm}    & 0         & \gA & \good 
                                  & \soso$^{++}$   & \bad   
                                  & $\sqrt{\sigma}=0.44\,\GeV$ 
                                  & $256(20)$
                                  & 0.686(54)                             \\
      Bali 92     & \cite{Bali:1992ru}    & 0         & \gA & \good 
                                  & \soso$^{++}$   & \bad 
                                  & $\sqrt{\sigma}=0.44\,\GeV$
                                  & $247(10)$                             
                                  & 0.661(27)                             \\
      & & & & & & & & & \\[-0.1cm]
      \hline
      \hline
\end{tabular*}
\begin{tabular*}{\textwidth}[l]{l@{\extracolsep{\fill}}lllllll}
\multicolumn{8}{l}{\vbox{\begin{flushleft} 
   $^\dagger$
   Since values of $\alpha_\mathrm{eff}$ within our designated range are used,
 we assign a \soso\ despite
   values of $\alpha_\mathrm{eff}$ up to $\alpha_\mathrm{eff}=0.5$ being used.  
   \\
   $^\#$     Since values of $2a/r$ within our designated range are used,
we assign a \soso\ although
   only values of $2a/r\geq1.14$ are used at $\alpha_\mathrm{eff}=0.3$.
\\
   $^\star$ Using results from \cite{Bazavov:2011nk}.  \\
   $^{\star\star}$
         $\alpha^{(3)}_{\overline{\rm MS}}(1.5\,\mbox{GeV}) = 0.326(19)$, 
         $\alpha^{(5)}_{\overline{\rm MS}}(M_Z) = 0.1156(^{+21}_{-22})$.  \\
   $^\S$ Both potential and $r_0/a$ are determined on a small 
   ($L=3.2r_0$) lattice.   \\
   $^{\dagger\dagger}$ Uses lattice results of \cite{Necco:2001xg}, 
   some of which have have very small lattice spacings where 
   according to more recent investigations a bias due to the freezing of
   topology may be present.  \\ 
   $^+$ Only $r_0\Lambda_{\overline{\rm MS}}$ is given.           \\ 
   $^{++}$ We give a $\soso$ because only a NLO formula is used and
       the error bars are very large; our criterion does not apply 
       well to these very early calculations.           
\end{flushleft}}}
\end{tabular*}
\vspace{-0.3cm}
\normalsize
\caption{Short distance potential results.}
\label{tab_short_dist}
\end{table}
of $r_0\Lambda_{\msbar}$ (together with $\Lambda_{\msbar}$
using the scale determination of the authors). 

The first determinations in the three-colour Yang Mills theory are by
UKQCD 92 \cite{Booth:1992bm} and Bali 92, \cite{Bali:1992ru} who used
$\alpha_\mathrm{qq}$ as explained above, but not in the tree-level
improved form. Rather a phenomenologically determined lattice artifact
correction was subtracted from the lattice potentials.  The comparison
with perturbation theory was on a more qualitative level on the basis
of a two-loop formula and a continuum extrapolation could not be
performed as yet. A much more precise computation of
$\alpha_\mathrm{qq}$ with continuum extrapolation was performed in
\cite{Necco:2001xg,Necco:2001gh}. Satisfactory agreement with
perturbation theory was found \cite{Necco:2001gh} but the stability of
the perturbative prediction was not considered sufficient to be able
to extract a $\Lambda$-parameter.

In Brambilla 10 \cite{Brambilla:2010pp} the same quenched lattice
results of \cite{Necco:2001gh} were used and a fit was performed to
the continuum potential, instead of the force, using three-loop
perturbation theory with the $\alpha_s^4 \ln\alpha_s$ term.  Close
agreement with perturbation theory was found when a renormalon
subtraction was performed.  Note that the renormalon subtraction
introduces a second scale into the perturbative formula which is
absent when the force is considered.

For the quenched calculation very small lattice spacings were available. 
For both ETM 11C \cite{Jansen:2011vv} and Bazavov 12 \cite{Bazavov:2012ka}
using dynamical fermions such small lattice spacings are not yet realized. 
They use the tree-level improved potential as described above. 
We note that the value of $\Lambda_\msbar$ in physical units by
 ETM 11C \cite{Jansen:2011vv} is based on a value of $r_0=0.42$~fm. 
 This is at least 10\% smaller than the large majority of
other values of $r_0$. Also the value of $r_0/a$ on the finest lattice
in that computation comes from a rather small lattice with 
$L\approx 3.2r_0 \approx 2.4/m_\pi$. 

One of the main issues for all these computations is whether the
perturbative running of the coupling constant
has been reached. While for quenched or $N_f=0$ fermions this seems
to be the case at the smallest distances, for dynamical fermions at present 
there is no consensus. While both Brambilla 10
\cite{Brambilla:2010pp} and Bazavov 12 \cite{Bazavov:2012ka} find
good agreement with perturbation theory after the renormalon is subtracted, 
Ref.~\cite{Knechtli:2011pz} uses the force, where no renormalon
contributes, and finds that far shorter distances are needed 
than are presently accessible for dynamical fermion simulations in order to 
match to perturbation theory.  
Further work is needed to clarify this point. 


\subsection{$\alpha_s$ from the vacuum polarisation at short distances}


\label{s:vac}


\subsubsection{General considerations}


The vacuum polarisation function for the flavour non-singlet 
currents $J^a_\mu$ ($a=1,2,3$) in the momentum representation is
parameterized as 
\begin{eqnarray}
   \langle J^a_\mu J^b_\nu \rangle 
      =\delta^{ab} [(\delta_{\mu\nu}Q^2 - Q_\mu Q_\nu) \Pi^{(1)}(Q) 
                                     - Q_\mu Q_\nu\Pi^{(0)}(Q)] \,,
\end{eqnarray}
where $Q_\mu$ is a space like momentum and $J_\mu\equiv V_\mu$
for a vector current and $J_\mu\equiv A_\mu$ for an axial-vector current. 
Defining $\Pi_J(Q)\equiv \Pi_J^{(0)}(Q)+\Pi_J^{(1)}(Q)$,
the operator product expansion (OPE) of the vacuum polarisation
function $\Pi_{V+A}(Q)=\Pi_V(Q)+\Pi_A(Q)$ is given by
\begin{eqnarray}
   \lefteqn{\Pi_{V+A}|_{\rm OPE}(Q^2,\alpha_s)}
      & &                                             \nonumber  \\
      &=& c + C_1(Q^2) + C_m^{V+A}(Q^2)
                       \frac{\bar{m}^2(Q)}{Q^2}
            + \sum_{q=u,d,s}C_{\bar{q}q}^{V+A}(Q^2)
                        \frac{\langle m_Q\bar{q}q \rangle}{Q^4}
                                                      \nonumber  \\
      & &   + C_{GG}(Q^2) 
                \frac{\langle \alpha_s GG\rangle}{Q^4}+{\rm O}(Q^{-6}) \,,
\label{eq:vacpol}
\end{eqnarray}
for large
$Q^2$. $C_X^{V+A}(Q^2)=\sum_{i\geq0}\left( C_X^{V+A}\right)^{(i)}\alpha^i(Q^2)$
are the perturbative coefficient functions for the
operators $X$ ($X=1$, $\bar{q}q$, $GG$).
Here $C_1$ is known up to four-loop
order in a continuum renormalization scheme such as the
$\overline{\rm MS}$ scheme \cite{Chetyrkin:2006xg, Boughezal:2006px}.
Non-perturbatively, there are terms in $C_X$ which do not have a 
series expansion in $\alpha_s$. For an example for the unit
operator see \cite{Balitsky:1993ki}.
The term $c$ is $Q$--independent and divergent in the limit of infinite
ultraviolet cutoff. However the Adler function defined as 
\begin{eqnarray}
   D(Q^2) \equiv - Q^2 { d\Pi(Q^2) \over dQ^2} \,,
\end{eqnarray}
is a scheme independent finite quantity. Therefore one can determine
the running coupling constant in the $\overline{\rm MS}$ scheme
from the vacuum polarisation function computed by a lattice QCD
simulation. 
In more detail, the lattice data of the vacuum polarization is fitted with the 
perturbative formula (\ref{eq:vacpol}) with fit parameter 
$\Lambda_{\overline{\rm MS}}$ parameterizing the running coupling 
$\alpha_{\overline{\rm MS}}(Q^2)$.  

While there is no problem in discussing the OPE at the
non-perturbative level, the `condensates' such as ${\langle \alpha_s
  GG\rangle}$ are ambiguous, since they mix with lower dimensional
operators including the unity operator.  Therefore one should work in
the high $Q^2$ regime where power corrections are negligible within
the given accuracy. Thus setting the renormalization scale as
$\mu\equiv \sqrt{Q^2}$, one should seek, as always, the window
$\Lambda_{\rm QCD} \ll \mu \ll a^{-1}$.


\subsubsection{Discussion of computations}


Results using this method are, to date, only available using
overlap fermions. These are collected in Table~\ref{tab_vac} for
\begin{table}[!htb]
   \vspace{3.0cm}
   \footnotesize
   \begin{tabular*}{\textwidth}[l]{l@{\extracolsep{\fill}}rllllllll}
   Collaboration & Ref. & $\Nf$ &
   \hspace{0.15cm}\begin{rotate}{60}{publication status}\end{rotate}
                                                    \hspace{-0.15cm} &
   \hspace{0.15cm}\begin{rotate}{60}{renormalization scale}\end{rotate}
                                                    \hspace{-0.15cm} &
   \hspace{0.15cm}\begin{rotate}{60}{perturbative behaviour}\end{rotate}
                                                    \hspace{-0.15cm} &
   \hspace{0.15cm}\begin{rotate}{60}{continuum extrapolation}\end{rotate}
      \hspace{-0.25cm} & 
                         scale & $\Lambda_\msbar[\MeV]$ & $r_0\Lambda_\msbar$ \\
   & & & & & & & & & \\[-0.1cm]
   \hline
   \hline
   & & & & & & & & & \\[-0.1cm]
   JLQCD 10 & \cite{Shintani:2010ph} & 2+1 &\gA & \bad 
            & \bad & \bad
            & $r_0 = 0.472\,\mbox{fm}$
            & $247(5)$$^\dagger$
            & $0.591(12)$              \\
   & & & & & & & & & \\[-0.1cm]
   \hline
   & & & & & & & & & \\[-0.1cm]
   JLQCD/TWQCD 08C & \cite{Shintani:2008ga} & 2 & \gA & \soso 
            & \bad & \bad
            & $r_0 = 0.49\,\mbox{fm}$
            & $234(9)(^{+16}_{-0})$
            & $0.581(22)(^{+40}_{-0})$    \\
            
   & & & & & & & & & \\[-0.1cm]
   \hline
   \hline
\end{tabular*}
\begin{tabular*}{\textwidth}[l]{l@{\extracolsep{\fill}}llllllll}
\multicolumn{8}{l}{\vbox{\begin{flushleft}
   {$^\dagger$ The paper cites $\alpha_\msbar^{(5)}(M_Z)=0.1181(3)(^{+14}_{-12})$.
   As a result of an inconsistency  found in this estimate by
   the FLAG working group, the number will be revised by JLQCD.}\\
\end{flushleft}}}
\end{tabular*}
\vspace{-0.3cm}
\normalsize
\caption{Vacuum polarisation results}
\label{tab_vac}
\end{table}
$N_f=2$, JLQCD/TWQCD 08C \cite{Shintani:2008ga} and for $N_f = 2+1$, JLQCD 10
\cite{Shintani:2010ph}.
At present, only one lattice spacing $a \approx 0.11\,\mbox{fm}$ 
has been simulated.

The fit to \eq{eq:vacpol} is done with the four-loop relation between
the running coupling and $\lms$.  It is found that without introducing
condensate contributions, the momentum scale where the perturbative
formula gives good agreement with the lattice results is very narrow,
$aQ \simeq 0.8-1.0$.  When condensate contributions are included the
perturbative formula gives good agreement with the lattice results for
the extended range $aQ \simeq 0.6-1.0$. Since there is only a single
lattice spacing there is a \bad\ for the continuum limit.  The
renormalization scale $\mu$ is in the range of $Q=1.6-2\,\mbox{GeV}$.
Choosing $\alpha_{\rm eff}=\alpha_{\overline{\rm MS}}(Q)$, we find
that $\alpha_{\rm eff}=0.25-0.30$ for $N_f=2$ and $\alpha_{\rm
  eff}=0.29-0.33$ for $N_f=2+1$. Thus we give a \soso\ and \bad\ for
$\Nf=2$ and $\Nf=2+1$ respectively for the renormalization scale and a
\bad\ for the perturbative behaviour.


\subsection{$\alpha_s$ from observables at the lattice spacing scale}


\label{s:WL}



\subsubsection{General considerations}


The general method is to evaluate a short distance quantity ${\cal O}$
at the scale of the lattice spacing $\sim 1/a$ and then determine
its relationship to $\alpha_{\overline{\rm MS}}$ via a power series expansion.

This is epitomized by the strategy of the HPQCD Collaboration
\cite{Mason:2005zx,Davies:2008sw}, discussed here for illustration,
which computes and then fits to a variety of short distance quantities, $Y$,
\begin{eqnarray}
   Y = \sum_{n=1}^{n_{\rm max}} c_n \alphah^n(q^*) \,.
\end{eqnarray}
$Y$ is taken as the logarithm of small Wilson loops (including some
non-planar ones), Creutz ratios, `tadpole-improved' Wilson loops and
the tadpole-improved or `boosted' bare coupling ($O(20)$ quantities in
total). $c_n$ are perturbative coefficients (each depending on the
choice of $Y$) known to $n = 3$ with additional coefficients up to
$n_{\rm max}$ being numerically fitted.  $\alphah$ is the running
coupling constant related to $\alphav$ from the static quark potential
(see sect.~\ref{s:qq}).\footnote{ $\alphah$ is defined by
  $\Lambda_\mathrm{V'}=\Lambda_\mathrm{V}$ and
  $b_i^\mathrm{V'}=b_i^\mathrm{V}$ for $i=0,1,2$ but $b_i=0$ for
  $i\geq3$.}

 The coupling
constant is fixed at a scale $q^* = d/a$.
This is chosen as the mean value of $\ln q$ with the one gluon loop
as measure,
\cite{Lepage:1992xa,Hornbostel:2002af}. (Thus a different result
for $d$ is found for every short distance quantity.)
A rough estimate yields $d \approx \pi$, and in general the
renormalization scale is always found to lie in this region.

For example for the Wilson loop $W_{mn} \equiv \langle W(ma,na) \rangle$
we have
\begin{eqnarray}
   \ln\left( \frac{W_{mn}}{u_0^{2(m+n)}}\right)
      = c_1 \alphah(q^*) +  c_2 \alphah^2(q^*)  + c_3 \alphah^3(q^*)
        + \cdots \,,
\label{short-cut}
\end{eqnarray}
for the tadpole-improved version, where $c_1$, $c_2\,, \ldots$
are the appropriate perturbative coefficients and $u_0 = W_{11}^{1/4}$.
Substituting the non-perturbative simulation value in the left hand side,
we can determine $\alphah(q^*)$, at the scale $q^*$.
Note that one finds empirically that perturbation theory for these
tadpole-improved quantities have smaller $c_n$ coefficients and so
the series has a faster apparent convergence.

Using the $\beta$ function in the $\rm V'$-scheme,
results can be run to a reference value, chosen as
$\alpha_0 \equiv \alphah(q_0)$, $q_0 = 7.5\,\mbox{GeV}$.
This is then converted perturbatively to the continuum
$\msbar$ scheme
\begin{eqnarray}
   \alpha_{\overline{\rm MS}}(q_0)
      = \alpha_0 + d_1 \alpha_0^2 + d_2 \alpha_0^3 + \cdots \,,
\end{eqnarray}
where $d_1, d_2$ are known one and two loop coefficients.

Other collaborations have focused more on the bare `boosted'
coupling constant and directly determined its relationship to
$\alpha_{\overline{\rm MS}}$. Specifically, the boosted coupling is
defined by 
\begin{eqnarray}
   \alphap(1/a) = {1\over 4\pi} {g_0^2 \over u_0^4} \,,
\end{eqnarray}
again determined at a scale $\sim 1/a$. As discussed previously
since the plaquette expectation value in the boosted coupling
contains the tadpole diagram contributions to all orders, which
are dominant contributions in perturbation theory,
there is an expectation that the perturbation theory using
the boosted coupling has 
smaller perturbative coefficients \cite{Lepage:1992xa}, and hence smaller 
perturbative errors.
 

\subsubsection{Continuum limit}


Lattice results always come along with discretisation errors,
which one needs to remove by a continuum extrapolation.
As mentioned previously, in this respect the present
method differs in principle from those in which $\alpha_s$ is determined
from physical observables. In the general case, the numerical
results of the lattice simulations at a value of $\mu$ fixed in physical 
units can be extrapolated to the continuum limit, and the result can be 
analysed as to whether it shows perturbative running as a function of 
$\mu$ in the continuum. For observables at the cutoff-scale ($q^*=d/a$),  
discretisation effects cannot easily be separated out
from perturbation theory, as the scale for the coupling
comes from the lattice spacing. 
Therefore the restriction  $a\mu  \ll 1$ (the `continuum extrapolation'
criterion) is not applicable here. Discretisation errors of 
order $a^2$ are, however, present. Since 
$a\sim \exp(-1/(2b_0 g_0^2)) \sim \exp(-1/(8\pi b_0 \alpha(q^*))$, 
these errors now appear as power corrections to the perturbative 
running, and have to be taken into account in the study of the 
perturbative behaviour, which is to be verified by changing $a$. 
One thus always should fit with power corrections in this method.

In order to keep a symmetry with the `continuum extrapolation' 
criterion for physical observables and to remember that discretisation 
errors are, of course, relevant, 
we replace it here by one for the lattice spacings used:
\begin{itemize}
   \item Lattice spacings
         \begin{itemize}
            \item[\good] 
               3 or more lattice spacings, at least 2 points below
               $a = 0.1\,\mbox{fm}$
            \item[\soso]
               2 lattice spacings, at least 1 point below
               $a = 0.1\,\mbox{fm}$
            \item[\bad]
               otherwise 
         \end{itemize}
\end{itemize}


\subsubsection{Discussion of computations}


\begin{table}[p]
   \vspace{3.0cm}
   \footnotesize
   \begin{tabular*}{\textwidth}[l]{l@{\extracolsep{\fill}}rllllllll}
   Collaboration & Ref. & $N_f$ &
   \hspace{0.15cm}\begin{rotate}{60}{publication status}\end{rotate}
                                                    \hspace{-0.15cm} &
   \hspace{0.15cm}\begin{rotate}{60}{renormalization scale}\end{rotate}
                                                    \hspace{-0.15cm} &
   \hspace{0.15cm}\begin{rotate}{60}{perturbative behaviour}\end{rotate}
                                                    \hspace{-0.15cm} &
   \hspace{0.15cm}\begin{rotate}{60}{lattice spacings}\end{rotate}
      \hspace{-0.25cm} & 
                         scale & $\Lambda_\msbar[\MeV]$ & $r_0\Lambda_\msbar$ \\
   & & & & & & & & \\[-0.1cm]
   \hline
   \hline
   & & & & & & & & \\[-0.1cm]
   HPQCD 10$^a$ $^\S$& \cite{McNeile:2010ji}& 2+1 & \gA & \soso
            & \good & \good
            & $r_1 = 0.3133(23)\, \mbox{fm}$
            & 340(9) 
            & 0.812(22)                                   \\ 
   HPQCD 08A$^a$& \cite{Davies:2008sw} & 2+1 & \gA & \soso
            & \good & \good
            & $r_1 = 0.321(5)\,\mbox{fm}$$^{\dagger\dagger}$
            & 338(12)$^\star$
            & 0.809(29)                                   \\
   Maltman 08$^a$& \cite{Maltman:2008bx}& 2+1 & \gA & \soso
            & \soso & \soso
            & $r_1 = 0.318\, \mbox{fm}$
            & 352(17)$^\dagger$
            & 0.841(40)                                   \\ 
   HPQCD 05A$^a$ & \cite{Mason:2005zx} & 2+1 & \gA & \soso
            & \soso & \soso
            & $r_1$$^{\dagger\dagger}$
            & 319(17)$^{\star\star}$
            & 0.763(42)                                   \\
   & & & & & & & & &  \\[-0.1cm]
   \hline
   & & & & & & & & &  \\[-0.1cm]
   QCDSF/UKQCD 05 & \cite{Gockeler:2005rv}  & 2 & \gA & \good
            & \bad  & \good
            & $r_0 = 0.467(33)\,\mbox{fm}$
            & 261(17)(26)
            & 0.617(40)(21)$^b$                           \\
   SESAM 99$^c$ & \cite{Spitz:1999tu} & 2 & \gA & \soso
            & \bad  & \bad
            & $c\bar{c}$(1S-1P)
            & 
            &                                             \\
   Wingate 95$^d$ & \cite{Wingate:1995fd} & 2 & \gA & \good
            & \bad  & \bad
            & $c\bar{c}$(1S-1P)
            & 
            &                                             \\
   Davies 94$^e$ & \cite{Davies:1994ei} & 2 & \gA & \good
            & \bad & \bad
            & $\Upsilon$
            & 
            &                                             \\
   Aoki 94$^f$ & \cite{Aoki:1994pc} & 2 & \gA & \good
            & \bad & \bad
            & $c\bar{c}$(1S-1P)
            & 
            &                                             \\
   & & & & & & & & &  \\[-0.1cm]
   \hline
   & & & & & & & & &  \\[-0.1cm]
   QCDSF/UKQCD 05 & \cite{Gockeler:2005rv}  & 0 & \gA & \good
            & \soso & \good
            & $r_0 = 0.467(33)\,\mbox{fm}$
            & 259(1)(20)
            & 0.614(2)(5)$^b$                              \\
   SESAM 99$^c$ & \cite{Spitz:1999tu} & 0 & \gA & \good
            & \bad  & \bad
            & $c\bar{c}$(1S-1P)
            & 
            &                                             \\
   Wingate 95$^d$ & \cite{Wingate:1995fd} & 0 & \gA & \good
            & \bad  & \bad
            & $c\bar{c}$(1S-1P)
            & 
            &                                             \\
   Davies 94$^e$ & \cite{Davies:1994ei}  & 0 & \gA & \good
            & \bad & \bad
            & $\Upsilon$
            & 
            &                                             \\
   El-Khadra 92$^g$ & \cite{ElKhadra:1992vn} & 0 & \gA & \good
            & \soso   & \soso
            & $c\bar{c}$(1S-1P)
            & 234(10)
            & 0.593(25)$^h$                                            \\
   & & & & & & & & &  \\[-0.1cm]
   \hline
   \hline
\end{tabular*}
\begin{tabular*}{\textwidth}[l]{l@{\extracolsep{\fill}}lllllll}
\multicolumn{8}{l}{\vbox{\begin{flushleft} 
   $^a$       The numbers for $\Lambda$ have been converted from the values for 
              $\alpha_s^{(5)}(M_Z)$. \\
   $^{\S}$     $\alpha_{\overline{\rm MS}}^{(3)}(5\ \mbox{GeV})=0.2034(21)$,
              $\alpha^{(5)}_{\overline{\rm MS}}(M_Z)=0.1184(6)$,
              only update of intermediate scale and $c$, $b$ quark masses,
              supersedes HPQCD 08A and Maltman 08. \\
   $^\dagger$ $\alpha^{(5)}_{\overline{\rm MS}}(M_Z)=0.1192(11)$. \\
   $^\star$    $\alpha_V^{(3)}(7.5\,\mbox{GeV})=0.2120(28)$, 
              $\alpha^{(5)}_{\overline{\rm MS}}(M_Z)=0.1183(8)$,
              supersedes HPQCD 05. \\
   $^{\dagger\dagger}$ Scale is originally determined from $\Upsilon$
              mass splitting. $r_1$ is used as an intermediate scale.
              In conversion to $r_0\Lambda_{\overline{\rm MS}}$, $r_0$ is
              taken to be $0.472\,\mbox{fm}$. \\
   $^{\star\star}$ $\alpha_V^{(3)}(7.5\,\mbox{GeV})=0.2082(40)$,
              $\alpha^{(5)}_{\overline{\rm MS}}(M_Z)=0.1170(12)$. \\
   $^b$       This supersedes 
              \cite{Gockeler:2004ad,Booth:2001uy,Booth:2001qp}.
              $\alpha^{(5)}_{\overline{\rm MS}}(M_Z)=0.112(1)(2)$.
              The $N_f=2$ results were based on values for $r_0 /a$
              which have later been found to be too small~\cite{Fritzsch:2012wq}. 
              The effect will  
              be of the order of 10--15\%, presumably an increase in 
              $\Lambda r_0$. \\
   $^c$       $\alpha^{(5)}_{\overline{\rm MS}}(M_Z)=0.1118(17)$. \\
   $^d$    
   $\alpha_V^{(3)}(6.48\,\mbox{GeV})=0.194(7)$ extrapolated from $\Nf=0,2$.
              $\alpha^{(5)}_{\overline{\rm MS}}(M_Z)=0.107(5)$.   \\
   $^e$       
              $\alpha_P^{(3)}(8.2\,\mbox{GeV})=0.1959(34)$ extrapolated
              from $N_f=0,2$. $\alpha^{(5)}_{\overline{\rm MS}}(M_Z)=0.115(2)$.
              \\
   $^f$       Estimated $\alpha^{(5)}_{\overline{\rm MS}}(M_Z)=0.108(5)(4)$. \\
   $^g$   This early computation violates our requirement that scheme conversions are 
             done at the two-loop level.  \\
   $^h$       Used $r_0=0.5$fm to convert to $r_0 \lms$.
              $\Lambda_{\overline{\rm MS}}^{(4)}=160(^{+47}_{-37})\mbox{MeV}$, 
              $\alpha^{(4)}_{\overline{\rm MS}}(5\mbox{GeV})=0.174(12)$.
              We converted this number to give
              $\alpha^{(5)}_{\overline{\rm MS}}(M_Z)=0.106(4)$.\\   
\end{flushleft}}}
\end{tabular*}
\vspace{-0.3cm}
\normalsize
\caption{Wilson loop results. }
\label{tab_wloops}
\end{table}


Note that due to $\mu \sim 1/a$ being relatively large the
results easily have a $\good$ or $\soso$ in the rating on 
renormalization scale.

The work of El-Khadra 92 \cite{ElKhadra:1992vn} employs a 1-loop
formula to relate $\alpha^{(0)}_{\overline{\rm MS}}(\pi/a)$
to the boosted coupling for three lattice spacings
$a^{-1} = 1.15$, $1.78$, $2.43\,\mbox{GeV}$. (The lattice spacing
is determined from the charmonium 1S-1P splitting.) They obtain
$\Lambda^{(0)}_{\overline{\rm MS}}=234\,\mbox{MeV}$, corresponding
to $\alpha_{\rm eff} = \alpha^{(0)}_{\overline{\rm MS}}(\pi/a)
\approx 0.15$ - $0.2$. The work of Aoki 94 \cite{Aoki:1994pc}
calculates $\alpha^{(2)}_V$ and $\alpha^{(2)}_{\overline{\rm MS}}$
for a single lattice spacing $a^{-1}\sim 2\,\mbox{GeV}$ again
determined from charmonium 1S-1P splitting in two-flavour QCD.
Using one-loop perturbation theory with boosted coupling,
they obtain $\alpha^{(2)}_V=0.169$ and $\alpha^{(2)}_{\overline{\rm MS}}=0.142$.
Davies 94 \cite{Davies:1994ei} gives a determination of $\alphav$
from the expansion 
\begin{equation}
   -\ln W_{11} \equiv \frac{4\pi}{3}\alphav^{(N_f)}(3.41/a)
        \times [1 - (1.185+0.070N_f)\alphav^{(N_f)} ]\,,
\end{equation}
neglecting higher order terms.  They compute the $\Upsilon$ spectrum
in $N_f=0$, $2$ QCD for single lattice spacings at $a^{-1} = 2.57$,
$2.47\,\mbox{GeV}$ and obtain $\alphav(3.41/a)\simeq 0.1$5, $0.18$
respectively.  Extrapolating the inverse coupling linearly in $N_f$, a
value of $\alphav^{(3)}(8.3\,\mbox{GeV}) = 0.196(3)$ is obtained.
SESAM 99 \cite{Spitz:1999tu} follows a similar strategy, again for a
single lattice spacing. They linearly extrapolated results for
$1/\alphav^{(0)}$, $1/\alphav^{(2)}$ at a fixed scale of
$9\,\mbox{GeV}$ to give $\alphav^{(3)}$, which is then perturbatively
converted to $\alpha_{\overline{\rm MS}}^{(3)}$. This finally gave
$\alpha_{\overline{\rm MS}}^{(5)}(M_Z) = 0.1118(17)$.  Wingate 95
\cite{Wingate:1995fd} also follow this method.  With the scale
determined from the charmonium 1S-1P splitting for single lattice
spacings in $N_f = 0$, $2$ giving $a^{-1}\simeq 1.80\,\mbox{GeV}$ for
$N_f=0$ and $a^{-1}\simeq 1.66\,\mbox{GeV}$ for $N_f=2$ they obtain
$\alphav^{(0)}(3.41/a)\simeq 0.15$ and $\alphav^{(2)}\simeq 0.18$
respectively. Extrapolating the coupling linearly in $N_f$, they
obtain $\alphav^{(3)}(6.48\,\mbox{GeV})=0.194(17)$.

The QCDSF/UKQCD Collaborations, QCDSF/UKQCD \cite{Gockeler:2005rv},
\cite{Gockeler:2004ad,Booth:2001uy,Booth:2001qp},
use the two-loop relation (re-written here in terms of $\alpha$)
\begin{eqnarray}
   {1 \over \alpha_{\overline{\rm MS}}(\mu)} 
      = {1 \over \alphap(1/a)} 
        + 4\pi(2b_0\ln a\mu - t_1^P) 
        + (4\pi)^2(2b_1\ln a\mu - t_2^P)\alphap(1/a) \,,
\label{gPtoMSbar}
\end{eqnarray}
where $t_1^P$ and $t_2^P$ are known. (A two-loop relation corresponds
to a three-loop lattice beta function.)  This was used to
directly compute $\alpha_{\rm \overline{\rm MS}}$, and the scale was
chosen so that the $O(\alpha^0_P)$ term vanishes, i.e.\
\begin{eqnarray}
   \mu^* = {1 \over a} \exp{[t_1^P/(2b_0)] } 
        \approx \left\{ \begin{array}{cc}
                           2.63/a  & N_f = 0 \\
                           1.4/a   & N_f = 2 \\
                        \end{array}
                 \right. \,.
\label{amustar}
\end{eqnarray}
The method is to first compute $\alphap(1/a)$ and from this using
eq.~(\ref{gPtoMSbar}) to find $\alpha_{\overline{\rm MS}}(\mu^*)$.
The RG equation, eq.~(\ref{eq:Lambda}), then determines
$\mu^*/\Lambda_{\overline{\rm MS}}$ and hence using
eq.~(\ref{amustar}) leads to the result for $r_0\Lambda_{\overline{\rm
    MS}}$.  This avoids giving the scale in $\mbox{MeV}$ until the
end.  In the $\Nf=0$ case $7$ lattice spacings were used,
\cite{Necco:2001xg}, giving a range $\mu^*/\Lambda_{\overline{\rm MS}}
\approx 24$ - $72$ (or $a^{-1} \approx 2$ - $7\,\mbox{GeV}$) and
$\alpha_{\rm eff} = \alpha_{\overline{\rm MS}}(\mu^*) \approx 0.14$ -
$ 0.11$. Neglecting higher order perturbative terms (see discussion
after eq.~(\ref{qcdsf:ouruncert}) below) in eq.~(\ref{gPtoMSbar}) this
is sufficient to allow a continuum extrapolation of
$r_0\Lambda_{\overline{\rm MS}}$.
A similar computation for $N_f = 2$ by QCDSF/UKQCD~05 \cite{Gockeler:2005rv}
gave $\mu^*/\Lambda_{\overline{\rm MS}} \approx 12$ - $17$
(or roughly $a^{-1} \approx 2$ - $3\,\mbox{GeV}$) 
and $\alpha_{\rm eff} = \alpha_{\overline{\rm MS}}(\mu^*)
\approx 0.20$ - $ 0.18$.
The $N_f=2$ results of QCDSF/UKQCD~05 are affected by an 
uncertainty which was not known at the time of publication: 
It has been realized that the values of $r_0/a$ of \cite{Gockeler:2005rv}
were significantly too low~\cite{Fritzsch:2012wq}. 
As this effect is expected to depend on $a$, it
influences the perturbative behaviour leading us to assign 
a \bad\ for that criterion.

The work of HPQCD 05A \cite{Mason:2005zx} (which supersedes
the original work \cite{Davies:2003ik}) uses three lattice spacings
$a^{-1} \approx 1.2$, $1.6$, $2.3\,\mbox{GeV}$ for $2+1$
flavour QCD. Typically the renormalization scale
$q \approx \pi/a \approx 3.50 - 7.10\,\mbox{GeV}$, corresponding to
$\alpha_{\rm eff} \equiv \alpha_\mathrm{V'} \approx 0.22-0.28$. 

In the later update HPQCD 08A \cite{Davies:2008sw} twelve data sets
(with six lattice spacings) are now used reaching up to $a^{-1}
\approx 4.4\,\mbox{GeV}$ corresponding to $\alpha_{\rm eff} \approx
0.18$. The values used for the scale $r_1$ were further updated in
HPQCD 10 \cite{McNeile:2010ji}.  Maltman 08~\cite{Maltman:2008bx}
uses most of the same lattice ensembles as HPQCD
08A~\cite{Davies:2008sw}, but considers a much smaller set of
quantities (three versus 22) that are less sensitive to condensates.
They also use different strategies for evaluating the condensates and
for the perturbative expansion, and a slightly different value for the
scale $r_1$.  The central values of the final results from Maltman 08
and HPQCD 08A differ by 0.0009 (which would be decreased to 0.0007
taking into account a reduction of 0.0002 in the value of the $r_1$
scale used by Maltman 08).
 
As mentioned before, the perturbative coefficients are computed
through $3$-loop order\cite{Mason:2004zt}, while the higher order
perturbative coefficients $c_n$ with $ n_{\rm max} \ge n > 3$ (with
$n_{\rm max} = 10$) are numerically fitted using the lattice
simulation data for the lattice spacings with the help of Bayesian
methods.  It turns out that corrections in \eq{short-cut} are of order
$|c_i/c_1|\alpha^i=$ 20\%, 5--15\% and 3--10\% for $i=1,2,3$,
respectively.  The inclusion of a fourth-order term is necessary to
obtain a good fit to the data, and leads to a shift of the result by
$1$ -- $2$ sigma. For all but one of the 22 quantities, central values
of $\approx 2-4$ were found, with errors from the fits of $\approx 2$.
 
For many of the quantities, the fitted central values of the ratios
$|c_4/c_1|$ appear to be larger than corresponding lower-order ratios
(which would be worrying for the application of perturbation theory),
but the coefficients $|c_5/c_1|$ are essentially undetermined by the
data and the errors on $|c_4/c_1|$ are sufficiently large that it is
premature to decide this issue.

Perturbative truncation errors are the largest source of uncertainty
in HPQCD~08A/10A, and a significant contribution in Maltman; both
estimate this error to be about $0.3$--$0.4$\%.  Maltman uses the
changes observed from fitting to data at the three finest versus
fitting to data at all lattice spacings, while HPQCD uses the
(correlated) errors in their fitted coefficients $c_4$ and $c_5$.  As
discussed in the introduction and conclusions, however, perturbative
truncation errors are notoriously difficult to estimate.  In the
concluding section (\ref{s:alpsumm}), we therefore also consider a
more conservative power-counting estimate of the perturbative error,
taking the estimated size of the $c_4$ term as the uncertainty.  With
$\alpha_1=\alpha_\msbar^{(3)}(5\,\GeV)$ and
$\alpha_2=\alpha_\msbar^{(5)}(M_Z)$ we have
\begin{eqnarray} 
  \Delta \alpha_1 = 
          \left|{c_4 \over c_1}\right|\alpha_1^4 \,,\quad
         \Delta \alpha_2 = 
         \left|{c_4 \over c_1}\right|\alpha_1^2 \alpha_2^2 \,,\quad
         {\Delta\Lambda \over \Lambda} 
         = {1\over 8\pi b_0 \alpha_1} { \Delta \alpha_1 \over \alpha_1}\,.
\label{qcdsf:ouruncert}
\end{eqnarray}
In order to obtain a numerical value we need 
$|c_4/c_1|$. It has been estimated as part of the fit by HPQCD.
Since the fit results are $|c_4/c_1| = 4\pm 2$
for the (log of the) plaquette and  unimproved 
Wilson-loops, the estimated four-loop
correction from eq.~(\ref{qcdsf:ouruncert}) is of order  2--6\%.   

As perturbative coefficients are fit parameters, it is important to
have isolated the perturbative piece of the short distance quantity,
or to show that non-perturbative effects are small. Checks were made
expanding the short distance quantity in a Taylor expansion in the
quark mass and adding `gluon condensate'-like terms.  This did not
change the fits perceptibly. With the $\alpha_{\rm eff}$ values given
above we assign a $\soso$ for the renormalization scale.  According to
our criterion the perturbative behaviour is verified.  However, one
should keep mind that it was necessary to include fitted higher-order
coefficients in order to describe the data.  The fact that these
fitted coefficients are not well-determined by the data makes the test
less stringent.  Table~\ref{tab_wloops} summarizes the results.


\subsection{$\alpha_s$ from current two-point functions}


\label{s:curr}


\subsubsection{General considerations}


The method has been introduced in \cite{Allison:2008xk}
and updated in \cite{McNeile:2010ji}, see also
\cite{Bochkarev:1995ai}. The basic observable is constructed
from a current $J(x) = i m_h\overline\psi_h(x)\gamma_5\psi_{h'}(x)$
of two mass-degenerate heavy valence quarks, $h$, $h^\prime$.
The pre-factor $m_h$ denotes the bare mass of the quark.
With a residual chiral symmetry, $J(x)$ is a renormalization group
invariant local field, i.e.\ it requires no renormalization.
Staggered fermions and twisted mass fermions have such a residual
chiral symmetry. The (Euclidean) time-slice correlation function
\begin{eqnarray}
   G(x_0) = a^3 \sum_{\vec{x}} \langle J^\dagger(x) J(0) \rangle \,,
\end{eqnarray}
($J^\dagger(x) = im_h\overline\psi_{h'}(x)\gamma_5\psi_{h}(x)$)
has a $\sim x_0^{-3}$ singularity at short distances and moments
\begin{eqnarray}
   G_n = a \sum_{t=-(T/2-a)}^{T/2-a} t^n \,G(t) \,,
\label{Gn_smu}
\end{eqnarray}
are finite for $n \ge 4$. Here $T$ is the time extent of the lattice.
The moments are dominated by contributions at $t$ of order $1/m_h$.
For large mass $m_h$ these are short distances and the moments
become increasingly perturbative for decreasing $n$.
Denoting the lowest order perturbation theory moments by $G_n^{(0)}$,
one defines the normalized moments
\begin{eqnarray}
   R_n = \left\{ \begin{array}{cc}
                    G_4/G_4^{(0)}          & \mbox{for $n=4$} \,, \\[0.5em]
                    {m_\mathrm{p} G_n^{1/(n-4)} \over
                     2m_h \left( G_n^{(0)} \right)^{1/(n-4)} }
                                        & \mbox{for $n \ge 6$} \,, \\
                 \end{array}
         \right.
\label{ratio_GG}
\end{eqnarray}
of even order $n$. 
The mass, $m_\mathrm{p}$, of the pseudoscalar flavoured $h h'$ state
is used to make $G_n$ dimensionless, while 
in the denominator the bare quark mass is used for this purpose.
In the continuum limit
the normalized moments can be parameterized in terms of functions
\begin{eqnarray}
   R_n \equiv \left\{ \begin{array}{cc}
                         r_4(\alpha_s(\mu), \mu/\bar{m}_h(\mu))
                                        & \mbox{for $n=4$} \,,     \\[0.5em]
                         z\cdot r_n(\alpha_s(\mu), \mu/\bar{m}_h(\mu))
                                        & \mbox{for $n \ge 6$} \,, \\
                      \end{array}
              \right.
\end{eqnarray}
where
\begin{eqnarray}
   z = { m_\mathrm{p} \over 2\bar{m}_h(\mu) } \,,
\end{eqnarray}
with $\bar{m}_h(\mu)$ being the renormalized quark mass.
The prefactor $z$ parameterizes the heavy quark mass and the quantities
$r_n$ have a perturbative expansion
\begin{eqnarray}
   r_n = 1 + r_{n,1}\alpha_s + r_{n,2}\alpha_s^2 + r_{n,3}\alpha_s^3 + \ldots\,,
\label{rn_expan}
\end{eqnarray}
where the written terms $r_{n,i}(\mu/\bar{m}_h(\mu))$, $i \le 3$ are known
for low $n$ from \cite{Chetyrkin:2006xg,Boughezal:2006px,Maier:2008he,
Maier:2009fz,Kiyo:2009gb}. In practice, the expansion is used in
the $\overline{\rm MS}$ scheme. Matching non-perturbative lattice results
for the moments to the perturbative expansion, one can determine an
approximation to $\alpha_{\overline{\rm MS}}(\mu)$ as well as $\bar m_h(\mu)$.
With the lattice spacing (scale) determined from some extra physical input,
this calibrates $\mu$.

A difficulty with this approach is that large masses are needed to enter
the perturbative domain. Lattice artefacts can then be sizeable and
have a complicated form. The ratios in eq.~(\ref{ratio_GG}) use the
tree level lattice results in the usual way for normalization.
This results in unity as the leading term in eq.~(\ref{rn_expan}),
suppressing some of the kinematical lattice artefacts.
We note that in contrast to e.g.\ the definition of $\alpha_\mathrm{qq}$,
here the cutoff effects are of order $a^n\alpha_s$, while there the
tree level term defines $\alpha_s$ and therefore the cutoff effects
after tree level improvement are of order $a^n\alpha_s^2$.

Furthermore finite size effects (FSE) due to the omission of
$|t| > T /2$ in eq.~(\ref{Gn_smu}) grow with $n$ as 
$(m_\mathrm{p}T/2)^n\, \exp{(-m_\mathrm{p} T/2)}$. 
In practice, however, since the (lower) moments
are short distance dominated, the FSE are expected to be irrelevant
at the present level of precision.  In the definitions above,
the mass of an artificial non-singlet pseudoscalar meson has 
been used, since this is done in the simulations. In the
determinations of the quark masses, this mass is approximated
by the mass of the $\eta$ (or $\eta_b$) in Nature. The difference,
due to quark-line disconnected diagrams is usually assumed to be small.
For the determination of $\alpha_s$, this approximation is actually
irrelevant, since one can consider the moments at arbitrary
(valence) quark masses.

Moments of correlation functions of the quark's electromagnetic
current can also be obtained from experimental data for $e^+e^-$
annihilation~\cite{Kuhn:2007vp,Chetyrkin:2009fv}.  This enables a
non-lattice determination of $\alpha_s$ using a similar analysis
method.  In particular, the same continuum perturbation theory enters
both the lattice and the phenomenological determinations.


\subsubsection{Discussion of computations}


The method has been applied in HPQCD 08B \cite{Allison:2008xk} and in
HPQCD 10 \cite{McNeile:2010ji}, based on the MILC ensembles with $2 +
1$ flavours of ASQTAD staggered quarks and HISQ valence quarks. The
scale was set using $r_1 = 0.321(5)\,\mbox{fm}$ in HPQCD 08B and the
updated value $r_1 = 0.3133(23)\,\mbox{fm}$ in HPQCD 10. The
effective range of couplings used is here given for $n = 4$, which is
the moment most dominated by short (perturbative) distances and
important in the determination of $\alpha_s$. The range is similar for
other ratios. With $r_{4,1} = 0.7427$ and $R_4 = 1.281(5)$ determined
in the continuum limit at the charm mass in \cite{Allison:2008xk}, we
have $\alpha_{\rm eff} = 0.38$ at the charm quark mass, which is the
mass value where HPQCD 08B carries out the analysis.  In HPQCD 10 a
set of masses is used, with $R_4 \in [1.090, 1.293]$ which corresponds
to $\alpha_{\rm eff} \in [0.121, 0.395]$.

The available data of HPQCD 10 is summarized in Fig.~\ref{hpqcd_alpha_eff}
where we plot $\alpha_\mathrm{eff}$ against $m_\mathrm{p} r_1$.
For the continuum limit criterion, we choose the scale $\mu = 2\bar m_h
\approx m_\mathrm{p}/1.1$, where we have taken $\bar m_h$ in the $\msbar$
scheme at scale $\bar m_h$ and the numerical value $1.1$ was determined in
HPQCD 10B.
\begin{figure}[!htb]
\hspace{-0.4cm}   \includegraphics[width=0.56\textwidth]{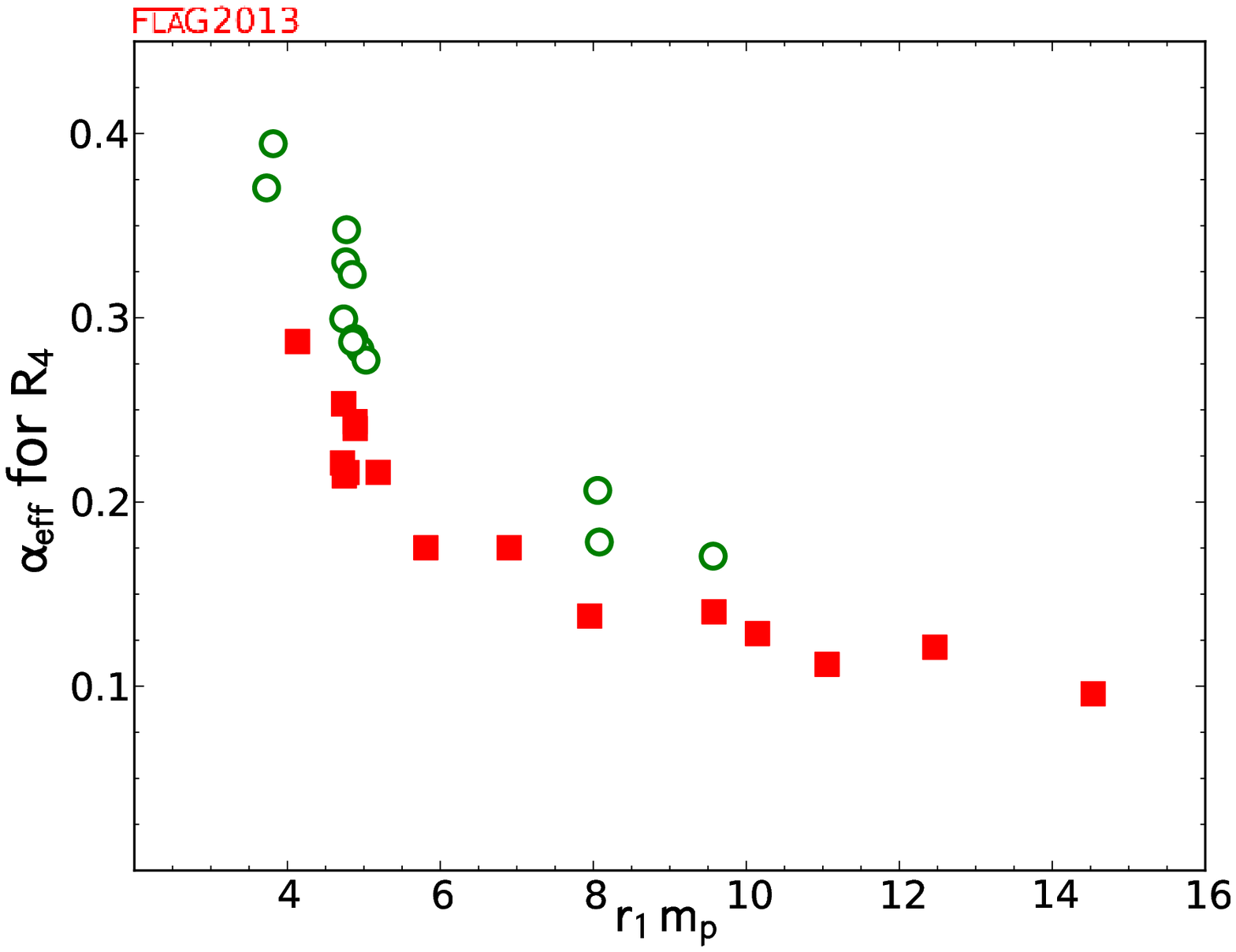}\hspace{-0.8cm} 
      \includegraphics[width=0.56\textwidth]{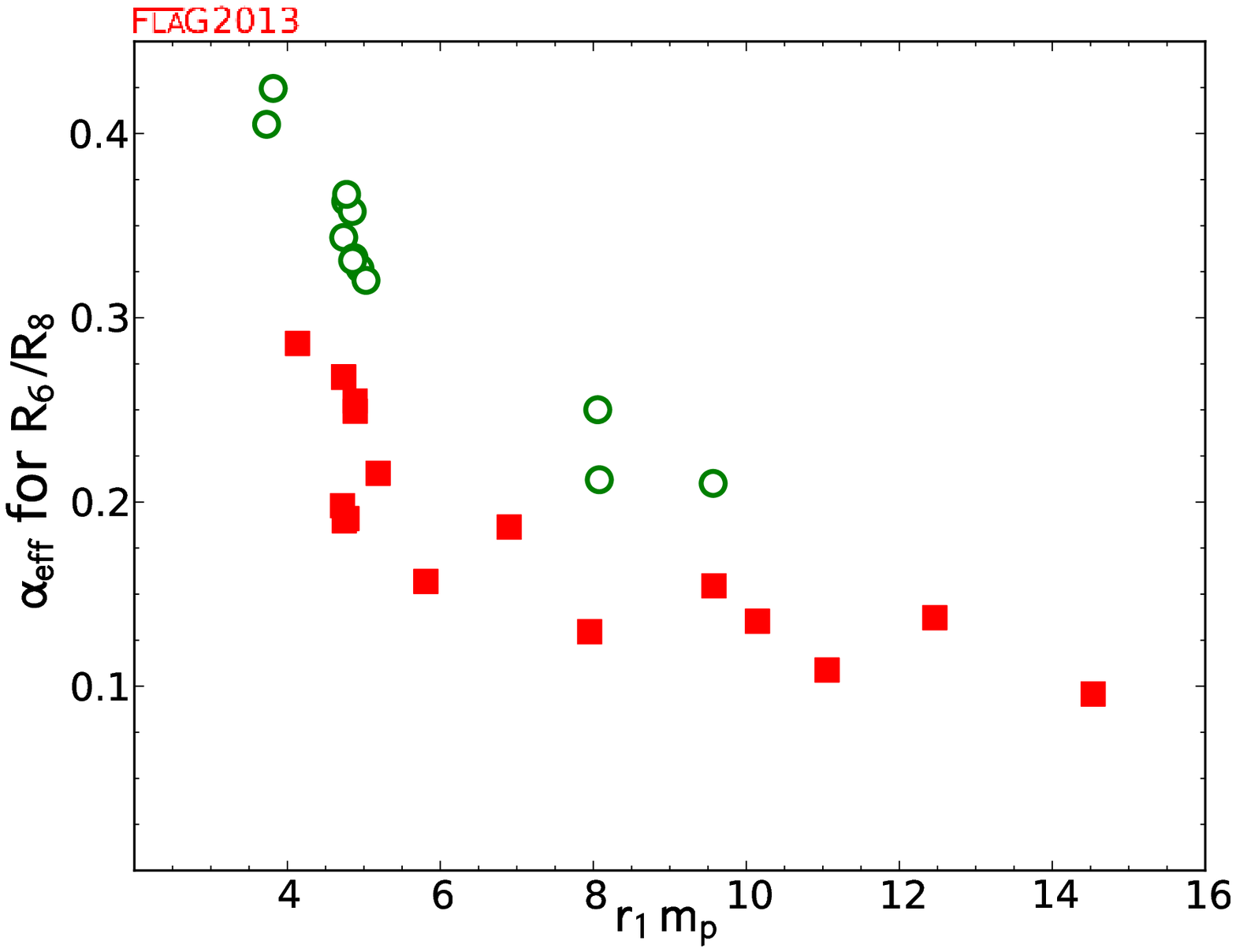}
\caption{$\alpha_{\rm eff}$ for $R_4$ (left) and
         $R_6/R_8$ (right) versus $r_1m_{\rm p}$.
         Symbols correspond to our continuum limit criterion,
         namely \protect\soso\ for data with $1\leq a\mu\leq 1.5$ and
         \protect\bad\ for $a\mu>1.5$, while \protect\good\ is not present. }
\label{hpqcd_alpha_eff}
\end{figure}
The data in Fig.~\ref{hpqcd_alpha_eff} are
grouped according to the range of $a\mu$ that they cover.
The vertical spread of the results for $\alpha_{\rm eff}$ 
at fixed $r_1m_\mathrm{p}$
in the figure measures the discretisation errors seen for
large masses: in the continuum we would expect all the points 
to lie on one universal curve. The plots illustrate the selection applied by
our quality criterion for the continuum limit with our choices for
$\mu$. Fig.~\ref{hpqcd_alpha_eff} gives reason for concern, since it shows
that the discretization errors that need to be removed in the continuum
extrapolation are not small.

With our choices for $\mu$, the continuum limit criterion is satisfied
for 3 lattice spacings when $\alpha_\mathrm{eff} \leq 0.3$ and $n=4$.
Larger $n$ moments are more influenced by non-perturbative effects.
For the $n$ values considered, adding a gluon condensate term, which
largely accounts for these effects, only changed error bars slightly.
We note that HPQCD in their papers perform a global fit to all data
using a joint expansion in powers of $\alpha_s^n$, $\left(
\Lambda/(m_\mathrm{p}/2) \right)^j$ to parameterize the heavy-quark
mass dependence, and $\left( am_\mathrm{p}/2 \right)^{2i}$ to
parameterize the lattice-spacing dependence.  To obtain a good fit,
they must exclude data with $am_\mathrm{p} > 1.95$ and include
lattice-spacing terms} $a^{2i}$ with $i$ greater than $10$.  Because
  these fits include many more fit parameters than data points, HPQCD
  uses their expectations for the sizes of coefficients as Bayesean
  priors.  The fits include data with masses as large as $am_{\textrm
    p}/2 \sim0.86$, so there is only minimal suppression of the many
  high order contributions for the heavier masses.  It is not clear,
  however, how sensitive the final results are to the larger
  $am_{\textrm p}/2$ values in the data.  The continuum limit of the
  fit is in agreement with a perturbative scale dependence (a
  five-loop running $\alpha_{\overline{\rm MS}}$ with a fitted
  five-loop coefficient in the beta-function is used).  Indeed, Fig.~2
  of Ref.~\cite{McNeile:2010ji} suggests that HPQCD's fit describes
  the data well.

In Table~\ref{tab_current_2pt} we list the current
two point function results.
Thus far, only one group has used this approach,
which models complicated and potentially large cutoff effects together
with a perturbative coefficient. We therefore are waiting to see confirmation
by other collaborations
of the small systematic errors obtained (cf.\ discussion in 9.9.2). 
We do however include the values of $\alpha_{\overline{\rm MS}}(M_Z)$
and $\Lambda_{\overline{\rm MS}}$ of HPQCD 10 in our final range.

\begin{table}[!htb]
   \vspace{3.0cm}
   \footnotesize
   \begin{tabular*}{\textwidth}[l]{l@{\extracolsep{\fill}}rlllllllll}
      Collaboration & Ref. & $N_f$ &
      \hspace{0.15cm}\begin{rotate}{60}{publication status}\end{rotate}
                                                       \hspace{-0.15cm} &
      \hspace{0.15cm}\begin{rotate}{60}{renormalization scale}\end{rotate}
                                                       \hspace{-0.15cm} &
      \hspace{0.15cm}\begin{rotate}{60}{perturbative behaviour}\end{rotate}
                                                       \hspace{-0.15cm} &
      \hspace{0.15cm}\begin{rotate}{60}{continuum extrapolation}\end{rotate}
      \hspace{-0.25cm} & 
                         scale & $\Lambda_\msbar[\MeV]$ & $r_0\Lambda_\msbar$ \\
      &&&&&&&& \\[-0.1cm]
      \hline
      \hline
      &&&&&&&& \\[-0.1cm]
      HPQCD 10    & \cite{McNeile:2010ji}  & 2+1       & \gA & \soso
                   & \soso  & \soso           
                   & $r_1 = 0.3133(23)\, \mbox{fm}$$^\dagger$
                   & 338(10)$^\star$           &  0.809(25)            \\
      HPQCD 08B    & \cite{Allison:2008xk}  & 2+1       & \gA & \bad 
                   & \bad  & \bad           
                   & $r_1 = 0.321(5)\,\mbox{fm}$$^\dagger$  
                   & 325(18)$^+$             &  0.777(42)  \\
      &&&&&&&& \\[-0.1cm]
      \hline
      \hline
\end{tabular*}
\begin{tabular*}{\textwidth}[l]{l@{\extracolsep{\fill}}lllllll}
\multicolumn{8}{l}{\vbox{\begin{flushleft} 
   $^\dagger$ Scale is determined from $\Upsilon$ mass splitting.    \\
   $^\star$  $\alpha^{(3)}_{\overline{\rm MS}}(5\,\mbox{GeV}) = 0.2034(21)$,
            $\alpha^{(5)}_{\overline{\rm MS}}(M_Z) = 0.1183(7)$.         \\
   $^+$     $\alpha^{(4)}_{\overline{\rm MS}}(3\,\mbox{GeV}) = 0.251(6)$,
            $\alpha^{(5)}_{\overline{\rm MS}}(M_Z) = 0.1174(12)$.         \\
\end{flushleft}}}
\end{tabular*}
\vspace{-0.3cm}
\normalsize
\caption{Current two point function results.}
\label{tab_current_2pt}
\end{table}


\subsection{$\alpha_s$ from QCD vertices}

\label{s:glu}


\subsubsection{General considerations}


The most intuitive and in principle direct way to determine the
coupling constant in QCD is to compute the appropriate
three or four point gluon vertices
or alternatively the
quark-quark-gluon vertex or ghost-ghost-gluon vertex (i.e.\ $
q\overline{q}A$ or $c\overline{c}A$ vertex respectively).  A suitable
combination of renormalization constants then leads to the relation
between the bare (lattice) and renormalized coupling constant. This
procedure requires the implementation of a non-perturbative
renormalization condition and the fixing of the gauge. For the study
of non-perturbative gauge fixing and the associated Gribov ambiguity,
we refer to \cite{Cucchieri:1997dx,Giusti:2001xf,Maas:2009ph} and
references therein.
In practice the Landau gauge is used and the
renormalization constants are defined 
by requiring that the vertex is equal to the tree level
value at a certain momentum configuration.
The resulting renormalization
schemes are called `MOM' scheme (symmetric momentum configuration)
or `$\rm \widetilde{MOM}$' (one momentum vanishes), 
which are then converted perturbatively
to the $\overline{\rm MS}$ scheme.

A pioneering work to determine the three gluon vertex in the $N_f = 0$
theory is Alles~96~\cite{Alles:1996ka} (which was followed by
\cite{Boucaud:2001qz} for two flavour QCD); a more recent $N_f = 0$
computation was \cite{Boucaud:2005gg} in which the three gluon vertex
as well as the ghost-ghost-gluon vertex was considered.  (This
requires in general a computation of the propagator of the
Faddeev--Popov ghost on the lattice.) The latter paper concluded that
the resulting $\Lambda_{\overline{\rm MS}}$ depended strongly on the
scheme used, the order of perturbation theory used in the matching and
also on non-perturbative corrections, \cite{Boucaud:2005xn}.

Subsequently in \cite{Sternbeck:2007br,Boucaud:2008gn} a specific
$\widetilde{\rm MOM}$ scheme with zero ghost momentum for the
ghost-ghost-gluon vertex was used. In this scheme, dubbed
the `MM' (Minimal MOM) or `Taylor' (T) scheme, the vertex
is not renormalized, and so the renormalized coupling reduces to
\begin{eqnarray}
   \alpha_{\rm T}(\mu) 
      = D^{\rm ghost}_{\rm lat}(\mu, a) D^{\rm gluon}_{\rm lat}(\mu, a)^2 \,
                      {g_0^2(a) \over 4\pi} \,,
\end{eqnarray}
where $D^{\rm ghost}_{\rm lat}$ and $D^{\rm gluon}_{\rm lat}$ are the
(bare lattice) dressed ghost and gluon `form factors' of these
propagator functions in the Landau gauge,
\begin{eqnarray}
   D^{ab}(p) = - \delta^{ab}\, {D^{\rm ghost}(p) \over p^2}\,, \qquad
   D_{\mu\nu}^{ab}(p) 
      = \delta^{ab} \left( \delta_{\mu\nu} - {p_\mu p_\nu \over p^2} \right) \,
        {D^{\rm gluon}(p) \over p^2 } \,,
\end{eqnarray}
and we have written the formula in the continuum with 
$D^{\rm ghost/gluon}(p)=D^{\rm ghost/gluon}_{\rm lat}(p, 0)$.
Thus there is now no need to compute the ghost-ghost-gluon vertex,
just the ghost and gluon propagators.


\subsubsection{Discussion of computations}


\begin{table}[!p]
   \vspace{3.0cm}
   \footnotesize
   \begin{tabular*}{\textwidth}[l]{l@{\extracolsep{\fill}}rllllllll}
   Collaboration & Ref. & $\Nf$ &
   \hspace{0.15cm}\begin{rotate}{60}{publication status}\end{rotate}
                                                    \hspace{-0.15cm} &
   \hspace{0.15cm}\begin{rotate}{60}{renormalization scale}\end{rotate}
                                                    \hspace{-0.15cm} &
   \hspace{0.15cm}\begin{rotate}{60}{perturbative behaviour}\end{rotate}
                                                    \hspace{-0.15cm} &
   \hspace{0.15cm}\begin{rotate}{60}{continuum extrapolation}\end{rotate}
      \hspace{-0.25cm} & 
                         scale & $\Lambda_\msbar[\MeV]$ & $r_0\Lambda_\msbar$ \\
      & & & & & & & & \\[-0.1cm]
      \hline
      \hline
      & & & & & & & & \\[-0.1cm]
      ETM 13D        & \cite{Blossier:2013ioa}   & {2+1+1} & {\gA}
                    & \soso & \soso  & \bad  
                    & $f_\pi$
                    & $314(7)(14)(10)$$^\S$
                    & $0.752(18)(34)(81)$$^\dagger$                               \\
       ETM 12C        & \cite{Blossier:2012ef}   & 2+1+1 & \gA 
                    & \soso & \soso  & \bad  
                    & $f_\pi$
                    & $324(17)$$^\S$
                    & $0.775(41)$$^\dagger$                                \\
      ETM 11D       & \cite{Blossier:2011tf}   & 2+1+1 & \gA 
                    & \soso & \soso  & \bad  
                    & $f_\pi$
                    & $316(13)(8)(^{+0}_{-9})$$^\star$
                    & $0.756(31)(19)(^{+0}_{-22})$$^\dagger$                 \\
      & & & & & & & & \\[-0.1cm]
      \hline
      & & & & & & & & \\[-0.1cm]
      Sternbeck 12  & \cite{Sternbeck:2012qs}  & 2+1  & \rC
                    &     &        & 
                    & \multicolumn{3}{l}{only running of 
                                         $\alpha_s$ in Fig.~4}            \\
      & & & & & & & & \\[-0.1cm]
      \hline
      & & & & & & & & \\[-0.1cm]
      Sternbeck 12  & \cite{Sternbeck:2012qs}  & 2  & \rC
                    &  &  & 
                    & \multicolumn{3}{l}{Agreement with $r_0\Lambda_\msbar$ value of \cite{Fritzsch:2012wq} }                              \\
      Sternbeck 10  & \cite{Sternbeck:2010xu}  & 2  & \rC 
                    & \soso  & \good & \bad
                    &
                    &
                    & $0.60(3)(2)$$^\#$                                   \\
      ETM 10F       & \cite{Blossier:2010ky}   & 2  & \gA 
                    & \soso  & \soso  & \soso 
                    & $f_\pi$
                    & $330(23)(22)(^{+0}_{-33})$\hspace{-2mm}
                    & $0.72(5)$$^+$                                       \\
      Boucaud 01B    & \cite{Boucaud:2001qz}    & 2 & \gA 
                    & \soso & \soso  & \bad
                    & $K^{\ast}-K$
                    & $264(27)$$^{\star\star}$
                    & 0.669(69)                              \\
      & & & & & & & & \\[-0.1cm]
      \hline
      & & & & & & & & \\[-0.1cm]
      Sternbeck 12  & \cite{Sternbeck:2012qs}   & 0 & \rC 
                    &  &  &
                    &  \multicolumn{3}{l}{Agreement with $r_0\Lambda_\msbar$ value of \cite{Brambilla:2010pp}}
\\
      Sternbeck 10  & \cite{Sternbeck:2010xu}   & 0 & \rC
                    & \good & \good & \bad
                    &
                    &
                    & $0.62(1)$$^\#$                                      \\
      Ilgenfritz 10 & \cite{Ilgenfritz:2010gu}  & 0 & \gA
                    &    \good    &  \good      & \bad 
                    & \multicolumn{2}{l}{only running of
                                         $\alpha_s$ in Fig.~13}           \\
{Boucaud 08}    & \cite{Boucaud:2008gn}       & 0         &\gA  
                    & \soso & \soso  & \bad 
                    & $\sqrt{\sigma} = 445\,\mbox{MeV}$
                    & $224(3)(^{+8}_{-5})$
                    & $0.59(1)(^{+2}_{-1})$         
 \\
{Boucaud 05}    & \cite{Boucaud:2005gg}       & 0       &\gA  
                    & \bad & \soso  & \bad 
                    & $\sqrt{\sigma} = 445\,\mbox{MeV}$
                    & 320(32)
                    & 0.85(9)           
 \\
   Soto 01        & \cite{DeSoto:2001qx}        & 0         & \gA  
                    & \soso & \soso  & \soso
                    & $\sqrt{\sigma} = 445\,\mbox{MeV}$
                    & 260(18)
                    & 0.69(5)          
 \\
{Boucaud 01A}    & \cite{Boucaud:2001st}      & 0         &\gA  
                    & \soso & \soso  & \soso
                    & $\sqrt{\sigma} = 445\,\mbox{MeV}$
                    & 233(28)~MeV
                    & 0.62(7)       
 \\
{Boucaud 00B}   & \cite{Boucaud:2000nd}      & 0         &\gA  
                    & \soso & \soso  & \soso
                    & 
                    & \multicolumn{2}{l}{only running of
                                         $\alpha_s$}
 \\
{Boucaud 00A}     &\cite{Boucaud:2000ey}     &  0    &\gA  
                    & \soso & \soso  & \soso
                    & $\sqrt{\sigma} = 445\,\mbox{MeV}$
                    & $237(3)(^{+~0}_{-10})$
                    & $0.63(1)(^{+0}_{-3})$            
 \\
{Becirevic 99B}  & \cite{Becirevic:1999hj} & 0 &\gA  
                    & \soso & \soso  & \bad 
                    & $\sqrt{\sigma} = 445\,\mbox{MeV}$
                    & $319(14)(^{+10}_{-20})$
                    & $0.84(4)(^{+3}_{-5})$   
 \\
{Becirevic 99A}  & \cite{Becirevic:1999uc} & 0 &\gA  
                    & \soso & \soso  & \bad 
                    & $\sqrt{\sigma} = 445\,\mbox{MeV}$
                    & $\lesssim 353(2)(^{+25}_{-15})$
                    & $\lesssim 0.93 (^{+7}_{-4})$         
 \\
{Boucaud 98B}  & \cite{Boucaud:1998xi} & 0 &\gA  
                    & \bad  & \soso  & \bad 
                    & $\sqrt{\sigma} = 445\,\mbox{MeV}$
                    & 295(5)(15)
                    & 0.78(4)           
 \\
{Boucaud 98A}    & \cite{Boucaud:1998bq} & 0 &\gA  
                    & \bad  & \soso  & \bad 
                    & $\sqrt{\sigma} = 445\,\mbox{MeV}$
                    & 300(5)
                    & 0.79(1)         
\\
{Alles 96}    & \cite{Alles:1996ka} & 0 &\gA  
                    & \bad  & \soso  & \bad 
                    & $\sqrt{\sigma} = 440\,\mbox{MeV}$\hspace{0.3mm}$^{++}$\hspace{-0.3cm}        
                    & 340(50)
                    & 0.91(13)   
\\
      & & & & & & & & \\[-0.1cm]
      \hline
      \hline
\end{tabular*}
\begin{tabular*}{\textwidth}[l]{l@{\extracolsep{\fill}}lllllll}
\multicolumn{8}{l}{\vbox{\begin{flushleft} 
   $^\dagger$ We use the 2+1 value $r_0=0.472$~fm.                        \\
   $^\S$ $\alpha_{\overline{\rm MS}}^{(5)}(M_Z)=0.1200(14)$.                   \\
   $^\star$ First error is statistical; second is due to the lattice
           spacing and third is due to the chiral extrapolation.
           $\alpha_{\overline{\rm MS}}^{(5)}(M_Z)=0.1198(9)(5)(^{+0}_{-5})$.    \\
   $^\#$ Only $r_0\Lambda_{\overline{\rm MS}}$ is given.                     \\
   $^+$ The determination of $r_0$
        from the $f_\pi$ scale is found in \cite{Baron:2009wt}.          \\
   $^{\star\star}$  $\alpha_{\overline{\rm MS}}^{(5)}(M_Z)=0.113(3)(4)$.         \\
   $^{++}$  The scale is taken from the string tension computation of \cite{Bali:1992ru}.
\end{flushleft}}}
\end{tabular*}
\vspace{-0.3cm}
\normalsize
\caption{Results for the gluon--ghost vertex.}
\label{tab_vertex}
\end{table}

For the calculations considered here, to match to perturbative
scaling, it was first necessary to reduce lattice artifacts by an
$H(4)$ extrapolation procedure (addressing $O(4)$ rotational
invariance), e.g.\ ETM 10F \cite{Blossier:2010ky} or lattice
perturbation theory, e.g.\ Sternbeck 12 \cite{Sternbeck:2012qs}.  To
match to perturbation theory, collaborations vary in their approach.
In ETM 10F \cite{Blossier:2010ky} it was necessary to include the
operator $A^2$ in the OPE of the ghost and gluon propagators, while in
{Sternbeck 12 \cite{Sternbeck:2012qs}} very large momenta are used and
$a^2p^2$ and $a^4p^4$ terms are included in their fit to the momentum
dependence.  A further later refinement was the introduction of
nonperturbative OPE power corrections in ETM 11D
\cite{Blossier:2011tf} and ETM 12C \cite{Blossier:2012ef}.  Although
the expected leading power correction, $1/q^4$, was tried, ETM finds
good agreement with their data only when they fit with the
next-to-leading order term, $1/q^6$.  The update ETM 13D
\cite{Blossier:2013ioa} investigates this point in more detail, using
better data with reduced statistical errors.  They find that after
again including the $1/q^6$ term they can describe their data over a
large momentum range from about 1.75~GeV to 7~GeV.

In all calculations except for Sternbeck 10 \cite{Sternbeck:2010xu},
Sternbeck 12 \cite{Sternbeck:2012qs} ,
the matching with the perturbative formula is performed including
power corrections in the form of 
condensates, in particular $\langle A^2 \rangle$. 
Three lattice spacings are present in almost all 
calculations with $N_f=0$, $2$, but the scales $ap$ are rather large.
This mostly results in a $\bad$ on the continuum extrapolation
(Sternbeck 10 \cite{Sternbeck:2010xu} ,
  Boucaud 01B \cite{Boucaud:2001qz} for $N_f=2$.
 Ilgenfritz 10 \cite{Ilgenfritz:2010gu},   
 Boucaud 08 \cite{Boucaud:2008gn} ,
 Boucaud 05 \cite{Boucaud:2005gg}, 
 Becirevic 99B \cite{Becirevic:1999hj},
  Becirevic 99A \cite{Becirevic:1999uc},
 Boucaud 98B \cite{Boucaud:1998xi},
 Boucaud 98A \cite{Boucaud:1998bq},
 Alles 96 \cite{Alles:1996ka} for $N_f=0$).
A \soso\ is reached in the $\Nf=0$ computations 
Boucaud 00A \cite{Boucaud:2000ey}, 00B \cite{Boucaud:2000nd},
01A \cite{Boucaud:2001st}, Soto 01 \cite{DeSoto:2001qx} due to
a rather small lattice spacing,  but this is done on a lattice
of a small physical size. 
The $N_f=2+1+1$ calculation, fitting with condensates, 
is carried out for two lattice spacings
and with $ap>1.5$ , giving $\bad$
for the continuum extrapolation as well. 
In { ETM 10F \cite{Blossier:2010ky}} we have
$0.25 < \alpha_{\rm eff} < 0.4$, while in ETM 11D, ETM 12C (and ETM 13)
we find
$0.24 < \alpha_{\rm eff} < 0.38$ which gives a green circle
in these cases for the renormalization scale.
In ETM~10F the values of $ap$ violate our criterion
for a continuum limit only slightly, and 
we give a \soso.

In {Sternbeck 10 \cite{Sternbeck:2010xu}}, the coupling ranges over
$0.07 \leq \alpha_{\rm eff} \leq 0.32$ for $N_f=0$ and $0.19 \leq
\alpha_{\rm eff} \leq 0.38$ for $N_f=2$ giving $\good$ and $\soso$ for
the renormalization scale respectively.  The fit with the perturbative
formula is carried out without condensates, giving a satisfactory
description of the data.  In {Boucaud 01A~\cite{Boucaud:2001st}},
depending on $a$, a large range of $\alpha_{\rm eff}$ is used which
goes down to $0.2$ giving a $\soso$ for the renormalization scale and
perturbative behaviour, and several lattice spacings are used leading
to $\soso$ in the continuum extrapolation.  The $\Nf=2$ computation
Boucaud 01B~\cite{Boucaud:2001st}, fails the continuum limit criterion
because both $a\mu$ is too large and an unimproved Wilson fermion
action is used.  Finally in the conference proceedings
Sternbeck~12~\cite{Sternbeck:2012qs}, the $N_f=0,2,3$ coupling
$\alpha_\mathrm{T}$ is studied.  Subtracting 1-loop lattice artefacts
and subsequently fitting with $a^2p^2$ and $a^4p^4$ additional lattice
artefacts, agreement with the perturbative running is found for large
momenta ($r_0^2p^2 > 600$) without the need for power corrections.  In
these comparisons, the values of $r_0\Lambda_\msbar$ from other
collaborations are used. As no numbers are given, we have not
introduced ratings for this study.

In Table~\ref{tab_vertex} we summarize the results. Presently there
are no $N_f \geq 3$ calculations of $\alpha_s$ from QCD vertices that
satisfy the FLAG criteria to be included in the range.


\subsection{Summary}


\label{s:alpsumm}
\begin{table}[!htb]
   \vspace{3.0cm}
   \footnotesize
   \begin{tabular*}{\textwidth}[l]{l@{\extracolsep{\fill}}rlllllllr}
   Collaboration & Ref. & $N_f$ &
   \hspace{0.15cm}\begin{rotate}{60}{publication status}\end{rotate}
                                                    \hspace{-0.15cm} &
   \hspace{0.15cm}\begin{rotate}{60}{renormalisation scale}\end{rotate}
                                                    \hspace{-0.15cm} &
   \hspace{0.15cm}\begin{rotate}{60}{perturbative behaviour}\end{rotate}
                                                    \hspace{-0.15cm} &
   \hspace{0.15cm}\begin{rotate}{60}{continuum extrapolation}\end{rotate}
      \hspace{-0.25cm} & 
       $\alpha_\msbar(M_\mathrm{Z})$ & Method  & Table \\
   & & & & & & & & \\[-0.1cm]
   \hline
   \hline
   & & & & & & & & \\[-0.1cm]
   ETM 13D    &  \cite{Blossier:2013ioa}   & 2+1+1& \gA
                    & \soso & \soso  & \bad 
                    & 0.1196(4)(8)(16)
                    & gluon-ghost vertex
                    & \ref{tab_vertex}                         \\
   ETM 12C    & \cite{Blossier:2012ef}   & 2+1+1 & \gA 
                    & \soso & \soso  & \bad  
                    & 0.1200(14)
		 & gluon-ghost vertex
                    & \ref{tab_vertex}                         \\
   ETM 11D   & \cite{Blossier:2011tf}   & 2+1+1 & \gA 
             & \soso & \soso & \bad  
                    & $0.1198(9)(5)(^{+0}_{-5})$
                    & gluon-ghost vertex
                    & \ref{tab_vertex}                         \\
   & & & & & & & & &  \\[-0.1cm]
   \hline
   & & & & & & & & & \\[-0.1cm]
   Bazavov 12  & \cite{Bazavov:2012ka}   & 2+1       & \gA & \soso
            & \soso  & \soso
            & $0.1156(^{+21}_{-22})$ 
            & $Q$-$\bar{Q}$ potential
            & \ref{tab_short_dist}                            \\
   HPQCD 10   & \cite{McNeile:2010ji}  & 2+1       & \gA & \soso
             & \soso  & \soso          
             & 0.1183(7)          
             & current two points
             & \ref{tab_current_2pt}                              \\
   HPQCD 10& \cite{McNeile:2010ji}& 2+1 & \gA & \soso
            & \good & \good
            & 0.1184(6)    
            & Wilson loops
            & \ref{tab_wloops}                              \\ 
  PACS-CS 09A& \cite{Aoki:2009tf} & 2+1 
            & \gA &\good &\good &\soso
            & $0.118(3)$$^\#$
            & Schr{\"o}dinger functional\hspace{-0.5cm}
            & \ref{tab_SF3}                                        \\
   Maltman 08& \cite{Maltman:2008bx}& 2+1 & \gA & \soso
            & \soso & \soso
            & $0.1192(11)$
            & Wilson loops
            & \ref{tab_wloops}                               \\ 
   HPQCD 08B  & \cite{Allison:2008xk}  & 2+1       & \gA & \bad
             & \bad  & \bad
             & 0.1174(12) 
             & current two points
             & \ref{tab_current_2pt}                               \\
   HPQCD 08A& \cite{Davies:2008sw} & 2+1 & \gA & \soso
            & \good & \good
            & 0.1183(8)
             & Wilson loops
            & \ref{tab_wloops}                                      \\
   HPQCD 05A & \cite{Mason:2005zx} & 2+1 & \gA & \soso
            & \soso & \soso
            & 0.1170(12)
            & Wilson loops
            & \ref{tab_wloops}                                       \\
   & & & & & & & & \\[-0.1cm]
   \hline
   & & & & & & & & \\[-0.1cm]
   QCDSF/UKQCD 05 & \cite{Gockeler:2005rv}  & $0,2 \to 3$ & \gA & \good 
            & \bad  & \good
            & 0.112(1)(2)
	   & Wilson loops
            & \ref{tab_wloops}                                       \\
   Boucaud 01B  & \cite{Boucaud:2001qz}    & $2\to 3$ & \gA 
            & \soso & \soso  & \bad
            & 0.113(3)(4)
	   & gluon-ghost vertex
            & \ref{tab_vertex}                                       \\
   SESAM 99 & \cite{Spitz:1999tu} & $0,2\to3$ & \gA & \good
            & \bad  & \bad
            & 0.1118(17)
            & Wilson loops
            & \ref{tab_wloops}                                       \\
   Wingate 95 & \cite{Wingate:1995fd} & $0,2\to3$  & \gA & \good
            & \bad  & \bad
            & 0.107(5)
            & Wilson loops
            & \ref{tab_wloops}                                       \\
   Davies 94& \cite{Davies:1994ei} & $0,2\to3$  & \gA & \good
            & \bad & \bad
            & 0.115(2)
            & Wilson loops
            & \ref{tab_wloops}                                       \\
   Aoki 94  & \cite{Aoki:1994pc} &  $2\to3$  & \gA & \good
            & \bad & \bad
            & 0.108(5)(4)
	   & Wilson loops
            & \ref{tab_wloops}                                       \\
   El-Khadra 92 & \cite{ElKhadra:1992vn} & $0\to3$ & \gA & \good
            & \soso  & \soso
            & 0.106(4)
            & Wilson loops
            & \ref{tab_wloops}                                      \\
   & & & & & & & & &  \\[-0.1cm]
   \hline
   \hline
\end{tabular*}
\begin{tabular*}{\textwidth}[l]{l@{\extracolsep{\fill}}lllllll}
\multicolumn{8}{l}{\vbox{\begin{flushleft} 
   $^\#$  Result with a linear continuum extrapolation in $a$.
\end{flushleft}}}
\end{tabular*}
\vspace{-0.3cm}
\caption{Results for $\alpha_\msbar(M_\mathrm{Z})$. 
         $N_f = 3$ results are matched at the
         charm and bottom thresholds and scaled to $M_Z$ to obtain
         the $N_f =5$ result.
         The arrows in the $N_f$ column indicates which $N_f$
         ($N_f = 0$, $2$ or a combination of both)
         were used to first extrapolate to $N_f = 3$ or estimate
         the $N_f = 3$ value through a model/assumption.
         The exact procedures used vary and are given in the
         various papers.}
\label{tab_alphmsbar}
\end{table}

\subsubsection{The present situation}

We first summarize the status of lattice-QCD calculations of the QCD
scale $\Lambda_\msbar$.  Fig.~\ref{r0LamMSbar} shows all results for
$r_0\Lambda_{\overline{\rm MS}}$ discussed in the previous sections.
\begin{figure}[!htb]\hspace{-2cm}\begin{center}
      \includegraphics[width=14.0cm]{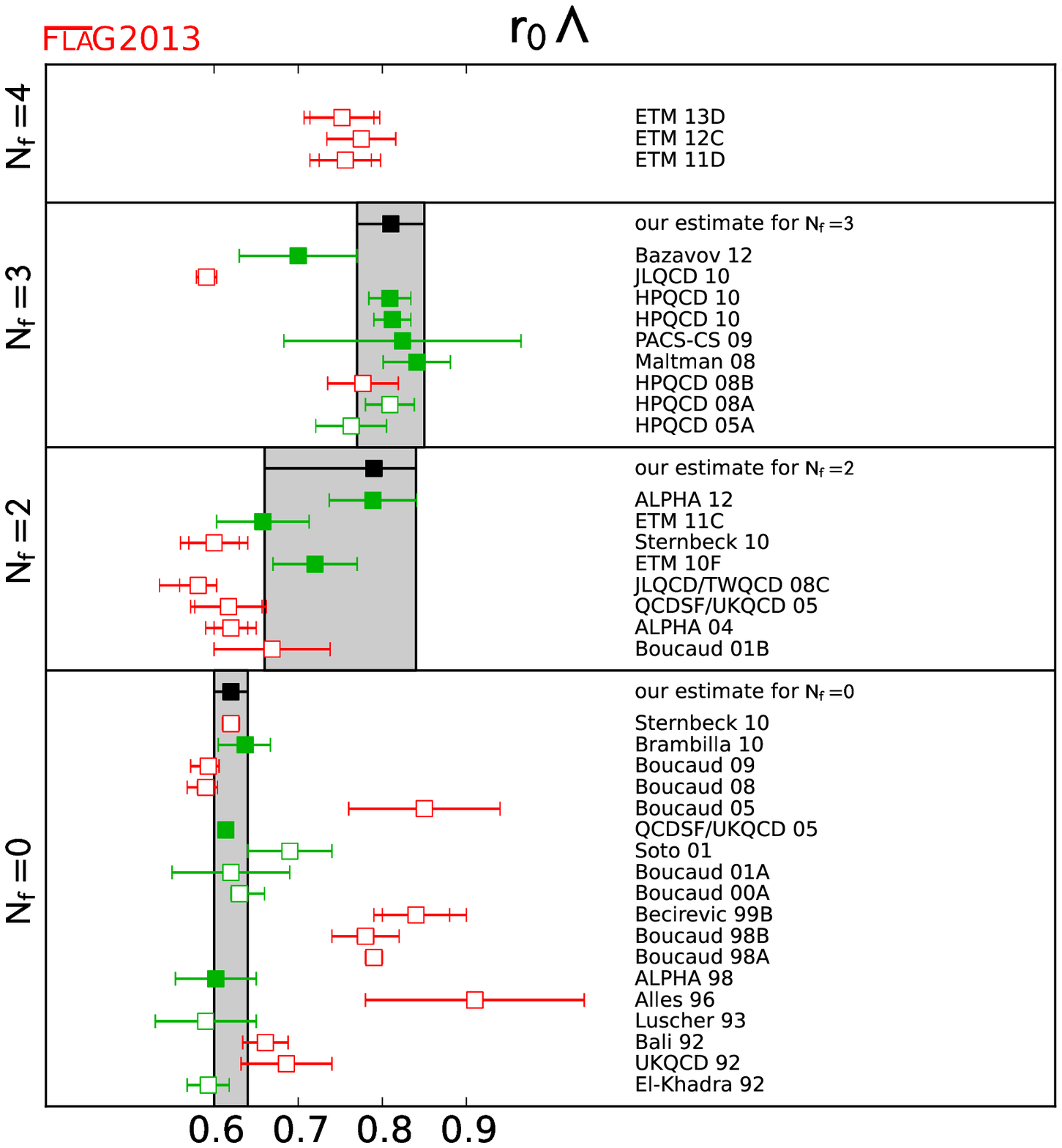}
      \end{center}
      
\vspace{-1cm}
\caption{$r_0\Lambda_{\overline{\rm MS}}$ estimates for
         $N_f = 0$, $2$, $3$, $4$ flavours.
         Full green squares are used in our final
         ranges, open green squares also indicate that there are no
         red squares in the colour coding but the computations were
         superseded by later more complete ones, while red open
         squares mean that there is at least one red square
         in the colour coding.}
\label{r0LamMSbar}
\end{figure}
Many of the numbers are the ones given directly in the papers. 
However, when only $\Lambda_{\overline{\rm MS}}$ in physical units
($\mbox{MeV}$) is available, we have converted them by multiplying
with the value of $r_0$ in physical units. The notation used
is full green squares for results used in our final average,
while an open green square indicates that there are no
red squares in the previous colour coding but the computation does
not enter the ranges because either it has been superseded by an update
or it is not published. Red open squares mean that there is at least
one red square in the colour coding.

For $N_f=0$ there is relatively little spread in the more recent
numbers, even in those which do not satisfy our quality
criteria. Clearly one could improve the statistical and many
systematic errors considerably nowadays, but the emphasis is on the
theory with quarks.

When two flavours of quarks are included, the numbers extracted 
by the various groups show a considerable spread, as in particular
older computations did not yet control the systematics sufficiently. 
This illustrates the difficulty of the problem and emphasizes the 
need for strict quality  criteria.  
The agreement among the more modern calculations with three or more flavours, 
however, is quite good.

We now turn to the status of the essential result for phenomenology,
$\alpha_{\overline{\rm MS}}^{(5)}(M_Z)$.  In Table~\ref{tab_alphmsbar}
and Fig.~\ref{alphasMSbarZ} we show all the results for
\begin{figure}[!htb]
   \begin{center}
      \includegraphics[width=14.0cm]{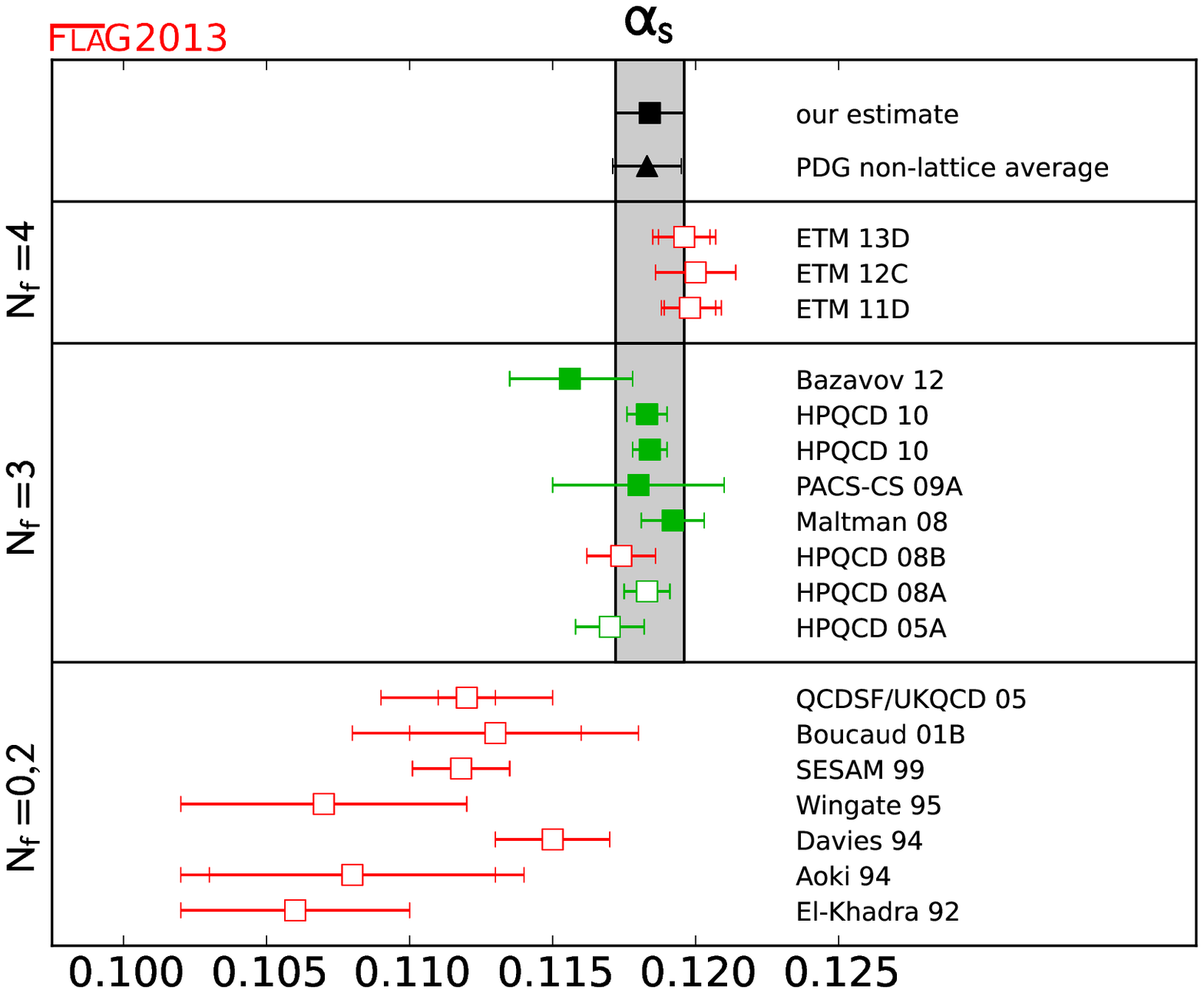}
   \end{center}
\caption{ $\alpha_{\overline{\rm MS}}^{(5)}(M_Z)$, the coupling
  constant in the $\overline{\rm MS}$ scheme at the $Z$ mass. The
  results labeled $N_f=0,2$ use estimates for $N_f=3$ obtained by
  first extrapolating in $N_f$ from $N_f=0,2$ results. Since this is
  not a theoretically justified procedure, these are not included in
  our final estimate and are thus given a red symbol. However, they
  are shown to indicate the progress made since these early
  calculations. The PDG entry indicates the outcome of their analysis
  excluding lattice results (see section
  \ref{subsubsec:alpha_s_Conclusions}).}
\label{alphasMSbarZ}
\end{figure}
$\alpha_{\overline{\rm MS}}^{(5)}(M_Z)$ (i.e.\ $\alpha_{\overline{\rm
    MS}}$ at the $M_Z$ mass) obtained from $N_f=2+1$ and $N_f = 2+1+1$
simulations.  For comparison, we also include results from $N_f = 0,2$
simulations, which are not relevant for phenomenology. For the $N_f
\geq 3$ simulations, the conversion from $\Nf = 3$ to $\Nf = 5$ is
made by matching the coupling constant at the charm and bottom quark
thresholds and using the scale as determined or used by the
authors. For $N_f = 0$,
$2$ the results for $\alpha_\msbar$ in the summary table come from
evaluations of $\alpha_\msbar$ at a low scale and are extrapolated in
$\Nf$ to $\Nf = 3$.

As can be seen from the tables and figures, at present there are
several computations satisfying the quality criteria to be included in
the FLAG average.  We note that none of those calculations of
$\alpha_{\overline{\rm MS}}^{(5)}(M_Z)$ satisfy all of our more
stringent criteria: a $\good$ for the renormalization scale,
perturbative behaviour and continuum extrapolation.  The results,
however, are obtained from four different methods that have different
associated systematics, and agree well within the stated
uncertainties.

\subsubsection{Our range for $\alpha_{\overline{\rm MS}}^{(5)}$}

We now explain the determination of our range.  We only include those
results without a red tag and that are published in a refereed journal.
We also do not include any numbers which were obtained by
extrapolating from theories with less than three flavours.  There is
no real basis for such extrapolations; rather they use ad hoc
assumptions on the low energy behaviour of the theories. One also
notices from the published results that the estimated numbers are
quite significantly below those with at least 2+1 flavours.

A general issue with most recent lattice calculations of
$\alpha_\msbar$ is that they are dominated by perturbative truncation
errors, which are difficult to estimate.  This concern also applies to
many non-lattice determinations.  Further, all results except for
those of sections \ref{s:SF},~\ref{s:WL} are based on extractions of
$\alpha_\msbar$ that are largely influenced by data with
$\alpha_\mathrm{eff}\geq 0.3$.  At smaller $\alpha$ the momentum scale
$\mu$ quickly is at or above $a^{-1}$. We have included computations
using $a\mu$ up to $1.5$ and $\alpha_\mathrm{eff}$ up to 0.4, but one
would ideally like to be significantly below that.  Accordingly we
wish at this stage to estimate the error ranges in a conservative
manner, and not simply perform weighted averages of the individual
errors estimated by each group.

Many of the methods have thus far only been applied by a single
collaboration, and with simulation parameters that could still be
improved.  We therefore think that the following aspects of the
individual calculations are important to keep in mind, and look
forward to additional clarification and/or corroboration in the
future.

\vspace{0.2em}
\noindent $\bullet\,$ The potential computations Brambilla 10~\cite{Brambilla:2010pp},
ETM 11C~\cite{Jansen:2011vv} and Bazavov 12~\cite{Bazavov:2012ka} give
evidence that they have reached distances where perturbation theory
can be used. However, in addition to $\Lambda_\mathrm{QCD}$, a scale
is introduced into the perturbative prediction by the process of
subtracting the renormalon contribution. The extractions of $\Lambda$
are dominated by data with $\alpha_\mathrm{eff}\geq 0.3$. In contrast,
Ref.~\cite{Knechtli:2011pz}, which studies the force instead of the
potential and therefore does not need a renormalon subtraction, finds
that significantly smaller lattice spacings would be needed in order
for perturbation theory to be reliable.  Further study is needed to
clarify the situation.

\vspace{0.2em}
\noindent $\bullet\,$ In the determination of $\alpha_s$ from observables at the lattice
spacing scale, there is an interplay of higher order perturbative
terms and lattice artefacts. In  HPQCD 05A \cite{Mason:2005zx}, HPQCD 08A \cite{Davies:2008sw}
and Maltman 08 \cite{Maltman:2008bx} both lattice artifacts (which are
power corrections in this approach) and higher order perturbative
terms are fitted.  We note that, Maltman 08~\cite{Maltman:2008bx} and
HPQCD 08A~\cite{Davies:2008sw} analyse largely the same data set but
use different versions of the perturbative expansion and treatments of
nonperturbative terms.  After adjusting for the slightly different
lattice scales used, the values of $\alpha_\msbar(M_Z)$ differ by
$0.0004$ to $0.0008$ for the three quantities considered.  In fact the
largest of these differences ($0.0008$) comes from a tadpole-improved
loop, which is expected to be best behaved perturbatively.

\vspace{0.2em}
\noindent $\bullet\,$ Another computation with very small errors is HPQCD 10
\cite{McNeile:2010ji}, where correlation functions of heavy quarks are
used to construct short-distance quantities. Due to the large quark
masses needed to reach the region of small coupling, considerable
discretisation errors are present, see \fig{hpqcd_alpha_eff}. These
are treated by fits to the perturbative running (a five-loop running
$\alpha_{\overline{\rm MS}}$ with a fitted five-loop coefficient in
the beta-function is used) with high order terms in a double expansion
in $a^2\Lambda^2$ and $a^2 m_\mathrm{h}^2$ supplemented by priors
which limit the size of the coefficients.  The priors play an
especially important role in these fits given the much larger number
of fit parameters than data points.  We note, however, that the size
of the coefficients does not prevent high-order terms from
contributing significantly, since the data includes values of
$am_{\textrm p}/2$ that are rather close to 1.  It is not clear how
sensitive the final results are to these large values of $am_{\textrm
  p}/2$.

As previously discussed $\alpha_{\overline{\rm MS}}^{(5)}(M_Z)$ is
summarized in Table~\ref{tab_alphmsbar} and Fig.~\ref{alphasMSbarZ}.
Early computations estimated the effect of the strange quark by
extrapolations from $N_f=0$ and $N_f=2$. They are included in the
table and figure but do not enter the final range. Indeed with our
present knowledge we see that such estimates were rather rough ones,
but also other systematic errors such as a lack of control of
discretisation errors presumably play a r\^ole in the differences seen
with today's results.  A number of calculations that include the
effect of the strange quark make up our final estimate.  These are
Bazavov 12, HPQCD~10A/10B, PACS-CS 09A, Maltman 08 while HPQCD~08A/05A
have been superseded by more complete calculations.  We obtain the
central value for our range,
\begin{eqnarray}
  \alpha_{\overline{\rm MS}}^{(5)}(M_Z) = 0.1184(12) \,, 
\label{eq:alpmz}
\end{eqnarray}
from the weighted average of the five results.  Of the results that
enter our range, those from Wilson loops (HPQCD~10A and Maltman 08)
and current two-point correlators (HPQCD~10B) presently have the
smallest quoted errors.  In both cases the uncertainties are dominated
by perturbative truncation errors.  Such errors are difficult to
estimate, and there is a considerable spread in opinion both in the
lattice and continuum phenomenology communities regarding how they
should be estimated.  We therefore choose to be conservative, and take
a larger range for $\alpha_{\overline{\rm MS}}^{(5)}(M_Z)$ than one
would obtain from the weighted average, or even from the most precise
individual calculation.  We make a conservative estimate of the
perturbative uncertainty in the calculation of $\alpha_s$ from small
Wilson loops, and take that estimate as the error range of the current
weighted average of all lattice results.  One approach for making such
an estimate would be to take the largest of the differences between
the calculations of Maltman 08~\cite{Maltman:2008bx} and HPQCD
08A~\cite{Davies:2008sw}, 0.0008, which comes from the quantity
computed by both groups that is expected to be best behaved
perturbatively.  This is somewhat larger than some of the estimates in
the individual papers.  An even more conservative estimate increases
this error further to make it commensurate with a power-counting
estimate of the truncation errors in the Wilson loop analyses.  Taking
the coefficient $|c_4/c_1|\approx 2$ in eq.~(\ref{qcdsf:ouruncert})
yields the estimate $\Delta \alpha_2 = 0.0012$ for
$\alpha^{(5)}_{\overline{\rm MS}}(M_Z)$.  This is what we adopt as our
final range.

The range for $\alpha_{\overline{\rm MS}}^{(5)}(M_Z)$ presented here
is based on results with rather different systematics (apart from the
matching across the charm threshold). We therefore believe that the
true value is quite likely to lie within this range.
 
We would like to emphasize once more that all computations which enter
this range rely on a perturbative inclusion of the charm and beauty
quarks.  While perturbation theory for the matching of $\gbar^2_{N_f}$
and $\gbar^2_{N_f-1}$ looks very well behaved even at the mass of the
charm, this scale is rather low and we have no reliable information
about the precision of perturbation theory. However, it seems unlikely
that the associated uncertainty is comparable with the present
errors. With future improved precision, this will become a relevant
issue. Note that this uncertainty is also present in some of the
phenomenological determinations, in particular from $\tau$ decays.

\subsubsection{Ranges for $[r_0 \Lambda]^{(\Nf)}$
and $\lms$}

In the present situation, we give ranges for $[r_0 \Lambda]^{(\Nf)}$
and $\lms$, discussing their determination case by case.  We include
results with $\Nf<3$ because it is interesting to see the
$\Nf$-dependence of the connection of low- and high-energy QCD.  This
aids our understanding of the field theory and helps in finding
possible ways to tackle it beyond the lattice approach. It is also of
interest in providing an impression on the size of the vacuum
polarisation effects of quarks, in particular with an eye on the still
difficult-to-treat heavier charm and beauty quarks. Even if this
information is rather qualitative, it may be valuable, given that it
is of a completely non-perturbative nature.

We emphasize that results for $[r_0 \Lambda]^{(0)}$
and $[r_0 \Lambda]^{(2)}$ are {\em not}\/ meant to be used
in phenomenology. 

For $\Nf=2+1+1$, we presently do not quote a range.  Our best estimate
is given by using the $\Nf=2+1$ result and converting it to
$\Nf=2+1+1$ perturbatively at the charm quark-mass threshold.

For $\Nf=2+1$, we take as a central value the weighted average of
Bazavov~12, HPQCD~10A, 10B, PACS-CS~09A and Maltman~08.  For the error
we take our own conservative estimate of the perturbative uncertainty
remaining in the determinations from small Wilson loops, HPQCD~10A and
Maltman~08.  From an estimate of $|c_4/c_1|\approx 2$ we obtain
(eq.~(\ref{qcdsf:ouruncert}) in section \ref{s:WL}) $\Delta
\Lambda/\Lambda = 0.05$. An independent estimate of the uncertainty
due to the fit to the $a$-dependence in the analysis of moments of
heavy quark correlators is much more difficult to make; as discussed
above, and in the absence of confirmation by other groups, we are not
yet ready to use the result of HPQCD 10 to reduce our conservative
estimate of the errors from other approaches.  Noting that the
statistical error is negligible, we thus assign the just mentioned 5\%
error to the overall range,
\begin{eqnarray}
   [r_0 \lms]^{(3)} = 0.81(4) \,.
   \label{eq:lms3}
\end{eqnarray}
It is in good agreement with all 2+1 results 
without red tags. 
In physical units, using $r_0=0.472$~fm, this means
\begin{eqnarray}
   \lms^{(3)} = 339(17)\,\mbox{MeV}\,.
\end{eqnarray}

For $N_f=2$, at present there is one computation with a \good\ rating
for all criteria, ALPHA 12.  We adopt it as our central value and
enlarge the error to cover the central values of the other two results
with filled green boxes. This results in an asymmetric error.  Our
present range is
\begin{eqnarray}
   [r_0 \lms]^{(2)} = 0.79(^{+~5}_{-13}) \,, \quad
   \label{eq:lms2}
\end{eqnarray}
and in physical units, using $r_0=0.472$fm,
\begin{eqnarray}
   \lms^{(2)} = 330(^{+21}_{-54}) \mbox{MeV}\,.  \quad 
\end{eqnarray}
A weighted average of the three eligible numbers would yield $[r_0
  \lms]^{(2)} = 0.725(30)$, not covering the best result and in
particular leading to a smaller error than we feel is justified, given
the issues discussed above.  Thus we believe that our estimate is a
conservative choice; the lower value of ETM 11C~\cite{Jansen:2011vv}
leads to the large downwards error.  We hope that future work will
improve the situation.

For $N_f=0$, ALPHA 98 has a \soso\ in the continuum limit since the
${\rm O}(a)$ improvement at the boundary was carried out only to
1-loop order. On the other hand, QCDSF/UKQCD~05 receives a \soso\ for
the perturbative behaviour since a power law correction was fitted to
the results, and additionally we note again that it is not obvious
that higher order perturbative terms are negligible; an estimate as
for HPQCD~10A (with $|c_4/c_1| \approx 2$) would be $\Delta [r_0
  \lms]^{(0)} = 0.018$.  A third result which enters our average is
Brambilla~10 but we exclude the older estimates shown in the graph.
They have a limited control of the systematic errors due to power law
corrections and discretisation errors.\footnote{We have assigned a
  \soso\ for the continuum limit, in Boucaud 00A\cite{Boucaud:2000ey},
  00B\cite{Boucaud:2000nd}, 01A\cite{Boucaud:2001st},
  Soto~01\cite{DeSoto:2001qx} but these results are from lattices of a
  very small physical size with finite size effects that are not
  easily quantified.}  Taking a weighted average of the three numbers,
we obtain $[r_0 \lms]^{(0)} = 0.615(5)$, dominated by the
QCDSF/UKQCD~05 result.  Since we are not yet convinced that such a
small uncertainty has been reached, we prefer to presently take a
range which encompasses all three central values and whose uncertainty
comes close to our estimate of the perturbative error:
\begin{eqnarray}
   [r_0 \lms]^{(0)} =  0.62(2) \,.  \quad
   \label{eq:lms0}
\end{eqnarray}
Converting to physical units, using $r_0=0.472$fm,
\begin{eqnarray}
   \lms^{(0)} =  260(7)\,\mbox{MeV}\,. \quad
\end{eqnarray}
While the conversion of the $\Lambda$-parameter to physical units is
quite unambiguous for $\Nf=2+1$, our choice of $r_0=0.472$~fm also for
smaller numbers of flavour amounts to a convention, in particular for
$\Nf=0$.  Indeed, in the tables ~\ref{tab_SF3}-\ref{tab_vertex}
somewhat different numbers in MeV are found.

 How sure are we about our ranges for $[r_0 \lms]$? In one case we
 have a result, \eq{eq:lms2} which easily passes our criteria, in
 another one (\eq{eq:lms0}) we have three compatible results which are
 close to that quality and agree. For $\Nf=2+1$ the range
 (\eq{eq:lms3}) takes account of results with rather different
 systematics (apart from the matching across the charm threshold). We
 therefore find it difficult to imagine that the ranges could be
 violated by much.

\subsubsection{Conclusions}
\label{subsubsec:alpha_s_Conclusions}
With the present results our range for the strong coupling is
(repeating eq.~(\ref{eq:alpmz}))
\begin{eqnarray*}
  \alpha_{\overline{\rm MS}}^{(5)}(M_Z) = 0.1184(12) \,. 
\end{eqnarray*}
As can be seen from Fig.~\ref{alphasMSbarZ}, when surveying the green
data points, the individual lattice results agree within their quoted
errors.  Further those points are based on different methods for
determining $\alpha_s$, each with its own difficulties and
limitations.  Thus the overall consistency of the lattice $\alpha_s$
results engenders confidence in our range.

While our range for $\alpha_{\overline{\rm MS}}(M_Z)$ in
eq.~(\ref{eq:alpmz}) has about the same central value as the PDG
average of lattice results, $\alpha_{\overline{\rm MS}}(M_Z) =
0.1185(5)$, our error estimate is more conservative, derived from an
estimate of perturbative uncertainties.  In contrast, in the PDG
review all published lattice results are taken with their errors at
face value and a $\chi$-squared weighted average is chosen because the
results are largely independent and compatible within errors.  We note
that there is a diversity of opinion over the size of our range for
$\alpha_{\overline{\rm MS}}(M_Z)$ in eq.~(\ref{eq:alpmz}) within FLAG.
Some members are sufficiently convinced by the overall consistency of
the results from various groups within their quoted errors, as well as
by the internal tests performed by individual groups, to take the
quoted errors at face value. Others prefer the more conservative error
estimate cited above, which aims to account for the difficulty
associated with estimating perturbative truncation errors, the largest
source of uncertainty in most of the calculations that enter the
range.  Given this diversity of opinion, we think it is appropriate to
choose the more conservative estimate for our quoted range.

It is also interesting to compare our result, eq.~(\ref{eq:alpmz}),
with the value quoted by the PDG for the average over all other
(non-lattice) sources, $\alpha_s=0.1183(12)$.  In the 2013 review, for
all subclasses of $\alpha_s$ determinations except for the lattice
results, the results disagree beyond those expected from the quoted
errors, presumably because of the challenges of evaluating systematic
uncertainties.  Thus the quoted range for each subclass is increased
to encompass the central values of all individual determinations.
This leads to subclass averages with errors that are larger than the
smallest error of individual determinations by factors between two and
four.

Our range for the lattice determination of $\alpha_{\overline{\rm
    MS}}(M_Z)$ in eq.~(\ref{eq:alpmz}) is in excellent agreement with
the PDG non-lattice average: the work done on the lattice provides an
entirely independent determination, which already reaches the same
precision even with our conservative estimate of the perturbative
error.

We finish by commenting on perspectives for the future.  In the next
few years we anticipate that a growing number of lattice calculations
of $\alpha_s$ from different quantities and by different
collaborations will enable increasingly precise determinations,
coupled with stringent cross-checks.  The determination of $\alpha_s$
from observables at the lattice spacing scale will improve due to a
further reduction of the lattice spacing. This reduces
$\alpha_\mathrm{eff}$ and thus the dominating error in
$\alpha_\msbar$.  Schr\"odinger functional methods for $N_f=2+1$ will
certainly reach the precision of the present $N_f=2$ results soon, as
this just requires an application of the presently known
techniques. Furthermore, we may expect a significant reduction of
errors due to new definitions of running couplings
\cite{Fodor:2012td,Fritzsch:2013je} using the Yang Mills gradient flow
\cite{Luscher:2010iy}. Factors of two and more in precision are
certainly possible. At this point it will then also be necessary to
include the charm quark in the computations such that the perturbative
matching of $N_f=2+1$ and $2+1+1$ theories at the charm quark
threshold is avoided. $N_f=2+1+1$ simulations are presently being
carried out.

\clearpage
\section*{Acknowledgments}
We are particularly indebted to J. Shigemitsu, who has actively
participated in the the production of the first version of the current
review: her invaluable contributions are gratefully acknowledged.  We
wish to thank Damir Be\v{c}irevi\'c, Nuria Carrasco, Christine Davies,
Rachel Dowdall, Petros Dimopoulos, Jochen Dingfelder, Elvira Gamiz,
Andreas Kronfeld, Peter Lepage, Matthew Moulson, Heechang Na, Jim
Simone, Doug Toussaint for correspondence and useful comments.
C.P. acknowledges the support of the Spanish MICINN under grant
FPA2009-08785, the Spanish MINECO under grant FPA2012-31686 and the
"Centro de excelencia Severo Ochoa Program" SEV-2012-0249, the
Community of Madrid under grant HEPHACOS S2009/ESP-1473, and the
European Union under the Marie Curie-ITN Program STRONGnet, grant
PITN-GA-2009-238353.  A.X.K.  thanks the Fermilab theory group for
hospitality while this review was finalized. The Albert Einstein
Center for Fundamental Physics at the University of Bern is supported
by the ``Innovations- und Kooperationsprojekt C-13'' of the
``Schweizerische Universit\"atskonferenz SUK/CRUS''. Fermilab is
operated by Fermi Research Alliance, LLC, under Contract
No. DE-AC02-07CH11359 with the U.S. Department of Energy. This work
was partially supported by Grant-in-Aid for Scientific Research
(25287046) and SPIRE (Strategic Program for Innovative Research) of
MEXT, Japan, by the JSPS Kakenhi Grant, No.~ 22224003, by MIUR (Italy)
under the contract PRIN 2010-2011, by the US Department of Energy
under grant numbers DE-FG02-91ER40628, DE-FG02-92ER40716, DOE
DE-FG02-13ER42001, DE-FG02-96ER40956, by the OCEVU Labex
(ANR-11-LABX-0060) and the A*MIDEX project (ANR-11-IDEX-0001-02)
funded by the "Investissements d'Avenir" French government program
managed by the ANR, by UK STFC grant ST/G000557/1, by the Helmholtz
Association through the virtual institute ``Spin and strong QCD''
(VH-VI-231), by the Swiss National Science Foundation, by the Spanish
Minister of Education and Science, project RyC-2011-08557 and by a URA
Visiting Scholarship.  The research leading to these results has also
received funding from the European Research Council under the European
Community's Seventh Framework Programme (FP7/2007-2013) ERC grant
agreement No 279757.

\appendix

\begin{appendix}
\clearpage
\section{Glossary}\label{comm}
\subsection{Lattice actions}\label{sec_lattice_actions}
In this appendix we give brief descriptions of the lattice actions
used in the simulations and summarize their main features.

\subsubsection{Gauge actions \label{sec_gauge_actions}}

The simplest and most widely used discretization of the Yang-Mills
part of the QCD action is the Wilson plaquette action\,\cite{Wilson:1974sk}:
\be
 S_{\rm G} = \beta\sum_{x} \sum_{\mu<\nu}\Big(
  1-\frac{1}{3}{\rm Re\,\Tr}\,W_{\mu\nu}^{1\times1}(x)\Big),
\label{eq_plaquette}
\ee
where $\beta \equiv 6/g_0^2$ (with $g_0$ the bare gauge coupling) and
the plaquette $W_{\mu\nu}^{1\times1}(x)$ is the product of
link variables around an elementary square of the lattice, i.e.
\be
  W_{\mu\nu}^{1\times1}(x) \equiv U_\mu(x)U_\nu(x+a\hat{\mu})
   U_\mu(x+a\hat{\nu})^{-1} U_\nu(x)^{-1}.
\ee
This expression reproduces the Euclidean Yang-Mills action in the
continuum up to corrections of order~$a^2$.  There is a general
formalism, known as the ``Symanzik improvement programme''
\cite{Symanzik:1983dc,Symanzik:1983gh}, which is designed to cancel
the leading lattice artifacts, such that observables have an
accelerated rate of convergence to the continuum limit.  The
improvement programme is implemented by adding higher-dimensional
operators, whose coefficients must be tuned appropriately in order to
cancel the leading lattice artifacts. The effectiveness of this
procedure depends largely on the method with which the coefficients
are determined. The most widely applied methods (in ascending order of
effectiveness) include perturbation theory, tadpole-improved
(partially resummed) perturbation theory, renormalization group
methods, and the nonperturbative evaluation of improvement
conditions.

In the case of Yang-Mills theory, the simplest version of an improved
lattice action is obtained by adding rectangular $1\times2$ loops to
the plaquette action, i.e.
\be
   S_{\rm G}^{\rm imp} = \beta\sum_{x}\left\{ c_0\sum_{\mu<\nu}\Big(
  1-\frac{1}{3}{\rm Re\,\Tr}\,W_{\mu\nu}^{1\times1}(x)\Big) +
   c_1\sum_{\mu,\nu} \Big(
  1-\frac{1}{3}{\rm Re\,\Tr}\,W_{\mu\nu}^{1\times2}(x)\Big) \right\},
\label{eq_Sym}
\ee
where the coefficients $c_0, c_1$ satisfy the normalization condition
$c_0+8c_1=1$. The {\sl Symanzik-improved} \cite{Luscher:1984xn},
{\sl Iwasaki} \cite{Iwasaki:1985we}, and {\sl DBW2}
\cite{Takaishi:1996xj,deForcrand:1999bi} actions are all defined
through \eq{eq_Sym} via particular choices for $c_0, c_1$. Details are
listed in Table\,\ref{tab_gaugeactions} together with the
abbreviations used in the summary tables.
\vspace{-0.07cm}
\begin{table}[!h]
\begin{center}
{\footnotesize
\begin{tabular*}{\textwidth}[l]{l @{\extracolsep{\fill}} c l}
\hline\hline \\[-1.0ex]
Abbrev. & $c_1$ & Description 
\\[1.0ex] \hline \hline \\[-1.0ex]
Wilson    & 0 & Wilson plaquette action \\[1.0ex] \hline \\[-1.0ex]
tlSym   & $-1/12$ & tree-level Symanzik-improved gauge action \\[1.0ex] \hline \\[-1.0ex]
tadSym  & variable & tadpole Symanzik-improved gauge
action \\[1.0ex] \hline \\[-1.0ex]
Iwasaki & $-0.331$ & Renormalization group improved (``Iwasaki'')
action \\[1.0ex] \hline \\[-1.0ex]
DBW2 & $-1.4088$ & Renormalization group improved (``DBW2'') action 
\\ [1.0ex] 
\hline\hline
\end{tabular*}
}
\caption{Summary of lattice gauge actions. The leading lattice
 artifacts are $O(a^2)$ or better for all
  discretizations. \label{tab_gaugeactions}} 
\end{center}
\end{table}

\clearpage

\subsubsection{Light-quark actions \label{sec_quark_actions}}

If one attempts to discretize the quark action, one is faced with the
fermion doubling problem: the naive lattice transcription produces a
16-fold degeneracy of the fermion spectrum. \\

\noindent
{\it Wilson fermions}\\
\noindent

Wilson's solution to the fermion doubling problem is based on adding a
dimension-5 (irrelevant) operator to the lattice action. The
Wilson-Dirac operator for the massless case reads
\cite{Wilson:1974sk,Wilson:1975id}
\be
     D_{\rm w} = \half\gamma_\mu(\nabla_\mu+\nabla_\mu^*)
   +a\nabla_\mu^*\nabla_\mu,
\ee
where $\nabla_\mu,\,\nabla_\mu^*$ denote the covariant forward and
backward lattice derivatives, respectively.  The addition of the
Wilson term $a\nabla_\mu^*\nabla_\mu$, results in fermion doublers
acquiring a mass proportional to the inverse lattice spacing; close to
the continuum limit these extra degrees of freedom are removed from
the low-energy spectrum. However, the Wilson term also results in an
explicit breaking of chiral symmetry even at zero bare quark mass.
Consequently, it also generates divergences proportional to the UV
cutoff (inverse lattice spacing), besides the usual logarithmic
ones. Therefore the chiral limit of the regularized theory is not
defined simply by the vanishing of the bare quark mass but must be
appropriately tuned. As a consequence quark mass renormalization
requires a power subtraction on top of the standard multiplicative
logarithmic renormalization.  The breaking of chiral symmetry also
implies that the nonrenormalization theorem has to be applied with
care~\cite{Karsten:1980wd,Bochicchio:1985xa}, resulting in a
normalization factor for the axial current which is a regular function
of the bare coupling.  On the other hand, vector symmetry is
unaffected by the Wilson term and thus a lattice (point split) vector
current is conserved and obeys the usual nonrenormalization theorem
with a trivial (unity) normalization factor. Thus, compared to lattice
fermion actions which preserve chiral symmetry, or a subgroup of it,
the Wilson regularization typically results in more complicated
renormalization patterns.

Furthermore, the leading order lattice artifacts are of order~$a$.
With the help of the Symanzik improvement programme, the leading
artifacts can be cancelled in the action by adding the so-called
``Clover'' or Sheikholeslami-Wohlert (SW) term~\cite{Luscher:1996sc}.
The resulting expression in the massless case reads
\be
   D_{\rm sw} = D_{\rm w}
   +\frac{ia}{4}\,\csw\sigma_{\mu\nu}\widehat{F}_{\mu\nu},
\label{eq_DSW}
\ee
where $\sigma_{\mu\nu}=\frac{i}{2}[\gamma_\mu,\gamma_\nu]$, and
$\widehat{F}_{\mu\nu}$ is a lattice transcription of the gluon field
strength tensor $F_{\mu\nu}$. The coefficient $\csw$ can be determined
perturbatively at tree-level ($\csw = 1$; tree-level improvement or
tlSW for short), via a mean field approach \cite{Lepage:1992xa}
(mean-field improvement or mfSW) or via a nonperturbative approach
\cite{Luscher:1996ug} (nonperturbatively improved or npSW).
Hadron masses, computed using $D_{\rm sw}$, with the coefficient
$\csw$ determined nonperturbatively, will approach the continuum
limit with a rate proportional to~$a^2$; with tlSW for $\csw$ the rate
is proportional to~$g_0^2 a$.

Other observables require additional improvement
coefficients~\cite{Luscher:1996sc}.  A common example consists in the
computation of the matrix element $\langle \alpha \vert Q \vert \beta
\rangle$ of a composite field $Q$ of dimension-$d$ with external
states $\vert \alpha \rangle$ and $\vert \beta \rangle$. In the
simplest cases, the above bare matrix element diverges logarithmically
and a single renormalization parameter $Z_Q$ is adequate to render it
finite. It then approaches the continuum limit with a rate
proportional to the lattice spacing $a$, even when the lattice action
contains the Clover term. In order to reduce discretization errors to
${\cal O}(a^2)$, the lattice definition of the composite operator $Q$
must be modified (or ``improved"), by the addition of all
dimension-$(d+1)$ operators with the same lattice symmetries as $Q$.
Each of these terms is accompanied by a coefficient which must be
tuned in a way analogous to that of $\csw$. Once these coefficients
are determined nonperturbatively, the renormalized matrix element of
the improved operator, computed with a npSW action, converges to the
continuum limit with a rate proportional to~$a^2$. A tlSW improvement
of these coefficients and $\csw$ will result in a rate proportional
to~$g_0^2 a$.

It is important to stress that the improvement procedure does not
affect the chiral properties of Wilson fermions; chiral symmetry
remains broken.

Finally, we mention ``twisted-mass QCD'' as a method which was
originally designed to address another problem of Wilson's
discretization: the Wilson-Dirac operator is not protected against the
occurrence of unphysical zero modes, which manifest themselves as
``exceptional'' configurations. They occur with a certain frequency in
numerical simulations with Wilson quarks and can lead to strong
statistical fluctuations. The problem can be cured by introducing a
so-called ``chirally twisted'' mass term. The most common formulation
applies to a flavour doublet $\bar \psi = ( u \quad d)$ of mass
degenerate quarks, with the fermionic part of the QCD action in the
continuum assuming the form
\cite{Frezzotti:2000nk}
\be
   S_{\rm F}^{\rm tm;cont} = \int d^4{x}\, \psibar(x)(\gamma_\mu
   D_\mu +
   m + i\mu_{\rm q}\gamma_5\tau^3)\psi(x).
\ee
Here, $\mu_{\rm q}$ is the twisted mass parameter, and $\tau^3$ is a
Pauli matrix in flavour space. The standard action in the continuum
can be recovered via a global chiral field rotation. The physical
quark mass is obtained as a function of the two mass parameters $m$
and $\mu_{\rm q}$. The corresponding lattice regularization of twisted-mass QCD (tmWil) for $\Nf=2$ flavours is defined through the fermion
matrix
\be
   D_{\rm w}+m_0+i\mu_{\rm q}\gamma_5\tau^3 \,\, .
\label{eq_tmQCD}
\ee
Although this formulation breaks physical parity and flavour
symmetries, resulting in nondegenerate neutral and charged pions,
is has a number of advantages over standard Wilson
fermions. Firstly, the presence of the twisted mass parameter
$\mu_{\rm q}$ protects the discretized theory against unphysical zero
modes. A second attractive feature of twisted-mass lattice QCD is the
fact that, once the bare mass parameter $m_0$ is tuned to its ``critical value"
(corresponding to massless pions in the standard Wilson formulation),
the leading lattice artifacts are of order $a^2$ without the
need to add the Sheikholeslami-Wohlert term in the action, or other
improving coefficients~\cite{Frezzotti:2003ni}. A third important advantage
is that, although the problem of explicit chiral
symmetry breaking remains, quantities computed with twisted fermions
with a suitable tuning of the mass parameter $\mu_{\rm q}$,
are subject to renormalization patterns which are simpler than the ones with
standard Wilson fermions. Well known examples are the pseudoscalar decay
constant  and $B_{\rm K}$.\\

\noindent
{\it Staggered Fermions}\\
\noindent

An alternative procedure to deal with the doubling problem is based on so-called ``staggered''
or Kogut-Susskind fermions \cite{Kogut:1974ag,Banks:1975gq, Banks:1976ia, Susskind:1976jm}.
Here the degeneracy is only lifted partially, from 16 down to 4.  It has become customary
to refer to these residual doublers as ``tastes'' in order to distinguish them from physical
flavours.  Taste changing interactions 
can occur via the exchange of gluons with one or more components
  of momentum near the cutoff $\pi/a$.  This leads to the breaking of the SU(4) vector symmetry among 
  tastes, thereby generating order $a^2$ lattice artifacts.

The residual doubling of staggered quarks (four tastes per
flavour) is removed by taking a fractional power of the fermion determinant \cite{Marinari:1981qf} --- the ``fourth-root 
procedure,'' or, sometimes, the ``fourth root trick.''  
This procedure would be unproblematic if
the action had full SU(4) taste symmetry, which would give a
Dirac operator that was block-diagonal in taste space.  
However, the breaking of taste symmetry at nonzero lattice spacing leads to a
variety of problems. In fact, the fourth root of the determinant is not equivalent
to the determinant of any local lattice Dirac operator \cite{Bernard:2006ee}.
This in turn leads 
to violations of unitarity 
on the lattice \cite{Prelovsek:2005rf,Bernard:2006zw,Bernard:2007qf,Aubin:2008wk}.

According to standard renormalization group lore, the taste
violations, which are associated with lattice operators of dimension
greater than four, might be expected go away in the continuum limit,
resulting in the restoration of locality and unitarity.  However,
there is a problem with applying the standard lore to this nonstandard
situation: the usual renormalization group reasoning assumes that the
lattice action is local.  Nevertheless, Shamir
\cite{Shamir:2004zc,Shamir:2006nj} shows that one may apply the
renormalization group to a ``nearby'' local theory, and thereby gives
a strong argument that that the desired local, unitary theory of QCD
is reproduced by the rooted staggered lattice theory in the continuum
limit.

A version of chiral perturbation that includes the lattice artifacts
due to taste violations and rooting (``rooted staggered chiral
perturbation theory'') can also be worked out
\cite{Lee:1999zxa,Aubin:2003mg,Sharpe:2004is} and shown to correctly
describe the unitarity-violating lattice artifacts in the pion sector
\cite{Bernard:2006zw,Bernard:2007ma}.  This provides additional
evidence that the desired continuum limit can be obtained. Further, it
gives a practical method for removing the lattice artifacts from
simulation results. Versions of rooted staggered chiral perturbation
theory exist for heavy-light mesons with staggered light quarks but
nonstaggered heavy quarks \cite{Aubin:2005aq}, heavy-light mesons with
staggered light and heavy quarks
\cite{Komijani:2012fq,Bernard:2013qwa}, staggered baryons
\cite{Bailey:2007iq}, and mixed actions with a staggered sea
\cite{Bar:2005tu,Bae:2010ki}, as well as the pion-only version
referenced above.

There is also considerable numerical evidence that the rooting
procedure works as desired.  This includes investigations in the
Schwinger model \cite{Durr:2003xs,Durr:2004ta,Durr:2006ze}, studies of
the eigenvalues of the Dirac operator in QCD
\cite{Follana:2004sz,Durr:2004as,Wong:2004nk,Donald:2011if}, and
evidence for taste restoration in the pion spectrum as $a\to0$
\cite{Aubin:2004fs,Bazavov:2009bb}.

Issues with the rooting procedure have led Creutz
\cite{Creutz:2006ys,Creutz:2006wv,Creutz:2007yg,Creutz:2007pr,Creutz:2007rk,Creutz:2008kb,Creutz:2008nk}
to argue that the continuum limit of the rooted staggered theory
cannot be QCD.  These objections have however been answered in
Refs.~\cite{Bernard:2006vv,Sharpe:2006re,Bernard:2007eh,Kronfeld:2007ek,Bernard:2008gr,Adams:2008db,Golterman:2008gt,
  Donald:2011if}. In particular, a claim that the continuum 't Hooft
vertex \cite{'tHooft:1976up,'tHooft:1976fv} could not be properly
reproduced by the rooted theory has been refuted
\cite{Bernard:2007eh,Donald:2011if}.

Overall, despite the lack of rigorous proof of the correctness of the
rooting procedure, we think the evidence is strong enough to consider staggered
QCD simulations on a par with simulations using other actions.
See the following reviews for further evidence and discussion:
\cite{Durr:2005ax,Sharpe:2006re,Kronfeld:2007ek,Golterman:2008gt,Bazavov:2009bb}.
\\

\noindent
{\it Improved Staggered Fermions}\\
\noindent

An improvement program can be used to suppress taste-changing
interactions, leading to ``improved staggered fermions,'' with the
so-called ``Asqtad'' \cite{Orginos:1999cr}, ``HISQ''
\cite{Follana:2006rc}, ``Stout-smeared'' \cite{Aoki:2005vt}, and
``HYP'' \cite{Hasenfratz:2001hp} actions as the most common versions.
All these actions smear the gauge links in order to reduce the
coupling of high-momentum gluons to the quarks, with the main goal of
decreasing taste-violating interactions. In the Asqtad case, this is
accomplished by replacing the gluon links in the derivatives by
averages over 1-, 3-, 5-, and 7-link paths.  The other actions reduce
taste changing even further by smearing more.  In addition to the
smearing, the Asqtad and HISQ actions include a three-hop term in the
action (the ``Naik term'' \cite{Naik:1986bn}) to remove order $a^2$
errors in the dispersion relation, as well as a ``Lepage term''
\cite{Lepage:1998vj} to cancel other order $a^2$ artifacts introduced
by the smearing.  In both the Asqtad and HISQ actions, the leading
taste violations are of order $\alpha_S^2 a^2$, and ``generic''
lattices artifacts (those associated with discretization errors other
than taste violations) are of order $\alpha_S a^2$.  The overall
coefficients of these errors are, however, significantly smaller with
HISQ than with Asqtad.  With the Stout-smeared and HYP actions, the
errors are formally larger (order $\alpha_S a^2$ for taste violations
and order $a^2$ for generic lattices artifacts).  Nevertheless, the
smearing seems to be very efficient, and the actual size of errors at
accessible lattice spacings appears to be at least as small as with
HISQ.

Although logically distinct from the light-quark improvement program
for these actions, it is customary with the HISQ action to include an
additional correction designed to reduce discretization errors for
heavy quarks (in practice, usually charm quarks)
\cite{Follana:2006rc}. The Naik term is adjusted to remove leading
$(am_c)^4$ and $\alpha_S(am_c)^2$ errors, where $m_c$ is the charm
quark mass and ``leading'' in this context means leading in powers of
the heavy-quark velocity $v$ ($v/c\sim 1/3$ for $D_s$).  With these
improvements, the claim is that one can use the staggered action for
charm quarks, although it must be emphasized that it is not obvious
{\it a priori}\/ how large a value of $am_c$ may be tolerated for a
given desired accuracy, and this must be studied in the simulations.
\\

\noindent
{\it Ginsparg-Wilson fermions}\\
\noindent

Fermionic lattice actions, which do not suffer from the doubling
problem whilst preserving chiral symmetry go under the name of
``Ginsparg-Wilson fermions''. In the continuum the massless Dirac
operator ($D$) anti-commutes with $\gamma_5$. At nonzero lattice spacing a 
chiral symmetry can be realized if this condition is relaxed
to \cite{Hasenfratz:1998jp,Hasenfratz:1998ri,Luscher:1998pqa}
\be
   \left\{D,\gamma_5\right\} = aD\gamma_5 D,
\label{eq_GWrelation}
\ee
which is now known as the Ginsparg-Wilson relation
\cite{Ginsparg:1981bj}. The Nielsen-Ninomiya
theorem~\cite{Nielsen:1981hk}, which states that any lattice
formulation for which $D$ anticommutes with $\gamma_5$ necessarily has
doubler fermions, is circumvented since $\{D,\gamma_5\}\neq 0$.

A lattice Dirac operator which satisfies \eq{eq_GWrelation} can be
constructed in several ways. The so-called ``overlap'' or
Neuberger-Dirac operator~\cite{Neuberger:1997fp} acts in four
space-time dimensions and is, in its simplest form, defined by
\be
   D_{\rm N} = \frac{1}{\abar} \left( 1-\epsilon(A)
   \right),\quad\mathrm{where}\quad\epsilon(A)\equiv A (A^\dagger A)^{-1/2}, \quad A=1+s-aD_{\rm w},\quad \abar=\frac{a}{1+s},
\label{eq_overlap}
\ee
$D_{\rm w}$ is the massless Wilson-Dirac operator and $|s|<1$
is a tunable parameter. The overlap operator $D_{\rm N}$ removes all
doublers from the spectrum, and can readily be shown to satisfy the
Ginsparg-Wilson relation. The occurrence of the sign function $\epsilon(A)$ in
$D_{\rm N}$ renders the application of $D_{\rm N}$ in a computer
program potentially very costly, since it must be implemented using,
for instance, a polynomial approximation.

The most widely used approach to satisfying the Ginsparg-Wilson
relation \eq{eq_GWrelation} in large-scale numerical simulations is
provided by \textit{Domain Wall Fermions}
(DWF)~\cite{Kaplan:1992bt,Shamir:1993zy,Furman:1994ky} and we
therefore describe this in some more detail. Following early
exploratory studies~\cite{Blum:1996jf}. this approach has been
developed into a practical formulation of lattice QCD with good chiral
and flavour symmetries leading to results which contribute
significantly to this review. In this formulation, the fermion fields
$\psi(x,s)$ depend on a discrete fifth coordinate $s=1,\ldots,N$ as well as
the physical 4-dimensional space-time coordinates $x_\mu,\,\mu=1\cdots
4$ (the gluon fields do not depend on $s$). The lattice on which the
simulations are performed, is therefore a five-dimensional one of size
$L^3\times T\times N$, where $L,\,T$ and $N$ represent the number of
points in the spatial, temporal and fifth dimensions respectively.
The remarkable feature of DWF is that for each flavour there exists a
physical light mode corresponding to the field $q(x)$:
\begin{eqnarray}
q(x)&=&\frac{1+\gamma^5}{2}\psi(x,1)+\frac{1-\gamma^5}{2}\psi(x,N)\\
\bar{q}(x)&=&\overline{\psi}(x,N)\frac{1+\gamma^5}{2} + \overline{\psi}(x,1)\frac{1-\gamma^5}{2}\,.
\end{eqnarray}
The left and right-handed modes of the physical field are located on
opposite boundaries in the 5th dimensional space which, for
$N\to\infty$, allows for independent transformations of the left and
right components of the quark fields, that is for chiral
transformations. Unlike Wilson fermions, where for each flavour the
quark mass parameter in the action is fine-tuned requiring a
subtraction of contributions of $O(1/a)$ where $a$ is the lattice
spacing, with DWF no such subtraction is necessary for the physical
modes, whereas the unphysical modes have masses of $O(1/a)$ and
decouple.

In actual simulations $N$ is finite and there are small violations of
chiral symmetry which must be accounted for. The theoretical framework
for the study of the residual breaking of chiral symmetry has been a
subject of intensive investigation (for a review and references to the
original literature see e.g.~\cite{Sharpe:2007yd}). The breaking
requires one or more \emph{crossings} of the fifth dimension to couple
the left and right-handed modes; the more crossings that are required
the smaller the effect.  For many physical quantities the leading
effects of chiral symmetry breaking due to finite $N$ are parameterized
by a \emph{residual} mass, $m_{\mathrm{res}}$.  For example, the PCAC
relation (for degenerate quarks of mass $m$) $\partial_\mu A_\mu(x) =
2m P(x)$, where $A_\mu$ and $P$ represent the axial current and
pseudoscalar density respectively, is satisfied with
$m=m^\mathrm{DWF}+m_\mathrm{res}$, where $m^\mathrm{DWF}$ is the bare
mass in the DWF action. The mixing of operators which transform under
different representations of chiral symmetry is found to be negligibly
small in current simulations. The important thing to note is that the
chiral symmetry breaking effects are small and that there are
techniques to mitigate their consequences.

The main price which has to be paid for the good chiral symmetry is
that the simulations are performed in 5 dimensions, requiring
approximately a factor of N in computing resources and resulting in
practice in ensembles at fewer values of the lattice spacing and quark
masses than is possible with other formulations. The current
generation of DWF simulations is being performed at physical quark
masses so that ensembles with good chiral and flavour symmetries are
being generated and analysed~\cite{Arthur:2012opa}. For a discussion
of the equivalence of DWF and overlap fermions
see~\cite{Borici:1999zw,Borici:1999da}.

A third example of an operator which satisfies the Ginsparg-Wilson
relation is the so-called fixed-point action
\cite{Bietenholz:1995cy,Hasenfratz:2000xz,Hasenfratz:2002rp}. This
construction proceeds via a renormalization group approach. A related
formalism are the so-called ``chirally improved'' fermions
\cite{Gattringer:2000js}.\\

\begin{table}
\begin{center}
{\footnotesize
\begin{tabular*}{\textwidth}[l]{l @{\extracolsep{\fill}} l l l l}
\hline \hline  \\[-1.0ex]
\parbox[t]{1.5cm}{Abbrev.} & Discretization & \parbox[t]{2.2cm}{Leading lattice \\artifacts} & Chiral symmetry &  Remarks
\\[4.0ex] \hline \hline \\[-1.0ex]
Wilson     & Wilson & $O(a)$ & broken & 
\\[1.0ex] \hline \\[-1.0ex]
tmWil   & twisted-mass Wilson &  \parbox[t]{2.2cm}{$O(a^2)$
at\\ maximal twist} & broken & \parbox[t]{5cm}{flavour symmetry breaking:\\ $(M_\text{PS}^{0})^2-(M_\text{PS}^\pm)^2\sim O(a^2)$}
\\[4.0ex] \hline \\[-1.0ex]
tlSW      & Sheikholeslami-Wohlert & $O(g^2 a)$ & broken & tree-level
impr., $\csw=1$
\\[1.0ex] \hline \\[-1.0ex]
\parbox[t]{1.0cm}{n-HYP tlSW}      & Sheikholeslami-Wohlert & $O(g^2 a)$ & broken & \parbox[t]{5cm}{tree-level
impr., $\csw=1$,\\
n-HYP smeared gauge links
}
\\[4.0ex] \hline \\[-1.0ex]
\parbox[t]{1.2cm}{stout tlSW}      & Sheikholeslami-Wohlert & $O(g^2 a)$ & broken & \parbox[t]{5cm}{tree-level
impr., $\csw=1$,\\
stout smeared gauge links
}
\\[4.0ex] \hline \\[-1.0ex]
\parbox[t]{1.2cm}{HEX tlSW}      & Sheikholeslami-Wohlert & $O(g^2 a)$ & broken & \parbox[t]{5cm}{tree-level
impr., $\csw=1$,\\
HEX smeared gauge links
}
\\[4.0ex] \hline \\[-1.0ex]
mfSW      & Sheikholeslami-Wohlert & $O(g^2 a)$ & broken & mean-field impr.
\\[1.0ex] \hline \\[-1.0ex]
npSW      & Sheikholeslami-Wohlert & $O(a^2)$ & broken & nonperturbatively impr.
\\[1.0ex] \hline \\[-1.0ex]
KS      & Staggered & $O(a^2)$ & \parbox[t]{3cm}{$\rm
  U(1)\otimes U(1)$ subgr.\\ unbroken} & rooting for $\Nf<4$
\\[4.0ex] \hline \\[-1.0ex]
Asqtad  & Staggered & $O(a^2)$ & \parbox[t]{3cm}{$\rm
  U(1)\otimes U(1)$ subgr.\\ unbroken}  & \parbox[t]{5cm}{Asqtad
  smeared gauge links, \\rooting for $\Nf<4$}  
\\[4.0ex] \hline \\[-1.0ex]
HISQ  & Staggered & $O(a^2)$ & \parbox[t]{3cm}{$\rm
  U(1)\otimes U(1)$ subgr.\\ unbroken}  & \parbox[t]{5cm}{HISQ
  smeared gauge links, \\rooting for $\Nf<4$}  
\\[4.0ex] \hline \\[-1.0ex]
DW      & Domain Wall & \parbox[t]{2.2cm}{asymptotically \\$O(a^2)$} & \parbox[t]{3cm}{remnant
  breaking \\exponentially suppr.} & \parbox[t]{5cm}{exact chiral symmetry and\\$O(a)$ impr. only in the limit \\
 $L_s\rightarrow \infty$}
\\[7.0ex] \hline \\[-1.0ex]
overlap    & Neuberger & $O(a^2)$ & exact
\\[1.0ex] 
\hline\hline
\end{tabular*}
}
\caption{The most widely used discretizations of the quark action
  and some of their properties. Note that in order to maintain the
  leading lattice artifacts of the action in nonspectral observables
  (like operator matrix elements)
  the corresponding nonspectral operators need to be improved as well. 
\label{tab_quarkactions}}
\end{center}
\end{table}

\noindent
{\it Smearing}\\
\noindent

A simple modification which can help improve the action as well as the
computational performance is the use of smeared gauge fields in the
covariant derivatives of the fermionic action. Any smearing procedure
is acceptable as long as it consists of only adding irrelevant (local)
operators. Moreover, it can be combined with any discretization of the
quark action.  The ``Asqtad'' staggered quark action mentioned above
\cite{Orginos:1999cr} is an example which makes use of so-called
``Asqtad'' smeared (or ``fat'') links. Another example is the use of
n-HYP smeared \cite{Hasenfratz:2001hp,Hasenfratz:2007rf}, stout smeared
\cite{Morningstar:2003gk,Durr:2008rw} or HEX (hypercubic stout) smeared \cite{Capitani:2006ni} gauge links in the tree-level clover improved
discretization of the quark action, denoted by ``n-HYP tlSW'',
``stout tlSW'' and ``HEX tlSW'' in the following.\\

\noindent
In Table \ref{tab_quarkactions} we summarize the most widely used
discretizations of the quark action and their main properties together
with the abbreviations used in the summary tables. Note that in order
to maintain the leading lattice artifacts of the actions as given in
the table in nonspectral observables (like operator matrix elements)
the corresponding nonspectral operators need to be improved as well.

\clearpage
\subsubsection{Heavy-quark actions}
\label{app:HQactions}

Charm and bottom quarks are often simulated with different
lattice-quark actions than up, down, and strange quarks because their
masses are large relative to typical lattice spacings in current
simulations; for example, $a m_c \sim 0.4$ and $am_b \sim 1.3$ at
$a=0.06$~fm.  Therefore, for the actions described in the previous
section, using a sufficiently small lattice spacing to control generic
$(am_h)^n$ discretization errors is computationally costly, and in
fact prohibitive at the physical $b$-quark mass.

One approach for lattice heavy quarks is direct application of
effective theory.  In this case the lattice heavy-quark action only
correctly describes phenomena in a specific kinematic regime, such as
Heavy-Quark Effective Theory
(HQET)~\cite{Isgur:1989vq,Eichten:1989zv,Isgur:1989ed} or
Nonrelativistic QCD (NRQCD)~\cite{Caswell:1985ui,Bodwin:1994jh}.  One
can discretize the effective Lagrangian to obtain, for example,
Lattice HQET~\cite{Heitger:2003nj} or Lattice
NRQCD~\cite{Thacker:1990bm,Lepage:1992tx}, and then simulate the
effective theory numerically.  The coefficients of the operators in
the lattice-HQET and lattice-NRQCD actions are free parameters that
must be determined by matching to the underlying theory (QCD) through
the chosen order in $1/m_h$ or $v_h^2$, where $m_h$ is the heavy-quark
mass and $v_h$ is the heavy-quark velocity in the the heavy-light
meson rest frame.

Another approach is to interpret a relativistic quark action such as
those described in the previous section in a manner suitable for heavy
quarks.  One can extend the standard Symanzik improvement program,
which allows one to systematically remove lattice cutoff effects by
adding higher-dimension operators to the action, by allowing the
coefficients of the dimension 4 and higher operators to depend
explicitly upon the heavy-quark mass.  Different prescriptions for
tuning the parameters correspond to different implementations: those
in common use are often called the Fermilab
action~\cite{ElKhadra:1996mp}, the relativistic heavy-quark action
(RHQ)~\cite{Christ:2006us}, and the Tsukuba
formulation~\cite{Aoki:2001ra}.  In the Fermilab approach, HQET is
used to match the lattice theory to continuum QCD at the desired order
in $1/m_h$.

More generally, effective theory can be used to estimate the size of
cutoff errors from the various lattice heavy-quark actions.  The power
counting for the sizes of operators with heavy quarks depends on the
typical momenta of the heavy quarks in the system.  Bound-state
dynamics differ considerably between heavy-heavy and heavy-light
systems.  In heavy-light systems, the heavy quark provides an
approximately static source for the attractive binding force, like the
proton in a hydrogen atom.  The typical heavy-quark momentum in the
bound-state rest frame is $|\vec{p}_h| \sim \Lambda_{\rm QCD}$, and
heavy-light operators scale as powers of $(\Lambda_{\rm QCD}/m_h)^n$.
This is often called ``HQET power-counting", although it applies to
heavy-light operators in HQET, NRQCD, and even relativistic
heavy-quark actions described below.  Heavy-heavy systems are similar
to positronium or the deuteron, with the typical heavy-quark momentum
$|\vec{p}_h| \sim \alpha_S m_h$.  Therefore motion of the heavy quarks
in the bound state rest frame cannot be neglected.  Heavy-heavy
operators have complicated power counting rules in terms of
$v_h^2$~\cite{Lepage:1992tx}; this is often called ``NRQCD power
counting."

Alternatively, one can simulate bottom or charm quarks with the same
action as up, down, and strange quarks provided that (1) the action is
sufficiently improved, and (2) the lattice spacing is sufficiently
fine.  These qualitative criteria do not specify precisely how large a
numerical value of $am_h$ can be allowed while obtaining a given
precision for physical quantities; this must be established
empirically in numerical simulations.  At present, both the HISQ and
twisted-mass Wilson actions discussed previously are being used to
simulate charm quarks.
Simulations with HISQ quarks have employed heavier quark masses than
those with twisted-mass Wilson quarks because the action is more
highly improved, but neither action can be used to simulate at the
physical $am_b$ for current lattice spacings.  Therefore calculations
of heavy-light decay constants with these actions still rely on
effective theory to reach the $b$-quark mass: the ETM Collaboration
interpolates between twisted-mass Wilson data generated near $am_c$
and the static point~\cite{Dimopoulos:2011gx}, while the HPQCD
Collaboration extrapolates HISQ data generated below $am_b$ up to the
physical point using an HQET-inspired series expansion in
$(1/m_h)^n$~\cite{McNeile:2011ng}.
\\


\noindent
{\it Heavy-quark effective theory}\\
\noindent

HQET was introduced by Eichten and Hill in
Ref.~\cite{Eichten:1989zv}. It provides the correct asymptotic
description of QCD correlation functions in the static limit
$m_{h}/|\vec{p}_h| \!\to\! \infty$. Subleading effects are described
by higher dimensional operators whose coupling constants are formally
of ${\mathcal O}((1/m_{h})^n)$.  The HQET expansion works well for
heavy-light systems in which the heavy-quark momentum is small
compared to the mass.

The HQET Lagrangian density at the leading (static) order in the rest
frame of the heavy quark is given by
\be
{\mathcal L}^{\rm stat}(x) = \overline{\psi}_{h}(x) \,D_0\, \psi_{h}(x)\;,
\ee
with
\be
P_+ \psi_{h} = \psi_{h} \; , \quad\quad \overline{\psi}_{h} P_+=\overline{\psi}_{h} \;,  
\quad\quad P_+={{1+\gamma_0}\over{2}} \;.
\ee
A bare quark mass $m_{\rm bare}^{\rm stat}$ has to be added to the energy  
levels $E^{\rm stat}$ computed with this Lagrangian to obtain the physical ones.
 For example, the mass of the $B$ meson in the static approximation is given by
\be
m_{B} = E^{\rm stat} + m_{\rm bare}^{\rm stat} \;.
\ee
At tree-level $m_{\rm bare}^{\rm stat}$ is simply the (static approximation of
the) $b$-quark mass, but in the quantized lattice formulation it has
to further compensate a divergence linear in the inverse lattice spacing.
Weak composite fields  are also rewritten in terms of the static fields, e.g.
\begin{equation}
A_0(x)^{\rm stat}=Z_{\rm A}^{\rm stat} \left( \overline{\psi}(x) \gamma_0\gamma_5\psi_h(x)\right)\;,
\end{equation}
where the renormalization factor of the axial current in the static
theory $Z_{\rm A}^{\rm stat}$ is scale-dependent.  Recent lattice-QCD
calculations using static $b$ quarks and dynamical light
quarks~\cite{Albertus:2010nm,Dimopoulos:2011gx} perform the operator
matching at one-loop in mean-field improved lattice perturbation
theory~\cite{Ishikawa:2011dd,Blossier:2011dg}.  Therefore the
heavy-quark discretization, truncation, and matching errors in these
results are of ${\mathcal O}(a^2 \Lambda_{\rm QCD}^2)$, ${\mathcal O}
(\Lambda_{\rm QCD}/m_h)$, and ${\mathcal O}(\alpha_s^2, \alpha_s^2
a \Lambda_{\rm QCD})$.

In order to reduce heavy-quark truncation errors in $B$-meson masses
and matrix elements to the few-percent level, state-of-the-art
lattice-HQET computations now include corrections of ${\mathcal
O}(1/m_h)$.  Adding the $1/m_{h}$ terms, the HQET Lagrangian reads
\begin{eqnarray}
{\mathcal L}^{\rm HQET}(x) &=&  {\mathcal L}^{\rm stat}(x) - \omegakin\Okin(x)
        - \omegaspin\Ospin(x)  \,, \\[2.0ex]
  \Okin(x) &=& \overline{\psi}_{h}(x){\bf D}^2\psi_{h}(x) \,,\quad
  \Ospin(x) = \overline{\psi}_{h}(x){\boldsymbol\sigma}\!\cdot\!{\bf B}\psi_{h}(x)\,.
\end{eqnarray}
At this order, two other parameters appear in the Lagrangian,
$\omegakin$ and $\omegaspin$. The normalization is such that the
tree-level values of the coefficients are
$\omegakin=\omegaspin=1/(2m_{h})$.  Similarly the operators are
formally expanded in inverse powers of the heavy-quark mass.  The time
component of the axial current, relevant for the computation of
mesonic decay constants is given by
\begin{eqnarray}
A_0^{\rm HQET}(x) &=& Z_{\rm A}^{\rm HQET}\left(A_0^{\rm stat}(x) +\sum_{i=1}^2 c_{\rm A}^{(i)} A_0^{(i)}(x)\right)\;, \\
A_0^{(1)}(x)&=&\overline{\psi}\frac{1}{2}\gamma_5 \gamma_k  (\nabla_k-\overleftarrow{\nabla}_k)\psi_h(x), \qquad k=1,2,3\\
A_0^{(2)} &=& -\partial_kA_k^{\rm stat}(x)\;, \quad A_k^{\rm stat}=\overline{\psi}(x) \gamma_k\gamma_5\psi_h(x)\;,
\end{eqnarray}
and depends on two additional parameters $c_{\rm A}^{(1)}$ and $c_{\rm A}^{(2)}$.

A framework for nonperturbative HQET on the lattice has been
introduced in~\cite{Heitger:2003nj,Blossier:2010jk}.  As pointed out
in Refs~\cite{Sommer:2006sj,DellaMorte:2007ny}, since $\alpha_s(m_h)$
decreases logarithmically with $m_h$, whereas corrections in the
effective theory are power-like in $\Lambda/m_h$, it is possible that
the leading errors in a calculation will be due to the perturbative
matching of the action and the currents at a given order
$(\Lambda/m_h)^l$ rather than to the missing ${\mathcal
O}((\Lambda/m_h)^{l+1})$ terms.  Thus, in order to keep matching
errors below the uncertainty due to truncating the HQET expansion, the
matching is performed nonperturbatively beyond leading order in
$1/m_{h}$. The asymptotic convergence of HQET in the limit
$m_h \to \infty$ indeed holds only in that case.

The higher dimensional interaction terms in the effective Lagrangian
are treated as space-time volume insertions into static correlation
functions.  For correlators of some multi-local fields ${\mathcal{O}}$
and up to the $1/m_h$ corrections to the operator, this means
\begin{equation}
\langle {\mathcal{O}} \rangle =\langle {\mathcal{O}} \rangle_{\rm stat} +\omegakin a^4 \sum_x
\langle {\mathcal{OO}}_{\rm kin}(x) \rangle_{\rm stat} + \omegaspin a^4 \sum_x
\langle {\mathcal{OO}}_{\rm spin}(x) \rangle_{\rm stat} \;, 
\end{equation}
where $\langle {\mathcal{O}} \rangle_{\rm stat}$ denotes the static
expectation value with ${\mathcal{L}}^{\rm stat}(x)
+{\mathcal{L}}^{\rm light}(x)$.  Nonperturbative renormalization of
these correlators guarantees the existence of a well-defined continuum
limit to any order in $1/m_h$.  The parameters of the effective action
and operators are then determined by matching a suitable number of
observables calculated in HQET (to a given order in $1/m_{h}$) and in
QCD in a small volume (typically with $L\simeq 0.5$ fm), where the
full relativistic dynamics of the $b$-quark can be simulated and the
parameters can be computed with good accuracy.
In~\cite{Blossier:2010jk,Blossier:2012qu} the Schr\"odinger Functional
(SF) setup has been adopted to define a set of quantities, given by
the small volume equivalent of decay constants, pseudoscalar-vector
splittings, effective masses and ratio of correlation functions for
different kinematics, that can be used to implement the matching
conditions.  The kinematical conditions are usually modified by
changing the periodicity in space of the fermions, i.e. by directly
exploiting a finite-volume effect.  The new scale $L$, which is
introduced in this way, is chosen such that higher orders in $1/m_hL$
and in $\Lambda_{\rm QCD}/m_h$ are of about the same size. At the end
of the matching step the parameters are known at lattice spacings
which are of the order of $0.01$ fm, significantly smaller than the
resolutions used for large volume, phenomenological, applications. For
this reason a set of SF-step scaling functions is introduced in the
effective theory to evolve the parameters to larger lattice spacings.
The whole procedure yields the nonperturbative parameters with an
accuracy which allows to compute phenomenological quantities with a
precision of a few percent
(see~\cite{Blossier:2010mk,Bernardoni:2012ti} for the case of the
$B_{(s)}$ decay constants).  Such an accuracy can not be achieved by
performing the nonperturbative matching in large volume against
experimental measurements, which in addition would reduce the
predictivity of the theory.  For the lattice-HQET action matched
nonperturbatively through ${\mathcal O}(1/m_h)$, discretization and
truncation errors are of ${\mathcal O}(a \Lambda^2_{\rm QCD}/m_h,
a^2 \Lambda^2_{\rm QCD})$ and ${\cal O}((\Lambda_{\rm QCD}/m_h )^2)$.

The noise-to-signal ratio of static-light correlation functions grows
exponentially in Euclidean time, $\propto e^{\mu x_0}$ . The rate
$\mu$ is nonuniversal but diverges as $1/a$ as one approaches the
continuum limit. By changing the discretization of the covariant
derivative in the static action one may achieve an exponential
reduction of the noise to signal ratio. Such a strategy led to the
introduction of the $S^{\rm stat}_{\rm HYP1,2}$
actions~\cite{DellaMorte:2005yc}, where the thin links in $D_0$ are
replaced by HYP-smeared links~\cite{Hasenfratz:2001hp}.  These actions
are now used in all lattice applications of HQET.
\\


\noindent
{\it Nonrelativistic QCD}\\
\noindent

Nonrelativistic QCD (NRQCD) \cite{Thacker:1990bm,Lepage:1992tx} is an
 effective theory that can be matched to full QCD order by order in
 the heavy-quark velocity $v_h^2$ (for heavy-heavy systems) or in
 $\Lambda_{\rm QCD}/m_h$ (for heavy-light systems) and in powers of
 $\alpha_s$.  Relativistic corrections appear as higher-dimensional
 operators in the Hamiltonian.
 
 As an effective field theory, NRQCD is only useful with an
 ultraviolet cutoff of order $m_h$ or less. On the lattice this means
 that it can be used only for $am_h>1$, which means that $O(a^n)$
 errors cannot be removed by taking $a\to0$ at fixed $m_h$. Instead
 heavy-quark discretization errors are systematically removed by
 adding additional operators to the lattice Hamiltonian.  Thus, while
 strictly speaking no continuum limit exists at fixed $m_h$, continuum
 physics can be obtained at finite lattice spacing to arbitrarily high
 precision provided enough terms are included, and provided that the
 coefficients of these terms are calculated with sufficient accuracy.
 Residual discretization errors can be parameterized as corrections to
 the coefficients in the nonrelativistic expansion, as shown in
 Eq.~(\ref{deltaH}).  Typically they are of the form
 $(a|\vec{p}_h|)^n$ multiplied by a function of $am_h$ that is smooth
 over the limited range of heavy-quark masses (with $am_h > 1$) used
 in simulations, and can therefore can be represented by a low-order
 polynomial in $am_h$ by Taylor's theorem (see
 Ref.~\cite{Gregory:2010gm} for further discussion).  Power-counting
 estimates of these effects can be compared to the observed lattice
 spacing dependence in simulations. Provided that these effects are
 small, such comparisons can be used to estimate and correct the
 residual discretization effects.

An important feature of the NRQCD approach is that the same action can
be applied to both heavy-heavy and heavy-light systems. This allows,
for instance, the bare $b$-quark mass to be fixed via experimental
input from $\Upsilon$ so that simulations carried out in the $B$ or
$B_s$ systems have no adjustable parameters left.  Precision
calculations of the $B_s$ meson mass (or of the mass splitting
$M_{B_s} - M_\Upsilon/2$) can then be used to test the reliability of
the method before turning to quantities one is trying to predict, such
as decay constants $f_B$ and $f_{B_s}$, semileptonic form factors or
neutral $B$ mixing parameters.

Given the same lattice-NRQCD heavy-quark action, simulation results
will not be as accurate for charm quarks as for bottom ($1/m_b <
1/m_c$, and $v_b < v_c$ in heavy-heavy systems).  For charm, however,
a more serious concern is the restriction that $am_h$ must be greater
than one.  This limits lattice-NRQCD simulations at the physical
$am_c$ to relatively coarse lattice spacings for which light-quark and
gluon discretization errors could be large.  Thus recent lattice-NRQCD
simulations have focused on bottom quarks because $am_b > 1$ in the
range of typical lattice spacings between $\approx$ 0.06 and 0.15~fm.

In most simulations with NRQCD $b$-quarks during the past decade one
has worked with an NRQCD action that includes tree-level relativistic
corrections through ${\cal O}(v_h^4)$ and discretization corrections
through ${\cal O}(a^2)$,
 \begin{eqnarray}
 \label{nrqcdact}
&&  S_{\rm NRQCD}  =
a^4 \sum_x \Bigg\{  {\Psi}^\dagger_t \Psi_t -
 {\Psi}^\dagger_t
\left(1 \!-\!\frac{a \delta H}{2}\right)_t
 \left(1\!-\!\frac{aH_0}{2n}\right)^{n}_t \nonumber \\
& \times &
 U^\dagger_t(t-a)
 \left(1\!-\!\frac{aH_0}{2n}\right)^{n}_{t-a}
\left(1\!-\!\frac{a\delta H}{2}\right)_{t-a} \Psi_{t-a} \Bigg\} \, ,
 \end{eqnarray}
where the subscripts $``t"$ and $``t-a"$ denote that the heavy-quark, gauge, $\bf{E}$,  and $\bf{B}$-fields are on time slices $t$ or $t-a$, respectively.
 $H_0$ is the nonrelativistic kinetic energy operator,
 \be
 H_0 = - {\delsq\over2m_h} \, ,
 \ee
and $\delta H$ includes relativistic and finite-lattice-spacing
corrections,
 \begin{eqnarray}
\delta H
&=& - c_1\,\frac{(\delsq)^2}{8m_h^3}
+ c_2\,\frac{i g}{8m_h^2}\left(\delv\cdot\Ev - \Ev\cdot\delv\right) \nl
& &
 - c_3\,\frac{g}{8m_h^2} \sigmav\cdot(\delvt\times\Ev - \Ev\times\delvt)\nl
& & - c_4\,\frac{g}{2m_h}\,\sigmav\cdot\Bv
  + c_5\,\frac{a^2\delfour}{24m_h}  - c_6\,\frac{a(\delsq)^2}
{16nm_h^2} \, .
\label{deltaH}
\end{eqnarray}
 $m_h$ is the bare heavy-quark mass, $\delsq$ the lattice Laplacian,
$\delv$ the symmetric lattice derivative and $\delfour$ the lattice
discretization of the continuum $\sum_i D^4_i$.  $\delvt$ is the
improved symmetric lattice derivative and the $\Ev$ and $\Bv$ fields
have been improved beyond the usual clover leaf construction. The
stability parameter $n$ is discussed in \cite{Lepage:1992tx}.  In most
cases the $c_i$'s have been set equal to their tree-level values $c_i
= 1$.  With this implementation of the NRQCD action, errors in
heavy-light meson masses and splittings are of ${\cal
O}(\alpha_S \Lambda_{\rm QCD}/m_h )$, ${\cal O}(\alpha_S (\Lambda_{\rm
QCD}/m_h)^2 )$, ${\cal O}((\Lambda_{\rm QCD}/m_h )^3)$, and ${\cal
O}(\alpha_s a^2 \Lambda_{\rm QCD}^2)$, with coefficients that are
functions of $am_h$.  One-loop corrections to many of the coefficients
in Eq.~(\ref{deltaH}) have now been calculated, and are starting to be
included in
simulations \cite{Morningstar:1994qe,Hammant:2011bt,Dowdall:2011wh}.

Most of the operator matchings involving heavy-light currents or
four-fermion operators with NRQCD $b$-quarks and AsqTad or HISQ light
quarks have been carried out at one-loop order in lattice perturbation
theory.  In calculations published to date of electroweak matrix
elements, heavy-light currents with massless light quarks have been
matched through ${\cal O}(\alpha_s, \Lambda_{\rm QCD}/m_h, \alpha_s/(a
m_h),
\alpha_s \Lambda_{\rm QCD}/m_h)$, and four-fermion operators through \\
 ${\cal O}(\alpha_s, \Lambda_{\rm QCD}/m_h, 
\alpha_s/(a m_h))$.
NRQCD/HISQ currents with massive HISQ quarks are also of interest,
e.g.  for the bottom-charm currents in $B \rightarrow D^{(*)}, l \nu$
semileptonic decays and the relevant matching calculations have been
performed at one-loop order in Ref.~\cite{Monahan:2012dq}.  Taking all
the above into account, the most significant systematic error in
electroweak matrix elements published to date with NRQCD $b$-quarks is
the ${\cal O}(\alpha_s^2)$ perturbative matching uncertainty.  Work is
therefore underway to use current-current correlator methods combined
with very high order continuum perturbation theory to do current
matchings nonperturbatively~\cite{Koponen:2010jy}.
\\


\noindent
{\it Relativistic heavy quarks}\\
\noindent

An approach for relativistic heavy-quark lattice formulations was
first introduced by El-Khadra, Kronfeld, and Mackenzie in
Ref.~\cite{ElKhadra:1996mp}.  Here they showed that, for a general
lattice action with massive quarks and non-Abelian gauge fields,
discretization errors can be factorized into the form $f(m_h
a)(a|\vec{p}_h|)^n$, and that the function $f(m_h a)$ is bounded to be
of ${\mathcal O}(1)$ or less for all values of the quark mass $m_h$.
Therefore cutoff effects are of ${\mathcal O}(a \Lambda_{\rm QCD})^n$
and ${\mathcal O}((a|\vec{p}_h|)^n)$, even for $am_h \gtapprox 1$, and
can be controlled using a Symanzik-like procedure.  As in the standard
Symanzik improvement program, cutoff effects are systematically
removed by introducing higher-dimension operators to the lattice
action and suitably tuning their coefficients.  In the relativistic
heavy-quark approach, however, the operator coefficients are allowed
to depend explicitly on the quark mass.  By including lattice
operators through dimension $n$ and adjusting their coefficients
$c_{n,i}(m_h a)$ correctly, one enforces that matrix elements in the
lattice theory are equal to the analogous matrix elements in continuum
QCD through $(a|\vec{p}_h|)^n$, such that residual heavy-quark
discretization errors are of ${\mathcal O}(a|\vec{p}_h|)^{n+1}$.

The relativistic heavy-quark approach can be used to compute the
matrix elements of states containing heavy quarks for which the
heavy-quark spatial momentum $|\vec{p}_h|$ is small compared to the
lattice spacing.  Thus it is suitable to describe bottom and charm
quarks in both heavy-light and heavy-heavy systems.  Calculations of
bottomonium and charmonium spectra serve as nontrivial tests of the
method and its accuracy.

At fixed lattice spacing, relativistic heavy-quark formulations
recover the massless limit when $(am_h) \ll 1$, recover the static
limit when $(am_h) \gg 1$, and smoothy interpolate between the two;
thus they can be used for any value of the quark mass, and, in
particular, for both charm and bottom.  Discretization errors for
relativistic heavy-quark formulations are generically of the form
$\alpha_s^k f(am_h)(a |\vec{p}_h|)^n$, where $k$ reflects the order of
the perturbative matching for operators of ${\mathcal O}((a
|\vec{p}_h|)^n)$.  For each $n$, such errors are removed completely if
the operator matching is nonperturbative. When $(am_h) \sim 1$, this
gives rise to nontrivial lattice-spacing dependence in physical
quantities, and it is prudent to compare estimates based on
power-counting with a direct study of scaling behaviour using a range
of lattice spacings.
At fixed quark mass, relativistic heavy-quark actions possess a smooth
continuum limit without power-divergences.  Of course, as
$m_h \to \infty$ at fixed lattice spacing, the power divergences of
the static limit are recovered (see, e.g. Ref.~\cite{Harada:2001fi}).

The relativistic heavy-quark formulations in use all begin with the
anisotropic Sheikholeslami-Wohlert (``clover")
action~\cite{Sheikholeslami:1985ij}:
\begin{equation}
S_\textrm{lat} = a^4 \sum_{x,x'} \bar{\psi}(x') \left( m_0 + \gamma_0 D_0 + \zeta \vec{\gamma} \cdot \vec{D} - \frac{a}{2} (D^0)^2 - \frac{a}{2} \zeta (\vec{D})^2+ \sum_{\mu,\nu} \frac{ia}{4} c_{\rm SW} \sigma_{\mu\nu} F_{\mu\nu} \right)_{x' x} \psi(x) \,,
\label{eq:HQAct}
\end{equation}
where $D_\mu$ is the lattice covariant derivative and $F_{\mu\nu}$ is
the lattice field-strength tensor.  Here we show the form of the
action given in Ref.~\cite{Christ:2006us}.  The introduction of a
space-time anisotropy, parameterized by $\zeta$ in
Eq.~(\ref{eq:HQAct}), is convenient for heavy-quark systems because
the characteristic heavy-quark four-momenta do not respect space-time
axis exchange ($\vec{p}_h < m_h$ in the bound-state rest frame).
Further, the Sheikoleslami-Wohlert action respects the continuum
heavy-quark spin and flavour symmetries, so HQET can be used to
interpret and estimate lattice discretization
effects~\cite{Kronfeld:2000ck,Harada:2001fi,Harada:2001fj}.  We
discuss three different prescriptions for tuning the parameters of the
action in common use below.  In particular, we focus on aspects of the
action and operator improvement and matching relevant for evaluating
the quality of the calculations discussed in the main text.

The meson energy-momentum dispersion relation plays an important role
in relativistic heavy-quark formulations:
\begin{equation}
	E(\vec{p}) = M_1 + \frac{\vec{p}^2}{2M_2} + {\mathcal O}(\vec{p}^4) \,,
\end{equation}
where $M_1$ and $M_2$ are known as the rest and kinetic masses,
respectively.  Because the lattice breaks Lorentz invariance, there
are corrections proportional to powers of the momentum.  Further, the
lattice rest masses and kinetic masses are not equal ($M_1 \neq M_2$),
and only become equal in the continuum limit.

The Fermilab interpretation~\cite{ElKhadra:1996mp} is suitable for
calculations of mass splittings and matrix elements of systems with
heavy quarks.  The Fermilab action is based on the hopping-parameter
form of the Wilson action, in which $\kappa_h$ parameterizes the
heavy-quark mass.  In practice, $\kappa_h$ is tuned such that the the
kinetic meson mass equals the experimentally-measured heavy-strange
meson mass ($m_{B_s}$ for bottom and $m_{D_s}$ for charm).  In
principle, one could also tune the anisotropy parameter such that $M_1
= M_2$.  This is not necessary, however, to obtain mass splittings and
matrix elements, which are not affected by
$M_1$~\cite{Kronfeld:2000ck}.  Therefore in the Fermilab action the
anisotropy parameter is set equal to unity.  The clover coefficient in
the Fermilab action is fixed to the value $c_{\rm SW} = 1/u_0^3$ from
mean-field improved lattice perturbation theory~\cite{Lepage:1992xa}.
With this prescription, discretization effects are of ${\mathcal
O}(\alpha_sa|\vec{p}_h|, (a|\vec{p}_h|)^2)$.  Calculations of
electroweak matrix elements also require improving the lattice current
and four-fermion operators to the same order, and matching them to the
continuum.  Calculations with the Fermilab action remove tree-level
${\mathcal O}(a)$ errors in electroweak operators by rotating the
heavy-quark field used in the matrix element and setting the rotation
coefficient to its tadpole-improved tree-level value (see e.g.
Eqs.~(7.8) and (7.10) of Ref.~\cite{ElKhadra:1996mp}).  Finally,
electroweak operators are typically renormalized using a mostly
nonperturbative approach in which the flavour-conserving light-light
and heavy-heavy current renormalization factors $Z_V^{ll}$ and
$Z_V^{hh}$ are computed nonperturbatively~\cite{ElKhadra:2001rv}.  The
flavour-conserving factors account for most of the heavy-light current
renormalization.  The remaining correction is expected to be close to
unity due to the cancellation of most of the radiative corrections
including tadpole graphs~\cite{Harada:2001fi}; therefore it can be
reliably computed at one-loop in mean-field improved lattice
perturbation theory with truncation errors at the percent to
few-percent level.

The relativistic heavy-quark (RHQ) formulation developed by Li, Lin,
and Christ builds upon the Fermilab approach, but tunes all the
parameters of the action in Eq.~(\ref{eq:HQAct})
nonperturbatively~\cite{Christ:2006us}.  In practice, the three
parameters $\{m_0a, c_{\rm SW}, \zeta\}$ are fixed to reproduce the
experimentally-measured $B_s$ meson mass and hyperfine splitting
($m_{B_s^*}-m_{B_s}$), and to make the kinetic and rest masses of the
lattice $B_s$ meson equal~\cite{Aoki:2012xaa}.  This is done by
computing the heavy-strange meson mass, hyperfine splitting, and ratio
$M_1/M_2$ for several sets of bare parameters $\{m_0a, c_{\rm
SW}, \zeta\}$ and interpolating linearly to the physical $B_s$ point.
By fixing the $B_s$-meson hyperfine splitting, one loses a potential
experimental prediction with respect to the Fermilab formulation.
However, by requiring that $M_1 = M_2$, one gains the ability to use
the meson rest masses, which are generally more precise than the
kinetic masses, in the RHQ approach.  The nonperturbative
parameter-tuning procedure eliminates ${\mathcal O}(a)$ errors from
the RHQ action, such that discretization errors are of ${\mathcal
O}((a|\vec{p}_h|)^2)$.  Calculations of $B$-meson decay constants and
semileptonic form factors with the RHQ action are in
progress~\cite{Witzel:2012pr,Kawanai:2012id}, as is the corresponding
one-loop mean-field improved lattice perturbation
theory~\cite{Lehner:2012bt}.  For these works, cutoff effects in the
electroweak vector and axial-vector currents will be removed through
${\mathcal O}(\alpha_s a)$, such that the remaining discretization
errors are of ${\mathcal O}(\alpha_s^2a|\vec{p}_h|,
(a|\vec{p}_h|)^2)$.  Matching the lattice operators to the continuum
will be done following the mostly nonperturbative approach described
above.

The Tsukuba heavy-quark action is also based on the
Sheikholeslami-Wohlert action in Eq.~(\ref{eq:HQAct}), but allows for
further anisotropies and hence has additional parameters: specifically
the clover coefficients in the spatial $(c_B)$ and temporal $(c_E)$
directions differ, as do the anisotropy coefficients of the $\vec{D}$
and $\vec{D}^2$ operators~\cite{Aoki:2001ra}.  In practice, the
contribution to the clover coefficient in the massless limit is
computed nonperturbatively~\cite{Aoki:2005et}, while the
mass-dependent contributions, which differ for $c_B$ and $c_E$, are
calculated at one-loop in mean-field improved lattice perturbation
theory~\cite{Aoki:2003dg}.  The hopping parameter is fixed
nonperturbatively to reproduce the experimentally-measured
spin-averaged $1S$ charmonium mass~\cite{Namekawa:2011wt}.  One of the
anisotropy parameters ($r_t$ in Ref.~\cite{Namekawa:2011wt}) is also
set to its one-loop perturbative value, while the other ($\nu$ in
Ref.~\cite{Namekawa:2011wt}) is fixed noperturbatively to obtain the
continuum dispersion relation for the spin-averaged charmonium $1S$
states (such that $M_1 = M_2$).  For the renormalization and
improvement coefficients of weak current operators, the contributions
in the chiral limit are obtained
nonperturbatively~\cite{Kaneko:2007wh,Aoki:2010wm}, while the
mass-dependent contributions are estimated using one-loop lattice
perturbation theory~\cite{Aoki:2004th}.  With these choices, lattice
cutoff effects from the action and operators are of ${\mathcal
O}(\alpha_s^2 a|\vec{p}|, (a|\vec{p}_h|)^2)$.
\\


\noindent
{\it Light-quark actions combined with HQET}\\
\noindent

The heavy-quark formulations discussed in the previous sections use
effective field theory to avoid the occurence of discretization errors
of the form $(am_h)^n$.  In this section we describe methods that use
improved actions that were originally designed for light-quark systems
for $B$ physics calculations. Such actions unavoidably contain
discretization errors that grow as a power of the heavy-quark mass. In
order to use them for heavy-quark physics, they must be improved to at
least ${\cal O}(am_h)^2$.  However, since $am_b > 1$ at the smallest
lattice spacings available in current simulations, these methods also
require input from HQET to guide the simulation results to the
physical $b$-quark mass.

The ETM collaboration has developed two methods, the ``ratio
method'' \cite{Blossier:2009hg} and the ``interpolation
method'' \cite{Guazzini:2006bn,Blossier:2009gd}. They use these
methods together with simulations with twisted-mass Wilson fermions,
which have discretization errors of $O(am_h)^2$.  In the interpolation
method $\Phi_{hs}$ and $\Phi_{h\ell}$ (or $\Phi_{hs}/\Phi_{h\ell}$)
are calculated for a range of heavy-quark masses in the charm region
and above, while roughly keeping $am_h \ltsim 0.5 $. The relativistic
results are combined with a separate calculation of the decay
constants in the static limit, and then interpolated to the physical
$b$ quark mass. In ETM's implementation of this method, the heavy
Wilson decay constants are matched to HQET using NLO in continuum
perturbation theory. The static limit result is renormalized using
one-loop mean-field improved lattice perturbation theory, while for
the relativistic data PCAC is used to calculate absolutely normalized
matrix elements. Both, the relativistic and static limit data are then
run to the common reference scale $\mu_b = 4.5 \GeV$ at NLO in
continuum perturbation theory.  In the ratio method, one constructs
physical quantities $P(m_h)$ from the relativistic data that have a
well-defined static limit ($P(m_h) \to$ const.~for $m_h \to \infty$)
and evaluates them at the heavy-quark masses used in the simulations.
Ratios of these quantities are then formed at a fixed ratio of heavy
quark masses, $z = P(m_h) / P(m_h/\lambda)$ (where $1 < \lambda \lsim
1.3$), which ensures that $z$ is equal to unity in the static limit.
Hence, a separate static limit calculation is not needed with this
method.  In ETM's implementation of the ratio method for the $B$-meson
decay constant, $P(m_h)$ is constructed from the decay constants and
the heavy-quark pole mass as $P(m_h) = f_{h\ell}(m_h) \cdot (m^{\rm
pole}_h)^{1/2}$. The corresponding $z$-ratio therefore also includes
ratios of perturbative matching factors for the pole mass to $\msbar$
conversion.  For the interpolation to the physical $b$-quark mass,
ratios of perturbative matching factors converting the data from QCD
to HQET are also included. The QCD-to-HQET matching factors improve
the approach to the static limit by removing the leading logarithmic
corrections. In ETM's implementation of this method (ETM 11 and 12)
both conversion factors are evaluated at NLO in continuum perturbation
theory. The ratios are then simply fit to a polynomial in $1/m_h$ and
interpolated to the physical $b$-quark mass.  The ratios constructed
from $f_{h\ell}$ ($f_{hs}$) are called $z$ ($z_s$).  In order to
obtain the $B$ meson decay constants, the ratios are combined with
relativistic decay constant data evaluated at the smallest reference
mass.

The HPQCD collaboration has introduced a method in
Ref.~\cite{McNeile:2011ng} which we shall refer to as the ``heavy
HISQ'' method.  The first key ingredient is the use of the HISQ action
for the heavy and light valence quarks, which has leading
discretization errors of ${\cal O} \left(\alpha_s (v/c) (am_h)^2,
(v/c)^2 (am_h)^4\right)$.  With the same action for the heavy and
light valence quarks it is possible to use PCAC to avoid
renormalization uncertainties.  Another key ingredient is the
availability of gauge ensembles over a large range of lattice
spacings, in this case in the form of the library of $N_f = 2+1$
asqtad ensembles made public by the MILC collaboration which includes
lattice spacings as small as $a \approx 0.045$~fm.  Since the HISQ
action is so highly improved and with lattice spacings as small as
$0.045$~fm, HPQCD is able to use a large range of heavy-quark masses,
from below the charm region to almost up to the physical $b$ quark
mass with $am_h \ltsim 0.85$. They then fit their data in a combined
continuum and HQET fit (i.e. using a fit function that is motivated by
HQET) to a polynomial in $1/m_H$ (the heavy pseudo scalar meson mass
of a meson containing a heavy ($h$) quark).

\bigskip

In Table~\ref{tab_heavy_quarkactions} we list the discretizations of
the quark action most widely used for heavy $c$ and $b$ quarks
together with the abbreviations used in the summary tables.  We also
summarize the main properties of these actions and the leading lattice
discretization errors for calculations of heavy-light meson matrix
quantities with them.  Note that in order to maintain the leading
lattice artifacts of the actions as given in the table in nonspectral
observables (like operator matrix elements) the corresponding
nonspectral operators need to be improved as well.

\begin{table}
\begin{center}
{\footnotesize
\begin{tabular*}{\textwidth}[l]{l @{\extracolsep{\fill}} l l l}
\hline \hline  \\[-1.0ex]
\parbox[t]{1.5cm}{Abbrev.} & Discretization & 
\parbox[t]{4cm}{Leading lattice artifacts\\and truncation errors\\for heavy-light mesons} &  
Remarks
\\[4.0ex] \hline \hline \\[-1.0ex]
tmWil   & twisted-mass Wilson &  ${\mathcal O}\big((am_h)^2\big)$ & \parbox[t]{4.cm}{PCAC relation for axial-vector current}  
\\[3.0ex] \hline \\[-1.0ex]
HISQ  & Staggered & \parbox[t]{4cm}{${\mathcal O}\big (\alpha_S (am_h)^2 (v/c), \\(am_h)^4 (v/c)^2 \big)$}  & \parbox[t]{4.cm}{PCAC relation for axial-vector current; Ward identity for vector current}  
\\[6.0ex] \hline \\[-1.0ex]
static  & static effective action &  \parbox[t]{4cm}{${\mathcal O}\big( a^2 \Lambda_{\rm QCD}^2, \Lambda_{\rm QCD}/m_h, \\ \alpha_s^2, \alpha_s^2 a \Lambda_{\rm QCD} \big)$}  & \parbox[t]{4.5cm}{implementations use APE, HYP1, and HYP2 smearing}  
\\[4.0ex] \hline \\[-1.0ex]
HQET  & Heavy-Quark Effective Theory &  \parbox[t]{4cm}{${\mathcal O}\big( a \Lambda^2_{\rm QCD}/m_h,  a^2 \Lambda^2_{\rm QCD},\\
 (\Lambda_{\rm QCD}/m_h)^2 \big)$}  & \parbox[t]{4.5cm}{Nonperturbative matching through ${\mathcal O}(1/m_h)$}  
\\[7.0ex] \hline \\[-1.0ex]
NRQCD  & Nonrelativistic QCD & \parbox[t]{4cm}{${\cal O}\big(\alpha_S \Lambda_{\rm QCD}/m_h, \\ \alpha_S (\Lambda_{\rm QCD}/m_h)^2 , \\ (\Lambda_{\rm QCD}/m_h )^3,  \alpha_s a^2 \Lambda_{\rm QCD}^2 \big)$}  & \parbox[t]{4.5cm}{Tree-level relativistic corrections through 
${\cal O}(v_h^4)$ and discretization corrections through ${\cal O}(a^2)$}  
\\[9.5ex] \hline \\[-1.0ex] 
Fermilab  & Sheikholeslami-Wohlert & ${\mathcal O}\big(\alpha_sa\Lambda_{\rm QCD}, (a\Lambda_{\rm QCD})^2\big)$  & \parbox[t]{4.5cm}{Hopping parameter tuned nonperturbatively; clover coefficient computed at tree-level in mean-field improved lattice perturbation theory}  
\\[12.0ex] \hline \\[-1.0ex] 
Tsukuba  & Sheikholeslami-Wohlert & ${\mathcal O}\big( \alpha_s^2 a\Lambda_{\rm QCD}, (a\Lambda_{\rm QCD})^2 \big)$  & \parbox[t]{4.5cm}{NP clover coefficient at $ma=0$ plus mass-dependent corrections calculated at one-loop in lattice perturbation theory; $\nu$ calculated NP from dispersion relation; $r_s$ calculated at one-loop in lattice perturbation theory}  
\\[18.0ex]
\hline\hline
\end{tabular*}
}
\caption{Discretizations of the quark action most widely used for heavy $c$ and $b$ quarks  and some of their properties.
\label{tab_heavy_quarkactions}}
\end{center}
\end{table}

\clearpage
\subsection{Setting the scale \label{sec_scale}}

In simulations of lattice QCD quantities such as hadron masses and
decay constants are obtained in ``lattice units'' i.e.~as
dimensionless numbers. In order to convert them into physical units
they must be expressed in terms of some experimentally known,
dimensionful reference quantity $Q$. This procedure is called
``setting the scale''. It amounts to computing the nonperturbative
relation between the bare gauge coupling $g_0$ (which is an input
parameter in any lattice simulation) and the lattice spacing~$a$
expressed in physical units. To this end one chooses a value for $g_0$
and computes the value of the reference quantity in a simulation: This
yields the dimensionless combination, $(aQ)|_{g_0}$, at the chosen
value of $g_0$. The calibration of the lattice spacing is then
achieved via
\be
 a^{-1}\,[{\rm MeV}] = \frac{Q|_{\rm{exp}}\,[{\rm MeV}]}{(aQ)|_{g_0}},
\ee
where $Q|_{\rm{exp}}$ denotes the experimentally known value of the
reference quantity. Common choices for $Q$ are the mass of the
nucleon, the $\Omega$ baryon or the decay constants of the pion and
the kaon. Vector mesons, such as the $\rho$ or $K^\ast$-meson, are
unstable and therefore their masses are not very well suited for
setting the scale, despite the fact that they have been used over many
years for that purpose.

Another widely used quantity to set the scale is the hadronic radius
$r_0$, which can be determined from the force between static quarks
via the relation \cite{Sommer:1993ce}
\be
   F(r_0)r_0^2 = 1.65.
\ee
If the force is derived from potential models describing heavy
quarkonia, the above relation determines the value of $r_0$ as
$r_0\approx0.5$\,fm. A variant of this procedure is obtained
\cite{Bernard:2000gd} by using the definition $F(r_1)r_1^2=1.00$,
which yields $r_1\approx0.32$\,fm. It is important to realize that
both $r_0$ and $r_1$ are not directly accessible in experiment, so
that their values derived from phenomenological potentials are
necessarily model-dependent. Inspite of the inherent ambiguity
whenever hadronic radii are used to calibrate the lattice spacing,
they are very useful quantities for performing scaling tests and
continuum extrapolations of lattice data. Furthermore, they can be
easily computed with good statistical accuracy in lattice simulations.

\subsection{Matching and running \label{sec_match}}

The lattice formulation of QCD amounts to introducing a particular
regularization scheme. Thus, in order to be useful for phenomenology,
hadronic matrix elements computed in lattice simulations must be
related to some continuum reference scheme, such as the
$\msbar$-scheme of dimensional regularization. The matching to the
continuum scheme usually involves running to some reference scale
using the renormalization group. 

In principle, the matching factors which relate lattice matrix
elements to the $\msbar$-scheme, can be computed in perturbation
theory formulated in terms of the bare coupling. It has been known for
a long time, though, that the perturbative expansion is not under good 
control. Several techniques have been developed which allow for a
nonperturbative matching between lattice regularization and continuum
schemes, and are briefly introduced here.\\

\newpage

\noindent
{\sl Regularization-independent Momentum Subtraction}\\
\noindent

In the {\sl Regularization-independent Momentum Subtraction}
(``RI/MOM'' or ``RI'') scheme \cite{Martinelli:1994ty} a
nonperturbative renormalization condition is formulated in terms of
Green functions involving quark states in a fixed gauge (usually
Landau gauge) at nonzero virtuality. In this way one relates
operators in lattice regularization nonperturbatively to the RI
scheme. In a second step one matches the operator in the RI scheme to
its counterpart in the $\msbar$-scheme. The advantage of this
procedure is that the latter relation involves perturbation theory
formulated in the continuum theory. The uncontrolled use of lattice
perturbation theory can thus be avoided. A technical complication is
associated with the accessible momentum scales (i.e. virtualities),
which must be large enough (typically several $\gev$) in order for the
perturbative relation to $\msbar$ to be reliable. The momentum scales
in simulations must stay well below the cutoff scale (i.e. $2\pi$ over
the lattice spacing), since otherwise large lattice artifacts are
incurred. Thus, the applicability of the RI scheme traditionally relies on the
existence of a ``window'' of momentum scales, which satisfy
\be
   \Lambda_{\rm QCD} \;\lesssim\; p \;\lesssim\; 2\pi a^{-1}.
\ee
However, solutions for mitigating this limitation, which involve
continuum limit, nonperturbative running to higher scales in the
RI/MOM scheme, have recently been proposed and implemented
\cite{Arthur:2010ht,Durr:2010vn,Durr:2010aw,Aoki:2010pe}.\\

\noindent
{\it Schr\"odinger functional}\\
\noindent

Another example of a nonperturbative matching procedure is provided
by the Schr\"odinger functional (SF) scheme \cite{Luscher:1992an}. It
is based on the formulation of QCD in a finite volume. If all quark
masses are set to zero the box length remains the only scale in the
theory, such that observables like the coupling constant run with the
box size~$L$. The great advantage is that the RG running of
scale-dependent quantities can be computed nonperturbatively using
recursive finite-size scaling techniques. It is thus possible to run
nonperturbatively up to scales of, say, $100\,\gev$, where one is
sure that the perturbative relation between the SF and
$\msbar$-schemes is controlled.\\

\noindent
{\sl Perturbation theory}\\
\noindent

The third matching procedure is based on perturbation theory in which
higher order are effectively resummed \cite{Lepage:1992xa}. Although
this procedure is easier to implement, it is hard to estimate the
uncertainty associated with it.\\

\bigskip

\noindent
{\sl Mostly nonperturbative renormalization}\\
\noindent

Some calculations of heavy-light and heavy-heavy matrix elements adopt a mostly nonperturbative matching approach.  Let us consider a weak 
decay process mediated by a current with quark flavours $h$ and $q$, where $h$ is the initial heavy quark (either bottom or charm) and 
$q$ can be a light ($\ell = u,d$), strange, or charm quark. The matrix elements of lattice current  $J_{hq}$ are matched to the 
corresponding continuum matrix elements with continuum current ${\cal J}_{hq}$ by calculating the renormalization factor $Z_{J_{hq}}$. 
The mostly nonperturbative renormalization method takes advantage of rewriting the current renormalization factor as the following product:
\begin{align}
Z_{J_{hq}} = \rho_{J_{hq}} \sqrt{Z_{V^4_{hh}}Z_{V^4_{qq}}} \,
\label{eq:Zvbl}
\end{align}
The flavour-conserving renormalization factors $Z_{V^4_{hh}}$ and $Z_{V^4_{qq}}$ can be obtained nonperturbatively from standard heavy-light 
and light-light meson charge normalization conditions.  $Z_{V^4_{hh}}$ and $Z_{V^4_{qq}}$ account for  the bulk of the renormalization. The remaining 
correction $\rho_{J_{hq}}$ is expected to be close to unity because most of the radiative corrections, including self-energy corrections and 
contributions from tadpole graphs, cancel in the ratio~\cite{Harada:2001fj,Harada:2001fi}.  The one-loop coefficients of $\rho_{J_{hq}}$  have been calculated for
heavy-light and heavy-heavy currents for Fermilab heavy and both (improved) Wilson light \cite{Harada:2001fj,Harada:2001fi} and 
asqtad light  \cite{ElKhadra:2007qe} quarks. In all cases the one-loop coefficients are found to be very small, yielding sub-percent to few percent level corrections.

\bigskip
\noindent
In Table~\ref{tab_match} we list the abbreviations used in the
compilation of results together with a short description.

\begin{table}[ht]
{\footnotesize
\begin{tabular*}{\textwidth}[l]{l @{\extracolsep{\fill}} l}
\hline \hline \\[-1.0ex]
Abbrev. & Description
\\[1.0ex] \hline \hline \\[-1.0ex]
RI  &  regularization-independent momentum subtraction scheme 
\\[1.0ex] \hline \\[-1.0ex]
SF  &  Schr\"odinger functional scheme
\\[1.0ex] \hline \\[-1.0ex]
PT1$\ell$ & matching/running computed in perturbation theory at one loop
\\[1.0ex] \hline \\[-1.0ex]
PT2$\ell$ & matching/running computed in perturbation theory at two loops 
\\[1.0ex] \hline \\[-1.0ex]
mNPR & mostly nonperturbative renormalization 
%
\\[1.0ex]
\hline\hline
\end{tabular*}
}
\caption{The most widely used matching and running
  techniques. \label{tab_match} 
}
\end{table}

\subsection{Chiral extrapolation\label{sec_ChiPT}}
As mentioned in the introduction, Symanzik's framework can be combined 
with Chiral Perturbation Theory. The well-known terms occurring in the
chiral effective Lagrangian are then supplemented by contributions 
proportional to powers of the lattice spacing $a$. The additional terms are 
constrained by the symmetries of the lattice action and therefore 
depend on the specific choice of the discretization. 
The resulting effective theory can be used to analyse the $a$-dependence of 
the various quantities of interest -- provided the quark masses and the momenta
considered are in the range where the truncated chiral perturbation series yields 
an adequate approximation. Understanding the dependence on the lattice spacing 
is of central importance for a controlled extrapolation to the continuum limit.
 
For staggered fermions, this program has first been carried out for a
single staggered flavour (a single staggered field) \cite{Lee:1999zxa}
at $O(a^2)$. In the following, this effective theory is denoted by
S{\Ch}PT. It was later generalized to an arbitrary number of flavours
\cite{Aubin:2003mg,Aubin:2003uc}, and to next-to-leading order
\cite{Sharpe:2004is}. The corresponding theory is commonly called
Rooted Staggered chiral perturbation theory and is denoted by
RS{\Ch}PT.

For Wilson fermions, the effective theory has been developed in
\cite{Sharpe:1998xm,Rupak:2002sm,Aoki:2003yv}
and is called W{\Ch}PT, while the theory for Wilson twisted-mass
fermions \cite{Sharpe:2004ny,Aoki:2004ta,Bar:2010jk} is termed tmW{\Ch}PT.

Another important approach is to consider theories in which the
valence and sea quark masses are chosen to be different. These
theories are called {\it partially quenched}. The acronym for the
corresponding chiral effective theory is PQ{\Ch}PT
\cite{Bernard:1993sv,Golterman:1997st,Sharpe:1997by,Sharpe:2000bc}.

Finally, one can also consider theories where the fermion
discretizations used for the sea and the valence quarks are different. The
effective chiral theories for these ``mixed action'' theories are
referred to as MA{\Ch}PT  \cite{Bar:2002nr,Bar:2003mh,Bar:2005tu,Golterman:2005xa,Chen:2006wf,Chen:2007ug,Chen:2009su}.
 
\newpage
\hspace{-1.5cm}
\subsection{Summary of simulated lattice actions}
In the following tables we summarize the gauge and quark actions used
in the various calculations with $N_f=2, 2+1$ and $2+1+1$ quark
flavours. The calculations with $N_f=0$ quark flavours mentioned in section \ref{sec:alpha_s} all used the Wilson gauge action and are not listed. Abbreviations are explained in section
\ref{sec_gauge_actions}, \ref{sec_quark_actions} and
\ref{app:HQactions}, and summarized in tables \ref{tab_gaugeactions},
\ref{tab_quarkactions} and \ref{tab_heavy_quarkactions}.
\begin{table}[h]
{\footnotesize
\begin{tabular*}{\textwidth}[l]{l @{\extracolsep{\fill}} c c c c}
\hline \hline \\[-1.0ex]
Collab. & Ref. & $\Nf$ & \parbox{1cm}{gauge\\action} & \parbox{1cm}{quark\\action} 
\\[2.0ex] \hline \hline \\[-1.0ex]
ALPHA 01A, 04, 05, 12 & \cite{Bode:2001jv,DellaMorte:2004bc,DellaMorte:2005kg,Fritzsch:2012wq} & 2 & Wilson & npSW \\
[2.0ex] \hline \\[-1.0ex]
Aoki 94 & \cite{Aoki:1994pc} & 2 & Wilson &  KS \\
[2.0ex] \hline \\[-1.0ex]
Bernardoni 10 & \cite{Bernardoni:2010nf} & 2 & Wilson & npSW ${}^\dagger$ \\
[2.0ex] \hline \\[-1.0ex]
Bernardoni 11 & \cite{Bernardoni:2011kd} & 2 & Wilson & npSW \\
[2.0ex] \hline \\[-1.0ex]
Brandt 13 & \cite{Brandt:2013dua} & 2 & Wilson & npSW \\
[2.0ex] \hline \\[-1.0ex]
Boucaud 01B & \cite{Boucaud:2001qz} & 2 & Wilson &  Wilson \\
[2.0ex] \hline \\[-1.0ex]
CERN-TOV 06 & \cite{DelDebbio:2006cn} & 2 & Wilson & Wilson/npSW \\
[2.0ex] \hline \\[-1.0ex]
CERN 08 & \cite{Giusti:2008vb} & 2 & Wilson & npSW \\
[2.0ex] \hline \\[-1.0ex]
{CP-PACS 01} & \cite{AliKhan:2001tx} & 2 & Iwasaki & mfSW  \\
[2.0ex] \hline \\[-1.0ex]
Davies 94  & \cite{Davies:1994ei} & 2 & Wilson  & KS \\
[2.0ex] \hline \\[-1.0ex]
D\"urr 11 & \cite{Durr:2011ed} & 2 & Wilson & npSW \\
[2.0ex] \hline \\[-1.0ex]
\parbox[t]{4.0cm}{ETM 07, 07A, 08, 09, 09A-D, 10B, 10D, 10F, 11C, 12, 13} & \parbox[t]{2.5cm}{
\cite{Blossier:2007vv,Boucaud:2007uk,Frezzotti:2008dr,Blossier:2009bx,Lubicz:2009ht,Jansen:2009tt,Baron:2009wt,Blossier:2009hg,Blossier:2010cr,Lubicz:2010bv,Blossier:2010ky,Jansen:2011vv,Burger:2012ti,Cichy:2013gja}} & 2 &  tlSym & tmWil \\
[7.0ex] \hline \\[-1.0ex]
{ETM 10A} & \cite{Constantinou:2010qv} & 2 &  tlSym & tmWil ${}^*$ \\
[2.0ex] \hline \\[-1.0ex]
G\"ulpers 13 & \cite{Gulpers:2013uca} & 2 & Wilson & npSW \\
[2.0ex] \hline \\[-1.0ex]
Hasenfratz 08 & \cite{Hasenfratz:2008ce} & 2 & tadSym &
n-HYP tlSW\\
[2.0ex] 
\hline\hline \\
\end{tabular*}
${}^\dagger$ The calculation uses overlap fermions in the valence quark sector.\\
${}^*$ The calculation uses Osterwalder-Seiler fermions \cite{Osterwalder:1977pc} in the valence quark sector.
}
\caption{Summary of simulated lattice actions with $\Nf=2$ quark
  flavours.}
\end{table}

\begin{table}[h]
\addtocounter{table}{-1}
{\footnotesize
\begin{tabular*}{\textwidth}[l]{l @{\extracolsep{\fill}} c c c c}
\hline \hline \\[-1.0ex]
Collab. & Ref. & $\Nf$ & \parbox{1cm}{gauge\\action} & \parbox{1cm}{quark\\action} 
\\[2.0ex] \hline \hline \\[-1.0ex]
JLQCD 08 & \cite{Aoki:2008ss} & 2 & Iwasaki & overlap \\
[2.0ex] \hline \\[-1.0ex]
{JLQCD 02, 05} & \cite{Aoki:2002uc,Tsutsui:2005cj} & 2 & Wilson & npSW \\
[2.0ex] \hline \\[-1.0ex]
JLQCD/TWQCD 07, 08A, 10 & \cite{Fukaya:2007pn,Noaki:2008iy,Fukaya:2010na} & 2 & Iwasaki & overlap \\
[2.0ex] \hline \\[-1.0ex]
QCDSF 07 & \cite{Brommel:2007wn} & 2 & Wilson &  npSW \\
[2.0ex] \hline \\[-1.0ex]
QCDSF/UKQCD 04, 06, 06A, 07 & \cite{Gockeler:2004rp,Gockeler:2006jt,Brommel:2006ww,QCDSFUKQCD} & 2 & Wilson &  npSW \\
[2.0ex] \hline \\[-1.0ex]
{RBC 04, 06, 07} & \cite{Aoki:2004ht,Dawson:2006qc,Blum:2007cy} & 2 & DBW2 & DW \\
[2.0ex] \hline \\[-1.0ex]
UKQCD/UKQCD 07 & \cite{Boyle:2007qe} & 2 & Wilson & npSW \\ 
[2.0ex]\hline \\ [-1.0ex]
RM123 11, 13 & \cite{deDivitiis:2011eh,deDivitiis:2013xla}& 2 &  tlSym & tmWil  \\
[2.0ex] \hline \\[-1.0ex]
Sesam 99 & \cite{Spitz:1999tu} & 2 & Wilson &  Wilson \\
[2.0ex] \hline \\[-1.0ex]
Sternbeck 10, 12 & \cite{Sternbeck:2010xu,Sternbeck:2012qs}& 2 &  Wilson & npSW  \\
[2.0ex] \hline \\[-1.0ex]
{SPQcdR 05} & \cite{Becirevic:2005ta} & 2 & Wilson & Wilson \\
[2.0ex] \hline \\ [-1.0ex]
TWQCD 11, 11A  & \cite{Chiu:2011bm,Chiu:2011dz} & 2 & Wilson & optimal DW \\
[2.0ex] \hline \\ [-1.0ex]
UKQCD 04 & \cite{Flynn:2004au,Boyle:2007qe} & 2 & Wilson & npSW \\ 
[2.0ex]\hline \\ [-1.0ex]
Wingate 95 & \cite{Wingate:1995fd} & 2 & Wilson &  KS \\
[2.0ex] 
\hline\hline \\
\end{tabular*}
}
\caption{(cntd.) Summary of simulated lattice actions with $\Nf=2$ quark
  flavours.}
\end{table}
%
%
\begin{table}[h]
{\footnotesize
\begin{tabular*}{\textwidth}[l]{l @{\extracolsep{\fill}} c c c c}
\hline \hline \\[-1.0ex]
Collab. & Ref. & $\Nf$ & \parbox{1cm}{gauge\\action} & \parbox{1cm}{quark\\action} 
\\[2.0ex] \hline \hline \\[-1.0ex]
ALPHA 10A & \cite{Tekin:2010mm} & $4$ & Wilson &  npSW \\
[2.0ex] \hline \\[-1.0ex]
Aubin 08, 09 & \cite{Aubin:2008ie,Aubin:2009jh} & $2+1$ & tadSym & Asqtad ${}^\dagger$\\
[2.0ex] \hline \\[-1.0ex]
Bazavov 12 & \cite{Bazavov:2012ka} & $2+1+1$ & tlSym & HISQ \\
[2.0ex] \hline \\[-1.0ex] 
Blum 10 & \cite{Blum:2010ym} & $2+1$ & Iwasaki & DW\\
[2.0ex] \hline \\[-1.0ex]
%
%
BMW 10A-C, 11, 13 & \cite{Durr:2010vn,Durr:2010aw,Portelli:2010yn,Durr:2011ap,Durr:2013goa} & $2+1$ &  tlSym & 2-level HEX tlSW \\
[2.0ex] \hline \\[-1.0ex]
BMW 10 & \cite{Durr:2010hr} & $2+1$ & tlSym & 6-level stout tlSW\\
[2.0ex] \hline \\[-1.0ex]
CP-PACS/JLQCD 07&  \cite{Ishikawa:2007nn} & $2+1$ & Iwasaki & npSW \\
[2.0ex] \hline \\[-1.0ex]
\parbox[t]{4.0cm}{ETM 10, 10E, 11, 11D, 12C, 13, 13D} &  \parbox[t]{2.5cm}{\cite{Baron:2010bv,Farchioni:2010tb,Baron:2011sf,Blossier:2011tf,Blossier:2012ef,Cichy:2013gja,Blossier:2013ioa}} & $2+1+1$ & Iwasaki & tmWil \\
[2.0ex] \hline \\[-1.0ex]
FNAL/MILC 12  & \cite{Bazavov:2012zs} & $2+1$ & tadSym & Asqtad \\
[2.0ex] \hline \\[-1.0ex]
FNAL/MILC 12B, 13 & \cite{Bazavov:2012dg,Bazavov:2013nfa} & $2+1+1$ & tadSym & HISQ \\
[2.0ex] \hline \\[-1.0ex] 
HPQCD 05, 05A, 08A, 13A&  \cite{Mason:2005bj,Mason:2005zx,Davies:2008sw,Dowdall:2013rya}& $2+1$ & tadSym &  Asqtad \\
[2.0ex] \hline \\[-1.0ex]
HPQCD 10 & \cite{McNeile:2010ji} & $2+1$ & tadSym & Asqtad ${}^*$\\ 
[2.0ex] \hline \\[-1.0ex]
HPQCD/UKQCD 06 & \cite{Gamiz:2006sq} & $2+1$ & tadSym & Asqtad \\ 
[2.0ex] \hline \\[-1.0ex]
HPQCD/UKQCD 07 & \cite{Follana:2007uv} & $2+1$ & tadSym & Asqtad ${}^*$\\ 
[2.0ex] \hline \\[-1.0ex]
HPQCD/MILC/UKQCD 04&  \cite{Aubin:2004ck}& $2+1$ & tadSym &  Asqtad \\
[2.0ex] \hline \\[-1.0ex]
JLQCD 09, 10 & \cite{Fukaya:2009fh,Shintani:2010ph} & $2+1$ & Iwasaki & overlap \\
[2.0ex] \hline \\[-1.0ex]
JLQCD 11, 12 & \cite{Kaneko:2011rp,Kaneko:2012cta}          & $2+1$ & \parbox[t]{3.5cm}{Iwasaki (fixed topology)} & overlap\\
[2.0ex] \hline \\[-1.0ex]
JLQCD/TWQCD 08B, 09A & \cite{Chiu:2008kt,JLQCD:2009sk} & $2+1$ & Iwasaki & overlap \\
[2.0ex] \hline \\[-1.0ex]
JLQCD/TWQCD 10 & \cite{Fukaya:2010na} & $2+1, 3$ & Iwasaki & overlap \\
[2.0ex] \hline \hline\\
\end{tabular*}
${}^\dagger$ The calculation uses domain wall fermions in the valence quark sector.\\
${}^*$ The calculation uses HISQ staggered fermions in the valence quark sector.\\
}
\caption{Summary of simulated lattice actions with $\Nf=2+1$ or $\Nf=2+1+1$ quark flavours.}
\end{table}

\begin{table}[h]
\addtocounter{table}{-1}
{\footnotesize
\begin{tabular*}{\textwidth}[l]{l @{\extracolsep{\fill}} c c c c}
\hline \hline \\[-1.0ex]
Collab. & Ref. & $\Nf$ & \parbox{1cm}{gauge\\action} & \parbox{1cm}{quark\\action} 
\\[2.0ex] \hline \hline \\[-1.0ex]
Laiho 11 & \cite{Laiho:2011np} & $2+1$ & tadSym & Asqtad ${}^\dagger$\\
[2.0ex] \hline \\[-1.0ex]
LHP 04 & \cite{Bonnet:2004fr} & $2+1$ & tadSym & Asqtad $^\dagger$ \\
[2.0ex] \hline \\[-1.0ex]
Maltman 08 & \cite{Maltman:2008bx} & $2+1$ & tadSym &  Asqtad \\
[2.0ex] \hline \\[-1.0ex]
MILC 04, 07, 09, 09A, 10, 10A & \parbox[t]{2.5cm}{\cite{Aubin:2004ck,Aubin:2004fs,Bernard:2007ps,Bazavov:2009bb,Bazavov:2010hj,Bazavov:2010yq}}& $2+1$ &
tadSym & Asqtad \\
[2.0ex] \hline \\[-1.0ex]
NPLQCD 06& \cite{Beane:2006kx}& $2+1$ & tadSym & Asqtad $^\dagger$ \\
[2.0ex] \hline \\[-1.0ex]
{PACS-CS 08, 08A, 09, 09A, 10, 12} & \cite{Aoki:2008sm,Kuramashi:2008tb,Aoki:2009ix,Aoki:2009tf,Aoki:2010wm} & $2+1$ & Iwasaki & npSW \\
[2.0ex] \hline \\[-1.0ex]
Perez 10 & \cite{PerezRubio:2010ke} & $4$ & Wilson &  npSW \\
[2.0ex] \hline \\[-1.0ex]
\parbox[t]{4.0cm}{RBC/UKQCD 07, 08, 08A, 10, 10A-B, 11, 12, 13}& \parbox[t]{2.5cm}{\cite{Antonio:2007pb,Allton:2008pn,Boyle:2008yd,Boyle:2010bh,Aoki:2010dy,Aoki:2010pe,Kelly:2012uy,Arthur:2012opa,Boyle:2013gsa}} & $2+1$ & \parbox{1.5cm}{Iwasaki, Iwasaki+DSDR} &  DW \\
[2.0ex] \hline \\[-1.0ex]
Sternbeck 12 & \cite{Sternbeck:2012qs} & $2+1$ & tlSym & npSW \\
[2.0ex] \hline \\[-1.0ex]
SWME 10, 11, 11A, 13 & \cite{Bae:2010ki,Kim:2011qg,Bae:2011ff,Bae:2013lja} & $2+1$ & tadSym & Asqtad ${}^+$ \\
[2.0ex] \hline \\[-1.0ex]
TWQCD 08 & \cite{Chiu:2008jq}  & $2+1$ & Iwasaki &  DW \\
[2.0ex] \hline\hline\\
\end{tabular*}
${}^\dagger$ The calculation uses domain wall fermions in the valence quark sector.\\
${}^+$ The calculation uses HYP smeared improved staggered fermions in the valence quark sector.
}
\caption{(cntd.) Summary of simulated lattice actions with $\Nf=2+1$ or $\Nf=2+1+1$ quark flavours.}
\end{table}

\begin{table}[!ht]
{\footnotesize
\begin{tabular*}{\textwidth}[l]{l @{\extracolsep{\fill}} c c c c c c}
\hline\hline \\[-1.0ex]
Collab. & Ref. & $\Nf$ & Gauge & \multicolumn{3}{c}{Quark actions}  \\
& & & action & sea & light valence & heavy \\[1.0ex] \hline \hline \\[-1.0ex]
ALPHA 11, 12A, 13 & \cite{Blossier:2011dk,Bernardoni:2012ti,Bernardoni:2013oda} & 2 &  plaquette & npSW  & npSW & HQET
\\[2.0ex] \hline \\[-1.0ex]
\parbox[t]{2.5cm}{Atoui 13} & \parbox[t]{2.5cm}{\cite{Atoui:2013zza}} & 2 &  tlSym & tmWil & tmWil & tmWil
\\[2.0ex] \hline \\[-1.0ex]
\parbox[t]{2.5cm}{ETM 09, 09D, 11B, 12A, 12B, 13B, 13C} & \parbox[t]{2.5cm}{\cite{Blossier:2009bx,Blossier:2009hg,DiVita:2011py,Carrasco:2012dd,Carrasco:2012de,Carrasco:2013zta,Carrasco:2013iba}} & 2 &  tlSym & tmWil & tmWil & tmWil
\\[2.0ex] \hline \\[-1.0ex]
ETM  11A & \cite{Dimopoulos:2011gx} & 2 &  tlSym & tmWil & tmWil & tmWil, static
\\[2.0ex] \hline \\[-1.0ex]
ETM 13E, 13F   & \cite{Carrasco:2013naa,Dimopoulos:2013qfa} & 2+1+1 &  Iwasaki & tmWil & tmWil & tmWil
\\[2.0ex] \hline \\[-1.0ex]
\parbox[t]{2.5cm}{FNAL/MILC 04, 04A, 05, 08, 08A, 10, 11, 11A, 12, 13B} &  \parbox[t]{2.5cm}{\cite{Aubin:2004ej,Okamoto:2004xg,Aubin:2005ar,Bernard:2008dn,Bailey:2008wp,Bailey:2010gb,Bazavov:2011aa,Bouchard:2011xj,Bazavov:2012zs,Qiu:2013ofa}}  & 2+1 & tadSym & Asqtad & Asqtad & Fermilab
\\[7.0ex] \hline \\[-1.0ex]
FNAL/MILC 12B, 13 & \cite{Bazavov:2012dg,Bazavov:2013nfa} & 2+1+1 & tadSym & HISQ & HISQ & HISQ
\\[2.0ex] \hline \\[-1.0ex]
HPQCD 06, 06A, 08B, 09& \cite{Dalgic:2006dt,Dalgic:2006gp,Allison:2008xk,Gamiz:2009ku} & 2+1 &  tadSym & Asqtad & Asqtad & NRQCD
\\[2.0ex] \hline \\[-1.0ex]
HPQCD 12 & \cite{Na:2012sp} & 2+1 &  tadSym & Asqtad & HISQ & NRQCD
\\[2.0ex] \hline \\[-1.0ex]
\parbox[t]{2.5cm}{HPQCD/UKQCD 07, HPQCD 10A,  10B, 11, 11A, 12A, 13C}  & \parbox[t]{2.5cm}{\cite{Follana:2007uv,Davies:2010ip,Na:2010uf,Na:2011mc,McNeile:2011ng,Na:2012iu,Koponen:2013tua}} & 2+1 &  tadSym & Asqtad & HISQ & HISQ
\\[10.0ex] \hline \\[-1.0ex]
\parbox[t]{2.5cm}{HPQCD 13} & \cite{Dowdall:2013tga} &  2+1+1 & tadSym & HISQ & HISQ & NRQCD
\\[7.0ex] \hline \\[-1.0ex]  
RBC/UKQCD 10C & \cite{Albertus:2010nm} & 2+1 & Iwasaki & DWF & DWF & static
\\[2.0ex] \hline \\[-1.0ex]
RBC/UKQCD 13A & \cite{Witzel:2013sla} & 2+1 & Iwasaki & DWF & DWF & RHQ
\\[2.0ex] \hline \\[-1.0ex]
PACS-CS 11 & \cite{Namekawa:2011wt} & 2+1 & Iwasaki & npSW & npSW & Tsukuba
\\[2.0ex] \hline\hline\\

\end{tabular*}
\caption{Summary of lattice simulations with $b$ and $c$ valence quarks.  
}
}
\end{table}

\clearpage
\section{Notes}
\subsection{Notes to section \ref{sec:qmass} on quark masses}

\begin{table}[!ht]
{\footnotesize
\begin{tabular*}{\textwidth}[l]{l @{\extracolsep{\fill}} c c c l}
\hline\hline \\[-1.0ex]
Collab. & Ref. & $\Nf$ & $a$ [fm] & Description
\\[1.0ex] \hline \hline \\[-1.0ex]
{RBC/UKQCD 12}& \cite{Arthur:2012opa} & 2+1 & 0.144, 0.113, 0.085  & \parbox[t]{4.5cm}{Scale set through $M_\Omega$. Coarsest lattice uses Iwasaki+DSDR gauge action.}\\
[5.0ex] \hline \\[-1.0ex]
PACS-CS 12 & \cite{Aoki:2012st} & 1+1+1 & 0.09
& \parbox[t]{4.5cm}{Reweighting of PACS-CS 08 $N_f=2+1$ QCD
  configurations with e.m. and $m_u\ne m_d$.} \\
[3.0ex] \hline \\[-1.0ex]
Laiho 11 & \cite{Laiho:2011np} & 2+1 & 0.15,0.09,0.06
& \parbox[t]{4.5cm}{MILC staggered ensembles~\cite{Bazavov:2010yq}, scale set using
  $r_1$ determined by HPQCD with $\Upsilon$ splittings, pseudoscalar
  decay constants, through $r_1$~\cite{Davies:2009tsa}.} \\ [1.0ex] \hline
\\[-1.0ex]
PACS-CS 10 & \cite{Aoki:2010wm} &  2+1 &  0.09
& \parbox[t]{4.5cm}{cf.~PACS-CS 08} \\
[1.0ex] \hline \\[-1.0ex]
{MILC 10A}& \cite{Bazavov:2010yq} & 2+1 & & cf.~MILC 09, 09A \\
[1.0ex] \hline \\[-1.0ex]
{BMW 10A, 10B}& \cite{Durr:2010vn,Durr:2010aw}
& 2+1 & \parbox[t]{2.3cm}{0.116,0.093,0.077, 0.065,0.054} &  Scale setting via
$M_\pi, M_K, M_\Omega$.\\
[4.0ex] \hline \\[-1.0ex]
{RBC/UKQCD 10A}& \cite{Aoki:2010dy} & 2+1 & 0.114, 0.087  & Scale set through $M_\Omega$.
\\
[1.0ex] \hline \\[-1.0ex]
{Blum 10}& \cite{Blum:2010ym}
& 2+1 & 0.11 &  \parbox[t]{4.5cm}{Relies on RBC/UKQCD 08 scale setting.} \\
[3.0ex] \hline \\[-1.0ex]
{PACS-CS 09}& \cite{Aoki:2009ix}& 2+1 & 0.09 & Scale setting via
$M_\Omega$.\\
[1.0ex] \hline \\[-1.0ex]
{HPQCD 09A, 10}&  \cite{Davies:2009ih,McNeile:2010ji}& 2+1 &  & \\
[1.0ex] \hline \\[-1.0ex]
{MILC 09A, 09} & \cite{Bazavov:2009fk,Bazavov:2009bb} & 2+1 & 0.045, 0.06, 0.09 & 
		\parbox[t]{4.5cm}{Scale set through $r_1$ and $\Upsilon$ and 
		continuum extrapolation based on RS{\Ch}PT.} \\ 
[6.0ex] \hline \\[-1.0ex]
PACS-CS 08 &\cite{Aoki:2008sm}& 2+1 &  0.09 & \parbox[t]{4.5cm}{Scale set 
		through $M_\Omega$. Non-perturbatively $O(a)$-improved.} \\
[4.0ex] \hline \\[-1.0ex]
RBC/UKQCD 08	&\cite{Allton:2008pn}	&2+1& 0.11 & \parbox[t]{4.5cm}{Scale
                 set through $M_\Omega$. Automatic  
		$O(a)$-improvement due to appoximate chiral symmetry. 
		$(\Lambda_{\rm QCD}a)^2\approx 4\%$ systematic error
		due to lattice artifacts added.} \\ 
[14.0ex] \hline \\[-1.0ex]
{CP-PACS/JLQCD 07}&  \cite{Ishikawa:2007nn} & 2+1 & 0.07,0.10,0.12 &  \parbox[t]{4.5cm}{Scale set 
		through $M_K$ or $M_\phi$. Non-perturbatively
                $O(a)$-improved.} \\
[6.0ex] \hline \\[-1.0ex]
{HPQCD 05}& \cite{Mason:2005bj}& 2+1 & 0.09,0.12 & \parbox[t]{4.5cm}{Scale set through the $\Upsilon-\Upsilon'$
mass difference.} \\
[4.0ex] \hline \\[-1.0ex]
\parbox [t]{4.5cm}{HPQCD/MILC/UKQCD 04,\\
 MILC 04}
& \cite{Aubin:2004ck,Aubin:2004fs}& 2+1 & 0.09,0.12 & \parbox[t]{4.5cm}{Scale
  set through $r_1$ and $\Upsilon$ and continuum extrapolation based on RS{\Ch}PT.} \\
[6.0ex] \hline \hline \\[-1.0ex]
\end{tabular*}
\caption{Continuum extrapolations/estimation of lattice artifacts in
  determinations of $m_{ud}$, $m_s$ and, in some cases $m_u$ and
  $m_d$, with $N_f=2+1$ quark flavours.}
}
\end{table}

\begin{table}[!ht]
{\footnotesize
\begin{tabular*}{\textwidth}[l]{l @{\extracolsep{\fill}} c c c l}
\hline\hline \\[-1.0ex]
Collab. & Ref. & $\Nf$ & $a$ [fm] & Description
\\[1.0ex] \hline \hline \\[-1.0ex]
RM123 13 & \cite{deDivitiis:2013xla} & 2 & \parbox[t]{2.3cm}{0.098,
  0.085,\\ 0.067, 0.054} & cf. ETM 10B\\
[4.0ex] \hline \\[-1.0ex]
ALPHA 12 & \cite{Fritzsch:2012wq} & 2 & \parbox[t]{2.3cm}{0.076, 0.066,\\ 0.049} & Scale set through $F_K$.\\
[4.0ex] \hline \\[-1.0ex]
RM123 11 & \cite{deDivitiis:2011eh} & 2 & \parbox[t]{2.3cm}{0.098,
  0.085,\\ 0.067, 0.054} & cf. ETM 10B\\
[4.0ex] \hline \\[-1.0ex]
D\"urr 11 & \cite{Durr:2011ed} & 2 & \parbox[t]{2.3cm}{0.076,
  0.072,\\ 0.060} & \parbox[t]{3cm}{Scale for light quark masses set through $m_c$.\\}\\
[4.0ex] \hline \\[-1.0ex]
ETM 10B & \cite{Blossier:2010cr} & 2 & \parbox[t]{2.3cm}{0.098, 0.085,\\ 0.067, 0.054} & Scale set through $F_\pi$.\\
[4.0ex] \hline \\[-1.0ex]
{JLQCD/TWQCD 08A} & \cite{Noaki:2008iy} & 2 & 0.12 & Scale set through $r_0$. \\
[1.0ex] \hline \\[-1.0ex]
{RBC 07} & \cite{Blum:2007cy} & 2 & 0.12 & Scale set through $M_\rho$.\\
[1.0ex] \hline \\[-1.0ex]
{ETM 07} & \cite{Blossier:2007vv} & 2 & 0.09 & Scale set through $F_\pi$. \\
[1.0ex] \hline \\[-1.0ex]
{QCDSF/UKQCD 06} & \cite{Gockeler:2006jt} & 2 & 0.065--0.09 & Scale set through $r_0$. \\
[2.0ex] \hline \\[-1.0ex]
{SPQcdR 05} & \cite{Becirevic:2005ta}
& 2 & 0.06,0.08 & Scale set through $M_{K^*}$. \\
[1.0ex] \hline \\[-1.0ex]
{ALPHA 05} & \cite{DellaMorte:2005kg}
& 2 & 0.07-0.12 & Scale set through $r_0$. \\
[1.0ex] \hline \\[-1.0ex]
{QCDSF/UKQCD 04} & \cite{Gockeler:2004rp}
& 2 &  0.07-0.12 & Scale set through $r_0$. \\
[1.0ex] \hline \\[-1.0ex]
{JLQCD 02} & \cite{Aoki:2002uc}
& 2 & 0.09 & Scale set through $M_\rho$. \\
[1.0ex] \hline \\[-1.0ex]
{CP-PACS 01} & \cite{AliKhan:2001tx}
& 2 & 0.11,0.16,0.22 & Scale set through $M_\rho$. \\
[1.0ex] \hline\hline
\end{tabular*}
\caption{Continuum extrapolations/estimation of lattice artifacts in
  determinations of $m_{ud}$, $m_s$ and, in some cases $m_u$ and
  $m_d$, with $N_f=2$ quark flavours.}
}
\end{table}

\newpage

\begin{table}[!ht]
{\footnotesize
\begin{tabular*}{\textwidth}[l]{l @{\extracolsep{\fill}} c c c l}
\hline \hline  \\[-1.0ex]
Collab. & Ref. & $\Nf$ & $M_{\pi,\text{min}}$ [MeV] & Description
\\[1.0ex] \hline \hline \\[-1.0ex]
{RBC/UKQCD 12}& \cite{Arthur:2012opa} & 2+1 & 170 & \parbox[t]{5.5cm}{Combined fit to Iwasaki and Iwasaki+DSDR gauge action ensembles.} \\
[7.0ex] \hline \\[-1.0ex]
PACS-CS 12 & \cite{Aoki:2012st} &  1+1+1 &  & cf.~PACS-CS 08 \\
[1.0ex] \hline \\[-1.0ex]
Laiho 11 & \cite{Laiho:2011np} & 2+1 & \parbox[t]{2cm}{210 (val.)\\ 280 (sea-RMS)} &
\parbox[t]{5.5cm}{NLO SU(3), mixed-action $\chi$PT \cite{Bar:2005tu}, with N$^2$LO-N$^4$LO
  analytic terms.} \\ [4.0ex] \hline \\[-1.0ex]

PACS-CS 10 & \cite{Aoki:2010wm} &  2+1 & & cf.~PACS-CS 08 \\
[1.0ex] \hline \\[-1.0ex]
{MILC 10A}& \cite{Bazavov:2010yq} & 2+1 & & \parbox[t]{5.5cm}{NLO SU(2)
S{\Ch}PT. Cf.~also MILC 09A,09.} \\
[3.0ex] \hline \\[-1.0ex]
{BMW 10A, 10B}& \cite{Durr:2010vn,Durr:2010aw}
& 2+1 & 135 & \parbox[t]{5.5cm}{Interpolation to the physical point.} \\
[1.0ex] \hline \\[-1.0ex]
{RBC/UKQCD 10A}& \cite{Aoki:2010dy} & 2+1 & 290 & \\
[1.0ex] \hline \\[-1.0ex]
{Blum 10}& \cite{Blum:2010ym,Allton:2008pn}
& 2+1 & \parbox[t]{2cm}{242 (valence), 330 (sea)} &
\parbox[t]{5.5cm}{Extrapolation done on the basis of PQ$\chi$PT formulae with
virtual photons.} \\
[4.0ex] \hline \\[-1.0ex]
{PACS-CS 09}& \cite{Aoki:2009ix}& 2+1 & 135 & \parbox[t]{5.5cm}{Physical point reached by
reweighting technique, no chiral extrapolation needed.}\\
[6.0ex] \hline \\[-1.0ex]
{HPQCD 09A, 10}&  \cite{Davies:2009ih,McNeile:2010ji}& 2+1 &  & \\
[1.0ex] \hline \\[-1.0ex]
{MILC 09A, 09} & \cite{Bazavov:2009fk,Bazavov:2009bb} & 2+1 & 177, 240 & \parbox[t]{5.5cm}{
		NLO SU(3) RS{\Ch}PT, continuum {\Ch}PT at NNLO and NNNLO and NNNNLO 
		analytic terms. The lightest Nambu-Goldstone mass
                is 177 MeV (09A) and 224 MeV (09)
		(at $a=$0.09fm) and the lightest RMS mass is 258MeV (at
                $a=$0.06fm).} \\ 
[17.0ex] \hline \\[-1.0ex]
PACS-CS 08 &\cite{Aoki:2008sm}& 2+1 &156& 
		\parbox[t]{5.5cm}{NLO SU(2) {\Ch}PT and SU(3)
                  (Wilson){\Ch}PT.}\\
[3.0ex] \hline \\[-1.0ex]
RBC/UKQCD 08	&\cite{Allton:2008pn}	&2+1& \parbox[t]{2cm}{242 (valence), 330 (sea)} & \parbox[t]{5.5cm}{SU(3) PQ{\Ch}PT 
and heavy kaon NLO SU(2) PQ{\Ch}PT fits.} \\
[4.0ex] \hline \\[-1.0ex]
{CP-PACS/JLQCD 07}&  \cite{Ishikawa:2007nn} & 2+1 & 620 & \parbox[t]{5.5cm}{NLO Wilson {\Ch}PT
fits to meson masses.} \\
[1.0ex] \hline \\[-1.0ex]
{HPQCD 05}& \cite{Mason:2005bj}& 2+1 & 240 & \parbox[t]{5.5cm}{PQ RS{\Ch}PT fits.} \\
[1.0ex] \hline \\[-1.0ex]
\parbox[t]{4.5cm}{HPQCD/MILC/UKQCD 04,\\ MILC 04}
& \cite{Aubin:2004ck,Aubin:2004fs}& 2+1 & 240 & \parbox[t]{5.5cm}{PQ RS{\Ch}PT fits.} \\
[4.0ex] \hline \hline \\[-1.0ex]
\end{tabular*}
\caption{Chiral extrapolation/minimum pion mass in determinations of
  $m_{ud}$, $m_s$ and, in some cases $m_u$ and
  $m_d$,  with $N_f=2+1$ quark flavours.}
}
\end{table}

\begin{table}[!ht]
{\footnotesize
\begin{tabular*}{\textwidth}[l]{l @{\extracolsep{\fill}} c c c l}
\hline \hline  \\[-1.0ex]
Collab. & Ref. & $\Nf$ & $M_{\pi,\text{min}}$ [MeV] & Description
\\[1.0ex] \hline \hline \\[-1.0ex]
RM123 13 & \cite{deDivitiis:2013xla} & 2 & 270 & \parbox[t]{6cm}{Fits based on NLO {\Ch}PT and Symanzik expansion up to $O(a^2)$. $O(\alpha)$ e.m. effects included.}\\
[7.0ex] \hline \\[-1.0ex]
ALPHA 12 & \cite{Fritzsch:2012wq} & 2 & 270 & \parbox[t]{6cm}{NLO
  SU(2) and SU(3) {\Ch}PT and $O(a^2)$ on LO LEC.}\\
[4.0ex] \hline \\[-1.0ex]
RM123 11 & \cite{deDivitiis:2011eh} & 2 & 270 & \parbox[t]{6cm}{Fits based on NLO {\Ch}PT and Symanzik expansion up to $O(a^2)$.}\\
[4.0ex] \hline \\[-1.0ex]
D\"urr 11 & \cite{Durr:2011ed} & 2 &  285 & \parbox[t]{6cm}{$m_c/m_s$
  determined by quadratic or cubic extrapolation in $M_\pi$.}\\
[4.0ex] \hline \\[-1.0ex]
ETM 10B & \cite{Blossier:2010cr} & 2 & 270 & \parbox[t]{6cm}{Fits based on NLO {\Ch}PT and Symanzik expansion up to $O(a^2)$.}\\
[4.0ex] \hline \\[-1.0ex]
{JLQCD/TWQCD 08A} & \cite{Noaki:2008iy} & 2 & 290 & NLO {\Ch}PT fits.\\
[1.0ex] \hline \\[-1.0ex]
{RBC 07} & \cite{Blum:2007cy}
& 2 & 440 & \parbox[t]{6cm}{NLO fit including $O(\alpha)$ effects.} \\
[1.0ex] \hline \\[-1.0ex]
{ETM 07} & \cite{Blossier:2007vv}
& 2 & 300 & Polynomial and PQ{\Ch}PT fits. \\
[1.0ex] \hline \\[-1.0ex]
{QCDSF/UKQCD 06} & \cite{Gockeler:2006jt} & 2 & \parbox[t]{2cm}{520 (valence), 620 (sea)} & NLO (PQ){\Ch}PT
fits. \\
[4.0ex] \hline \\[-1.0ex]
{SPQcdR 05} & \cite{Becirevic:2005ta}
& 2 & ~600 & Polynomial fit. \\
[1.0ex] \hline \\[-1.0ex]
{ALPHA 05} & \cite{DellaMorte:2005kg}
& 2 & ~560 & LO {\Ch}PT fit. \\
[1.0ex] \hline \\[-1.0ex]
{QCDSF/UKQCD 04} & \cite{Gockeler:2004rp}
& 2 &  \parbox[t]{2cm}{520 (valence), 620 (sea)} & NLO (PQ){\Ch}PT fits. \\
[4.0ex] \hline \\[-1.0ex]
{JLQCD 02} & \cite{Aoki:2002uc}
& 2 & 560 & Polynomial and {\Ch}PT fits. \\
[1.0ex] \hline \\[-1.0ex]
{CP-PACS 01} & \cite{AliKhan:2001tx}
& 2 & 430 & \parbox{6cm}{Polynomial fits.} \\
[1.0ex]
\hline\hline
\end{tabular*}
\caption{Chiral extrapolation/minimum pion mass in determinations of
  $m_{ud}$, $m_s$ and, in some cases $m_u$ and
  $m_d$,  with $N_f=2$ quark flavours.}
}
\end{table}

\newpage

\begin{table}
{\footnotesize
\begin{tabular*}{\textwidth}[l]{l @{\extracolsep{\fill}} c c c c l}
\hline\hline  \\[-1.0ex]
Collab. & Ref. & $\Nf$ & $L$ [fm] & $M_{\pi,\text{min}} L$ & Description
\\[1.0ex] \hline \hline \\[-1.0ex]
{RBC/UKQCD 12}& \cite{Arthur:2012opa} & 2+1 & 2.7, 4.6 &  $\gtrsim 4.0$
& \parbox[t]{3.7cm}{Uses FV chiral perturbation theory to estimate the error.} \\
[6.0ex] \hline \\[-1.0ex]

PACS-CS 12 & \cite{Aoki:2012st} &  1+1+1 & & & cf.~PACS-CS 08 \\
[1.0ex] \hline \\[-1.0ex]

Laiho 11 & \cite{Laiho:2011np} & 2+1 &
\parbox[t]{2cm}{2.5, 2.9, 3.0,\\ 3.6, 3.8, 4.8}&\parbox[t]{0.5cm}{4.1 (val.)\\
4.1 (sea)}&  \parbox[t]{3.7cm}{Data corrected using NLO SU(3) $\chi$PT finite-V formulae.} \\
[4.0ex] \hline \\[-1.0ex]

PACS-CS 10 & \cite{Aoki:2010wm} &  2+1 & & & cf.~PACS-CS 08 \\
[1.0ex] \hline \\[-1.0ex]
{MILC 10A}& \cite{Bazavov:2010yq} & 2+1 & & & cf.~MILC 09A,09 \\
[1.0ex] \hline \\[-1.0ex]
{BMW 10A, 10B}& \cite{Durr:2010vn,Durr:2010aw}
& 2+1 &  $\gtrsim 5.0$ & $\gtrsim 4.0$ & \parbox[t]{3.7cm}{FS corrections
  below 5 per mil on the largest lattices.}\\
[4.0ex] \hline \\[-1.0ex]
{RBC/UKQCD 10A}& \cite{Aoki:2010dy} & 2+1 & 2.7 &  $\gtrsim 4.0$ & \\
[4.0ex] \hline \\[-1.0ex]
{Blum 10}& \cite{Blum:2010ym}
& 2+1 & 1.8, 2.7 & --- & \parbox[t]{3.7cm}{Simulations done with quenched photons; large
finite volume effects analytically corrected for, but not related to
$M_\pi L$.} \\
[12.0ex] \hline \\[-1.0ex]
{PACS-CS 09}& \cite{Aoki:2009ix}& 2+1 & 2.9 & 2.0 & \parbox[t]{3.7cm}{Only one volume.} \\
[1.0ex] \hline \\[-1.0ex]
{HPQCD 09A, 10}&  \cite{Davies:2009ih,McNeile:2010ji}& 2+1 &  &  &  \\
[1.0ex] \hline \\[-1.0ex]
{MILC 09A, 09} & \cite{Bazavov:2009fk,Bazavov:2009bb} & 2+1 &
\parbox[t]{2cm}{2.5, 2.9, 3.4,\\ 3.6, 3.8, 5.8}&4.1, 3.8&  \parbox[t]{3.7cm}{} \\
[4.0ex] \hline \\[-1.0ex]
PACS-CS 08 &\cite{Aoki:2008sm}& 2+1 &2.9&2.3& 
		\parbox[t]{3.7cm}{Correction for FSE from {\Ch}PT using 
		\cite{Colangelo:2005gd}.} \\
[4.0ex] \hline \\[-1.0ex]
RBC/UKQCD 08	&\cite{Allton:2008pn}	&2+1& 1.8, 2.7&4.6&
		\parbox[t]{3.7cm}{Various volumes for comparison and 
		correction for FSE from {\Ch}PT 
		\cite{Gasser:1986vb,Gasser:1987ah,Colangelo:2005gd}.}\\
[6.0ex] \hline \\[-1.0ex]
{CP-PACS/JLQCD 07}&  \cite{Ishikawa:2007nn} & 2+1 & 2.0 & 6.0&
\parbox[t]{3.7cm}{Estimate based on the comparison to a $L=1.6$ fm volume
  assuming powerlike dependence on $L$.} \\
[9.0ex] \hline \\[-1.0ex]
{HPQCD 05}& \cite{Mason:2005bj}& 2+1 & 2.4, 2.9 & 3.5 &  \\
[1.0ex] \hline \\[-1.0ex]
\parbox[t]{4.5cm}{HPQCD/MILC/UKQCD 04,\\ MILC 04}
& \cite{Aubin:2004ck,Aubin:2004fs}& 2+1 & 2.4, 2.9 & 3.5 & NLO S{\Ch}PT. \\
[4.0ex] \hline \hline \\[-1.0ex]
\end{tabular*}
\caption{Finite volume effects in determinations of $m_{ud}$, $m_s$
  and, in some cases $m_u$ and $m_d$, with $N_f=2+1$ quark flavours.}
}
\end{table}

\begin{table}
{\footnotesize
\begin{tabular*}{\textwidth}[l]{l @{\extracolsep{\fill}} c c c c l}
\hline\hline  \\[-1.0ex]
Collab. & Ref. & $\Nf$ & $L$ [fm] & $M_{\pi,\text{min}} L$ & Description
\\[1.0ex] \hline \hline \\[-1.0ex]
RM123 13 & \cite{deDivitiis:2013xla} & 2 & $\gtrsim 2.0$ & 3.5 & \parbox[t]{4.2cm}{One volume
$L=1.7$ fm at $m_\pi=495, a=0.054$fm.} \\
[4.0ex] \hline \\[-1.0ex]
ALPHA 12 & \cite{Fritzsch:2012wq} & 2 & 2.1--3.2 & 4.2
& \parbox[t]{4.2cm}{Roughly 2 distinct volumes; no analysis of FV effects.} \\
[4.0ex] \hline \\[-1.0ex]
RM123 11 & \cite{deDivitiis:2011eh} & 2 & $\gtrsim 2.0$ & 3.5 & \parbox[t]{4.2cm}{One volume
$L=1.7$ fm at $m_\pi=495, a=0.054$fm.} \\
[4.0ex] \hline \\[-1.0ex]
D\"urr 11 & \cite{Durr:2011ed} & 2 & 1.22-2.30 & 2.8
& \parbox[t]{4.2cm}{A number of volumes in determination of $m_c/m_s$, but
  all but one have $L<2\,\fm$.}\\
[6.0ex] \hline \\[-1.0ex]
ETM 10B & \cite{Blossier:2010cr} & 2 & $\gtrsim 2.0$ & 3.5 & \parbox[t]{4.2cm}{One volume
$L=1.7$ fm at $m_\pi=495, a=0.054$fm.} \\
[4.0ex] \hline \\[-1.0ex]
{JLQCD/TWQCD 08A} & \cite{Noaki:2008iy} & 2 & 1.9 & 2.8 &
\parbox[t]{4.2cm}{Corrections for FSE based on NLO {\Ch}PT.}\\
[4.0ex] \hline \\[-1.0ex]
{RBC 07} & \cite{Blum:2007cy}
& 2 & 1.9 & 4.3 &\parbox[t]{4.2cm}{Estimate of FSE based on a model.}  \\
[4.0ex] \hline \\[-1.0ex]
{ETM 07} & \cite{Blossier:2007vv}
& 2 &  2.1 & 3.2 & NLO PQ{\Ch}PT \\
[1.0ex] \hline \\[-1.0ex]
{QCDSF/UKQCD 06} & \cite{Gockeler:2006jt} & 2 & 1.4--1.9 & 4.7 & \\
[1.0ex] \hline \\[-1.0ex]
{SPQcdR 05} & \cite{Becirevic:2005ta}
& 2 & 1.0--1.5 & 4.3 & \parbox[t]{4.2cm}{Comparison between 1.0 and 1.5 fm.}\\
[4.0ex] \hline \\[-1.0ex]
{ALPHA 05} & \cite{DellaMorte:2005kg}
& 2 & 2.6 & 7.4 & \\
[1.0ex] \hline \\[-1.0ex]
{QCDSF/UKQCD 04} & \cite{Gockeler:2004rp}
& 2 &  1.7--2.0 & 4.7 & \\
[1.0ex] \hline \\[-1.0ex]
{JLQCD 02} & \cite{Aoki:2002uc}
& 2 & 1.8 & 5.1 & \parbox[t]{4.2cm}{Numerical study with three volumes.} \\
[4.0ex] \hline \\[-1.0ex]
{CP-PACS 01} & \cite{AliKhan:2001tx}
& 2 & 2.0--2.6 & 5.7 &  \\
[1.0ex]
\hline\hline
\end{tabular*}
\caption{Finite volume effects in determinations of $m_{ud}$, $m_s$
  and, in some cases $m_u$ and $m_d$, 
   with $N_f=2$ quark flavours.}
}
\end{table}

\newpage

\begin{table}[!ht]
{\footnotesize
\begin{tabular*}{\textwidth}[l]{l @{\extracolsep{\fill}} c c l}
\hline\hline \\[-1.0ex]
Collab. & Ref. & $\Nf$  & Description
\\[1.0ex] \hline \hline \\[-1.0ex]
{RBC/UKQCD 12}& \cite{Arthur:2012opa} & 2+1
& \parbox[t]{7cm}{Non-perturbative renormalization (RI/SMOM).}\\
[4.0ex] \hline \\[-1.0ex]

PACS-CS 12 & \cite{Aoki:2012st} &  1+1+1 & \parbox[t]{7cm}{cf.~PACS-CS 10} \\
[4.0ex] \hline \\[-1.0ex]

Laiho 11 & \cite{Laiho:2011np} & 2+1 &  \parbox[t]{7cm}{$Z_A$ from AWI
  and $Z_A/Z_S-1$ from 1-loop, tadpole-improved, perturbation theory.}\\
[4.0ex] \hline \\[-1.0ex]

PACS-CS 10 & \cite{Aoki:2010wm} &  2+1
& \parbox[t]{7cm}{Non-perturbative renormalization and running; Schr\"odinger functional method.} \\
[4.0ex] \hline \\[-1.0ex]
{MILC 10A}& \cite{Bazavov:2010yq} & 2+1 & cf.~MILC 09A,09\\
[1.0ex] \hline \\[-1.0ex]
{BMW 10A, 10B}& \cite{Durr:2010vn,Durr:2010aw}
& 2+1 & \parbox[t]{7cm}{Non-perturbative renormalization (tree-level
  improved RI-MOM), non-perturbative running.} \\
[4.0ex] \hline \\[-1.0ex]
{RBC/UKQCD 10A}& \cite{Aoki:2010dy} & 2+1
& \parbox[t]{7cm}{Non-perturbative renormalization (RI/SMOM).}\\
[1.0ex] \hline \\[-1.0ex]
{Blum 10}& \cite{Blum:2010ym}
& 2+1 & \parbox[t]{7cm}{Relies on non-perturbative renormalization factors calculated
by RBC/UKQCD 08; no QED renormalization.} \\
[6.0ex] \hline \\[-1.0ex]
{PACS-CS 09}& \cite{Aoki:2009ix}& 2+1 & \parbox[t]{7cm}{Non-perturbative
renormalization; Schr\"odinger functional method.} \\
[4.0ex] \hline \\[-1.0ex]
{HPQCD 09A, 10}&  \cite{Davies:2009ih,McNeile:2010ji}& 2+1 & \parbox[t]{7cm}{Lattice
  calculation of $m_s/m_c$: $m_s$ derived from a perturbative
  determination of $m_c$.} \\ 
[4.0ex] \hline \\[-1.0ex]
{MILC 09A, 09} & \cite{Bazavov:2009fk,Bazavov:2009bb} & 2+1 & \parbox[t]{7cm}{2-loop
perturbative renormalization.}  \\
[1.0ex] \hline \\[-1.0ex]
PACS-CS 08 &\cite{Aoki:2008sm}& 2+1 & \parbox[t]{7cm}{1-loop perturbative
  renormalization.} \\ 
[1.0ex] \hline \\[-1.0ex]
RBC/UKQCD 08	&\cite{Allton:2008pn}	&2+1& \parbox[t]{7cm}{Non-perturbative
renormalization, 3-loop perturbative matching.} \\
[4.0ex] \hline \\[-1.0ex]
{CP-PACS/JLQCD 07}&  \cite{Ishikawa:2007nn} & 2+1 & \parbox[t]{7cm}{1-loop
  perturbative renormalization, tadpole improved.} \\ 
[4.0ex] \hline \\[-1.0ex]
{HPQCD 05}& \cite{Mason:2005bj}& 2+1 & \parbox[t]{7cm}{2-loop perturbative
  renormalization.} \\ 
[1.0ex] \hline \\[-1.0ex]
\parbox[t]{4.5cm}{HPQCD/MILC/UKQCD 04,\\ MILC 04}
& \cite{Aubin:2004ck,Aubin:2004fs}& 2+1 &  \parbox[t]{7cm}{1-loop perturbative
renormalization.} \\ 
[4.0ex] \hline \hline \\[-1.0ex]
\end{tabular*}
\caption{Renormalization in determinations of $m_{ud}$, $m_s$
  and, in some cases $m_u$ and $m_d$, with $N_f=2+1$ quark flavours.}
}
\end{table}

\begin{table}[!ht]
{\footnotesize
\begin{tabular*}{\textwidth}[l]{l @{\extracolsep{\fill}} c c l}
\hline\hline \\[-1.0ex]
Collab. & Ref. & $\Nf$  & Description
\\[1.0ex] \hline \hline \\[-1.0ex]
RM123 13 & \cite{deDivitiis:2013xla} & 2 &\parbox[t]{7cm}{Non-perturbative
renormalization.} \\
[1.0ex] \hline \\[-1.0ex]
ALPHA 12 & \cite{Fritzsch:2012wq} & 2 &\parbox[t]{7cm}{Non-perturbative
renormalization.} \\
[1.0ex] \hline \\[-1.0ex]
RM123 11 & \cite{deDivitiis:2011eh} & 2 &\parbox[t]{7cm}{Non-perturbative
renormalization.} \\
[1.0ex] \hline \\[-1.0ex]
D\"urr 11 & \cite{Durr:2011ed} & 2 & \parbox[t]{7cm}{Lattice
  calculation of $m_s/m_c$: $m_s$ derived from a perturbative
  determination of $m_c$.}\\
[4.0ex] \hline \\[-1.0ex]
ETM 10B & \cite{Blossier:2010cr} & 2 &\parbox[t]{7cm}{Non-perturbative
renormalization.} \\
[1.0ex] \hline \\[-1.0ex]
{JLQCD/TWQCD 08A} & \cite{Noaki:2008iy} & 2 & \parbox[t]{7cm}{Non-perturbative
renormalization.} \\
[1.0ex] \hline \\[-1.0ex]
{RBC 07} & \cite{Blum:2007cy}
& 2 & \parbox[t]{7cm}{Non-perturbative renormalization.}  \\
[1.0ex] \hline \\[-1.0ex]
{ETM 07} & \cite{Blossier:2007vv}
& 2 & \parbox[t]{7cm}{Non-perturbative renormalization.}  \\
[1.0ex] \hline \\[-1.0ex]
{QCDSF/UKQCD 06} & \cite{Gockeler:2006jt} & 2 & \parbox[t]{7cm}{Non-perturbative
renormalization.} \\
[1.0ex] \hline \\[-1.0ex]
{SPQcdR 05} & \cite{Becirevic:2005ta}
& 2 &  \parbox[t]{7cm}{Non-perturbative renormalization.} \\
[1.0ex] \hline \\[-1.0ex]
{ALPHA 05} & \cite{DellaMorte:2005kg}
& 2 & \parbox[t]{7cm}{Non-perturbative renormalization.} \\
[1.0ex] \hline \\[-1.0ex]
{QCDSF/UKQCD 04} & \cite{Gockeler:2004rp}
& 2 & \parbox[t]{7cm}{Non-perturbative
renormalization.} \\
[1.0ex] \hline \\[-1.0ex]
{JLQCD 02} & \cite{Aoki:2002uc}
& 2 & \parbox[t]{7cm}{1-loop perturbative
renormalization.} \\
[1.0ex] \hline \\[-1.0ex]
{CP-PACS 01} & \cite{AliKhan:2001tx}
& 2 & \parbox[t]{7cm}{1-loop perturbative
renormalization.} \\
[1.0ex] \hline \hline \\[-1.0ex]
\end{tabular*}
\caption{Renormalization in determinations of $m_{ud}$, $m_s$
  and, in some cases $m_u$ and $m_d$, with $N_f=2$ quark flavours.}
}
\end{table}


\clearpage


\subsection{Notes to section \ref{sec:vusvud} on $|V_{ud}|$ and  $|V_{us}|$}
\label{app:VusVud}

\begin{table}[!h]
{\footnotesize
\begin{tabular*}{\textwidth}[l]{l @{\extracolsep{\fill}} c c c l}
\hline\hline \\[-1.0ex]
Collab. & Ref. & $\Nf$ & $a$ [fm] & Description
\\[1.0ex] \hline \hline \\[-2.5ex]
FNAL/MILC 13C &\cite{Gamiz:2013xxa}			&2+1+1&0.09, 0.12, 0.15&
	\parbox[t]{6cm}{Relative scale through $r_1$, physical scale from 
	         $f_\pi$ calculated by MILC 09A at $N_f = 2 + 1$.}\\
[6.0ex] \hline \\[-2.5ex]
FNAL/MILC 12		&\cite{Bazavov:2012cd}			&2+1&0.09, 0.12&
	\parbox[t]{6cm}{Relative scale $r_1$, physical scale determined from a 
		mixture of $f_\pi$, $f_K$, radial excitation of $\Upsilon$
		and $m_{D_s}-\frac 12 m_{ {\eta}_c}$.}\\
[6.0ex] \hline \\[-2.5ex]
RBC/UKQCD 13		&\cite{Boyle:2013gsa}			&2+1&0.09, 0.11, 0.14&
	\parbox[t]{6cm}{Scale set through $\Omega$ mass.}\\
[0.0ex] \hline \\[-2.5ex]
JLQCD 12	&\cite{Kaneko:2012cta}			&2+1&0.112&
	\parbox[t]{6cm}{Scale set through $\Omega$ mass.}\\
[0.0ex] \hline \\[-2.5ex]
JLQCD 11	&\cite{Kaneko:2011rp}			&2+1&0.112&
	\parbox[t]{6cm}{Scale set through $\Omega$ mass.}\\
[0.0ex] \hline \\[-2.5ex]
RBC/UKQCD 07,10 &\cite{Boyle:2007qe,Boyle:2010bh} 	&2+1& 0.114(2) & 
		\parbox[t]{6cm}{
		Scale fixed through $\Omega$ baryon mass. Add 
		$(\Lambda_{\rm QCD}a)^2\approx4$\%
		systematic error for lattice artifacts.
		Fifth dimension with extension $L_s=16$, therefore small
		residual chiral symmetry breaking and approximate $O(a)$-improvement.}\\
[15.0ex] \hline \\[-2.5ex]
ETM 10D		&\cite{Lubicz:2010bv}		&2  & 
		\parbox[t]{2cm}{\centering 0.05, 0.07, 0.09, 0.10} & \parbox[t]{6cm}{ Scale set through $F_\pi$. Automatic $O(a)$ 
		impr., flavour symmetry
		breaking:\\ $(M_{PS}^{0})^2 - (M_{PS}^\pm)^2\sim O(a^2)$.} 
\\[7.0ex] \hline \\[-2.5ex]
ETM 09A		&\cite{Lubicz:2009ht}	&2  & 0.07, 0.09, 0.10&
		\parbox[t]{6cm}{
		Scale set through $F_\pi$. Automatic $O(a)$ impr., flavour symmetry
		breaking:\\ $(M_{PS}^{0})^2 - (M_{PS}^\pm)^2\sim O(a^2)$. \\
  		Three lattice spacings only for pion mass 470\,MeV.}\\
[12.0ex] \hline \\[-2.5ex]
QCDSF 07	&\cite{Brommel:2007wn}	&2  &0.075& 
		\parbox[t]{6cm}{Scale set with $r_0$. Non-perturbatively
		$O(a)$-improved Wilson fermions, not clear whether 
		currents improved.}\\
[6.0ex] \hline \\[-2.5ex]
RBC 06		&\cite{Dawson:2006qc}	&2  &0.12& 
		\parbox[t]{6cm}{Scale set through $M_\rho$.
		Automatic $O(a)$-improvement due to approximate 
		chiral symmetry of the action. }\\
[6.0ex] \hline \\[-2.5ex]
JLQCD 05	&\cite{Tsutsui:2005cj}	&2  &0.0887&
		\parbox[t]{6cm}{Scale set through $M_\rho$. Non-perturbatively
		$O(a)$-improved Wilson fermions.}\\
[1.0ex]
\hline\hline
\end{tabular*}
\caption{Continuum extrapolations/estimation of lattice artifacts in
  determinations of $f_+(0)$.}
}
\end{table}

\noindent
\begin{table}[!h]
{\footnotesize
\begin{tabular*}{\textwidth}[l]{l @{\extracolsep{\fill}} c c c l}
\hline \hline  \\[-1.0ex]
Collab. & Ref. & $\Nf$ & $M_{\pi,\text{min}}$ [MeV] & Description
\\[1.0ex] \hline \hline \\[-2.5ex]
FNAL/MILC 13C &\cite{Gamiz:2013xxa}			&2+1+1&$173_{\rm RMS}(128_{\rm \pi,5})$&
		\parbox[t]{6cm}{NLO SU(3) PQ staggered {\Ch}PT with
		        continuum {\Ch}PT at NNLO. Lightest Nambu-Goldstone 
		        mass is 128~MeV and lightest RMS mass is 173~MeV
			for the same gauge ensemble with $a \simeq 0.09$~fm.
			}\\
[12.0ex] \hline \\[-2.5ex]
FNAL/MILC 12		&\cite{Bazavov:2012cd}			&2+1&$378_{\rm RMS}(263_{\rm \pi,5})$&
		\parbox[t]{6cm}{NLO SU(3) PQ staggered {\Ch}PT with either 
			phenomenological NNLO ansatz or NNLO {\Ch}PT.
			Lightest Nambu-Goldstone mass is 263~MeV with 
			$a=0.12$~fm and lightest RMS mass is 378~MeV
			with $a=0.09$~fm.
			}\\
[12.0ex] \hline \\[-2.5ex]
RBC/UKQCD 13		&\cite{Boyle:2013gsa}			&2+1&170&
		\parbox[t]{6cm}{NLO SU(3) {\Ch}PT with phenomenological ansatz 
			for higher orders.}\\
[1.0ex] \hline \\[-2.5ex]
JLQCD 12	&\cite{Kaneko:2012cta}			&2+1&290&
		\parbox[t]{6cm}{NLO SU(3) {\Ch}PT with phenomenological ansatz 
			for higher orders.}\\
[1.0ex] \hline \\[-2.5ex]
JLQCD 11	&\cite{Kaneko:2011rp}			&2+1&290&
		\parbox[t]{6cm}{NLO SU(3) {\Ch}PT with phenomenological ansatz 
			for higher orders.}\\
[1.0ex] \hline \\[-2.5ex]
RBC/UKQCD 07,10 &\cite{Boyle:2007qe,Boyle:2010bh} 	&2+1&330& 
		\parbox[t]{6cm}{NLO SU(3) {\Ch}PT with phenomenological ansatz 
			for higher orders.}\\
[0.0ex] \hline \\[-2.5ex]
ETM 10D         & \cite{Lubicz:2010bv}  &2  &$210_{\pi^0}(260_{\pi^\pm})$&
		\parbox[t]{6cm}{NLO heavy kaon SU(2) {\Ch}PT and
		NLO SU(3) {\Ch}PT and phenomenological ansatz for higher orders.
		Average of $f_+(0)$-fit and joint $f_+(0)$-$f_K/f_\pi$-fit.}\\[9.0ex] \hline \\[-2.5ex]
ETM 09A		&\cite{Lubicz:2009ht}	&2  &$210_{\pi^0}(260_{\pi^\pm})$& 
		\parbox[t]{6cm}{NLO heavy kaon SU(2) {\Ch}PT and
		NLO SU(3) {\Ch}PT and phenomenological ansatz for higher orders.}\\
[0.0ex] \hline \\[-2.5ex]
QCDSF 07	&\cite{Brommel:2007wn}	&2  &591& 
		\parbox[t]{6cm}{Only one value for the pion mass.}\\
[0.0ex] \hline \\[-2.5ex]
RBC 06		&\cite{Dawson:2006qc}	&2  &490& 
		\parbox[t]{6cm}{NLO SU(3) {\Ch}PT and phenomenological ansatz 
			for higher orders.}\\
[0.0ex] \hline \\[-2.5ex]
JLQCD 05	&\cite{Tsutsui:2005cj}	&2  &550& 
		\parbox[t]{6cm}{NLO SU(3) {\Ch}PT and phenomenological ansatz 
			for higher orders.}\\
[3.0ex]
\hline\hline\\
\end{tabular*}
\caption{Chiral extrapolation/minimum pion mass in determinations of $f_+(0)$.
The subscripts RMS and $\pi,5$ in the case of staggered fermions indicate
the root-mean-square mass and the Nambu-Goldstone boson mass, respectively. 
In the case
of twisted-mass fermions $\pi^0$ and $\pi^\pm$ indicate the neutral and
charged pion mass where applicable.}
}
\end{table}

\noindent
\begin{table}[!h]
{\footnotesize
\begin{tabular*}{\textwidth}[l]{l @{\extracolsep{\fill}} c c c c l}
\hline \hline  \\[-1.0ex]
Collab. & Ref. & $\Nf$ &$L$ [fm]& $M_{\pi,\text{min}}L$ & Description
\\[1.0ex] \hline \hline \\[-2.5ex]
FNAL/MILC 13C &\cite{Gamiz:2013xxa} 		&2+1+1&2.9--5.8&$4.9_{\rm RMS}(3.6_{\pi,5})$& \parbox[t]{5cm}{The values correspond to $M_{\pi,\rm RMS}=173$~MeV and 
		$M_{\pi,5}=128$~MeV, respectively.}
		\parbox[t]{5cm}{}\\
[6.0ex] \hline \\[-2.5ex]
FNAL/MILC 12		&\cite{Bazavov:2012cd} 		&2+1&2.4--3.4&$6.2_{\rm RMS}(3.8_{\pi,5})$& \parbox[t]{5cm}{The values correspond to $M_{\pi,\rm RMS}=378$~MeV and 
		$M_{\pi,5}=263$~MeV, respectively.}
		\parbox[t]{5cm}{}\\
[6.0ex] \hline \\[-2.5ex]
RBC/UKQCD 13		&\cite{Boyle:2013gsa} 	&2+1&2.7, 4.6&3.9& 
		\parbox[t]{5cm}{}\\
[0.0ex] \hline \\[-2.5ex]

JLQCD 12	&\cite{Kaneko:2012cta} 	&2+1&1.8, 2.7&4.1& 
		\parbox[t]{5cm}{}\\
[0.0ex] \hline \\[-2.5ex]
JLQCD 11	&\cite{Kaneko:2011rp} 	&2+1&1.8, 2.7&4.1& 
		\parbox[t]{5cm}{}\\
[0.0ex] \hline \\[-2.5ex]
RBC/UKQCD 07,10 &\cite{Boyle:2007qe,Boyle:2010bh} 	&2+1&1.8,2.7&4.7& 
		\parbox[t]{5cm}{Two volumes for all but the lightest
			pion mass.}\\
[0.0ex] \hline \\[-2.5ex]
ETM 10D		&\cite{Lubicz:2010bv}		&2  &2.1--2.8 &$3.0_{\pi^0}(3.7_{\pi^\pm})$ & \\
[0.0ex] \hline \\[-2.5ex]
ETM 09A		&\cite{Lubicz:2009ht}	&2  &2.1, 2.8&$3.0_{\pi^0}(3.7_{\pi^\pm})$& 
		\parbox[t]{5cm}{Two volumes at $M_\pi=300$\,MeV and {\Ch}PT-motivated 
		estimate of the error due to FSE.}\\
[6.0ex] \hline \\[-2.5ex]
QCDSF 07	&\cite{Brommel:2007wn}	&2  &1.9&5.4& 
		\parbox[t]{5cm}{}\\
[0.0ex] \hline \\[-1.0ex]
RBC 06		&\cite{Dawson:2006qc}	&2
&1.9&4.7& \parbox[t]{5cm}{}\\
[0.0ex] \hline \\[-1.0ex]
JLQCD 05	&\cite{Tsutsui:2005cj}	&2  &1.8&4.9& \parbox[t]{5cm}{}\\
[0.0ex]
\hline\hline
\end{tabular*}
\caption{Finite volume effects in determinations of $f_+(0)$. The subscripts RMS and $\pi,5$ in the case of staggered fermions indicate
the root-mean-square mass and the Nambu-Goldstone boson mass, respectively. 
In the case
of twisted-mass fermions $\pi^0$ and $\pi^\pm$ indicate the neutral and
charged pion mass where applicable.}
}
\end{table}

\begin{table}[!h]
{\footnotesize
\begin{tabular*}{\textwidth}[l]{l @{\extracolsep{\fill}} c c c l}
\hline \hline  \\[-1.0ex]
Collab. & Ref. & $\Nf$ & $a$ [fm] & Description
\\[2.5ex] \hline \hline \\[-2.5ex]
{HPQCD 13A}& \cite{Dowdall:2013rya} & 2+1+1 & 0.09,\,0.12,\,0.15  &  
	\parbox[t]{5.7cm}{ 
	Relative scale through Wilson flow and absolute scale through $f_\pi$.}\\
[3.0ex] \hline  \\[-2.5ex]
{MILC 13A}& \cite{Bazavov:2013cp} & 2+1+1 & 0.06,\,0.09,\,0.12,\,0.15  &  
	\parbox[t]{5.7cm}{ 
	Absolute scale though $f_\pi$.}\\
 \hline  \\[-2.5ex]
{ETM 13F}& \cite{Dimopoulos:2013qfa} & 2+1+1 & 0.062, 0.082, 0.089  &
		    \parbox[t]{5.7cm}{ 
		    Scale set through $f_\pi$. Automatic $Ow(a)$ improvement, flavour symmetry
		    breaking: $(M_{PS}^{0})^2-(M_{PS}^\pm)^2\sim O(a^2)$. Discretization and volume effects
		    due to the $\pi^0 - \pi^\pm$ mass splitting are taken into account through {\Ch}PT for
		    twisted-mass fermions.} \\
\hline  \\[-2.5ex]
{ETM 10E}& \cite{Farchioni:2010tb} & 2+1+1 & 0.061, 0.078  &
		\parbox[t]{5.7cm}{ 
		Scale set through $f_\pi/m_\pi$. 
		Two lattice spacings but
		$a$-dependence ignored in all fits. Finer lattice spacing
		from \cite{Baron:2011sf}.} \\
\hline  \\[-2.5ex]
{MILC 11}& \cite{Bazavov:2011fh} & 2+1+1 & 0.12,\,0.09  &  
	\parbox[t]{5.7cm}{ 
	Relative scale through
	$f_{\rm PS}/m_{\rm PS}=\textrm{fixed}$, absolute scale though $f_\pi$.}\\
[3.0ex] \hline  \\[-2.5ex]
{RBC/UKQCD 12}& \cite{Arthur:2012opa} & 2+1 & 0.09,\, 0.11,\, 0.14  &  
	\parbox[t]{5.7cm}{ 
	Scale set through $m_\Omega$.}\\
\hline  \\[-2.5ex]
{LAIHO 11}& \cite{Laiho:2011np} & 2+1 & 0.125, 0.09, 0.06 &  
	\parbox[t]{5.7cm}{Scale set through $r_1$ and $\Upsilon$ and 
	continuum extrapolation based on MA{\Ch}PT.}\\
[3.0ex] \hline  \\[-2.5ex]
{JLQCD/TWQCD 10}& \cite{Noaki:2010zz} & 2+1 & 0.112  & Scale set
through $M_\Omega$. \\
 \hline  \\[-2.5ex]
{RBC/UKQCD 10A}& \cite{Aoki:2010dy} & 2+1 & 0.114, 0.087  & Scale set
through $M_\Omega$. \\
 \hline  \\[-2.5ex]
MILC 10 & \cite{Bazavov:2010hj} & 2+1 & 0.09, 0.06, 0.045 & \parbox[t]{5.7cm}{3 lattice
spacings, continuum extrapolation by means of RS{\Ch}PT.}\\
[2.0ex] \hline  \\[-2.5ex]
BMW 10		&\cite{Durr:2010hr}		&2+1&
		\parbox[t]{2.5cm}{\centering 0.07, 0.08,\\ 0.12}&
		\parbox[t]{5.7cm}{Scale set through $M_{\Omega,\Xi}$. Perturbative
		$O(a)$-improvement.}\\
\hline \\[-2.5ex]
JLQCD/TWQCD 09A &\cite{Noaki:2008iy}	&2+1& 0.1184(3)(21) & 
		\parbox[t]{5.7cm}{Scale set through $F_\pi$. Automatic 
		$O(a)$-improvement due to chiral symmetry of action.}\\[6.0ex]
 \hline \\[-2.5ex]
{PACS-CS 09}& \cite{Aoki:2009ix} & 2+1 & 0.0900(4)  & Scale set
through $M_\Omega$. \\
 \hline  \\[-2.5ex]
MILC 09A	&\cite{Bazavov:2009fk}	&2+1& 0.045, 0.06, 0.09 & 
		\parbox[t]{5.7cm}{Scale set through $r_1$ and $\Upsilon$ and 
		continuum extrapolation based on RS{\Ch}PT.}\\
 \hline \\[-2.5ex]
MILC 09		&\cite{Bazavov:2009bb}	&2+1& 	\parbox[t]{2.5cm}{\centering 0.045, 0.06, \\0.09, 0.12} & 
		\parbox[t]{5.7cm}{Scale set through $r_1$ and $\Upsilon$ and 
		continuum extrapolation based on RS{\Ch}PT.}\\
[2.0ex] \hline \\[-2.5ex]
Aubin 08	&\cite{Aubin:2008ie}	&2+1& 0.09, 0.12 & 	
		\parbox[t]{5.7cm}{Scale set through $r_1$ and $\Upsilon$ and 
		continuum extrapolation based on MA{\Ch}PT.}\\
 \hline \\[-2.5ex]
PACS-CS 08, 08A &\cite{Aoki:2008sm,Kuramashi:2008tb}& 2+1 & 0.0907(13) & 
		\parbox[t]{5.7cm}{Scale set 
		through $M_\Omega$. Non-perturbatively $O(a)$-improved. }\\
 \hline \\[-2.5ex]
HPQCD/UKQCD 07	&\cite{Follana:2007uv}	&2+1& 0.09, 0.12, 0.15 & 
		\parbox[t]{5.7cm}{Scale set through $r_1$ and $\Upsilon$ and 
		continuum extrapolation on continuum-{\Ch}PT motivated ansatz. Taste
		breaking of sea quarks ignored.}\\
 \hline \\[-2.5ex]
RBC/UKQCD 08	&\cite{Allton:2008pn}	&2+1& 0.114(2) & 
		\parbox[t]{5.7cm}{Scale set through $M_\Omega$. Automatic 
		$O(a)$-improvement due to appoximate chiral symmetry. 
		$(\Lambda_{\rm QCD}a)^2\approx 4\%$ systematic error
		due to lattice artifacts added.}\\[8.0ex]
 \hline \\[-2.5ex]
NPLQCD 06	&\cite{Beane:2006kx}	&2+1& 0.125 & 
		\parbox[t]{5.7cm}{Scale set through $r_0$ and $F_\pi$. Taste breaking
		of sea quarks ignored.}\\
\hline\hline
\end{tabular*}
\caption{Continuum extrapolations/estimation of lattice artifacts in
  determinations of $f_K/f\pi$ for $N_f=2+1$ and  $N_f=2+1+1$ simulations.}
}
\end{table}

\begin{table}[!h]
{\footnotesize
\begin{tabular*}{\textwidth}[l]{l @{\extracolsep{\fill}} c c c l}
\hline \hline  \\[-1.0ex]
Collab. & Ref. & $\Nf$ & $a$ [fm] & Description
\\[2.5ex] \hline \hline \\[-2.5ex]
ALPHA 13		&\cite{Lottini:2013rfa}		&2  & 	\parbox[t]{2.5cm}{\centering 0.05, 0.065, 0.075} &\parbox[t]{5.7cm}{Scale set through $F_\pi$.	$O(a)$-improved Wilson action.}\\
[3.0ex] \hline \\[-2.5ex]
BGR 11        &\cite{Engel:2011aa}                &2  & 0.135 &
                   \parbox[t]{5.7cm}{Scale set through $r_0 = 0.48$ fm.
Chirally improved Dirac operator.}\\
[3.0ex] \hline \\[-2.5ex]
ETM 10D		&\cite{Lubicz:2010bv}		&2  & 	\parbox[t]{2.5cm}{\centering 0.05, 0.07,\\ 0.09, 0.10} &\parbox[t]{5.7cm}{Scale set through $F_\pi$.	Automatic $O(a)$ impr., flavour symmetry breaking: $(M_{PS}^{0})^2-(M_{PS}^\pm)^2\sim O(a^2)$.}\\
[6.0ex] \hline \\[-2.5ex]
ETM 09		&\cite{Blossier:2009bx}		&2  & 0.07, 0.09, 0.10 & 
		\parbox[t]{5.7cm}{Scale set through $F_\pi$.
		Automatic $O(a)$ impr., flavour symmetry
		breaking: $(M_{PS}^{0})^2-(M_{PS}^\pm)^2\sim O(a^2)$. 
		  }\\
[6.0ex] \hline \\[-2.5ex]
QCDSF/UKQCD 07 	&\cite{QCDSFUKQCD}	&2  & 0.06, 0.07 & 
		\parbox[t]{5.7cm}{Scale set through $F_\pi$. Non-perturbative
		$O(a)$-improvement.}\\
\hline\hline
\end{tabular*}
\caption{Continuum extrapolations/estimation of lattice artifacts in
  determinations of $f_K/f\pi$ for $N_f=2$ simulations.}
}
\end{table}

\noindent
\begin{table}[!h]
{\footnotesize
\begin{tabular*}{\textwidth}[l]{l @{\extracolsep{\fill}} c c c l}
\hline \hline  \\[-1.0ex]
Collab. & Ref. & $\Nf$ & $M_{\pi,\text{min}}$ [MeV] & Description
\\[2.5ex] \hline \hline \\[-2.5ex]
{HPQCD 13A}& \cite{Dowdall:2013rya} & 2+1+1 &$173_{\rm RMS}(128_{\pi,5})$  &  
	\parbox[t]{6.2cm}{ 
	NLO $\chi$PT supplemented by model for NNLO. Both the lightest RMS and 
	the lightest
	Nambu-Goldstone mass are from the $a=0.09$~fm ensemble.}\\
[9.0ex] \hline  \\[-2.5ex]
{MILC 13A}& \cite{Bazavov:2013cp} & 2+1+1 & $143_{\rm RMS}(128_{\pi,5})$  &  
	\parbox[t]{6.2cm}{ 
	Linear interpolation to physical point. The lightest RMS mass is from the 
	$a=0.06$~fm ensemble and the lightest Nambu-Goldstone mass is from the
	$a=0.09$~fm ensemble.}\\ 
[9.0ex]\hline  \\[-2.5ex]
{ETM 13F}& \cite{Dimopoulos:2013qfa} & 2+1+1 & $155_{\rm \pi^0}(220_{\pi^\pm})$  & 
          \parbox[t]{6.2cm}{Chiral extrapolation performed through SU(2) {\Ch}PT or polynomial fit.}\\
\hline  \\[-2.5ex]
{ETM 10E}& \cite{Farchioni:2010tb} & 2+1+1 & $215_{\rm \pi^0}(265_{\pi^\pm})$  & \parbox[t]{6.2cm}{}\\
\hline  \\[-2.5ex]
{MILC 11}& \cite{Bazavov:2011fh} & 2+1+1 &$173_{\rm RMS}(128_{\pi,5})$ &\parbox[t]{6.2cm}{
		Quoted result from polynomial interpolation to the physical 
		point. The lightest RMS mass is from the 
	$a=0.06$~fm ensemble and lightest the Nambu-Goldstone mass is from the
	$a=0.09$~fm ensemble.}\\[12.0ex]
\hline\hline\\
\end{tabular*}
\caption{Chiral extrapolation/minimum pion mass in determinations of $f_K / f\pi$ for $N_f=2+1+1$ simulations. 
The subscripts RMS and $\pi,5$ in the case of staggered fermions indicate
the root-mean-square mass and the Nambu-Goldstone boson mass. In the case
of twisted-mass fermions $\pi^0$ and $\pi^\pm$ indicate the neutral and charged pion mass 
and, where applicable, ``val'' and ``sea'' indicate valence and sea pion masses.}
}
\end{table}

\noindent
\vspace{-3cm}
\begin{table}[!h]
{\footnotesize
\begin{tabular*}{\textwidth}[l]{l @{\extracolsep{\fill}} c c c l}
\hline \hline  \\[-1.0ex]
Collab. & Ref. & $\Nf$ & $M_{\pi,\text{min}}$ [MeV] & Description
\\[2.5ex] \hline \hline \\[-2.5ex]
{RBC/UKQCD 12}& \cite{Arthur:2012opa} & 2+1 & 171$_{\rm sea}$, 143$_{\rm val}$ &\parbox[t]{6.2cm}{
		NLO PQ SU(2) $\chi$PT as well as analytic ans\"atze.}\\
[3.0ex] \hline \\[-2.5ex]
{LAIHO 11}& \cite{Laiho:2011np} & 2+1 & 250$_{\rm RMS}(220_{\pi,5})$ &
	\parbox[t]{6.2cm}{NLO MA{\Ch}PT }\\[1mm]
\hline \\[-2.5ex]
{JLQCD/TWQCD 10}& \cite{Noaki:2010zz} & 2+1 & 290 &
	\parbox[t]{6.2cm}{NNLO {\Ch}PT}\\
\hline \\[-2.5ex]
{RBC/UKQCD 10A}& \cite{Aoki:2010dy} & 2+1 & 290 &\parbox[t]{6.2cm}{
		Results are based on 
		heavy kaon NLO SU(2) PQ{\Ch}PT. }\\
\hline \\[-2.5ex]
MILC 10       & \cite{Bazavov:2010hj}  & 2+1 & $258_{\rm RMS}(177_{\pi,5})$ &  \parbox[t]{6.2cm}{Lightest
Nambu-Goldstone mass is 177\,MeV (at 0.09\,fm) and
 lightest RMS mass is 258\,MeV (at 0.06\,fm). NLO rS{\Ch}PT and NNLO {\Ch}PT.} \\
\hline \\[-2.5ex]
BMW 10		&\cite{Durr:2010hr}		&2+1& 190 & 
		\parbox[t]{6.2cm}{Comparison of various fit-ans\"atze:
		SU(3) {\Ch}PT, heavy kaon SU(2) {\Ch}PT, polynomial. }\\
[2.5ex] \hline \\[-2.5ex]
JLQCD/TWQCD 09A &\cite{Noaki:2008iy}	&2+1&290& 
		\parbox[t]{6.2cm}{NNLO SU(3) {\Ch}PT.}\\
\hline \\[-2.5ex]
{PACS-CS 09}& \cite{Aoki:2009ix} & 2+1 & 156 &
	\parbox[t]{6.2cm}{NNLO {\Ch}PT}\\
\hline \\[-2.5ex]
MILC 09A	&\cite{Bazavov:2009fk}	&2+1&$258_{\rm RMS}(177_{\pi,5})$& 
		\parbox[t]{6.2cm}{
		NLO SU(3) RS{\Ch}PT, continuum {\Ch}PT at NNLO and up to NNNNLO 
		analytic terms. Heavy kaon SU(2) RS{\Ch}PT with 
		NNLO continuum chiral logs on a sub-set of the
                lattices. The lightest Nambu-Goldstone mass is 177\,MeV  
		(at $a=0.09$ fm)
		and the lightest RMS mass is 258\,MeV (at $a=0.06$ fm).}\\
[2.5ex] \hline \\[-2.5ex]
MILC 09		&\cite{Bazavov:2009bb}	&2+1&$258_{\rm RMS}(224_{\pi,5})$& 
		\parbox[t]{6.2cm}{
		NLO SU(3) RS{\Ch}PT with continuum {\Ch}PT NNLO and NNNLO analytic
		terms added. According to \cite{Bazavov:2009fk}
		the lightest sea Nambu-Goldstone mass is 224\,MeV
		and the lightest RMS mass is 258\,MeV (at $a=0.06$ fm).}\\
[2.5ex] \hline \\[-2.5ex]
Aubin 08	&\cite{Aubin:2008ie}	&2+1&$329_{\rm RMS}(246_{\pi,5})$& 
		\parbox[t]{6.2cm}{NLO MA{\Ch}PT. According to \cite{Bazavov:2009fk}
		the lightest sea Nambu-Goldstone mass is 246\,MeV (at $a=0.09$ fm)
		and the lightest RMS mass is 329\,MeV (at $a=0.09$ fm).}\\
[2.5ex] \hline \\[-2.5ex]
PACS-CS 08, 08A &\cite{Aoki:2008sm,Kuramashi:2008tb}& 2+1 &156& 
		\parbox[t]{6.2cm}{NLO SU(2) {\Ch}PT and SU(3) (Wilson){\Ch}PT.}\\
\hline \\[-2.5ex]
HPCD/UKQCD 07	&\cite{Follana:2007uv}	&2+1&$375_{\rm RMS}(263_{\pi,5})$& 
		\parbox[t]{6.2cm}{NLO SU(3) chiral perturbation theory
		with NNLO and NNNLO analytic terms. 
		The lightest RMS mass is from the $a=0.09$~fm ensemble and 
		the lightest Nambu-Goldstone mass is from the 
		$a=0.12$~fm ensemble.}\\
[11.0ex] \hline \\[-2.5ex]
RBC/UKQCD 08	&\cite{Allton:2008pn}	&2+1&$330_{\rm sea}$, $242_{\rm val}$& 
		\parbox[t]{6.2cm}{While SU(3) PQ{\Ch}PT fits were studied,
		final results are based on 
		heavy kaon NLO SU(2) PQ{\Ch}PT. }\\
[6.0ex] \hline \\[-2.5ex]
NPLQCD 06	&\cite{Beane:2006kx}	&2+1&300& 
		\parbox[t]{6.2cm}{NLO SU(3) {\Ch}PT and some NNLO terms. 
		The sea RMS mass for the employed lattices is heavier.}\\
\hline\hline\\
\end{tabular*}
\vskip -0.5cm
\caption{Chiral extrapolation/minimum pion mass in determinations of $f_K/f\pi$ for $N_f=2+1$ simulations. The subscripts RMS and $\pi,5$ in the case of staggered fermions indicate
the root-mean-square mass and the Nambu-Goldstone boson mass. In the case
of twisted-mass fermions $\pi^0$ and $\pi^\pm$ indicate the neutral and
charged pion mass and where applicable, ``val'' and ``sea'' indicate valence
and sea pion masses.}
}
\end{table}

\begin{table}[!h]
{\footnotesize
\begin{tabular*}{\textwidth}[l]{l @{\extracolsep{\fill}} c c c l}
\hline \hline  \\[-1.0ex]
Collab. & Ref. & $\Nf$ & $M_{\pi,\text{min}}$ [MeV] & Description
\\[2.5ex] \hline \hline \\[-2.5ex]
ALPHA 13		&\cite{Lottini:2013rfa}	&2  &190& 
                 \parbox[t]{6.2cm}{NLO SU(3) {\Ch}PT and phenomenological ansatz for higher orders.}\\
[3.0ex] \hline \\[-2.5ex]
BGR 11        &\cite{Engel:2011aa}        &2  &250&
                  \parbox[t]{6.2cm}{NLO SU(2) {\Ch}PT. Strange quark
mass fixed by reproducing the $\Omega$ mass.}\\
[0.0ex] \hline \\[-2.5ex]
ETM 10D         & \cite{Lubicz:2010bv}  &2  &$210_{\pi^0}(260_{\pi^\pm})$&\parbox[t]{6.2cm}{
		NLO SU(3) {\Ch}PT and phenomenological ansatz for higher orders.
		Joint $f_+(0)$-$f_K/f_\pi$-fit.}\\
[2.5ex] \hline \\[-2.5ex]
ETM 09		&\cite{Blossier:2009bx}		&2  &$210_{\pi^0}(260_{pi^\pm})$& 
		\parbox[t]{6.2cm}{NLO heavy meson SU(2) {\Ch}PT and NLO
		SU(3) {\Ch}PT.}\\
[2.5ex] \hline \\[-2.5ex]
QCDSF/UKQCD 07 	&\cite{QCDSFUKQCD}	&2  &300& 
		\parbox[t]{6.2cm}{Linear extrapolation of lattice data.}\\
\hline\hline\\
\end{tabular*}
\vskip -0.5cm
\caption{Chiral extrapolation/minimum pion mass in determinations of $f_K/f\pi$ for $N_f=2$ simulations. The subscripts RMS and $\pi,5$ in the case of staggered fermions indicate
the root-mean-square mass and the Nambu-Goldstone boson mass. In the case
of twisted-mass fermions $\pi^0$ and $\pi^\pm$ indicate the neutral and
charged pion mass and where applicable, ``val'' and ``sea'' indicate valence
and sea pion masses.}
}
\end{table}

\noindent
\begin{table}[!h]
{\footnotesize
\begin{tabular*}{\textwidth}[l]{l @{\extracolsep{\fill}} c c c c l}
\hline \hline  \\[-1.0ex]
Collab. & Ref. & $\Nf$ &$L$ [fm]& $M_{\pi,\text{min}}L$ & Description
\\ \hline \hline \\[-2.5ex]
{HPQCD 13A}& \cite{Dowdall:2013rya} & 2+1+1 &2.5-5.8& $4.9_{\rm RMS}(3.7_{\pi,5})$  &  
	\parbox[t]{4.7cm}{ }\\
[0.0ex] \hline  \\[-2.5ex]
{MILC 13A}& \cite{Bazavov:2013cp} & 2+1+1 &2.8-5.8 & $3.9_{\rm RMS}(3.7_{\pi,5}) $ &  
	\parbox[t]{4.7cm}{ }\\
[0.0ex] \hline  \\[-2.5ex]
{ETM 13F}& \cite{Dimopoulos:2013qfa} & 2+1+1 & 2.0 - 3.0 & $1.7_{\pi^0}(3.3_{\pi^\pm})$ &
        \parbox[t]{4.7cm}{FSE for the pion is corrected through resummed NNLO {\Ch}PT for twisted-mass fermions,
        which takes into account the effects due to the $\pi^0 - \pi^\pm$ mass splitting.} \\
[0.0ex] \hline \\[-2.5ex]
{ETM 10E}& \cite{Farchioni:2010tb} & 2+1+1 & 1.9 - 2.9 &  $
3.1_{\pi^0}(3.9_{\pi^\pm})$ &\parbox[t]{4.7cm}{Simulation parameters from \cite{Baron:2011sf,
Herdoiza_private}.} \\
[0.0ex] \hline \\[-2.5ex]
{MILC 11}& \cite{Bazavov:2011fh} & 2+1+1 & 5.6,\,5.7 &  $4.9_{\rm RMS}(3.7_{\pi,5})$ & \\
[0.0ex] \hline \\[-2.5ex]
{RBC/UKQCD 12}& \cite{Arthur:2012opa} & 2+1 & 2.7,\,4.6 &  $3.3$ & 
\parbox[t]{4.7cm}{For partially quenched $M_\pi=143$MeV, $M_\pi L=3.3$ and for
unitary $M_\pi=171$MeV, $M_\pi L=4.0$.}\\
[0.0ex] \hline \\[-2.5ex]
{LAIHO 11}& \cite{Laiho:2011np} & 2+1 & {2.5--4.0} & $4.9_{\rm RMS}(4.3_{\pi,5})$ &\parbox[t]{4.7cm}{} \\
[0.0ex] \hline \\[-2.5ex]
{JLQCD/TWQCD 10}& \cite{Noaki:2010zz} & 2+1 & 1.8, 2.7 &  4.0 & \\
[0.0ex] \hline \\[-2.5ex]
{RBC/UKQCD 10A}& \cite{Aoki:2010dy} & 2+1 & 2.7 &  $4.0$ &
\parbox[t]{4.7cm}{$M_\pi L=4.0$ for lightest sea quark mass and $M_\pi L=3.1$ for lightest partially quenched quark mass.} \\
[0.0ex] \hline \\[-2.5ex]
MILC 10       & \cite{Bazavov:2010hj}                 & 2+1 & 2.5-3.8 &
$7.0_{\rm RMS}(4.0_{\pi,5})$  & \parbox[t]{4.7cm}{$L\!\geq\!2.9\,\fm$ for the lighter masses.}\\
[0.0ex] \hline \\[-2.5ex]
BMW 10		&\cite{Durr:2010hr}		&2+1& 
{2.0--5.3} &4.0& 
		\parbox[t]{4.7cm}{Various volumes for comparison and
		correction for FSE from {\Ch}PT using \cite{Colangelo:2005gd}.}\\
[0.0ex] \hline \\[-2.5ex]
JLQCD/TWQCD 09A &\cite{Noaki:2008iy}	&2+1&1.9&2.8& 
		\parbox[t]{4.7cm}{Estimate of FSE using {\Ch}PT 
		\cite{Colangelo:2005gd,Brower:2003yx}}\\
[0.0ex] \hline \\[-2.5ex]
{PACS-CS 09}& \cite{Aoki:2009ix} & 2+1 & 2.9& 2.28 & 
	\parbox[t]{4.7cm}{after reweighting to 
	the physical point $M_{\pi,\rm min}L=1.97$}\\
[0.0ex] \hline \\[-2.5ex]
MILC 09A	&\cite{Bazavov:2009fk}	&2+1&
{2.5--5.8}&$7.0_{\rm RMS}(4.1_{\pi,5})$& 
		\parbox[t]{4.7cm}{}\\
[0.0ex] \hline \\[-2.5ex]
MILC 09		&\cite{Bazavov:2009bb}	&2+1&
{2.4--5.8}&$7.0_{\rm RMS}(4.8_{\pi,5})$& 
		\parbox[t]{4.7cm}{Various volumes for comparison and
		correction for FSEs from (RS){\Ch}PT \cite{Colangelo:2005gd}.}\\
[0.0ex] \hline \\[-2.5ex]
Aubin 08	&\cite{Aubin:2008ie}	&2+1&2.4--3.6&4.0& 
		\parbox[t]{4.7cm}{Correction for FSE from MA{\Ch}PT.}\\
[0.0ex] \hline \\[-2.5ex]
PACS-CS 08, 08A &\cite{Aoki:2008sm,Kuramashi:2008tb}& 2+1 &2.9&2.3& 
		\parbox[t]{4.7cm}{Correction for FSE from {\Ch}PT using 
		\cite{Colangelo:2005gd}.}\\
[0.0ex] \hline \\[-2.5ex]
HPCD/UKQCD 07	&\cite{Follana:2007uv}	&2+1&
{2.4--2.9}&$4.1_{\rm RMS}(3.8_{\pi,5})$& 
		\parbox[t]{4.7cm}{Correction for FSE from {\Ch}PT using 
		\cite{Colangelo:2005gd}.}\\
[0.0ex] \hline \\[-2.5ex]
RBC/UKQCD 08	&\cite{Allton:2008pn}	&2+1&1.8, 2.7&$4.6_{\rm sea}$, $3.4_{\r val}$&
		\parbox[t]{4.7cm}{Various volumes for comparison and 
		correction for FSE from {\Ch}PT 
		\cite{Gasser:1986vb,Gasser:1987ah,Colangelo:2005gd}.
		}\\
[0.0ex] \hline \\[-2.5ex]
NPLQCD 06	&\cite{Beane:2006kx}	&2+1&2.5&3.8& 
		\parbox[t]{4.7cm}{Correction for FSE from S{\Ch}PT 
		\cite{Aubin:2003uc,Aubin:2003mg}}\\
[0.0ex] \hline \\[-2.5ex]
ALPHA 13		&\cite{Lottini:2013rfa}		&2  & {2.1, 2.4, 3.1}& $4.0$ & \\
[0.0ex] \hline \\[-2.5ex]
BGR 11        &\cite{Engel:2011aa}                 &2 & {2.1, 2.2} & $2.7$ & \\
[0.0ex] \hline \\[-2.5ex]
ETM 10D		&\cite{Lubicz:2010bv}		&2  & {2.1--2.8}& $3.0_{\pi^0}(3.7_{\pi^\pm})$ & \\
[0.0ex] \hline \\[-2.5ex]
ETM 09		&\cite{Blossier:2009bx}		&2  &2.0--2.7&$3.0_{\pi^0}(3.7_{\pi^\pm})$& 
		\parbox[t]{4.7cm}{Correction for FSE from {\Ch}PT 
		\cite{Gasser:1986vb,Gasser:1987ah,Colangelo:2005gd}.}\\
[0.0ex] \hline \\[-2.5ex]
QCDSF/UKQCD 07 	&\cite{QCDSFUKQCD}	&2  &1.4,\dots,2.6&4.2& 
		\parbox[t]{4.7cm}{Correction for FSE from {\Ch}PT}\\
[0.0ex] \hline\hline
\end{tabular*}
\caption{Finite volume effects in determinations of $f_K/f\pi$.
The subscripts RMS and $\pi,5$ in the case of staggered fermions indicate
the root-mean-square mass and the Nambu-Goldstone boson mass. In the case
of twisted-mass fermions $\pi^0$ and $\pi^\pm$ indicate the neutral and
charged pion mass and where applicable, ``val'' and ``sea'' indicate valence
and sea pion masses.}
}
\end{table}


\clearpage
\newpage
\subsection{Notes to section \ref{sec:LECs} on Low-Energy Constants}

\enlargethispage{20ex}

\begin{table}[!h]
{\footnotesize
\begin{tabular*}{\textwidth}[l]{l @{\extracolsep{\fill}} c c l l}
\hline\hline
\\[-1.2ex]
Collab. & Ref. & $\Nf$ & $a$ [fm]& Description
\\[1.0ex]
\hline \hline
\\[-1.2ex]
HPQCD 13A & \cite{Dowdall:2013rya} & 2+1+1 & 0.09--0.15 & \parbox[t]{6.0cm}{Configurations are shared with MILC.}\\
[1.0ex] \hline \\[-1.2ex]%
ETM 13 & \cite{Cichy:2013gja} & 2+1+1 & 0.0607--0.0863 & \parbox[t]{6.0cm}{Configurations are shared with ETM 11.}\\
[1.0ex] \hline \\[-1.2ex]%
ETM 11 & \cite{Baron:2011sf} & 2+1+1 & 0.0607--0.0863 & \parbox[t]{6.0cm}{Three lattice spacings fixed through $F_\pi/M_\pi$.}\\
[1.0ex] \hline\\[-1.2ex]
ETM 10 & \cite{Baron:2010bv} & 2+1+1 & 0.078, 0.086 & \parbox[t]{6.0cm}{Two lattice spacings fixed through $F_\pi/M_\pi$.}
\\[1.0ex]
\hline\hline
\\[-1.2ex]
BMW 13 & \cite{Durr:2013goa} & 2+1 & 0.054--0.093  & Scale set through Omega baryon mass.\\
[1.0ex] \hline  \\[-1.2ex]
RBC/UKQCD 12 & \cite{Arthur:2012opa} & 2+1 & \parbox[t]{2.5cm}{0.086, 0.114 and 0.144 for $\Mpi^{\rm min}$} & Scale set through Omega baryon mass.\\
[3.0ex] \hline  \\[-1.2ex]
 Borsanyi 12 & \cite{Borsanyi:2012zv} & 2+1 & 0.097--0.284 & Scale fixed through $F_\pi/M_\pi$.\\
[1.0ex] \hline  \\[-1.2ex]
 NPLQCD 11 & \cite{Beane:2011zm} & 2+1 & 0.09, 0.125 & \parbox[t]{6.0cm}{Configurations are shared with MILC 09 \cite{Bazavov:2009bb}.}\\
[1.0ex] \hline  \\[-1.2ex]
 MILC 10, 10A & \cite{Bazavov:2010hj,Bazavov:2010yq} & 2+1 & 0.045--0.09 & \parbox[t]{6.0cm}{3 lattice
spacings, continuum extrapolation by means of RS{\Ch}PT.}\\
[3.0ex] \hline  \\[-1.2ex]
JLQCD/TWQCD 10 & \cite{Fukaya:2010na} & 2+1, 3 & 0.11 & \parbox[t]{6.0cm}{One lattice spacing, scale fixed through $m_\Omega$.}
  \\
[1.0ex] \hline\\[-1.2ex]
RBC/UKQCD 9, 10A & \cite{Boyle:2009xi, Aoki:2010dy} & 2+1 & \parbox[t]{1.7cm}{0.1106(27), 0.0888(12)} & \parbox[t]{6.0cm}{Two lattice spacings. Data combined in global chiral-continuum fits.}
  \\
[3.0ex] \hline\\[-1.2ex]
JLQCD 09 & \cite{Fukaya:2009fh} & 2+1 & 0.1075(7) &  \parbox[t]{6.0cm}{Scale fixed through $r_0$.
} \\
[1.0ex] \hline \\[-1.2ex]
MILC 09, 09A   & \cite{Bazavov:2009bb,Bazavov:2009fk}  & 2+1 & 0.045--0.18 & \parbox[t]{6.0cm}
 {Total of 6 lattice spacings, continuum extrapolation by means of RS{\Ch}PT.} \\
[3.0ex] \hline \\[-1.2ex]
TWQCD 08 & \cite{Chiu:2008jq} & 2+1 & 0.122(3) & \parbox[t]{6.0cm}{Scale fixed through $m_\rho$, $r_0$. 
}  \\
[1.0ex] \hline\\[-1.2ex]
JLQCD/TWQCD 08B & \cite{Chiu:2008kt} & 2+1 &  0.1075(7) & \parbox[t]{6.0cm}{Scale fixed through $r_0$. 
}   \\
[1.0ex] \hline \\[-1.2ex]
PACS-CS 08   & \cite{Aoki:2008sm}     & 2+1 & 0.0907       & \parbox[t]{6.0cm}
 {One lattice spacing.
} \\
[1.0ex] \hline \\[-1.2ex]
RBC/UKQCD 08 & \cite{Allton:2008pn}   & 2+1 & 0.114        & \parbox[t]{6.0cm}
 {One lattice spacing, attempt to estimate cut-off effects via
 formal argument.} \\
[3.0ex]\hline \\[-1.2ex]
RBC/UKQCD 08A & \cite{Boyle:2008yd} & 2+1 &0.114 & \parbox[t]{6.0cm}
 {Only one lattice spacing, attempt to estimate size of cut-off effects via
 formal argument.} \\
[3.0ex]\hline \\[-1.2ex]
NPLQCD 06 & \cite{Beane:2006kx} & 2+1 & 0.125 & \parbox[t]{6.0cm}{One lattice spacing, continuum {\Ch}PT used.} \\
[1.0ex]\hline \\[-1.2ex]
LHP 04 & \cite{Bonnet:2004fr} & 2+1 & $\simeq 0.12 $ & \parbox[t]{6.0cm}{Only one lattice spacing, mixed discretization approach.}\\
[3.0ex]\hline \\[-1.2ex]
\end{tabular*}
\caption{Continuum extrapolations/estimation of lattice artifacts in
  $\Nf=2+1+1$ and $2+1$ determinations of the Low-Energy Constants.}
}
\end{table}

\clearpage

\begin{table}[!h]
{\footnotesize
\begin{tabular*}{\textwidth}[l]{l @{\extracolsep{\fill}} c c l l}
\hline\hline \\[-1.2ex]
Collab. & Ref. & $\Nf$ & $a$ [fm]& Description
\\[1.0ex] \hline \hline \\[-1.2ex]
G\"ulpers 13& \cite{Gulpers:2013uca} & 2 & 0.063 & \parbox[t]{7.5cm}{Scale fixed through omega Baryon mass.
}\\
[1.0ex] \hline \\[-1.2ex]%
Brandt 13& \cite{Brandt:2013dua} & 2 & 0.05--0.08 & \parbox[t]{7.5cm}{Configurations are shared with CLS.
}\\
[1.0ex] \hline \\[-1.2ex]%
QCDSF 13& \cite{Horsley:2013ayv} & 2 & 0.06--0.076 & \parbox[t]{7.5cm}{Scale fixed through $r_0 = 0.50(1)$ fm.
}\\
[1.0ex] \hline \\[-1.2ex]%
ETM 13& \cite{Cichy:2013gja} & 2 & 0.05--0.1 & \parbox[t]{7.5cm}{Configurations are shared with ETM 09C.
}\\
[1.0ex] \hline \\[-1.2ex]%
ETM 12 & \cite{Burger:2012ti} & 2 & 0.05--0.1 & \parbox[t]{7.5cm}{Configurations are shared with ETM 09C.
}\\
[1.0ex] \hline \\[-1.2ex]
Bernardoni 11 & \cite{Bernardoni:2011kd} & 2 & 0.0649(10) & \parbox[t]{7.5cm}{Configurations are shared with CLS.
}\\
[1.0ex] \hline \\[-1.2ex]
TWQCD 11 & \cite{Chiu:2011bm} & 2 & 0.1034(1)(2) & \parbox[t]{7.5cm}
{
Scale fixed through $r_0$.
}\\
[1.0ex] \hline \\[-1.2ex]
TWQCD 11A & \cite{Chiu:2011dz} & 2 & 0.1032(2) & \parbox[t]{7.5cm}{
Scale fixed through $r_0$.
}\\
[1.0ex] \hline \\[-1.2ex]
Bernardoni 10 & \cite{Bernardoni:2010nf} & 2 & 0.0784(10) & \parbox[t]{7.5cm}{Scale fixed through $M_K$. Non-perturbative $O(a)$ improvement.  No estimate of systematic error. }\\
[3.0ex] \hline \\[-1.2ex]
JLQCD/TWQCD 09, 10 & \cite{Fukaya:2010na} & 2 & 0.11
& \parbox[t]{7.5cm}{One lattice spacing fixed
through $r_0$.}
 \\
[1.0ex] \hline  \\[-1.2ex]
%
%
ETM 09B & \cite{Jansen:2009tt} & 2 & 0.063, 0.073 & \parbox[t]{7.5cm}{Automatic $O(a)$ impr.
$r_0=0.49$\,fm used.}\\ 
[1.0ex] \hline  \\[-1.2ex]
ETM 09C & \cite{Baron:2009wt}  & 2 &  0.051--0.1 & \parbox[t]{7.5cm}{Automatic $O(a)$ impr. Scale fixed through $F_\pi$. 4 lattice spacings, continuum extrapolation.}\\
[3.0ex] \hline \\[-1.2ex]
ETM 08 & \cite{Frezzotti:2008dr} & 2 & 0.07-0.09             & \parbox[t]{7.5cm}{Automatic $O(a)$ impr. Two lattice spacings. Scale fixed through $F_\pi$.   }              \\
[3.0ex] \hline \\[-1.2ex]
\parbox[t]{3.5 cm}{JLQCD/TWQCD 08A \\JLQCD 08A} & \parbox[t]{0.7
  cm}{\cite{Noaki:2008iy} \\ \cite{Shintani:2008qe}} & 2 & 0.1184(3)(21)
& \parbox[t]{7.5cm}{Automatic $O(a)$ impr., exact chiral symmetry. Scale fixed through $r_0$. 
}\\
[3.0ex] \hline \\[-1.2ex]
CERN 08 & \cite{Giusti:2008vb} & 2 & 0.0784(10) & \parbox[t]{7.5cm}{Scale fixed through $M_K$. Non-perturbative $O(a)$ improvement.
}\\
[3.0ex] \hline \\[-1.2ex]
Hasenfratz 08 & \cite{Hasenfratz:2008ce} & 2 & 0.1153(5) & \parbox[t]{7.5cm}{Tree level $O(a)$
improvement. Scale fixed through $r_0$. Estimate of lattice artifacts via W{\Ch}PT \cite{Bar:2008th}.}\\
[3.0ex] \hline \\[-1.2ex]
JLQCD/TWQCD 07  & \cite{Fukaya:2007pn} & 2 & 0.1111(24) & \parbox[t]{7.5cm}{Automatic $O(a)$ impr., exact chiral symmetry. Scale fixed through $r_0$.
} \\
[3.0ex] \hline \\[-1.2ex]
JLQCD/TWQCD 07A & \cite{Aoki:2007pw} & 2 & $\simeq$ 0.12 & \parbox[t]{7.5cm}{Automatic $O(a)$ impr., exact chiral symmetry. Scale fixed through $r_0$.
} \\
[3.0ex] \hline  \\[-1.2ex]
 CERN-TOV 06 & \cite{DelDebbio:2006cn} & 2 & \parbox[t]{1.0cm}{0.0717(15), 0.0521(7),  0.0784(10)} &\parbox[t]{7.5cm}{Scale fixed through $M_K$. The lattice with $a=0.0784(10)$ is obtained with non-perturbative $O(a)$ improvement.
}\\
[7.0ex] \hline \\[-1.2ex]
QCDSF/UKQCD 06A & \cite{Brommel:2006ww} & 2& 0.07-0.115 &  \parbox[t]{7.5cm}{5 lattice spacings. Non-perturbative $O(a)$ improvement. Scale fixed through $r_0$.}\\
[3.0ex] \hline \hline \\[-1.2ex]
\end{tabular*}
\caption{Continuum extrapolations/estimation of lattice artifacts in $\Nf=2$
  determinations of the Low-Energy Constants.}
}
\end{table}

\clearpage


\begin{table}[!h]
{\footnotesize
\begin{tabular*}{\textwidth}[l]{l @{\extracolsep{\fill}} c c c l}
\hline\hline \\[-1.2ex]
Collab. & Ref. & $\Nf$ & $M_{\pi,\text{min}}$ [MeV] & Description
\\[1.0ex] \hline \hline \\[-1.2ex]
HPQCD 13A& \cite{Dowdall:2013rya} & 2+1+1 & 128 & \parbox[t]{7.5cm}{NLO chiral fit.
}\\
[1.0ex] \hline \\[-1.2ex]%
ETM 13& \cite{Cichy:2013gja} & 2+1+1 & 270 & \parbox[t]{7.5cm}{Linear fit in the quark mass.
}\\
[1.0ex] \hline \\[-1.2ex]%
ETM 11 & \cite{Baron:2011sf} & 2+1+1 & 270 & \parbox[t]{7.0cm}{NLO SU(2) chiral fit.}\\
[2.0ex] \hline\\[-1.2ex]
ETM 10 & \cite{Baron:2010bv} & 2+1+1 &  $270$ & \parbox[t]{5.9cm}{SU(2) NLO and NNLO fits. }\\
[1.0ex] \hline\hline\\[-1.2ex]
BMW 13 & \cite{Durr:2013goa} & 2+1 & 120 & \parbox[t]{5.9cm}{NLO and NNLO SU(2) fits tested with x and $\xi$ expansion.}\\
[3.0ex] \hline  \\[-1.2ex]
RBC/UKQCD 12 & \cite{Arthur:2012opa} & 2+1 & \parbox[t]{2.2cm}{293 plus run at 171, 246} & \parbox[t]{5.9cm}{NLO SU(2) ChPT incl.\ finite-V and some discr.\ effects} \\
[3.0ex] \hline  \\[-1.2ex]
Borsanyi 12 & \cite{Borsanyi:2012zv} & 2+1 & 135 &  NNLO SU(2) chiral fit.\\
[1.0ex] \hline  \\[-1.2ex]
NPLQCD 11 & \cite{Beane:2011zm} & 2+1 & 235 & NNLO SU(2) mixed action $\chi$PT.\\
[1.0ex] \hline  \\[-1.2ex]
MILC 10, 10A       & \cite{Bazavov:2010hj,Bazavov:2010yq}  & 2+1 &     &  Cf.~MILC 09A. \\
[1.0ex] \hline \\[-1.2ex]
JLQCD/TWQCD 09, 10 & \cite{Fukaya:2010na} & 2+1,3 &
\parbox[t]{1.8cm}{ 100($\epsilon$-reg.),\\ 290($p$-reg.)}
& \parbox[t]{5.9cm}{$N_f=2+1$ runs both in $\epsilon$- and $p$-regime; $N_f=3$ runs only in $p$-regime. NLO $\chi$PT fit of the spectral density interpolating the two regimes.}
\\
[9.0ex] \hline\\[-1.2ex]
RBC/UKQCD 9, 10A & \cite{Boyle:2009xi, Aoki:2010dy} & 2+1 & 290--420 & \parbox[t]{5.9cm}{Valence pions mass is 225-420 MeV. NLO SU(2) $\chi$PT fit.} \\
[3.0ex] \hline\\[-1.2ex]
%
%
MILC 09, 09A      & \cite{Bazavov:2009bb,Bazavov:2009fk}  & 2+1 &  $258$ & \parbox[t]{5.9cm}
 {
 Lightest Nambu-Goldstone mass is 224\,MeV and
 lightest RMS mass is 258\,MeV (at 0.06\,fm). }\\
[5.0ex] \hline \\[-1.2ex]
%
%
TWQCD 08 & \cite{Chiu:2008jq} & 2+1 &  \parbox[t]{2.2cm}{{$m_{ud}=m_s/4$},  $m_s\sim$ phys.}  & \parbox[t]{5.9cm}{Quark condensate extracted from topological susceptibility, LO chiral fit.}  \\
[3.0ex] \hline\\[-1.2ex]
JLQCD/TWQCD 08B & \cite{Chiu:2008kt} & 2+1 & \parbox[t]{2.2cm}{$m_{ud}\geq m_s/6$, \\ $m_s\sim$ phys.}  &  \parbox[t]{5.9cm}{Quark condensate extracted from topological susceptibility, LO chiral fit.} \\
[3.0ex] \hline \\[-1.2ex]
PACS-CS 08   & \cite{Aoki:2008sm}     & 2+1 &  $156$ & \parbox[t]{5.9cm}
 {To date, lightest published quark mass reached in a direct simulation.} \\
[3.0ex] \hline \\[-1.2ex]
RBC/UKQCD 08 & \cite{Allton:2008pn}   & 2+1 &  $330$ & \parbox[t]{5.9cm}
 {
 Lightest velence pion mass is $242\MeV$.} \\
[1.0ex] \hline \\[-1.2ex]
RBC/UKQCD 08A & \cite{Boyle:2008yd} & 2+1 & 330 & \parbox[t]{5.9cm}{Pion electromagnetic form factor computed at one pion mass.
}\\
[3.0ex] \hline \\[-1.2ex]
NPLQCD 06 & \cite{Beane:2006kx} & 2+1 & $460$  & \parbox[t]{5.9cm}{Value refers to lightest RMS mass at $a=0.125$\,fm as quoted in \cite{Bazavov:2009fk}.} \\
[3.0ex]\hline \\[-1.2ex]
LHP 04 & \cite{Bonnet:2004fr} & 2+1 &  $318$ &  \parbox[t]{5.9cm}{Pion form factor extracted from vector meson dominance fit.}\\
[3.0ex]\hline \hline\\[-1.2ex]
\end{tabular*}
\caption{Chiral extrapolation/minimum pion mass in $\Nf=2+1+1$ and $2+1$
  determinations of the Low-Energy Constants.}
}
\end{table}

\clearpage

\begin{table}[!h]
{\footnotesize
\begin{tabular*}{\textwidth}[l]{l @{\extracolsep{\fill}} c c c l}
\hline\hline
\\[-1.2ex]
Collab. & Ref. & $\Nf$ & $M_{\pi,\text{min}}$ [MeV] & Description \\
[1.0ex] \hline \hline \\[-1.2ex]
G\"ulpers 13& \cite{Gulpers:2013uca} & 2 & 280 & \parbox[t]{7.5cm}{NLO ChPT fit.
}\\
[1.0ex] \hline \\[-1.2ex]%
Brandt 13& \cite{Brandt:2013dua} & 2 & 280 & \parbox[t]{7.5cm}{Configurations are shared with CLS.
}\\
[1.0ex] \hline \\[-1.2ex]%
QCDSF 13& \cite{Horsley:2013ayv} & 2& 130 & \parbox[t]{7.5cm}{Fit with ChPT + analytic.
}\\
[1.0ex] \hline \\[-1.2ex]%
ETM 13& \cite{Cichy:2013gja} & 2 & 260 & \parbox[t]{7.5cm}{Configurations are shared with ETM 09C.
}\\
[1.0ex] \hline \\[-1.2ex]%
ETM 12 & \cite{Burger:2012ti} & 2 & 260 & \parbox[t]{7.5cm}{Configurations are shared with ETM 09C.
}\\
[1.0ex] \hline \\[-1.2ex]
Bernardoni 11 & \cite{Bernardoni:2011kd} & 2 & 312 & \parbox[t]{5.5cm}{Overlap valence + O(a) improved Wilson sea, mixed regime $\chi$PT.}
\\
[3.0ex] \hline \\[-1.2ex]
TWQCD 11 & \cite{Chiu:2011bm} & 2 & 230 & \parbox[t]{5.5cm}
{
NLO SU(2) $\chi$PT fit.
}\\
[1.0ex] \hline \\[-1.2ex]
TWQCD 11A & \cite{Chiu:2011dz} & 2 & 220 & \parbox[t]{5.5cm}{
NLO $\chi$PT (infinite $V$) for topological susceptibility $\chi_{\rm top}$.
}\\
[3.0ex] \hline \\[-1.2ex]
Bernardoni 10 & \cite{Bernardoni:2010nf}&  2& 297, 377, 426 & \parbox[t]{5.5cm}{NLO SU(2) fit of $\chi_{\rm top}$.}\\
[1.0ex] \hline \\[-2.0ex]
JLQCD/TWQCD 10 & \cite{Fukaya:2010na} & 2 &  \parbox[t]{3cm}{$\sqrt{2m_{\rm min}\Sigma}/F$=120 ($\epsilon$-reg.), 290 ($p$-reg.)}
& \parbox[t]{5.5cm}{Data both in the $p$ and $\epsilon$-regime. NLO chiral fit of the spectral density interpolating the two regimes.
} \\
[3.0ex]\hline \\[-2.0ex]
JLQCD/TWQCD 09 & \cite{JLQCD:2009qn} & 2 & 290  &  \parbox[t]{5.5cm}{LECs extracted from NNLO chiral fit of vector and scalar radii $\langle r^2\rangle^\pi_{V,S}$.}\\
[3.0ex]\hline \\[-1.2ex]
ETM 09B & \cite{Jansen:2009tt} & 2 & \parbox{3.5cm}{
$\sqrt{2m_{\rm min}\Sigma}/F$=85
} &  \parbox[t]{5.5cm}{NLO SU(2) $\epsilon$-regime fit.}\\
[1.0ex]\hline \\[-1.2ex]
ETM 09C & \cite{Baron:2009wt} & 2 & $280$    & \parbox[t]{5.5cm}{NNLO SU(2) fit. 
} \\
[1.0ex] \hline \\[-1.2ex]
ETM 08  & \cite{Frezzotti:2008dr} & 2 &   $260$  &   \parbox[t]{5.5cm}{From pion form factor using NNLO {\Ch}PT and exp. value of $\langle r^2\rangle^\pi_{S}$.   }        \\
[3.0ex] \hline \\[-1.2ex]
\parbox[t]{3.5 cm}{JLQCD/TWQCD 08A \\JLQCD 08A} & \parbox[t]{0.7
  cm}{\cite{Noaki:2008iy} \\ \cite{Shintani:2008qe}} & 2 & 290 &
\parbox[t]{5.5cm}{NNLO SU(2) fit.
}\\
[1.0ex] \hline \\[-1.2ex]
CERN 08 & \cite{Giusti:2008vb} &  2 & $m_{q,\text{min}}$=13 MeV & \parbox[t]{5.5cm}{NLO SU(2) fit for the mode number.
}\\
[1.0ex] \hline \\[-1.2ex]
Hasenfratz 08 & \cite{Hasenfratz:2008ce} & 2 &  \parbox[t]{3.5cm}{$\sqrt{2m_{\rm min}\Sigma}/F$=220
}
&  NLO SU(2) $\epsilon$-regime fit.
\\
[1.0ex] \hline \\[-1.2ex]
JLQCD/TWQCD 07  & \cite{Fukaya:2007pn} & 2 & $\sqrt{2m_{\rm min}\Sigma}/F$=120 & \parbox[t]{5.5cm}{NLO SU(2) $\epsilon$-regime fit.} \\
[1.0ex]\hline \\[-1.2ex]
JLQCD/TWQCD 07A & \cite{Aoki:2007pw} & 2 & $m_{ud}= m_s/6-m_s$ & \parbox[t]{5.5cm}{Quark condensate from topological susceptibility, LO chiral fit.}\\
[3.0ex] \hline \\[-1.2ex]
 CERN-TOV 06 & \cite{DelDebbio:2006cn} & 2 & 403, 381, 377 & \parbox[t]{5.5cm}{NLO SU(2) fit.
}\\
[1.0ex] \hline \\[-1.2ex]
QCDSF/UKQCD 06A & \cite{Brommel:2006ww} & 2 & 400 & \parbox[t]{5.5cm}{Several  fit functions to extrapolate the pion form factor.} \\
[3.0ex] \hline \hline \\[-1.2ex]
\end{tabular*}
\caption{Chiral extrapolation/minimum pion mass in $\Nf=2$ determinations of the Low-Energy Constants.}
}
\end{table}

\clearpage


\begin{table}[!h]
{\footnotesize
\begin{tabular*}{\textwidth}[l]{l @{\extracolsep{\fill}} c c c c l}
\hline\hline \\[-1.2ex]
Collab.\ & Ref.\ & $\Nf$ & $L$ [fm] & $M_{\pi,\text{min}}L$ & Description
\\[1.0ex] \hline \hline \\[-1.2ex]
HPQCD 13A& \cite{Dowdall:2013rya} & 2+1+1 & 4.8--5.5 & 3.3 & \parbox[t]{7.5cm}{
3 volumes are compared.
}\\
[1.0ex] \hline \\[-1.2ex]%
ETM 13& \cite{Cichy:2013gja} & 2+1+1 & 1.9--2.8 & 3.0 &\parbox[t]{7.5cm}{
4 volumes compared.
}\\
[1.0ex] \hline \\[-1.2ex]%
ETM 11 & \cite{Baron:2011sf} & 2+1+1 & 1.9--2.8 & 3.0 & See \cite{Baron:2010bv}.\\
[1.0ex] \hline\\[-1.2ex]
ETM 10 & \cite{Baron:2010bv} & 2+1+1 & 1.9-2.8 & 3.0 &
\parbox[t]{4.25cm}{FSE estimate using \cite{Colangelo:2005gd}.
  $M_{\pi^+} L\gtrsim 4$, but $M_{\pi^0} L \sim 2$. }\\
[3.0ex] \hline\hline\\[-1.2ex]
BMW 13 & \cite{Durr:2013goa} & 2+1 & 2.1 & 3.0 & 3 volumes are compared.\\
[1.0ex] \hline  \\[-1.2ex]
RBC/UKQCD 12 & \cite{Arthur:2012opa} & 2+1 & 2.7--4.6 & $>4$ & FSE seem to be very small. \\
[1.0ex] \hline  \\[-1.2ex]
 Borsanyi 12 & \cite{Borsanyi:2012zv} & 2+1 & 3.9 & 3.3 & Expected to be less than 1\%. \\
[1.0ex] \hline  \\[-1.2ex]
 NPLQCD 11 & \cite{Beane:2011zm} & 2+1 & 2.5--3.5 & 3.6 & Expected to be less than 1\%. \\
[1.0ex] \hline  \\[-1.2ex]
MILC 10, 10A       & \cite{Bazavov:2010hj,Bazavov:2010yq}                 & 2+1 & 2.52 &
4.11  & \parbox[t]{4.8cm}{$L\!\geq\!2.9\,\fm$ for lighter masses.}\\
[1.0ex] \hline \\[-1.2ex]
JLQCD/TWQCD 09, 10 & \cite{Fukaya:2010na} & 2+1, 3 & 1.9, 2.7 &
& \parbox[t]{4.25cm}{
2 volumes are compared for a fixed quark mass.
}\\
[2.0ex]\hline\\[-1.2ex]
RBC/UKQCD 9, 10A & \cite{Boyle:2009xi, Aoki:2010dy} & 2+1 &  2.7   & $\simeq$ 4  & \parbox[t]{4.25cm}{
FSE estimated using $\chi$PT.} \\
[1.0ex] \hline\\[-1.2ex]
%
%
MILC 09, 09A  & \cite{Bazavov:2009bb,Bazavov:2009fk}  & 2+1 & 2.4/2.9 & 3.5/4.11 & \parbox[t]{4.8cm}
{$L\!\geq\!2.9\,\fm$ for lighter masses.}
\\
[1.0ex] \hline \\[-1.2ex]
TWQCD 08 & \cite{Chiu:2008jq} & 2+1 &  1.95  &  - &  \parbox[t]{4.25cm}{No estimate of FSE.} \\
[1.0ex] \hline\\[-1.2ex]
JLQCD/TWQCD 08B & \cite{Chiu:2008kt} & 2+1 &1.72  &  - &  \parbox[t]{4.25cm}{Fixing topoloical charge (to $\nu=0$) gives FSE \protect\cite{Aoki:2007ka}.} \\
[3.0ex]\hline\\[-1.2ex]
 PACS-CS 08   & \cite{Aoki:2008sm}     & 2+1 & 2.9     & 2.3      & \parbox[t]{4.25cm}
 {FSE is the main concern of the authors.} \\
[3.0ex] \hline \\[-1.2ex]
RBC/UKQCD 08 & \cite{Allton:2008pn}   & 2+1 & 2.74    & 4.6      & \parbox[t]{4.25cm}{FSE by means of {\Ch}PT.} \\
[1.0ex]\hline \\[-1.2ex]
RBC/UKQCD 08A & \cite{Boyle:2008yd} & 2+1 & 2.74 & 4.6 & \parbox[t]{4.25cm}{FSE estimated to be  $< 1 \%$ using $\chi$PT.}\\
[3.0ex] \hline \\[-1.2ex]
NPLQCD 06 & \cite{Beane:2006kx} & 2+1 &  2.5    & 3.7     & \parbox[t]{4.25cm}{Value refers to lightest valence pion mass.} \\
[3.0ex]\hline \\[-1.2ex]
 LHP 04 & \cite{Bonnet:2004fr} & 2+1 & $\simeq 2.4$ & 3.97 &  \parbox[t]{4.25cm}{Value refers to domain-wall valence pion mass.}\\
[3.0ex] \hline \hline\\[-1.2ex]
\end{tabular*}
\caption{Finite volume effects in $\Nf=2+1+1$ and $2+1$ determinations of the Low-Energy Constants.}
}
\end{table}

\clearpage

\begin{table}[!h]
{\footnotesize
\begin{tabular*}{\textwidth}[l]{l @{\extracolsep{\fill}} c c c c l}
\hline\hline \\[-1.2ex]
Collab.\ & Ref.\ & $\Nf$ & $L$ [fm] & $M_{\pi,\text{min}}L$ & Description
\\[1.0ex] \hline \hline \\[-1.2ex]
G\"ulpers 13& \cite{Gulpers:2013uca}& 2 & 4--6 & 4.3 &  \parbox[t]{7.5cm}{Configs. shared with CLS.
}\\
[1.0ex] \hline \\[-1.2ex]%
Brandt 13& \cite{Brandt:2013dua} & 2 & $\sim 5$ & 4 &  \parbox[t]{7.5cm}{Configs. shared with CLS.
}\\
[1.0ex] \hline \\[-1.2ex]%
QCDSF 13& \cite{Horsley:2013ayv} & 2& 1.8--2.4 & 2.7 &  \parbox[t]{7.5cm}{ NLO ChPT is used for FSE.
}\\
[1.0ex] \hline \\[-1.2ex]%
ETM 13& \cite{Cichy:2013gja} & 2 & 2.0--2.5 & 3--4 & \parbox[t]{7.5cm}{Configs. shared with ETM 09C.
}\\
[1.0ex] \hline \\[-1.2ex]%
ETM 12 & \cite{Burger:2012ti} & 2 & 2.0--2.5 & 3--4 &\parbox[t]{7.5cm}{Configs. shared with ETM 09C.
}\\
[1.0ex] \hline \\[-1.2ex]
Bernardoni 11 & \cite{Bernardoni:2011kd} & 2 & 1.56 & 2.5  & \parbox[t]{5.25cm}{
Mixed regime $\chi$PT for FSE used.
}\\
[1.0ex] \hline \\[-1.2ex]
TWQCD 11 & \cite{Chiu:2011bm} & 2 & 1.65 & 1.92 & \parbox[t]{5.25cm}
{
SU(2) $\chi$PT is used for FSE.
}\\
[1.0ex] \hline \\[-1.2ex]
TWQCD 11A & \cite{Chiu:2011dz} & 2 & 1.65 & 1.8 & \parbox[t]{5.25cm}{
No estimate of FSE.
}\\
[1.0ex] \hline \\[-1.2ex]
Bernardoni 10 & \cite{Bernardoni:2010nf} & 2 & 1.88 & 2.8 & \parbox[t]{5.25cm}{FSE included in the NLO chiral fit.}  \\
[1.0ex] \hline \\[-2.0ex]
JLQCD/TWQCD 10 & \cite{Fukaya:2010na} & 2 & 1.8-1.9 &  & \parbox[t]{5.25cm}{
FSE estimated from
different topological sectors.
} \\
[3.0ex] \hline  \\[-1.2ex]
JLQCD/TWQCD 09 & \cite{JLQCD:2009qn} & 2 & 1.89 & 2.9
&  \parbox[t]{5.25cm}{
FSE by NLO  $\chi$PT,
Additional FSE for fixing topology \protect\cite{Aoki:2007ka}.
} \\
[3.0ex] \hline  \\[-2.0ex]
ETM 09B & \cite{Jansen:2009tt} & 2 & 1.3, 1.5 &  $\epsilon$-regime
&  \parbox[t]{5.25cm}{Topology: not fixed. 
2 volumes.
}\\
[1.0ex] \hline  \\[-2.0ex]
ETM 09C & \cite{Baron:2009wt} & 2 & 2.0-2.5   & 3.2--4.4 
& \parbox[t]{5.25cm}{Several volumes. Finite-volume effects estimated through \protect\cite{Colangelo:2005gd}.}\\
[3.0ex] \hline \\[-2.0ex]
ETM 08 & \cite{Frezzotti:2008dr} & 2 & 2.1, 2.8  & 3.4, 3.7 &\parbox[t]{5.25cm}{
Only data with $M_\pi L\gtrsim 4$ are considered.
.}\\
[3.0ex] \hline \\[-2.0ex]
\parbox[t]{3.5 cm}{JLQCD/TWQCD 08A \\JLQCD 08A} & \parbox[t]{0.7
  cm}{\cite{Noaki:2008iy} \\ \cite{Shintani:2008qe}} & 2 & 1.89 & 2.9
&  \parbox[t]{5.25cm}{
FSE estimates through
  \protect\cite{Colangelo:2005gd}. Additional FSE 
for fixing topology \protect\cite{Aoki:2007ka}.} \\
[3.0ex] \hline \\[-1.2ex]
CERN 08 & \cite{Giusti:2008vb}  & 2  & 1.88, 2.51  & - & \parbox[t]{5.25cm}{
Two volumes compared.}  \\
[1.0ex] \hline \\[-1.2ex]
Hasenfratz 08 & \cite{Hasenfratz:2008ce} & 2 & 1.84, 2.77 &  $\epsilon$-regime &  \parbox[t]{5.25cm}{Topology: not fixed, 2 volumes.}\\
[1.0ex] \hline \\[-2.0ex]
JLQCD/TWQCD 07  & \cite{Fukaya:2007pn} & 2 & 1.78 & $\epsilon$-regime & \parbox[t]{5.25cm}{Topology: fixed to $\nu=0$.}\\
[1.0ex]\hline \\[-1.2ex]
JLQCD/TWQCD 07A & \cite{Aoki:2007pw} & 2 & 1.92 & - & \parbox[t]{5.25cm}{Topology fixed to $\nu=0$ \protect\cite{Aoki:2007ka}.
} \\
[1.0ex] \hline  \\[-1.2ex]
CERN-TOV 06 & \cite{DelDebbio:2006cn} & 2 & 1.72, 1.67, 1.88 & 3.5, 3.2, 3.6 & \parbox[t]{5.25cm}{No estimate for FSE.}  \\
[1.0ex] \hline \\[-1.2ex]
QCDSF/UKQCD 06A & \cite{Brommel:2006ww} & 2 & 1.4-2.0 & 3.8 & \parbox[t]{5.25cm}{NLO $\chi$PT estimate for FSE \protect\cite{Bunton:2006va}.} \\
[1.0ex] \hline \hline \\[-1.2ex]
\end{tabular*}
\caption{Finite volume effects in $\Nf=2$ determinations of the Low-Energy Constants.}
}
\end{table}

\clearpage


\begin{table}[!h]
{\footnotesize
\begin{tabular*}{\textwidth}[l]{l @{\extracolsep{\fill}} c c l}
\hline\hline \\[-1.2ex]
Collab. & Ref. & $\Nf$ &  Description
\\[1.0ex] \hline \hline \\[-1.2ex]
HPQCD 13A& \cite{Dowdall:2013rya} & 2+1+1 & ---  \\
[1.0ex] \hline \\[-1.2ex]%
ETM 13& \cite{Cichy:2013gja} & 2+1+1 & Non-perturbative\\
[1.0ex] \hline \\[-1.2ex]%
ETM 11 & \cite{Baron:2011sf} & 2+1+1 & Non-perturbative\\
[1.0ex] \hline\\[-1.2ex]
ETM 10 & \cite{Baron:2010bv}  & 2+1+1 & Non-perturbative\\
[1.0ex] \hline\hline \\[-1.2ex]
BMW 13 & \cite{Durr:2013goa} & 2+1 & Non-perturbative\\
[1.0ex] \hline  \\[-1.2ex]
RBC/UKQCD 12 & \cite{Arthur:2012opa} & 2+1 & Non-perturbative (RI/SMOM)\\
[1.0ex] \hline  \\[-1.2ex]
Borsanyi 12 & \cite{Borsanyi:2012zv} & 2+1 & \parbox[t]{7cm}{Indirectly non-perturbative
              through \cite{Durr:2010vn} for $\Sigma$; no renormalization needed for $F$,
              since only $\Fpi/F$ computed and scale set through $\Fpi$.}\\
[6.0ex] \hline  \\[-1.2ex]
NPLQCD 11 & \cite{Beane:2011zm} & 2+1 & Not needed (no result for $\Sigma$).\\
[1.0ex] \hline  \\[-1.2ex]
JLQCD/TWQCD 10 & \cite{Fukaya:2010na} & 2+1, 3 &  Non-perturbative\\
[1.0ex]\hline \\[-1.2ex]
MILC 10, 10A & \cite{Bazavov:2010hj, Bazavov:2010yq}  & 2+1 &  2 loop \\
[1.0ex] \hline \\[-1.2ex]
RBC/UKQCD 10A & \cite{Aoki:2010dy}   & 2+1 &  Non-perturbative\\
[1.0ex]\hline \\[-1.2ex]
JLQCD 09 & \cite{Fukaya:2009fh} & 2+1 & Non-perturbative\\
[1.0ex] \hline \\[-1.2ex]
MILC 09, 09A & \cite{Bazavov:2009bb,Bazavov:2009fk}  & 2+1 &  2 loop \\
[1.0ex] \hline \\[-1.2ex]
TWQCD 08 & \cite{Chiu:2008jq} & 2+1 &  Non-perturbative\\
[1.0ex] \hline \\[-1.2ex]
JLQCD/TWQCD 08B & \cite{Chiu:2008kt} & 2+1 & Non-perturbative\\
[1.0ex]\hline\\[-1.2ex]
PACS-CS 08   & \cite{Aoki:2008sm}     & 2+1 &  1 loop \\
[1.0ex] \hline \\[-1.2ex]
RBC/UKQCD 08, 08A & \cite{Allton:2008pn,Boyle:2008yd}   & 2+1 &  Non-perturbative\\
[1.0ex]\hline \\[-1.2ex]
NPLQCD 06 & \cite{Beane:2006kx} & 2+1 & --- \\
[1.0ex] \hline \\[-1.2ex]
LHP 04 & \cite{Bonnet:2004fr} & 2+1 & --- \\
[1.0ex] \hline \hline\\[-1.2ex]
 All collaborations  &  & 2 & Non-perturbative\\
[1.0ex] \hline\hline \\[-1.2ex]
\end{tabular*}
\caption{Renormalization in determinations of the Low-Energy Constants.}
}
\end{table}

\clearpage

\clearpage
\subsection{Notes to section \ref{sec:BK} on Kaon $B$-parameter $B_K$}
\label{app-BK}

In the following, we summarize the characteristics (lattice actions,
pion masses, lattice spacings, etc.) of the recent $\Nf = 2+1$ and
$\Nf = 2$ runs. We also provide brief descriptions of how systematic
errors are estimated by the various authors.

\begin{table}[!ht]

{\footnotesize
\begin{tabular*}{\textwidth}{l c c c l}
\hline\hline \\[-1.0ex]
Collab. & Ref. & $\Nf$ & $a$ [fm] & Description 
\\[1.0ex] \hline \hline \\[-1.0ex]
%
SWME 13 & \cite{Bae:2013lja} & 2+1 & 0.12, 0.09, 0.06, 0.045  &
\parbox[t]{5.5cm}{Continuum extrapolation with the coarsest lattice spacing
  omitted; residual combined discretization and sea-quark extrapolationg
error of 1.1\% from difference between linear
  fit in $a^2$, $m_{\rm sea}$ and a constrained nine-parameter extrapolation.}
\\[1.0ex] \hline \\[-1.0ex]
RBC/UKQCD 12 & 
\cite{Arthur:2012opa} & 2+1 & 0.146, 0.114, 0.087  &
\parbox[t]{5.5cm}{Coarsest lattice spacing uses different action.
Combined continuum and chiral fits.}
\\[1.0ex] \hline \\[-1.0ex]
Laiho 11 & \cite{Laiho:2011np} & 2+1 & 0.12, 0.09, 0.06  &
\parbox[t]{5.5cm}{Combined continuum and chiral extrapolation based on SU(3)
  mixed-action partially quenched {\Ch}PT.}
\\[1.0ex] \hline \\[-1.0ex]
SWME 11, 11A & \cite{Kim:2011qg,Bae:2011ff} & 2+1 & 0.12, 0.09, 0.06, 0.045  &
\parbox[t]{5.5cm}{Continuum extrapolation with the coarsest lattice spacing
  omitted; residual discretization error of 1.9\% from difference between
  fit to a constant and a constrained five-parameter extrapolation.}
\\[1.0ex] \hline \\[-1.0ex]
BMW 11 & \cite{Durr:2011ap} & 2+1 & 0.093, 0.077, 0.065, 0.054  &
\parbox[t]{5.5cm}{Combined continuum and chiral extrapolation;
  discretization error of 0.1\% from comparison of $O({\alpha_s}a)$
  and $O(a^2)$ extrapolations.} 
\\[1.0ex] \hline \\[-1.0ex]
RBC/UKQCD 10B & \cite{Aoki:2010pe} & 2+1 & 0.114, 0.087  &
\parbox[t]{5.5cm}{Two lattice spacings. Combined chiral and continuum
fits.} 
\\[1.0ex] \hline \\[-1.0ex]
SWME 10   & \cite{Bae:2010ki} & 2+1 &  0.12, 0.09, 0.06 &
\parbox[t]{5.5cm}{Continuum extrapolation of results obtained at four
lattice spacings; residual discretization error of 0.21\% from
difference to result at smallest lattice spacing.}   
\\[1.0ex] \hline \\[-1.0ex]
Aubin 09 & \cite{Aubin:2009jh} & 2+1 & 0.12, 0.09 & \parbox[t]{5.5cm}{Two
   lattice spacings; quote 0.3\% discretization error, estimated from
   various $a^2$-terms in fit function}
\\[6.0ex] \hline \\[-1.0ex]
RBC/UKQCD 07A, 08 & \cite{Antonio:2007pb,Allton:2008pn} & 2+1 &
0.114(2) & \parbox[t]{5.5cm}{Single lattice spacing; quote 4\%
  discretization error, estimated from the difference between
  computed and experimental values of $f_\pi$.}
\\ [10.0ex] \hline \\[-1.0ex]
HPQCD/UKQCD  06 & \cite{Gamiz:2006sq} & {2+1} & 0.12 & \parbox[t]{5.5cm}{Single
  lattice spacing; 3\% discretization error quoted without providing
  details.}
\\[4.0ex] \hline \\[-1.0ex]
ETM 10A      & \cite{Constantinou:2010qv} & 2 & 0.1, 0.09, 0.07 & 
 \parbox[t]{5.5cm}{Three lattice spacings; 1.2\% error quoted.}
 \\[1.0ex] \hline \\[-1.0ex] 
JLQCD 08 & \cite{Aoki:2008ss} & 2 & 0.118(1) & \parbox[t]{5.5cm}{Single
  lattice spacing; no error quoted.}
\\[1.0ex] \hline \\[-1.0ex]
RBC 04 & \cite{Aoki:2004ht} & 2 & 0.117(4) & \parbox[t]{5.5cm}{Single
   lattice spacing; no error quoted.}
\\[1.0ex] \hline \\[-1.0ex]
UKQCD 04 & \cite{Flynn:2004au} & 2 & 0.10 & \parbox[t]{5.5cm}{Single
   lattice spacing; no error quoted.}
\\[1.0ex]
\hline\hline
\end{tabular*}
\caption{Continuum extrapolations/estimation of lattice artifacts in
  determinations of $B_K$.}
}
\end{table}

\begin{table}[!ht]
{\footnotesize
\begin{tabular*}{\textwidth}{l @{\extracolsep{\fill}} c c c l}
\hline\hline \\[-1.0ex]
{Collab.} & {Ref.} & {$\Nf$} & {$M_{\pi,\rm min}\,[\mev]$} & {Description}  
\\[1.0ex] \hline \hline \\[-1.0ex]
%
SWME 13 & \cite{Bae:2013lja} & 2+1 & 
\parbox[t]{1.8cm}{442/445,\, 299/273,\, 237/256,\, 222/334}
 & \parbox[t]{6cm}{Valence/sea RMS $M_{\pi,\rm min}$ entries correspond to the
four lattice spacings. Chiral extrapolations based on SU(2) staggered
{\Ch}PT at NNLO (with some coefficients fixed by Bayesian priors), and
also including one analytic NNNLO term. Residual error of 0.33\% error from
doubling the widths of Bayesian priors.}
\\[1.0ex] \hline \\[-1.0ex]
RBC/UKQCD 12 & \cite{Arthur:2012opa} & 2+1 & 
\parbox[t]{1.8cm}{140/170,\, 240/330,\, 220/290}  &
\parbox[t]{6.0cm}{Valence/sea
$M_{\pi,\rm min}$ entries correspond to the three lattice
spacings. Combined chiral \& continuum extrapolation, using
$M_\pi<350\;$MeV.}
\\[1.0ex] \hline \\[-1.0ex]
Laiho 11 & \cite{Laiho:2011np} & 2+1 & \parbox[t]{1.8cm}{210/280}
 & \parbox[t]{6cm}{$M_{\pi,\rm min}$ entries correspond to the smallest
  valence/sea quark masses. Chiral \& continuum fits based on NLO mixed action
  {\Ch}PT, including a subset of NNLO terms. Systematic error estimated
  from spread arising from variations in the fit function.}
\\[1.0ex] \hline \\[-1.0ex]
SWME 11, 11A & \cite{Kim:2011qg,Bae:2011ff} & 2+1 & 
\parbox[t]{1.8cm}{442/445,\, 299/325,\, 237/340,\, 222/334}
 & \parbox[t]{6cm}{Valence/sea RMS $M_{\pi,\rm min}$ entries correspond to the
four lattice spacings. Chiral extrapolations based on SU(2) staggered
{\Ch}PT at NNLO (with some coefficients fixed by Bayesian priors), and
also including one analytic NNNLO term. Residual error of 0.33\% error from
doubling the widths of Bayesian priors.}
\\[1.0ex] \hline \\[-1.0ex]
BMW 11 & \cite{Durr:2011ap} & 2+1 & 
\parbox[t]{1.8cm}{219,\, 182,\\ 120,\, 131}
 & \parbox[t]{6cm}{$M_{\pi,\rm min}$ entries correspond to the
four lattice spacings used in the final result. Combined fit to the chiral and continuum
behaviour. Systematics investigated by applying cuts to the maximum pion
mass used in fits. Uncertainty of 0.1\% assigned to chiral fit.} 
\\[1.0ex] \hline \\[-1.0ex]
RBC/UKQCD 10B & \cite{Aoki:2010pe} & 2+1 & 
\parbox[t]{1.8cm}{240/330,\, 220/290}  &
\parbox[t]{6cm}{Valence/sea $M_{\pi,\rm min}$ entries correspond to the two
lattice spacings. Combined chiral and continuum extrapolations.}
\\[1.0ex] \hline \\[-1.0ex]
SWME 10   & \cite{Bae:2010ki} & 2+1 &  
\parbox[t]{1.8cm}{442/445,\, 299/325,\, 237/340} &
\parbox[t]{6cm}{Valence/sea $M_{\pi,\rm min}$ entries correspond to the three
lattice spacings. Chiral extrapolations based on SU(2) staggered {\Ch}PT
at NLO, including some analytic NNLO terms. SU(3) staggered {\Ch}PT as
cross-check. Combined 1.1\% error from various different variations in
the fit procedure.} \\
\hline\hline
\end{tabular*}
\caption{Chiral extrapolation/minimum pion mass in
  determinations of $B_K$.}
}
\end{table}

\begin{table}[!ht]
\addtocounter{table}{-1}
{\footnotesize
\begin{tabular*}{\textwidth}{l @{\extracolsep{\fill}} c c c l}
\hline\hline \\[-1.0ex]
{Collab.} & {Ref.} & {$\Nf$} & {$M_{\pi,\rm min}\,[\mev]$} & {Description}  
\\[1.0ex] \hline \hline \\[-1.0ex]
Aubin 09 & \cite{Aubin:2009jh} & 2+1 & \parbox[t]{1.8cm}{240/370} &
  \parbox[t]{6cm}{ $M_{\pi,\rm min}$ entries correspond to the
  smallest valence/sea quark masses. Chiral \& continuum fits based on
  NLO mixed action {\Ch}PT at NLO, including a subset of NNLO
  terms. Systematic error estimated from spread arising from
  variations in the fit function.}
\\[1.0ex] \hline \\[-1.0ex]
RBC/UKQCD 07A, 08 & {\cite{Antonio:2007pb,Allton:2008pn}} & {2+1} & 
 \parbox[t]{1.8cm}{330} & {\parbox[t]{6cm}{Fits based on SU(2) PQ{\Ch}PT
 at NLO. Effect of neglecting higher orders estimated at 6\% via
 difference between fits based on LO and NLO expressions.}} 
\\[10.0ex] \hline \\[-1.0ex]
HPQCD/UKQCD 06 & {\cite{Gamiz:2006sq}} & {2+1} & \parbox[t]{1.8cm}{360}
    & {\parbox[t]{6cm}{3\% uncertainty from chiral extrapolation 
    quoted, without giving further details.}}
\\[4.0ex] \hline \\[-1.0ex]
ETM 10A & \cite{Constantinou:2010qv} & 2 & \parbox[t]{1.8cm}{400,
  270,\\ 300} & 
  \parbox[t]{6cm}{Each $M_{\pi,\rm min}$ entry corresponds to a
  different lattice spacing. Simultaneous chiral \& continuum
  extrapolations, based on {\Ch}PT at NLO, are carried out. Systematic
  error from several sources, including lattice calibration, quark
  mass calibration, chiral and continuum extrapolation etc., estimated
  at 3.1\%.} 
\\ [21.0ex] \hline \\[-1.0ex]
JLQCD 08 & \cite{Aoki:2008ss} & 2 & 290 & \parbox[t]{6cm}{Fits based on
  NLO PQ{\Ch}PT. Range of validity investigated. Fit error included in
  statistical uncertainty.}  
\\[7.0ex] \hline \\[-1.0ex]
RBC 04 & \cite{Aoki:2004ht} & 2 & 490 & \parbox[t]{6cm}{Fits based on NLO
     PQ{\Ch}PT. Fit error included in statistical uncertainty.} 
\\[4.0ex] \hline \\[-1.0ex]
UKQCD 04 & \cite{Flynn:2004au} & 2 & 780 & \parbox[t]{6cm}{Fits to
  continuum chiral behaviour at fixed sea quark mass. Separate
  extrapolation in sea quark mass. Fit error included in overall
  uncertainty.} 
\\[10.0ex]
\hline\hline
\end{tabular*}
\caption{(cntd.) Chiral extrapolation/minimum pion mass in
  determinations of $B_K$ in two-flavour QCD.}
}
\end{table}

\begin{table}[!ht]
{\footnotesize
\begin{tabular*}{\textwidth}{l @{\extracolsep{\fill}} c c c c l}
\hline\hline \\[-1.0ex]
Collab. & Ref. & $\Nf$ & $L$ [fm] & ${M_{\pi,\rm min}}L$ & Description 
\\[1.0ex] \hline \hline \\[-1.0ex]
%
SWME 13 & \cite{Bae:2013lja} & 2+1
& \parbox[t]{1.8cm}{2.4-3.3, 2.4-5.5, 2.8-3.8, 2.8} & 
$\gtrsim 3.2$ 
& \parbox[t]{5cm}{$L$ entries correspond to the four lattice spacings,
with several volumes in most cases. Finite-volume effects
estimated using NLO {\Ch}PT.}
\\[1.0ex] \hline \\[-1.0ex]
RBC/UKQCD 12 & \cite{Arthur:2012opa} 
& 2+1 & \parbox[t]{1.8cm}{4.6,\,2.7,\,2.8} &
$\gtrsim 3.2$ 
& \parbox[t]{5cm}{
$L$ entries correspond to the three lattice
spacings. Finite volume effects estimated using NLO {\Ch}PT.}
\\[1.0ex] \hline \\[-1.0ex]
Laiho 11 & \cite{Laiho:2011np} & 2+1 & \parbox[t]{1.8cm}{2.4,\,3.4,\,3.8} &
$\gtrsim 3.5$
 & \parbox[t]{5cm}{$L$ entries correspond to the three lattice spacings.
Finite volume effects estimated using NLO {\Ch}PT.}
\\[1.0ex] \hline \\[-1.0ex]
SWME 11, 11A & \cite{Kim:2011qg,Bae:2011ff} & 2+1
& \parbox[t]{1.8cm}{2.4/3.3, 2.4, 2.8, 2.8} & 
$\gtrsim 3.2$ 
& \parbox[t]{5cm}{$L$ entries correspond to the four lattice spacings,
with two volumes at the coarsest lattice. Finite-volume effects
estimated using NLO {\Ch}PT.}
\\[1.0ex] \hline \\[-1.0ex]
BMW 11 & \cite{Durr:2011ap} & 2+1 & \parbox[t]{1.8cm}{6.0, 4.9,
4.2, 3.5} & $\gtrsim 3.8, 3.0$ &
\parbox[t]{5cm}{$L$ entries correspond to the four lattice
spacings, and are the largest of several volumes at each $a$.
${M_{\pi,\rm min}}L\approx3.0$ for the ensemble
at $a\approx0.08\,\fm$. Finite volume effects estimated in {\Ch}PT and
by combined fit to multiple volumes.}
\\[1.0ex] \hline \\[-1.0ex]
RBC/UKQCD 10B & \cite{Aoki:2010pe} & 2+1 & \parbox[t]{1.8cm}{2.7,\,2.8} &  
$\gtrsim 3.1$ 
& \parbox[t]{5cm}{$L$ entries correspond to the three lattice spacings.
Finite volume effects estimated using NLO {\Ch}PT.}
\\[1.0ex] \hline \\[-1.0ex]
SWME 10   & \cite{Bae:2010ki} & 2+1 & \parbox[t]{1.8cm}{2.4/3.3, 2.4,2.8} & 
$\gtrsim 3.4$ 
& \parbox[t]{5cm}{$L$ entries correspond to the three lattice spacings,
with two volumes for the coarsest spacing. 
Finite-volume error of 0.9\% estimated from difference
obtained these two volumes.}
\\[1.0ex] \hline \\[-1.0ex]
Aubin 09 & \cite{Aubin:2009jh} & 2+1 & \parbox[t]{1.8cm}{2.4,\,3.4} & 3.5 &
   \parbox[t]{5cm}{$L$ entries correspond to the two lattice spacings.
Keep $m_{\pi}L\;\gtaeq\;3.5$; no
   comparison of results from different volumes; 0.6\% error
   estimated from mixed action {\Ch}PT correction.}
\\[15.0ex] \hline \\[-1.0ex]
RBC/UKQCD 07A, 08 & \cite{Antonio:2007pb,Allton:2008pn} & 2+1 &
  \parbox[t]{1.8cm}{1.83/2.74} & 4.60 & \parbox[t]{5cm}{Each $L$ entry
  corresponds to a different volume at the same lattice spacing; 1\%
  error from difference in results on two volumes.} 
\\[10.0ex] \hline \\[-1.0ex]
HPQCD/UKQCD 06 & \cite{Gamiz:2006sq} & 2+1 & \parbox[t]{1.8cm}{2.46} &
  \parbox[t]{1.8cm}{4.49} & {Single volume; no error quoted.}
\\
\hline\hline
\end{tabular*}
\caption{Finite volume effects in determinations of $B_K$.
If partially-quenched fits are used,
the quoted $M_{\pi,\rm min}L$ is for lightest valence (RMS) pion.}
}
\end{table}

\begin{table}[!ht]
\addtocounter{table}{-1}
{\footnotesize
\begin{tabular*}{\textwidth}{l @{\extracolsep{\fill}} c c c c l}
\hline\hline \\[-1.0ex]
Collab. & Ref. & $\Nf$ & $L$ [fm] & ${M_{\pi,\rm min}}L$ & Description 
\\[1.0ex] \hline \hline \\[-1.0ex]
ETM 10A & \cite{Constantinou:2010qv} & 2 & \parbox[t]{1.8cm}{2.1,\\
  2.2/2.9,\\ 2.2}& \parbox[t]{1.8cm}{5,\\ 3.3/4.3,\\ 3.3} &  
  \parbox[t]{5cm}{Each $L$ entry corresponds to a different
  lattice spacing, with two volumes at the intermediate lattice
  spacing. Results from these two volumes at $M_{\pi} \sim 300$ MeV are
  compatible.} 
\\ [15.0ex] \hline \\[-1.0ex]
JLQCD 08 & \cite{Aoki:2008ss} & 2 & 1.89 & 2.75 & \parbox[t]{5cm}{Single
  volume; data points with $m_{\rm val} < m_{\rm sea}$ excluded; 5\%
  error quoted as upper bound of PQ{\Ch}PT estimate of the 
  effect.}
\\[10.0ex] \hline \\[-1.0ex]
RBC 04 & \cite{Aoki:2004ht} & 2 & 1.87 & 4.64 & Single volume; no
   error quoted.
\\[1.0ex] \hline \\[-1.0ex]
UKQCD 04 & \cite{Flynn:2004au} & 2 & 1.6 & 6.51 & Single volume; no
   error quoted.
\\[1.0ex] 
\hline\hline
\end{tabular*}
\caption{(cntd.) Finite volume effects in determinations of $B_K$ in
   two-flavour QCD.}
}
\end{table}

\begin{table}[!ht]
{\footnotesize
\begin{tabular*}{\textwidth}{l @{\extracolsep{\fill}} c c c c l}
\hline\hline \\[-1.0ex]
& & & & running & \\
\rb{Collab.} & \rb{Ref.} & \rb{$\Nf$} & \rb{Ren.} & match. & \rb{Description} 
\\[1.0ex] \hline \hline \\[-1.0ex]
%
SWME 13 & \cite{Bae:2013lja} & 2+1 & PT1$\ell$ & PT1$\ell$ &
\parbox[t]{5cm}{Uncertainty from neglecting higher orders estimated at
4.4\% by identifying the unknown 2-loop coefficient with result at the
  smallest lattice spacing.}
\\[1.0ex] \hline \\[-1.0ex]
RBC/UKQCD 12 & \cite{Arthur:2012opa} & 2+1 & RI & PT1$\ell$ &
\parbox[t]{5cm}{Two different RI-SMOM schemes used to estimate
2\% systematic error in conversion to $\msbar$.}
\\[1.0ex] \hline \\[-1.0ex]
Laiho 11 & \cite{Laiho:2011np} & 2+1 & RI & PT1$\ell$ &
\parbox[t]{5cm}{Total uncertainty in matching \& running of 3\%.
  Perturbative truncation error in the conversion to $\msbar$, RGI schemes is
  dominant uncertainty.} 
\\[1.0ex] \hline \\[-1.0ex]
SWME 11, 11A & \cite{Kim:2011qg,Bae:2011ff} & 2+1 & PT1$\ell$ & PT1$\ell$ &
\parbox[t]{5cm}{Uncertainty from neglecting higher orders estimated at
4.4\% by identifying the unknown 2-loop coefficient with result at the
  smallest lattice spacing.}
\\[1.0ex] \hline \\[-1.0ex]
BMW 11 & \cite{Durr:2011ap} & 2+1 & RI & PT1$\ell$ &
\parbox[t]{5cm}{Uncertainty of 0.05\% in the determination of the
  renormalization factor included. 1\% error estimated due to truncation of perturbative
  matching to $\msbar$ and RGI schemes at NLO.}
\\[1.0ex] \hline \\[-1.0ex]
RBC/UKQCD 10B & \cite{Aoki:2010pe} & 2+1 & RI & PT1$\ell$ &
\parbox[t]{5cm}{Variety of different RI-MOM schemes including
non-exceptional momenta. Residual uncertainty of 2\% uncertainty in
running \& matching.}\\[1.0ex] \hline \\[-1.0ex] 
SWME 10  & \cite{Bae:2010ki} & 2+1 & PT1$\ell$ & PT1$\ell$ &
\parbox[t]{5cm}{Uncertainty from neglecting higher orders estimated at
5.5\% by identifying the unknown 2-loop coefficient with result at the
  smallest lattice spacing.}
\\
\hline\hline
\end{tabular*}
\caption{Running and matching in determinations of $B_K$.}
}
\end{table}

\begin{table}[!ht]
\addtocounter{table}{-1}
{\footnotesize
\begin{tabular*}{\textwidth}{l @{\extracolsep{\fill}} c c c c l}
\hline\hline \\[-1.0ex]
& & & & running & \\
\rb{Collab.} & \rb{Ref.} & \rb{$\Nf$} & \rb{Ren.} & match. & \rb{Description} 
\\[1.0ex] \hline \hline \\[-1.0ex]
Aubin 09 & \cite{Aubin:2009jh} & 2+1 & RI & PT1$\ell$ &
  \parbox[t]{5cm}{Total uncertainty in matching \& running of 3.3\%,
  estimated from a number of sources, including chiral extrapolation
  fit ansatz for n.p. determination, strange sea quark mass
  dependence, residual chiral symmetry breaking, perturbative matching
  \& running.}
\\[21.0ex] \hline \\[-1.0ex]
RBC/UKQCD 07A, 08 & {\cite{Antonio:2007pb,Allton:2008pn}} & 
 {2+1} & {RI} & {PT1$\ell$} & {\parbox[t]{5cm}{Uncertainty
 from n.p. determination of ren. factor included in statistical error; 
 2\% systematic error from perturbative matching to $\msbar$
 estimated via size of correction itself.}}
\\[15.0ex] \hline \\[-1.0ex]
HPQCD/UKQCD 06 & {\cite{Gamiz:2006sq}} & {2+1} & {PT1$\ell$}&
    {PT1$\ell$} & {\parbox[t]{5cm}{Uncertainty due to neglecting 2-loop
    order in perturbative matching and running estimated by
    multiplying result by $\alpha^2$.}}
\\[15.0ex] \hline \\[-1.0ex]
ETM 10A & \cite{Constantinou:2010qv} & 2 & RI & PT1$\ell$ &
  \parbox[t]{5cm}{Uncertainty from RI renormalization estimated at 2.5\%.}
\\ [4.0ex] \hline \\[-1.0ex]
JLQCD 08 & \cite{Aoki:2008ss} & 2 & RI & PT1$\ell$ &
 \parbox[t]{5cm}{Uncertainty from n.p. determination of ren. factor 
 included in statistical error; 2.3\% systematic error from
 perturbative matching to $\msbar$ estimated via size of correction
 itself.} 
\\[15.0ex] \hline \\[-1.0ex]
RBC 04 & \cite{Aoki:2004ht} & 2 & RI & PT1$\ell$ &
   \parbox[t]{5cm}{Uncertainty from n.p. determination of ren. factor
 included.}
\\[4.0ex] \hline \\[-1.0ex]
UKQCD 04 & \cite{Flynn:2004au} & 2 & PT1$\ell$ & PT1$\ell$ & No error quoted.
\\[1.0ex] 
\hline\hline
\end{tabular*}
\caption{(cntd.) Running and matching in determinations of $B_K$ in
two-flavour QCD.}
}
\end{table}

\clearpage

\clearpage
\subsection{Notes to section \ref{sec:DDecays} on $D$-meson decay constants and form factors}
\label{app:DDecays}

In the following, we summarize the characteristics (lattice actions,
pion masses, lattice spacings, etc.) of the recent $\Nf = 2+1$ and
$\Nf = 2$ runs. We also provide brief descriptions of how systematic
errors are estimated by the various authors.  We focus on calculations
with either preliminary or published quantitative results.

\subsubsection{$D_{(s)}$-meson decay constants}
\label{app:fD_Notes}
\begin{table}[!htb]

{\footnotesize
\begin{tabular*}{\textwidth}{l @{\extracolsep{\fill}} c c l l l}
\hline\hline \\[-1.65ex]
Collab. & Ref. & $\Nf$ & $a$ [fm]  & Continuum extrapolation & Scale Setting
\\[.2ex] \hline \hline \\[-1.65ex]
\parbox[t]{2cm}{ETM~13F} & \cite{Dimopoulos:2013qfa} & 2+1+1 & \parbox[t]{.9cm}{0.09, 0.08, 0.06}  & \parbox[t]{3.8cm}{Chiral and continuum extrapolations  
performed simultaneously by adding an $O(a^2)$ term to the chiral fit.} &
\parbox[t]{3.5cm}{Relative scale set through $M_{c's'}$, the mass of a fictitious meson made of valence quarks of mass $r_0m_{s'}\!=0.22$ and
$r_0m_{c'}=2.4$. Absoulte scale through $f_\pi$.} 
\\[9.0ex] \hline \\[-1.65ex]
\parbox[t]{2cm}{FNAL/MILC~12B \\ FNAL/MILC~13} & \cite{Bazavov:2012dg,Bazavov:2013nfa} & 2+1+1 & 
	     \parbox[t]{.9cm}{0.15, 0.12, 0.09, 0.06} &
\parbox[t]{3.8cm}{Chiral and continuum extrapolations  performed simultaneously. Central values 
 produced  using a fit function quadratic in $a^2$ and linear in the sea quark mass. In FNAL/MILC~13 terms of $O(a^4)$ are included.} &
 \parbox[t]{3.4cm}{Absolute scale set through $f_\pi$; the uncertainty is propagated into the final error.\\} 
\\
\hline\hline
\end{tabular*}
\caption{Lattice spacings and description of actions used in $N_f=2+1+1$ determinations of the $D$ and $D_s$ meson decay constants.} 
}
\end{table}

\begin{table}[!htb]

{\footnotesize
\begin{tabular*}{\textwidth}{l @{\extracolsep{\fill}} c c l l l}
\hline\hline \\[-1.65ex]
Collab. & Ref. & $\Nf$ & $a$ [fm]  & Continuum extrapolation & Scale Setting
\\[.2ex] \hline \hline \\[-1.65ex]
HPQCD 12A& \cite{Na:2012iu} & 2+1 & \parbox[t]{.9cm}{0.12, 0.09} &
\parbox[t]{3.8cm}{
Chiral and continuum extrapolations  peformed 
simultaneously  using PQHM$\chi$PT augmentd by $a$ dependent terms: $c_0(am_c)^2 + c_1(am_c)^4$.
} & \parbox[t]{3.4cm}{Relative scale set through $r_1$; absolute scale from $f_\pi$, $f_{\rm K}$ and the $\Upsilon$ splitting.
Uncertainties from both $r_1$ and $r_1/a$  propagated.}
\\[9.0ex] \hline \\[-1.65ex]
FNAL/MILC 11 & \cite{Bazavov:2011aa} & 2+1 &
             \parbox[t]{.9cm}{0.15, 0.12, 0.09} &
\parbox[t]{3.8cm}{
Chiral and continuum extrapolations  peformed 
simultaneously using one-loop HM$\chi$PT for rooted staggered quarks. Effects of hyperfine and flavour splittings
are also included. \\
} & \parbox[t]{3.4cm}{Relative scale set through $r_1=0.3117(22)$. The error in $r_1$
comes from the spread of different absolute scale determinations using $f_\pi$, $f_{\rm K}$ and the $\Upsilon$ splitting.}
\\[9.0ex] \hline \\[-1.65ex]
PACS-CS 11 & \cite{Namekawa:2011wt} & 2+1 &
             \parbox[t]{.9cm}{$0.09$} & \parbox[t]{3.8cm}{
Cutoff effects from the heavy-quark action estimated by naive power counting to be at the percent level.\\} 
& \parbox[t]{3.4cm}{Scale set through $m_\Omega$.}
\\[9.0ex] \hline \\[-1.65ex]
HPQCD 10A & \cite{Davies:2010ip} & 2+1 &
\parbox[t]{.9cm}{0.15, 0.12, 0.09, 0.06, 0.044} &
\parbox[t]{3.8cm}{
Chiral and continuum 
extrapolations  performed simultaneously. Polynomials 
up to $am_{c}^8$ are kept (even powers only).
} & \parbox[t]{3.4cm}{See the discussion for HPQCD 12A.}
\\[9.0ex] \hline \\[-1.65ex]
HPQCD/UKQCD 07 & \cite{Follana:2007uv} & 2+1 & 
\parbox[t]{.9cm}{0.15, 0.12, 0.09} &
\parbox[t]{3.8cm}{
Combined chiral and continuum extrapolations using
HM$\chi$PT at NLO augmented by second and third-order polynomial terms in $m_q$ and terms up to $a^4$.
} & \parbox[t]{3.4cm}{Scale set through $r_1$ obtained from the $\Upsilon$ spectrum using the 
non-relativistic QCD action for $b$ quarks. Uncertainty propagated among the systematics.}
\\[9.0ex] \hline \\[-1.65ex]
FNAL/MILC 05 & \cite{Aubin:2005ar} & 2+1 &
\parbox[t]{.9cm}{0.175, 0.121, 0.086} &
\parbox[t]{3.8cm}{Most light-quark cutoff effects are removed through
 NLO HM$\chi$PT for rooted staggered quarks. Continuum values are then obtained 
by averaging the $a \approx 0.12$ and $a \approx 0.09$ fm results.\\}
 &  \parbox[t]{3.4cm}{Scale set through $r_1$ obtained from the $\Upsilon$ spectrum using the 
non-relativistic QCD action for $b$ quarks.}
\\[9.0ex] \hline \\[-2.1ex]
\hline\hline
\end{tabular*}
\caption{Lattice spacings and description of actions used in $N_f=2+1$ determinations of the $D$ and $D_s$ meson decay constants.} 
}
\end{table}

\begin{table}[!htb]


{\footnotesize
\begin{tabular*}{\textwidth}{l @{\extracolsep{\fill}} c c l l l}
\hline\hline \\[-1.65ex]
Collab. & Ref. & $\Nf$ & $a$ [fm]  & Continuum extrapolation & Scale Setting
\\[.2ex] \hline \hline \\[-1.65ex]
\parbox[t]{2.0cm}{ETM~09\\ ETM~11A\\ ETM~13B} & \cite{Blossier:2009bx,Dimopoulos:2011gx,Carrasco:2013zta} & 2 &
	     \parbox[t]{.9cm}{0.10, 0.085, 0.065, 0.054 } &
\parbox[t]{3.8cm}{
NLO SU(2) HM$\chi$PT 
supplemented by terms linear in $a^2$ and in $m_D a^2$ is used in the 
combined chiral/continuum extrapolation.
\\} & \parbox[t]{3.4cm}{Scale set through $f_\pi$.}
\\
\hline\hline
\end{tabular*}
\caption{Lattice spacings and description of actions used in $N_f=2$ determinations of the $D$ and $D_s$ meson decay constants.} 
}
\end{table}


\begin{table}[!htb]

\vspace{-.8cm}

{\footnotesize
\begin{tabular*}{\textwidth}{l @{\extracolsep{\fill}} c c c   l}
\hline\hline \\[-1.0ex]
{Collab.} & {Ref.} & {$\Nf$} & {$M_{\pi,\rm min}\,[\mev]$}  & {Description}  
\\[1.0ex] \hline \hline \\[-1.0ex]
\parbox[t]{2cm}{ETM~13F} & \cite{Dimopoulos:2013qfa} & 2+1+1 &  \parbox[t]{2cm}{245, 238, 211}  & \parbox[t]{5.8cm}{$f_{D_s}$ is extrapolated
using both a quadratic and a linear fit in $m_l$ plus $O(a^2)$ terms. Then the double ratio $(f_{D_s}/f_D)/(f_K/f_\pi)$ is fitted in continuum HM$\chi$PT,
as no lattice spacing dependnce is visible within statistical errors.\\[-1.7ex]}  
\\[5.0ex] \hline \\[-2.0ex]
\parbox[t]{2cm}{FNAL/MILC~12B \\FNAL/MILC~13} & \cite{Bazavov:2012dg,Bazavov:2013nfa} & 2+1+1 & \parbox[t]{2cm}{310, 245, 179, 145} &
\parbox[t]{5.8cm}{Chiral and continuum extrapolations are peformed simultaneously. Central values 
are produced  using a fit function quadratic in $a^2$ and linear in the sea-quark mass.
In FNAL/MILC~13 terms of $O(a^4)$ are included.}
\\[5.0ex] \hline \\[-2.0ex]

HPQCD 12A& \cite{Na:2012iu} & 2+1 & \parbox[t]{2cm}{460, 329} & \parbox[t]{5.8cm}{Chiral and continuum extrapolations
 are peformed simultaneously using PQHM$\chi$PT augmented by $a$ dependent terms: $c_0(am_c)^2 + c_1(am_c)^4$.}
\\[5.0ex] \hline \\[-2.0ex]

FNAL/MILC 11 & \cite{Bazavov:2011aa} & 2+1 & \parbox[t]{2cm}{570, 440, 320} &
\parbox[t]{5.8cm}{Chiral and continuum extrapolations are peformed 
simultaneously using HM$\chi$PT for rooted staggered quarks.
Effects of hyperfine and flavour splittings are also included.}
\\[5.0ex] \hline \\[-2.0ex]
PACS-CS 11 & \cite{Namekawa:2011wt} & 2+1 & \parbox[t]{2cm}{152} & \parbox[t]{5.8cm}{Simulations are  
reweighted in the light- and strange-quark masses to the physical point.}
\\[5.0ex] \hline \\[-2.0ex]
HPQCD 10A & \cite{Davies:2010ip} & 2+1 & \parbox[t]{2cm}{542, 460, 329, 258, 334} & \parbox[t]{5.8cm}{Chiral and continuum 
extrapolations are performed simultaneously. Polynomials up to  $\left( \frac{m_{q,sea}-m_{q,phys}}{m_{q,phys}}\right)^2$ for $q=s,l$ and
up to  $(am_{c})^8$ are kept.}
\\[5.0ex] \hline \\[-2.0ex]
HPQCD/UKQCD 07 & \cite{Follana:2007uv} & 2+1 & \parbox[t]{2cm}{542, 460, 329} & \parbox[t]{5.8cm}{Combined chiral and continuum extrapolations using
HM$\chi$PT at NLO augmented by second and third-order polynomial terms in $m_q$ and terms up to $a^4$.}
\\[5.0ex] \hline \\[-2.0ex]
FNAL/MILC 05 & \cite{Aubin:2005ar} & 2+1 &
\parbox[t]{2cm}{$>~440\;, 440\;, 400$} & \parbox[t]{5.8cm}{Chiral extrapolations are first performed at each lattice spacing
uisng NLO HM$\chi$PT for rooted staggered quarks. Lattice artefacts are then extrapolated linearly in $a^2$. }
\\[5.0ex] \hline \\[-2.0ex]
\parbox[t]{2cm}{ETM 09\\ ETM 11A\\ ETM~13B} & \cite{Blossier:2009bx,Dimopoulos:2011gx,Carrasco:2013zta}  & 2 & \parbox[t]{2cm}{410, 270, 310, 270}
& \parbox[t]{5.8cm}{$M_{\pi,{\rm min}}$ refers to the charged pions.  NLO SU(2) HM$\chi$PT 
supplemented by terms linear in $a^2$ and in $m_D a^2$ is used in the 
combined chiral/continuum extrapolation. To estimate the systematic due to chiral extrapolation,
once $f_{D_s}\sqrt{m_{D_s}}$ and $f_{D_s}\sqrt{m_{D_s}}/(f_{D}\sqrt{m_{D}})$ 
and once $f_{D_s}\sqrt{m_{D_s}}/f_K$ and 
$f_{D_s}\sqrt{m_{D_s}}/f_K \; \times \;  f_\pi/(f_{D}\sqrt{m_{D}})$
are fitted. In ETM~13 the double ratio $(f_{D_s}/f_D)/(f_K/f_\pi)$ is fitted in HM$\chi$PT.\\ }
\\[5.0ex]  \\[-2.0ex]

\hline\hline
\end{tabular*}
\caption{Chiral extrapolation/minimum pion mass in determinations of the
  $D$ and $D_s$ meson decay constants.  For actions with multiple
 species of pions, masses quoted are the RMS pion masses.    The
 different $M_{\pi,\rm min}$ entries correspond to the different lattice
 spacings.} 
}
\end{table}


\begin{table}[!htb]
{\footnotesize
\begin{tabular*}{\textwidth}{l @{\extracolsep{\fill}} c c c c l}
\hline\hline \\[-1.0ex]
Collab. & Ref. & $\Nf$ & $L$ [fm] & ${M_{\pi,\rm min}}L$ & Description 
\\[1.0ex] \hline \hline \\[-1.0ex]
\parbox[t]{2cm}{ETM~13F} & \cite{Dimopoulos:2013qfa} & 2+1+1 & \parbox[t]{1.75cm}{2.13/2.84, 1.96/2.61, 2.97}   & 
\parbox[t]{1.6cm}{$3.5, 3.2, 3.2 $}  &  \parbox[t]{4.7cm}{The comparison of two different volumes at the two largest lattice spacings indicates that FV effects
are below the statistical errors.} 
\\[8.0ex] \hline \\[-1.0ex]
\parbox[t]{2cm}{FNAL/MILC~12B \\ FNAL/MILC~13} & \cite{Bazavov:2012dg,Bazavov:2013nfa} & 2+1+1 & 
	     \parbox[t]{1.75cm}{2.4/4.8, 2.88/5.76, 2.88/5.76, 2.88/5.76} & \parbox[t]{1.6cm}{$3.3, 3.9$, $3.7, 4$}  &
\parbox[t]{4.7cm}{FV errors  estimated in $\chi$PT at NLO and, in FNAL/MILC~12B, by analysing otherwise identical ensembles with three different spatial sizes at $a=0.12$ fm and $m_l/m_s=0.1$.}  
\\[8.0ex] \hline \\[-1.0ex]
HPQCD 12A& \cite{Na:2012iu} & 2+1 &
\parbox[t]{1.75cm}{2.4/2.8, 2.4/3.4}
& \parbox[t]{1.6cm}{3.8, 4.2}  &
\parbox[t]{4.7cm}{FV errors  estimated by comparing finite and infinite volume $\chi$PT.}  
\\[8.0ex] \hline \\[-1.0ex]
FNAL/MILC 11 & \cite{Bazavov:2011aa} & 2+1 & 
	     \parbox[t]{1.75cm}{2.4, 2.4/2.88, 2.52/3.6} & \parbox[t]{1.6cm}{3.9, 3.8, 4,2 }  &
\parbox[t]{4.7cm}{FV errors  estimated using  finite-volume $\chi$PT.}
\\[8.0ex] \hline \\[-1.0ex]
PACS-CS 11 & \cite{Namekawa:2011wt} & 2+1 &
	     \parbox[t]{1.75cm}{$2.88$} & \parbox[t]{1.6cm}{2.2 (before reweighting)}  &
\parbox[t]{4.7cm}{No discussion of FSE.}
\\[8.0ex] \hline \\[-1.0ex]
HPQCD 10A & \cite{Davies:2010ip} & 2+1 &
             \parbox[t]{1.75cm}{2.4, 2.4/2.88/3.36, $2.52, 2.88$, 2.82} & \parbox[t]{1.6cm}{$3.9, 3.8$, $4.1, 4.5$, 4.6 }  &
\parbox[t]{4.7cm}{FV errors  estimated using finite- vs infinite-volume $\chi$PT.}
\\[8.0ex] \hline \\[-1.0ex]
HPQCD/UKQCD 07 & \cite{Follana:2007uv} & 2+1 & 
	     \parbox[t]{1.75cm}{2.4, 2.4/2.88, 2.52} & \parbox[t]{1.6cm}{3.9, 3.8, 4.1}  &
\parbox[t]{4.7cm}{FV errors  estimated using finite- vs infinite-volume $\chi$PT.}
\\[8.0ex] \hline \\[-1.0ex]
FNAL/MILC 05 & \cite{Aubin:2005ar} & 2+1 & \parbox[t]{1.75cm}{2.8, 2.9, 2.5} & \parbox[t]{1.6cm}{$3.8, 3.8,$ $4.1$ }  &
\parbox[t]{4.7cm}{FV errors estimated to be  $1.5\%$ or less from $\chi$PT.} 
\\[8.0ex] \hline \\[-1.0ex]
\parbox[t]{2cm}{ETM~09\\ ETM~11A\\ ETM~13B} & \cite{Blossier:2009bx,Dimopoulos:2011gx,Carrasco:2013zta} & 2 &
	     \parbox[t]{1.75cm}{2.4, 2.0/2.7, 2.1, 2.6 } & \parbox[t]{1.6cm}{5, 3.3, 3.3, 3.5}  &
\parbox[t]{4.7cm}{FV errors are found to be negligible by comparing results at $m_\pi L = 3.3$ and $m_\pi L=4.3$ for $m_\pi \simeq 310$ MeV.}
\\[8.0ex]  \\[-1.0ex]
\hline\hline
\end{tabular*}
\caption{Finite volume effects in determinations of the $D$ and $D_s$ meson
 decay constants.  Each $L$-entry corresponds to a different
 lattice spacing, with multiple spatial volumes at some lattice
 spacings. For actions with multiple species of pions, the lightest masses are
quoted.} 
}
\end{table}


\begin{table}[!htb]
{\footnotesize
\begin{tabular*}{\textwidth}{l @{\extracolsep{\fill}} c c c l}
\hline\hline \\[-1.0ex]
Collab. & Ref. & $\Nf$ &  Ren. & Description
\\[1.0ex] \hline \hline \\[-1.0ex]
\parbox[t]{2cm}{ETM~13F} & \cite{Dimopoulos:2013qfa} & 2+1+1 & $-$ & \parbox[t]{6.9cm}{The axial current is absolutely normalized.} 
\\[3.0ex] \hline \\[-1.0ex]
\parbox[t]{2cm}{FNAL/MILC~12B \\ FNAL/MILC~13} & \cite{Bazavov:2012dg,Bazavov:2013nfa} & 2+1+1 & 
$-$ &
\parbox[t]{6.9cm}{The axial current is absolutely normalized.} 
\\[3.0ex] \hline \\[-1.0ex]
HPQCD 12A& \cite{Na:2012iu} & 2+1 & 
$-$ &
\parbox[t]{6.9cm}{The axial current is absolutely normalized.}
\\[3.0ex] \hline \\[-1.0ex]
FNAL/MILC 11 & \cite{Bazavov:2011aa} & 2+1 &
mNPR &
\parbox[t]{6.9cm}{Two-loop and higher-order perturbative truncation errors estimated to be the full size of the one-loop term.}
\\[3.0ex] \hline \\[-1.0ex]
PACS-CS 11 & \cite{Namekawa:2011wt} & 2+1 &
PT1$\ell$+NP &
\parbox[t]{6.9cm}{Mass dependent part of the renormalization constant of the axial current computed at one-loop; the NP contribution
is added in the chiral limit.\\}
\\[3.0ex] \hline \\[-1.ex]
HPQCD 10A & \cite{Davies:2010ip} & 2+1 &
$-$ &
\parbox[t]{6.9cm}{The axial current is absolutely normalized.}
\\[3.0ex] \hline \\[-1.0ex]
HPQCD/UKQCD 07 & \cite{Follana:2007uv} & 2+1 & 
$-$&
\parbox[t]{6.9cm}{The axial current is absolutely normalized.}
\\[3.0ex] \hline \\[-1.0ex]
FNAL/MILC 05 & \cite{Aubin:2005ar} & 2+1 &
mNPR &
\parbox[t]{6.9cm}{Errors due to higher order corrections in the perturbative part are estimated to be $1.3\%$.}
\\[3.0ex] \hline \\[-1.0ex]
\parbox[t]{2cm}{ETM 09\\ ETM 11A\\ ETM~13B} & \cite{Blossier:2009bx,Dimopoulos:2011gx,Carrasco:2013zta} & 2 &
$-$ &
\parbox[t]{6.9cm}{The axial current is absolutely normalized.}
\\[3.0ex]  \\[-1.0ex]
\hline\hline
\end{tabular*}
\caption{Operator renormalization in  determinations of the $D$ and $D_s$ meson decay constants.} 
}
\end{table}


\begin{table}[!htb]
{\footnotesize
\begin{tabular*}{\textwidth}{l @{\extracolsep{\fill}} c c c l}
\hline\hline \\[-1.0ex]
Collab. & Ref. & $\Nf$ &  Action & Description
\\[1.0ex] \hline \hline \\[-1.0ex]
\parbox[t]{2cm}{ETM~13F} & \cite{Dimopoulos:2013qfa} & 2+1+1 & tmWil  &
\parbox[t]{3.9cm}{$0.15 \lsim am_c \lsim 0.20$. $D(a_{\rm min})  \geq 2\%$ also when the relative scale is set through $M_{c's'}$.}
\\[4.0ex] \hline \\[-1.0ex]
\parbox[t]{2cm}{FNAL/MILC~12B \\ FNAL/MILC~13} & \cite{Bazavov:2012dg,Bazavov:2013nfa} & 2+1+1 & 
HISQ (on HISQ)&
\parbox[t]{3.9cm}{$0.29<am_c<0.7$. Discretization errors estimated using different fit ans\"atze to be~$\approx 1.5\%$ for $f_{D_{(s)}}$.} 
\\[4.0ex] \hline \\[-1.0ex]
HPQCD 12A& \cite{Na:2012iu} & 2+1 & 
HISQ &
\parbox[t]{3.9cm}{$0.41<am_c<0.62$. Heavy-quark discretization errors estimated using different fit ans\"atze to be $\approx~1.2\%$.\\}
\\[4.0ex] \hline \\[-1.0ex]
FNAL/MILC 11 & \cite{Bazavov:2011aa} & 2+1 &
Fermilab  &
\parbox[t]{3.9cm}{Discretization errors from charm quark estimated through a combination of Heavy Quark and Symanzik Effective Theories to be around 3\% for $f_{D_{(s)}}$ and negligible for the ratio.\\}
\\[4.0ex] \hline \\[-1.0ex]
PACS-CS 11 & \cite{Namekawa:2011wt} & 2+1 &
Tsukuba &
\parbox[t]{3.9cm}{$am_c\approx 0.57$. Heavy-quark discretization errors estimated to be at the percent level by power counting.}
\\[4.0ex] \hline \\[-1.0ex]
HPQCD 10A & \cite{Davies:2010ip} & 2+1 &
HISQ &
\parbox[t]{3.9cm}{$0.193< am_c<0.825$. Heavy-quark discretization errors estimated by changing the fit-inputs to be $\approx~0.4\%$.\\}
\\[4.0ex] \hline \\[-1.0ex]
HPQCD/UKQCD 07 & \cite{Follana:2007uv} & 2+1 & 
HISQ &
\parbox[t]{3.9cm}{$0.43<am_c<0.85$.  Heavy-quark discretization errors estimated from the chiral/continuum fits to be $\approx~0.5\%$.
$\delta(a_{\rm min})$ slightly $>\!1\!$ for $f_{D_s}$.\\}
\\[4.0ex] \hline \\[-1.0ex]
FNAL/MILC 05 & \cite{Aubin:2005ar} & 2+1 &
Fermilab  &
\parbox[t]{3.9cm}{Discretization errors from charm quark estimated via heavy-quark power-counting at 4.2\% for $f_{D_{(s)}}$ and 0.5\% for the ratio.\\}
\\[4.0ex] \hline \\[-1.0ex]
\parbox[t]{2cm}{ETM~09\\ ETM~11A\\ ETM~13B} & \cite{Blossier:2009bx,Dimopoulos:2011gx,Carrasco:2013zta} & 2 &
tmWil &
\parbox[t]{3.9cm}{$0.16<am_c<0.23$. $D(a_{\rm min})\approx 5\%$ in ETM~09.}
\\[4.0ex]  \\[-1.0ex]
\hline\hline
\end{tabular*}
\caption{Heavy-quark treatment in  determinations of the $D$ and $D_s$ meson decay constants.} 
}
\end{table}

\clearpage

\subsubsection{$D\to\pi\ell\nu$ and $D\to K\ell\nu$ form factors}
\label{app:DtoPi_Notes}

\begin{table}[!ht]

{\footnotesize
\begin{tabular*}{\textwidth}{l c c c l l}
\hline\hline \\[-1.0ex]
Collab. & Ref. & $\Nf$ & $a$ [fm] & Continuum extrapolation & Scale setting 
\\[1.0ex] \hline \hline \\[-1.0ex]
HPQCD 13C & \cite{Koponen:2013tua} & 2+1 & \parbox[t]{1.3cm}{0.09, 0.12}  &
\parbox[t]{3.8cm}{Modified z-expansion fit combining the continuum and chiral extrapolations and the momentum trans- fer dependence.} & \parbox[t]{3.8cm}{Relative scale $r_1/a$ set from the static-quark potential.  Absolute scale $r_1$ set from several quantities including $f_\pi$, $f_K$, and $\Upsilon$ $2S-1S$ splitting c.f. HPQCD~09B~\cite{Davies:2009tsa}.}
\\[18.0ex] \hline \\[-2.0ex]
HPQCD 10B, 11 & \cite{Na:2010uf,Na:2011mc} & 2+1 & \parbox[t]{1.3cm}{0.09, 0.12}  &
\parbox[t]{3.8cm}{Modified $z$-expansion fit combining the continuum and chiral extrapolations and the momentum transfer dependence. Leading discretization errors from $(am_c)^n$ charm-mass effects (see Table~\ref{tab:DtoPiKHQ}). 
    Subleading $(aE)^n$ discretization corrections 
     estimated to be 1.0\% for both 
    $D\to\pi$ and $D\to K$.} & \parbox[t]{3.8cm}{Relative scale $r_1/a$ set from the static-quark potential.  Absolute scale $r_1$ set from several quantities including $f_\pi$, $f_K$, and $\Upsilon$ $2S-1S$ splitting c.f. HPQCD~09B~\cite{Davies:2009tsa}.  Scale uncertainty estimated to be 0.7\% in $D\to\pi$ and and 0.2\% in $D\to K$.}
\\[32.0ex] \hline \\[-2.0ex]
\parbox[t]{1cm}{FNAL/MILC~04} & \cite{Aubin:2004ej} & 2+1 & 0.12  &
\parbox[t]{3.8cm}{Discretization effects from light-quark sector estimated to be 4\% by power counting. Discretization effects from final-state pion and kaon energies estimated to be 5\%.} & \parbox[t]{3.8cm}{Scale set through $\Upsilon$ $2S-1S$ splitting c.f. HPQCD 03~\cite{Wingate:2003gm}.  Error in $a^{-1}$ estimated to be 1.2\%, but scale error in dimensionless form factor negligible compared to other uncertainties.}
\\[21.0ex] \hline \\[-2.0ex]
ETM 11B & \cite{DiVita:2011py} & 2 & \parbox[t]{1.3cm}{0.068, 0.086, 0.102} &
\parbox[t]{3.8cm}{Discretization errors estimated to be 5\% for $D\to\pi$ and 3\% for $D\to K$ from comparison of results in the continuum limit to those at the finest lattice spacing.}
& \parbox[t]{3.8cm}{Scale set through $f_\pi$ c.f. ETM~07A~\cite{Boucaud:2007uk} and ETM~09C~\cite{Baron:2009wt}.}
\\[19.0ex]
\hline\hline
\end{tabular*}
\caption{Continuum extrapolations/estimation of lattice artifacts in determinations of the $D\to\pi\ell\nu$ and $D\to K \ell\nu$ form factors.}
}
\end{table}
\begin{table}[!ht]
{\footnotesize
\begin{tabular*}{\textwidth}{l @{\extracolsep{\fill}} c  c c  l}
\hline\hline \\[-1.0ex]
{Collab.} & {Ref.} & {$\Nf$} & {$M_{\pi,\rm min}\,[\mev]$}  & {Description}  
\\[1.0ex] \hline \hline \\[-1.0ex]
HPQCD 13C & \cite{Koponen:2013tua} & 2+1 & \parbox[t]{2cm}{390, 390}
& \parbox[t]{6cm}{Modified z-expansion fit combining the continuum and chiral extrapolations and the momentum transfer dependence.}
\\[7.0ex] \hline \\[-2.0ex]
HPQCD 10B, 11 & \cite{Na:2010uf, Na:2011mc} & 2+1 & \parbox[t]{2cm}{390, 390}
& \parbox[t]{6cm}{Modified $z$-expansion fit combining the continuum and chiral extrapolations and the momentum transfer dependence. Contributions to error budget from light valence and sea-quark mass dependence estimated to be 2.0\% for $D \to\pi$ and 1.0\% for $D \to K$. }
\\[18.0ex] \hline \\[-2.0ex]
FNAL/MILC~04 & \cite{Aubin:2004ej} & 2+1 & \parbox[t]{2cm}{510}
 & \parbox[t]{6cm}{Fit to S$\chi$PT, combined with the Becirevic-Kaidalov ansatz for the momentum
 transfer dependence of form factors. Error estimated to be 3\% for $D \to\pi$ and 2\% for $D \to K$ by comparing fits with and without one extra analytic term.}
\\[15.0ex] \hline \\[-2.0ex]
ETM 11B & \cite{DiVita:2011py} & 2 & \parbox[t]{2cm}{270}
& \parbox[t]{6cm}{$SU(2)$ tmHM$\chi$PT plus Becirevic-Kaidalov ansatz for fits to the momentum transfer dependence of form factors.  Fit uncertainty estimated to be 7\% for $D \to\pi$ and 5\% for $D \to K$
by considering fits with and without NNLO corrections of order $\cO(m_\pi^4)$ and/or higher-order terms through $E^5$, and by excluding data with $E \gtapprox 1$~GeV.}
\\[24.0ex]
\hline\hline
\end{tabular*}
\caption{Chiral extrapolation/minimum pion mass in
  determinations of the $D\to \pi\ell\nu$ and $D\to K \ell\nu$ form factors.  For actions with multiple species of pions, masses quoted are the RMS pion masses.  The different $M_{\pi,\rm min}$ entries correspond to the different lattice spacings.}
}
\end{table}
\begin{table}[!ht]
{\footnotesize
\begin{tabular*}{\textwidth}{l @{\extracolsep{\fill}} c c  c c l}
\hline\hline \\[-1.0ex]
Collab. & Ref. & $\Nf$ & $L$ [fm] & ${M_{\pi,\rm min}}L$ & Description 
\\[1.0ex] \hline \hline \\[-1.0ex]
HPQCD 13C & \cite{Koponen:2013tua} & 2+1 & \parbox[t]{1.8cm}{2.4, 2.4/2.9} & $\gtrsim 3.8$ & 
\parbox[t]{5cm}{No explicit estimate of FV error, but expected to be small for simulation masses and volumes.}
\\[7.0ex] \hline \\[-1.0ex]
HPQCD 10B, 11 & \cite{Na:2010uf,Na:2011mc} & 2+1 & \parbox[t]{1.8cm}{2.4, 2.4/2.9} & $\gtrsim 3.8$ & 
\parbox[t]{5cm}{Finite volume effects estimated to be 0.04\% for $D\to\pi$ and 0.01\% for $D\to K$ by comparing the ``$m_\pi^2 {\rm log} (m_\pi^2)$" term in infinite and finite volume.}
\\[12.0ex] \hline \\[-1.0ex]
FNAL/MILC 04 & \cite{Aubin:2004ej} & 2+1 & \parbox[t]{1.8cm}{2.4/2.9} & $\gtrsim 3.8$ & 
\parbox[t]{5cm}{No explicit estimate of FV error, but expected to be small for simulation masses and volumes.}
\\[7.0ex] \hline \\[-1.0ex]
ETM 11B & \cite{DiVita:2011py} & 2 & \parbox[t]{1.8cm}{2.2, 2.1/2.8, 2.4} & $\gtrsim 3.7$ &
\parbox[t]{5cm}{Finite volume uncertainty estimated to be at most 2\% by considering
fits with and without the lightest pion mass point at $m_\pi L \approx 3.7$.}
\\[12.0ex]
\hline\hline
\end{tabular*}
\caption{Finite volume effects in determinations of the $D\to \pi\ell\nu$ and $D\to K\ell\nu$ form factors.  Each $L$-entry corresponds to a different lattice
spacing, with multiple spatial volumes at some lattice spacings.  For actions with multiple species of pions, the lightest pion masses are quoted.}
}
\end{table}
\begin{table}[!ht]
{\footnotesize
\begin{tabular*}{\textwidth}{l @{\extracolsep{\fill}} c c c l}
\hline\hline \\[-1.0ex]
Collab. & Ref. & $\Nf$ & Ren. & Description 
\\[1.0ex] \hline \hline \\[-1.0ex]
HPQCD 13C & \cite{Koponen:2013tua} & 2+1 & --- &
\parbox[t]{6cm}{Scalar form factor extracted from absolutely normalized scalar current. Vector form factor extracted from both point-split spatial (normalized by requiring $Zf_+(0) = 1$ for flavourless currents and checking mass independence) and local temporal vector currents (normalized by matching to the scalar case using kinematic constraint at zero momentum transfer $f_+(0) = f_0(0)$).}
\\[20.0ex] \hline \\[-1.0ex]
HPQCD 10B, 11 & \cite{Na:2010uf, Na:2011mc} & 2+1 & --- &
\parbox[t]{6cm}{Form factor extracted from absolutely normalized scalar-current matrix element then using kinematic constraint at zero momentum-transfer $f_+(0) = f_0(0)$.}
\\[10.0ex] \hline \\[-1.0ex]
FNAL/MILC~04 & \cite{Aubin:2004ej} & 2+1 & mNPR &
\parbox[t]{6cm}{Size of two-loop correction to current renormalization factor assumed to be negligible.}
\\[7.0ex] \hline \\[-1.0ex]
ETM 11B & \cite{DiVita:2011py} & 2 & --- &
\parbox[t]{6cm}{Form factors extracted from double ratios insensitive to current normalization.}
\\[4.0ex]
\hline\hline
\end{tabular*}
\caption{Operator renormalization in determinations of the $D\to\pi\ell\nu$ and $D\to K \ell\nu$ form factors.}
}
\end{table}
\begin{table}[!ht]
{\footnotesize
\begin{tabular*}{\textwidth}{l @{\extracolsep{\fill}} c c c l}
\hline\hline \\[-1.0ex]
Collab. & Ref. & $\Nf$ & Action & Description 
\\[1.0ex] \hline \hline \\[-1.0ex]
HPQCD 13C & \cite{Koponen:2013tua} & 2+1 & HISQ &
\parbox[t]{6cm}{Bare charm-quark mass $am_c \sim 0.41$--0.63. No explicit estimate of $(am_c)^n$ errors.}
\\[4.0ex] \hline \\[-1.0ex]
HPQCD 10B, 11 & \cite{Na:2010uf,Na:2011mc} & 2+1 & HISQ &
\parbox[t]{6cm}{Bare charm-quark mass $am_c \sim 0.41$--0.63.  Errors of $(am_c)^n$ estimated within modified $z$-expansion to be 1.4\% for $D\to K$ and 2.0\% for $D \to\pi$.  Consistent with expected size of dominant one-loop cutoff effects on the finest lattice spacing, ${\mathcal O}(\alpha_S (am_c)^2 (v/c) )\sim 1.6$\%.}
\\[18.0ex] \hline \\[-1.0ex]
FNAL/MILC~04 & \cite{Aubin:2004ej} & 2+1 & Fermilab &
\parbox[t]{6cm}{Discretization errors from charm quark estimated via heavy-quark power-counting to be 7\%.}
\\[7.0ex] \hline \\[-1.0ex]
ETM 11B & \cite{DiVita:2011py} & 2 & tmWil &
\parbox[t]{6cm}{Bare charm-quark mass $am_c \sim 0.17$--0.30.   Expected size of ${\mathcal O}((am_c)^2)$ cutoff effects on the finest lattice spacing consistent with quoted 5\% continuum-extrapolation uncertainty.}
\\[12.0ex]
\hline\hline
\end{tabular*}
\caption{Heavy quark treatment in determinations of the $D\to\pi\ell\nu$ and $D\to K \ell\nu$ form factors. \label{tab:DtoPiKHQ} 
}}
\end{table}

\clearpage
\subsection{Notes to section \ref{sec:BDecays} on $B$-meson decay constants, mixing parameters, and form factors}
\label{app:BDecays}

In the following, we summarize the characteristics (lattice actions,
pion masses, lattice spacings, etc.) of the recent $\Nf = 2+1$ and
$\Nf = 2$ runs. We also provide brief descriptions of how systematic
errors are estimated by the various authors.  We focus on calculations
with either preliminary or published quantitative results.

\subsubsection{$B_{(s)}$-meson decay constants}
\label{app:fB_Notes}


\begin{table}[!htb]
{\footnotesize
\begin{tabular*}{\textwidth}{l @{\extracolsep{\fill}} c c c l l}
\hline\hline \\[-1.0ex]
Collab. & Ref. & $\Nf$ & $a$ [fm]  & Continuum extrapolation & Scale setting
\\[1.0ex] \hline \hline \\[-2.0ex]
\parbox[t]{2.0cm}{ALPHA 13 \\ ALPHA 12A \\ ALPHA 11} & \cite{Bernardoni:2013oda,Bernardoni:2012ti,Blossier:2011dk} & 2 &
\parbox[t]{0.9cm}{0.075, 0.065, 0.048} & 
\parbox[t]{3.8cm}{Combined continuum and chiral extrapolation with linear in  
$a^2$ term. Continuum extrapolation 
errors estimated to be 5~MeV in ALPHA 11. 
}
& \parbox[t]{3.6cm}{Relative scale set from $r_0$. Absolute scale set from $f_K$. Scale setting uncertainty included in combined statistical and extrapolation error.}
\\[14.0ex] \hline \\[-2.0ex]
\parbox[t]{1.7cm}{ETM 13B, 13C \\ ETM 12B  \\ ETM 11A}  & 
\cite{Carrasco:2013zta,Carrasco:2013iba,Carrasco:2012de,Dimopoulos:2011gx} & 2 &
	     \parbox[t]{0.9cm}{0.098, 0.085, 0.067, 0.054} &
\parbox[t]{3.8cm}{Combined continuum and chiral extrapolation, with a term linear in $a^2$.
ETM 12 and 13 include a heavier masses than ETM 11A. Discretization error included in combined
statistical and systematic error, estimated by dropping the data at the coarsest lattice spacing
as $\sim 0.5 - 1\%$. \\
}
& \parbox[t]{3.6cm}{Scale set from $f_\pi$. Scale setting uncertainty included in combined 
statistical and systematic error. }
\\[-2.0ex] \hline \\[-2.0ex]
ETM 09D & \cite{Blossier:2009hg} & 2 &
	     \parbox[t]{0.9cm}{0.098, 0.085, 0.067} &
\parbox[t]{3.8cm}{Combined continuum and chiral extrapolation with a term linear in 
$a^2$. 
}
& \parbox[t]{3.6cm}{Scale set from $f_\pi$. Scale setting uncertainty included in 
combined statistical and systematic error.}
\\[4.0ex] \\[-1.0ex]
\hline\hline
\end{tabular*}
\caption{Continuum extrapolations/estimation of lattice artifacts in determinations of 
the $B$ and $B_s$ meson decay constants for $N_f=2$ simulations.} 
}
\end{table}

\begin{table}[!htb]
\vskip-2cm
{\footnotesize
\begin{tabular*}{\textwidth}{l @{\extracolsep{\fill}} c c c l l}
\hline\hline \\[-1.0ex]
Collab. & Ref. & $\Nf$ & $a$ [fm]  & Continuum extrapolation & Scale setting
\\[1.0ex] \hline \hline \\[-2.0ex]
ETM 13E &  \cite{Carrasco:2013naa} & 2+1+1 & \parbox[t]{0.9cm}{0.89, 0.82, 0.62}
& \parbox[t]{3.8cm}{Combined  continuum and chiral extrapolation, linear in $a^2$.  
}
& \parbox[t]{3.8cm}{Scale set from $f_\pi$. Scale setting uncertainty included in combined 
statistical and systematic error. }
\\[2.0ex] \hline \\[-2.0ex]
HPQCD 13 & \cite{Dowdall:2013tga} & 2+1+1 & \parbox[t]{0.9cm}{0.15, 0.12, 0.09}
& \parbox[t]{3.8cm}{Combined  continuum and chiral extrapolation. 
Continuum extrapolation errors estimated to be $0.7\%$. 
}
& \parbox[t]{3.6cm}{Scale set from $\Upsilon$(2S-1S)
splitting, see Ref.~\cite{Dowdall:2011wh}.  Scale uncertainty included in statistical error.}
\\[9.0ex] \hline \\[-2.0ex]
RBC/UKQCD 13A & \cite{Witzel:2013sla} & 2+1 & \parbox[t]{0.9cm}{0.11, 0.086}
& \parbox[t]{3.8cm}{Combined  continuum and chiral extrapolation with linear in $a^2$ term. 
No systematic error estimate. 
}
& \parbox[t]{3.6cm}{Scale set by the $\Omega$ baryon mass.}
\\[10.0ex] \hline \\[-2.0ex]
HPQCD 12 & \cite{Na:2012sp} & 2+1 & \parbox[t]{0.9cm}{0.12, 0.09}
& \parbox[t]{3.8cm}{Combined  continuum and chiral extrapolation. 
Continuum extrapolation errors estimated to be $0.9\%$. 
}
& \parbox[t]{3.6cm}{Relative scale $r_1/a$ from the static-quark potential.  
Absolute scale $r_1$  from $f_\pi$, $f_K$, and $\Upsilon$(2S-1S) 
splitting.  Scale uncertainty estimated to be $1.1\%$.}
\\[15.0ex] \hline \\[-2.0ex]
HPQCD 11A & \cite{McNeile:2011ng} & 2+1 & \parbox[t]{0.9cm}{0.15, 0.12, 0.09, 0.06, 0.045}
& \parbox[t]{3.8cm}{ $am_Q \approx 0.2 - 0.85$. Combined  continuum and HQET fit. 
 Continuum extrapolation error estimated by varying the fit ansatz and the included data
 points to be $0.63\%$. Discretization errors appear to decrease with increasing heavy-meson 
 mass. 
 } 
& \parbox[t]{3.6cm}{Relative scale $r_1/a$ from the static-quark potential.  
Absolute scale $r_1$ from $f_\pi$, $f_K$, and 
$\Upsilon$(2S-1S) splitting.  Scale uncertainty estimated to be $0.74\%$.}
\\[21.0ex] \hline \\[-2.0ex]
FNAL/MILC 11 &\cite{Bazavov:2011aa}  & 2+1 & \parbox[t]{0.9cm}{0.15, 0.12, 0.09}
& \parbox[t]{3.8cm}{Combined continuum and chiral extrapolation. Continuum extrapolation 
errors estimated to be 1.3\%.
}
& \parbox[t]{3.6cm}{Relative scale $r_1/a$ from the static-quark potential.  
Absolute scale $r_1$ from $f_\pi$, $f_K$, and $\Upsilon$(2S-1S) splitting.  
Scale uncertainty estimated to be 1~MeV.}
\\[15.0ex] \hline \\[-2.0ex]
RBC/UKQCD 10C &\cite{Albertus:2010nm}  & 2+1 & \parbox[t]{0.9cm}{0.11}
& \parbox[t]{3.8cm}{One lattice spacing with discretization errors estimated 
by power counting as  3\%.}
& \parbox[t]{3.6cm}{Scale set by the $\Omega$ baryon mass. Combined scale and mass tuning
uncertainties on $f_{B_s}/f_B$ estimated as 1\%}
\\[8.0ex] \hline \\[-2.0ex]
HPQCD 09 & \cite{Gamiz:2009ku} & 2+1 & \parbox[t]{0.9cm}{0.12, 0.09}
& \parbox[t]{3.8cm}{Combined continuum and chiral extrapolation. Continuum 
extrapolation errors estimated to be 3\%. 
}
& \parbox[t]{3.6cm}{Relative scale $r_1/a$ from the static-quark potential.  
Absolute scale $r_1$  from the $\Upsilon$(2S-1S) splitting.  Scale uncertainty
 estimated to be $2.3\%$. }
\\[14.0ex] \\[-1.0ex]
\hline\hline
\end{tabular*}
\caption{Continuum extrapolations/estimation of lattice artifacts in determinations of 
the $B$ and $B_s$ meson decay constants for $N_f=2+1+1$ and $N_f=2+1$ simulations.} 
}
\end{table}

\begin{table}[!htb]
\vskip-2cm
{\footnotesize
\begin{tabular*}{\textwidth}{l @{\extracolsep{\fill}} c c c   l}
\hline\hline \\[-1.0ex]
{Collab.} & {Ref.} & {$\Nf$} & {$M_{\pi,\rm min}\,[\mev]$}  & {Description}  
\\[1.0ex] \hline \hline \\[-2.0ex]
\parbox[t]{1.8cm}{ETM 13B, 13C \\ ETM 12B \\ ETM 11A} & 
\cite{Carrasco:2013zta,Carrasco:2013iba,Carrasco:2012de,Dimopoulos:2011gx} & 2 & 
\parbox[t]{1.3cm}{410, 275, 300, 270} &
 \parbox[t]{7.0cm}{$M_{\pi,{\rm min}}$ refers to the charged pions. Linear and NLO (full QCD) 
HM$\chi$PT supplemented by an $a^2$ term is used.The chiral fit error is estimated from the difference between
the NLO HM$\chi$PT and linear fits with half the difference
used as estimate of the systematic error. 
For the static limit calculation in ETM 11A, $\Phi_s^{\rm stat}$ is extrapolated 
assuming a constant in light quark mass.
The ratio $\Phi_s^{\rm stat}/\Phi_{\ell}^{\rm stat}$ is fit using three 
different chiral fit forms (NLO HM$\chi$PT, linear, and quadratic) to estimate the chiral fir error. }
\\[30.0ex] \hline \\[-2.0ex]
ETM 09D & \cite{Blossier:2009hg} & 2 & \parbox[t]{1.3cm}{410, 275, 300}
& \parbox[t]{7.0cm}{$M_{\pi,{\rm min}}$ refers to the charged pions.  
Linear and NLO (full QCD) 
HM$\chi$PT is used. The final result given by the average of NLO HMChiPT and linear
{\it Ans\"atze} $\pm$ half the difference).}
\\[10.0ex] \hline \\[-2.0ex]
\parbox[t]{1.8cm}{ALPHA 13 \\ALPHA 12A } & \cite{Bernardoni:2013oda,Bernardoni:2012ti} & 
2 & \parbox[t]{1.3cm}{270, 190, 270}
& \parbox[t]{7.0cm}{LO and NLO HMChPT supplemented by a term linear in $a^2$ are 
used. The final result is an average between LO and NLO with half the difference
used as estimate of the systematic error. }
\\[9.0ex]\hline \\[-2.0ex]
ALPHA 11 & \cite{Blossier:2011dk} & 
2 & \parbox[t]{1.3cm}{331, 268, 267}
& \parbox[t]{7.0cm}{Linear and NLO (full QCD) HMChPT supplemented by a term linear in $a^2$ are 
used. The final result is an average between linear and NLO fits with half the difference
used as estimate of the systematic error. }
\\[9.0ex] 
\hline\hline
\end{tabular*}
\caption{Chiral extrapolation/minimum pion mass in determinations of the
  $B$ and $B_s$ meson decay constantsfor $N_f=2$ simulations.  
  For actions with multiple species of pions, masses quoted are the RMS pion masses.    
  The different $M_{\pi,\rm min}$ entries correspond to the different lattice
 spacings.} 
}
\end{table}

\begin{table}[!htb]
\vskip-2cm
{\footnotesize
\begin{tabular*}{\textwidth}{l @{\extracolsep{\fill}} c c c   l}
\hline\hline \\[-1.0ex]
{Collab.} & {Ref.} & {$\Nf$} & {$M_{\pi,\rm min}\,[\mev]$}  & {Description}  
\\[1.0ex] \hline \hline \\[-2.0ex]
ETM 13E & \cite{Carrasco:2013naa} & 2+1+1 & \parbox[t]{1.3cm}{245, 239, 211}
& \parbox[t]{7.0cm}{$M_{\pi,{\rm min}}$ refers to the charged pions.  Linear and NLO (full QCD) 
HM$\chi$PT supplemented by an $a^2$ term  is used for the SU(3) breaking ratios.The chiral fit 
error is estimated from the difference between
the NLO HM$\chi$PT and linear fits with half the difference
used as estimate of the systematic error. The ratio $z_s$ is fit using linear light quark mass
dependence supplemented by an $a^2$ term. }
\\[14.0ex] \hline \\[-2.0ex]
HPQCD 13 &  \cite{Dowdall:2013tga} & 2+1+1 & \parbox[t]{1.3cm}{310, 294, 173} & 
\parbox[t]{7.0cm}{Two or three pion masses at each lattice spacing, one each 
with a physical mass GB pion. NLO (full QCD) HM$\chi$PT supplemented by 
generic $a^2$ and $a^4$ terms is used to
interpolate to the physical pion mass.}
\\[9.0ex] \hline \\[-2.0ex]
RBC/UKQCD 13A & \cite{Witzel:2013sla} & 2+1 & \parbox[t]{1.3cm}{329, 289} &
\parbox[t]{7.0cm}{Three (two) light quark masses per lattice spacing. 
NLO SU(2) HM$\chi$PT is used. No systematic error estimate.}
\\[8.0ex] \hline \\[-2.0ex]
HPQCD 12 &  \cite{Na:2012sp} & 2+1 & \parbox[t]{1.3cm}{390, 390} & 
\parbox[t]{7.0cm}{Two or three pion masses at each lattice spacing. NLO (full QCD) 
HM$\chi$PT supplemented by NNLO analytic terms and 
generic $a^2$ and $a^4$ terms is used. The systematic error is estimated by 
varying the fit Ansatz, in particular for the NNLO analytic terms and the $a^{2n}$ terms. }
\\[12.0ex] \hline \\[-2.0ex]
HPQCD 11A & \cite{McNeile:2011ng} & 2+1 & \parbox[t]{1.3cm}{570, 450, 390, 330, 330}  &
\parbox[t]{7.0cm}{One light sea quark mass only at each lattice spacing. The sea-quark mass 
dependence is assumed to be negligible, based on the calculation of $f_{D_s}$ in 
Ref.~\cite{Davies:2010ip}, where the sea quark extrapolation error is estimated as $0.34\%$. } 
\\[9.0ex] \hline \\[-2.0ex]
FNAL/MILC 11 & \cite{Bazavov:2011aa} & 2+1 & \parbox[t]{1.3cm}{570, 440, 320} &
\parbox[t]{7.0cm}{Three to five sea-quark masses per lattice spacing, 
and  $9-12$ valence light quark masses per ensemble. NLO partially quenched HMrS$\chi$PT 
including $1/m$ terms and  supplemented by NNLO analytic and $\alpha_s^2a^2$ terms is used. 
The systematic error is estimated by varying the fit Ansatz,  in particular the NNLO analytic 
terms and the chiral scale. }
\\[17.0ex] \hline \\[-2.0ex]
RBC/UKQCD 10C & \cite{Albertus:2010nm} & 2+1 & \parbox[t]{1.3cm}{430} &
\parbox[t]{7.0cm}{Three light quark masses at one lattice spacing. 
NLO SU(2) $\chi$PT is used. The systematic error is estimated
 from the difference between NLO $\chi$PT and linear fits as $\sim 7\%$.}
\\[8.0ex] \hline \\[-2.0ex]
HPQCD 09 &  \cite{Gamiz:2009ku} & 2+1 & \parbox[t]{1.3cm}{440, 400} & 
\parbox[t]{7.0cm}{Four or two pion masses per lattice spacing. NLO (full QCD) 
HMrS$\chi$PT supplemented by NNLO analytic terms and 
 $\alpha_s a^2, a^4$ terms is used. The chiral fit error is estimated by varying the fit Ansatz,
in particular, by adding or removing NNLO and discretization terms. }
\\[12.0ex]
\hline\hline
\end{tabular*}
\caption{Chiral extrapolation/minimum pion mass in determinations of the
  $B$ and $B_s$ meson decay constants for $N_f=2+1+1$ and $N_f=2+1$ simulations.  
  For actions with multiple species of pions, masses quoted are the RMS pion masses.    
  The different $M_{\pi,\rm min}$ entries correspond to the different lattice
 spacings.} 
}
\end{table}

\begin{table}[!htb]
{\footnotesize
\begin{tabular*}{\textwidth}{l @{\extracolsep{\fill}} c c c c l}
\hline\hline \\[-1.0ex]
Collab. & Ref. & $\Nf$ & $L$ [fm] & ${M_{\pi,\rm min}}L$ & Description 
\\[1.0ex] \hline \hline \\[-2.0ex]
ETM 13E & \cite{Carrasco:2013naa} & 2+1+1 & 
\parbox[t]{1.8cm}{2.84/2.13, 2.61/1.96, 2.97} &
\parbox[t]{1.5cm}{3.53, 3.16, 3.19} &
\parbox[t]{4.9cm}{ FV error estimated how?}
\\[6.0ex] \hline \\[-2.0ex]
HPQCD 13 & \cite{Dowdall:2013tga} & 2+1+1 & 
\parbox[t]{1.8cm}{2.4/3.5/4.7, 2.9/3.8/5.8, 2.8/5.6}  & 
\parbox[t]{1.5cm}{3.30, 3.88, 3.66} &
\parbox[t]{4.9cm}{ The analysis uses finite-volume $\chi$PT.}
\\[7.0ex] \hline \\[-2.0ex]
RBC/UKQCD 13A & \cite{Witzel:2013sla} & 2+1 & 
\parbox[t]{1.8cm}{2.6,2.75} & \parbox[t]{1.5cm}{4.54, 4.05}& 
\parbox[t]{4.9cm}{No FV error estimate. }
\\[6.0ex] \hline \\[-2.0ex]
HPQCD 12 & \cite{Na:2012sp} & 2+1 & \parbox[t]{1.8cm}{2.4/2.9, 2.5/3.6}  & 
\parbox[t]{1.5cm}{3.84, 4.21} &
\parbox[t]{4.9cm}{FV error is taken from Ref.~\cite{Follana:2007uv} for HPQCD's $D$ 
meson analysis, where it was estimated using finite volume $\chi$PT .}
\\[9.0ex] \hline \\[-2.0ex]
HPQCD 11A & \cite{McNeile:2011ng} & 2+1 & \parbox[t]{1.8cm}{2.4, 2.4, 2.5, 2.9, 2.9} &
\parbox[t]{1.5cm}{3.93, 4.48, 4.14, 4.49, 4.54}  & 
\parbox[t]{4.9cm}{FV error is assumed to negligible.}
\\[6.0ex] \hline \\[-2.0ex]
FNAL/MILC 11 & \cite{Bazavov:2011aa} & 2+1 & \parbox[t]{1.8cm}{2.4, 2.4/2.9, 2.5/3.6} &
\parbox[t]{1.5cm}{3.93, 3.78, 4.14}  & 
\parbox[t]{4.9cm}{FV error is estimated using  finite-volume $\chi$PT.}
\\[4.0ex] \hline \\[-2.0ex]
RBC/UKQCD 10C & \cite{Albertus:2010nm} & 2+1 & 
\parbox[t]{1.8cm}{1.8} & $3.9$ & 
\parbox[t]{4.9cm}{FV error estimated using  finite-volume $\chi$PT to be $1\%$
 for SU(3) breaking ratios. }
\\[6.0ex] \hline \\[-2.0ex]
HPQCD 09 & \cite{Gamiz:2009ku} & 2+1 & \parbox[t]{1.8cm}{2.4/2.9, 2.5} &
\parbox[t]{1.5cm}{3.78, 4.14}  & 
\parbox[t]{4.9cm}{FV error is assumed to negligible.}
\\[2.0ex] \hline \\[-2.0ex]
\parbox[t]{1.8cm}{ETM 13B, 13C \\ETM 12B \\ETM 11A} & 
\parbox[t]{1.0cm}{ \cite{Carrasco:2013zta,Carrasco:2013iba}\\ \cite{Carrasco:2012de}\\ \cite{Dimopoulos:2011gx}} & 2 &
	     \parbox[t]{1.8cm}{2.4, 2.0/2.7, 2.1, 1.7/2.6} &
	      \parbox[t]{1.5cm}{5, 3.7, 3.3, 3.5}  &
\parbox[t]{4.9cm}{FV errors are found to be negligible by comparing results at $m_\pi L = 3.3$ and $m_\pi L=4.3$ for $m_\pi \simeq 310$ MeV.}
\\[10.0ex] \hline \\[-2.0ex]
\parbox[t]{1.8cm}{ALPHA 13\\ALPHA 12A \\ALPHA 11} & 
\parbox[t]{1.0cm}{ \cite{Bernardoni:2013oda}\\ \cite{Bernardoni:2012ti}\\ \cite{Blossier:2011dk}} & 2 &
\parbox[t]{1.8cm}{2.4, 2.1/4.2 [3.1], 2.3/3.1} & \parbox[t]{1.5cm}{5.2, 4.1, 4.2}  & 
\parbox[t]{4.9cm}{No explicit estimate of FV errors, but expected to be much smaller than other uncertainties.}
\\[7.0ex]
\hline\hline
\end{tabular*}
\caption{Finite volume effects in determinations of the $B$ and $B_s$ meson
 decay constants.  Each $L$-entry corresponds to a different
 lattice spacing, with multiple spatial volumes at some lattice
 spacings.} 
}
\end{table}


\begin{table}[!htb]
{\footnotesize
\begin{tabular*}{\textwidth}{l @{\extracolsep{\fill}} c c c l}
\hline\hline \\[-1.0ex]
Collab.  &  Ref. & $\Nf$ & Ren. & Description 
\\[1.0ex] \hline \hline \\[-2.0ex]
ETM 13E  & \cite{Carrasco:2013naa} & 2+1+1 & --, PT1$\ell$ &
\parbox[t]{7.5cm}{The current used for the relativistic decay constants is absolutely normalized. 
 The ratio is constructed from the relativistic decay constant data
 and the heavy-quark pole masses. 
 Ratios of pole-to-$\msbar$ mass  conversion factors are included 
at NLO in continuum perturbation theory.}
\\[7.0ex] \hline \\[-2.0ex]
HPQCD 13 & \cite{Dowdall:2013tga} & 2+1+1 & PT1$\ell$
& \parbox[t]{7.5cm}{The NRQD effective current is matched through $O(1/m$) and renormalized using one-loop PT.  Included are all terms though
$O(\alpha_s)$,  $O(\alpha_s \, a)$, $O(\Lambda_{\rm QCD}/M)$, $O(\alpha_s/a M)$ ,  $O(\alpha_s \, \Lambda_{\rm QCD}/M)$. The dominant error is due unknown $O(\alpha_s^2)$ contributions to the current renormalization. The perturbation theory used in this work is the same as in 
HPQCD 09 and 12, but is rearranged to match the mNPR method. Using the fact that the heavy-heavy temporal vector current is normalized, and that the light-light HISQ vector current receives a small one-loop correction, the error is estimated as  $\sim1.4$\%. }
\\[29.0ex] \hline \\[-2.0ex]
RBC/UKQCD 13A & \cite{Witzel:2013sla}  & 2+1 & mNPR 
& \parbox[t]{7.5cm}{No systematic error estimate. }
\\[4.0ex] \hline \\[-2.0ex]
HPQCD 12/09 & \cite{Na:2012sp,Gamiz:2009ku} & 2+1 & PT1$\ell$
& \parbox[t]{7.5cm}{The NRQD effective current is matched through $O(1/m$) and renormalized using one-loop PT.  Included are all terms though
$O(\alpha_s)$,  $O(\alpha_s \, a)$, $O(\Lambda_{\rm QCD}/M)$, $O(\alpha_s/a M)$ ,  $O(\alpha_s \, \Lambda_{\rm QCD}/M)$. The dominant error is due unknown $O(\alpha_s^2)$ contributions to the current renormalization. The authors take the perturbative error as $\sim 2  \rho_0 \, \alpha_s^2$, where $\rho_0$ is the coefficient of the one-loop correction to the leading term, which yields an
error of $\sim4$\%. }
\\[20.0ex] \hline \\[-2.0ex]
HPQCD 11A & \cite{McNeile:2011ng} & 2+1 & -- 
& \parbox[t]{7.5cm}{This work uses PCAC together with an absolutely normalized current.}
\\[4.0ex] \hline \\[-2.0ex]
FNAL/MILC 11 & \cite{Bazavov:2011aa}  & 2+1 & mNPR 
& \parbox[t]{7.5cm}{The authors' estimate of the perturbative errors is comparable in size to the actual 
one-loop corrections. }
\\[4.0ex] \hline \\[-2.0ex]
RBC/UKQCD 10C & \cite{Albertus:2010nm} & 2+1 & PT1$\ell$ & 
\parbox[t]{7.5cm}{The static-light current is matched through
 $O(\alpha_s a, \alpha_s)$ and renormalized using one-loop
tadpole improved PT. For massless light quarks, the renormalization factors
cancel in the ratio of decay constants.}
\\
\hline\hline
\end{tabular*}
\caption{Description of the renormalization/matching procedure adopted in the determinations of the $B$ and $B_s$ meson decay constants for $N_f=2+1$ simulations.} 
}
\end{table}

\begin{table}[!htb]
{\footnotesize
\begin{tabular*}{\textwidth}{l @{\extracolsep{\fill}} c c c l}
\hline\hline \\[-1.0ex]
Collab.  &  Ref. & $\Nf$ & Ren. & Description  
\\[1.0ex] \hline \hline \\[-2.0ex]
\parbox[t]{1.8cm}{ALPHA 13 \\ALPHA 12A \\ALPHA 11} & 
\parbox[t]{1.0cm}{ \cite{Bernardoni:2013oda}\\ \cite{Bernardoni:2012ti}\\ \cite{Blossier:2011dk}} & 
2 & NPR & 
\parbox[t]{7.5cm}{The authors use the Schr\"dingier functional for the NP matching. }
\\[4.0ex] \hline \\[-2.0ex]
\parbox[t]{1.8cm}{ETM 13B, 13C \\ETM 12B \\ETM 11A} & 
\parbox[t]{1.0cm}{ \cite{Carrasco:2013zta,Carrasco:2013iba}\\ \cite{Carrasco:2012de}\\ \cite{Dimopoulos:2011gx}} & 
 2 & --, PT1$\ell$ & 
\parbox[t]{7.5cm}{The current used for the relativistic decay constants is absolutely normalized. 
{\bf Interpolation method:} The static limit current renormalization is calculated in one-loop 
mean field improved perturbation theory, there half the correction is used to estimate the
 error.  {\bf Ratio method:} The ratio is constructed from the relativistic decay constant data
 and the heavy-quark pole masses. 
 Ratios of pole-to-$\msbar$ mass  conversion factors are included 
at NLO in continuum perturbation theory.}
\\[7.0ex]  \\[-2.0ex]
\hline\hline
\end{tabular*}
\caption{Description of the renormalization/matching procedure adopted in the determinations of the $B$ and $B_s$ meson decay constants for $N_f=2$ simulations.} 
}
\end{table}

\begin{table}[!htb]
\vskip-1cm
{\footnotesize
\begin{tabular*}{\textwidth}{l @{\extracolsep{\fill}} c c c l}
\hline\hline \\[-2.0ex]
Collab. & Ref. & $\Nf$ &  Action  &  Description
\\[1.0ex] \hline \hline \\[-1.0ex]
ETM 13E  & \cite{Carrasco:2013naa} & 2+1+1 & tmWil &
\parbox[t]{7.5cm}{The estimate of the discretization effects is described in the 
continuum table. The relativistic data are 
matched to HQET using NLO continuum PT in an intermediate step, and 
converted back to QCD at the end. 
The error due to HQET matching (estimated by replacing the NLO expressions  with LO)
 is a very small contribution to the systematic error due to the 
heavy quark mass dependence.}
\\[8.0ex] \hline  \\[-2.0ex]
HPQCD 13 & \cite{Dowdall:2013tga} & 2+1+1 & NRQCD 
& \parbox[t]{7.5cm}{HQ truncation effects estimated as in HPQCD 09  to be $1.0\%$}
\\[4.0ex] \hline \\[-2.0ex]
RBC/UKQCD 13A & \cite{Witzel:2013sla} & 2+1 & RHQ & 
\parbox[t]{7.5cm}{No estimate of HQ discretization errors.}
\\[6.0ex] \hline \\[-2.0ex]
HPQCD 12 & \cite{Na:2012sp} & 2+1 & NRQCD 
& \parbox[t]{7.5cm}{HQ truncation effects estimated as in HPQCD 09  to be $1.0\%$}
\\[4.0ex] \hline \\[-2.0ex]
HPQCD 11A & \cite{McNeile:2011ng} & 2+1 & HISQ
& \parbox[t]{7.5cm}{ The analysis uses a combined continuum and $1/m$ extrapolation. }
\\[4.0ex] \hline \\[-2.0ex]
FNAL/MILC 11 & \cite{Bazavov:2011aa} & 2+1 & Fermilab
& \parbox[t]{7.5cm}{HQ discretization effects are  
included in the combined chiral and continuum fits, and are estimated by varying the fit Ansatz
and excluding the data at the coarsest lattice spacing to be $\sim 2\%$, consistent with
simple power counting estimates but larger than the residual discretization errors observed
in the data.}
\\[17.0ex] \hline \\[-2.0ex]
RBC/UKQCD 10C & \cite{Albertus:2010nm} & 2+1 & Static & 
\parbox[t]{7.5cm}{Truncation effects of ${\cal O}(1/m_h)$ on the SU(3) breaking ratios are
estimated by power counting to be 2\%.}
\\[6.0ex] \hline \\[-2.0ex]
HPQCD 09 & \cite{Gamiz:2009ku} & 2+1 & NRQCD
& \parbox[t]{7.5cm}{The leading HQ truncation effects are of 
${\cal O}(\alpha_s \Lambda_{\rm QCD}/m_h)$ due to the
tree-level coefficient of the ${\bm \sigma}\cdot{\rm\bf B}$ term. 
The error is estimated by calculating the $B^*-B$ hyperfine
splitting and comparing with experiment as $1\%$.}
\\[10.0ex] \hline \\[-2.0ex]
\parbox[t]{1.8cm}{ALPHA 13 \\ALPHA 12A \\ALPHA 11} & 
\parbox[t]{1.0cm}{ \cite{Bernardoni:2013oda}\\ \cite{Bernardoni:2012ti}\\ \cite{Blossier:2011dk}} & 2 & HQET &
\parbox[t]{7.5cm}{NP improved through ${\cal O}(1/m_h)$. Truncation errors of 
${\cal O}(\Lambda_{\rm QCD}/m_h)^2$ are not included. \\}
\\[8.0ex] \hline \\[-2.0ex]
\parbox[t]{1.8cm}{ETM 13B, 13C \\ETM 12B \\ETM 11A} & 
\parbox[t]{1.0cm}{ \cite{Carrasco:2013zta,Carrasco:2013iba}\\ \cite{Carrasco:2012de}\\ \cite{Dimopoulos:2011gx}} & 
2 & tmWil &
\parbox[t]{7.5cm}{The estimate of the discretization effects is described in the 
continuum table. In both methods the relativistic data are 
matched to HQET using NLO continuum PT in an intermediate step, and 
converted back to QCD at the end. 
The error due to HQET matching (estimated by replacing the NLO expressions  with LO)
 is a very small contribution to the systematic error due to the 
heavy quark mass dependence. The variation observed from adding 
heavier masses to their data and/or including $1/m_h^3$ terms is 
$0.4-1.3\%$.}
\\[8.0ex]
\hline\hline
\end{tabular*}
\caption{Heavy quark treatment in determinations of the $B$ and $B_s$ meson decay constants.} 
}
\end{table}

\clearpage
\subsubsection{$B_{(s)}$-meson mixing matrix elements}
\label{app:BMix_Notes}
\begin{table}[!ht]
{\footnotesize
\begin{tabular*}{\textwidth}{l c c c l l}
\hline\hline \\[-1.0ex]
Collab. & Ref. & $\Nf$ & $a$ [fm] & Continuum extrapolation & Scale setting 
\\[1.0ex] \hline \hline \\[-1.0ex]
FNAL/MILC 12 & \cite{Bazavov:2012zs} & 2+1 & \parbox[t]{1.1cm}{0.12, 0.09} & 
\parbox[t]{3.5cm}{Combined continuum and chiral extrapolation with
		 NLO rHMS$\chi$PT, NNLO analytic and generic 
		 ${\cal O}(\alpha_s^2 a^2, a^4)$ terms. 
		 Combined statistical, chiral and light-quark discretization error 
		 is estimated, by examining the variation with different fit Ans\"atze
		 to be 3.7\% on $\xi$.} &
\parbox[t]{3.5cm}{Relative scale $r_1/a$ is set via static quark
		  potential. Absolute scale $r_1=0.3117(22)$ fm is
		  determined \cite{Bazavov:2011aa} through averaging the
		  $f_\pi$ input and the estimate of HPQCD collaboration 
		  \cite{Davies:2009tsa}. The scale uncertainty on $\xi$ is estimated as 0.2\%.} 
\\[32.0ex] \hline \\[-2.0ex]
FNAL/MILC 11A & \cite{Bouchard:2011xj} & 2+1 & \parbox[t]{1.1cm}{0.12, 0.09, 0.06} & 
\parbox[t]{3.5cm}{Combined continuum and chiral extrapolation with
		 NLO rHMS$\chi$PT, NNLO analytic and and generic 
		 ${\cal O}(\alpha_s^2 a^2, a^4)$ terms.} & 
\parbox[t]{3.5cm}{See above. The error in $r_1$ yields a 3\% 
		 uncertainty on $f^2_B B_B$.} 
\\[14.0ex] \hline \\[-2.0ex]
RBC/UKQCD 10C & \cite{Albertus:2010nm} & 2+1 & \parbox[t]{1.5cm}{0.11} &
\parbox[t]{3.5cm}{Only one lattice spacing is used. Discretization error
		 is estimated to be 4\% on $\xi$ by power counting.}&
\parbox[t]{3.5cm}{Scale is set using the $\Omega^-$ mass as input 
		 \cite{Allton:2008pn}.The error on $\xi$ due to the combined scale and 
		 light quark mass uncertainties is estimated as 1\%.} 
\\[14.0ex] \hline \\[-2.0ex]
HPQCD 09 & \cite{Gamiz:2009ku} & 2+1 & \parbox[t]{1.1cm}{0.12, 0.09} &
\parbox[t]{3.5cm}{Combined continuum and chiral extrapolation with
		 NLO rHMS$\chi$PT and NNLO analytic terms. Light-quark
		 discretization error
		 is estimated as 3, 2 and 0.3\% for $f_B\sqrt{B_B}$,
		 $f_{B_s}\sqrt{B_{B_s}}$ and $\xi$ respectively.} &
\parbox[t]{3.5cm}{Relative scale $r_1/a$ is set via static quark
		  potential. Absolute scale $r_1=0.321(5)$ fm is
		  determined through $\Upsilon$ mass \cite{Gray:2005ur}.
		  The error on $f_B\sqrt{B_B}$ due to the scale uncertainty
		  is estimated as 2.3\%.} 
\\[23.0ex] \hline \\[-2.0ex]
HPQCD 06A & \cite{Dalgic:2006gp} & 2+1 & \parbox[t]{1.1cm}{0.12} & 
\parbox[t]{3.5cm}{Only one lattice spacing is used. Light-quark discretization error
		 on $f_{B_s}^2B_{B_s}$ is estimated as 4\% 
		 by power counting.} &
\parbox[t]{3.5cm}{Scale is set using the $\Upsilon$ $2S-1S$ splitting as 
		 input \cite{Gray:2005ur}. The error on $f^2_B B_B$ due to the scale uncertainty
		  is estimated as 5\%.} 
\\[14.0ex]
\hline\hline
\end{tabular*}
\caption{Continuum extrapolations/estimation of lattice artifacts 
 in determinations of the neutral $B$-meson mixing matrix elements for
 $N_f=2+1$ simulations.}
\label{tab:BBcontinum}
}
\end{table}

\begin{table}[!ht]
{\footnotesize
\begin{tabular*}{\textwidth}{l c c c l l}
\hline\hline \\[-1.0ex]
Collab. & Ref. & $\Nf$ & $a$ [fm] & Continuum extrapolation & Scale setting 
\\[1.0ex] \hline \hline \\[-1.0ex]
ETM 13B  & \cite{Carrasco:2013zta} & 2 & \parbox[t]{1.1cm}{0.098,
	     0.085, 0.067, 0.054} & 
\parbox[t]{3.5cm}{
	     Combined chiral and continuum extrapolation, with a term linear in $a^2$.
	     Discretization error is estimated by omitting the
	     coarsest lattice as 0.5, 1.7, 1.3 and 1.0 \% for 
	     $B_{B_s}$, $B_B$, $B_{B_s}/B_B$ and $\xi$ respectively.
	     The heavy-quark masses vary in the range $0.13 \lsim  a m_h \lsim 0.85$.} &
\parbox[t]{3.5cm}{See below.} 
\\[32.0ex] 
\hline \\[-2.0ex]
ETM 12A,12B  & \cite{Carrasco:2012dd,Carrasco:2012de} & 2 & \parbox[t]{1.1cm}{0.098,
	     0.085, 0.067} & 
\parbox[t]{3.5cm}{Combined chiral and continuum extrapolation, with a term linear in $a^2$.
		Discretization error included in combined statistical, chiral and continuum extrapolation error
		and estimated as $4.5$\%. 
		The heavy-quark masses vary in the range $0.25 \lsim  a m_h \lsim 0.6$.} &
\parbox[t]{3.5cm}{Relative scale $r_0/a$ set from the static quark
		     potential. Absolute scale set from $f_\pi$. Scale setting uncertainty 
		     included in combined statistical and systematic error.} 
\\[30.0ex] 
\hline\hline
\end{tabular*}
\caption{Continuum extrapolations/estimation of lattice artifacts 
 in determinations of the neutral $B$-meson mixing matrix elements for
 $N_f=2$ simulations.}
\label{tab:BBcontinum2}
}
\end{table}

\begin{table}[!ht]
{\footnotesize
\begin{tabular*}{\textwidth}{l @{\extracolsep{\fill}} c c c  l}
\hline\hline \\[-1.0ex]
Collab. & Ref. & $\Nf$ & {$M_{\pi,\rm min}\,[\mev]$}  & {Description}
\\[1.0ex] \hline \hline \\[-1.0ex]
FNAL/MILC 12 & \cite{Bazavov:2012zs} & 2+1 & \parbox[t]{2cm}{440, 320}
	     &  \parbox[t]{5cm}{Combined continuum and chiral extrapolation with
		 NLO rHMS$\chi$PT and NNLO analytic terms.
		 See the entry in Table~\ref{tab:BBcontinum}.
		 The omission of wrong-spin contributions \cite{Bernard:2013dfa} 
		 in the HMrS$\chi$PT is treated as a systematic error and
		 estimated to be 3.2\% for $\xi$.}
\\[14.0ex] \hline \\[-2.0ex]
FNAL/MILC 11A & \cite{Bouchard:2011xj} & 2+1 & \parbox[t]{2cm}{440, 320, 250} & 
\parbox[t]{5cm}{Combined continuum and chiral extrapolation with
		NLO rHMS$\chi$PT and NNLO analytic terms.}
\\[8.0ex] \hline \\[-2.0ex]
RBC/UKQCD 10C & \cite{Albertus:2010nm} & 2+1 & \parbox[t]{2cm}{430} &
\parbox[t]{5cm}{Linear fit matched with $SU(2)$ NLO HM$\chi$PT at the lightest
		$ud$ mass point is used as the preferred fit. Many different fit Ans\"atze
		are considered. The systematic
		 error is estimated from the difference between the $SU(2)$ 
		 HM$\chi$PT fit described above and a linear fit. }
\\[20.0ex]\hline \\[-2.0ex] 
HPQCD 09 & \cite{Gamiz:2009ku} & 2+1 & \parbox[t]{2cm}{440, 400} &
\parbox[t]{5cm}{Combined continuum and chiral extrapolation with
		NLO rHMS$\chi$PT and NNLO analytic terms.} 
\\[8.0ex] \hline \\[-2.0ex]
HPQCD 06A & \cite{Dalgic:2006gp} & 2+1 & \parbox[t]{2cm}{510} & 
\parbox[t]{5cm}{Two sea $ud$ quark masses $m_{ud}/m_s=0.25$ and $0.5$
		 are used to calculate the matrix element for $B_s$
		 meson at the predetermined value of the strange quark
		 mass. No significant sea quark mass dependence is observed
		 and the value at the lighter sea $ud$ mass is taken as the result.} 
\\[24.0ex] \hline \\[-2.0ex]
\parbox[t]{2.2cm}{ETM 13B\\ ETM 12A,12B}  & \parbox[t]{1cm}{\cite{Carrasco:2013zta}\\ \cite{Carrasco:2012dd,Carrasco:2012de}} & 2 & \parbox[t]{2cm}{410, 275, 300, 270} & 
\parbox[t]{5cm}{$M_{\pi,{\rm min}}$ refers to the charged pions, where 270 MeV
	on the finest lattice only included in ETM 13. Linear and NLO (full
	QCD) HM$\chi$PT supplemented by an $a^2$ term is used. The chiral fit error is
	estimated from the difference between the NLO HM$\chi$PT and linear fits. }
\\[20.0ex] 
%
\hline\hline
\end{tabular*}
\caption{Chiral extrapolation/minimum pion mass
 in determinations of the neutral $B$-meson mixing matrix elements.
 For actions with multiple species of pions, masses quoted are the RMS
 pion masses.  The different $M_{\pi,\rm min}$ entries correspond to the
 different lattice spacings.}
}
\end{table}

\begin{table}[!ht]
{\footnotesize
\begin{tabular*}{\textwidth}{l @{\extracolsep{\fill}} c c c c l}
\hline\hline \\[-1.0ex]
Collab. & Ref. & $\Nf$ & $L$ [fm] & ${M_{\pi,\rm min}}L$ & Description
\\[1.0ex] \hline \hline \\[-1.0ex]
FNAL/MILC 12 & \cite{Bazavov:2012zs} & 2+1 & 
\parbox[t]{1.8cm}{2.4/2.9, 2.5} & $\gtrsim 3.8$ & 
\parbox[t]{5.8cm}{FV error is estimated to be less than $0.1\%$ for
		     $SU(3)$ breaking ratios from FV HMrS$\chi$PT.} 
\\[6.0ex] \hline \\[-2.0ex]
FNAL/MILC 11A & \cite{Bouchard:2011xj} & 2+1 & 
\parbox[t]{1.8cm}{2.4/2.9, 2.5/2.9/3.6, 3.8} & $\gtrsim 3.8$ &
\parbox[t]{5.8cm}{FV error on $f_B\sqrt{B_B}$ is estimated to be less
		  than $1\%$, 
		  which is inferred from the study of the $B$-meson decay
		  constant using FV HM$\chi$PT \cite{Bazavov:2011aa}.} 
\\[6.0ex] \hline \\[-2.0ex]
RBC/UKQCD 10C & \cite{Albertus:2010nm} & 2+1 & 
\parbox[t]{1.8cm}{1.8} & $\gtrsim 3.9$ & 
\parbox[t]{5.8cm}{FV error estimated through FV HM$\chi$PT as $1\%$ for $SU(3)$ breaking ratios. }
\\[4.0ex]\hline \\[-2.0ex] 
HPQCD 09 & \cite{Gamiz:2009ku} & 2+1 & 
\parbox[t]{1.8cm}{2.4/2.9, 2.5} & $\gtrsim 3.8$ &
\parbox[t]{5.8cm}{No explicit estimate of FV error, but expected to be much smaller than other uncertainties.}
\\[6.0ex] \hline \\[-2.0ex]
HPQCD 06A & \cite{Dalgic:2006gp} & 2+1 & 
\parbox[t]{1.8cm}{2.4} & $\gtrsim 4.5$ & 
\parbox[t]{5.8cm}{No explicit estimate of FV error, but expected to be much smaller than other uncertainties.}
\\[6.0ex] \hline \\[-2.0ex]
\parbox[t]{2.2cm}{ETM 13B\\ ETM 12A,12B}  &
     \parbox[t]{1cm}{\cite{Carrasco:2013zta}\\ \cite{Carrasco:2012dd,Carrasco:2012de}} & 2 & 
\parbox[t]{1.8cm}{2.4, 2.0/2.7, 2.1, 1.7/2.6} & $\gtrsim 3.2$  & 
\parbox[t]{5.8cm}{$L=1.7/2.6\, {\rm fm}$ only included in ETM~13. FV error is assumed to be negligible based on the study of $D$-meson decay constants in 
Ref.~\cite{Blossier:2009bx}.}
\\[9.0ex] 
%
%
\hline\hline
\end{tabular*}
\caption{Finite volume effects 
 in determinations of the neutral $B$-meson mixing matrix elements.
 Each $L$-entry corresponds to a different
 lattice spacing, with multiple spatial volumes at some lattice
 spacings.  For actions with multiple species of pions, the lightest
 masses are quoted.}
}
\end{table}

\begin{table}[!ht]
{\footnotesize
\begin{tabular*}{\textwidth}{l @{\extracolsep{\fill}} c c c l}
\hline\hline \\[-1.0ex]
Collab. & Ref. & $\Nf$ & Ren. & Description 
\\[1.0ex] \hline \hline \\[-1.0ex]
FNAL/MILC 12 & \cite{Bazavov:2012zs} & 2+1 & PT$1l$ & 
\parbox[t]{6cm}{One-loop mean-field improved PT
		 is used to renormalize the four-quark operators
		 with heavy quarks rotated to eliminate tree-level $O(a)$
		 errors. The error from neglecting higher order corrections 
		 is estimated to be $0.5$\% on $\xi$.} 
\\[16.0ex] \hline \\[-2.0ex]
FNAL/MILC 11A & \cite{Bouchard:2011xj} & 2+1 & PT$1l$ &
\parbox[t]{6cm}{One-loop mean-field improved PT is used to 
                renormalize the four-quark operators
		 with heavy quarks rotated to eliminate tree-level $O(a)$
		 errors. The error from neglected higher order corrections 
		 is estimated to be $4$\% on $f_B\sqrt{B_B}$.} 
\\[16.0ex] \hline \\[-2.0ex]
RBC/UKQCD 10C & \cite{Albertus:2010nm} & 2+1 & PT$1l$ & 
\parbox[t]{6cm}{Static-light four-quark operators are 
		 renormalized with one-loop
		 mean field improved PT.
		 The error due to neglected higher order effects 
		 is estimated to be $2.2$\% on $\xi$.}
\\[10.0ex]\hline \\[-2.0ex]
HPQCD 09 & \cite{Gamiz:2009ku} & 2+1 & PT$1l$ &
\parbox[t]{6cm}{Four-quark operators in lattice NRQCD are matched to
		 QCD through order $\alpha_s$, $\Lambda_{\rm QCD}/M$ 
		 and  $\alpha_s/(aM)$ \cite{Gamiz:2008sk} using one-loop PT. 
		 The error due to neglected higher order effects is estimated to 
		 be 4\% on $f_B\sqrt{B_B}$ and 0.7\% on $\xi$.}
\\[16.0ex] \hline \\[-2.0ex]
HPQCD 06A & \cite{Dalgic:2006gp} & 2+1 & PT$1l$ &
\parbox[t]{6cm}{Four-quark operators in lattice NRQCD are matched to
		 full QCD through order $\alpha_s$, $\Lambda_{\rm QCD}/M$ 
		 and  $\alpha_s/(aM)$ \cite{Gamiz:2008sk}. The error  
		 is estimated as $\sim 1\cdot \alpha_s^2$ to be 9\% on
		 $f_{B_s}^2 B_{B_s}$}
\\[10.0ex] \hline \\[-2.0ex]
ETM 13B, 12A, 12B & \cite{Carrasco:2013zta,Carrasco:2012dd,Carrasco:2012de} & 2 & NPR & 
\parbox[t]{6cm}{The bag parameters are nonperturbatively 
	         renormalized in the RI'-MOM scheme. They are calculated
	         as functions of the ($\msbar$) heavy-quark mass (renormalized
	         nonperturbatively in RI/MOM). }
\\[13.0ex] 
%
%
%
\hline\hline
\end{tabular*}
\caption{Operator renormalization 
 in determinations of the neutral $B$-meson mixing matrix elements.
}}
\end{table}

\begin{table}[!ht]
{\footnotesize
\begin{tabular*}{\textwidth}{l @{\extracolsep{\fill}} c c c l}
\hline\hline \\[-1.0ex]
Collab. & Ref. & $\Nf$ & Action & Description
\\[1.0ex] \hline \hline \\[-1.0ex]
FNAL/MILC 12 & \cite{Bazavov:2012zs} & 2+1 & Fermilab & 
\parbox[t]{6cm}{The heavy-quark discretization error on $\xi$ is estimated
		to be 0.3 \%. The error on $\xi$ 
		 due to the uncertainty in the $b$-quark mass is
		 are estimated to be 0.4 \%.} 
\\[9.0ex] \hline \\[-2.0ex]
FNAL/MILC 11A & \cite{Bouchard:2011xj} & 2+1 & Fermilab &
\parbox[t]{6cm}{The 
		 heavy-quark discretization error on $f_B\sqrt{B_B}$
		 is estimated as 4\% using power-counting.} 
\\[6.0ex] \hline \\[-2.0ex]
RBC/UKQCD 10C & \cite{Albertus:2010nm} & 2+1 & Static & 
\parbox[t]{6cm}{Two different static-quark actions with Ape and HYP
		 smearings are used. The discretization error
		 on $\xi$ is estimated as  $\sim 4$\% and  the error due to the missing
		 $1/m_b$ corrections  as
		 $\sim 2$\%, both using power-counting.}
\\[13.0ex]\hline \\[-2.0ex]
HPQCD 09 & \cite{Gamiz:2009ku} & 2+1 & NRQCD &
\parbox[t]{6cm}{Heavy-quark truncation errors due to relativistic corrections
		 are estimated to be 2.5, 2.5 and 0.4 \% for $f_B\sqrt{B_B}$,
		 $f_{B_s}\sqrt{B_{B_s}}$ and $\xi$ respectively.}
\\[9.0ex] \hline \\[-2.0ex]
HPQCD 06A & \cite{Dalgic:2006gp} & 2+1 & NRQCD &
\parbox[t]{6cm}{Heavy-quark truncation errors due to relativistic corrections
		 are estimated to be 3\% for $f_{B_s}^2{B_{B_s}}$.}
\\[6.0ex] \hline \\[-2.0ex]
\parbox[t]{2.2cm}{ETM 13B\\ ETM 12A,12B}  &
     \parbox[t]{1cm}{\cite{Carrasco:2013zta}\\ \cite{Carrasco:2012dd,Carrasco:2012de}} & 2 & 
	 tmWilson & 
\parbox[t]{6cm}{The ratio method is used to perform an interpolation to
	the physical $b$ quark mass from the simulated
	heavy mass and the known static limit.
	In an intermediate step, the ratios include HQET matching
	factors calculated to tree-level, leading-log, 
	and next-to-leading-log (ETM 13 only)
	in continuum PT.
	The interpolation uses a polynomial up to
	quadratic in the inverse quark-mass. 
	The systematic errors added together with those of the chiral
	fit are estimated as $1.3-1.6$\% for bag parameters for ETM 13,
	while they are estimated from
	changing the interpolating polynomial as $2$\% and from changing the order of HQET 
	matching factors as $3$\% for ETM 12A and 12B.}
\\[46.0ex]
%
%
\hline\hline
\end{tabular*}
\caption{Heavy-quark treatment
 in determinations of the neutral $B$-meson mixing matrix elements.
}}
\end{table}

\clearpage
\subsubsection{$B\to\pi\ell\nu$ form factor}
\label{app:BtoPi_Notes}

\begin{table}[!ht]

{\footnotesize
\begin{tabular*}{\textwidth}{l c c c l l}
\hline\hline \\[-1.0ex]
Collab. & Ref. & $\Nf$ & $a$ [fm] & Continuum extrapolation & Scale setting 
\\[1.0ex] \hline \hline \\[-1.0ex]
FNAL/MILC 08A & \cite{Bailey:2008wp} & 2+1 & 0.09, 0.12   &
\parbox[t]{4.2cm}{Fit to rHMS$\chi$PT to remove light-quark discretization errors.  Residual heavy-quark discretization errors estimated with power-counting to be 3.4\%.} & \parbox[t]{4.0cm}{Relative scale $r_1/a$ set from the static-quark potential.  Absolute scale $r_1$ set through $f_\pi$ cf.~MILC 07~\cite{Bernard:2007ps}; error in scale taken to be difference from scale set through $\Upsilon$ $2S-1S$ splitting c.f.~HPQCD 05B~\cite{Gray:2005ur}.  Scale-uncertainty estimated at between 1\% and 1.5\% in the range of $q^2$ explored.}
\\[29.0ex] \hline \\[-2.0ex]
HPQCD 06 & \cite{Dalgic:2006dt} & 2+1 & 0.09,0.12  &
\parbox[t]{4.2cm}{Central values obtained from data at $a=0.12$~fm.  Discretization errors observed to be within the statistical error by comparison with data at $a=0.09$~fm.} &  \parbox[t]{4.0cm}{Relative scale $r_1/a$ set from the static-quark potential.  Absolute scale $r_1$ set through $\Upsilon$ $2S-1S$ splitting c.f.~HPQCD 05B~\cite{Gray:2005ur}.}
\\[15.0ex]
\hline\hline
\end{tabular*}
\caption{Continuum extrapolations/estimation of lattice artifacts in
  determinations of the $B\to\pi\ell\nu$ form factor.}
}
\end{table}
\begin{table}[!ht]
{\footnotesize
\begin{tabular*}{\textwidth}{l @{\extracolsep{\fill}} c c c   l}
\hline\hline \\[-1.0ex]
{Collab.} & {Ref.} & {$\Nf$} & {$M_{\pi,\rm min}\,[\mev]$}  & {Description}  
\\[1.0ex] \hline \hline \\[-1.0ex]
FNAL/MILC 08A & \cite{Bailey:2008wp} & 2+1 & \parbox[t]{2cm}{400, 440}
& \parbox[t]{5cm}{Simultaneous chiral-continuum extrapolation and $q^2$ interpolation using $SU(3)$ rHMS$\chi$PT.  Systematic error estimated by adding higher-order analytic terms and varying the $B^*$-$B$-$\pi$ coupling.}
\\[15.0ex] \hline \\[-2.0ex]
HPQCD 06 & \cite{Dalgic:2006dt} & 2+1 & \parbox[t]{2cm}{400, 440}
& \parbox[t]{5cm}{First interpolate data at fixed quark mass to fiducial values of $E_\pi$ using the Becirevic-Kaidalov and Ball-Zwicky ans\"atze, then extrapolate data at fixed $E_\pi$ to physical quark masses using $SU(3)$ rHMS$\chi$PT. Systematic error estimated by varying interpolation and extrapolation fit functions.}
\\[23.0ex]
\hline\hline
\end{tabular*}
\caption{Chiral extrapolation/minimum pion mass in
  determinations of the $B\to\pi\ell\nu$ form factor.  For actions with multiple species of pions, masses quoted are the RMS pion masses.    The different $M_{\pi,\rm min}$ entries correspond to the different lattice spacings.}
}
\end{table}
\begin{table}[!ht]
{\footnotesize
\begin{tabular*}{\textwidth}{l @{\extracolsep{\fill}} c c c c l}
\hline\hline \\[-1.0ex]
Collab. & Ref. & $\Nf$ & $L$ [fm] & ${M_{\pi,\rm min}}L$ & Description 
\\[1.0ex] \hline \hline \\[-1.0ex]
FNAL/MILC 08A & \cite{Bailey:2008wp} & 2+1 & \parbox[t]{1.8cm}{2.4, 2.4/2.9} & $\gtrsim 3.8$  &
\parbox[t]{5cm}{Estimate FV error to be $0.5\%$ using 1-loop rHMS$\chi$PT.
}
\\[4.0ex] \hline \\[-1.0ex]
HPQCD 06 & \cite{Dalgic:2006dt} & 2+1 & \parbox[t]{1.8cm}{2.4/2.9} & $\gtrsim 3.8$ & 
\parbox[t]{5cm}{No explicit estimate of FV error, but expected to be much smaller than other uncertainties.}
\\[7.0ex]
\hline\hline
\end{tabular*}
\caption{Finite volume effects in determinations of the $B\to\pi\ell\nu$ form factor.  Each $L$-entry corresponds to a different lattice
spacing, with multiple spatial volumes at some lattice spacings.  For actions with multiple species of pions, the lightest masses are quoted.}
}
\end{table}
\begin{table}[!ht]
{\footnotesize
\begin{tabular*}{\textwidth}{l @{\extracolsep{\fill}} c c c l}
\hline\hline \\[-1.0ex]
Collab. & Ref. & $\Nf$ & Ren. & Description 
\\[1.0ex] \hline \hline \\[-1.0ex]
FNAL/MILC 08A & \cite{Bailey:2008wp} & 2+1 & mNPR &
\parbox[t]{6cm}{Perturbative truncation error estimated at 3\% with size of 1-loop correction on finer ensemble.}
\\[7.0ex] \hline \\[-1.0ex]
HPQCD 06 & \cite{Dalgic:2006dt} & 2+1 & PT1$\ell$ &
\parbox[t]{6cm}{Currents included through ${\mathcal O}(\alpha_S \Lambda_{\rm QCD}/M$, $\alpha_S/(aM)$, $\alpha_S\, a \Lambda_{\rm QCD})$. Perturbative truncation error estimated from power-counting.}
\\[10.0ex]
\hline\hline
\end{tabular*}
\caption{Operator renormalization in determinations of the $B\to\pi\ell\nu$ form factor.}
}
\end{table}
\begin{table}[!ht]
{\footnotesize
\begin{tabular*}{\textwidth}{l @{\extracolsep{\fill}} c c c l}
\hline\hline \\[-1.0ex]
Collab. & Ref. & $\Nf$ & Action & Description 
\\[1.0ex] \hline \hline \\[-1.0ex]
FNAL/MILC 08A & \cite{Bailey:2008wp} & 2+1 & Fermilab &
\parbox[t]{8cm}{Discretization errors in $f+(q^2)$ from heavy-quark action estimated to be 3.4\% by heavy-quark power-counting.}
\\[4.0ex] \hline \\[-1.0ex]
HPQCD 06 & \cite{Dalgic:2006dt} & 2+1 & NRQCD &
\parbox[t]{8cm}{Discretization errors in $f_+(q^2)$ estimated to be ${\mathcal O} (\alpha_s (a \Lambda_{\rm QCD})^2) \sim  3\%$.  Relativistic errors estimated to be ${\mathcal O}((\Lambda_{\rm QCD}/M)^2) \sim 1\%$.}
\\[7.0ex]
\hline\hline
\end{tabular*}
\caption{Heavy quark treatment in determinations of the $B\to\pi\ell\nu$ form factor.}
}
\end{table}

\clearpage
\subsubsection{$B\to D \ell\nu$ and $B \to D^* \ell \nu$ form factors and $R(D)$}
\label{app:BtoD_Notes}

\begin{table}[!ht]

{\footnotesize
\begin{tabular*}{\textwidth}{l c c c l l}
\hline\hline \\[-1.0ex]
Collab. & Ref. & $\Nf$ & $a$ [fm] & Continuum extrapolation & Scale setting 
\\[1.0ex] \hline \hline \\[-1.0ex]
\parbox[t]{2.8cm}{FNAL/MILC 13B\\ $B \to D^*$} & \cite{Qiu:2013ofa} & 2+1 & \parbox[t]{1.5cm}{ 0.045, 0.06, 0.09, 0.12,\\0.15} &
\parbox[t]{3.6cm}{See FNAL/MILC~10.}
& \parbox[t]{3.6cm}{See below.} 
\\[9ex] \hline \\[-2.0ex]
%
\parbox[t]{2.8cm}{FNAL/MILC 13B\\ $B \to D$} & \cite{Qiu:2013ofa} & 2+1 & \parbox[t]{1.5cm}{0.045, 0.06, 0.09, 0.12} &
\parbox[t]{3.6cm}{Continuum extrapolation using rHMS$\chi$PT to remove light-quark discretization errors.  Residual discretization errors estimated from power-counting to be 2\% in $h_+$ (the dominant contributor to $f_+$) and 10\% in $h_-$.}
& \parbox[t]{3.6cm}{Relative scale $r_1/a$ set from the static-quark potential.  Absolute scale $r_1$ set through combination of $f_\pi$ from MILC 09B~\cite{Bazavov:2009tw} and several quantities including $\Upsilon$ splittings from HPQCD~\cite{Davies:2009tsa}, as described in~\cite{Bazavov:2011aa}.} 
\\[27.ex] \hline \\[-2.0ex]
FNAL/MILC 12A & \cite{Bailey:2012jg} & 2+1 & \parbox[t]{1.5cm}{ 0.09, 0.12} &
\parbox[t]{3.6cm}{Continuum extrapolation using rHMS$\chi$PT to remove light-quark discretization errors.  Residual discretization errors estimated to be 
very small for the ratio of branching fractions $R(D)$ at 0.2\% }
& \parbox[t]{3.6cm}{See below.} 
\\[23.5ex] \hline \\[-2.0ex]
FNAL/MILC 10 & \cite{Bailey:2010gb} & 2+1 & \parbox[t]{1.5cm}{0.06, 0.09, 0.12, 0.15} &
\parbox[t]{3.6cm}{Continuum extrapolation using rHMS$\chi$PT to remove light-quark discretization errors.  Residual discretization errors estimated to be 1.0\% from power-counting.  Further, the data displays no observable trend with lattice spacing.}
& \parbox[t]{3.6cm}{Relative scale $r_1/a$ set from the static-quark potential.  Absolute scale $r_1$ set through $f_\pi$ c.f.~MILC 09B~\cite{Bazavov:2009tw}.  Comparison with $r_1$ set via other quantities by HPQCD~\cite{Davies:2009tsa} shows negligible change.} 
\\[29.5ex] \hline \\[-2.0ex]
FNAL/MILC 08 & \cite{Bernard:2008dn} & 2+1 & 0.09, 0.12, 0.15 &
\parbox[t]{3.6cm}{Continuum extrapolation using rHMS$\chi$PT to remove light-quark discretization errors.  Residual discretization errors estimated to be 1.5\% from power-counting and by comparison of data at 0.12 and 0.09~fm.}
& \parbox[t]{3.6cm}{Relative scale $r_1/a$ set from the static-quark potential.  Absolute scale $r_1$ set through $f_\pi$ c.f.~MILC 07~\cite{Bernard:2007ps}.  Comparison with scale set through $\Upsilon$ $2S-1S$ shows negligible change.} 
\\[24ex]
\hline\hline
\end{tabular*}
\caption{Continuum extrapolations/estimation of lattice artifacts in
  determinations of the $B\to D \ell\nu$, $B\to D^* \ell\nu$,  $B_s\to D_s \ell\nu$ form factors and of $R(D)$.}
}
\end{table}

\begin{table}[!ht]

{\footnotesize
\begin{tabular*}{\textwidth}{l c c c l l}
\hline\hline \\[-1.0ex]
Collab. & Ref. & $\Nf$ & $a$ [fm] & Continuum extrapolation & Scale setting 
\\[1.0ex] \hline \hline \\[-1.0ex]
FNAL/MILC 04A & \cite{Okamoto:2004xg} & 2+1 & 0.12  &
\parbox[t]{3.6cm}{Central value obtained from data at a single lattice spacing. 
Comparison with quenched simulations at different lattice spacings interpreted as indication
of small discretization effects.}
& \parbox[t]{3.6cm}{Relative scale $r_1/a$ set from the static-quark potential.  Absolute scale $r_1$ set through $\Upsilon$ $2S-1S$ splitting c.f. HPQCD 03~\cite{Wingate:2003gm}.}
\\[21.0ex] \hline \\[-2.0ex]
Atoui  13 & \cite{Atoui:2013zza} & 2& \parbox[t]{1.5cm}{0.054, 0.067, 0.085, 0.098} &
\parbox[t]{3.6cm}{Combined continuum and chiral extrapolation, with linear terms in $a^2$ and $m_{\rm sea}$.
No dependence on $a$ or $m_{\rm sea}$ observed within errors. Stability of results vs fits with no $m_{\rm sea}$
dependence checked.}
& \parbox[t]{3.6cm}{Scale set through $F_\pi$.} 
\\[24ex]
\hline\hline
\end{tabular*}
\caption{(cntd.) Continuum extrapolations/estimation of lattice artifacts in
  determinations of the $B\to D \ell\nu$, $B\to D^* \ell\nu$,  $B_s\to D_s \ell\nu$ form factors and of $R(D)$.}
}
\end{table}

\begin{table}[!ht]
{\footnotesize
\begin{tabular*}{\textwidth}{l @{\extracolsep{\fill}} c c c  l}
\hline\hline \\[-1.0ex]
{Collab.} & {Ref.} & {$\Nf$} & {$M_{\pi,\rm min}\,[\mev]$}  & {Description}  
\\[1.0ex] \hline \hline \\[-1.0ex]
FNAL/MILC 13B $B \to D^*$ & \cite{Qiu:2013ofa} & 2+1 & \parbox[t]{2cm}{330, 260, 280, 470, 590}
 & \parbox[t]{5cm}{Simultaneous chiral-continuum extrapolation using $SU(3)$ rHMS$\chi$PT.  Systematic error estimated by adding higher-order analytic terms, varying the $D^*$-$D$-$\pi$ coupling, and comparison with continuum $HM\chi$PT.}
\\[18.0ex] \hline \\[-2.0ex]
FNAL/MILC 13B $B \to D$ & \cite{Qiu:2013ofa} & 2+1 & \parbox[t]{2cm}{330, 260, 280, 470}
 & \parbox[t]{5cm}{Simultaneous chiral-continuum extrapolation using $SU(3)$ rHMS$\chi$PT supplemented by terms analytic in $(w-1)$ to interpolate at nonzero recoil.  Systematic error included in statistical errors via inclusion of NNLO analytic terms and $D^*$-$D$-$\pi$ coupling with Bayesian priors.}
\\[21.0ex] \hline \\[-2.0ex]
FNAL/MILC 12A & \cite{Bailey:2012jg} & 2+1 & \parbox[t]{2cm}{ 400, 440}
 & \parbox[t]{5cm}{See below.}
\\[2.0ex] \hline \\[-2.0ex]
FNAL/MILC 10 & \cite{Bailey:2010gb} & 2+1 & \parbox[t]{2cm}{ 340, 320, 440, 570}
 & \parbox[t]{5cm}{See below.}
\\[3.5ex] \hline \\[-2.0ex]
FNAL/MILC 08 & \cite{Bernard:2008dn} & 2+1 & \parbox[t]{2cm}{320, 440, 570}
 & \parbox[t]{5cm}{Simultaneous chiral-continuum extrapolation using $SU(3)$ rHMS$\chi$PT.  Systematic errors estimated by adding higher-order analytic terms and varying the $D^*$-$D$-$\pi$ coupling.}
\\[15.0ex] \hline \\[-2.0ex]
FNAL/MILC 04A & \cite{Okamoto:2004xg} & 2+1 & \parbox[t]{2cm}{510}
 & \parbox[t]{5cm}{Linear extrapolation in the light-quark mass.}
\\[4.0ex] \hline \\[-2.0ex]
Atoui  13 & \cite{Atoui:2013zza} & 2 & \parbox[t]{2cm}{270, 300, 270, 410}
 & \parbox[t]{5cm}{Combined continuum and chiral extrapolation, with linear terms in $a^2$ and $m_{\rm sea}$.
No dependence on $a$ or $m_{\rm sea}$ observed within errors. Stability of results vs fits with no $m_{\rm sea}$
dependence checked.}
\\[4.0ex]
\hline\hline
\end{tabular*}
\caption{Chiral extrapolation/minimum pion mass in
  determinations of the $B\to D \ell\nu$, $B\to D^* \ell\nu$,  $B_s\to D_s \ell\nu$ form factors and of $R(D)$.  For actions with multiple species of pions, masses quoted are the RMS pion masses.  The different $M_{\pi,\rm min}$ entries correspond to the different lattice spacings.}
}
\end{table}
\begin{table}[!ht]
{\footnotesize
\begin{tabular*}{\textwidth}{l @{\extracolsep{\fill}} c c c c l}
\hline\hline \\[-1.0ex]
Collab. & Ref. & $\Nf$ & $L$ [fm] & ${M_{\pi,\rm min}}L$ & Description 
\\[1.0ex] \hline \hline \\[-1.0ex]
\parbox[t]{2.8cm}{FNAL/MILC 13B\\ $B \to D^*$}  & \cite{Qiu:2013ofa} & 2+1 & \parbox[t]{1.8cm}{2.9, 2.9/3.4/3.8, 2.4/2.7/3.4/5.5, 2.4/2.9,  2.4} & $\gtrsim 3.8$ & 
\parbox[t]{5cm}{Estimate FV error to be negligible using 1-loop rHMS$\chi$PT.}
\\[10.0ex] \hline \\[-1.0ex]
\parbox[t]{2.8cm}{FNAL/MILC 13B\\ $B \to D$}  & \cite{Qiu:2013ofa} & 2+1 & \parbox[t]{1.8cm}{2.9, 2.9/3.4/3.8, 2.4/2.7/3.4/5.5, 2.4/2.9} & $\gtrsim 3.8$ & 
\parbox[t]{5cm}{ FV error estimated to be negligible.}
\\[10.0ex] \hline \\[-1.0ex]
FNAL/MILC 12A & \cite{Bailey:2012jg} & 2+1 & \parbox[t]{1.8cm}{2.5, 2.4} & $\gtrsim 3.8$ & 
\parbox[t]{5cm}{ FV error estimated to be negligible in \cite{Bailey:2012rr}.}
\\[4.0ex] \hline \\[-1.0ex]
FNAL/MILC 10 & \cite{Bailey:2010gb} & 2+1 & \parbox[t]{1.8cm}{2.8, 2.4/3.4, 2.4/2.9,
2.4} & $\gtrsim 3.8$ & 
\parbox[t]{5cm}{See below.}
\\[4.0ex] \hline \\[-1.0ex]
FNAL/MILC 08 & \cite{Bernard:2008dn} & 2+1 & \parbox[t]{1.8cm}{2.4/3.4, 2.4/2.9,
2.4} & $\gtrsim 3.8$ & 
\parbox[t]{5cm}{Estimate FV error to be negligible using 1-loop rHMS$\chi$PT.}
\\[4.0ex] \hline \\[-1.0ex]
FNAL/MILC 04A & \cite{Okamoto:2004xg} & 2+1 & \parbox[t]{1.8cm}{2.4/2.9} & $\gtrsim 4.5$ & 
\parbox[t]{5cm}{No estimate of FV error quoted.}
\\[2.0ex] \hline \\[-1.0ex]
Atoui  13 & \cite{Atoui:2013zza} & 2 & \parbox[t]{1.8cm}{1.7/2.6, 2.1, 2.0/2.7, 2.4} & $\gtrsim 3.6$ & 
\parbox[t]{5cm}{No volume dependence observed within errors.}
\\[4.0ex]
\hline\hline
\end{tabular*}
\caption{Finite volume effects in determinations of the $B\to D \ell\nu$, $B\to D^* \ell\nu$,  $B_s\to D_s \ell\nu$ form factors and of $R(D)$.  Each $L$-entry corresponds to a different lattice spacing, with multiple spatial volumes at some lattice spacings.  For actions with multiple species of pions, the lightest pion masses are quoted.}
}
\end{table}
\begin{table}[!ht]
{\footnotesize
\begin{tabular*}{\textwidth}{l @{\extracolsep{\fill}} c c c l}
\hline\hline \\[-1.0ex]
Collab. & Ref. & $\Nf$ & Ren. & Description 
\\[1.0ex] \hline \hline \\[-1.0ex]
FNAL/MILC 13B $B \to D^*$ & \cite{Qiu:2013ofa} & 2+1  & mNPR &
\parbox[t]{6cm}{See error estimate for $h_+$ below.} 
\\[3.0ex] \hline \\[-1.0ex]
FNAL/MILC 13B $B \to D$ & \cite{Qiu:2013ofa} & 2+1 & mNPR &
\parbox[t]{6cm}{See FNAL/MILC 08A.  A 0.4\% perturbative truncation error for $h_+$ (the dominant contributor to $f_+$) is estimated from $\alpha_s^2$ times the largest 1-loop coefficient observed in the relevant mass range for all similar currents, while a conservative 20\% error is taken for $h_-$.} 
\\[18.0ex] \hline \\[-1.0ex]
FNAL/MILC 12A & \cite{Bailey:2012jg} & 2+1 & mNPR &
\parbox[t]{6cm}{Only the relative matching of the spatial and 
temporal components of currents is relevant for the ratio $R(D)$. 
Uncertainty for this is estimated to be 0.4\%.} 
\\[10.0ex] \hline \\[-1.0ex]
FNAL/MILC 10 & \cite{Bailey:2010gb} & 2+1 & mNPR &
\parbox[t]{6cm}{See below.  A 0.3\% perturbative truncation error is estimated.}
\\[5.0ex] \hline \\[-1.0ex]
FNAL/MILC 08 & \cite{Bernard:2008dn} & 2+1 & mNPR &
\parbox[t]{6cm}{Majority of current renormalization factor cancels in double ratio of lattice correlation functions.  Remaining correction calculated with 1-loop tadpole-improved lattice perturbation theory. 0.3\% perturbative truncation 
error estimated from size of one-loop correction on finest ensemble.}
\\[18.0ex] \hline \\[-1.0ex]
FNAL/MILC 04A & \cite{Okamoto:2004xg} & 2+1 & mNPR &
\parbox[t]{6cm}{ No explicit estimate of perturbative truncation error.}
\\[5.0ex] \hline \\[-1.0ex]
Atoui  13 & \cite{Atoui:2013zza} & 2 & --- &
\parbox[t]{6cm}{Observables obtained from ratios that do not require renormalization. Checks performed by comparing with results coming from currents that are renormalized separately with non-perturbative $Z_{\rm V}$.}
\\[12.0ex]
\hline\hline
\end{tabular*}
\caption{Operator renormalization in determinations of the $B\to D \ell\nu$, $B\to D^* \ell\nu$,  $B_s\to D_s \ell\nu$ form factors and of $R(D)$.}
}
\end{table}
\begin{table}[!ht]
{\footnotesize
\begin{tabular*}{\textwidth}{l @{\extracolsep{\fill}} c c c l}
\hline\hline \\[-1.0ex]
Collab. & Ref. & $\Nf$ & Action & Description 
\\[1.0ex] \hline \hline \\[-1.0ex]
\parbox[t]{2.8cm}{FNAL/MILC 13B\\ $B \to D^*$}  & \cite{Qiu:2013ofa} & 2+1 & Fermilab &
\parbox[t]{6cm}{See FNAL/MILC~10.}
\\[4.5ex] \hline \\[-1.0ex]
\parbox[t]{2.8cm}{FNAL/MILC 13B\\ $B \to D$}  & \cite{Qiu:2013ofa} & 2+1 & Fermilab &
\parbox[t]{6cm}{Heavy-quark discretization errors estimated from power-counting to be 2\% in $h_+$ (the dominant contributor to $f_+$) and 10\% in $h_-$.}
\\[10.0ex] \hline \\[-1.0ex]
FNAL/MILC 12A & \cite{Bailey:2012jg} & 2+1 & Fermilab &
\parbox[t]{6cm}{Discretization errors of form factors estimated 
via power counting which leads to negligible ($\sim$0.2\%) errors 
in the ratio $R(D)$.}
\\[10.0ex] \hline \\[-1.0ex]
FNAL/MILC 10 & \cite{Bailey:2010gb} & 2+1 & Fermilab &
\parbox[t]{6cm}{Discretization errors from heavy quark action estimated to be 
1.1\% from power counting and a more detailed theory of cutoff effects.}
\\[10.0ex] \hline \\[-1.0ex]
FNAL/MILC 08 & \cite{Bernard:2008dn} & 2+1 & Fermilab &
\parbox[t]{6cm}{Heavy-quark discretization errors estimated 
to be 1.5\% from power counting and comparisons of data at different lattice spacings.}
\\[10.0ex] \hline \\[-1.0ex]
FNAL/MILC 04A & \cite{Okamoto:2004xg} & 2+1 & Fermilab &
\parbox[t]{6cm}{ No explicit estimate of heavy-quark discretization errors.}
\\[5.0ex] \hline \\[-1.0ex]
Atoui  13 & \cite{Atoui:2013zza} & 2 & tmWil &
\parbox[t]{6cm}{Results obtained from step-scaling in heavy quark mass via the ratio method. Separate continuum limit extrapolations with mild $a^2$ dependence carried out for each mass point separately. Result at physical value of $m_b$ obtained by interpolation between data region and known exact HQET limit.}
\\[16.0ex]
\hline\hline
\end{tabular*}
\caption{Heavy quark treatment in determinations of the $B\to D \ell\nu$, $B\to D^* \ell\nu$,  $B_s\to D_s \ell\nu$ form factors and of $R(D)$.}
}
\end{table}

\clearpage

\subsection{Notes to section \ref{sec:alpha_s} on the strong coupling $\alpha_{\rm s}$}

\subsubsection{Renormalization scale and perturbative behaviour}




\begin{table}[!htb]
   \footnotesize
   \begin{tabular*}{\textwidth}[l]{l@{\extracolsep{\fill}}rlllllllll}
\hline\hline \\[-1.0ex]
      Collab. & Ref. & $\Nf$ & $\alpha_{\rm eff}$ & Description
     & & & &  \\
      \hline
      \hline
      & & & &  \\[-0.1cm]
Sternbeck 12  & \cite{Sternbeck:2012qs}       & 0        &  
  \parbox[t]{0.1\textwidth}{0.11-0.18}
& \parbox[t]{0.6\textwidth}{$\alpha_{\mathrm{T}}(p)$ for $p = 5 - 40$\,GeV.
  Fitted with 4-loop formulae without power corrections.
  ($\beta$ = 6.0, 6.4, 6.7, 6.92.)}
\\
 {Ilgenfritz 10} & \cite{Ilgenfritz:2010gu}   & 0        &
  \parbox[t]{0.1\textwidth}{0.07-0.9.}
& \parbox[t]{0.6\textwidth}{$\alpha_{\rm T}(p)$ for $p = 1 - 240$\,GeV.
  ($\beta$ = 5.8, 6.0, 6.2, 6.4, 9.0.)}
\\
Sternbeck 10  & \cite{Sternbeck:2010xu}       & 0        & 
  \parbox[t]{0.1\textwidth}{0.07-0.32} 
& \parbox[t]{0.6\textwidth}{$\alpha_{\mathrm{T}}$ for 
  $p = 2.5 - 140$\, GeV, fitted with 4-loop formula partially on
  very small lattices.} 
\\
Brambilla 10  & \cite{Brambilla:2010pp}       & 0        & 
  \parbox[t]{0.1\textwidth}{0.22-0.47} 
& \parbox[t]{0.6\textwidth}{$\alpha_\mathrm{qq}(1/r)$ for
  the range $r/r_0 = 0.15 - 0.5$. Fit of $V(r)$ with 3-loop formula
  (NNNLO) with renormalon subtraction and resummation reproduces
  the static potential for $r/r_0 = 0.15 - 0.45$ well.} 
\\
{Boucaud~08}  & \cite{Boucaud:2008gn}         & 0        & 
  \parbox[t]{0.1\textwidth}{0.18-0.35}
& \parbox[t]{0.6\textwidth}{$\alpha_\mathrm{T}(p)$ with 
  $p = 3 - 6$\, GeV. Fitted to 4-loop perturbation formula
   with $1/p^2$ correction.}
\\
{Boucaud 05}  & \cite{Boucaud:2005gg}         & 0        &
  \parbox[t]{0.1\textwidth}{0.22-0.55}
& \parbox[t]{0.6\textwidth}{$\Lambda_{\widetilde{\rm MOM}_{g,c}}$ using 
  gluon and ghost propagators with $2 \leq \mu \leq 6$\, GeV. 
  Fitted to 4-loop perturbation theory.}
\\
\parbox[t]{0.12\textwidth}{ {QCDSF-\\ UKQCD~05}}
              & \cite{Gockeler:2005rv}        & 0        &
  \parbox[t]{0.1\textwidth}{0.10-0.15}
& \parbox[t]{0.6\textwidth}
  {$\alpha_{\overline{MS}}(2.63/a)$ computed from the boosted coupling.}
\\        
CP-PACS 04     & \cite{Takeda:2004xha}         & 0        &      
  \parbox[t]{0.12\textwidth}{0.08-0.28}
& \parbox[t]{0.60\textwidth}{$\alpha_{\rm SF}(1/L)$ step scaling
  functions at $\alpha_{\rm eff} = 0.08, 0.19$, study of continuum limit.
  Agreement of continuum limit with ALPHA 98.}
\\
{Boucaud 01A} & \cite{Boucaud:2001st}         &  0       &
  \parbox[t]{0.1\textwidth}{0.18-0.45}
& \parbox[t]{0.6\textwidth}{$\alpha_{\rm MOM}$ with
  $p = 2.5 - 10$\, GeV. Consistency check of 3-loop perturbation
  formula with gluon condensate. $\langle A^2 \rangle$ from
  $\alpha_{\rm MOM}$ and gluon propagator are consistent.}
\\
{Soto 01}     & \cite{DeSoto:2001qx}          & 0        &
  \parbox[t]{0.1\textwidth}{ {0.25-0.36, 0.3-0.36, 0.19-0.24 }}
& \parbox[t]{0.6\textwidth}{{$\alpha_{\widetilde{\rm MOM}}$ for
  $p = 3 - 10$\, GeV. Fit with 3-loop formula with gluon condensate.
  3-loop formula without condensate does not fit the lattice data.
  ($\beta$ = 6.0, 6.2, 6.8.)}}
\\
{Boucaud 00A} & \cite{Boucaud:2000ey}         &  0       &
  \parbox[t]{0.1\textwidth}{0.35-0.55, 0.25-0.45, 0.22-0.28, 0.18-0.22}
& \parbox[t]{0.6\textwidth}{$\alpha_{\widetilde{\rm MOM}}$ with
  $p = 2 - 10$\, GeV. Fitted to 3-loop perturbation theory with power
  correction. 4-loop coefficient is strongly correlated to the power
  correction coefficient. ($\beta$ = 6.0, 6.2, 6.4, 6.8.)}
\\
{Boucaud 00B} & \cite{Boucaud:2000nd}         &  0       &
  \parbox[t]{0.1\textwidth}{0.35-0.55, 0.25-0.45, 0.22-0.28, 0.18-0.22}
& \parbox[t]{0.6\textwidth}{$\alpha_{\rm MOM, \widetilde{MOM}}$ with
  $2 \leq \mu \leq 10$\, GeV.  Consistency check of 3-loop
  perturbation formula with gluon condensate.
 $\beta_2^{MOM}=1.5\times \beta_2^{\widetilde{MOM}}$ is needed.
 ($\beta$ = 6.0, 6.2, 6.4, 6.8.)}
\\
{Becirevic 99A}& \cite{Becirevic:1999uc}      &  0       &
  \parbox[t]{0.1\textwidth}{0.25-0.4} 
& \parbox[t]{0.6\textwidth}{$\alpha_{\widetilde{\rm MOM}}$ with
  $p$ = 2.5 - 5.5\, GeV.}
\\
{Becirevic 99B}  & \cite{Becirevic:1999hj}    &  0       &
  \parbox[t]{0.1\textwidth}{0.18-0.25}
& \parbox[t]{0.6\textwidth}{$\alpha_{\widetilde{\rm MOM}}$ from
a single lattice spacing with $p$ = 5.6 - 9.5\, GeV.}
\\
{SESAM 99}    & \cite{Spitz:1999tu}           & 0        &
  \parbox[t]{0.1\textwidth}{0.15}
& \parbox[t]{0.6\textwidth}{$\alpha_{V}(3.41/a)$ computed
  from the boosted coupling.}
\\
ALPHA 98      & \cite{Capitani:1998mq}        & 0        & 
  \parbox[t]{0.1\textwidth}{0.07-0.28}
& \parbox[t]{0.60\textwidth}{$\alpha_{\rm SF}(1/L)$ step scaling,
  agreement with 3-loop running for $\alpha_{\rm eff} < 0.15$.}
\\
{Boucaud 98A} & \cite{Boucaud:1998bq}         & 0        &
  \parbox[t]{0.1\textwidth}{0.35-0.5} 
& \parbox[t]{0.6\textwidth}{$\alpha_\mathrm{MOM}$,  
  with $2.1 \leq \mu \leq 3.9$\, GeV. Fitted to 3-loop
  perturbation theory without power correction.}
\\
{Boucaud 98B} & \cite{Boucaud:1998xi}         & 0        &
  \parbox[t]{0.1\textwidth}{0.27-0.50} 
& \parbox[t]{0.6\textwidth}{$\alpha_{\widetilde{\rm MOM}}$
  with $\mu = 2.2 - 4.5$\, GeV.}
\\
Alles 96      & \cite{Alles:1996ka}           &  0       & 
  \parbox[t]{0.1\textwidth}{0.35-0.71}
& \parbox[t]{0.6\textwidth}{$\alpha_{\widetilde{\rm MOM}}(p)$
  with $p=1.8-3.0$\, GeV.}
\\
{Wingate 95}  & \cite{Wingate:1995fd}         & 0        &
  \parbox[t]{0.1\textwidth}{0.15}
& \parbox[t]{0.6\textwidth}{$\alpha_{V}(3.41/a)$ computed
  from the boosted coupling.}
\\
{Davies 94}   & \cite{Davies:1994ei}          & 0        &
  \parbox[t]{0.1\textwidth}{0.15}
& \parbox[t]{0.6\textwidth}{$\alpha_{V}(3.41/a)$ computed from
  the boosted coupling.}
\\
L\"uscher 93 & \cite{Luscher:1993gh}          & 0        & 
  \parbox[t]{0.1\textwidth}{0.09-0.28}
& \parbox[t]{0.60\textwidth}{$\alpha_{\rm SF}(1/L)$ step scaling,
  agreement with 3-loop running for $\alpha_{\rm eff} < 0.17$.}
\\      
UKQCD 92     & \cite{Booth:1992bm}            & 0        & 
  \parbox[t]{0.1\textwidth}{0.17-0.40} 
& \parbox[t]{0.6\textwidth}{$\alpha_\mathrm{qq}(1/r)$ for a single
  lattice spacing. Fit of $\alpha_\mathrm{qq}(1/r)$ to a NLO formula.}
\\
Bali 92     & \cite{Bali:1992ru}              & 0        & 
  \parbox[t]{0.1\textwidth}{0.15-0.35} 
& \parbox[t]{0.6\textwidth}{$\alpha_\mathrm{qq}(1/r)$ for the
  lattice spacing used in the analysis. Box size $L\approx1.05$\,fm.
  Fit of $\alpha_\mathrm{qq}(1/r)$ to a NLO formula. $\Lambda_\msbar$ is
  found to depend on the fit-range.}
\\
El-Khadra 92& \cite{ElKhadra:1992vn}          & 0        &
  \parbox[t]{0.1\textwidth}{0.15, 0.13, 0.12}
& \parbox[t]{0.6\textwidth}{$\alpha_{\overline{MS}}(\pi/a)$ from 1-loop
  boosted perturbation theory.}
\\
   &&&&&&&&& \\[-0.1cm]
   \hline
   \hline
\end{tabular*}
\caption{Renormalization scale and perturbative behaviour
         of $\alpha_s$ determinations for $N_f=0$.}
\label{tab_Nf=0_renormalisation}
\end{table}

\begin{table}[!htb]
   \vspace{3.0cm}
   \footnotesize
   \begin{tabular*}{\textwidth}[l]{l@{\extracolsep{\fill}}rlllllllll}
\hline\hline \\[-1.0ex]
      Collab. & Ref. & $\Nf$ &  $\alpha_{\rm eff}$ & Description   \\
      &&&&&&& & \\[-0.1cm]
      \hline
      \hline
      &&&&&&&&& \\[-0.1cm]
ALPHA 12 & \cite{Fritzsch:2012wq} & 2  & 
\parbox[t]{0.1\textwidth}{see \\ ALPHA 04}
    & 
\parbox[t]{0.60\textwidth}{Determination of $\Lambda_\msbar / f_K$ using ALPHA 04} 
\\ \\[-0.1cm] \hline\\[-1.5ex] 
 {Sternbeck 12}  & \cite{Sternbeck:2012qs}       & 2        &  
 {0.17-0.23}
& \parbox[t]{0.6\textwidth}{ {
$\alpha_\mathrm{T}$ for $(r_0 p)^2=200-2000$. 
Fit with 4-loop formula without condensate. Deviation at higher energy is observed.}}
\\ \\[-0.1cm] \hline\\[-1.5ex] 
ETM 11C   & \cite{Jansen:2011vv}   & 2 & 
0.26-0.96 & 
\parbox[t]{0.6\textwidth}{$\alpha_{\rm qq}(1/r)$ as computed by us from $\Lambda_\msbar=315$MeV.
\newline Fit of $V(r)$ with 3-loop formula (NNNLO) with renormalon subtraction
         and resummation reproduces the static potential for $r/r_0=0.2-0.6$ well. One fit-range, using $r/a=2-4$ at the smallest lattice spacing corresponds to 
         $\alpha_{\rm eff}=0.26-0.40$. 
         In the $\msbar$ scheme one has 
         $\alpha_\msbar(1/r)=0.24 - 0.63$ and for the restricted
         fit $\alpha_\msbar(1/r)=0.24 - 0.36$.
         Central values taken from $a=0.042$\,fm lattice with $L=1.3$\,fm
         and $m_\pi=350$\,MeV. 
}
\\ \\[-0.1cm] \hline\\[-1.5ex] 
 {ETM 10F}       & \cite{Blossier:2010ky}   & 2              &  
0.24-0.45 & \parbox[t]{0.60\textwidth}{$\alpha_T$ for momentum up to
2.6 - 5.6\,GeV. Fitted with 4 loop formula with gluon condensate}
\\ \\[-0.1cm] \hline\\[-1.5ex]  
 {Sternbeck 10}  & \cite{Sternbeck:2010xu}       & 2       & 
 {0.19-0.38} & {$\alpha_\mathrm{T}$ for $1\leq(ap)^2\leq 10$.
Fitted with 4-loop formula.}
\\ \\[-0.1cm] \hline\\[-1.5ex] 
 {JLQCD 08} & \cite{Shintani:2008ga} & 2                  
&  {0.25 -0.30}
& \parbox{0.6\textwidth}{ {$\alpha_{\overline{\rm MS}}(Q)$ for $0.65<(aQ)^2<1.32$. Fit with the perturbative formula 
with power corrections. }}
\\ \\[-0.1cm] \hline\\[-1.5ex] 
\parbox[t]{0.11\textwidth}{QCDSF-\\UKQCD 05}
   & \cite{Gockeler:2005rv}
   & 2
   & \parbox[t]{0.1\textwidth}
        {0.20-0.18}
   & \parbox[t]{0.6\textwidth}
        {$\alpha_\msbar(1.4/a)$ computed from the boosted coupling.}
\\ \\[-0.1cm] \hline\\[-1.5ex] 
ALPHA 04 & \cite{DellaMorte:2004bc} & 2 & 
0.078-0.44 & 
\parbox[t]{0.60\textwidth}{$\alpha_{\rm SF}(1/L)$  step scaling, agreement with 
3-loop running for $\alpha_s < 0.2$}
\\ \\[-0.1cm] \hline\\[-1.5ex] 
ALPHA 01 & \cite{Bode:2001jv} & 2 & 
0.078-0.44 & 
\parbox[t]{0.60\textwidth}{$\alpha_{\rm SF}(1/L)$  step scaling, agreement with 
3-loop running for $\alpha_s < 0.2$}
\\ \\[-0.1cm] \hline\\[-1.5ex] 
{Boucaud 01B}    & \cite{Boucaud:2001qz}    & 2  & 
0.25-0.5   &  
\parbox[t]{0.60\textwidth}{$\alpha_{\widetilde{MOM}}$ for momentum up to 7\,GeV.
Fitted with 4-loop formula with and without power correction,
leading to different results for $\Lambda_\msbar^{(2)}$. 
Extrapolation of $\alpha_s(1.3\,\GeV)$ in $N_f$ from $N_f=0, 2$ to
$N_f=3$ is made.}
\\ \\[-0.1cm] \hline\\[-1.5ex] 
 {SESAM 99}
   & \cite{Spitz:1999tu}
   & 2
   & \parbox[t]{0.1\textwidth}
        {0.17}
   & \parbox[t]{0.6\textwidth}
        {The boosted coupling $\alpha_P(3.41/a)$.}
\\ \\[-0.1cm] \hline\\[-1.5ex] 
 {Wingate 95}
   & \cite{Wingate:1995fd}
   & 2
   & \parbox[t]{0.1\textwidth}
        {0.18}
   & \parbox[t]{0.6\textwidth}
        {$\alpha_{V}(3.41/a)$ computed from the boosted coupling.}
\\ \\[-0.1cm] \hline\\[-1.5ex] 
 {Aoki 94}
   & \cite{Aoki:1994pc}
   & 2                   
   & \parbox[t]{0.1\textwidth}
        {0.14}
   & \parbox[t]{0.6\textwidth}
        {$\alpha_\msbar(\pi/a)$ computed from the boosted coupling.}
\\ \\[-0.1cm] \hline\\[-1.5ex] 
 {Davies 94} 
   & \cite{Davies:1994ei}  
   & 2
   & \parbox[t]{0.1\textwidth}
        {0.18}
   & \parbox[t]{0.6\textwidth}
        {$\alpha_{V}(3.41/a)$ computed from the boosted coupling.}
\\                            
[0.5ex]
     &&&&&&&&& \\[-0.1cm]
\hline\hline
\end{tabular*}
\caption{Renormalization scale and perturbative behaviour 
of $\alpha_s$ determinations for $N_f=2$.}
\label{tab_Nf=2_renormalisation}
\end{table}


\begin{table}[!htb]
   \vspace{3.0cm}
   \footnotesize
   \begin{tabular*}{\textwidth}[l]{l@{\extracolsep{\fill}}rlllllllll}
\hline\hline \\[-1.0ex]
      Collab. & Ref. & $\Nf$ & $\alpha_{\rm eff}$ & Description \\
      &&&&& \\[-0.1cm]
      \hline
      \hline
      &&&&& \\[-0.1cm]
Bazavov 12   & \cite{Bazavov:2012ka}  & 2+1  & 
0.23-0.57 & 
\parbox[t]{0.60\textwidth}{$\alpha_{\rm qq}$ computed by us from $\Lambda_\msbar r_0=0.70 $. 
Fit of $V(r)$ with 3-loop formula (NNNLO) with renormalon subtraction
and resummation reproduces the static potential for 
$r/r_0=0.135-0.5$ well. 
}
\\ \\[-0.1cm] \hline\\[-1.5ex]
 {Sternbeck 12}  & \cite{Sternbeck:2012qs}  & 2+1       &  
 {0.19-0.25}  &   {$\alpha_\mathrm{T}$ for $(p r_0)^2=200-2000$. 
Comparison with 4-loop formula.}
\\ \\[-0.1cm] \hline\\[-1.5ex]
 {JLQCD 10} & \cite{Shintani:2010ph} & 2+1                  
 & {0.29-0.35}
 &\parbox{0.6\textwidth}{ {$\alpha_{\overline{\rm MS}}(Q)$ for $0.4<(aQ)^2<1.0$. Fit with the perturbative formula 
with power corrections.}}
\\ \\[-0.1cm] \hline\\[-1.5ex] 
 {HPQCD 10}
   & \cite{McNeile:2010ji}
   & 2+1
   & 
   & \parbox[t]{0.6\textwidth}
        {Update of $r_1$ and $r_1/a$ in HPQCD 08A.}
\\ \\[-0.1cm] \hline\\[-1.5ex]
HPQCD 10    & \cite{McNeile:2010ji}  & 2+1 &  
0.12-0.42  &\parbox[t]{0.60\textwidth}{Combined range given for 
$\alpha_\mathrm{eff}$ from $R_4$ and $R_6/R_8$. 
Fit of $R_n, \;n=4\ldots10$ to NNNLO of the ratios (meaning NNLO for $\alpha_s$)
including $(a m)^{2i}$ terms with $i\leq 10$; coefficients constrained by priors. } 
\\ \\[-0.1cm] \hline\\[-1.5ex]
PACS-CS 09A& \cite{Aoki:2009tf} & 2+1 & 
0.08-0.27 & 
\parbox[t]{0.60\textwidth}{$\alpha_{\rm SF}(1/L)$  step scaling, agreement with 
3-loop running for $\alpha_s \leq 0.27$}
\\
HPQCD 08B    & \cite{Allison:2008xk}  & 2+1 &  
0.378  &\parbox[t]{0.60\textwidth}{Fit to NNNLO of the ratios 
(meaning NNLO for $\alpha_s$) at the charm mass
including $(a m)^{2i}$ terms with $i\leq 2\ldots 4$; coefficients constrained by priors. } 
\\ \\[-0.1cm] \hline\\[-1.5ex]
 {HPQCD 08A}
   & \cite{Davies:2008sw}
   & 2+1
   & \parbox[t]{0.1\textwidth}
        {0.15-0.4}
   & \parbox[t]{0.6\textwidth}
        {$\alphav(q^*)$ for a variety of short distance quantities,
         using same method as in HPQCD 05A.}
\\ \\[-0.1cm] \hline\\[-1.5ex]
 {Maltman 08}
   & \cite{Maltman:2008bx}
   & 2+1
   & 
   & \parbox[t]{0.6\textwidth}
        {Re-analysis of HPQCD 05A for a restricted set of
         short distance quantities with similar results.}
\\ \\[-0.1cm] \hline\\[-1.5ex]
 {HPQCD 05A}
   & \cite{Mason:2005zx}
   & 2+1
   & \parbox[t]{0.1\textwidth}
        {0.2-0.4}
   & \parbox[t]{0.6\textwidth}
        {$\alphav(q^*)$ for a variety of short distance quantities.
        }
\\
      &&&&& \\[-0.1cm]
      \hline
      \hline
\end{tabular*}
\caption{Renormalization scale for $N_f=3$.}
\label{tab_Nf=3_renormalisation}
\end{table}


\begin{table}[!htb]
   \vspace{3.0cm}
   \footnotesize
   \begin{tabular*}{\textwidth}[l]{l@{\extracolsep{\fill}}rlllllllll}
\hline\hline \\[-1.0ex]
      Collab. & Ref. & $\Nf$ & $\alpha_{\rm eff}$ & Description \\
      &&&&& \\[-0.1cm]
      \hline
      \hline
      &&&&& \\[-0.1cm]
 { {ETM 13D}}       &  \cite{Blossier:2013ioa} & 2+1+1 &  0.26-0.7    & 
	\parbox[t]{0.60\textwidth}{$\alpha_\mathrm{T}(p)$= for $p=1.6-6.5$ GeV. 
	Update of \cite{Blossier:2012ef}
                                   with improved power law determination.}
 \\ \\[-0.1cm] \hline\\[-1.5ex]
 { ETM 12C}       & \cite{Blossier:2012ef}   & 2+1+1 &
0.24-0.38 & 
	\parbox[t]{0.60\textwidth}{$\alpha_\mathrm{T}(p)$ for $p=1.7-6.8$ GeV. 
	Fit with 4-loop formula with gluon condensate or higher power.}
	 \\ \\[-0.1cm] \hline\\[-1.5ex]
      ALPHA 10A & \cite{Tekin:2010mm} & 4                  & 0.07-0.28
        & \parbox[t]{0.6\textwidth}{$\alpha_\mathrm{SF}(1/L)$.
        Comparison with 2-, 3-loop $\beta$ function.}
 \\ \\[-0.1cm] \hline\\[-1.5ex]
 {ETM 11D}      & \cite{Blossier:2011tf}   & 2+1+1 & 
	0.24-0.4& 
	\parbox[t]{0.60\textwidth}{$\alpha_\mathrm{T}(p)$ for $p=3.8-7.1$\,GeV
        with H(4)-procedure. Fit with 4-loop formula with gluon condensate.}
 \\ \\[-0.1cm] \hline\\[-1.5ex]
      Perez 10 & \cite{PerezRubio:2010ke} & 4           &0.06-0.28  
        & \parbox[t]{0.6\textwidth}{$\alpha_\mathrm{SF}(1/L)$.
          Comparison with 1-, 2-, 3-loop $\beta$ function.}\\
      &&&&&&\\[-0.1cm]
      \hline
      \hline
\end{tabular*}
\caption{Renormalization scale of $\alpha_s$ determinations 
for $N_f=4$.}
\label{tab_Nf=4_renormalisation}
\end{table}

\clearpage


\subsubsection{Continuum limit}

\hspace{-1.5cm}

\begin{table}[!htb]
   \footnotesize
   \begin{tabular*}{\textwidth}[l]{l@{\extracolsep{\fill}}rlllllllll}
\hline\hline \\[-1.0ex]
      Collaboration & Ref. & $\Nf$ & $a\,\mu$  & Description\\
     & & & &  \\
      \hline
      \hline
      & & & &  \\[-0.1cm]
      Sternbeck 12  & \cite{Sternbeck:2012qs}       & 0        &  
      {4 lattice spacings $a\leq 0.1$\,fm} & 
      {At $\alpha_s$ = 0.18, $ap$ = 2.7, 1.5 for $\beta = 6.0, 6.4$.}\\
      Brambilla 10 & \cite{Brambilla:2010pp} & 0  & 
      \parbox[t]{0.25\textwidth}{ At least 3 lattice spacings 
          with  $0.2 \leq 2a/r \leq 1.1 $ } &
      \parbox[t]{0.4\textwidth}{Extrapolation of potential differences 
          $V(r)-V(0.51r_0)$ linear in $a^2$ performed in \cite{Necco:2001xg}
          with several lattice spacings.}
\\ 
      Ilgenfritz 10 & \cite{Ilgenfritz:2010gu}              & 0       & 
      \parbox[t]{0.25\textwidth}{{$a$ = 0.136, 0.093, 0.068, 0.051\,fm
      ($\beta=$ 5.8, 6.0, 6.2, 6.4), while no value of $a$ is given
      for $\beta = 9.0$}} &
      \parbox[t]{0.4\textwidth}{ {At $\alpha_s = 0.3$, 
      $ap =$ 2.0, 1.4, 1.0, 0.8 ($\beta$ = 5.8, 6.0, 6.2, 6.4). 
      For $\beta = 9.0$ at $ap = 1.4$, $\alpha_s = 0.082$.}
     }\\
      Sternbeck 10  & \cite{Sternbeck:2010xu}       & 0        & 
     \parbox[t]{0.25\textwidth}{{8 lattice spacings $a$ = 0.004 - 0.087\,fm
     ($r_0 = 0.467$\,fm)}}
&  \parbox[t]{ 0.4\textwidth}{
       {$\sqrt{3} < ap < \sqrt{12}$}.} \\
     Boucaud 08    & \cite{Boucaud:2008gn}         
     & 0           & {  $a = 0.1, 0.07, 0.05$\,fm}  & At $\alpha_s = 0.3$
     the data have $ap = 2.6, 1.9, 1.5$.\\
     {QCDSF/UKQCD 05}
     & \cite{Gockeler:2005rv}
     & 0                   
     & \parbox[t]{0.25\textwidth}
        {7 lattice spacings with $a = 0.10$ - $0.028\,\mbox{fm}$.}
     & \parbox[t]{0.6\textwidth}
        {$r_0/a$, together with $r_0 = 0.467\,\mbox{fm}$.}
\\
    {Boucaud 05}    & \cite{Boucaud:2005gg}         & 0       &
    {$a = 0.1, 0.07, 0.05$\,fm} & 
    {At $\alpha_s \leq 0.3$ $ap = 1.9, 1.4, 1.0$}.\\
    CP-PACS 04 & \cite{Takeda:2004xha} & 0  & 
    \parbox[t]{0.25\textwidth}{ 4 spacings, $a/L = 1/12 - 1/4$.} &
    \parbox[t]{0.4\textwidth}{ Iwasaki and  
    L\"uscher Weisz tree-level improved bulk actions; boundary improvement at 
    tree-level, 1-loop and with two different choices of implementation.}
\\
    Soto 01            & \cite{DeSoto:2001qx}       & 0       &  
    {$a = 0.07, 0.05, 0.03$\,fm} & {At $\alpha_s \leq 0.3$,
    the data have $ap = 1.4, 1.0,  0.6$.}\\
    {Boucaud 01A}      & \cite{Boucaud:2001st}      & 0       &   
    {$a = 0.1, 0.07, 0.05, 0.03$\,fm}  & 
    {At $\alpha_s \leq 0.3$ $ap = 1.9, 1.4, 1.0, 0.6$.} \\
    {Boucaud 00A}      &\cite{Boucaud:2000ey}           &  0    & 
    {$a = 0.1, 0.07, 0.05, 0.03$~fm}  &
    {At $\alpha_s \leq 0.3$ $ap = 1.9, 1.4, 1.0,   0.6$.}\\
    {Boucaud 00B}   & \cite{Boucaud:2000nd}      & 0         & 
    {$a = 0.1, 0.07, 0.05, 0.03$\,fm} & 
    {At $\alpha_s \leq 0.3$ $ap = 1.9, 1.4, 1.0, 0.6$.}
\\
 {SESAM 99}
   & \cite{Spitz:1999tu}
   & 0                   
   & \parbox[t]{0.25\textwidth}
        {1 lattice spacing with $a = 0.086\,\mbox{fm}$}
   & \parbox[t]{0.4\textwidth}
        {$\Upsilon$ spectrum splitting.}
\\
{Becirevic 99A}  & \cite{Becirevic:1999uc} & 0 &
{$a = 0.07, 0.05$\,fm} & {At $\alpha_s \leq 0.3$ $ap = 1.4, 1.0$.}
\\
{Becirevic 99B}  & \cite{Becirevic:1999hj} & 0 &
{$a = 0.1, 0.07, 0.03$\,fm} & 
\parbox[t]{ 0.4\textwidth}{Only $a = 0.03$fm used to extract
$\alpha_s$. At $\alpha_s \leq 0.3$, $ap = 0.6 - 1.5$.}
\\
ALPHA 98 & \cite{Capitani:1998mq} & 0  & 
\parbox[t]{0.25\textwidth}{ 4 to 6 spacings, $a/L = 1/12 - 1/5$ 
in step scaling functions (SSF)} &  
\parbox[t]{0.4\textwidth}{1-loop O($a$) boundary improvement,
linear extrapolation in $a/L$.
\newline $a/L = 1/8 - 1/5$ for $\alpha_s\leq 0.11$ SSF,
\newline $a/L = 1/12 - 1/5$ for $0.12 \leq \alpha_s\leq 0.20$ SSF,
\newline  $L_{\rm max}/r_0$ from \cite{Guagnelli:1998ud},
where several lattice spacings were used.}
\\
{Boucaud 98A}    & \cite{Boucaud:1998bq} & 0 &
{$a = 0.1, 0.07, 0.05$\,fm} & {At $\alpha_s \leq 0.3$, $ap = 1.9, 1.4, 1.0$}.\\
Boucaud 98B  & \cite{Boucaud:1998xi} & 0 & 
{$a = 0.1, 0.07, 0.05$\,fm} & {At $\alpha_s \leq 0.3$, $ap = 1.9, 1.4, 1.0$}.
\\
Alles 96           & \cite{Alles:1996ka}                &  0        & 
{$a \leq 0.1$\,fm} & {At $\alpha_s = 0.35$, $ap = 1.5$}. \\
{Wingate 95}
   & \cite{Wingate:1995fd}
   & 0                   
   & \parbox[t]{0.25\textwidth}
        {1 lattice spacing with $a = 0.11\,\mbox{fm}$}
   & \parbox[t]{0.4\textwidth}
        {Charmonium $1S$-$1P$ splitting.}
\\
 {Davies 94} 
   & \cite{Davies:1994ei}  
   & 0                   
   & \parbox[t]{0.25\textwidth}
        {1 lattice spacing with $a = 0.077\,\mbox{fm}$}
   & \parbox[t]{0.4\textwidth}
        {$\Upsilon$ spectrum splitting.}
\\
L\"uscher 93 & \cite{Luscher:1993gh} & 0  & 
\parbox[t]{0.25\textwidth}{4 or 5 lattice spacings, $a/L = 1/12 - 1/5$ 
in step scaling functions} &  
\parbox[t]{0.4\textwidth}{ 1-loop O($a$) boundary improvement,
linear extrapolation in $a/L$.
\newline $a/L = 1/8 - 1/5$ for $\alpha_s\leq 0.11$ SSF,
\newline $a/L = 1/10 - 1/5$ for $0.11 \leq \alpha_s\leq 0.22$ SSF,
\newline $a/L = 1/12 - 1/6$ for $0.22 \leq \alpha_s\leq 0.28$ SSF,
\newline $a/L = 1/8.5 - 1/4.5$ for continuum extrapolation of 
$L_{\rm max}/ \sqrt{K}$.}
\\
UKQCD 92     & \cite{Booth:1992bm}    & 0  &
\parbox[t]{0.25\textwidth}{ One lattice spacing
          with  \newline $0.44 \leq 2a/r \leq 1.6$ } &
\parbox[t]{0.4\textwidth}{No continuum limit.}
\\ 
Bali 92     & \cite{Bali:1992ru}    & 0  &
\parbox[t]{0.25\textwidth}{One lattice spacing
          with \newline $0.4 \leq 2a/r \leq 1.6 $ } &
\parbox[t]{0.4\textwidth}{No continuum limit.}
\\ 
{El-Khadra 92}   
   & \cite{ElKhadra:1992vn} 
   & 0       
   & \parbox[t]{0.25\textwidth}
        {3 lattice spacings with $a = 0.17,\ 0.11,\ 0.08\,\mbox{fm}$}
   & \parbox[t]{0.6\textwidth}
        {Charmonium $1S$-$1P$ splitting.}
\\
[-0.1cm]
     &&&&&&&&& \\[-0.1cm]
             \hline
      \hline
\end{tabular*}
\caption{Continuum limit for $\alpha_s$ determinations with $N_f=0$.}
\label{tab_Nf=0_continuumlimit}
\end{table}


\begin{table}[!htb]
   \vspace{0.0cm}
   \footnotesize
   \begin{tabular*}{\textwidth}[l]{l@{\extracolsep{\fill}}rlllllllll}
\hline\hline \\[-1.0ex]
      Collab. & Ref. & $\Nf$ &  $a\,\mu$ & Description   \\
      &&&&&&& & \\[-0.1cm]
      \hline
      \hline
      &&&&&&&&& \\[-0.1cm]
ALPHA 12 & \cite{Fritzsch:2012wq} & 2 &  
\parbox[t]{0.25\textwidth}{$a=0.049, 0.066, 0.076$fm from $f_K$} & 
\parbox[t]{0.45\textwidth}{2-loop O($a$) boundary improvement, 
linear extrapolation of $L_\mathrm{max} f_\mathrm{K}$ in $a^2$.}
 \\ \\[-0.1cm] \hline\\[-1.5ex]
      Sternbeck 12  & \cite{Sternbeck:2012qs}  & 2        &
     {$a=0.073,0.07,0.06$\,fm}  &  
     {At $\alpha_s = 0.23$, $ap = 2.1, 2.0, 1.7$.}
 \\ \\[-0.1cm] \hline\\[-1.5ex]
ETM 11C   & \cite{Jansen:2011vv}    & 2    & 
\parbox[t]{0.25\textwidth}{$0.30 \leq 2a/r \leq 1.0 $ 
\newline $0.67 \leq 2a/r \leq 1.26 $ when $\alpha_s = 0.3$} &
\parbox[t]{0.45\textwidth}{Four lattice spacings; continuum limit
studied with a particular range in $r$; central result
from the smallest lattice spacing, $a = 0.042$fm.}
 \\ \\[-0.1cm] \hline\\[-1.5ex]
 {ETM 10F}       & \cite{Blossier:2010ky}   & 2              &  
\parbox[t]{0.25\textwidth}{
 {$a=0.05,0.07,0.08$~fm. Different lattice spacings are patched together.}}
&  \parbox[t]{0.45\textwidth}{  At $\alpha_s=0.3$, $ap=1.6, 1.3, 1.1$.} 
 \\ \\[-0.1cm] \hline\\[-1.5ex]
     Sternbeck 10  & \cite{Sternbeck:2010xu}  & 2        &  
     $a=0.068,0.076,0.082$~fm
     &  {At $\alpha_s \leq 0.3$, $ap \geq 1.7$.} 
 \\ \\[-0.1cm] \hline\\[-1.5ex]
 {JLQCD 08} & \cite{Shintani:2008ga} & 2                  
& \parbox[t]{0.25\textwidth}{$a=0.12$ fm  from $r_0=0.49$\,fm} 
& \parbox[t]{0.45\textwidth}{{Single lattice spacing, $0.64 < (aQ)^2 < 1.32$.
At $\alpha_s = 0.3$, $ap = 0.81$.}}
 \\ \\[-0.1cm] \hline\\[-1.5ex]
\parbox[t]{0.11\textwidth}{QCDSF-\\ UKQCD 05}
   & \cite{Gockeler:2005rv}
   & 2
   & \parbox[t]{0.25\textwidth}
        {4 lattice spacings with $a = 0.10$ - $0.066\,\mbox{fm}$}
   & \parbox[t]{0.45\textwidth}
        {$r_0$, together with $r_0 = 0.467\,\mbox{fm}$.}
 \\ \\[-0.1cm] \hline\\[-1.5ex]
ALPHA 04 & \cite{DellaMorte:2004bc} & 2  & 
$a/L=1/8,1/6,1/5,1/4$ & 
\parbox[t]{0.45\textwidth}{ 1-loop (at weak coupling) and
2-loop O($a$) boundary improvement,
linear extrapolation of 
SSF in $(a/L)^2$}
 \\ \\[-0.1cm] \hline\\[-1.5ex]
ALPHA 01A & \cite{Bode:2001jv} & 2 & 
$a/L = 1/6, 1/5, 1/4$ & 
\parbox[t]{0.45\textwidth}{1-loop (at weak coupling) and
2-loop O($a$) boundary improvement,
weighted average of 
SSF with $a/L=1/5,1/6$.}
 \\ \\[-0.1cm] \hline\\[-1.5ex]
 {Boucaud 01B}    & \cite{Boucaud:2001qz}    & 2         & 
\parbox[t]{0.25\textwidth}{
 {$a = 0.05, 0.07, 0.09$\,fm. Data at different lattice spacings 
 are patched together}} &
  \parbox[t]{0.45\textwidth}{ 
  {At $\alpha_s = 0.3$, $ap = 1.6, 1.3, 0.9$; plain Wilson action
  with ${\rm O}(a)$ errors}.}
 \\ \\[-0.1cm] \hline\\[-1.5ex]
 {SESAM 99}
   & \cite{Spitz:1999tu}
   & 2
   & \parbox[t]{0.25\textwidth}
        {1 lattice spacing with $a = 0.079\,\mbox{fm}$}
   & \parbox[t]{0.45\textwidth}
        {$\Upsilon$ spectrum splitting.}
 \\ \\[-0.1cm] \hline\\[-1.5ex]
 {Wingate 95}
   & \cite{Wingate:1995fd}
   & 2
   & \parbox[t]{0.25\textwidth}
        {1 lattice spacing with $a = 0.11\,\mbox{fm}$}
   & \parbox[t]{0.45\textwidth}
        {Charmonium $1S$-$1P$ splitting.}
 \\ \\[-0.1cm] \hline\\[-1.5ex]
 {Aoki 94}
   & \cite{Aoki:1994pc}
   & 2                   
   & \parbox[t]{0.25\textwidth}
        {1 lattice spacing with $a = 0.10\,\mbox{fm}$}
   & \parbox[t]{0.6\textwidth}
        {Charmonium $1P-1S$ splitting}
 \\ \\[-0.1cm] \hline\\[-1.5ex]
 {Davies 94} 
   & \cite{Davies:1994ei}  
   & 2                   
   & \parbox[t]{0.25\textwidth}
        {1 lattice spacing with $a = 0.08\,\mbox{fm}$}
   & \parbox[t]{0.45\textwidth}
        {$\Upsilon$ spectrum splitting.}
    \\  \\[-0.1cm]
      \hline
      \hline
\end{tabular*}
\caption{Continuum limit for $\alpha_s$ determinations with $N_f=2$.}
\label{tab_Nf=2_continuumlimit}
\end{table}


\begin{table}[!htb]
   \vspace{0.0cm}
   \footnotesize
   \begin{tabular*}{\textwidth}[l]{l@{\extracolsep{\fill}}rlllllllll}
\hline\hline \\[-1.0ex]
      Collab. & Ref. & $\Nf$ & $a\,\mu$ & Description \\
      &&&&&&& & \\[-0.1cm]
      \hline
      \hline
      &&&&&&& \\[-0.1cm]
Bazavov 12   & \cite{Bazavov:2012ka} & 2+1  & 
$2a/r=0.6 - 2.0 $  &
\parbox[t]{0.4\textwidth}{7 lattice spacings; 4 lattice spacings with 
\newline $1.14 \leq 2a/r \leq 1.5$ when $\alpha_s(1/r) = 0.3$.
\newline $2a/r = 2$ when $\alpha_s(1/r) = 0.23$ (on the finest lattice).} 
 \\ \\[-0.1cm] \hline\\[-1.5ex]
Sternbeck 12& \cite{Sternbeck:2012qs} & {2+1}       
 & {$a=0.07$\,fm}  &  
   {At $\alpha_s = 0.23$, $ap = 2.1$.}
 \\ \\[-0.1cm] \hline\\[-1.5ex]
HPQCD 10    & \cite{McNeile:2010ji}  & 2+1 & 
\parbox[t]{0.25\textwidth}{$a\mu=2 a\bar m_h = 0.61 - 1.75$}& 
\parbox[t]{0.4\textwidth}{5 lattice spacings;
3 lattice spacings with $1.0 \leq a\mu  \leq 1.5$ 
when $\alpha_{R_4}(\mu)  \leq 0.3$;
3 lattice spacings with $1.0 \leq a\mu  \leq 1.5$ 
when $\alpha_{R_6/R_8}(\mu) \leq 0.33$.
}
 \\ \\[-0.1cm] \hline\\[-1.5ex]
  JLQCD 10 & \cite{Shintani:2010ph} & 2+1                  
 &  $a = 0.11$\,fm from $r_0 = 0.49$\,fm  
 &\parbox[t]{0.4\textwidth}{{Single lattice spacing, $0.4 < (aQ)^2 < 1.0$
 for the momentum fit range. At $\alpha_s = 0.3$, $ap = 0.89$.}}
 \\ & \\[-0.1cm] \hline\\[-1.5ex]
 {HPQCD 10}
   & \cite{McNeile:2010ji}
   & 2+1
   &
   & \parbox[t]{0.6\textwidth}
        {Update of $r_1$ and $r_1/a$ in HPQCD 08A.}
 \\ \\[-0.1cm] \hline\\[-1.5ex]
PACS-CS 09A& \cite{Aoki:2009tf} & 2+1 & 
$a/L=1/8,1/6,1/4$ & 
\parbox[t]{0.4\textwidth}{Tree-level
O($a$) boundary improvement, which has been seen to behave better
than 1-loop in simulations \cite{Takeda:2004xha}; 
weighted average of $a/L=1/8,1/6$ for step scaling function
which agrees with a
linear extrapolation in $a/L$ of all data points of the 
SSF. Linear extrapolation  in $a/L$ of $L_\mathrm{max}m_\rho$ with 
$a/L_\mathrm{max} = 1/8, 1/6, 1/4$.}
 \\ \\[-0.1cm] \hline\\[-1.5ex]
{HPQCD 08B}   & \cite{Allison:2008xk}  & 2+1 &  
\parbox[t]{0.25\textwidth}{ $a\mu=2 a\bar m_h$ = 0.8,\, 1.2,\, 1.7,\, 2.1} & 
\parbox[t]{0.4\textwidth}{4 lattice spacings with heavy quark mass
approximately the charm mass, where $\alpha_{R_4}(\mu) = 0.38$.}
 \\ \\[-0.1cm] \hline\\[-1.5ex]
 {HPQCD 08A}
   & \cite{Davies:2008sw}
   & 2+1
   & \parbox[t]{0.25\textwidth}
        {6 lattice spacings with $a = 0.18$ - $0.045\,\mbox{fm}$}
   & \parbox[t]{0.6\textwidth}
        {$r_1$ using $\Upsilon$ spectrum splitting.}
 \\ \\[-0.1cm] \hline\\[-1.5ex]
 {Maltman 08}
   & \cite{Maltman:2008bx}
   & 2+1
   & \parbox[t]{0.25\textwidth}
        {5 lattice spacings with $a = 0.18$ - $0.06\,\mbox{fm}$}
   & \parbox[t]{0.40\textwidth}
        {Re-analysis of HPQCD 05A with additional lattice
         spacings $a = 0.06$, $0.15\,\mbox{fm}$.}
 \\ \\[-0.1cm] \hline\\[-1.5ex]
 {HPQCD 05A}
   & \cite{Mason:2005zx}
   & 2+1
   & \parbox[t]{0.25\textwidth}
        {3 lattice spacings with $a = 0.18$ - $0.09\,\mbox{fm}$}
   & \parbox[t]{0.6\textwidth}
        {$r_1$ using $\Upsilon$ spectrum splitting.}
\\ 
        &&&&&&&&& \\[-0.1cm]
   \hline
      \hline
\end{tabular*}
\caption{Continuum limit for $\alpha_s$ determinations with $N_f=3$.}
\label{tab_Nf=3_continuumlimit}
\end{table}


\begin{table}[!htb]
   \vspace{3.0cm}
   \footnotesize
   \begin{tabular*}{\textwidth}[l]{l@{\extracolsep{\fill}}rlllllllll}
\hline\hline \\[-1.0ex]
      Collab. & Ref. & $\Nf$ & $a\,\mu$ & Description \\
      &&&& \\[-0.1cm]
      \hline
      \hline
      &&&& \\[-0.1cm]
{ETM 13B}
                &  \cite{Blossier:2013ioa} & 2+1+1 &    
  $a=0.060, 0.068$ fm   from $f_\pi$            &For $\alpha_s \leq 0.3$, $ap=1.5,1.7$.
	\parbox[t]{0.60\textwidth}{Update of \cite{Blossier:2012ef}.}
\\ \\[-0.1cm] \hline\\[-1.5ex]      
{ETM 12C}       & \cite{Blossier:2012ef}   & 2+1+1 &  
      $a=0.061, 0.078$ from $f_\pi$ & 
      \parbox[t]{0.4\textwidth}{Global fit with $(ap)^2$ discretisation effects.
       For $\alpha_s \leq 0.3$, $ap = 1.5, 2.2$.}       
\\ \\[-0.1cm] \hline\\[-1.5ex]      
{ETM 11D}      & \cite{Blossier:2011tf}   & 2+1+1 & 
      $a = 0.061, 0,078$\,fm & {For $\alpha_s \leq 0.3$, $ap = 1.5, 2.0$.}
\\ \\[-0.1cm] \hline\\[-1.5ex]      
      ALPHA 10A & \cite{Tekin:2010mm} & 4                  & 
      $a/L = 1/4, 1/6, 1/8$ & Constant or global linear fit in $(a/L)^2$.
\\ \\[-0.1cm] \hline\\[-1.5ex]      
      Perez 10 & \cite{PerezRubio:2010ke} & 4 & $a/L=1/4, 1/6, 1/8$ &
      \parbox[t]{0.4\textwidth}{Linear extrapolation in $(a/L)^2$.
      1-loop improvement at the boundary.}
\\  &&&&&&&&& \\[-0.1cm]
   \hline
   \hline
\end{tabular*}
\caption{Continuum limit for $\alpha_s$ determinations with $N_f=4$.}
\label{tab_Nf=4_continuumlimit}
\end{table}


\clearpage


\end{appendix}

\clearpage
\bibliography{FLAG} 
\bibliographystyle{JHEP_FLAG}

\end{document}